\documentclass[a4paper,11pt, twoside]{book}
\usepackage[top=2.5cm,bottom=2.5cm,left=2.5cm,right=2.5cm, heightrounded, bindingoffset=0.5cm]{geometry}
\usepackage[utf8]{inputenc}
\usepackage[T1]{fontenc}
\usepackage[english]{babel}
\usepackage{amssymb, amsmath, mathrsfs}
\usepackage{cancel}
\usepackage{slashed}
\usepackage{mathtools}
\usepackage{amsfonts}
\usepackage{nicefrac}
\usepackage{empheq}
\usepackage{bbold}
\usepackage{amsthm}
\usepackage{hyperref}
\usepackage{enumitem}
\usepackage{bm}
\usepackage{xcolor}
\definecolor{oceanboatblue}{rgb}{0.0, 0.47, 0.75}
\definecolor{forestgreen}{rgb}{0.13, 0.55, 0.13}
\definecolor{darkgreen}{rgb}{0.2, 0.56, 0.2}
\hypersetup{
colorlinks,
    linkcolor={blue},
    citecolor={darkgreen},
    urlcolor={darkgreen}
}
\usepackage[T1]{fontenc}
\usepackage{blindtext}
\usepackage[pdftex]{graphicx}
\usepackage{tabularx}
\usepackage{indentfirst}
\usepackage{subfigure}
\usepackage[small]{caption}
\usepackage{eucal}
\usepackage{eso-pic}
\usepackage{url}
\usepackage{appendix}
\usepackage{booktabs}
\usepackage{afterpage}
\usepackage{parskip}
\usepackage{listings}
\usepackage{fancyhdr}
\usepackage{textcomp}
\usepackage{cite}
\usepackage{multirow}
\usepackage{setspace}
\usepackage{afterpage}
\usepackage{indentfirst}
\usepackage{emptypage}
\usepackage{siunitx}
\usepackage{tikz}
\numberwithin{equation}{section}
\usepackage{pdfpages}
\usepackage{graphicx}
\usepackage[symbol]{footmisc}

\newcommand{\beq}{\begin{equation}}
\newcommand{\eeq}{\end{equation}}
\renewcommand{\[}{\left[}
\renewcommand{\]}{\right]}
\renewcommand{\(}{\left(}
\renewcommand{\)}{\right)}
\newcommand{\bra}[1]{\left\langle #1 \right |}
\newcommand{\ket}[1]{\left| #1 \right \rangle}
\newcommand{\braket}[2]{\left\langle #1 | #2 \right\rangle}

\DeclarePairedDelimiter{\abs}{\lvert}{\rvert}
\DeclarePairedDelimiter{\norm}{\lVert}{\rVert}

\newcommand{\N}{\mathcal{N}}
\newcommand{\M}{\mathcal{M}}
\newcommand{\C}{\mathcal{C}}
\renewcommand{\P}{\mathcal{P}}
\newcommand{\T}{\mathcal{T}}
\renewcommand{\L}{\mathcal{L}}
\renewcommand{\H}{\mathcal{H}}
\newcommand{\Q}{\mathcal{Q}}
\renewcommand{\S}{\mathcal{S}}
\newcommand{\E}{\mathcal{E}}
\newcommand{\F}{\mathcal{F}}
\newcommand{\A}{\mathcal{A}}
\newcommand{\G}{\mathcal{G}}
\newcommand{\Z}{\mathcal{Z}}
\newcommand{\K}{\mathcal{K}}

\newcommand{\bQ}{\bar{Q}}
\newcommand{\dalpha}{\dot{\alpha}}
\newcommand{\dbeta}{\dot{\beta}}
\newcommand{\bepsilon}{\bar{\epsilon}}
\newcommand{\bsigma}{\bar{\sigma}}
\newcommand{\btheta}{\bar{\theta}}
\newcommand{\bpsi}{\bar{\psi}}
\newcommand{\bPsi}{\bar{\Psi}}
\newcommand{\bchi}{\bar{\chi}}
\newcommand{\bV}{\overline{V}}
\newcommand{\MP}{M_{\scalebox{0.6}{P}}}
\newcommand{\fgamma}{\bar{\gamma}}
\newcommand{\mk}{\vec{k}^2}
\newcommand{\vk}{\vec{k}}
\newcommand{\vx}{\vec{x}}
\newcommand{\vfgamma}{\vec{\bar{\gamma}}}
\newcommand{\vpsi}{\vec{\psi}}

\newcommand{\kg}{\left(\vec{k}\cdot\vec{\fgamma}\right)}
\newcommand{\hO}{\hat{O}}
\newcommand{\g}{\sqrt{-g}}

\newcommand{\cR}{C_\textup{R}}
\newcommand{\cI}{C_\textup{I}}
\newcommand{\dr}{d_\textup{R}}
\newcommand{\di}{d_\textup{I}}
\newcommand{\ta}{\tilde{a}}

\newcommand{\halpha}{\hat{\alpha}}
\newcommand{\hbeta}{\hat{\beta}}
\newcommand{\mpsi}{\lambda}
\newcommand{\bp}{\bm{p}}

\newcommand{\balpha}{\bar{\alpha}}
\newcommand{\bL}{\bar{L}}
\newcommand{\bN}{\bar{N}}
\newcommand{\td}{\tilde{d}}
\newcommand{\tb}{\tilde{b}}

\newcommand{\lnabla}{\overset{\leftarrow}{\nabla}}

\newcommand*{\bfrac}[2]{\genfrac{\lbrace}{\rbrace}{0pt}{}{#1}{#2}}
\newcommand*{\sfrac}[2]{\genfrac{[}{]}{0pt}{}{#1}{#2}}

\DefineFNsymbols{sym}{\textdagger \textdaggerdbl \textparagraph %
{\textdagger\textdagger} {\textdaggerdbl\textdaggerdbl} %
{\textparagraph\textparagraph}}
\setfnsymbol{sym}

\title{Linear and nonlinear supersymmetry \\ in field and string theory}
\author{Gabriele Casagrande\footnote{Email address: \href{mailto:gabriele.casagrande@polytechinque.edu}{\texttt{gabriele.casagrande@polytechinque.edu}}.}\footnote{Address from October $1^\textup{st}$ 2025: Ben-Gurion University of the Negev, Department of Physics.}
}
\date{28$^\text{th}$ March 2025}

\begin{document}

\frontmatter

\includepdf[pages=-]{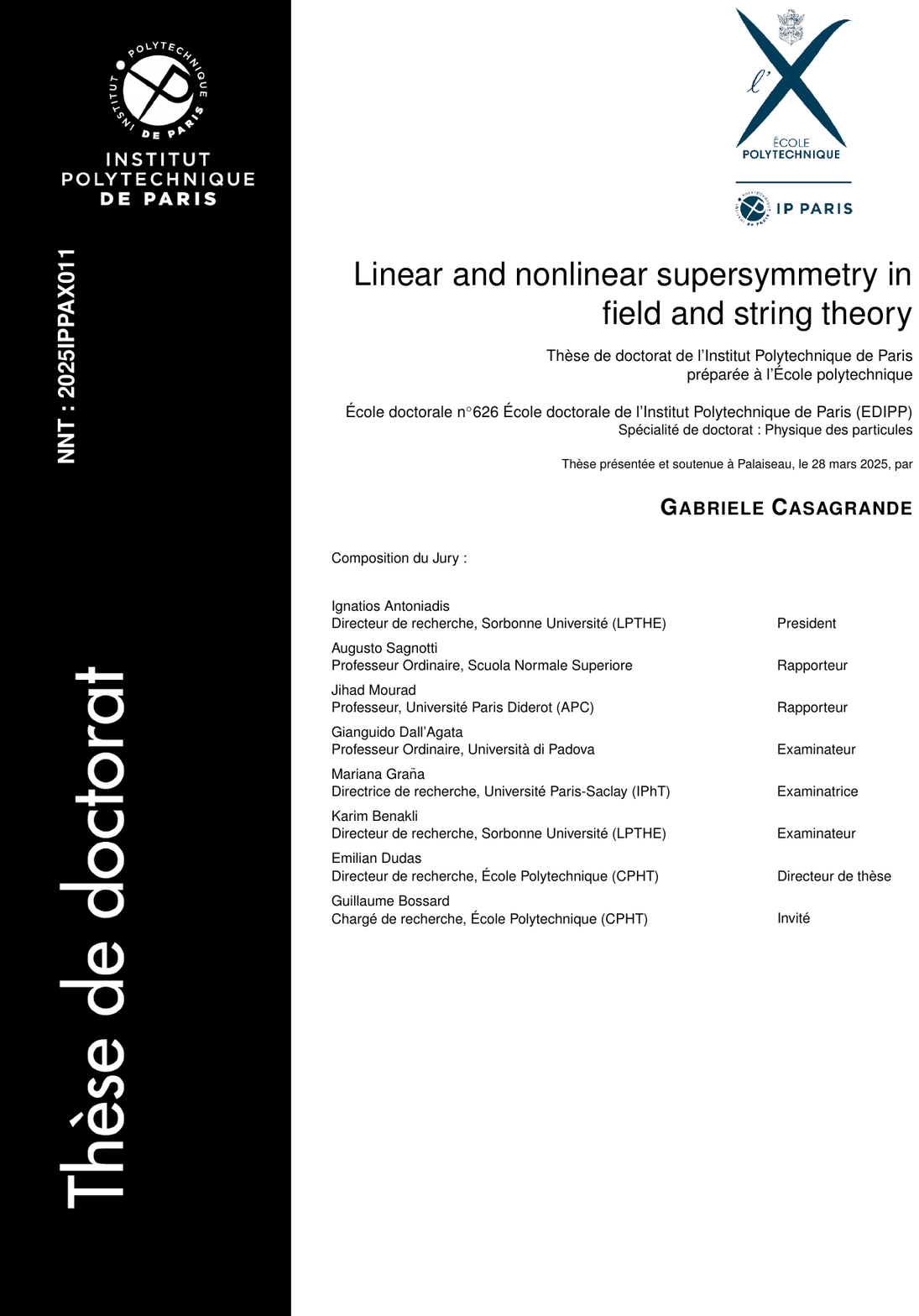}

\maketitle



\chapter*{Acknowledgments}

As for most of the people in the academic world, the completion of the PhD represents a moment of transition. Therefore, I take this opportunity to express my deepest and most sincere gratitude and affection to all the people I have been lucky to have alongside me so far, for truly having made the difference in this experience and continuing to do so in what will come next. \\


First of all, thank you Emilian. For the unconditional trust you have had in me since the very beginning, for everything you taught me, and for your constant care, empathy and respect. I am extremely grateful for your guidance throughout these years. \\

With the same spirit, thank you Guillaume for having taken me under your supervision as well. Working with you is extremely stimulating, and the insights, the support and the understanding I have received from you are really invaluable. \\

I would like to acknowledge the members of my doctoral committee. Thanks to Ignatios Antoniadis, Karim Benakli and Mariana Gra$\tilde{\text{n}}$a for evaluating my work and for the instructive questions about it. Thanks to Jihad Mourad for having been a \textit{rapporteur} as kind as precise. Thanks to Gianguido Dall'Agata, for having supported me through my entire academic life, from my time in Padova to today, and for having set the stage of my PhD by putting me in contact with Emilian in the first place. Finally, thanks to Augusto Sagnotti, for his rich comments and notes on this thesis as \textit{rapporteur}, and, mostly, for his precious advice and his unexpected and flattering support.\\


I also thank very much Quentin Bonnefoy and Marco Peloso, for the work done together and for having made me understand what a good researcher is. Thank you also to Adrien Loty and Matteo Morittu for working side by side with me on our projects, both as collaborators and as friends.\\

A big thank you goes to the Centre de Physique Th\'eorique (CPHT) and all its members. Thank you to Jean-R\'en\'e, Marios, Balt, Blaise, Dario, Olga, Christoph and Cedric, as well as to Fadila, Florence, Malika and Cynthia, for making the lab the warm, stimulating, and pleasant place that it is. For the same reasons, thank you to all the fellow students and postdocs of the CPHT. Thank you Filippo for having been a big brother to me even before I arrived in Paris, for all the discussions we had in our office and for our sincere friendship. Thank you to Adrien and Mathieu (aka \textit{Nino \& Lele}) for having welcomed me to France and made me feel included in the best possible way. Thank you Mikel for our morning chats, our tennis matches and for being the best of friends. Thank you to Erik, Vaios, Adi, David, Maddalena and Amedeo, for everything we share and have experienced together. Thank you Fanny for your decisive contribution to my \textit{pot de th\`ese} and, mostly, for being a kind friend. Many thanks also to Victor, Magali, Francesco, Edoardo, Rajah, Yorgo, Cl\'ement, Mathieu~B., David~R.~B., Emilio, Blagoje, Thomas, Simon, Adrien~F., Junchen and Evgeny. \\

I would also like to thank the broader Paris physics community, in particular Veronica, Marcello, Dimitrios, Tom and Jules, as well as to the whole high-energy theory people I had the chance to meet during this time. \\

Ringrazio tantissimo tutti i miei amici, con i quali so di essere sempre nel posto giusto. Grazie alla \textit{banda} di Montaner: Chiara, Dimitri, Elena, Martina, Oriana, Silvia, Vittoria ed Enrico~M. Grazie agli \textit{Amici di Mauro}: Alessandro, Elena, Enrico~T., Francesca, Checco, Nenni, Giovanni, Giorgio, Margherita e Tommaso, nonch\'e a tutto il progetto \textit{FWB}. Grazie ad Alberto, Begonia, Luca, Michell, Pier e Vanessa per essere venuti a Parigi a festeggiare. Grazie a Riccardo per continuare a capirsi l'un l'altro anche dopo tanto tempo. Grazie a Matteo per la sua amicizia spontanea e sincera. \\

Un ringraziamento speciale, per quanto sempre insufficiente, va alla mia famiglia, i cui meriti vanno ben oltre queste poche righe. Grazie a nonna Ines, per il bene che vuole a tutti. Grazie a zia Maru e zio Robi, per la loro ironia e saggezza. Grazie a zia Virna e Rachele, per l'affetto che ci lega. Grazie a zio Gory, per il suo spirito, il suo esempio e la cura che ha verso tutti noi. Grazie ad Andrea, per insegnarmi tante cose pur essendo il più giovane. Grazie a Cristiana, per pensare meglio le stesse cose che penso io. Grazie a mamma Serena e pap\`a Rudy, per avermi insegnato tutto quello che serve davvero sapere. \\

Infine, grazie a Gloria. Per starci vicini qualsiasi sia la distanza, per aiutarmi a capire meglio le cose, vecchie e nuove, e per tutto ciò che di speciale ci unisce.


\chapter*{Abstract}

The current state of high-energy physics is characterized by a division into two distinct branches. The Standard Model, validated with high precision, governs phenomena at the $\si{TeV}$ scale, but it must be modified in order to reach higher energy scales. Key steps in this respect are the coupling to gravity, following the principles of general relativity, and the quantization of such gravitational interaction. String theory, operating near the Planck scale, remains the only consistent framework achieving such unification. Within the effective field theory approach, one expects a series of intermediate models connecting the Standard Model in the infrared to String Theory in the ultraviolet, progressively disclosing the features of quantum gravity. However, the concrete realization of this connection remains elusive.

A major role in bridging this gap is played by supersymmetry, as it is compellingly motivated at both ends of the energy spectrum. In field theory, it is the sole extension of the Poincar\'e group and leads naturally to gravity when realized locally. In string theory, supersymmetry is necessary for describing fermionic excitations and ensuring vacuum stability, with supergravity emerging as its low-energy limit. Furthermore, spontaneous supersymmetry breaking provides a coherent framework for transitioning from string theory to the non-supersymmetric Standard Model, with the associated breaking scale acting as a vehicle for high-energy physics to manifest at lower energies.

This Ph.D. thesis investigates effective field and string theories in which supersymmetry is realized and broken in various ways. First, we address effective theories with nonlinearly realized supersymmetry, constructed using the formalism of constrained superfields. We establish a criterion for identifying inconsistent constraints, formulate their improved versions, and propose, along these lines, improved supergravity models of inflation. Then, we discuss a peculiar application of this framework: the gravitational production of massive gravitinos in time-dependent backgrounds. We emphasize some physical subtleties in the standard description of this mechanism, which become particularly severe in the case of the gravitino. Next, the focus shifts to linear supersymmetry, and we investigate the leading-order coupling of the massive spin-2 field supermultiplet to pure supergravity in four dimensions. This analysis is carried out using the supercurrent superfield formalism, and it shows that only a single class of such couplings can be consistently defined. This is fundamentally distinct from the coupling obtained by compactifying higher-dimensional supergravities and it has interesting connections to string theory. Finally, we turn to string theory and present a novel orientifold projection of the type IIB string. This is a Scherk--Schwarz orientifold in which supersymmetry is entirely broken, and the O-planes couple only to the twisted sector of the theory. We argue that this model is invariant under S-duality and can be formulated as an F-theory compactification. We also analyze the D-brane spectrum of the theory and compare the results with earlier similar constructions.

\chapter*{R\'esum\'e en français}

L’état actuel de la physique des hautes énergies se caractérise par une division en deux branches distinctes. Le Modèle Standard, validé avec une grande précision, décrit les phénomènes à l’échelle du $\si{TeV}$, mais il doit être modifié pour atteindre des énergies plus élevées. Les étapes clés dans cette perspective sont le couplage à la gravité, selon les principes de la relativité générale, ainsi que la quantification de cette interaction gravitationnelle. La théorie des cordes, opérant à proximité de l’échelle de Planck, demeure le seul cadre cohérent permettant une telle unification. Dans le cadre de la théorie des champs efficace, on s’attend dès lors à une série de modèles intermédiaires reliant le Modèle Standard dans l’infrarouge à la théorie des cordes dans l’ultraviolet, révélant progressivement les caractéristiques de la gravité quantique. Toutefois, la réalisation concrète de cette connexion reste insaisissable.

Un rôle majeur dans la tentative de combler cet écart est joué par la supersymétrie, car elle est solidement motivée aux deux extrémités du spectre énergétique. En théorie des champs, elle constitue la seule extension du groupe de Poincaré et conduit naturellement à la gravité lorsqu’elle est réalisée localement. Dans la théorie des cordes, la supersymétrie est nécessaire pour décrire les excitations fermioniques et garantir la stabilité du vide, la supergravité apparaissant comme sa limite à basse énergie. En outre, la rupture spontanée de la supersymétrie offre un cadre cohérent pour la transition de la théorie des cordes vers le Modèle Standard non-supersymétrique, l’échelle de rupture jouant alors le rôle de vecteur permettant à la physique des hautes énergies de se manifester à basse énergie.

Cette thèse de doctorat étudie des théories des champs et de cordes efficaces dans lesquelles la supersymétrie est réalisée et brisé de diverses manières. On commençe par examiner des théories efficaces avec une supersymétrie réalisée de manière non linéaire, construites à l’aide du formalisme des superchamps contraints. Nous établissons un critère permettant d’identifier les contraintes inconsistantes, formulons leurs versions améliorées, et proposons, dans cette optique, des modèles améliorés de supergravité pour l’inflation. Ensuite, on discute une application particulière de ce cadre : la production gravitationnelle de gravitinos massifs dans des fonds dépendant du temps. Nous soulignons certaines subtilités physiques dans la description standard de ce mécanisme, qui deviennent particulièrement sévères dans le cas du gravitino. Par la suite, l’attention se porte sur la supersymétrie linéaire, et nous étudions le couplage au premier ordre du supermultiplet de spin-2 massif à la supergravité pure en quatre dimensions. Cette analyse est réalisée à l’aide du formalisme du superchamp de supercourant, et elle montre qu’une seule classe de tels couplages peut être définie de manière cohérente. Cela est fondamentalement distinct de celui obtenu en compactifiant la supergravité en dimensions supérieures, et il est d'une manière interessante lié à la théorie des cordes. Enfin, nous nous tournons vers la théorie des cordes et présentons une nouvelle projection orientifold de la corde de type~IIB. Il s’agit d’un orientifold de Scherk--Schwarz dans lequel la supersymétrie est complètement brisé, et les O-plans ne se couplent qu’au secteur twisté de la théorie. Nous soutenons que ce modèle est invariant sous la S-dualité et peut être formulé comme une compactification en théorie F. Nous analysons également les D-branes de cette théorie et comparons les résultats à ceux obtenus dans des constructions similaires antérieures.

\chapter*{Publications}\addcontentsline{toc}{chapter}{Publications}

\noindent The present manuscript is based on the following published papers:

\vspace{0.5cm}
\begin{description}
   
    \item[\cite{Bonnefoy:2022rcw}]  Q.~Bonnefoy, G.~Casagrande and E.~Dudas, \textit{Causality constraints on nonlinear supersymmetry}, \href{https://link.springer.com/article/10.1007/JHEP11(2022)113}{\textit{JHEP} \textbf{11} (2022), 113} \href{https://arxiv.org/pdf/2206.13451}{[\texttt{arXiv:2206.13451}]}. 

    \vspace{0.3cm}
    \item[\cite{Casagrande:2023fjk}]  G.~Casagrande, E.~Dudas and M.~Peloso, \textit{On energy and particle production in cosmology: the particular case of the gravitino}, \href{https://link.springer.com/article/10.1007/JHEP06(2024)003}{\textit{JHEP} \textbf{06} (2024), 003} \href{https://arxiv.org/pdf/2310.14964}{[\texttt{arXiv:2310.14964}]}. 

    \vspace{0.3cm}
    \item[\cite{Bossard:2024mls}]  G.~Bossard, G.~Casagrande and E.~Dudas, \textit{Twisted Orientifold planes and S-duality without supersymmetry}, \href{https://link.springer.com/article/10.1007/JHEP02(2025)062?utm_source=rct_congratemailt&utm_medium=email&utm_campaign=oa_20250213&utm_content=10.1007/JHEP02(2025)062}{\textit{JHEP} \textbf{02} (2025), 062}  \href{https://arxiv.org/pdf/2411.00955}{\texttt{[arXiv:2411.00955]}}. 

    \vspace{0.3cm}
    \item[\cite{ms2}]  G.~Bossard, G.~Casagrande, E.~Dudas and A.~Loty, \textit{A unique coupling of
    the massive spin-2 field to supergravity}, \href{https://arxiv.org/pdf/2502.09599}{\texttt{[arXiv:2502.09599]}}. 
\end{description}


\begingroup
 \hypersetup{linkcolor=black} 
 \tableofcontents 
\endgroup


\renewcommand{\thefootnote}{\arabic{footnote}}


\chapter{Introduction}

The modern perspective on quantum field theory is to view every model as an \textit{effective field theory}, describing physics consistently only up to a certain energy scale. An effective theory successfully describes the physical processes whose typical scale lies within the appropriate energy range, whereas beyond this energy threshold the spectrum of phenomena expands and the theory eventually fails to be predictive. Famous examples of effective field theories are the Fermi theory of weak interactions and the chiral lagrangian describing the low-energy dynamics of QCD in terms of pions. In both cases, we know that the underlying physics is more complex: weak interactions are mediated by the $Z_0$ and $W_\pm$ bosons, and the pions in the chiral lagrangian are bound states of quarks. However, these fields are fine structures that cannot be resolved at the energy scale at which the effective field theory is defined, and the relevant physics is more easily described by this effective theory. More precisely, within the energy range where it applies, the effective field theory is as predictive as the "complete" theory from which it is derived.

Beyond the purely phenomenological meaning, quantum field theories are now regarded as effective field theories in the sense that we do not expect a given theory to be descriptive of all phenomena at all energies. We expect instead to have to deal with different theories at different energy scales, all seen as low-energy realizations of the same, unifying theory. Taking this perspective, the spectrum of energy scales is divided into infrared (IR) and ultraviolet (UV) regions which have traditionally been critically disconnected. On the infrared side there is the Standard Model of particle physics \cite{Glashow:1961tr,Weinberg:1967tq,Salam:1968rm}, which has been tested
with great precision and confirmed to be the theory describing physics at the $\si{TeV}$ scale \cite{Englert:1964et,Higgs:1964pj,ATLAS:2012yve,CMS:2012qbp,Aoyama:2012wj,Fan:2022eto}. However, it is evident that this theory is incomplete, as there are several crucial questions that the Standard Model fails to address, highlighting the need for its extension. For example, the Standard Model does not predict any neutrino masses at the renormalizable level, and it fails to explain why the mass of the Higgs boson is so small and protected by radiative corrections. Also, it does not provide any viable dark matter candidate and suffers from the well-known CP problem. In addition, recent precision tests have challenged some of its predictions -- for instance, the muon magnetic moment \cite{Muong-2:2004fok}. 

From a broader point of view, the main issue with the Standard model is that it does not include the gravitational interaction. Moreover, gravity does not only need to be coupled to the Standard Model fields following Einstein's general relativity principles, but it must also be properly quantized in order to be truly unified in a single picture with all other forces and fields. However, this task is as crucial as it is difficult to accomplish, since Einstein's general relativity is inherently non-renormalizable \cite{Goroff:1985sz,Goroff:1985th}. This failure of the standard quantum field theory paradigm in describing the gravitational interaction is actually a major indication that our understanding of the laws of nature needs to evolve and in which direction. To date, the only framework that has achieved a quantum description of gravity is String Theory \cite{Veneziano:1968yb,Virasoro:1969me,Shapiro:1969km,Scherk:1974ca,Yoneya:1974jg,Green:1981yb,Gross:1984dd,Witten:1995ex}, which lives close to the Planck scale and represents the UV side of the theories' energy spectrum. According to the effective field theory principles, we expect String Theory and the Standard Model to be connected through a series of effective models living between the $\si{TeV}$ and Planck scales, gradually revealing the quantum features of gravity. However, how to concretely realize such a connection is neither obvious nor unique. Following a top-down approach, one finds an extraordinarily large number of possible four-dimensional effective theories that can be derived from string theory via compactification \cite{Candelas:1985en}, which is known as the string landscape, and no vacuum selection mechanism leading uniquely to the Standard Model has yet been identified \cite{Douglas:2003um}.\\

Bridging this IR--UV gap and making the quantum description of gravity at high energies consistent with the low-scale physics are the main challenges of modern theoretical physics. In this respect, a crucial role is played by \textit{supersymmetry}, which has been a highly productive framework and a prominent example of unification since its first formulation around fifty years ago \cite{Wess:1973kz,Wess:1974tw} and is the central focus of the present thesis. In fact, supersymmetry has strong motivations to be realized on both sides of the energy spectrum and can therefore help to concretely establish a bridge between the two IR and UV regions we long to connect. 

On the field theory side, supersymmetry is the maximal possible extension of the Poincar\'e group of special relativity, as it is the only way to evade the famous Coleman--Mandula theorem \cite{Coleman:1967ad,Haag:1974qh}. Moreover, following this general symmetry principle, when realized as a local symmetry, it automatically yields gravitational interactions in what are known as supergravity theories \cite{Freedman:1976xh, Deser:1976eh}, which realize a first unification of gravity and the other interactions in a unified theoretical picture. Some notable advantages of supersymmetry from the quantum field theory perspective include a natural solution to the Standard Model's hierarchy problem, potential dark matter candidates, improved renormalizability (including for gravity \cite{Bern:2006kd}) and the unification of gauge interactions at high energy scales. 

On the String Theory side, supersymmetry must necessarily be part of the worldsheet theory in order to describe fermionic
excitations and to ensure the stability of string vacua by removing the tachyonic closed string mode. Furthermore, the low-energy limits of superstring theories coincide with the consistent supergravity theories in ten dimensions. From this point of view, one may say that supersymmetry is actually predicted by String Theory, assuming its validity.  

Furthermore, supersymmetry offers a well-defined framework in which to embed the transition from String Theory to low-energy field theories. This is the framework of the \textit{supersymmetry breaking} mechanisms, which can occur both in field and string theory. Supersymmetry must ultimately be broken at some point in this evolution, as the Standard Model is clearly non-supersymmetric. As a result, the scale of supersymmetry breaking acts as the channel through which the UV physics propagates at lower energies. \\

The research I conducted during my doctoral training takes place within this broader context and focuses on the study of string and field theories characterized by different amounts of supersymmetry. The main outcomes of this work are the articles \cite{Bonnefoy:2022rcw,Casagrande:2023fjk,Bossard:2024mls,ms2}, which are the core of the present thesis:
\vspace{5mm}
\begin{itemize}
    \item the subject of \cite{Bonnefoy:2022rcw} is nonlinear rigid supersymmetry and its implementation through the so-called constrained superfields.  We prove the inconsistency of a certain class of superfield constraints, whose associated goldstino lagrangians exhibit nontrivial causality constraints that are in tension with the effective field theory setup. We propose an alternative, consistent version of these constraints and use them to formulate improved minimal models of inflation in supergravity. \\

    \item  In \cite{Casagrande:2023fjk} we revisit the theoretical aspects of what is known as  gravitational particle production mechanism. This phenomenon takes place when fields are quantized in a time-dependent background, whose expansion drives the creation of particles from the vacuum. More specifically, we analyze the physical consequences of some ambiguities of the formalism used to describe this phenomenon, which we found to be particularly relevant in the case of the massive gravitino in cosmological supergravity models. \\
    
    \item In \cite{Bossard:2024mls} we construct a novel orientifold projection of the type~IIB string. It is a Scherk--Schwarz deformation that lives in nine dimensions and that has no leftover supersymmetry. The resulting theory contains orientifold planes which couple only to the massive twisted states and is conjectured to exhibit S-duality invariance.\\
    
    \item Finally, in \cite{ms2} we investigate the supersymmetric coupling of a single massive spin-2 field to $D=4$, $\N=1$ supergravity at leading order in the Planck mass. Using the supercurrent superfield formalism, we demonstrate that this coupling is unique, that it involves a non-minimal, higher-derivative coupling to the metric field, and that it matches the coupling arising from the first oscillator mode of the open superstring.

\end{itemize}
\vspace{5mm}

The content of the thesis is organized in the following way. In Chapter \ref{ch_susy} we give a general introduction to rigid supersymmetry and the superspace formulation. We then discuss the nonlinear realization of supersymmetry, the constrained superfield formalism and present the results of \cite{Bonnefoy:2022rcw}. Chapter \ref{ch2_sugra} is instead centered on supergravity. We first introduce supergravity as the gauge theory of the Super-Poincar\'e group, describing its coupling to matter fields and the supersymmetric Higgs mechanism. Focusing on this setup, we discuss the dynamics of the massive gravitino in cosmological models based on nonlinear supergravity and present the results of \cite{Casagrande:2023fjk} on their gravitational production. Next, we discuss the various off-shell formulations of $D=4$, $\N=1$ supergravity, in relation to the different types of supercurrent superfields. Then, we present in detail the analysis and results of \cite{ms2} on the couplings of the massive spin-2 supermultiplet to supergravity. In the end, Chapter \ref{ch3_st} is devoted to string theory. We give a comprehensive review of the bosonic string and the superstring theories, with specific emphasis on the one-loop vacuum amplitudes and the orientifold projections. We then specify the discussion to orientifold compactifications and describe in detail the new such construction put forward in \cite{Bossard:2024mls}. Finally, Chapter \ref{ch_conclusions} summarizes the results of my work presented in this manuscript and gives some outlook for its future developments. The content of the various chapters is supplemented and integrated by four appendices \ref{app_conventions}--\ref{app_branes}.


\mainmatter

\chapter{Rigid supersymmetry and constrained superfields}\label{ch_susy}

The first chapter of this manuscript is devoted to the description of the general framework of rigid supersymmetry. The main focus will be on the case of $D=4$, $\N=1$ supersymmetry, its spontaneous breaking, and the resulting nonlinear realization. The original results discussed in this chapter were presented in \cite{Bonnefoy:2022rcw} and concern the construction of low-energy effective theories of nonlinear supersymmetry using the so-called constrained superfield formalism. More specifically, we identify and motivate a criterion for distinguishing between superfield constraints which define consistent effective theories and those which do not. 

The chapter is organized as follows. In Section \ref{ch1_sec_SPA} we review the basic elements of the four-dimensional \textit{Super-Poincar\'e algebra} and its representations. In Section \ref{ch1_sec_ss_sf} we introduce the \textit{superspace} formalism and the general elements of supersymmetric field theories. In Section \ref{ch1_sec_ssb} we discuss the mechanism of \textit{spontaneous supersymmetry breaking}, present explicit models in which it occurs, and examine the \textit{nonlinear realization} of supersymmetry that emerges and characterizes the dynamics of the goldstino, the Goldstone fermion of supersymmetry breaking. In Section \ref{ch1_sec_nls} we move to the more advanced topic of \textit{constrained superfields} and the construction of effective theories of nonlinear supersymmetry, presenting the original results of \cite{Bonnefoy:2022rcw}.\\

In this chapter we follow the conventions of \cite{Wess:1992cp}, which we collect, together with useful formulae, in Appendix \ref{app_conventions}. Further references are \cite{Bertolini:2024xny,Bilal:2001nv,Martin:1997ns,Gates:1983nr,Freedman:2012zz,DallAgata:2021uvl,Antoniadis:2024hvw}.

\section{The Super-Poincar\'e algebra}\label{ch1_sec_SPA}

The core idea and best-known feature of supersymmetry is the relationship it establishes between bosonic and fermionic degrees of freedom:
\beq
    \text{Bosons} \ \xleftrightarrow{\text{ \ SUSY \ }} \ \text{Fermions} \ .
\eeq

\noindent Thus, the field content of a supersymmetric theory will organize into multiplets of bosons and fermions, joined together by supersymmetry. The fact that it relates particles of different spin implies that supersymmetry must be a \textit{spacetime} symmetry (\textit{i.e.} not an internal one). 


This property is actually very deep and has powerful implications. The Coleman--Mandula theorem \cite{Coleman:1967ad} states that the largest continuous symmetry that the S-matrix of a given quantum field theory can posses is given by the Poincar\'e group times possible internal symmetries. The Poincar\'e group is generated by the translations $P_\mu$ and the Lorentz transformations $M_{\mu\nu}$, which satisfy the standard algebra
\beq
\begin{aligned}
    [P_\mu,P_\nu]&=0 \ , \\
    [P_\mu,M_{\rho\sigma}]&=-i\(\eta_{\mu\rho}P_{\sigma}-\eta_{\mu\sigma}P_{\rho}\) \ , \\
    [M_{\mu\nu},M_{\rho\sigma}]&=i\(\eta_{\mu\rho}M_{\nu\sigma}-\eta_{\mu\sigma}M_{\nu\rho}-\eta_{\nu\rho}M_{\mu\sigma}+\eta_{\nu\sigma}M_{\mu\rho}\) \ . \\
\end{aligned}
\label{eqn:ch1_P_alg}
\eeq

\noindent The Coleman--Mandula theorem holds provided some physical assumptions (\textit{e.g.} locality, causality, positivity of the energy, \dots) are satisfied. One of these assumptions is that the generators of the S-matrix symmetry algebra are bosonic, \textit{i.e.} satisfy standard commutation relation, like those of eq.~\eqref{eqn:ch1_P_alg}. Thus, the Coleman--Mandula theorem can be extended by allowing the algebra to contain also \textit{fermionic} generators, satisfying anticommutation relations. It was then proven by Haag, Lopuszanski and Sohnius \cite{Haag:1974qh} that the specific combination of Poincar\'e and supersymmetry, namely the Super-Poincar\'e group, is the only possible fermionic extension of the S-matrix symmetry algebra. Therefore, the Super-Poincar\'e group is the largest symmetry group that a quantum field theory can enjoy. This is the main theoretical motivation in favor of supersymmetry and its realisation in nature.

The algebra of the super Super-Poincar\'e group is an example of \textit{graded} Lie algebra, which is a vector space $L$ given by 
\beq
    L=\oplus_{i=0}^{i=n}L_i \ ,
\eeq

\noindent with $L_i$ also being vector spaces, endowed with a product
\beq
    [ \;\; , \;\; \}: \ L\times L \to L \ ,
\eeq

\noindent such that
\beq
\begin{aligned}
    &[L_i,L_j\}\in L_{i+j} \qquad  \text{mod} \ n+1 \ , \\
    &[L_i,L_j\}=-\(-1\)^{\eta_i \eta_j}[L_j,L_i\} \ , \\
    &[L_i,[L_j,L_k\}\}\(-1\)^{\eta_i \eta_k}+[L_j,[L_k,L_i\}\}\(-1\)^{\eta_i \eta_j}+[L_k,[L_i,L_j\}\}\(-1\)^{\eta_j \eta_k}=0 \ ,
\end{aligned}
\label{eqn:ch1_gla}
\eeq

\noindent where $\eta_i$=0 for bosonic generators and $\eta_i=1$ for fermionic ones. The first equation in \eqref{eqn:ch1_gla} implies that the first element $L_0$ must be a Lie algebra, while all the others are not. The second equation is instead known as supersymmetrization condition, while the third one is the generalization of the standard Jacobi identity \cite{Bertolini:2024xny}. The Super-Poincar\'e algebra is then given by
\beq
    L_\textup{SP}=L_0\oplus L_1 \ ,
\eeq

\noindent with $L_0$ being the Poincar\'e group \eqref{eqn:ch1_P_alg} and $L_1$ the actual supersymmetry algebra. For $\N$ supersymmetries, this is generated by a set of $2\N$ fermionic generators, $Q_\alpha^I$ and $\bQ_{\dalpha}^I$, with $I=1,\dots,\N$ and $\bQ_{\dalpha}^I=\(Q_\alpha^I\)^\dagger$, transforming respectively in the $\(\frac12,0\)$ and $\(0,\frac12\)$ representations of the Lorentz group, to which the indices $\alpha=1,2$ and $\dalpha=\dot{1},\dot{2}$ are associated. In addition to \eqref{eqn:ch1_P_alg}, the Super-Poincar\'e algebra involves the following commutation relations:
\beq
\begin{aligned}
    [P_\mu,Q_\alpha^I]&= [P_\mu,\bQ_{\dalpha}^I]=0 \ , &&&&& \{Q_\alpha^I,\bQ_{\dalpha}^J\}&=2\delta^{IJ}\sigma^\mu_{\alpha\dalpha}P_\mu \ , \\
     [M_{\mu\nu},Q_\alpha^I]&=-i\(\sigma_{\mu\nu}\)_\alpha{}^\beta Q_\beta^I \ , &&&&&  \{Q_\alpha^I,Q_\beta^J\}&=\epsilon_{\alpha\beta}Z^{IJ} \ , \\
     [M_{\mu\nu},\bQ_{\dalpha}^I]&= -i\(\bar{\sigma}_{\mu\nu}\)_{\dalpha\dbeta}\bQ^{I}{}^{\dbeta}\ , &&&&& \{\bQ_{\dalpha}^I,\bQ_{\dbeta}^J\}&=\epsilon_{\dalpha\dbeta}\(Z^{IJ}\)^* \ .
\end{aligned}
\label{eqn:ch1_SP_alg}
\eeq

\noindent The antisymmetric matrix $Z^{IJ}=-Z^{JI}$ is the central extension of the algebra. It commutes in fact with all the generators and it is usually given by some combinations of the generators of the internal symmetries, which we did not include explicitly in eq.~\eqref{eqn:ch1_SP_alg}.
 An important example of such symmetries is the so-called R-symmetry, which is the largest internal symmetry group that can have nontrivial commutators with supersymmetry generators $\(Q_\alpha^I,\bQ_{\dalpha}^I\)$. In the case of $\N$ supersymmetries, it can be shown that this R-symmetry group is actually $\mathrm{U}(\N)$.
 
The Super-Poincar\'e algebra \eqref{eqn:ch1_SP_alg} allows one to prove explicitly some of the main properties of all supersymmetric theories. First, we can see that the action of supersymmetry relates bosonic and fermionic degrees of freedom. Let's take for instance the generators $Q_1^I$ and an eigenstate $\ket{j,m}$ of the angular momentum $J_i=\varepsilon_{ijk}M^{jk}$, namely
\begin{align}
    J^2\ket{j,m}=j\(j+1\)\ket{j,m} \ , && J_3\ket{j,m}=m\ket{j,m} \ .    
\end{align}

\noindent Then, the state $Q_1^I\ket{j,m}$ is such that 
\beq
    J_3\(Q_1^I\ket{j,m}\)=\(m-\frac12\)Q_1^I\ket{j,m} \ ,
\eeq

\noindent which means that the supersymmetry generator $Q_1^I$ decreases the angular momentum along the $z$-axis by $\frac12$. The same is done by the generator $\bQ_{\dot{2}}^I$, while the generators $Q_2^I$ and $\bQ_{\dot{1}}^I$ instead increase the angular momentum along $z$ by $\frac12$. Hence, the action of the supersymmetry generators on a state is to shift its spin by $\frac12$, mapping therefore bosonic states into fermionic ones and vice versa.
On the contrary, the operator $P^2$ is a Casimir invariant of the Super-Poincar\'e algebra, since the supersymmetry generators commute with the momentum $P_\mu$, and therefore the states belonging to a supersymmetric representation will have the same mass.

Another consequence of the algebra \eqref{eqn:ch1_SP_alg} is that the energy of any supersymmetric state $\ket{\psi}$ is positive. This follows from
\beq
\begin{aligned}
    \bra{\psi} \left\{Q_\alpha^I,\bar{Q}_{\dot{\alpha}}^I\right\}\ket{\psi}&=2\N\sigma^{\mu}_{\alpha\dot{\alpha}}\bra{\psi}P_\mu\ket{\psi} \\
    &\,\,\, \Big | \\
    &=\bra{\psi}\[Q_\alpha^I\(Q_\alpha^I\)^\dagger+\(Q_\alpha^I\)^\dagger Q_\alpha^I\]\ket{\psi}=\norm{Q_\alpha^I\ket{\psi}}^2+\norm{\(Q_\alpha^I\)^\dagger\ket{\psi}}^2 \ ,
\end{aligned}
\eeq

\noindent which, taking the trace on the $(\alpha,\dalpha)$ indices, yields the semi-positivity condition
\beq
    \bra{\psi}P_0\ket{\psi}\ge 0 \ .
\label{eqn:ch1_E_pos}
\eeq

Finally, we can show how any representation of the Super-Poincar\'e group contains the same number of bosonic and fermionic degrees of freedom. Taking $N_\textup{F}$ to be the fermion number operator, the operator $\(-1\)^{N_\textup{F}}$ is equal to 1 for bosonic states and to $-1$ for fermionic ones, so that its trace over a finite supersymmetric representation is equal to the number of bosons $n_\textup{B}$ minus the number of fermions $n_\textup{F}$ contained in such representation:
\beq
    \text{Tr}\(-1\)^{N_\textup{F}}=n_\textup{B}-n_\textup{F} \ .
\eeq

\noindent The operator $\(-1\)^{N_\textup{F}}$ satisfies the relation 
\beq
    \left\{\(-1\)^{N_\textup{F}},Q_\alpha^I\right\}=0 \ ,
\eeq

\noindent which combined with the cyclicity property of the trace and the algebra \eqref{eqn:ch1_SP_alg} gives the identity
\beq
\begin{aligned}
    0&=\text{Tr}\[\(-1\)^{N_\textup{F}}Q_\alpha^I\bQ_{\dalpha}^I-\bQ_{\dalpha}^I\(-1\)^{N_\textup{F}}Q_\alpha^I\]=\text{Tr}\[\(-1\)^{N_\textup{F}}\left\{Q_\alpha^I,\bQ_{\dalpha}^I\right\}\]=\\
    &=2\N\[\text{Tr}\(-1\)^{N_\textup{F}}\]\sigma^\mu_{\alpha\dalpha}P_\mu=2\N\(n_\textup{B}-n_\textup{F}\)\sigma^\mu_{\alpha\dalpha}P_\mu \ ,
\end{aligned}
\eeq

\noindent which implies that a finite representation of the Super-Poincar\'e algebra \eqref{eqn:ch1_SP_alg} contains an equal number of bosonic and fermionic degrees of freedom $n_\textup{B}=n_\textup{F}$.

\subsection{Massless representations}\label{ch1_sec_m=0_rep}

After having introduced the Super-Poincar\'e algebra \eqref{eqn:ch1_SP_alg} and its general properties, we can now discuss in detail its representations. As said before, such representations contain particle states with different spins but equal mass. We then distinguish between massless and massive representations.

We start from the case of massless representations, \textit{i.e.} those for which $P^2=0$. The associated 4-momentum is written, in the rest frame, as $P_\mu=\(-E,0,0,E\)$, with $E$ being the (positive) energy, so that the algebra \eqref{eqn:ch1_SP_alg} gives
\beq
    \left\{Q_\alpha^I,\bQ_{\dalpha}^J\right\}=2\delta^{IJ}\sigma^\mu_{\alpha\dalpha}P_\mu=\begin{pmatrix}4E&0\\0&0\end{pmatrix}\delta^{IJ} \ .
\eeq

\noindent Thus, the generators $Q_2^I$ and $\bQ_{\dot{2}}^I$ are totally anticommuting and can therefore be set consistently to zero. In addition, also the central charges $Z^{IJ}$ must vanish, as implied by eq.~\eqref{eqn:ch1_SP_alg}. The generators $Q_1^I$ and $\bQ_{\dot{1}}^I$ define instead a set of $\N$ pairs of creation and annihilation operators:
\begin{align}
    a_I=\frac{1}{2\sqrt{E}}Q_1{}_I \ , && a_I^\dagger=\frac{1}{2\sqrt{E}}\bQ_{\dot{1}}{}_I=\(a_I\)^\dagger \ ,
\end{align}

\noindent satisfying
\begin{align}
    \left\{a_I,a_J^\dagger\right\}=\delta_{IJ} \ , && \left\{a_I,a_J\right\}=\left\{a^\dagger_I,a^\dagger_J\right\}=0 \ .
\end{align}

\noindent The annihilation operators $a^I$ lower the helicity of a state by a factor $\frac12$, the creation operators $a^\dagger{}^I$ raise it by $\frac12$. A given representation is built starting from a state, called the Clifford vacuum $\ket{\lambda_0}$, of energy $E$ and helicity $\lambda_0$, which is annihilated by all the annihilation operators, namely
\beq
    a_I\ket{\lambda_0}=0 \quad \forall \,\, I \ ,
\label{eqn:ch1_cv_def_1}
\eeq

\noindent and then acting repeatedly on it with the creation operators. The representation, which is called supermultiplet, will thus contain the states
\begin{align}
    \ket{\lambda_0}, && a^\dagger{}_I\ket{\lambda_0}=\ket{\lambda_0+\textstyle{\frac12}}_I, && \dots && a_{\N}{}^\dagger\dots a_1{}^\dagger\ket{\lambda_0}=\ket{\lambda_0+\textstyle{\frac{\N}{2}}} \ ,
\label{eqn:ch1_cv_multiplet_1}
\end{align}

\noindent carrying helicities from $\lambda_0$ up to $\lambda_0+\frac{\N}{2}$. Consistent theories involve fields of helicity not greater than 2, so that $\N=8$ is the highest possible number of supersymmetries. For $4<\N\le8$, the supermultiplets necessarily involve the spin-2 state, \textit{i.e.} the graviton, and/or its spin-$\frac32$ partner, the gravitino. These are therefore gravitational theories, where supersymmetry must be realized as a local symmetry. Coherent theories of rigid supersymmetry can therefore be defined up to $\N=4$. 

This is not yet enough to have a well-defined irreducible representation of the Super-Poincar\'e algebra \eqref{eqn:ch1_SP_alg}. The representations have in fact to be consistent with the $\C\P\T$ theorem. The $\C\P\T$ transformations flip the helicity of a given state, so that consistent representations must have a symmetric helicity spectrum. It is often the case that one has to combine "by hand" set of states of the type \eqref{eqn:ch1_cv_multiplet_1} with mirrored helicity spectra to make up a good supermultiplet.

\subsubsection{\texorpdfstring{$\N=1$}{N=1} massless multiplets}\label{ch1_sub_N=1_m=0}

We now focus on the $\N=1$ case and analyze concretely the possible massless representations \eqref{eqn:ch1_cv_multiplet_1}. In this case we have one creation and one annihilation operators, $a$ and $a^\dagger$, so that for every Clifford vacuum $\ket{\lambda_0}$, such that $a\ket{\lambda_0}=0$, there exist one single superpartner state $a^\dagger\ket{\lambda_0}=\ket{\lambda_0+\frac12}$. Thus, the $\N=1$ supermultiplets are given by the couples
\beq
    \left\{\lambda_0,\lambda_0+\textstyle{\frac12}\right\}\oplus\left\{-\(\lambda_0+\textstyle{\frac12}\),-\lambda_0\right\} \ , 
\label{eqn:ch1_N=1_cv_mult}
\eeq

\noindent where $\oplus$ denotes the addition of the $\C\P\T$-conjugate states. The possible $\N=1$ multiplets are then the following:

\paragraph{$\bullet$ Chiral multiplet:} the chiral multiplet is constructed starting from the Clifford vacuum $\ket{\lambda_0=0}$ and contains the degrees of freedom
\beq
    \left\{0,\frac12\right\}\oplus\left\{-\frac12,0\right\} \ ,
\label{eqn:ch1_m0_chiral}
\eeq

\noindent which correspond to a complex scalar and a Weyl (chiral) fermion.

\paragraph{$\bullet$ Vector multiplet:} The gauge multiplet is instead constructed from the Clifford vacuum $\ket{\lambda_0=\textstyle{\frac12}}$, which yields
\beq
    \left\{\frac12, 1 \right\}\oplus\left\{-1,-\frac12\right\} \ ,
\label{eqn:ch1_m0_vec}
\eeq

\noindent which are the degrees of freedom of a gauge field and a Weyl fermion, usually called gaugino. Standing in the same multiplet, the gaugino must sit in the same representation of the gauge group as the vector field, namely the adjoint one.

Chiral and vector multiplets are the $\N=1$ supersymmetric generalization of matter and gauge fields and allow one to describe supersymmetric renormalizable theories in four dimensions. In principle there exist two more $\N=1$ multiplets which do not involve higher-helicity states, the ones that are constructed from the Clifford vacua $\ket{\lambda_0=1}$ and $\ket{\lambda_0=\textstyle{\frac32}}$. These multiplets contain though the helicity states corresponding to the gravitino and the graviton, meaning that they should be embedded in the framework of local supersymmetry rather than global, hence we do not discuss them further here.

\subsubsection{Extended supermultiplets}

In this work we will mainly be concerned with $\N=1$ supersymmetry and so we do not discuss extended representations in detail. We though present, for completeness, one such example, the $\N=2$ matter multiplets, which are known as hypermultiplets. They are built from the Clifford vacuum $\ket{\lambda_0=-\frac12}$ and contain
\beq
    \left\{-\frac12,0,0,\frac12\right\}\oplus\left\{-\frac12,0,0,\frac12\right\} \ ,
\eeq

\noindent which are the degrees of freedom of two complex scalars and two Weyl fermions. The reason why a seemingly identical $\C\P\T$-conjugate was added is the underlying R-symmetry of extended representations, which in this case is given by the group $\mathrm{SU}(2)_\textup{R}$. Under such R-symmetry, the fermions states are singlets but the helicity-0 ones must come into a doublet, which requires then the addition of the $\C\P\T$-conjugate to be formed. In the language of $\N=1$ supersymmetry, they correspond to two chiral multiplets. However, belonging to the same group, the two Weyl fermions must transform in the same representation of the given internal symmetry group, so that the $\N=2$ hypermultiplet is not chiral.

\subsection{Massive \texorpdfstring{$\N=1$}{N=1} representations}\label{ch1_sec_m_rep}

The procedure to build massive representations, \textit{i.e.} those for which $P^2=-m^2$, is completely analogous to the massless case discussed in Section \ref{ch1_sec_m=0_rep}. The substantial difference is that now there are no trivial supersymmetry generator, which means a larger set of ladder operators and therefore longer supermultiplets. Taking the rest-frame momentum $P_\mu=\(-M,0,0,0\)$, the $\N=1$ non-trivial\footnote{Recall that in the case of $\N=1$ supersymmetry there is no central extension. This is not the case for massive representations of extended supersymmetry, which we will briefly discuss at the end of this section.}
anticommutator of eq.~\eqref{eqn:ch1_SP_alg} becomes
\beq
    \left\{Q_\alpha,\bQ_{\dalpha}\right\}=2M \delta_{\alpha\dalpha} \ ,
\eeq

\noindent which allows one to define the system of massive ladder operators
\begin{align}
    b_\alpha=\frac{1}{\sqrt{2M}}Q_\alpha \ , &&  b^\dagger_{\dalpha}{}=\frac{1}{\sqrt{2M}}\bQ_{\dalpha} \ ,
\end{align}

\noindent satisfying
\begin{align}
    \left\{ b_\alpha,b^\dagger_{\dalpha}{}\right\}=\delta_{\alpha\dalpha} \ , && \left\{ b_\alpha,b_{\beta}{}\right\}=\left\{ b^{\dagger}_{\dalpha},b^\dagger_{\dbeta}\right\}=0 \ .
\end{align}

\noindent Notice that, while the creation operator $b_{\dot1}^\dagger$ raises the spin by a factor $\frac12$, as we saw previously, the new one $b_{\dot2}^\dagger$ instead decreases it by the same amount. 

The Clifford vacuum $\ket{j_0}$ for massive representations carries now spin $j_0$ and it is annihilated by both $b_1$ and $b_2$, as in eq.~\eqref{eqn:ch1_cv_def_1}. The massive version of the chiral and vector massless multiplets of eq.~\eqref{eqn:ch1_m0_chiral} and \eqref{eqn:ch1_m0_vec} are given by 
\begin{align}
    &\bullet\,\, \textbf{chiral multiplet:} && \ket{j_0=0} && \longrightarrow&&  \left\{-\frac12,0,0,\frac12\right\} \ , \label{eqn:ch1_m_chiral}\\
    &\bullet \,\,\textbf{vector multiplet:} && \ket{j_0=\textstyle{\frac12}} && \longrightarrow && \left\{0,2\times\frac12,1\right\}\oplus\left\{-1,2\times-\frac12,0\right\} \ . \label{eqn:ch1_m_vec}
\end{align}

\noindent The massive chiral multiplet \eqref{eqn:ch1_m_chiral} contains the same degrees of freedom of the massless one \eqref{eqn:ch1_m0_chiral}, although it is $\C\P\T$-invariant by itself. The states described by the massive vector multiplet \ref{eqn:ch1_m_vec} are those of one massive vector, one massive Dirac fermion and one massive real scalar. In terms of massless multiplets, it corresponds to one massless vector \eqref{eqn:ch1_m0_vec} and one massless chiral \eqref{eqn:ch1_m0_chiral} multiplet. This is the supersymmetric version of the St\"uckelberg mechanism for the massive Proca field, which can be consistently written as the sum of a massless gauge field and a scalar mode. 

\subsubsection{Massive representations of extended supersymmetries}

Massive representations of extended supersymmetry are richer because of the possible nontrivial central extension of the algebra \eqref{eqn:ch1_SP_alg}. Taking the number $\N$ of supersymmetries to be even, the non-trivial central charges are $Z_r$, with $r=1,\dots,\frac{\N}{2}$. Then, the algebra \eqref{eqn:ch1_SP_alg} allows one to define a set of $2\N$ pairs of creation and annihilation operators, respectively $\(c_\alpha^r, \(c_\alpha^r\)^\dagger\)$ and $\(d_\alpha^r, \(d_\alpha^r\)^\dagger\)$, such that the non-trivial anticommutators of the algebra they define are
\begin{align}
    \left\{c_\alpha^r, \(c_\beta^s\)^\dagger\right\}=\(2m+Z_r\)\delta^{rs}\delta_{\alpha\beta} \ , && \left\{d_\alpha^r, \(d_\beta^s\)^\dagger\right\}=\(2m-Z_r\)\delta^{rs}\delta_{\alpha\beta} \ .
\end{align}

From the positivity of the Hilbert space defined by such ladder operators, it follows that the central charges are bounded as
\beq
    \abs{Z_r}\le 2m \quad \forall \,\, r \ ,
\label{eqn:ch1_cg_bound}
\eeq

\noindent from which we find back that massless representations cannot be associated with nontrivial central charges. Moreover, we see that if some of the central charges saturate this bound, the correspondent $d$-ladder operators are trivially realized and can be set to zero. This is an example of BPS condition and the multiplets associated with such BPS configurations are shorter. For instance, the $\N=2$ massive short multiplets contain the same degrees of freedom as the massless ones. Thus, the massive representations of extended supersymmetry can be realized as long multiplets, short multiplets or ultra-short multiplets, based on the number of central charges that satisfy the BPS condition $Z_r=2m$ that follows from the bound \eqref{eqn:ch1_cg_bound}.

\subsection{Representations on fields and supersymmetry in components}

The pattern described to build the representations of the Super-Poincar\'e algebra \eqref{eqn:ch1_SP_alg} in terms of spin states can be applied to construct the supermultiplets of fields. The starting point is a bottom-component field, analogous to the Clifford vacuum, on which half of the supersymmetry generators act trivially. Its superpartners are then obtained as the (nontrivial) iterative action of the supersymmetry generators. Each step of the procedure is controlled by the Super-Poincar\'e algebra \eqref{eqn:ch1_SP_alg}, in particular by the generalized Jacobi identity of eq.~\eqref{eqn:ch1_gla}, and it goes on as long as the supersymmetry transformations result in new degrees of freedom.

We illustrate this in the case of the $\N=1$ chiral multiplet of eq.~\eqref{eqn:ch1_m0_chiral}. Its bottom component is a complex scalar field $\phi(x)$ such that
\beq
    \[\bQ_{\dalpha},\phi\]=0 \ .
\label{eqn:ch1_cv_field}
\eeq

\noindent Then, the second element of the multiplet associated with such a field $\phi$ is given by\footnote{Note that if $\phi$ were real, it would follow from eq.~\eqref{eqn:ch1_cv_field} that the action of all supersymmetry generators would be trivial.}
\beq
    \[Q_{\alpha},\phi\]\equiv \psi_\alpha \ .
\eeq

\noindent It follows from the algebra \eqref{eqn:ch1_SP_alg} that the supersymmetry generators have mass dimension equal to $\frac12$, so that the field $\psi_\alpha$ has mass dimension $\frac32$ and it carries a Weyl index: it is therefore a spin-$\frac12$ Weyl fermion. The next combinations in line are given by
\begin{align}
    \left\{Q_{\alpha},\psi_\beta\right\}=F_{\alpha\beta} \ , &&   \left\{\bQ_{\dalpha},\psi_\beta\right\}=H_{\dalpha\beta}\ .
\end{align}

\noindent Making use of the generalised Jacobi identity of eq.~\eqref{eqn:ch1_gla}, one can see that
\begin{align}
    \begin{aligned}F_{\alpha\beta}+F_{\beta\alpha}=0 && \longleftrightarrow && F_{\alpha\beta}=F\epsilon_{\alpha\beta} \end{aligned} \ , && H_{\dalpha\beta}=2\sigma^\mu_{\beta\dalpha}\[P_\mu,\phi\]\sim\partial_\mu\phi \ ,
\end{align}

\noindent so the only new degree of freedom emerging at this step is the complex scalar field $F$, of mass dimension $2$. Proceeding in the same way, namely acting on $F$ with the supersymmetry generators $Q_\alpha$ and $\bQ_{\dalpha}$ and using the Jacobi identity \eqref{eqn:ch1_gla}, one obtains
\begin{align}
    \[Q_\alpha,F\]=0 \ , && \[\bQ_{\dalpha},F\]=-2\(\partial_\mu\psi\sigma^\mu\)_{\dalpha} \ .
\end{align}

Thus, since no new degrees of freedom arose from this last step, the supersymmetric multiplet defined by the field $\phi$ of eq.~\eqref{eqn:ch1_cv_field} is given by
\beq
    \left\{\phi,\psi, F\right\} \ .
\label{eqn:ch1_chiral_offshell}
\eeq

Compared to the one in eq.~\eqref{eqn:ch1_m0_chiral}, this multiplet contains the complex scalar field $F$ and accommodates therefore twice the degrees of freedom, four bosonic and four fermionic. The reason is that the multiplet \eqref{eqn:ch1_chiral_offshell} is an \textit{off-shell} multiplet. The field $F$ is an auxiliary field, carrying no actual dynamical degree of freedom. Going on-shell, it is fixed to a given function $F(\phi,\psi)$, depending on the specific model considered, while the Dirac equation for $\psi$ halves its degrees of freedom, giving back the (on-shell) chiral multiplet of eq.~\eqref{eqn:ch1_m0_chiral}. Despite seemingly being a redundant description, off-shell supermultiplets are very useful and will come naturally in the superspace formalism, as we will see in Section \ref{ch1_sec_ss_sf}. 

The field components of the supermultiplet are related, as described, by supersymmetry transformations, which are usually given at the infinitesimal level, depending on a parameter $\epsilon$ of mass dimension $-\frac12$. The infinitesimal action of supersymmetry is given by
\beq
    \delta_\epsilon=i\(\epsilon Q+\bepsilon \bQ\) \ .
\label{eqn:ch1_d_susy_inf}
\eeq

\noindent The infinitesimal supersymmetry transformations of the off-shell are, in the standard normalization convention,
\beq
\begin{aligned}
    \delta\phi &=\sqrt{2}\epsilon\psi \ , \\
    \delta\psi &=i\sqrt{2}\(\sigma^\mu\bepsilon\)\partial_\mu\phi+\sqrt{2}F \epsilon \, \\
    \delta F &=i\sqrt{2}\bepsilon\bsigma^\mu\partial_\mu\psi \ ,
\end{aligned}
\label{eqn:ch1_wz_free_susy}
\eeq

\noindent in agreement with the computation above. This set of transformations closes on the spacetime translations as
\beq
    \[\delta_{\epsilon_1},\delta_{\epsilon_2}\]=-2i\(\epsilon_1\sigma^\mu\bepsilon_2-\epsilon_2\sigma^\mu\bepsilon_1\)\partial_\mu \ ,
\eeq

\noindent as prescribed by the Super-Poincar\'e algebra \eqref{eqn:ch1_SP_alg}. The transformations \eqref{eqn:ch1_wz_free_susy} characterize the so-called free Wess--Zumino model \cite{Wess:1974tw,Wess:1973kz}.

\section{Superspace and superfields}\label{ch1_sec_ss_sf}

In Section \ref{ch1_sec_SPA} we described how supersymmetry emerges naturally as the maximal extension of the Poincar\'e group, resulting in the so-called Super-Poincar\'e algebra given in eq.~\eqref{eqn:ch1_SP_alg}. We have then discussed its main properties and its representations, the supersymmetric multiplets, first in terms of spin states and finally in terms of component fields.

From now on, we restrict to $\N=1$ supersymmetry. The most elegant, general as well as effective way to construct and study $\N=1$ supersymmetric field theories is the framework of \textit{superspace} \cite{Salam:1974yz,Salam:1974jj}. The idea behind superspace is the following. As the standard Poincar\'e group \eqref{eqn:ch1_P_alg}, describing translations $P_\mu$ and rotations $M_{\mu\nu}$ of the ordinary spacetime coordinates $x^\mu$, has been extended to the Super-Poincar\'e group \eqref{eqn:ch1_SP_alg}, one can think of associating also the new supersymmetry generators $Q_\alpha$ and $\bQ_{\dalpha}$ to some additional coordinates. By construction, these are $2+2$ Grassmann coordinates,\footnote{See Section \ref{app_conv_grassmann} for a summary of properties and useful formulae on Grassmann algebras.} which are usually noted as $\(\theta_\alpha,\btheta_{\dalpha}\)$. A point in superspace is thus given by the set of eight coordinates $\(x^\mu,\theta_\alpha,\btheta_{\dalpha}\)$. In this setup, a supersymmetry transformation of parameter $\epsilon$ corresponds to the following translation of the superspace coordinates:
\begin{align}
    \theta\to\theta+\epsilon \ , && \btheta\to\btheta+\bepsilon \ , && \delta_\epsilon x^\mu=i\(\epsilon\sigma^\mu\btheta-\theta\sigma\bepsilon\) \ ,
\label{eqn:ch1_susy_translation}
\end{align}

\noindent where the spacetime coordinate transformation $\delta_\epsilon x^\mu$ follows by consistency with $\left\{Q,\bQ\right\}\sim P_\mu$ \eqref{eqn:ch1_SP_alg}. A full "supertranslation" is described by the Super-Poincar\'e group element
\beq
    G\(x,\theta,\btheta\)=e^{i\(-x^\mu P_\mu+\epsilon Q+\bepsilon\bQ\)} \ .
\label{eqn:ch1_ss_g_element}
\eeq


A field defined on such superspace, \textit{i.e.} depending on the full set of superspace coordinates, \textit{e.g.} $F\(x,\theta,\btheta\)$, is called \textit{superfield}. Since Grassmann numbers are nilpotent, a superfield has a finite expansion in the superspace coordinates, of the type
\begin{equation}
    \begin{aligned}
        F\(x^\mu,\theta_\alpha,\btheta_{\dalpha}\)=&f(x)+\theta\psi(x)+\btheta\bar{\chi}(x)+\theta^2M(x)+\btheta^2\bar{N}(x)+\theta\sigma^\mu\btheta A_\mu(x)\\
        &+\theta^2\btheta\bar{\lambda}(x)+\btheta^2\theta\zeta(x)+\theta^2\btheta^2D(x) \ .
    \end{aligned}
\label{eqn:ch1_sf_gen}
\end{equation}

\noindent The action of a supersymmetry transformation of parameter $\epsilon$ on such a superfield is computed as
\beq
    e^{i\(\epsilon Q+\bepsilon\bQ\)}F\(x,\theta,\btheta\)e^{-i\(\epsilon Q+\bepsilon\bQ\)} \ ,
\eeq

\noindent in agreement with eq.~\eqref{eqn:ch1_d_susy_inf}, and yields, once expanded, the representation of the infinitesimal supersymmetry transformations as differential operators in superspace:
\begin{align}
    Q_\alpha=-i\partial_\alpha -\(\sigma^\mu\btheta\)_\alpha\partial_\mu \ , && \bQ_{\dalpha}=i\bar{\partial}_{\dalpha}+\(\theta\sigma^\mu\)_{\dalpha}\partial_\mu ,
\label{eqn:ch1_Q_bQ_diff}
\end{align}

\noindent which explicitly verify the Super-Poincar\'e algebra condition
\beq
    \left\{Q_\alpha,\bQ_{\dalpha}\right\}=-2i\sigma^\mu_{\alpha\dalpha}\partial_\mu \ .
\label{eqn:ch1_comm_susy_derivative}
\eeq

\noindent of eq.~\eqref{eqn:ch1_SP_alg}. In addition to $Q_\alpha$ and $\bQ_{\dalpha}$, we introduce two further differential operators, the covariant derivatives
\begin{align}
    D_\alpha=\partial_\alpha+i\(\sigma^\mu\btheta\)_\alpha\partial_\mu \ , && \bar{D}_{\dalpha}=-\bar{\partial}_{\dalpha}-i\(\theta\sigma^\mu\)_{\dalpha}\partial_\mu \ ,
\label{eqn:ch1_ss_cov_dev}
\end{align}

\noindent which satisfy the anticommutation relations
\begin{equation}
    \begin{aligned}
         \left\{D,D\right\}&=\left\{\bar{D},\bar{D}\right\}=\left\{D,Q\right\}=\left\{D,\bQ\right\}=\left\{\bar{D},Q\right\}=\left\{\bar{D},\bQ\right\}=0 \ , \\
        \left\{D_\alpha,\bar{D}_{\dalpha}\right\}&=-2i\sigma^\mu_{\alpha\dalpha}\partial_\mu \ .
    \end{aligned}
\end{equation}

The crucial advantage of the superspace formalism is that it allows one to construct supersymmetric lagrangians, including in particular interactions among the component fields, in a straightforward manner. The reason is that a superspace action constructed in terms of superfields \eqref{eqn:ch1_sf_gen}, given by\footnote{A product of superfields is, by definition \eqref{eqn:ch1_sf_gen}, itself a superfield, so that \eqref{eqn:ch1_ss_action} is, for a real $F$, the most general form of a superspace action.}
\beq
    \int d^4x d^2\theta d^2\btheta \,\, F\(x,\theta,\btheta\) \ ,
\label{eqn:ch1_ss_action}
\eeq

\noindent is automatically a supersymmetric object. Acting on \eqref{eqn:ch1_ss_action} and using eq.~\eqref{eqn:ch1_d_susy_inf}, \eqref{eqn:ch1_susy_translation} and \eqref{eqn:ch1_Q_bQ_diff}, the integration measure is by definition invariant, so that 
\begin{equation}
    \begin{aligned}
        \delta_\epsilon\int d^4x d^2\theta d^2\btheta &\,\,F\(x,\theta,\btheta\)=\int d^4x d^2\theta d^2\btheta \,\,\delta_\epsilon F\(x,\theta,\btheta\)= \\
        &=\int d^4x d^2\theta d^2\btheta \(i\epsilon Q+i\bepsilon\bQ\)F\(x,\theta,\btheta\)= \\
        &=\int d^4x d^2\theta d^2\btheta \[\epsilon^\alpha \partial_\alpha +\bepsilon_{\dalpha}\partial^{\dalpha} +i\(\epsilon\sigma^\mu\btheta-\theta\sigma^\mu\bepsilon\)\partial_\mu\]F\(x,\theta,\btheta\) \ .
    \end{aligned}
\end{equation}

\noindent The first two terms vanish because of the Grassmann integration, while the last term is a total spacetime derivative and thus vanishes as well. Therefore, starting from a superspace action of the type \eqref{eqn:ch1_ss_action} and performing the integration in the fermionic coordinates yields a standard action which is, by construction, invariant under supersymmetry.

Moreover, irreducible representations of the Super-Poincar\'e algebra are obtained in superspace in a simple and unified way, starting from the general superfield \eqref{eqn:ch1_sf_gen} and reducing its components through supersymmetry-invariant conditions. We'll see this explicitly for the chiral \eqref{eqn:ch1_m0_chiral}-\eqref{eqn:ch1_chiral_offshell} and vector \eqref{eqn:ch1_m0_vec} multiplets in what follows.

\subsection{Chiral superfields}

Chiral multiplets \eqref{eqn:ch1_m0_chiral} contain a complex scalar a spin-$\frac12$ Weyl fermion, together with a second complex scalar in its off-shell formulation \eqref{eqn:ch1_chiral_offshell}. In superspace, this multiplet is described by a superfield $\Phi$ that satisfies the condition
\beq
    \bar{D}_{\dalpha}\Phi=0 \ .
\label{eqn:ch1_chiral_sf_def}
\eeq

\noindent In order to see that this constraint reduces the components of $\Phi$ to those of the off-shell chiral multiplet \eqref{eqn:ch1_chiral_offshell}, it is convenient to use a modified superspace coordinates system, where the spacetime coordinate is redefined into
\begin{align}
    x^\mu \quad \longrightarrow \quad y^\mu= x^\mu+i\theta\sigma^\mu\btheta \ .
\end{align}

\noindent In the basis $\(y,\theta,\btheta\)$, the superspace covariant derivatives \eqref{eqn:ch1_ss_cov_dev} are given by\footnote{We will use, with abuse of notation and the help of the context, the symbol $\partial_\mu$ to indicate derivatives with respect to both $x^\mu$ and $y^\mu$.}
\begin{align}
    D_\alpha=\partial_\alpha+2i\(\sigma^\mu\btheta\)_\alpha\partial_\mu \ , && \bar{D}_{\dalpha}=-\bar{\partial}_{\dalpha} \ .
\end{align}

\noindent Since we have by construction that $\bar{D}_{\dalpha}y^\mu=0$ and $\bar{D}_{\dalpha}\theta_\alpha=0$, a superfield that is function only of the coordinates $y^\mu$ and $\theta$ will be automatically chiral, solving the condition \eqref{eqn:ch1_chiral_sf_def} directly. The superspace expansion of a chiral superfield is therefore
\beq
    \Phi\(y,\theta\)=\phi(y)+\sqrt{2}\theta\psi(y)+\theta^2F(y) \ ,
\label{eqn:ch1_chiral_exp_y}
\eeq

\noindent in which we recognize the off-shell field components $\{\phi,\psi,F\}$ of eq.~\eqref{eqn:ch1_chiral_offshell}. In terms of the original basis $\(y,\theta,\btheta,\)$, it reads
\beq
\begin{aligned}
    \Phi\(x,\theta,\btheta\)=&\phi(x)+\sqrt{2}\theta\psi(x)+\theta^2 F(x)+i\theta\sigma^\mu\btheta\partial_\mu\phi(x)\\
    &-\frac{i}{\sqrt{2}}\theta^2\partial_\mu\psi(x)\sigma^\mu\btheta+\frac{1}{4}\theta^2\btheta^2\Box\phi(x) \ .
\end{aligned}
\eeq

\noindent Then, following eq.~\eqref{eqn:ch1_d_susy_inf} and \eqref{eqn:ch1_Q_bQ_diff}, the supersymmetry transformations of the field components of $\Phi$ are precisely those given in eq.~\eqref{eqn:ch1_wz_free_susy}.

Similarly to eq.~\eqref{eqn:ch1_chiral_sf_def}, a superfield $\bar{\Phi}$ satisfying $D_\alpha\bar{\Phi}=0$ is an antichiral superfield and it is the complex conjugate of \eqref{eqn:ch1_chiral_exp_y}. The simplest superspace action describing such chiral superfields is then
\begin{equation}
    S=\int d^4x d^2\theta d^2\btheta\,\, \bar{\Phi}\Phi \ ,
\label{eqn:ch1_csf_kin_1}
\end{equation}

\noindent and the superspace integration yields the kinetic lagrangian
\beq
    \L=-\partial_\mu\bar{\phi}\partial^\mu\phi-i\psi\sigma^\mu\partial_\mu\bpsi+\bar{F}F \ ,
\label{eqn:ch1_wz_kin_free}
\eeq

\noindent which is indeed invariant (up to total spacetime derivatives) under the supersymmetry transformations \eqref{eqn:ch1_wz_free_susy}.

\subsubsection{K\"ahler potential and superpotential}

Following the construction of chiral superfields just presented, we can now go on and study their interactions in a general fashion. Again, the superspace formalism allows one to do so in a compact and effective manner. The first step in this direction concerns the kinetic action \eqref{eqn:ch1_csf_kin_1}. This can be generalized, to 
\beq
    \int d^4x d^2\theta d^2\btheta\,\, K\(\bar{\Phi},\Phi\) \ ,
\label{eqn:ch1_csf_kin_KP_1}
\eeq

\noindent where $K\(\bar{\Phi},\Phi\)$ is a real superfield, of mass dimension 2, known as \textit{K\"ahler potential}, and is a polynomial in the chiral superfields $\Phi$ and $\bar{\Phi}$. The action \eqref{eqn:ch1_csf_kin_KP_1} it defines is symmetric under
\beq
    K\(\bar{\Phi},\Phi\) \quad \longrightarrow \quad K\(\bar{\Phi},\Phi\)+\Lambda\(\Phi\)+\bar{\Lambda}\(\bar{\Phi}\) \ ,
\label{eqn:ch1_K_tr}
\eeq

\noindent which is called K\"ahler transformation. Considering  $n$ chiral superfields $\Phi_i$, $i=1,\dots,n$, of the type \eqref{eqn:ch1_chiral_exp_y} and employing the notation
\begin{align}
    K_i\equiv\partial_i K\equiv\frac{\partial K}{\partial\phi^i} \ , 
\end{align}

\noindent the contribution to the component action of the K\"ahler potential term \eqref{eqn:ch1_csf_kin_KP_1} is equal to
\beq
\begin{aligned}
    \int d^2\theta d^2\btheta \,\, K\(\Phi,\bar{\Phi}\)=&K_{i\bar{j}}\[-\partial_\mu\phi^i\partial^\mu\bar{\phi}^{\bar{j}}-\(\frac{i}{2}\bpsi^{\bar{j}}\bsigma^\mu\partial_\mu\psi^{i}+h.c.\)+F^i\bar{F}^{\bar{j}}\] \\
    &-\[\frac{i}{2}K_{ij\bar{k}}\(\psi^i\sigma^\mu\bpsi^{\bar{k}}\partial_\mu\phi^j-i\psi^i\psi^j\bar{F}^{\bar{k}}\)+h.c.\]\\
    &+\frac{1}{4}K_{ij\bar{k}\bar{l}}\,\psi^i\psi^j\bpsi^{\bar{k}}\bpsi^{\bar{l}} \ .
\end{aligned}
\label{eqn:ch1_KP_action_1}
\eeq

\noindent The first line of this formula is the generalization of the kinetic lagrangian \eqref{eqn:ch1_wz_kin_free}, which is characterised by the scalar field-dependent kinetic coefficient $K_{i\bar{j}}$, known as K\"ahler metric. Supersymmetric theories of this type have in fact a geometrical description in terms of the so-called K\"ahler geometry \cite{Zumino:1979et}. A K\"ahler manifold is a Riemannian manifold, parameterised here by the complex scalars $\phi^i$, whose metric is defined in terms of a real function $K(\phi,\bar{\phi})$, the K\"ahler potential, as
\beq
    K_{i\bar{j}}=\partial_i\partial_{\bar{j}}K \ ,
\label{eqn:ch1_K_metric}
\eeq

\noindent in such a way to be invariant under the K\"ahler transformations \eqref{eqn:ch1_K_tr}. The K\"ahler geometry characterizes naturally the supersymmetric theories under consideration, since the general chiral superfield action \eqref{eqn:ch1_csf_kin_KP_1} is invariant under K\"ahler transformations by construction. 
Starting from the K\"ahler metric \eqref{eqn:ch1_K_metric}, of inverse $K^{i\bar{j}}$, one can define the associated connection and Riemann tensor as
\begin{align}
    \Gamma^i_{jk}=K^{i\bar{l}}K_{jk\bar{l}} \ , && \Gamma^{\bar{i}}_{\bar{j}\bar{k}}=K^{l\bar{i}}K_{\bar{j}\bar{k}l} \ , && R_{i\bar{j}k\bar{l}}=K_{m\bar{l}}\partial_{\bar{j}}\Gamma^m_{ik} \ ,
\label{eqn:ch1_K_geom_formulae_1}
\end{align}

\noindent as well as covariant derivatives
\begin{align}
    D_\mu \psi^i=\partial_\mu \psi^i+\Gamma^i_{jk}\(\sigma^\mu\psi^j\)\partial_\mu\phi^k \ , && D_i V_j=\partial_i V_j-\Gamma^l_{ij}V_l \ .
\label{eqn:ch1_K_geom_formulae_2}
\end{align}

The K\"ahler term \eqref{eqn:ch1_KP_action_1} already contains some interactions for the chiral supermultiplets. However, chiral superfields allows for a further extension of the supersymmetric action \eqref{eqn:ch1_ss_action}-\eqref{eqn:ch1_KP_action_1}, given in terms of an holomorphic function $W\(\Phi\)$ as
\beq
    \int d^2\theta\,\,W\(\Phi\) +h.c. \ .
\label{eqn:ch1_W_term_1}
\eeq

\noindent This superspace integral is supersymmetric because the $\theta^2$ component of $W\(\Phi\)$, being a chiral superfield itself, transforms as a total spacetime derivative. The function $W\(\Phi\)$ is called \textit{superpotential} and describes the interactions of chiral superfields. The associated component lagrangian is given by
\beq
    \int d^2\theta\,\,W\(\Phi\)=W_iF^i-\frac{1}{2}W_{ij}\psi^i\psi^j \ .
\eeq

Therefore, a chiral superfield theory is completely specified by the K\"ahler potential $K\(\Phi,\bar{\Phi}\)$ and the superpotential $W\(\Phi\)$. The complete superspace action is given by
\beq
\begin{aligned}
    S_\textup{chiral}=\int d^4x\left\{\int d^2\theta d^2\btheta \,\, K\(\Phi,\bar{\Phi}\)+\(\int d^2\theta\,\,W\(\Phi\) +h.c.\)\right\} \ ,
\end{aligned}
\label{eqn:ch1_chiral_K_W_ss}
\eeq

\noindent which turns into the off-shell lagrangian
\beq
\begin{aligned}
    \L_\textup{chiral}=& K_{i\bar{j}}\[-\partial_\mu\phi^i\partial^\mu\bar{\phi}^{\bar{j}}-\(\frac{i}{2}\bpsi^{\bar{j}}\bsigma^\mu\partial_\mu\psi^{i}+h.c.\)+F^i\bar{F}^{\bar{j}}\]\\ 
    &-\[\frac{i}{2}K_{ij\bar{k}}\(\psi^i\sigma^\mu\bpsi^{\bar{k}}\partial_\mu\phi^j-i\psi^i\psi^j\bar{F}^{\bar{k}}\)+h.c.\]\\
    &+\frac{1}{4}K_{ij\bar{k}\bar{l}}\,\psi^i\psi^j\bpsi^{\bar{k}}\bpsi^{\bar{l}} +\(W_iF^i-\frac{1}{2}W_{ij}\psi^i\psi^j+h.c.\) \ .
\end{aligned}
\label{eqn:ch1_chiral_K_W_lag}
\eeq

\noindent The auxiliary fields $F^i$ can be integrated out explicitly in such theory. Their equation of motion fixes
\beq
    F^i=-K^{i\bar{j}}\overline{W}_{\bar{j}}+\Gamma^i_{jk}\psi^j\psi^k \ ,
\label{eqn:ch1_F_int_out}    
\eeq

\noindent and therefore the on-shell supersymmetric lagrangian describing $n$ chiral superfields and their interactions is
\begin{equation}
    \begin{aligned}
        \L_\textup{chiral}=&-K_{i\bar{j}}\[\partial_\mu\phi^i\partial^\mu\bar{\phi}^{\bar{j}}+\(\frac{i}{2}\bpsi^{\bar{j}}\bsigma^\mu D_\mu\psi^{i}+h.c.\)\]-K^{i\bar{j}}D_iW\bar{D}_{\bar{j}}\overline{W} \\
        &-\frac12\(D_iD_jW\psi^i\psi^j+h.c.\)+\frac14R_{i\bar{j}k\bar{l}}\psi^i\psi^k\bpsi^{\bar{j}}\bpsi^{\bar{l}} \ .
    \end{aligned}
\label{eqn:ch1_on_shell_lag_KW}
\end{equation}

\noindent All the terms in this lagrangian have been written in a K\"haler-geometrical form, according to eq.~\eqref{eqn:ch1_K_geom_formulae_1}-\eqref{eqn:ch1_K_geom_formulae_2}, so that $D_iW=W_i$ and $D_iD_jW=W_{ij}-\Gamma^l_{ij}W_l$. In particular, the second term in the first line of \eqref{eqn:ch1_on_shell_lag_KW} is the scalar potential of supersymmetric theories:
\beq
    V\(\phi,\bar{\phi}\)=K^{i\bar{j}}D_iW\bar{D}_{\bar{j}}\overline{W} \ ,
\label{eqn:ch1_scal_p_F}
\eeq

\noindent which is again completely specified by the K\"ahler and superpotential. 

An example of chiral superfield theory is the famous Wess--Zumino model \cite{Wess:1974tw,Wess:1973kz}, which is the supersymmetric realisation of a Yukawa-type lagrangian. This model contains one chiral superfield $\Phi$, of K\"ahler and scalar potentials
\begin{align}
    K_\textup{WZ}\(\Phi\)=\bar{\Phi}\Phi \ , && W_\textup{WZ}\(\Phi\)=\frac{m}{2}\Phi^2+\frac{\lambda}{3}\Phi^3 \ ,
\label{eqn:ch1_wz_KW}
\end{align}

\noindent and results in the lagrangian
\beq
    \L_\textup{WZ}=-\partial_\mu\phi\partial^\mu\bar{\phi}-\(\frac{i}{2}\bpsi\bsigma^\mu\partial_\mu\psi+h.c.\)-\frac{m}{2}\(\psi\psi+\bpsi\bpsi\)-\abs{m\phi+\lambda\phi^2}^2 -\lambda\(\phi\psi\psi+h.c.\)\ ,
\label{eqn:ch1_L_wz}
\eeq

\noindent which is left invariant by the transformations
\begin{align}
    \delta\phi=\sqrt{2}\epsilon\psi \ , && \delta\psi=i\sqrt{2}\(\sigma^\mu\bepsilon\)\partial_\mu\phi-\sqrt{2}\(m\bar{\phi}+\lambda\bar{\phi}^2\)\epsilon \ ,
\end{align}

\noindent in agreement with \eqref{eqn:ch1_wz_free_susy}.

\subsection{Vector superfields}

A vector superfield $V$ is defined by the reality condition
\beq
    V=\bV \ .
\label{eqn:ch1_vec_sf_def}
\eeq

\noindent The resulting superfield can be expanded as
\beq
    \begin{aligned}
        V\(x,\theta,\btheta\)=& C(x)+i\theta\chi(x)-i\btheta\bchi(x)-\theta\sigma^\mu\btheta A_\mu(x)\\
        &+\frac{i}{2}\theta^2\[M(x)+iN(x)\]-\frac{i}{2}\btheta^2\[M(x)-iN(x)\]\\
        &+i\theta^2\btheta\[\bar{\lambda}(x)+\frac{i}{2}\bsigma^\mu\partial_\mu\chi(x)\]-i\btheta^2\theta\[\lambda(x)+\frac{i}{2}\sigma^\mu\partial_\mu\bchi(x)\]\\
        &+\frac{1}{2}\theta^2\btheta^2\[D(x)+\Box C(x)\] \ .
    \end{aligned}
\eeq

The supersymmetric version of the $\mathrm{U}(1)$ gauge transformation is given by
\beq
    V \quad \longrightarrow \quad V+i\Lambda-i\bar{\Lambda} \ ,
\label{eqn:ch1_susy_gauge_tr}
\eeq

\noindent with $\Lambda\(y,\theta\)=\alpha\(y\)+\sqrt{2}\theta\psi(y)+\theta^2F(y)$ being a chiral superfield. The $\theta\sigma^\mu\btheta$-term of this transformation reproduces in fact the standard gauge transformations on the field $A_\mu$
\beq
    A_\mu \quad \longrightarrow \quad A_\mu+\partial_\mu\(\alpha+\bar{\alpha}\) \ .
\label{eqn:ch1_gauge_tr_field}
\eeq

\noindent Moreover, there exists a choice of $\Lambda\(y,\theta\)$, called Wess--Zumino gauge, for which the component fields $C$, $\chi$, $M$ and $N$ can be removed. In this parameterization the vector superfield $V$ takes the form
\beq
    V\(x,\theta,\btheta\)=-\theta\sigma^\mu\btheta A_\mu(x)+i\theta^2\btheta\bar{\lambda}(x)-i\btheta^2\theta\lambda(x)+\frac{1}{2}\theta^2\btheta^2 D(x) \ ,
\label{eqn:ch1_vec_sf_wz_gauge}
\eeq

\noindent where $\lambda$ and $D$ are invariant under the gauge transformation \eqref{eqn:ch1_susy_gauge_tr}. The fields $A_\mu$ and $\lambda$ are the degrees of freedom of the vector multiplet we introduced in eq.~\eqref{eqn:ch1_m0_vec}, while $D$ is the auxiliary field of its off-shell formulation, analogous the one of the chiral multiplet. The supersymmetry transformations relating the degrees of freedom of this vector multiplet are\footnote{The supersymmetry transformations \eqref{eqn:ch1_vec_sf_susy} actually do not preserve the Wess--Zumino gauge at the level of the superfield V of eq.~\eqref{eqn:ch1_vec_sf_wz_gauge} and need to be accompanied by a compensating gauge transformation of parameter $\Lambda=-i\theta\sigma^\mu\bepsilon A_\mu-\theta^2\bepsilon\bar{\lambda}$.}
\beq
\begin{aligned}
    &\delta A_\mu = i\(\bepsilon\bsigma_\mu\lambda-\bar{\lambda}\bsigma_\mu\epsilon\) \ , \\
    &\delta\lambda = iD\epsilon +F_{\mu\nu}\sigma^{\mu\nu}\epsilon \ , \\
    &\delta D = -\epsilon\sigma^\mu\partial_\mu\bar{\lambda}-\partial_\mu\lambda\sigma^\mu\bepsilon \ ,
\end{aligned}
\label{eqn:ch1_vec_sf_susy}
\eeq

\noindent where $F_{\mu\nu}=\partial_\mu A_\nu-\partial_\nu A_\mu$. Notice that in the Wess--Zumino gauge the vector superfield $V$ of eq.~\eqref{eqn:ch1_vec_sf_wz_gauge} can appear at most quadratically, since $V^2=-\frac12\theta^2\btheta^2 A^\mu A_\mu$ and $V^3=0$.

In order to write a superspace lagrangian for the vector superfield \eqref{eqn:ch1_vec_sf_def}, it is natural to define the superfield version of the gauge field strength $F_{\mu\nu}$. This object is given by
\beq
    W_\alpha=-\frac{1}{4}\overline{D}^2D_\alpha V \ .
\eeq

\noindent By construction, the super-field strength $W_\alpha$ is invariant under the supersymmetric gauge transformations \eqref{eqn:ch1_susy_gauge_tr} and it is a chiral superfield, \textit{i.e.} $\overline{D}_{\dalpha}W_\alpha=0$. It is also such that $D^\alpha W_\alpha=\overline{D}_{\dalpha}\overline{W}^{\dalpha}$. In the Wess--Zumino gauge it takes the form
\beq
    W_\alpha \(y,\theta\)=-i\lambda_\alpha(y)+\[D(y)\delta_\alpha{}^\beta-iF_{\mu\nu}(y)\(\sigma^{\mu\nu}\)_\alpha{}^\beta\]+\theta^2\(\sigma^\mu\partial_\mu\bar{\lambda}\)_\alpha \ ,
\eeq

\noindent and the lagrangian
\beq
    \L=\frac14\int d^2\theta\,\, W^\alpha W_\alpha +\frac14\int d^2\btheta\,\,\overline{W}_{\dalpha}\overline{W}^{\dalpha}=-\frac14 F^{\mu\nu}F_{\mu\nu}-\(\frac{i}{2}\lambda\sigma^\mu\partial_\mu\bar{\lambda}+h.c.\)+\frac12 D^2 \ ,
\eeq

\noindent is the supersymmetric (off-shell) extension of the free Maxwell lagrangian. 

There is yet another possible real, supersymmetric and gauge invariant action that one can write for abelian gauge groups (or abelian factors of non-abelian groups which are not semi-simple). The $\theta^2\btheta^2$-component of the real vector superfield \eqref{eqn:ch1_vec_sf_wz_gauge}, \textit{i.e.} the auxiliary field $D$, is gauge invariant and transforms as a total spacetime derivative under supersymmetry \eqref{eqn:ch1_vec_sf_susy}. The associated lagrangian contribution
\beq
    \L_\textup{FI}=2\xi\int d^2\theta d^2\btheta\,\, V =\xi D \ ,
\label{eqn:ch1_FI_term}
\eeq

\noindent is called Fayet--Iliopoulos term \cite{Fayet:1974jb} and will play an important role in the context of supersymmetry breaking mechanisms.

\subsubsection{Super-Yang--Mills theories}

The non-abelian generalization of the above discussion is obtained extending the definition of vector superfield \eqref{eqn:ch1_vec_sf_def} and supersymmetric gauge transformations \eqref{eqn:ch1_susy_gauge_tr} to
\begin{align}
    V=V_a T^a \ , && \Lambda=\Lambda_a T^a \ , && \[T^a,T^b\]=f^a{}_{bc}T^c \ , && \text{Tr}\(T^aT^b\)=\delta^{ab} \ ,
\end{align}

\noindent where $T^a$ are the generators of the non-abelian gauge group and $f_{abc}$ the associated with the structure constants. In the Wess--Zumino gauge, $V_a$ takes the form of eq.~\eqref{eqn:ch1_vec_sf_wz_gauge} for every $a$. Calling $g$ the gauge coupling, the correct non-abelian super-field strength is
\beq
    W_\alpha=-\frac{1}{8g}\overline{D}^2\(e^{-2gV}D_\alpha e^{2gV}\) \ . 
\eeq

\noindent Such field-strength $W_\alpha$ transforms as $W_\alpha\to e^{-i\Lambda}W_\alpha e^{i\Lambda}$ under gauge transformations, so that the Yang--Mills term $\text{Tr}\(W^\alpha W_\alpha\)$ is indeed gauge invariant. In the Wess--Zumino gauge it reduces to
\beq
    W_\alpha=-\frac14\overline{D}^2D_\alpha V+\frac{g}{4}\overline{D}^2\[V,D_\alpha V\] \ ,
\eeq

\noindent which turns into
\beq
    W^a_\alpha \(y,\theta\)=-i\lambda^a_\alpha(y)+\[D^a(y)\delta_\alpha{}^\beta-iF^a_{\mu\nu}(y)\(\sigma^{\mu\nu}\)_\alpha{}^\beta\]\theta_\beta+\theta^2\(\sigma^\mu\mathcal{D}_\mu\bar{\lambda}^a\)_\alpha \ ,
\eeq

\noindent where, consistently, we have
\begin{align}
    F_{\mu\nu}^a=\partial_\mu A^a_\nu-\partial_\nu A^a_\mu-gf^a{}_{bc} A^b_\mu A^c_\nu \ , && \mathcal{D}_\mu\lambda^a=\partial_\mu \lambda^a-gf^a{}_{bc}A^b_\mu\lambda^c \ .
\end{align}

\noindent Notice that the gauginos $\lambda^a$ transform in the adjoint representation of the non-abelian gauge group, as they belong to the same multiplet as the gauge fields.

The supersymmetric coupling of gauge and matter fields, \textit{i.e.} of vector \eqref{eqn:ch1_vec_sf_def} to chiral superfields \eqref{eqn:ch1_chiral_sf_def} is obtained as follows. The supersymmetric gauge transformations of chiral superfields say in the fundamental representation is
\beq
    \Phi \quad \longrightarrow \quad e^{-i\Lambda}\Phi \ ,
\eeq

\noindent which respects the chiral structure of the superfield, and so a gauge invariant kinetic term has the form
\beq
    \bar{\Phi}e^{2gV}\Phi \ .
\label{eqn:ch1_vec_chiral_min_kin}
\eeq

\noindent Thus, an example of supersymmetric Yang--Mills theory, where a non-abelian vector multiplet is coupled to $n$ chiral superfields of minimal kinetic term \eqref{eqn:ch1_vec_chiral_min_kin} and superpotential $W\(\Phi\)$ (which must be gauge invariant), is given by 
\beq
    \L_\textup{SYM}=\int d^2\theta d^2\btheta\,\,\bar{\Phi}e^{2gV}\Phi+\(\int d^2\theta\[\frac14\text{Tr}\(W^\alpha W_\alpha\) +W\(\Phi\)\]+ h.c.\) \ .
\label{eqn:ch1_sym_theory}
\eeq

\noindent The off-shell supersymmetry transformations characterizing this theory are
\beq
\begin{aligned}
    &\delta\phi^i=\sqrt{2}\bepsilon\psi^i \ , &&&&\delta A^a_\mu = i\(\bepsilon\bsigma_\mu\lambda^a-\bar{\lambda}^a\bsigma_\mu\epsilon\) \ , \\
    &\delta\psi^i=i\sqrt{2}\(\sigma^\mu\bepsilon\)\mathcal{D}_\mu\phi^i+\sqrt{2}F^i\epsilon \ , &&&&\delta\lambda^a = iD^a\epsilon +F^a_{\mu\nu}\sigma^{\mu\nu}\epsilon \ , \\
    &\delta F^i=\sqrt{2}\bepsilon\bsigma^\mu\mathcal{D}_\mu\psi+2ig\(T_a\)^i{}_j\phi^j\bepsilon\bar{\lambda}^a &&&&\delta D^a = -\epsilon\sigma^\mu\mathcal{D}_\mu\bar{\lambda}^a-\mathcal{D}_\mu\lambda^a\sigma^\mu\bepsilon \ ,
\end{aligned}
\label{eqn:ch1_sym_susy_tr}
\eeq

\noindent where
\begin{align}
    \mathcal{D}_\mu\phi^i=\partial_\mu\phi^i+igA^a_\mu\(T_a\)^i{}_j\phi^j \ , && \mathcal{D}_\mu\psi^i=\partial_\mu\psi^i+igA^a_\mu\(T_a\)^i{}_j\psi^j \, 
\end{align}

\noindent are the fundamental representation covariant derivatives. A very important element of theories of this type is again the scalar potential. The coupling to vector multiplets introduces a new term in the scalar potential, associated with the $D^a$ auxiliary fields and thus called D-term, which adds the so-called $F$-term coming from the superpotential $W\(\Phi\)$ \eqref{eqn:ch1_scal_p_F}. In the case of the theory \eqref{eqn:ch1_sym_theory}, it is given by
\beq
    V\(\phi,\bar{\phi}\)=\sum_i\abs{F^i}^2+\frac12\sum_aD^a D_a \ ,
\label{eqn:ch1_vec_chiral_sp_1}
\eeq

\noindent and the on-shell equations for the auxiliary fields are
\begin{align}
    F^i=-\overline{W}_{\bar{i}} \ , && D^a=-g\(T^a\)_{ij}\phi^i\bar{\phi}^{\bar j} \ .
\label{eqn:ch1_vec_chiral_af_on_shell_1}
\end{align}

The scalar potential in eq.~\eqref{eqn:ch1_vec_chiral_sp_1}-\eqref{eqn:ch1_vec_chiral_af_on_shell_1} can easily be further generalized by including a non-minimal K\"ahler potential, as well as Fayet--Iliopoulos terms \eqref{eqn:ch1_FI_term} (see eq.~\eqref{eqn:ch1_F_D_term_eq}), and it is the key element to study the phenomenon of spontaneous supersymmetry breaking, as we do in the next section.

\section{Spontaneous breaking and nonlinear realisations of supersymmetry}\label{ch1_sec_ssb}

The breaking of supersymmetry is a central feature of supersymmetric theories. It gives rise to rich and complex dynamics and, most importantly, is essential for supersymmetry to be phenomenologically viable. Since the Standard Model is clearly non-supersymmetric, supersymmetry breaking plays a key role in the physics beyond the Standard Model and, more generally, in mechanisms that connect the observed low-energy world to higher energy scales.

\begin{figure}
\centering
\includegraphics[scale=0.5]{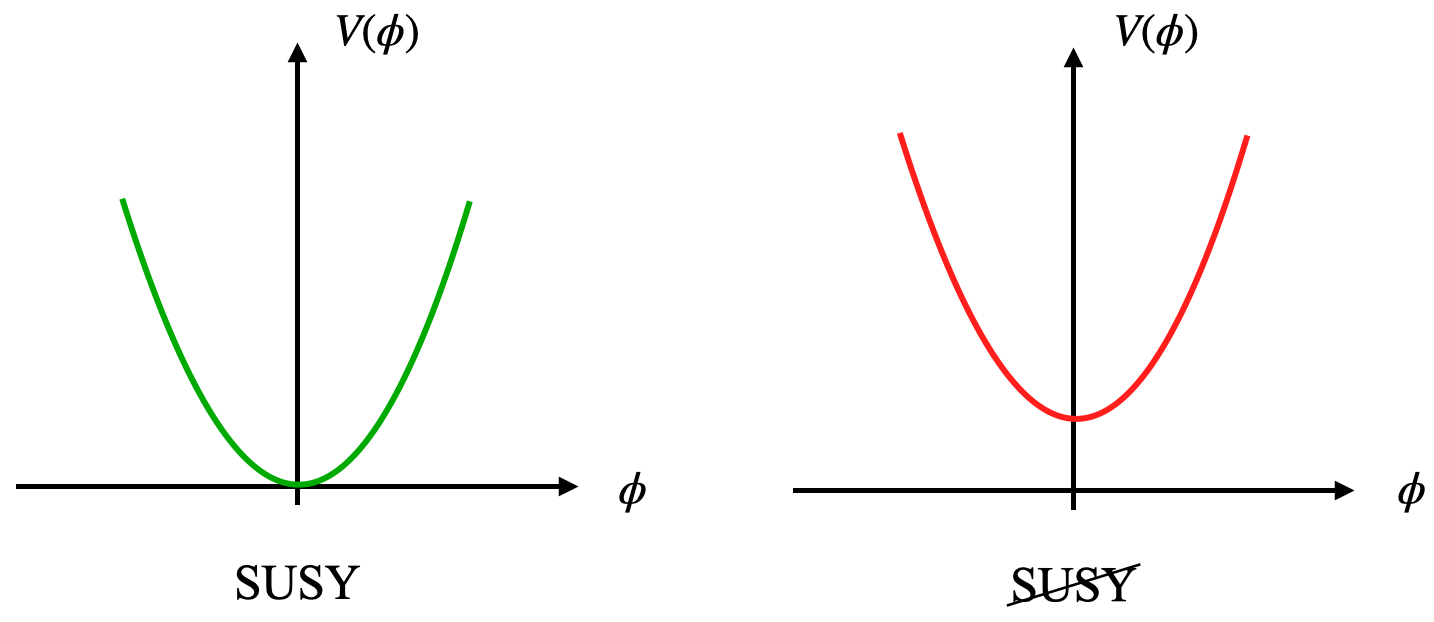}
\caption{Basic representation of supersymmetry preserving versus supersymmetry breaking vacua.}
\label{fig_susy_breaking}
\end{figure}

Here we focus on the case where supersymmetry is broken \textit{spontaneously}. 
This is a natural scenario from a high-energy perspective, where supersymmetry is expected to be realized as a local symmetry, namely as supergravity. Restricting ourselves to the $\N=1$ (rigid) case, the spontaneous breaking of supersymmetry is entirely governed by the scalar potential. Since the vacuum state of the theory must be a Lorentz-invariant quantity, only scalar fields can acquire a non-trivial vacuum expectation value. Furthermore, as discussed in Section~\ref{ch1_sec_SPA}, a general consequence of the Super-Poincar\'e algebra is that the energy in a supersymmetric theory is semi-positive definite, as stated in Eq.~\eqref{eqn:ch1_E_pos}. It follows that supersymmetric vacua correspond to the zeros of the scalar potential, while any non-zero minimum defines a vacuum in which supersymmetry is spontaneously broken. This is illustrated in Figure~\ref{fig_susy_breaking}.

The general form of the scalar potential in a supersymmetric field theory that couples vector and chiral multiplets was given in eq.~\eqref{eqn:ch1_vec_chiral_sp_1}. Off-shell, it is a function of the auxiliary fields $F^i$ and $D^a$, which is indeed semi-positive definite, in agreement with eq.~\eqref{eqn:ch1_E_pos}. Therefore, supersymmetric vacua are determined by the conditions
\begin{align}
    F^i=-K^{i\bar j}\overline{W}_{\bar j}=0 \ , && D^a=-g\(T^a\)_{ij}\phi^i\bar{\phi}^{\bar j}-\xi^a=0 \ ,
\label{eqn:ch1_F_D_term_eq}
\end{align}

\noindent where we included a non-minimal K\"ahler metric\footnote{Although it does modify the shape of the scalar potential, a (non-singular) non-minimal K\"ahler metric does not influence the F-term vanishing condition in eq.~\eqref{eqn:ch1_F_D_term_eq}, which is any case given by $W_i=0$.} and the Fayet--Iliopoulos terms \eqref{eqn:ch1_FI_term}. The scalar field values for which both the F and D-term equations are solved is called the moduli space. If the F-term and/or the D-term equations cannot be solved, the theory breaks supersymmetry spontaneously. We now discuss simple models which realize the two possibilities.

\subsubsection{D-term breaking and the Fayet--Iliopoulos term}

Fayet--Iliopoulos terms \cite{Fayet:1974jb} generally introduce a positive contribution to the vacuum energy and realize the spontaneous breaking of supersymmetry in supersymmetric gauge theories. For instance, we take the following theory of an abelian vector superfield $V$ and two chiral superfields $\Phi_1$ and $\Phi_2$:
\beq
    \L=\int d^2\theta d^2\btheta\(\bar{\Phi}_1e^{2gV}\Phi_1+\bar{\Phi}_2e^{-2gV}\Phi_2+2\xi V\)+\[\int d^2\theta\(\frac14W^\alpha W_\alpha+m\Phi_1\Phi_2\)+h.c.\] \ .
\eeq

\noindent The scalar potential \eqref{eqn:ch1_vec_chiral_sp_1} in this model is given by
\beq
    V=m^2\(\abs{\phi_1}^2+\abs{\phi_2}^2\)+\frac12\[g\(\abs{\phi_1}^2-\abs{\phi_2}^2\)+2\xi\] \ ,
\eeq

\noindent and the F-term and D-term equations \eqref{eqn:ch1_F_D_term_eq} are
\begin{align}
    F_1=-m\bar{\phi}_2=0 \ , &&  F_2=-m\bar{\phi}_1=0 \ , && D=-\xi-g\(\abs{\phi_1}^2-\abs{\phi_2}^2\)=0 \ .
\end{align}

\noindent These equations cannot be solved simultaneously because of the nontrivial Fayet--Iliopoulos term, and supersymmetry is thus broken spontaneously.

\subsubsection{F-term breaking}

We now turn to the case where supersymmetry is broken through an F-term. A generic superpotential $W\(\Phi\)$ depending on $n$ chiral superfields $\Phi_i$, can be written as the polynomial sum
\beq
    W\(\Phi\)=a_i\Phi_i+\frac12 b_{ij}\Phi^i\Phi^j+\frac13 c_{ijk}\Phi^i\Phi^j\Phi^k+\dots \ .
\label{eqn:ch1_W_exp_poly}
\eeq

\noindent For a minimal K\"ahler potential, the F-term equation \eqref{eqn:ch1_F_D_term_eq} takes then the form
\beq
    a_i+b_{ij}\phi^j+c_{ijk}\phi^i\phi^j+\dots=0 \ .
\eeq

\noindent Hence, spontaneous supersymmetry breaking is possible if at least one of the coefficients $a_i$ is different from zero, namely if there is at least a linear term in the superpotential \eqref{eqn:ch1_W_exp_poly}. Otherwise, there always exist a supersymmetric vacuum, for some/all the scalar fields set to zero.

\paragraph{The Polonyi model}

The so-called Polonyi model is the simplest instance of F-term supersymmetry breaking. It describes a single chiral superfield $\Phi$ with linear superpotential:
\begin{align}
    K\(\Phi,\bar{\Phi}\)=\bar{\Phi}\Phi \ , && W\(\Phi\)=f\Phi \ .
\end{align}

\noindent The scalar potential is constant $V=\abs{f}^2$ and supersymmetry is clearly broken in the whole field space.

\paragraph{The O'Raifeartaigh model}

A more sophisticated and widely studied setup for spontaneous supersymmetry breaking is the so-called The O'Raifeartaigh model \cite{ORaifeartaigh:1975nky,Intriligator:2007py}. The simplest O'Raifeartaigh setup amounts to three chiral superfields $\Phi_1$, $\Phi_2$ and $X$, with canonical K\"ahler potential and interacting through the superpotential
\beq
    W_\text{O'R}=\frac{h}{2}X\Phi_1^2+m\Phi_1\Phi_2+f X \ ,
\label{eqn:ch1_OR_W}
\eeq

\noindent with $h$, $m$ and $f$ positive coefficients. The F-term equations are\footnote{We employ the same notation, namely $X$, for the superfield and its bottom scalar field component.}
\beq
\begin{aligned}
    \bar{F}_1&=-hX\phi_1-m\phi_2=0 \ , \\
    \bar{F}_2&=-m\phi_1=0 \ , \\
    \bar{F}_X&=-\frac{h}{2}\phi_1^2-f=0 \ ,
\end{aligned}
\eeq

\noindent and cannot be solved all at the same time. O'Raifeartaigh models are characterized by interaction terms engineered precisely to give rise to a set of F-term equations whose solutions are in conflict with each other. The model \eqref{eqn:ch1_OR_W} is actually characterized by a (pseudo)moduli space of supersymmetry breaking vacua. Defining the parameter
\beq
    y\equiv\left\lvert\frac{hf}{m^2}\right\rvert \ ,
\eeq

\noindent for $y<1$ this (pseudo)moduli space is parameterized by $\phi_1=\phi_2=0$ and arbitrary $X$. In the $y>1$ case, the (pseudo)moduli spaces are instead two, given by
\begin{align}
    \phi_2=-\frac{h}{m}X\phi \ , && \phi_1=\pm i\sqrt{\frac{2f}{h}\(\frac{y-1}{y}\)}\ ,
\end{align}

\noindent and $X$ still being a flat direction. 

This vacuum degeneracy is actually lifted by 1-loop corrections \cite{Intriligator:2007py}, which are generally given by the Coleman--Weinberg potential \cite{Coleman:1973jx}. Moreover, supersymmetry can be restored by deforming the superpotential \eqref{eqn:ch1_OR_W} with a term breaking the R-symmetry it originally enjoys.\footnote{The R-charges of the superfields entering the superpotential \eqref{eqn:ch1_OR_W} are $\text{R}_X=\text{R}_{\Phi_2}=2$ and $\text{R}_{\Phi_1}=0$.} For (parametrically) small enough R-symmetry breaking, the new supersymmetric vacuum is pushed far away with respect to the old one, which survives in this setup as a metastable vacuum \cite{Intriligator:2007py}.

\subsection{The Goldstino and nonlinear supersymmetry}\label{ch1_sub_goldstino}

When a symmetry is spontaneously broken, the Goldstone theorem predicts the existence of massless particles, associated with the number of broken generators, \textit{i.e.} that do not leave the vacuum invariant \cite{Nambu:1961tp,Goldstone:1961eq,Goldstone:1962es}. Supersymmetry makes, of course, no exception, and having anticommuting generators the Goldstone particle must be a spin-$\frac12$ fermion, known as the \textit{goldstino}. 

Let us focus on the model \eqref{eqn:ch1_sym_theory}. When supersymmetry is broken spontaneously, either via an F-term $\langle F^i\rangle\ne 0$ or a D-term $\langle D^a\rangle\ne 0$, the transformations of the fermions \eqref{eqn:ch1_sym_susy_tr} become inhomogeneous:
\begin{align}
    \langle\delta\psi^i\rangle\sim\langle F^i\rangle \epsilon \ , &&  \langle\delta\lambda^a\rangle\sim \langle D^a\rangle \epsilon \ ,
\label{eqn:ch1_ferm_susy_inhom}
\end{align}

\noindent and the goldstino mode $G$ is given as the following combinations of fermionic fields
\beq
    G=\frac{1}{f}\(\langle F_i\rangle\psi^i+\langle D_a\rangle\lambda^a\) \ ,
\label{eqn:ch1_G_def_ferm}
\eeq

\noindent where $f=\langle\abs{F^i}^2\rangle+\frac12\langle D^aD_a\rangle$ is the scale of supersymmetry breaking. The combination \eqref{eqn:ch1_G_def_ferm} is in fact an eigenvector of vanishing eigenvalue of the fermionic mass matrix, 
\beq
    -\frac12\begin{pmatrix} \psi^i & \lambda^a \end{pmatrix}\begin{pmatrix}
        W_{ij} & -i\sqrt{2}g\bar{\phi}^{\bar j}\(T_b\)_{i\bar{j}}\\ -i\sqrt{2}g\bar{\phi}^{j}\(T_a\)_{ij} & 0 
    \end{pmatrix}
    \begin{pmatrix} \psi^j \\ \lambda^b \end{pmatrix} \ ,
\eeq

\noindent originating from the lagrangian in eq.~\eqref{eqn:ch1_sym_theory}.

The inhomogeneous supersymmetry transformations of the fermions \eqref{eqn:ch1_ferm_susy_inhom} and the goldstino dynamics that emerges can be elegantly understood in terms of a \textit{nonlinear realization} of the supersymmetry algebra. Nonlinear supersymmetry relates a fermion to a scalar combination of the fermion itself, while still satisfying the supersymmetry algebra \eqref{eqn:ch1_SP_alg}. To develop such formalism, one starts from a fermionic field $\chi\(x\)$ whose supersymmetry transformation is taken to be \cite{Volkov:1973ix}
\beq
    \chi'(x')=\chi\(x\)+f\epsilon \ ,
\label{eqn:ch1_nl_susy_tr_1}
\eeq

\noindent in which $f$ is a real coefficient of mass dimension 2, while $x'^\mu=x^\mu+i\(\epsilon\sigma^\mu\btheta-\theta\sigma^\mu\bepsilon\)$ is the transformation of the spacetime coordinate $x^\mu$ in superspace, as in eq.~\eqref{eqn:ch1_susy_translation}. The transformation \eqref{eqn:ch1_nl_susy_tr_1} is actually the generalization of the other coordinate transformation $\theta'=\theta+\epsilon$ \eqref{eqn:ch1_susy_translation} --- which are similar in form to the transformations in eq.~\eqref{eqn:ch1_ferm_susy_inhom} --- to an arbitrary spinor field. The infinitesimal field variation at equal spacetime point is then
\beq
    \delta_\epsilon\chi\(x\)=\chi'\(x\)-\chi\(x\)=f\epsilon+\frac{i}{f}\[\chi\(x\)\sigma^\mu\bepsilon-\epsilon\sigma^\mu\bchi\(x\)\]\partial_\mu\chi\(x\) \ ,
\label{eqn:ch1_nl_susy_tr_2}
\eeq

\noindent which satisfies the supersymmetry algebra \eqref{eqn:ch1_SP_alg}:
\beq
    \[\delta_1,\delta_2\]\chi=-2i\(\epsilon_1\sigma^\mu\bepsilon_2-\epsilon_2\sigma^\mu\bepsilon_1\)\partial_\mu\chi \ .
\eeq

\noindent Therefore, the nonlinear transformation \eqref{eqn:ch1_nl_susy_tr_1} is a supersymmetry transformation, relating the spinor $\chi$ to a scalar combination of $\chi$ itself. Invariant lagrangians are then constructed in a geometric way, starting from the generalized vielbein
\beq
    a_\mu{}^\alpha=\delta_\mu{}^\alpha+\frac{i}{f^2}\(\chi\sigma^\alpha\partial_\mu\bchi-\partial_\mu\chi\sigma^\alpha\bchi\) \ .
\label{eqn:ch1_VA_veilbein}
\eeq

\noindent The determinant $\det a$ transforms in fact as a total spacetime derivative under the supersymmetry transformation \eqref{eqn:ch1_nl_susy_tr_2} --- which is interpreted as a diffeomorphism for the vielbein $a_\mu{}^\alpha$ \eqref{eqn:ch1_VA_veilbein} --- and thus defines the supersymmetric lagrangian
\beq
    \L_\text{VA}=-\frac{f^2}{2}\det a=-\frac{f^2}{2}-\frac{i}{2}\(\chi\sigma^\mu\partial_\mu\bchi-\partial_\mu\chi\sigma^\mu\bchi\)+\dots \ ,
\label{eqn:ch1_VA_lag}
\eeq

\noindent where the dots stand for higher-derivative self-interactions of the spinor $\chi$. This lagrangian is the famous Volkov--Akulov lagrangian of nonlinear supersymmetry \cite{Volkov:1973ix}. The positive vacuum energy, proportional to the constant $f$, spontaneously breaks linear supersymmetry, leaving though a residual nonlinear supersymmetry \eqref{eqn:ch1_nl_susy_tr_2}, realized in terms of the massless fermion $\chi$, which is then identified with the goldstino.

\section{Constrained superfields}\label{ch1_sec_nls}

Theories with spontaneously broken supersymmetry present a fermionic Goldstone mode, the goldstino \eqref{eqn:ch1_G_def_ferm}, on which the original supersymmetry is now nonlinearly realized \eqref{eqn:ch1_nl_susy_tr_2}. Such nonlinear supersymmetry completely determines the dynamics of the goldstino, which is of the Volkov--Akulov type \eqref{eqn:ch1_VA_lag} \cite{Volkov:1973ix}. In particular, the residual nonlinear supersymmetry should also characterize the interactions of the goldstino with the other fields in the theory.

As we will see more in detail in Section \ref{ch2_sec_sugra_Higgs}, in supergravity, that is, when supersymmetry becomes a local symmetry, the fermionic version of the Higgs mechanism takes place \cite{Deser:1977uq}: the massless goldstino is eaten by the gauge field of local supersymmetry, the spin-$\frac32$ \textit{gravitino}, which becomes massive. The original goldstino corresponds now to the longitudinal polarizations of the massive gravitino. However, if the supersymmetry breaking scale $f$ is well below the Planck mass $M_\textup{P}$, the low-energy (\textit{i.e.} at a scale below $f$) interactions of such longitudinal modes of the massive gravitino dominate over those of the original transverse ones, since the former scale as $\nicefrac{1}{f}$ while the latter as $\nicefrac{1}{M_\textup{P}}$. This property is expressed in general terms by the gravitino equivalence theorem \cite{Fayet:1986zc,Casalbuoni:1988kv,Casalbuoni:1988qd}, so that at energies below the supersymmetry breaking scale --- but above the gravitino mass --- the couplings of the gravitino are controlled by the longitudinal mode and can be studied in terms of the goldstino interactions in the rigid limit. From this point of view, nonlinear supersymmetry and the associated goldstino dynamics are a direct manifestation of the high-energy physics at lower scales, in this case even with gravity decoupled. 

A particularly simple and effective tool to construct effective theories invariant under nonlinear supersymmetry is the so-called \textit{constrained superfield} formalism \cite{Rocek:1978nb, Lindstrom:1979kq,Komargodski:2009rz,DallAgata:2015zxp,DallAgata:2016syy}. The idea is that extra constraints are imposed on superfields, which effectively remove some of their components, simulating the decoupling of a massive degree of freedom while allowing one to use the standard supersymmetric lagrangians. In this framework, the goldstino $G$ is embedded into a chiral superfield $S$ subject to the nilpotency condition \cite{Rocek:1978nb,Ivanov:1978mx,Ivanov:1982bpa,Komargodski:2009rz}
\beq 
    S^2=0 \ .
\label{eqn:ch1_G_nil_sf}
\eeq

\noindent This constraint fixes $S$ to be of the form
\beq
    S=\frac{G^2}{2F_S}+\sqrt{2}\theta G+\theta^2 F_S \ ,
\label{eqn:ch1_nil_sf_sol}
\eeq

\noindent namely it determines the bottom scalar component of the superfield to be a function of the fermion $G$ and the auxiliary field $F_S$. Notice that $F_S$ must be different from zero in order for the nilpotent constraint \eqref{eqn:ch1_G_nil_sf} to be solved, signaling already the breaking of supersymmetry. Moreover, the supersymmetry transformation of $S$ now relates the fermion $G$ to the bilinear $s\sim G^2$ of the fermion itself. Thus, supersymmetry is realized nonlinearly on the nilpotent superfield $S$, so that its fermionic component $G$ is identified with the goldstino of supersymmetry breaking. The most general 2-derivative theory that one can write with $S$ is given by
\begin{align}
    K\(S,\bar{S}\)=\bar{S}S \ , && W\(S\)=f S \ .
\end{align}

\noindent The associated on-shell lagrangian, obtained after integrating out the auxiliary field as
\beq
    F_S=-f\(1+\frac{\bar{G}^2}{4f^4}\Box G^2-\frac{3}{16 f^8}G^2\bar{G}^2\Box G^2\Box \bar{G}^2\) \ ,
\label{eqn:ch1_F_S_sol}
\eeq

\noindent is found to be
\beq
    \L_S=-f^2-iG\sigma^\mu\partial_\mu\bar{G}+\frac{1}{4f^2}\bar{G}^2\Box G^2-\frac{1}{16f^6}G^2\bar{G}^2\Box G^2\Box \bar{G}^2 \ ,
\eeq

\noindent which can be shown to be equivalent to the Volkov--Akulov lagrangian \eqref{eqn:ch1_VA_lag} \cite{Volkov:1973ix} by means of a nonlinear field redefinition \cite{Ivanov:1978mx,Ivanov:1982bpa,Kuzenko:2010ef}. 

Therefore, the nilpotent superfield $S$ correctly describes the goldstino field and the associated nonlinear supersymmetry. Its coupling to other fields allows then to construct in a very direct way effective theories of spontaneously broken supersymmetry. Following these lines, the superfield $S$ also allows for the definition of further superfield constraints, which decouple superpartners in different ways, providing a richer spectrum of effective field theories of nonlinearly realized supersymmetry.\cite{Komargodski:2009rz}. This constraining procedure through the nilpotent goldstino superfield \eqref{eqn:ch1_G_nil_sf} can be generalised into a prescription that allows one to remove specific, single superfield components \cite{DallAgata:2016syy}.

However, despite being algebraically clear, this procedure is not always physically well-defined \cite{Bonnefoy:2022rcw}. This is the case when the given superfield constraint also affects the auxiliary field of the constrained multiplet. In fact, whereas the removal by the constraints of a scalar or fermion can naturally fit with the microscopic picture of the standard decoupling of heavy particles, the same interpretation does not apply to auxiliary fields --- which are not dynamical and, hence, are not associated with any physical mass --- and their removal requires higher-derivative couplings directly in the UV, as proposed already in \cite{DallAgata:2016syy}. This feature manifests in low-energy goldstino lagrangians in terms of higher-derivative operators which are subject to positivity bounds in order to be consistent \cite{Adams:2006sv,Bellazzini:2016xrt}, and therefore cannot originate from a 2-derivative action in the UV. On the other hand, any constraint that does not affect the auxiliary fields does not suffer from any causality issue of this type and so, as expected from the physical point of view, there is no obstruction to an uplift to a 2-derivative UV theory. 

This is the main result of the paper \cite{Bonnefoy:2022rcw}, subject of the present thesis, which we now discuss in detail.
We begin by analyzing two examples of superfield constraints, proposed in \cite{Komargodski:2009rz} and further studied in \cite{Bonnefoy:2022rcw}. The theories defined by these constraints describe the (on-shell) coupling of the goldstino to a real and a complex scalar field, respectively. After introducing the constraints, we compute the most general lagrangian they determine and show that they involve operators that may lead to superluminal goldstino propagation unless appropriate positivity bounds on the scattering amplitudes are taken into account. Then, we provide evidence that the obstruction to a UV completion of such effective theories is related to the non-trivial solution imposed by such constraints on the auxiliary fields. We do so first studying a third superfield constraint, that couples the goldstino to a fermion and which leaves untouched the auxiliary field, and show that the resulting lagrangian does not contain any dangerous operator. Second, we refine the models that suffer from potential acausal behavior through the generalized constraints of \cite{DallAgata:2016syy}, by constructing effective models with the same field spectrum of the ill-defined ones but without touching the auxiliary fields. The resulting theories are found to be completely causal in the whole field space. 

\subsection{Coupling to a real scalar -- the orthogonal constraint}\label{sub_ch1_oc}

The first superfield constraint induced by the goldstino superfield \eqref{eqn:ch1_G_nil_sf} is the so-called \textit{orthogonal constraint}, which acts on a chiral superfield as
\beq
    S\(\Phi-\bar{\Phi}\)=0 \ ,
\label{eqn:ch1_oc}
\eeq    

\noindent which implies also $\(\Phi-\bar{\Phi}\)^3=0$. Calling the $\Phi$ components
\begin{align}
    \Phi=\phi+\sqrt{2}\theta\chi_\phi+\theta^2 F_\phi \ , && \phi=A+iB \ ,
\label{eqn:ch1_oc_phi_comp}
\end{align}

\noindent the orthogonal constraint \eqref{eqn:ch1_oc} is solved by \cite{Komargodski:2009rz}
\beq
\begin{aligned}
    B&=-\frac{G\sigma^\mu\bar G} {2\abs{F_S}^2}\partial_\mu A \ , \\
    \chi_\phi&=-i\sigma^\mu\frac{\bar G}{\bar{F}_S}\partial_\mu\phi \ , \\
    F_\phi&=-\partial_\nu\left(\frac{\bar G}{\bar{F}_S}\right)\bar{\sigma}^\mu\sigma^\nu \frac{\bar G} {\bar{F}_S} \partial_\mu A+\frac{1}{2}\left(\frac{\bar G}{\bar{F}_S}\right)^2\Box A \ . 
\end{aligned}
\label{eqn:ch1_oc_sol}
\eeq

\noindent The resulting on-shell degrees of freedom are the goldstino $G$ and a real scalar $A$. Because of such minimal spectrum, the orthogonal constraint played a significant role in defining minimal models of inflation in supergravity \cite{Ferrara:2015tyn,Carrasco:2015iij,Kolb:2021xfn,Dudas:2021njv,Kahn:2015mla}. We will discuss this in Section \ref{ch2_sugra_grav_cosmology}.

The most general theories coupling the nilpotent superfield $S$ \eqref{eqn:ch1_G_nil_sf} and such orthogonally-constrained superfield $\Phi$ \eqref{eqn:ch1_oc} involves the following K\"ahler and superpotential \cite{Ferrara:2015tyn}:
\begin{align}
    K=h\(\mathcal{A}\)\mathcal{B}^2+\bar{S}S \ , && W=f\(\Phi\)S+g\(\Phi\) \ ,
\label{eqn:ch1_oc_KW}
\end{align}

\noindent where the function $h$ is real, while $f$ and $g$ are holomorphic, and where we employed the definition
\begin{align}
    \Phi=\mathcal{A}+i\mathcal{B} && \longleftrightarrow && \mathcal{A}=\frac12\(\Phi+\bar{\Phi}\) \ , && \mathcal{B}=\frac{1}{2i}\(\Phi-\bar{\Phi}\) \ ,
\end{align}

\noindent Then, making use of the general formula \eqref{eqn:ch1_chiral_K_W_lag}, one obtains the following off-shell lagrangian
\begin{equation}
    \begin{aligned}
    \mathcal{L}=&\, \abs{F_S}^2+\frac{h}{2}\partial_\mu A\partial^\mu A-\left(\frac{i}{2}\partial_\mu G \sigma^\mu\bar{G}+h.c.\right)+\frac{1}{4}\frac{{\bar G}^2}{\bar{F}_S}\Box\left(\frac{G^2}{F_S}\right)\\
     &+\frac{h}{2}\left[i\, \partial^\mu\left(\frac{G}{F_S}\right)\sigma^\nu\left(\frac{\bar{G}}{\bar{F}_S}\right)\partial_\mu A\partial_\nu A-\frac{i}{2}\partial_\mu\left(\frac{G}{F_S}\right)\sigma^\mu\left(\frac{\bar{G}}{\bar{F}_S}\right)(\partial A)^2+h.c.\right]\\
     &+\left[fF_S+\frac{i}{2}f'\frac{G\sigma^\mu\bar{G}}{\bar{F}_S}\partial_\mu A+\frac{g''}{2}\left(\frac{\bar{G}}{\bar{F}_S}\right)^2(\partial A)^2+\frac{g'}{2}\left(\frac{\bar{G}}{\bar{F}_S}\right)^2\Box A\right.\\
    &\left.\qquad-g'\partial_\nu\left(\frac{\bar{G}}{\bar{F}_S}\right)\bar{\sigma}^\mu\sigma^\nu\frac{\bar{G}}{\bar{F}_S}\partial_\mu A +h.c.\right].
    \end{aligned}
\label{eqn:ch1_oc_lag_off}
\end{equation}

Next, one has to integrate out the goldstino auxiliary field $F_S$. The solution to its equations of motion can be found iteratively, expanding in the number of fermions. Truncating such solution to the quadratic order in goldstinos --- which corresponds to quartic operators in the lagrangian --- one finds the on-shell spacetime lagrangian \cite{Bonnefoy:2022rcw}
\begin{equation}
    \begin{aligned}
    \mathcal{L}=&-\abs{f}^2-\frac{h}{2}\partial_\mu A\partial^\mu A+\left(\frac{i}{2}\partial_\mu G \sigma^\mu\bar{G}+h.c.\right)-\(i\frac{f'}{2f}+h.c.\)G\sigma^\mu\bar{G}\partial_\mu A\\
    &+\frac{h}{2\abs{f}^2}\left[i\,\partial^\mu G \sigma^\nu\bar{G}\partial_\mu A\partial_\nu A-\frac{i}{2}\partial_\mu G \sigma^\mu\bar{G}\partial_\nu A\partial^\nu A+h.c.\right]\\
    &-\left[\frac{\bar{g}'}{\bar{f}^2}\left(\frac{1}{2} G^2\Box A-\partial_\mu G \sigma^\mu\bar{\sigma}^\nu G \partial_\nu A\right)+\xi\, G^2\,\partial_\mu A\partial^\mu A+h.c.\right] \\
    &+\frac{1}{4\abs{f}^2}\bar{G}^2\Box G^2\ ,
    \end{aligned}
\label{eqn:ch1_oc_lag_1}
\end{equation}

\noindent in which
\begin{equation}
    \xi\equiv \frac{\bar{g}''}{2\bar{f}^2}-\frac{\bar{g}'\bar{f}'}{\bar{f}^3} \ .
\label{eqn:ch1_oc_xi_def}
\end{equation}

\noindent The lagrangian \eqref{eqn:ch1_oc_lag_1} captures exactly all the terms quadratic in the goldstino. An alternative form of this lagrangian is obtained performing the following field redefinitions:
\begin{align}
     \delta G^\alpha =\frac{h}{4\abs{f}^2}(\partial A)^2\,G^\alpha \ , && \delta G^\alpha=-i\frac{g'}{f^2}(\bar{G}\bar{\sigma}^\mu)^\alpha\,\partial_\mu A \ , && \delta A=\frac{\bar{g}'}{2h\bar{f}^2}G^2+h.c \ ,
\end{align}

\noindent which in the quartic limit yield
\begin{equation}
    \begin{aligned}
        \mathcal{L}=&-\abs{f}^2-\frac{h}{2}\partial_\mu A\partial^\mu A+\(\frac{i}{2}\partial_\mu G \sigma^\mu\bar{G}+h.c.\)+\(mG^2+h.c.\)+\lambda G^2\bar{G}^2\\
        &-\(i\frac{f'}{2f}G\sigma^\mu\bar{G}\partial_\mu A+\tilde{\xi}G^2\partial_\mu A\partial^\mu A+h.c.\)-\frac{1}{8h}\left(\frac{\bar g'^2}{\bar f^4}G^2\Box G^2+h.c.\right)\\
        &+\frac{1}{2\abs{f}^2}\left(h-\frac{2\abs{g'}^2}{\abs{f}^2}\right)\left(i\,\partial^\mu G \sigma^\nu\bar{G}\partial_\mu A\partial_\nu A+h.c.\right)\\
        &+\frac{1}{4\abs{f}^2}\left(1-\frac{\abs{g'}^2}{h\abs{f}^2}\right)\bar{G}^2 \Box G^2 \ ,
    \end{aligned}
\label{eqn:ch1_oc_lag_2}
\end{equation}

\noindent where the coefficients $m$, $\lambda$ and $\tilde{\xi}$ are short-hand notations for
\begin{align}
    m=\frac{g'f'}{h f} \ , && \lambda=\frac{\(f'{}^2+ff''\)g'{}^2}{2h^2f^4} \ , && \tilde{\xi}=\xi-\frac{g'h'}{4hf^2} \ .
\label{eqn:ch1_oc_coeff_l2}
\end{align}

\noindent In this field basis, the dependence on the quadratic goldstino operators is not exact anymore, but it makes manifest some peculiar quartic interactions, which will be subject to positivity bounds.

\subsubsection{Causality constraints}

The lagrangians \eqref{eqn:ch1_oc_lag_1}-\eqref{eqn:ch1_oc_lag_2} involve peculiar derivative self-interactions of the goldstino $G$. This is directly related to the fact that the orthogonal constraint \eqref{eqn:ch1_oc} implies a nontrivial solution for the auxiliary field $F_\phi$, as in \eqref{eqn:ch1_oc_sol}. Focusing on the quartic-limit lagrangian \eqref{eqn:ch1_oc_lag_2}, such operators are directly related to nontrivial positivity bounds on the scattering amplitudes \cite{Adams:2006sv,Bellazzini:2016xrt,Dine:2009sw}. The relevant $2\to2$ scattering amplitudes, twice subtracted and in the forward limit, are
\beq
    \begin{aligned}
        \mathcal{A}(G,\bar G\to G,\bar G)&=
        \frac{h\abs{f}^2-\abs{g'}^2}{h\abs{f}^4}s^2 \\
    \mathcal{A}(G, G\to \bar G, \bar G)&=
    \frac{2\bar g'^2}{h\bar f^4}
    s^2 \ , \\
    \mathcal{A}(G,A\to G,A)&=
    \frac{h\abs{f}^2-2\abs{g'}^2}{2\abs{f}^4}s^2 \ ,
    \end{aligned}
\label{eqn:ch1_oc_amp}
\eeq

\noindent and they are found, following \cite{Bellazzini:2016xrt,Dine:2009sw,Bellazzini:2014waa,Trott:2020ebl}, to be subject to the following positivity bounds:
\beq
\begin{aligned}
    \frac{d^2}{ds^2}\mathcal{A}\(G,\bar G\to G,\bar G\)_{\big|_{s=t=0}}&\ge 0 &&\iff&& h\abs{f}^2\ge \abs{g'}^2 \ , \\
    \frac{d^2}{ds^2}\mathcal{A}\(G,A\to G,A\)_{\big|_{s=t=0}}&\ge 0 &&\iff&&  h\abs{f}^2\ge 2\abs{g'}^2 \ , \\
    \frac{d^2}{ds^2}\[\mathcal{A}\(G,\bar G\to G,\bar G\)-\frac12\abs{\mathcal{A}\(G,G\to \bar G,\bar G\)}\]_{\big|_{s=t=0}}&\ge0 && \iff && h\abs{f}^2\ge 2\abs{g'}^2 \ ,
\end{aligned}
\label{eqn:ch1_oc_pb_all}
\eeq

\noindent which thus yield \cite{Bonnefoy:2022rcw}
\beq
    h\(0\)\abs{f\(0\)}^2\ge 2\abs{g'\(0\)}^2 \ .
\label{eqn:ch1_oc_pb_fin}
\eeq

\noindent The positivity bounds \eqref{eqn:ch1_oc_pb_all}-\eqref{eqn:ch1_oc_pb_fin} actually hold in the vacuum expectation value for the scalar field $A$, which can be taken to be $A=0$ in full generality. The higher-order terms in the expansions around such vacuum are actually operators of higher-order in the number of fields and do not contribute to the amplitudes of eq.~\eqref{eqn:ch1_oc_amp}.

However, we argue in \cite{Bonnefoy:2022rcw} that the vacuum positivity bound \eqref{eqn:ch1_oc_pb_fin} must nevertheless be extended to the whole functions domain in order to have a causal propagation of the goldstino $G$. This is related to the sound speed parameter, which parameterize the propagation of the goldstino and in presence of non-minimal couplings and a nontrivial background can develop peculiar features like the possibility to vanish and also to become superluminal \cite{Kolb:2021xfn,Dudas:2021njv,Bonnefoy:2022rcw}. This property was actually studied for the massive gravitino in spontaneously broken supergravity \cite{Benakli:2014bpa,Kahn:2015mla,Hasegawa:2017hgd,Kolb:2021xfn,Kolb:2021nob,Terada:2021rtp,Dudas:2021njv,Antoniadis:2021jtg}, but it directly relates to the goldstino in the rigid limit by means of the equivalence theorem \cite{Fayet:1986zc,Casalbuoni:1988kv,Casalbuoni:1988qd}. In the case at hand, the sound speed can be computed exactly from the quadratic-limit lagrangian \eqref{eqn:ch1_oc_lag_1}, since it contains all the operators quadratic in the goldstino, and assuming an homogeneous background solution for the scalar field $A=A\(t\)$.\footnote{Derivatives with respect to the time $t$ will be denoted with a dot, \textit{i.e.} $\partial_0 A\equiv\dot{A}$.} 

Working in momentum space, parameterized for the goldstino by $k^\mu=\(k_0,0,0,k_3\)$, the quadratic equation of motion for $G$ that follows from the lagrangian \eqref{eqn:ch1_oc_lag_1} is found to be \cite{Bonnefoy:2022rcw}
\beq
    \left\{\(1+\frac{h\dot A{}^2}{2\abs{f}^2}\)^2k_0^2  -\[\(1-\frac{h\dot A{}^2}{2\abs{f}^2}\)^2+4 \dot A{}^2\frac{\abs{g'}^2}{\abs{f}^4}\]k_3^3\right\} G_\alpha = 0 \ ,
\label{eqn:ch1_oc_G_eom}
\eeq

\noindent which yields the time-dependent sound speed
\beq
    c_s^2= 1 - \frac{4\dot A{}^2}{\(\abs{f(A)}^2+\frac{h(A)\dot A{}^2}{2}\)^2} \(\frac{h(A)}{2}\abs{f(A)}^2 -\abs{g'(A)}^2\) \ .
\label{eqn:ch1_oc_sound_speed}
\eeq

\noindent Therefore, in order to have subluminal goldstino propagation, \textit{i.e.} $c_s^2\le 1$, at all times one must require
\beq
     h\(A\)\abs{f\(A\)}^2\ge 2\abs{g'\(A\)}^2 \ ,
\label{eqn:ch1_oc_sound_speed_bound}
\eeq

\noindent which is indeed the extension of the positivity bound \eqref{eqn:ch1_oc_pb_fin} to all field space. This result is coherent with \cite{Dine:2009sw,Kahn:2015mla} and with the longitudinal gravitino dynamics in supergravity \cite{Kolb:2021nob,Kolb:2021xfn,Dudas:2021njv}.

The causality bound \eqref{eqn:ch1_oc_sound_speed_bound} implies that the theory \eqref{eqn:ch1_oc_KW}, despite being the most general one with respect to the nonlinear supersymmetry setup defined by the orthogonal constraint \eqref{eqn:ch1_oc}, is not automatically well-defined. From the low-energy effective field theory point of view, we interpret this as an obstruction to a completion into a microscopic 2-derivative theory, which instead requires higher-derivative operators already in the UV \cite{DallAgata:2016syy}. We will substantiate this claim in Section \ref{sub_ch1_imp_const}.

\subsection{Coupling to a complex scalar}

The second model we examine, very similar to the orthogonal constraint one, involves again a chiral superfield
\beq
    \mathcal{H}=H+\sqrt{2}\theta\chi_H+\theta^2F_H \ ,
\eeq

\noindent subject now to the constraint
\beq
    S \overline{\mathcal{H}}=\text{chiral} \ .
\label{eqn:ch1_csc}
\eeq

\noindent This constraint is solved by \cite{Komargodski:2009rz}
\beq
\begin{aligned}
    \chi_H&=-i\sigma^\mu\frac{\bar{G}}{\bar{F}_S}\partial_\mu H \ , \\
    F_H&=-\partial_\nu\left(\frac{\bar{G}}{\bar{F}_S}\right)\bar{\sigma}^\mu\sigma^\nu \frac{\bar{G}}{\bar{F}_S} \partial_\mu H+\frac{1}{2}\left(\frac{\bar{G}}{\bar{F}_S}\right)^2\Box H \ ,
\end{aligned}
\label{eqn:ch1_csc_sol}
\eeq

\noindent and thus describes the coupling of the goldstino to a complex scalar $H$. The most general K\"ahler and superpotential for such constrained superfields are\footnote{The goldstino kinetic term in the K\"ahler potential \eqref{eqn:ch1_csc_KW} is canonical since any non-minimal coefficient $\xi\(\H,\overline{\H}\)$ can actually be reabsorbed into the goldstino superfield as $S\to\xi^{-\frac12}S$, which preserves the chiral nature of $S$ because of the constraints \eqref{eqn:ch1_G_nil_sf} and \eqref{eqn:ch1_csc} \cite{DallAgata:2015zxp}. The whole dependence on $\xi$ is then redefined away up to terms in the lagrangian which are beyond the quartic order in the fields discussed here.}
\begin{align}
    K=\kappa\(\H,\overline{\H}\) +\bar{S}S  \ , && W\(\H\)=f\(\H\)S+g\(\H\) \ ,
\label{eqn:ch1_csc_KW}
\end{align}

\noindent where $f$ and $g$ are holomorphic functions. Proceeding then in the same way as for the orthogonal constraint, the on-shell lagrangian is determined by \eqref{eqn:ch1_csc_KW} is \cite{Bonnefoy:2022rcw}
\begin{equation}
\begin{aligned}
	\mathcal{L}=&-\abs{f
	}^2+\left(\frac{i}{2}\partial_\mu G \sigma^\mu {\bar G}+h.c.\right)-\kappa_{H\bar H}\partial_\mu H\partial^\mu {\bar H}\\
	&+\frac{1}{4\abs{f}^2} {\bar G}^2 \Box G^2+\frac{\kappa_{H \bar H}}{2\abs{f}^2}\left(i\partial^\mu G \sigma^\nu {\bar G}+h.c.\right)\left(\partial_\mu H\partial_\nu {\bar H} +\partial_\nu H \partial_\mu {\bar H} \right)\\
	&-\frac{\kappa_{H\bar{H}}}{2\abs{f}^2}\(i\partial_\mu G\sigma^\mu\bar{G}+h.c.\)\partial_\nu H\partial^\nu\bar{H}-\left[i\frac{f
	_H}{f} G \sigma^\mu {\bar G} \partial_\mu H+h.c.\right]\\
	&+\frac{\kappa_{H \bar H}}{\abs{f}^2}\[\frac{1}{2}G \sigma^\mu {\bar G} \left(i\,\partial_\mu H\Box {\bar H} +h.c.\right)-\epsilon^{\mu\nu\alpha\beta}\partial_\mu G \sigma_\nu {\bar G}  \partial_\alpha H\partial_\beta {\bar H}\]\\
	&+\left[\frac{g'}{f^2} {\bar G} \bar{\sigma}^\mu\sigma^\nu\partial_\nu {\bar G}  \partial_\mu H -\frac{g''}{2f^2} {\bar G}^2(\partial H)^2-\frac{g'}{2f^2} {\bar G}^2\Box H+h.c.\right] \ .
\end{aligned}
\label{eqn:ch1_csc_lag_1}
\end{equation}

\noindent This is the lagrangian in what we refer to as the quadratic limit, \textit{i.e.} it is exact in the operators quadratic in the goldstino $G$, as the orthogonal constraint lagrangian \eqref{eqn:ch1_oc_lag_1}. One can switch to the quartic-limit basis through the field redefinitions
\begin{equation}
\begin{aligned}
    \delta G^\alpha&=-\frac{\kappa_{H\bar{H}}}{2 \abs{f}^2}\partial_\mu H\partial^\mu\bar{H}G^\alpha \ ,  &&&&& \delta H&=-\frac{i}{2\abs{f}^2}G\sigma^\mu\bar{G}\partial_\mu H \ , \\
    \delta G^\alpha&= i\frac{g'}{f^2}({\bar G} \bar{\sigma}^\mu)^\alpha\,\partial_\mu H \ , &&&&& \delta H&=\frac{\bar{g}'}{2\kappa_{H \bar H}\bar{f}^2} G^2 \ ,
\end{aligned}
\end{equation}

\noindent which yield
\begin{equation}
\begin{aligned}
	\mathcal{L}=&-\abs{f}^2-\kappa_{H \bar H}\partial_\mu H\partial^\mu {\bar H} +\left(\frac{i}{2}\partial_\mu G \sigma^\mu {\bar G} +h.c.\right)-\left(m_1\, G^2+h.c.\right)\\
	&+2\left[\bar{m}_1({\bar k} {\bar H}-kH) {\bar G}^2+h.c.\right]+\frac{1}{4\abs{f}^2}\left(1-\frac{\abs{g'}^2}{\abs{f}^2\kappa_{H\bar H}}\right) {\bar G}^2\Box G^2\\
	&+\frac{1}{2}\left(k^2H^2+{\bar k}^2 {\bar H}^2-2\abs{k}^2\abs{H}^2\right)\left(m_1 G^2+\bar{m}_1 {\bar G}^2\right)-m_2\,G^2 {\bar G}^2
	\\
	&+\frac{1}{2\abs{f}^2}\left(\kappa_{H \bar H}+\frac{\abs{g'}^2}{\abs{f}^2}\right)\(i\partial^\mu G \sigma^\nu\bar{G}\partial_\mu H\partial_\nu {\bar H}+h.c.\)\\
	&+\frac{1}{2\abs{f}^2}\left(\kappa_{H \bar H}-3\frac{\abs{g'}^2}{\abs{f}^2}\right)\(i\partial^\mu G\sigma^\nu\bar{G}\partial_\mu {\bar H} \partial_\nu H+h.c.\)\\
	&-\left[\lambda_1 G^2\partial_\mu H\partial^\mu {\bar H}+\lambda_2 {\bar G}^2(\partial H)^2+h.c.\right]\\
    &-\[\frac{i\(2f_H\bar f-\partial_H\abs{f}^2\)}{2\abs{f}^2}G\sigma^\mu {\bar G}\partial_\mu H+h.c. \] \\
	&-\frac{1}{\abs{f}^2}\left(\kappa_{H \bar H}+\frac{\abs{g'}^2}{\abs{f}^2}\right)\epsilon^{\mu\nu\alpha\beta}\partial_\mu G \sigma_\nu {\bar G} \partial_\alpha H\partial_\beta {\bar H},
\end{aligned}
\label{eqn:ch1_csc_lag_2}
\end{equation}

\noindent where $m_1$, $m_2$, $\lambda_1$ and $\lambda_2$ are again functions of the starting ones $f$ and $g$, similar to eq.~\eqref{eqn:ch1_oc_coeff_l2}, whose expression is not relevant.

Therefore, also in the case of the constraint \eqref{eqn:ch1_csc} the resulting lagrangians \eqref{eqn:ch1_csc_lag_1}-\eqref{eqn:ch1_csc_lag_2} are characterised by peculiar quartic, derivative self-couplings of the goldstino. Their source is, as for the orthogonal constraint \eqref{eqn:ch1_oc}-\eqref{eqn:ch1_oc_sol}, the nontrivial condition imposed by the constraint \eqref{eqn:ch1_csc} on the auxiliary field $F_H$ in eq.~\eqref{eqn:ch1_csc_sol}.

\subsubsection{Causality constraints}

Analogously to the orthogonal constraint \eqref{eqn:ch1_oc} and the discussion of Section \ref{sub_ch1_oc}, the resulting self-interactions of the goldstino are subject to positivity bounds \cite{Bonnefoy:2022rcw}. The $2\to2$ scattering amplitudes, computed from the quartic-limit lagrangian \eqref{eqn:ch1_csc_lag_2}, that are nontrivially bounded are the elastic ones, involving the real part of $H$, which all yield the bound
\beq
    \kappa_{H,\bar H}\(0\)\abs{f\(0\)}^2\ge\abs{g'\(0\)} \ .
\label{eqn:ch1_csc_pb}
\eeq

As in eq.~\eqref{eqn:ch1_oc_pb_fin}, this positivity bound holds in the vacuum $H=0$, but it is extended to the whole fields domain by requiring a causal goldstino propagation. The sound speed parameter is in fact computed from the quadratic-limit lagrangian \eqref{eqn:ch1_csc_lag_1} to be
\beq
    c_s^2=1-\frac{4\abs{\dot H}^2}{\(\abs{f}^2+\kappa_{H\bar{H}}\abs{\dot H}^2\)^2}\[\kappa_{H\bar{H}}\abs{f}^2-\abs{g'}^2\] \ ,
\label{eqn:ch1_csc_sound_speed}
\eeq

\noindent so that one must have
\beq
    \kappa_{H,\bar H}\(H,\bar H\)\abs{f\(H\)}^2\ge\abs{g'\(H\)} \ ,
\label{eqn:ch1_csc_sound_speed_bound}
\eeq

\noindent in order to have subluminal propagation. The sound speed \eqref{eqn:ch1_csc_sound_speed} matches again its supergravity version in the rigid limit\footnote{In this case, the sound speed \eqref{eqn:ch1_csc_sound_speed} differs from the supergravity one by $1/M_\text{P}^2$-suppressed terms. These gravitational corrections are still consistent with causality, which should be either be satisfied or be unresolvable \cite{Alberte:2020jsk,Tokuda:2020mlf,Alberte:2020bdz,Caron-Huot:2021rmr,Arkani-Hamed:2021ajd}.} \cite{Bonnefoy:2022rcw}.

The appearance of such nontrivial causality constraints \eqref{eqn:ch1_csc_pb}-\eqref{eqn:ch1_csc_sound_speed_bound} suggest that, like the orthogonal constraint \eqref{eqn:ch1_oc}, also the complex scalar constrain \eqref{eqn:ch1_csc} does not have a well-defined 2-derivative UV origin. This is linked to the solution imposed on the auxiliary fields by both constraints \eqref{eqn:ch1_oc} and \eqref{eqn:ch1_csc}, which has no obvious microscopic interpretation in terms of decoupling of physical degrees of freedom in a two-derivative lagrangian.

\subsection{Consistent superfield constraints}\label{sub_ch1_imp_const}

The orthogonal constraint \eqref{eqn:ch1_oc} and the complex field constraint \eqref{eqn:ch1_csc}, despite being consistent with (nonlinear) supersymmetry, as well as from the algebraic point of view, give rise to couplings of the goldstino that are subject to nontrivial causality bounds \eqref{eqn:ch1_oc_sound_speed_bound}-\eqref{eqn:ch1_csc_sound_speed_bound}. These bounds prevent an uplifting of the associated effective theories into 2-derivative UV completions. One of the main aspects of \cite{Bonnefoy:2022rcw} --- which we now discuss more in detail --- is that the constraints generating these effective constructions have the subtle feature of imposing a set of conditions on the auxiliary fields, as in eq.~\eqref{eqn:ch1_oc_sol}-\eqref{eqn:ch1_csc_sol}. These may in fact require higher-derivative operators directly at the microscopic level, as already suggested in \cite{DallAgata:2016syy} for the orthogonal constraint \eqref{eqn:ch1_oc}. 

In this section we consider superfield constraints which do not affect the auxiliary fields and show that they are indeed causal throughout the entire field space. The first direct example in this sense is the nilpotent constraint \eqref{eqn:ch1_G_nil_sf} defining the goldstino superfield itself. Indeed, such constraint does not act on the auxiliary field (see eq.~\eqref{eqn:ch1_nil_sf_sol}) and the associated lagrangian is the Volkov--Akulov one \eqref{eqn:ch1_VA_lag} \cite{Volkov:1973ix,Rocek:1978nb,Ivanov:1978mx,Komargodski:2009rz,Kuzenko:2010ef}, which is perfectly causal and has a clear UV origin. 

A second model, more similar to the two constraints \eqref{eqn:ch1_oc} and \eqref{eqn:ch1_csc} but not acting on the auxiliary field, is given by
\beq
    S\Q=0 \ ,
\label{eqn:ch1_fc}
\eeq

\noindent where $\Q$ is the chiral superfield
\beq
    \Q=Q+\sqrt{2}\theta\chi_Q+\theta^2 F_Q \ .
\eeq   

\noindent The constraint \eqref{eqn:ch1_fc}, which also implies $\Q^3=0$ like the orthogonal constraint \eqref{eqn:ch1_oc}, fixes the complex scalar $Q$ to be \cite{Komargodski:2009rz}
\beq
    Q=\frac{\chi_Q G}{F_S}-\frac{F_Q}{2F_S^2} G^2 \ .
\eeq

\noindent and thus describes a coupling between the goldstino $G$ and a second fermion $\chi_Q$. The K\"ahler and superpotential are taken to be \cite{Bonnefoy:2022rcw}
\begin{align}
    K=S\bar{S}+\Q\bar{\Q}+\frac{1}{2}\Q\bar{\Q}\left(h_1\Q+\bar{h}_1\bar{\Q}\right)+\frac{h_2}{4}\left(\Q\bar{\Q}\right)^2 \ , && W=f(\Q)S+g(\Q) \ ,
\label{eqn:ch1_fc_KW}
\end{align}

\noindent in which $f$ and $g$ are holomorphic functions, as in the previous cases, while $h_1$ and $h_2$ are real constants. In the resulting off-shell lagrangian, the auxiliary field $F_Q$, being unconstrained, has to be integrated out together with $F_S$, giving
\begin{equation}
\begin{aligned}
	\mathcal{L}=&\left(\frac{i}{2}\partial_\mu G \sigma^\mu\bar{G}+h.c.\right)+\left(\frac{i}{2}\partial_\mu\chi_q\sigma^\mu\bchi_q+h.c.\right)+\frac{1}{2\abs{f}^2}\left[\frac{g'}{f}\partial_\mu\left(\chi_q G \right)\partial^\mu\bar{G}^2+h.c.\right]+\\
	&+\frac{1}{\abs{f}^2}\left(\bchi_q\bar{G}\right)\Box\left(\chi_q G \right)+\frac{1}{4}\abs{f}^2\left(1+\frac{\abs{g'}^2}{\abs{f}^2}\right)\bar{G}^2\Box G ^2+\dots \ ,
	\end{aligned}
\label{eqn:ch1_fc_lag}
\end{equation}

\noindent where we omitted four-fermion contact terms, which are not relevant for the present discussion. 

The lagrangian \eqref{eqn:ch1_fc_lag} does involve nontrivial quartic, derivative interactions of the goldstino and the fermion $\chi_Q$ but these operators satisfy automatically the associated positivity bounds \cite{Bonnefoy:2022rcw,Bellazzini:2016xrt}, namely without the need to impose any extra condition on the general model \eqref{eqn:ch1_fc_KW}. Decoupling a scalar field has a well-defined physical meaning and can indeed be done from a 2-derivative UV theory. The outcome is a well-defined effective lagrangian \eqref{eqn:ch1_fc_lag}, \textit{i.e.} that does not suffer from the causality issues that arise when also an auxiliary field gets constrained in this procedure.  

\subsubsection{Improved, generalized constraints}

The strongest evidence supporting the claim on constrained auxiliary fields as sources for the potential inconsistency of the goldstino effective theories is the construction of an improved version of the orthogonal constraint \eqref{eqn:ch1_oc}, which maintains the minimal field spectrum but without constraining the auxiliary field. This is done in \cite{Bonnefoy:2022rcw} by means of the generalized superfield constraints derived in \cite{DallAgata:2016syy}. This generalized constraint has the form
\beq
    \bar{S}SQ_L=0 \ ,
\label{eqn:ch1_gsc}
\eeq

\noindent where $S$ is the goldstino nilpotent superfield \eqref{eqn:ch1_G_nil_sf} and $Q_L$ is a generic superfield which may carry a Lorentz index $L$. This constraint has the implicit solution
\beq
    Q_\textup{L}=-2\frac{\bar{D}_{\dot{\alpha}}\bar{S}\bar{D}^{\dot{\alpha}}Q_\textup{L}}{\bar{D}^2\bar{S}}-\frac{\bar{S}\bar{D}^2Q_\textup{L}}{\bar{D}^2\bar{S}}-2\frac{D^\alpha SD_\alpha\bar{D}^2\left(\bar{S}Q_\textup{L}\right)}{D^2S\bar{D}^2\bar{S}}-S\frac{D^2\bar{D}^2\left(\bar{S}Q_\textup{L}\right)}{D^2S\bar{D}^2\bar{S}} \ ,
\label{eqn:ch1_gsc_sol}
\eeq

\noindent which is nontrivial only on the lowest component of $Q_L$ \cite{DallAgata:2016syy}. 

The power of such constraint is that it can be used to remove single components of the superfields it acts on. It follows in particular that standard constraints can be expressed in terms of combinations of such generalized constraint \eqref{eqn:ch1_gsc}. This is the case of the orthogonal constraint \eqref{eqn:ch1_oc}, which was shown in \cite{DallAgata:2016syy} to be equivalent to the set of three constraints
\begin{align}
    \bar{S}S\(\Phi-\bar{\Phi}\)&=0 \ , \label{eqn:ch1_gsc_oc_sc}\\
    \bar{S}SD_\alpha \Phi &= 0 \ , \label{eqn:ch1_gsc_oc_f}\\
    \bar{S}S D^2\Phi &=0 \ , \label{eqn:ch1_gsc_oc_aux}
\end{align}

\noindent each one removing, respectively, the imaginary scalar, the fermion, and the auxiliary component of $\Phi$, as in eq.~\eqref{eqn:ch1_oc_sol}. Note in particular that the condition \eqref{eqn:ch1_gsc_oc_aux} removing the auxiliary field is a higher-derivative constraint.

Thus, following the previous discussion about ill-defined goldstino couplings and constraints on the auxiliary fields, the decomposition above suggest considering the effective theory generated only by the first two constraints \eqref{eqn:ch1_gsc_oc_sc}-\eqref{eqn:ch1_gsc_oc_f}
\begin{align}
    \bar{S}S\(\Phi-\bar{\Phi}\)=0 \ , &&    \bar{S}SD_\alpha \Phi= 0 \ ,
\label{eqn:ch1_gsc_oc_imp}
\end{align}

\noindent without the third one \eqref{eqn:ch1_gsc_oc_aux}: the resulting model will be an improved version of the orthogonal constraint \eqref{eqn:ch1_oc}, with the same minimal field spectrum but without touching the auxiliary field \cite{Bonnefoy:2022rcw}. The two constraints of eq.~\eqref{eqn:ch1_gsc_oc_imp} bound the imaginary scalar $B$ and the fermion $\chi_\phi$ of the chiral superfield $\Phi$ \eqref{eqn:ch1_oc_phi_comp} to be
\begin{align}
    B&=\frac{i}{4}\[\(\frac{\bar{G}}{\bar{F}_S}\)^2\bar{F}_\phi-\(\frac{G}{F_S}\)^2F_\phi\]-\frac{G\sigma^\mu\bar{G}}{2|F_S|^2}\partial_\mu A +\dots \ ,\\
    \psi_\alpha&=\frac{F_\phi}{F_S}G_\alpha-i\frac{\(\sigma^\mu\bar{G}\)_\alpha}{\bar{F}_S}\partial_\mu A+\frac{i}{4}\frac{\(\sigma^\mu\bar{G}\)_\alpha}{\bar{F}_S}\partial_\mu\[\(\frac{G}{F_S}\)^2F_\phi\]+\dots \ ,
\end{align}

\noindent in which we neglect terms higher-order in the number of fields with respect to the quartic one under consideration. The most general K\"ahler and superpotential that can be defined in this improved setup are
\begin{align}
    K=\bar{S}S+\kappa\(\Phi,\bar{\Phi}\) \ , && W\(\Phi\)=f\(\Phi\)S+g\(\Phi\) \ , 
\label{eqn:ch1_gsc_KW}
\end{align}

\noindent where the function $\kappa$ is real and $f$ and $g$ are holomorphic. The associated on-shell lagrangian in the quadratic limit, obtained integrating out both $F_S$ and $F_\phi$ in the standard way, is equal to
\begin{equation}
    \begin{aligned}
        \mathcal{L}=&-\[\abs{f}^2+\frac{\abs{g'}^2}{\kappa_{\phi\bar{\phi}}}\]-\kappa_{\phi\bar{\phi}}\partial_\mu A\partial^\mu A+Z\(\frac{i}{2}\partial_\mu G\sigma^\mu\bar{G}+h.c.\)+\(mG^2+h.c.\)\\
        &+c\,G^2\bar{G}^2-Z\(i\frac{f'}{2f}+h.c.\) G\sigma^\mu\bar{G}\partial_\mu A+\(d\,G^2+h.c.\)\(\partial A\)^2\\
        &+\[\frac{\bar{g}'}{\bar{f}^2}\partial_\mu G\sigma^\mu\bar{\sigma}^\nu G\partial_\nu A+\frac{\kappa_{\phi\bar{\phi}}}{\abs{f}^2}i\partial^\mu G\sigma^\nu\bar{G}\partial_\mu A\partial_\nu A-\frac{\kappa_{\phi\bar{\phi}}}{2\abs{f}^2}i\partial_\mu G\sigma^\mu\bar{G}\(\partial A\)^2+h.c.\] \\
        &+\frac{1}{4\abs{f}^2}\(1+\frac{\abs{g'}^2}{2\kappa_{\phi\bar{\phi}}\abs{f}^2}\)\bar{G}^2\Box G^2-\frac{1}{16\kappa_{\phi\bar{\phi}}}\(\frac{\bar{g}'^2}{\bar{f}^4}G^2\Box G^2+h.c.\)\ ,
    \end{aligned}
\label{eqn:ch1_gsc_lag_1}
\end{equation}

\noindent where 
\begin{equation}
    Z\equiv 1+\frac{\abs{g'}^2}{\kappa_{\phi\bar{\phi}}\abs{f}^2} \ ,
\label{eqn:ch1_gsc_Z_coeff}
\end{equation}

\noindent while the coefficients $m$, $c$ and $d$ are functions of the original $\kappa$, $f$ and $g$ of which we do not need the explicit form. The main differences with the lagrangian \eqref{eqn:ch1_oc_lag_1} produced by the orthogonal constraint are the scalar potential, which now includes also the term coming from $F_\phi$, and the non-minimal goldstino kinetic term, parameterized in this case by the function $Z$ \eqref{eqn:ch1_gsc_Z_coeff}. Instead, the quartic limit lagrangian is obtained through the field redefinitions
\begin{align}
    \delta G^\alpha=\frac{\kappa_{\phi\bar{\phi}}}{2Z\abs{f}^2}G^\alpha\(\partial A\)^2 \ , && \delta G^\alpha= -\frac{i}{Z}\frac{g'}{f^2}\(\bar{G}\bar{\sigma}^\mu\)^\alpha\partial_\mu A \ ,
\end{align}

\noindent which yield
\begin{equation}
    \begin{aligned}
        \mathcal{L}=&-\[\abs{f}^2+\frac{\abs{g'}^2}{\kappa_{\phi\bar{\phi}}}\]-\kappa_{\phi\bar{\phi}}\partial_\mu A\partial^\mu A+Z\(\frac{i}{2}\partial_\mu G\sigma^\mu\bar{G}+h.c.\)\\
        &+\(mG^2+h.c.\)+\lambda G\sigma^\mu\bar{G}\partial_\mu A+c\,G^2\bar{G}^2+\(\tilde{d}\,G^2+h.c.\)\partial_\mu A\partial^\mu A\\
        &+\(\frac{\kappa_{\phi\bar{\phi}}}{\abs{f}^2+\frac{\abs{g'}^2}{\kappa_{\phi\bar{\phi}}}}\)\(i\partial^\mu G\sigma^\nu\bar{G}\partial_\mu A\partial_\nu A+h.c.\)+\frac{1}{4\abs{f}^2}\(1+\frac{\abs{g'}^2}{2\kappa_{\phi\bar{\phi}}\abs{f}^2}\)\bar{G}^2\Box G^2 \\
        &-\frac{1}{16\kappa_{\phi\bar{\phi}}}\(\frac{\bar{g}'^2}{\bar{f}^4}G^2\Box G^2+h.c.\)\ ,
    \end{aligned}
\label{eqn:ch1_gsc_lag_2}
\end{equation}

\noindent where, again, we do not need to specify the explicit expression of the functions $\lambda$ and $\tilde{d}$.

Proceeding in the same way as for the orthogonal constraint in Section \ref{sub_ch1_oc}, we now analyze the goldstino propagation and the causality of the effective theory defined by the constraints \eqref{eqn:ch1_gsc_oc_imp}. It is easy to see that the potentially dangerous operators appearing in the quartic-limit lagrangian \eqref{eqn:ch1_gsc_lag_2} automatically meet the associated positivity requirements. They constrain in fact the same scattering amplitudes as in eq.~\eqref{eqn:ch1_oc_amp}, but the improved setup is such that these conditions are now identically satisfied by the new couplings. Furthermore, we can also compute the goldstino sound speed parameter from the quadratic-limit lagrangian \eqref{eqn:ch1_gsc_lag_1}, evaluated on a time-dependent scalar field background $A(t)$. The new superfield constraints setup thus yield 
\beq
    c_s^2=1-\frac{4{\dot A}^2\abs{f}^2\kappa_{\phi\bar{\phi}}}{\(\abs{f}^2+\frac{\abs{g'}^2}{\kappa_{\phi\bar{\phi}}}+\kappa_{\phi\bar{\phi}}{\dot A}^2\)^2} \ ,
\label{eqn:ch1_gsc_sound_speed}
\eeq

\noindent which is subluminal at all times \cite{Bonnefoy:2022rcw}.

Therefore, improving the orthogonal constraint \eqref{eqn:ch1_oc} by imposing only the conditions \eqref{eqn:ch1_gsc_oc_imp}, which have a standard physical interpretation in terms of decoupling of degrees of freedom, and without touching the auxiliary fields, yields a completely causal effective field theory, with the same minimal field content of the orthogonal constraint \eqref{eqn:ch1_oc} but resolving all its inconsistencies. The algebraic side of this argument is that decoupling an auxiliary field requires higher-derivative operators already in the UV, as the one in eq.~\eqref{eqn:ch1_gsc_oc_aux} for the orthogonal constraint \cite{DallAgata:2016syy}, and thus do not define a priori a consistent effective field theory. The main result of \cite{Bonnefoy:2022rcw} is then that consistent effective field theories of nonlinear supersymmetry must be constructed through superfield constraints that only remove physical, and not auxiliary, degrees of freedom.

From this perspective, the improved models constructed in \cite{Bonnefoy:2022rcw} through the generalized superfield constraint \eqref{eqn:ch1_gsc} are indeed well-defined effective field theories and admit therefore a UV completion into a microscopic 2-derivative theory. In \cite{Bonnefoy:2022rcw} we take a step in this direction by studying the coupling of the improved orthogonal constraint \eqref{eqn:ch1_gsc_oc_imp} to (nonlinear) supergravity, where the goldstino $G$ is now identified with the longitudinal component of the massive gravitino \cite{Deser:1977uq}. The gravitino sound speed in such supergravity model is found to match exactly the rigid one \eqref{eqn:ch1_gsc_sound_speed}, so that subluminal propagation persist also at this level, and therefore the improved constraint \eqref{eqn:ch1_gsc_oc_imp} defines improved minimal models of inflation in supergravity with nonlinearly realized supersymmetry.

\chapter{Linear and nonlinear supergravity}\label{ch2_sugra}

In the first chapter of this thesis, we introduced the general framework of supersymmetry, arising as the maximal possible extension of the global Poincar\'e group. In this second chapter, we take a further step forward and promote global supersymmetry to a local symmetry. The striking outcome of this standard, but rather deep and powerful procedure is the emergence of Einstein gravity as a necessary and essential element of the spectrum of the resulting theory, which is thus a theory of \textit{supergravity}. The beauty of this framework is that gravity and the other forces and matter fields get unified\footnote{Such a unification holds, at the field theory level, only classically, since gravity is not consistent as quantum field theory. The unification of matter and forces at the quantum level is realized by string theory, which we will discuss in Chapter \ref{ch3_st}.} in a single, elegant picture and by a unique symmetry principle.

As in the previous chapter, the main focus will be on $D=4$, $\N=1$ supergravity and its couplings to matter fields. We begin in Section \ref{ch2_sec_gauged_susy} introducing pure supergravity as the gauge theory of the local Super-Poincar\'e group. We then proceed to discuss its coupling to chiral multiplets, the spontaneous breaking of local supersymmetry, and the corresponding super-Higgs mechanism. In this framework, we present in Section \ref{ch2_sugra_grav_cosmology} the results of the work \cite{Casagrande:2023fjk}, concerning effective theories of nonlinear local supersymmetry, in continuation with Chapter \ref{ch_susy}, and the dynamics of massive gravitinos in cosmological scenarios, in particular the phenomenon of \textit{gravitational particle production}. Then, we restore supersymmetry and discuss in Section \ref{ch2_sec_off_shell_sugra} the possible off-shell formulations of $D=4$, $\N=1$ pure supergravity in relation to the supercurrent superfield formalism. This sets the stage for presenting the findings of the work \cite{ms2}, in which we use this formalism to construct and investigate possible couplings to supergravity of the supermultiplet of the \textit{massive spin-2 field}. In this final section, along with Appendix \ref{app_ms2_kk}, we briefly cover some aspects of pure supergravity in five dimensions. \\

The main review references for this chapter are the books \cite{Freedman:2012zz} and \cite{DallAgata:2021uvl}. 

\section{Supergravity as the gauge theory of Super-Poincar\'e}\label{ch2_sec_gauged_susy}

A theory of supergravity is a theory where supersymmetry has been promoted from global to local symmetry. This means that the supersymmetry parameter $\epsilon$ acquires a spacetime dependence:
\beq
    \epsilon \,\, \longrightarrow \,\, \epsilon(x) \ .
\label{eqn:ch2_local_epsilon}
\eeq

\noindent The outcomes of this promotion can be understood qualitatively starting from a rigid supersymmetric theory and studying the deformations implied by \eqref{eqn:ch2_local_epsilon}. This is the so-called Noether procedure, which we can apply following the standard algorithm of gauge interactions, to which local supersymmetry makes no exception. Thus, the gauging \eqref{eqn:ch2_local_epsilon} requires the introduction of a gauge field for local supersymmetry, which we call $\psi_\mu$. Being a gauge field, it must transform with the derivative of the symmetry parameter:
\beq
    \delta\psi_\mu\sim\partial_\mu\epsilon \ ,
\label{eqn:ch2_delta_psi_noether}
\eeq

\noindent and couple in the lagrangian to the supersymmetry conserved current, which we call $\Xi^\mu$. This current is a Majorana vector-spinor\footnote{While in Chapter \ref{ch_susy} we used 2-components Weyl spinor, here we employ instead 4-components spinors, in this case of the Majorana type. See Appendix \ref{app_conventions} for more details and for the conventions on the Clifford algebra.} of mass dimension $\[\Xi^\mu\]=\frac{7}{2}$, and so the gauge field $\psi_\mu$ must also be a Majorana vector-spinor. Its minimal coupling to $\Xi^\mu$ has the form
\beq
    \frac{1}{\MP}\bpsi_\mu\Xi^\mu \ ,
\label{eqn:ch2_noether_psi_Xi}
\eeq

\noindent where $\MP$ must be the Planck mass, as will be clear in a moment. The gauge field $\psi_\mu$ is known as the \textit{gravitino}, which is a Majorana vector-spinor field, of spin-$\frac32$. The coupling of the gravitino $\psi_\mu$ to the supercurrent $\Xi^\mu$ of eq.~\eqref{eqn:ch2_noether_psi_Xi} is the first deformation of the theory and it requires further ones in order to have full local supersymmetry invariance. Following this Noether procedure, the supersymmetry transformation of the supercurrent $\Xi^\mu$ contains the stress-energy tensor $T^{\mu\nu}$ \cite{Ferrara:1974pz}:
\beq
    \delta\Xi^\mu\sim T^{\mu\nu}\gamma_\nu\epsilon \ .
\label{eqn:ch2_delta_Xi_noether}
\eeq

\noindent In order to embed consistently the stress-energy tensor in the lagrangian, there must exist its own associated gauge field, which is nothing but the metric field $g_{\mu\nu}$:
\beq
    g_{\mu\nu}T^{\mu\nu} \ .
\label{eqn:ch2_g_emt_noether}
\eeq

\noindent Therefore, we see explicitly from this Noether procedure argument that promoting supersymmetry to a local symmetry requires, by consistency, the existence of the gravitational interaction. The graviton field $g_{\mu\nu}$ and the spin-$\frac32$ gravitino form what is known as the pure supergravity multiplet $\left\{g_{\mu\nu},\psi_\mu\right\}$. At the linear level, the supersymmetry transformations characterising this multiplet have the schematic form
\begin{align}
     \delta\psi_\mu\sim\MP\partial_\mu\epsilon \ , && \delta g_{\mu\nu}\sim\frac{1}{\MP}\bepsilon\gamma_{(\mu}\psi_{\nu)} \ .
\label{eqn:ch2_delta_sugra_noether}
\end{align}

Going beyond the linear order through the Noether procedure highlighted above is rather involved and not much transparent. On the contrary, the complete supersymmetry transformations of the pure supergravity multiplet --- which also determine the associated lagrangian --- can be elegantly derived in full generality by performing the gauging of the full Super-Poincar\'e algebra of eq.~\eqref{eqn:ch1_SP_alg}. In this way, the connection between local supersymmetry and gravitational interactions will be manifest. We denote collectively the generators of the ($\N=1$) Super-Poincar\'e algebra \eqref{eqn:ch1_SP_alg}, namely the translations $P_a$, the Lorentz rotations $M_{ab}$ and the supersymmetry transformations $Q_\alpha$ as\footnote{From now on we set the Planck mass $\MP$ to 1 and reinstate it explicitly when needed.}
\beq
    T_A=\left\{P_a,M_{ab},Q_\alpha\right\} \ ,
\eeq

\noindent in which the small latin letters denote four-dimensional local Lorentz indices, whereas $\alpha$ is a 4-component spinor index. The gauging of this algebra can be realized introducing a set of gauge fields $A_\mu^A$ for each of these generators, which must have the form
\beq
    A_\mu^AT_A=e_\mu{}^aP_a+\frac12\omega_\mu^{ab}M_{ab}+\bpsi_\mu Q \ .
\eeq

\noindent It is clear from this formula that the gauge fields $e_\mu{}^a$ and $\omega_\mu^{ab}$ are the vierbein and the associated spin-connection, while $\psi_\mu$ is the gravitino field introduced above. These gauge fields should transform under local supersymmetry as
\beq
    \delta A_\mu^A=\partial_\mu\epsilon^A+f^A{}_{BC}A_\mu^B\epsilon^C \ ,
\eeq

\noindent where $f^A{}_{BC}$ are the structure constants of the local Super-Poincar\'e algebra, which can be read from eq.~\eqref{eqn:ch1_SP_alg}. The resulting supersymmetry transformations of the vierbein and the gravitino are
\begin{align}
    \delta e_\mu{}^a=\frac{1}{2}\bepsilon\gamma^a\psi_\mu \ , && \delta \psi_\mu= D_\mu\epsilon \ ,
\end{align}

\noindent where $D_\mu$ is the Lorentz-covariant derivative
\beq
    D_\mu\epsilon=\partial_\mu \epsilon+\frac{1}{4}\omega_\mu^{ab}\gamma_{ab}\epsilon \ .
\eeq

\noindent Moreover, the structure constants $f^A{}_{BC}$ imply that the vierbein $e_\mu{}^a$ and the spin-connection $\omega_\mu^{ab}$ satisfy a nontrivial torsion equation
\beq
    T^a_{\mu\nu}=2\(\partial_{[\mu}e_{\nu]}{}^a+\omega_{[\mu}^a{}_be_{\nu]}{}^b\)-\frac{1}{2}\bpsi_{[\mu}\gamma^a\psi_{\nu]} \ ,
\eeq

\noindent which involves a gravitino bilinear contribution. The torsionless condition yields therefore not the standard connection $\omega_\mu^{ab}(e)$, but rather a connection with torsion:
\beq
    \hat{\omega}_\mu^{ab}\equiv\omega_\mu^{ab}\(e,\psi\)=\omega_\mu^{ab}(e)-\frac{1}{4}\(\bpsi^{[a}\gamma_\mu\psi^{b]}-\bpsi_\mu\gamma^{[a}\psi^{b]}-\bpsi^{[a}\gamma^{b]}\psi_\mu\) \ . 
\label{eqn:ch2_conn_torsion}
\eeq   

\noindent This torsion term is a peculiar feature of supergravity theories and summarizes in an elegant way their non-linearities, which is required by general consistency with local supersymmetry.

The pure supergravity lagrangian is therefore given by \cite{Freedman:1976xh,Deser:1976eh}
\beq
   e^{-1}\L_\textup{SUGRA}=\frac{1}{2} R(\hat{\omega})-\frac12\bpsi_\mu\gamma^{\mu\nu\rho}D_\nu\(\hat{\omega}\)\psi_\rho \ ,
\label{eqn:ch2_sugra_lag_1}
\eeq

\noindent and it is invariant under the supersymmetry transformations
\beq
\begin{aligned}
    \delta e_\mu{}^a&=\frac12\bepsilon\gamma^a\psi_\mu \ , \\
    \delta\psi_\mu&=D_\mu\(\hat{\omega}\)\epsilon \ .
\end{aligned}
\label{eqn:ch2_pure_sugra_susy}
\eeq

\noindent Unpacking the torsion terms of the connection $\hat{\omega}_\mu^{ab}$ \eqref{eqn:ch2_conn_torsion}, which are left implicit in eq.~\eqref{eqn:ch2_sugra_lag_1}-\eqref{eqn:ch2_pure_sugra_susy}, amounts to a four-gravitino contact term in the action, which results in 
\begin{equation}
    e^{-1}\L_\textup{SUGRA}=\frac12 R-\frac12\psi_\mu\gamma^{\mu\nu\rho}D_\nu\psi_\rho+\L_\textup{torsion} \ ,
\label{eqn:ch2_sugra_lag_2}
\end{equation}

\noindent where $R\equiv R(e)$ and $D_\mu\equiv D_\mu(\omega)$ are the standard torsion-free quantities, and 
\beq
    e^{-1}\L_\textup{torsion}=-\frac{1}{32}\[\(\bpsi^\mu\gamma^\nu\psi^\rho\)\(\bpsi_\mu\gamma_\nu\psi_\rho+\bpsi_\mu\gamma_\rho\psi_\nu\)-4\(\bpsi_\mu\gamma^\rho\psi_\rho\)\(\bpsi^\mu\gamma^\rho\psi_\rho\)\]\ .
\label{eqn:ch2_sugra_L_tor}
\eeq

The closure of the algebra of local supersymmetry is now modified with respect to the global case \eqref{eqn:ch1_SP_alg}-\eqref{eqn:ch1_comm_susy_derivative}, and it is realized on a covariant general coordinate transformation:
\beq
    \[\delta_1,\delta_2\]=\delta_\xi-\delta_{\hat{\Lambda}}-\delta_{\hat{\epsilon}} \ ,
\label{eqn:ch2_sugra_closure}
\eeq

\noindent where $\delta_\xi$ is a general coordinate transformation, $\delta_{\hat{\Lambda}}$ is a local Lorentz transformation and $\delta_{\hat{\epsilon}}$ is a local supersymmetry transformation, all given by field-dependent parameters:
\begin{align}
    \xi^a=-\frac{1}{2}\bepsilon_1\gamma^a\epsilon_2 \ , && \hat{\Lambda}^{ab}=\xi^\rho\omega_\rho^{ab} \ , && \hat{\epsilon}=\xi^\rho\psi_\rho \ .
\end{align}

\noindent The closure \eqref{eqn:ch2_sugra_closure} is exact on the vierbein $e_\mu{}^a$ and it holds on-shell (as expected) on the gravitino $\psi_\mu$.

\subsection{The coupling to chiral multiplets}\label{eqn:ch2_sec_chiral_mult}

After having introduced the pure supergravity multiplet \eqref{eqn:ch2_pure_sugra_susy}, 
we now discuss its coupling to $n$ chiral multiplets $\{\phi^i, \chi^i\}$, $i=1,\dots,n$, of the type we introduced in eq.~\eqref{eqn:ch1_m0_chiral}-\eqref{eqn:ch1_chiral_sf_def} of Chapter \ref{ch_susy}. As we saw previously, chiral multiplets are inherently associated with K\"ahler geometry: the scalar fields $\phi^i$ parameterize in fact a K\"ahler manifold, whose characteristic isometries are the K\"ahler transformations 
\beq
    K\(\phi,\bar{\phi}\)\quad \longrightarrow \quad K\(\phi,\bar{\phi}\)+\Lambda(\phi)+\bar{\Lambda}\(\bar \phi\)\ ,
\label{eqn:ch2_K_tr}
\eeq

\noindent as in eq.~\eqref{eqn:ch1_K_tr}. When coupled to supergravity, the requirement of K\"ahler invariance gives rise to new structures and symmetry relations among the elements of the theory, which turn into new interaction terms in the resulting action. These additional couplings will appear suppressed by appropriate powers of the Planck mass $\MP$, so that the global supersymmetry action \eqref{eqn:ch1_chiral_K_W_lag} is indeed recovered in the decoupling limit of gravity.

The first new element of this construction is that the gravitino $\psi_\mu$ transforms nontrivially under K\"ahler transformations \eqref{eqn:ch2_K_tr}, in the form of a chiral rotation:
\beq
    \psi_\mu \,\, \to \,\, e^{-\frac{i}{2}\mathfrak{Im}\(\Lambda\)\gamma_5}\psi_\mu \ .
\label{eqn:ch2_gravitino_K_tr}
\eeq

\noindent The invariance of the action under K\"ahler reparameterization implies that the derivatives acting on the gravitino should include an appropriate covariantization with respect to this K\"ahler transformations \eqref{eqn:ch2_K_tr}. This K\"ahler-covariant derivative takes the form
\beq
    \mathscr{D}_\mu\psi_\nu=D_\mu\psi_\nu+\frac{i}{2}Q_{\mu}\gamma_5\psi_\nu \ ,
\label{eqn:ch2_gravitino_K_cov_der}
\eeq

\noindent where $Q_\mu$ is the composite operator
\beq
    Q_\mu=\frac{i}{2}\[K_{\bar i}\partial_\mu \bar{\phi}^{\bar i}-K_i\partial_\mu\phi^i\] \ . 
\eeq

\noindent Note that $Q_\mu$ transforms as $Q_\mu\,\,\to\,\,Q_\mu+\partial_\mu\Lambda$ under K\"ahler transformations \eqref{eqn:ch2_K_tr}, like a $\mathrm{U}(1)$ gauge field, and thus the gravitino derivative $\mathscr{D}_\mu\psi_\nu$ \eqref{eqn:ch2_gravitino_K_cov_der} transforms indeed covariantly under \eqref{eqn:ch2_K_tr}, \textit{i.e.} like $\psi_\mu$ in eq.~\eqref{eqn:ch2_gravitino_K_tr}. Moreover, consistency of the supersymmetry transformation \eqref{eqn:ch2_pure_sugra_susy} requires first the supersymmetry parameter $\epsilon$ to transform in the same way as the gravitino \eqref{eqn:ch2_gravitino_K_tr}, and consequently also the chiral fermions $\chi^i$:
\begin{align}
    \chi_{\mu L}^i \,\, \to \,\, e^{\frac{i}{2}\mathfrak{Im}\(\Lambda\)\gamma_5}\chi^i_{\mu L} \ , && \mathscr{D}_\mu\chi^i_{\nu L}=D_\mu\chi^i_\nu+\Gamma^i_{jk}\partial_\mu\phi^j\chi^k_{L}-\frac{i}{2}Q_{\mu}\gamma_5\chi^i_L \ ,
\label{eqn:ch2_ferm_K_cov}
\end{align}

\noindent for which the K\"ahler-covariant derivative includes also the K\"ahler connection term, as we defined in eq.~\eqref{eqn:ch1_K_geom_formulae_1}-\eqref{eqn:ch1_K_geom_formulae_2}.

Thus, when coupling chiral multiplets to supergravity, consistency with local supersymmetry and K\"ahler geometry implies that the fermionic fields of the theory must also transform under K\"ahler transformations, in the form of a chiral $\mathrm{U}(1)$-like symmetry \eqref{eqn:ch2_gravitino_K_tr}-\eqref{eqn:ch2_ferm_K_cov}, with respect to which the corresponding lagrangians must be properly covariantized. As a consequence, also the coupling of the superpotential $W\(\phi\)$ is modified. First, it must come through the combination
\beq
    e^{\frac{K}{2}}W \ ,
\label{eqn:ch2_sugra_exp_K_W}
\eeq

\noindent and should transform as
\beq
    W\(\phi\) \,\, \to \,\, e^{-\Lambda\(\phi\)}W\(\phi\)\ ,
\label{eqn:ch2_W_K_tr}
\eeq

\noindent under K\"ahler transformations \eqref{eqn:ch2_K_tr}. The associated K\"ahler-covariantization of the derivatives is given by
\beq
    e^{\frac{K}{2}}\mathscr{D}_iW=e^{\frac{K}{2}}\(\partial_i+K_i\)W \ .
\eeq

\noindent This new structure linking the superpotential $W$ and the K\"ahler potential implies that the two objects are no longer independent, as in global supersymmetry, and terms can be moved between the two without changing the final result, thanks to the K\"ahler transformations of eq.~\eqref{eqn:ch2_K_tr}, \eqref{eqn:ch2_sugra_exp_K_W} and \eqref{eqn:ch2_W_K_tr}. This motivates the definition of a generalized potential $\mathscr{G}$, as
\beq
    \mathscr{G}\equiv K+\log\abs{W}^2 \ ,
\label{eqn:ch2_gen_pot_KW}
\eeq

\noindent which is manifestly invariant under K\"ahler transformations \eqref{eqn:ch2_K_tr}-\eqref{eqn:ch2_W_K_tr}.

Collecting all these ingredients, we obtain the following Lagrangian describing the coupling of $n$ chiral multiplets to pure $D=4$, $\N=1$ supergravity \cite{Cremmer:1982wb,Cremmer:1982en,Bagger:1982ab}:
\beq
\begin{aligned}
    e^{-1}\L=&\frac12R-\frac12\bpsi_\mu\gamma^{\mu\nu\rho}\mathscr{D}_\mu\psi_\nu-K_{i\bar j}\(\partial_\mu \phi^i\partial^\mu\bar{\phi}^{\bar j}+\bchi_L^{i}\slashed{\mathscr{D}}\chi_R^{\bar j}+\bchi_R^{\bar j}\slashed{\mathscr{D}}\chi_L^{i}\) \\
    &+\[\frac12 e^{\frac{K}{2}}W\bpsi_{\mu R}\gamma^{\mu\nu}\psi_{\nu R}-e^{\frac{K}{2}}\mathscr{D}_i\mathscr{D}_jW\bchi^i_L\chi^j_L+h.c.\]\\
    &+\[K_{i\bar j}\bpsi_{\mu L} \gamma^\nu\gamma^\mu\chi^i_L\partial_\nu{\bar \phi}^{\bar j}+\bpsi_{\mu R}\gamma^\mu\chi^i_Le^{\frac{K}{2}}\mathscr{D}_iW+h.c.\] -V\(\phi,\bar{\phi}\) \\
    &+\L_\textup{4F}\ .
\end{aligned}
\label{eqn:ch2_sugra_chiral_lag}
\eeq

\noindent The first line contains the kinetic terms of all fields, while the second line contains mass-like terms for both the gravitino and the chiral fermions. The third line contains instead the coupling of the gravitino to the supersymmetry current, as in eq.~\eqref{eqn:ch2_noether_psi_Xi}, and the scalar potential, which is now given by
\beq
    V\(\phi,\bar \phi\)=e^K\(K^{i\bar j}\mathscr{D}_i W\bar{\mathscr{D}}_{\bar j}\overline{W}-3\abs{W}^2\) \ ,
\label{eqn:ch2_sugra_scal_pot_F}
\eeq

\noindent where the first term appeared already in global supersymmetry, whereas the second is a consequence of the coupling to pure supergravity. In terms of the generalized potential \eqref{eqn:ch2_gen_pot_KW}, this scalar potential can be expressed as
\beq
    V=e^{\mathscr G}\(\mathscr{G}^{i\bar j}\mathscr{G}_i\mathscr{G}_{\bar j}-3\) \ .
\eeq

\noindent The last line of the lagrangian \eqref{eqn:ch2_sugra_chiral_lag} contains instead quartic fermionic terms and reads
\beq
    \L_\text{4F}=\L_\textup{torsion}-\frac12 K_{i\bar j}\bpsi_\mu\chi^i\bpsi^\mu\chi^{\bar j}+\frac14\(R_{i\bar{j}k\bar{l}}-K_{i\bar j}K_{k \bar l}\)\bchi^i\chi^{k}\bchi^{\bar j}\chi^{\bar l} \ .
\eeq

\noindent where $\L_\textup{torsion}$ is the quartic gravitino term of pure supergravity \eqref{eqn:ch2_conn_torsion} and $R_{i\bar{j}k\bar{l}}$ was given in eq.~\eqref{eqn:ch1_K_geom_formulae_1}. 

At quadratic order in the number of fermions, the local supersymmetry transformations that leave invariant the lagrangian \eqref{eqn:ch2_sugra_chiral_lag} are
\beq
\begin{aligned}
    \delta e_\mu{}^a&=\frac12\bepsilon\gamma^a\psi_\mu \ , &&& \delta\psi_{\mu L} &=\mathscr{D}_\mu\epsilon_L+\frac12 e^{\frac{K}{2}}W\gamma_\mu\epsilon_R \ , \\
    \delta\phi^i&=\bepsilon_L\chi^i_L \ , &&& \delta\chi^i_L&=\frac12\slashed{\partial}\phi^i\epsilon_R-\frac12e^{\frac{K}{2}}K^{i\bar j}\bar{\mathscr{D}}_{\bar j}\overline{W}\epsilon_L \ .    
\end{aligned}
\label{eqn:ch2_sugra_chiral_susy}
\eeq

\subsubsection{Comments on the coupling of vector multiplets}

We end this section with a few additional comments on the coupling to supergravity of vector multiplets \eqref{eqn:ch1_m0_vec}-\eqref{eqn:ch1_vec_sf_def}. Similarly to the inclusion of chiral multiplets, the global supersymmetry setup needs to be modified with appropriate covariantization of the terms and the addition of the gravitino coupling to the supercurrent \eqref{eqn:ch2_noether_psi_Xi}, as well as further quartic fermionic interactions. Moreover, further interactions and covariantizations need to be included if some of the isometries of the scalar manifold are gauged through the newly added vector fields. 

We do not display the full lagrangian here. We though highlight that, as in global supersymmetry, adding gauge interactions between the matter fields also give rise in the local case to a new term in the scalar potential $V\(\phi,\bar\phi \)$, which takes the general form
\beq
   V\(\phi,\bar \phi\)=e^K\(K^{i\bar j}\mathscr{D}_i W\bar{\mathscr{D}}_{\bar j}\overline{W}-3\abs{W}^2\)+\frac12 \(\mathfrak{Re}f_{IJ}\)D^ID^J \ ,
\label{eqn:ch2_sugra_scal_pot_F_D}
\eeq

\noindent where $f_{IJ}$ is the gauge-kinetic function \cite{Cremmer:1982wb,Cremmer:1982en,Bagger:1982ab} and $D^I$ are the D-terms that already appeared in global supersymmetry in eq.~\eqref{eqn:ch1_vec_chiral_sp_1}-\eqref{eqn:ch1_vec_chiral_af_on_shell_1}. For more details, also on the inclusion of Fayet--Iliopoulos terms \eqref{eqn:ch1_FI_term}, see \cite{Cremmer:1982wb,Cremmer:1982en,Bagger:1982ab,Elvang:2006jk,DeRydt:2007vg,Binetruy:2004hh}.

\subsection{Local supersymmetry breaking}\label{ch2_sec_sugra_Higgs}

The scalar potential \eqref{eqn:ch2_sugra_scal_pot_F_D} is the starting point to discuss the spontaneous breaking of local supersymmetry and the consequent supersymmetric Higgs mechanism, in which the goldstino of supersymmetry breaking, which we introduced in Section \ref{ch1_sec_ssb}, is eaten by the gravitino, which in this way acquires a mass \cite{Zumino:1979et}.

With respect to the global case, supersymmetric and non-supersymmetric vacua can no longer be identified by the corresponding value of the energy, as depicted in Figure \ref{fig_susy_breaking}. In local supersymmetry, the semi-positivity of the energy \eqref{eqn:ch1_E_pos} does not hold anymore. Indeed, the coupling to pure supergravity generates a negative contribution to the scalar potential, the term $-3e^{K}\abs{W}^2$ in eq.~\eqref{eqn:ch2_sugra_scal_pot_F_D}. However, the vacua of a supergravity theory can still be classified based on the homogeneous structure of the vacuum expectation values of the fermionic supersymmetry transformations. They should in fact vanish identically, since the vacuum must be a Lorentz-invariant state, and this translates into consistency conditions on their explicit expressions, which are directly related to the terms in the scalar potential. This structure was also present in the rigid case, as shown in eq.~\eqref{eqn:ch1_ferm_susy_inhom}. To illustrate this, let us consider pure supergravity $\left\{e_\mu{}^a,\psi_\mu\right\}$ coupled to $n_\textup{C}$ chiral multiplets $\left\{\phi^i,\chi^i\right\}$ and $n_\textup{V}$ vector multiplets $\left\{A_\mu^I,\lambda^I\right\}$. The vacuum expectation value of the fermionic supersymmetry transformations are
\begin{align}
    \langle\delta\psi_{\mu L}\rangle&=\langle\mathscr{D}_\mu\epsilon_L\rangle+\frac12\left\langle e^{\frac{K}{2}}W\right\rangle\gamma_\mu\epsilon_R=0 \ , \label{eqn:ch2_vev_susy_psi}\\
    \langle \delta\chi^i_L \rangle&=-\frac12\left\langle K^{i\bar j}e^{\frac{K}{2}}\bar{\mathscr{D}}_{\bar j}\overline{W}\right\rangle\epsilon_L=0 \ , \label{eqn:ch2_vev_susy_chi}\\
    \langle \delta\lambda^I_L\rangle &=\frac{i}{2}\left\langle D^I\right\rangle\epsilon_L=0 \ . \label{eqn:ch2_vev_susy_lambda}
\end{align}

\noindent Thus, a vacuum is supersymmetric if these conditions are satisfied identically, while it breaks supersymmetry if they are not. From the matter fermions equations \eqref{eqn:ch2_vev_susy_chi}-\eqref{eqn:ch2_vev_susy_lambda} we have that a supersymmetric vacuum is determined by the conditions
\begin{align}
    \left\langle\mathscr{D}_iW\right\rangle=0 \ , && \left\langle D^I\right\rangle=0 \ .
\label{eqn:ch2_susy_vev_cond}
\end{align}

\noindent Note, however, that there is no a priori condition on the vacuum expectation value of the superpotential $\left\langle W\right\rangle$. The gravitino equation \eqref{eqn:ch2_vev_susy_psi} can in fact still be satisfied by a nontrivial $\left\langle\mathscr{D}_\mu\epsilon_L\right\rangle$. This is a purely gravitational effect and and can occur only in anti-de Sitter (AdS) solutions, \textit{i.e.} with negative vacuum energy. On the contrary, the condition $\left\langle W\right\rangle=0$ identifies Minkowski vacua, consistently with the global supersymmetry case. The energy of a local supersymmetric vacuum $V_0$ is in fact given by eq.~\eqref{eqn:ch2_sugra_scal_pot_F_D}-\eqref{eqn:ch2_susy_vev_cond} to be
\beq
    V_0=-3\left\langle e^K\abs{W}^2\right\rangle \ .
\label{eqn:ch2_sugra_vac_energy}
\eeq

\noindent We see therefore that while positive vacuum energy (namely, de Sitter) solutions breaks supersymmetry, the energy of supersymmetric vacua is semi-negative definite. Also note that while the energy scale of supersymmetry breaking $\Lambda_{\slashed{\textup{susy}}}$ is identified by eq.~\eqref{eqn:ch2_susy_vev_cond} to be the sum of the F and D-terms in the scalar potential:
\beq
    \Lambda_{\slashed{\textup{susy}}}^4=\left\langle e^K K^{i\bar j}\mathscr{D}_i W\bar{\mathscr{D}}_{\bar j}\overline{W}\right\rangle+\frac12\left\langle\(\mathfrak{Re}f_{IJ}\)D^ID^J\right\rangle \ ,
\label{eqn:ch2_sugra_susy_breaking_scale}
\eeq

\noindent the associated vacuum energy $V_0$ is lower, because of the gravitational term \eqref{eqn:ch2_sugra_vac_energy}:
\beq
    V_0 = \Lambda_{\slashed{\textup{susy}}}^4-3\left\langle e^K\abs{W}^2\right\rangle \ .
\eeq

Let us now consider a vacuum that spontaneously breaks supersymmetry. For simplicity of presentation and relevance for the present manuscript, we focus on the case of a Minkowski vacuum, \textit{i.e.} $V_0=0$, with non-vanishing F-terms only. The general case with also D-terms and a nontrivial cosmological constant $V_0\ne 0$ follows anyway the same pattern \cite{Ferrara:2016ntj}. As we discussed in Section \ref{ch1_sec_ssb}, the Goldstone theorem predicts the existence in this vacuum of a massless state, the goldstino \eqref{eqn:ch1_G_def_ferm}, which is given by the following combination of chiral fermions:
\beq
    G_L=\left\langle\frac{\mathscr{D}_i W}{W}\right\rangle\chi^i_L \ .
\label{eqn:ch2_sugra_G}
\eeq

\noindent Similarly to the rigid case, its supersymmetry transformations is of the type
\beq
    \delta G_L = -\sqrt{3}\left\langle e^{\frac{K}{2}}\overline{W}\right\rangle \epsilon_L \ ,
\label{eqn:ch2_sugra_G_susy}
\eeq

\noindent indicating the residual nonlinear realization \eqref{eqn:ch1_nl_susy_tr_2}. However, in the local case supersymmetry has an associated gauge field, the gravitino $\psi_\mu$, which reabsorbs the goldstino \eqref{eqn:ch2_sugra_G} through the standard Higgs mechanism, becoming a massive field \cite{Zumino:1979et}. This can be seen explicitly from the lagrangian \eqref{eqn:ch2_sugra_chiral_lag}. Taking into account the Minkowski background, the relevant sector is the gravitino-goldstino lagrangian 
\beq
\begin{aligned}
    \L=&-\frac12\bpsi_\mu\gamma^{\mu\nu\rho}\partial_\nu\psi_\rho-\(\bar{G}_L\slashed{\partial}G_R+\bar{G}_R\slashed{\partial}G_L\) \\
    &+\frac12\[\left\langle e^{\frac{K}{2}}W\right\rangle\(\bar{G}_LG_L+2\sqrt{3}\bpsi_{\mu R}\gamma^\mu G_L+\bpsi_{\mu R}\gamma^{\mu\nu}\psi_{\nu R}\)+h.c.\]\ .
\end{aligned}
\label{eqn:ch2_psi_G_lag}
\eeq

\noindent This lagrangian suggests that the goldstino dependence can be completely canceled by performing the redefinition
\beq
    \psi_{\mu L} \,\, \to \,\, \psi_{\mu L}+\frac{2}{\sqrt{3}}\left\langle e^{\frac{K}{2}}\overline{W}\right\rangle^{-1}\partial_\mu G_L+\frac{1}{\sqrt{3}}\gamma_\mu G_R \ ,
\label{eqn:ch2_psi_G_redef}
\eeq

\noindent which yields the lagrangian of a massive gravitino
\beq
    \L=-\frac12\bpsi_\mu\gamma^{\mu\nu\rho}\partial_\nu\psi_\rho+\frac{1}{2}\(m_{3/2}\bpsi_{\mu R}\gamma^{\mu\nu}\psi_{\nu R}+h.c.\) \ .
\label{eqn:ch2_sugra_massive_psi_lag_1}
\eeq

\noindent Comparing the gravitino redefinition \eqref{eqn:ch2_psi_G_redef} with its supersymmetry transformation in eq.~\eqref{eqn:ch2_sugra_chiral_susy}, which evaluated on the Minkowski vacuum under consideration becomes
\beq
    \delta\psi_{\mu L}=\partial_\mu \epsilon_L+\left\langle e^{\frac{K}{2}}\abs{W}\right\rangle\gamma_\mu \epsilon_R \ ,
\eeq

\noindent one realizes that such a redefinition \eqref{eqn:ch2_psi_G_redef} is a special instance of supersymmetry transformation, of composite parameter
\beq
    \epsilon=\frac{2}{\sqrt 3}\left\langle e^{\frac{K}{2}}\abs{W}\right\rangle^{-1}G \ .
\eeq

\noindent Therefore, the difference between the gravitino-goldstino lagrangian \eqref{eqn:ch2_psi_G_lag} and the massive gravitino one \eqref{eqn:ch2_sugra_massive_psi_lag_1} is nothing but a gauge transformation. The goldstino $G$ is then a pure gauge degree of freedom in (spontaneously broken) supergravity, and can be completely reabsorbed in the gravitino through \eqref{eqn:ch2_psi_G_redef}. This identifies the unitary gauge, which can be effectively selected by setting $G=0$.  

This is the supersymmetric version of the Higgs mechanism, stating that a spontaneously supersymmetry breaking vacuum yields a massive gravitino, of mass
\beq
    \abs{m_{3/2}}=\frac{1}{\MP^2}\left\langle e^{\frac{K}{2}}\abs{W}\right\rangle\sim\frac{\Lambda_{\slashed{\textup{susy}}}^2}{\MP} \ ,
\label{eqn:ch2_m_32}
\eeq

\noindent in which we reinstated the Planck mass $\MP$ for clarity. The massive gravitino contains four polarization states: two transverse $\pm\frac32$ polarizations, corresponding to the original massless gravitino, and two longitudinal $\pm\frac12$ polarizations, corresponding to the goldstino degrees of freedom. In the lagrangian they both couple to the current of supersymmetry $\Xi^\mu$ as in eq.~\eqref{eqn:ch2_noether_psi_Xi}, but the resulting dynamics is substantially different for the two states. While the transverse polarizations interact with gravitational strength, the longitudinal states interact with the scale of supersymmetry breaking \eqref{eqn:ch2_sugra_susy_breaking_scale}. This is clear from the gauge transformation \eqref{eqn:ch2_psi_G_redef}. Therefore, the couplings of the longitudinal gravitino are enhanced with respect to the transverse one, which is instead gravitationally suppressed. 

This property is expressed by the gravitino equivalence theorem \cite{Fayet:1986zc,Casalbuoni:1988kv,Casalbuoni:1988qd}, which states that at energies well above of the gravitino mass $m_{3/2}$ the scattering amplitudes of the longitudinal gravitino polarizations can be computed in terms of the goldstino $G$ through the replacement rule
\beq
    \psi_\mu \,\, \to \,\, \frac{1}{m_{3/2}}\partial_\mu G \ . 
\label{eqn:ch2_sugra_eq_theorem}
\eeq

\noindent The corresponding interaction terms are then given by
\beq
\begin{aligned}
    \frac{1}{\MP}\overline{\Xi}^\mu\psi_\mu\,\,\longrightarrow\,\,\frac{1}{m_{3/2}\MP}\overline{\Xi}^\mu \partial_\mu G\sim\frac{1}{\Lambda_{\slashed{\textup{susy}}}^2}\partial_\mu \overline{\Xi}^\mu G\ ,
\end{aligned}
\eeq

\noindent where we see explicitly that the coupling scale is indeed the scale of supersymmetry breaking. Following this perspective, one can take in particular the decoupling limit of gravity $\MP\to\infty$ while keeping the $\Lambda_{\slashed{\textup{susy}}}$ scale fixed. At low-energies, where some of the superpartners of the original theory may have decoupled as well, one obtains in this way an effective theory for the goldstino $G$ and the remaining light fields. These are the theories of nonlinear supersymmetry we discussed in Chapter \ref{ch_susy}. In particular, one finds that the low-energy interactions of the goldstino are described by the Volkov--Akulov lagrangian \eqref{eqn:ch1_VA_lag} \cite{Volkov:1973ix}. From the point of view of the constrained superfield formalism and their consistency, that we discuss in Section \ref{ch1_sec_nls}, this shows explicitly how the nilpotent superfield constraint \eqref{eqn:ch1_G_nil_sf} describing the goldstino arises as an effective field theory operator from a 2-derivative microscopic theory.

\section{Massive gravitino in cosmology}\label{ch2_sugra_grav_cosmology}

We now focus on supergravity theories with spontaneously broken supersymmetry, discussed in Section \ref{ch2_sec_sugra_Higgs}, and present the results we obtained in \cite{Casagrande:2023fjk}, which concern the so-called \textit{gravitational production} of massive gravitinos in cosmological settings. 

As we saw already in Section \ref{ch1_sec_nls} for the global case, supersymmetry breaking is a crucial element to investigate the phenomenological consequences of supergravity, not only as a consistency requirement to match observations, but, more importantly, because it drives and fully characterizes the effective low-energy dynamics. The formalism of constrained superfields presented in Section \ref{ch1_sec_nls} is very useful for building effective theories with spontaneous supersymmetry breaking also in the local case \cite{Lindstrom:1979kq,Ivanov:1984hs,Ivanov:1989bh,Samuel:1982uh,Farakos:2013ih,Antoniadis:2014oya,Ferrara:2014kva,Kallosh:2014via,DallAgata:2014qsj,Kallosh:2014hxa,Dudas:2015eha,Bergshoeff:2015tra,Hasegawa:2015bza,Ferrara:2015gta,Kuzenko:2015yxa,Antoniadis:2015ala,Kallosh:2015tea,Kallosh:2015sea,DallAgata:2015pdd,Bandos:2015xnf,Ferrara:2015tyn,DallAgata:2015zxp,DallAgata:2016syy,Bandos:2016xyu}. The main difference with respect to rigid supersymmetry is that now the super-Higgs mechanism takes place, so that one ends up with effective theories for the massive gravitino (in curved spacetime) rather than goldstino ones.

A specific application of this type is the construction of supergravity models for inflation \cite{Antoniadis:2014oya,Kallosh:2014via,DallAgata:2014qsj,Carrasco:2015iij,Ferrara:2015tyn,Hasegawa:2017hgd,Kolb:2021xfn,Dudas:2021njv,Terada:2021rtp,Antoniadis:2021jtg,Bonnefoy:2022rcw}. In particular, the nilpotent superfield \eqref{eqn:ch1_G_nil_sf} and the orthogonal superfield \eqref{eqn:ch1_oc} yield, once coupled to gravity, models of inflation with a minimal spectrum, made of the graviton, the massive gravitino and a real scalar field, which is identified with the inflaton \cite{Kallosh:2014via,Ferrara:2015tyn,Carrasco:2015iij}. Although the associated rigid goldstino lagrangian, given in eq.~\eqref{eqn:ch1_oc_lag_off}, is rather involved, in supergravity the goldstino gets absorbed, as said, by the massive gravitino, so that --- after the constraints have been solved for --- one can work directly in the unitary gauge and set the goldstino field consistently to zero. Thus, starting from a canonical K\"ahler potential\footnote{With respect to eq.~\eqref{eqn:ch1_oc_KW}, we take here $h=1$.} and the most general superpotential with this superfield constraint setup, given in eq.~\eqref{eqn:ch1_oc_KW}, one obtains the very simple on-shell lagrangian \cite{Ferrara:2015tyn}
\beq
\begin{aligned}
    e^{-1}\L =\frac{1}{2}R-\frac12\bPsi_\mu\gamma^{\mu\nu\rho}D_\nu\Psi_\rho+\frac{m\(\phi\)}{2}\bPsi_\mu\gamma^{\mu\nu}\Psi_\nu-\frac12\partial_\mu\phi\partial^\mu\phi-V\(\phi\) +e^{-1}\L_\textup{torsion}\ ,
\end{aligned}
\label{eqn:ch2_sugra_inflation_oc_lag}
\eeq

\noindent where $\Psi_\mu$ is the massive gravitino, $\phi$ the real scalar field left independent by the orthogonal constraint \eqref{eqn:ch1_oc}-\eqref{eqn:ch1_oc_sol} and $\L_\textup{torsion}$ is the standard supergravity torsion term \eqref{eqn:ch2_sugra_L_tor}. The field-dependent gravitino mass parameter $m(\phi)$ and the scalar potential $V(\phi)$ are parameterized by the two holomorphic functions $f(\phi)$ and $g(\phi)$ that enter the superpotential \eqref{eqn:ch1_oc_KW}. Taking these functions to be real, the gravitino mass is given, according to eq.~\eqref{eqn:ch2_sugra_chiral_lag}-\eqref{eqn:ch2_m_32}, by
\beq
    m\(\phi\)=g\(\phi\) \ ,
\label{eqn:ch2_sugra_oc_gr_mass}
\eeq

\noindent while the scalar potential is equal to
\beq
    V\(\phi\)=f^2\(\phi\)-3g^2\(\phi\).
\label{eqn:ch2_sugra_oc_sc_pot}
\eeq

\noindent Notice that, with respect to the general formula \eqref{eqn:ch2_sugra_scal_pot_F}, only the F-term of the nilpotent superfield appears in the scalar potential. The orthogonal superfield one is constrained to be a function of the fermions, according to \eqref{eqn:ch1_oc_sol}, and therefore does not contribute to the scalar potential. Thus, as anticipated, the lagrangian \eqref{eqn:ch2_sugra_inflation_oc_lag} contains a single scalar field coupled to spontaneously broken supergravity and constitutes a minimal supergravity model for inflation. The known inflationary models can be embedded in this setup by engineering the functions $f\(\phi\)$ and $g\(\phi\)$ that determine the scalar potential \eqref{eqn:ch2_sugra_oc_sc_pot} \cite{Ferrara:2015tyn,Carrasco:2015iij}. 

\subsection{Gravitino in FLRW}\label{ch2_sec_grav_flrw}

More generally, the theory \eqref{eqn:ch2_sugra_inflation_oc_lag} is an explicit and simplified supergravity framework in which to further investigate cosmological applications. One can in fact assume a time-dependent scalar field background solution of the Friedmann--Lemaitre--Robertson--Walker (FLRW) type:
\beq
    g_{\mu\nu}=a(\tau)^2\eta_{\mu\nu} \ ,
\label{eqn:ch2_flrw}
\eeq

\noindent where $a(\tau)$ is the scale factor and $\tau$ is the so-called conformal time, related to the standard cosmological time $t$ as $dt=a(\tau)d\tau$. The Einstein equations that describe this background are
\beq
    G_{\mu\nu}=T_{\mu\nu}^{(\phi)} \ ,
\label{eqn:ch2_EE_gen}
\eeq  

\noindent where the gravity side is given by\footnote{Derivatives with respect to the conformal time will be denoted as $a^{\prime}\equiv\frac{da}{d\tau}$, while derivatives with respect to the cosmological time $t$ will be denoted as $\dot{a}\equiv\frac{da}{dt}$.}
\begin{align}
    G_{\mu\nu}=a^2\left[\left(4H^2-\frac{R}{3}\right)\delta^0_\mu\delta^0_\nu+\left(H^2-\frac{R}{3}\right)\eta_{\mu\nu}\right]  \ , && R=&6\frac{a^{\prime\prime}}{a^3} \ ,
\label{eqn:ch2_flrw_R_G}
\end{align}

\noindent in which $H$ is the Hubble parameter
\beq
    H=\frac{a^{\prime}}{a^2} \ ,
\eeq

\noindent whereas the energy-momentum tensor of $\phi$ is
\begin{equation}
\begin{aligned}
	T^{(\phi)}_{\mu\nu}&=\partial_\mu\phi\partial_\nu\phi-g_{\mu\nu}\(\frac{1}{2}\partial_\rho\phi\partial^\rho\phi+V(\phi)\)=\\
	&=a^2\delta^0_\mu\delta^0_\nu\(\frac{{\phi^{\prime}}{}^2}{2a^2}+V(\phi)\)+a^2\delta_{ij}\delta^i_\mu\delta^j_\nu\(\frac{{\phi^{\prime}}{}^2}{2a^2}-V(\phi)\) \ .
\end{aligned}
\label{eqn:ch2_EE_T_scalar}
\end{equation}

\noindent Recalling eq.~\eqref{eqn:ch2_sugra_oc_gr_mass}-\eqref{eqn:ch2_sugra_oc_sc_pot}, the Einstein equations \eqref{eqn:ch2_EE_gen} for the FLRW ansatz \eqref{eqn:ch2_flrw} result
\begin{equation}
\begin{aligned}
	3\(H^2+m^2\)=&\frac{{\phi^{\prime}}{}^2}{2a^2}+f^2 \ , \\
	\(H^2-3m^2-\frac{R}{3}\)=&\frac{{\phi^{\prime}}{}^2}{2a^2}-f^2 \ .
\end{aligned}
\label{eqn:ch2_EE_flrw}
\end{equation}

In such a cosmological setup, the dynamics of the massive gravitino turns out to be quite peculiar \cite{Kallosh:1999jj,Kallosh:2000ve,Giudice:1999yt, Giudice:1999am, Nilles:2001fg, Nilles:2001ry,Hasegawa:2017hgd,Kolb:2021xfn,Kolb:2021nob,Dudas:2021njv}, and we now discuss it in detail. In a curved background, the gravitino sector of the lagrangian \eqref{eqn:ch2_sugra_inflation_oc_lag} is\footnote{We neglect the quartic terms $\L_\textup{torsion}$ in \eqref{eqn:ch2_sugra_inflation_oc_lag} since they do not play a significant role in the present discussion.}
\beq
    e^{-1}\L_\Psi=-\frac12\bPsi_\mu\gamma^{\mu\nu\rho}\nabla_\nu\Psi_\rho+\frac{m}{2}\bPsi_\mu\gamma^{\mu\nu}\Psi_\nu \ ,
\label{eqn:ch2_m_grav_lag}
\eeq

\noindent where $m$ is the background-dependent mass, as in eq.~\eqref{eqn:ch2_sugra_inflation_oc_lag}-\eqref{eqn:ch2_sugra_oc_gr_mass}, and $\nabla_\mu$ is the full covariant derivative, including also the Levi--Civita connection.\footnote{Despite redundant at the level of eq.~\eqref{eqn:ch2_m_grav_lag}, the full covariant derivative $\nabla_\mu$ is necessary to derive the gravitino equations of motion and constraints, given in eq.~\eqref{eqn:ch2_gr_eom_1},\eqref{eqn:ch2_gr_const} and \eqref{eqn:ch2_gr_eom_2}, consistently with diffeomorphism invariance.} The equations of motion of the massive gravitino that follow from the lagrangian \eqref{eqn:ch2_m_grav_lag} are
\begin{equation}
	\gamma^{\mu\nu\rho}\nabla_\nu\Psi_\rho=m\gamma^{\mu\nu}\Psi_\nu \ .
\label{eqn:ch2_gr_eom_1}
\end{equation}

\noindent Contracting this equation with $\gamma^\mu$ and $\nabla_\mu$ yields two constraint equations, respectively
\beq
\begin{aligned}
    \gamma^{\mu\nu}\nabla_\mu\Psi_\nu=&\frac{3}{2}
    m\gamma^\mu\Psi_\mu \ , \\
    G_{\mu\nu}\gamma^\mu\Psi^\nu=&3m^2\gamma^\mu\Psi_\mu+2\(\partial_\mu m\)\gamma^{\mu\nu}\Psi_\nu \ ,
\end{aligned}
\label{eqn:ch2_gr_const}
\eeq

\noindent where $G_{\mu\nu}$ is the Einstein tensor. 
These constraints allow one to rewrite the equations of motion (\ref{eqn:ch2_gr_eom_1}) in alternative form
\begin{equation}
	\cancel{\nabla}\Psi^\mu-\nabla^\mu\(\gamma^\rho\Psi_\rho\)=-\frac{m}{2}\(\gamma^{\mu\rho}\Psi_\rho+3\Psi^\mu\) \ .
\label{eqn:ch2_gr_eom_2}
\end{equation}

Let us now specify these general equations to the FLRW background \eqref{eqn:ch2_flrw} under consideration. The gravitino lagrangian \eqref{eqn:ch2_m_grav_lag} is typically written in canonical form via the redefinition
\beq
    \Psi_\mu=\frac{1}{\sqrt{a}}\psi_\mu \ ,
\label{eqn:ch2_psi_cn}
\eeq

\noindent and after having expanded the whole dependence on the metric \eqref{eqn:ch2_flrw}, one obtains\footnote{For convenience, we employ in this section the following notation for the flat space Dirac matrices, which will carry the same Greek indices as the curved ones but denoted with a bar, \textit{i.e.} $$\gamma^\mu=e^\mu{}_a\gamma^a=a(\tau)^{-1}\delta^\mu{}_a\gamma^a\equiv a(\tau)^{-1}\fgamma^\mu$$.}
\beq
    \L_\psi=-\frac12\bpsi_\mu\fgamma^{\mu\nu\rho}\partial_\nu\psi_\rho+\frac{am}{2}\bpsi_\mu\fgamma^{\mu\nu}\psi_\nu-aH\bpsi_\rho\fgamma^\rho\psi_0 \ .
\label{eqn:ch2_grav_lag_cn}
\eeq

\noindent The equations of motion \eqref{eqn:ch2_gr_eom_1} become
\begin{equation}
	\fgamma^{\mu\nu\rho}\partial_\nu\psi_\rho-am\fgamma^{\mu\rho}\psi_\rho+aH\left(\fgamma^\mu\psi_0-\delta^{\mu}_0\fgamma^\rho\psi_\rho\right)=0 \ , 
\label{eqn:ch2_gr_eom_flrw_1}
\end{equation}

\noindent while the constraints \eqref{eqn:ch2_gr_const} are instead given by
\beq
    \partial_\mu\psi^\mu-\cancel{\partial}\left(\fgamma^\rho\psi_\rho\right)-2aH\psi_0+\frac{1}{2}\left(3am-aH\fgamma^0\right)\left(\fgamma^\rho\psi_\rho\right) =0 \ ,
\label{eqn:ch2_gr_const_flrw_1}
\eeq

\noindent and
\begin{equation}
    \fgamma^0\psi_0= C(\tau)\,\fgamma^i\psi_i \ . 
\label{eqn:ch2_gr_const_flrw_2}
\end{equation}

\noindent This second constraint implies that the component $\psi_0$ is actually not an independent field. The function $C(\tau)$ is equal to \cite{Kallosh:1999jj,Kallosh:2000ve,Kolb:2021xfn}
\begin{equation}
	C(\tau)=C_\textup{R}(\tau)+C_\textup{I}(\tau)\fgamma_0=\frac{H^2-3m^2-\nicefrac{R}{3}}{3\left(H^2+m^2\right)}+\frac{2m^{\prime}}{3a\left(H^2+m^2\right)}\fgamma_0  \ , \label{eqn:ch2_const_coeff}
\end{equation}

\noindent it reduces to $C\(\tau\)\to -1$ in the flat space limit and its square is equal to the sound speed parameter $\abs{C}^2=c_s^2$ we introduced in Section \ref{sub_ch1_oc}.

In the context of the low-energy effective theories of nonlinear rigid supersymmetry studied in Section \ref{ch1_sec_nls}, the sound speed parameter $c_s^2$ characterized the motion of the goldstino field $G$ (see eq.~\eqref{eqn:ch1_oc_sound_speed}, \eqref{eqn:ch1_csc_sound_speed} and \eqref{eqn:ch1_gsc_sound_speed}). In the local case, the sound speed indeed parameterizes the propagation of the longitudinal polarization states of the massive gravitino. The transverse states propagate instead at the speed of light (\textit{i.e.} $c_s=1$). This can be seen explicitly through a proper identification of the gravitino field $\psi_\mu$ into transverse and longitudinal modes. This is done in momentum space, in which one works via the expansion
\beq
    \psi_\mu(\tau,\vec{x})=\int\frac{d^3k}{(2\pi)^3}\psi_{\mu,k}(\tau)e^{i\vec{k}\cdot\vec{x}} \ , 
\label{eqn:ch2_psi_mom_space}
\eeq

\noindent where the independent space components $\psi_{i,k}$ can be decomposed as\footnote{More details about the notation and the algebraic rules for the momentum space operators are collected in Appendix \ref{app:hed_algebra}.}
\cite{Kallosh:1999jj,Kallosh:2000ve,Nilles:2001fg}
\begin{equation}
	\psi_{i,k}=\psi_{i,k}^\textup{t}+\left(P_\gamma\right)_i\vec{\fgamma}\cdot\vec{\psi}_k+\left(P_k\right)_i\vec{k}\cdot\vec{\psi}_k \ ,
\label{eqn:ch2_psi_i_dec_1}
\end{equation}

\noindent in terms of the two projectors
\begin{align}
	\left(P_\gamma\right)_i=\frac{1}{2}\left[\frac{k_i}{\mk}\left(\vec{k}\cdot\vec{\fgamma}\right)-\fgamma_i\right] \ , &&
	\left(P_k\right)_i=\frac{1}{2\mk}\left[\fgamma_i\left(\vk\cdot\vfgamma\right)-3k_i\right] \ .
\label{eqn:ch2_psi_proj}
\end{align}

\noindent The decomposition \eqref{eqn:ch2_psi_i_dec_1} is such that
\begin{equation}
  \left(P_\gamma\right)^i\psi_{i,k}^\textup{t}=\left(P_k\right)^i\psi_{i,k}^\textup{t}=0 \ ,
\end{equation}

\noindent which characterizes $\psi_{i,k}^\textup{t}$ as being the transverse gravitino component. Moreover, the two constraints \eqref{eqn:ch2_gr_const_flrw_1}-\eqref{eqn:ch2_gr_const_flrw_2} become, in momentum space, 
\begin{align}
	\vk\cdot\vpsi_k=\[\vk\cdot\vfgamma+ia\(m-H\fgamma^0\)\]\vfgamma\cdot\vpsi_k \ , && \fgamma^0\psi_{0,k}=C(\tau)\(\vfgamma\cdot\vpsi_k\) \ ,
\label{eqn:ch2_const_k}
\end{align}

\noindent so that the decomposition \eqref{eqn:ch2_psi_i_dec_1} writes as
\beq
    \psi_{i,k}=\psi_{i,k}^\textup{t}+\hO_i\theta_k \ ,
\label{eqn:ch2_psi_dec}
\eeq

\noindent where $\theta_k\equiv\fgamma^i\psi_{i,k}$ specifies the longitudinal gravitino components, which are extracted through the projective operator
\beq
    \hO_i=\frac{1}{\mk}\left\{-k_i\left(\vk\cdot\vfgamma\right)+\frac{ia}{2}\left[\fgamma_i\left(\vk\cdot\vfgamma\right)-3k_i\right]\left(m-H\fgamma^0\right)\right\} \ .
\label{eqn:ch2_psi_dec_O}
\eeq

Thus, making use of this decomposition, together with the redefinition
\begin{equation}
    \theta_k=\sqrt{\frac{3\(m^2+H^2\) a^2}{2\vk^2}}\fgamma^0\vartheta_k \ ,
\label{eqn:ch2_theta_cn}
\end{equation}

\noindent that canonically normalizes the field $\theta$, one finds that the massive gravitino lagrangian in FRLW \eqref{eqn:ch2_m_grav_lag} splits into transverse and longitudinal sectors \cite{Kallosh:1999jj,Kallosh:2000ve,Nilles:2001fg}
\beq
    \mathcal{L}_\psi=\mathcal{L}_\textup{t}+\mathcal{L}_\vartheta \ ,
\eeq

\noindent which are completely decoupled from each other. The transverse sector is equal to
\beq
    \mathcal{L}_\textup{t}=-\frac{1}{2}\sum_{i=1}^3\bpsi_{i,k}^\textup{t}\left[\fgamma^0\partial_0+i\kg+am\right]\psi_{i,k}^\textup{t} \ ,
\label{eqn:ch2_L_psi_t}
\eeq

\noindent and thus has the form of a standard spin-$\frac12$ fermion in FLRW, propagating with unitary sound speed and canonical time-dependent rescaling of the mass. On the contrary, the longitudinal sector is given by 
\begin{equation}
	 \mathcal{L}_\vartheta=-\frac{1}{2}\bar{\vartheta}_k\[\fgamma^0\partial_0-i\kg C+aM\]\vartheta_k \ ,
\label{eqn:ch2_L_theta}
\end{equation}

\noindent which shows that the dynamics of the longitudinal gravitino $\vartheta_k$ it describes is characterized, as anticipated around eq.~\eqref{eqn:ch2_const_coeff}, by a nontrivial sound speed
\beq
    c_s^2=\abs{C^2(\tau)}=C_\textup{R}^2(\tau)+C_\textup{I}^2(\tau) \ ,
\label{eqn:ch2_grav_sound_speed}
\eeq

\noindent as well as a nontrivial time-dependent mass parameter
\begin{equation}
    M \equiv -\[\frac{m}{2}+\frac{3}{2}\(mC_\textup{R}+HC_\textup{I}\)\] \ .
\label{eqn:ch2_grav_mass_M}
\end{equation}

The nontrivial sound speed is what makes the longitudinal gravitino dynamics in FLRW very peculiar. Going back to the starting model \eqref{eqn:ch2_sugra_inflation_oc_lag}, which we recall is built coupling pure supergravity to one nilpotent and one orthogonal superfield \cite{Ferrara:2015tyn}, we can write the sound speed \eqref{eqn:ch2_grav_sound_speed} in terms of the scalar field background by means of the associated Einstein equations \eqref{eqn:ch2_EE_flrw}. For a general background, one obtains\footnote{Note that eq.~\eqref{eqn:ch2_sound_speed_flrw} is written using the cosmological time $t$, for convenience of writing.}
\beq
    c_s^2=1-\frac{2\dot{\phi}^2}{\(\frac{\dot{\phi}^2}{2}+3m^2+V\)^2}\[V+3m^2-2\(m^{\prime}\)^2\] \ ,
\eeq
\label{eqn:ch2_grav_sound_speed_gen}

\noindent so that in the case of the theory \eqref{eqn:ch2_sugra_inflation_oc_lag}, of gravitino mass $m$ and scalar potential $V$ given in eq.~\eqref{eqn:ch2_sugra_oc_gr_mass} and \eqref{eqn:ch2_sugra_oc_sc_pot}, it results in
\beq
    c_s^2=1-\frac{2\dot{\phi}^2}{\(\frac{\dot{\phi}^2}{2}+f(\phi)^2\)^2}\[f(\phi)^2-2g^{\prime}(\phi)^2\] \ .
\label{eqn:ch2_sound_speed_flrw}
\eeq

\noindent This sound speed displays two critical limits, associated with equally critical behaviors. The first issue is that there is no condition, at this level, that prevents the sound speed to become superluminal during the scalar field evolution \cite{Dudas:2021njv,Bonnefoy:2022rcw} and this is clearly not acceptable from the point of view of the consistency of the theory. The second issue is instead phenomenological. It was in fact argued in \cite{Kolb:2021xfn,Kolb:2021nob} that the limit in which the sound speed goes to zero is associated with a divergent gravitational production of longitudinal gravitinos, signaling the breakdown of the theory before the Planck scale. 

We first encountered the superluminal sound speed problem in Chapter \ref{ch_susy}, while presenting the results from \cite{Bonnefoy:2022rcw}. In the next paragraph, we expand and complete the discussion of this topic from Section \ref{ch1_sec_nls}. In the following paragraph we instead elaborate more on the gravitational overproduction of gravitinos for vanishing sound speed, as discussed in \cite{Kolb:2021xfn}. Gravitational particle production of gravitinos, on the other hand, is also the focus of the work \cite{Casagrande:2023fjk}, and we will address it in detail in the next section.

\subsubsection{The \texorpdfstring{$\bm{c_s>1}$}{$c_s>1$} limit}

As reported in Section \ref{ch1_sec_nls}, we analyzed and solved the problem of superluminal sound speed in nonlinear supergravity in \cite{Bonnefoy:2022rcw}. The nontrivial causality constraint
\beq
    c_s\le 1 \,\, \longleftrightarrow\,\,f(\phi)^2\le 2g^{\prime}(\phi)^2 \ ,
\label{eqn:ch2_sugra_caus_const}
\eeq

\noindent that follows from eq.~\eqref{eqn:ch2_sound_speed_flrw} is understood as a consequence of the badly defined orthogonal constraint \eqref{eqn:ch1_oc}, which is used to construct the model \eqref{eqn:ch2_sugra_inflation_oc_lag}. This constraint requires in fact higher-derivative operators already in the UV \cite{DallAgata:2016syy}, so that the associated effective field theory is not guaranteed to be well-defined a priori. By means of the gravitino equivalence theorem \cite{Fayet:1986zc,Casalbuoni:1988kv,Casalbuoni:1988qd} \eqref{eqn:ch2_sugra_eq_theorem}, in \cite{Bonnefoy:2022rcw} we address this problem in terms of the goldstino in the rigid limit. The goldstino sound speed, which we wrote in eq.~\eqref{eqn:ch1_oc_sound_speed}, actually matches the longitudinal gravitino one \eqref{eqn:ch2_sound_speed_flrw}, since no term scales with the Planck mass $\MP$ in \eqref{eqn:ch2_sound_speed_flrw}. Then, from this goldstino perspective the causality constraint \eqref{eqn:ch1_oc_sound_speed_bound}-\eqref{eqn:ch2_sugra_caus_const} is related to positivity bounds on the scattering amplitudes \cite{Bonnefoy:2022rcw,Adams:2006sv,Bellazzini:2016xrt} and we interpret it, from the bottom-up point of view, as an obstruction to a UV completion into a microscopic two-derivative theory. This is the main result of \cite{Bonnefoy:2022rcw}, which we described in detail in Section \ref{ch1_sec_nls} of this manuscript.

In \cite{Bonnefoy:2022rcw} we also formulate improved theories like \eqref{eqn:ch2_sugra_inflation_oc_lag}, based on the generalized constraint \eqref{eqn:ch1_gsc} of \cite{DallAgata:2016syy}. These theories, defined by the constraints of eq.~\eqref{eqn:ch1_gsc_oc_sc}-\eqref{eqn:ch1_gsc_oc_f}, maintain the minimality of the spectrum of \eqref{eqn:ch2_sugra_inflation_oc_lag} but are not subject to any causality constraint. This can be seen explicitly from the associated goldstino sound speed, given in eq.~\eqref{eqn:ch1_gsc_sound_speed}, which is subluminal all along the evolution of the background scalar field. Moving then to the associated supergravity model, following the equivalence theorem backwards, the longitudinal gravitino sound speed \eqref{eqn:ch2_grav_sound_speed_gen} again matches its rigid goldstino counterpart. The supergravity scalar potential \eqref{eqn:ch2_sugra_scal_pot_F} for this improved constraints setup is in fact equal to
\beq
    V_\textup{improved}=f(\phi)^2+2g^{\prime}(\phi)^2-3g(\phi)^2 \ ,
\eeq

\noindent so that the sound speed of this model is computed from eq.~\eqref{eqn:ch2_grav_sound_speed_gen} to be
\beq
    c_s^2=1-\frac{2\dot{\phi}^2f(\phi)^2}{\(\frac{\dot{\phi}^2}{2}+f(\phi)^2+g^{\prime}(\phi)^2\)^2} \ ,
\label{eqn:ch2_sugra_sound_speed_improved}
\eeq

\noindent which is indeed exactly equal to the goldstino one of eq.~\eqref{eqn:ch1_csc_sound_speed}, and thus it is always subluminal. Therefore, the construction we put forward in \cite{Bonnefoy:2022rcw} defines consistent effective field theories of nonlinear supergravity and represent an improved class of minimal supergravity models for inflation with respect to \eqref{eqn:ch2_sugra_inflation_oc_lag}. 

Note, however, that also the improved sound speed \eqref{eqn:ch1_csc_sound_speed}-\eqref{eqn:ch2_sugra_sound_speed_improved} can vanish during the scalar field evolution, so that the problem of the longitudinal gravitino overproduction \cite{Kolb:2021xfn} does affect, from this point of view, also the improved models of \cite{Bonnefoy:2022rcw}.

\subsubsection{The \texorpdfstring{$\bm{c_s\to 0}$}{$c_s\to 0$} limit and gravitational particle production}

The production of gravitinos, which becomes divergent in the $c_s\to0$ limit for the orthogonal constraint model \eqref{eqn:ch2_sugra_inflation_oc_lag}, exemplifies a well-known and significant phenomenon in cosmology --- and more generally in quantum field theory in expanding spacetime backgrounds --- namely, gravitational particle production \cite{Parker:1969au,Zeldovich:1971mw,Ford:1986sy,Birrell:1982ix,Gorbunov:2011zzc}. This is a quantum, non-perturbative process in which particles are produced from the vacuum as a consequence of the expanding spacetime background. This phenomenon plays a crucial role in the theory of reheating of the universe after inflation \cite{Kofman:1994rk,Kofman:1997yn}. In supergravity, it has been applied to study the production of gravitinos in the early universe, in order to test its compatibility with primordial nucleosynthesis \cite{Kallosh:1999jj, Kallosh:2000ve, Giudice:1999yt, Giudice:1999am, Nilles:2001fg, Nilles:2001ry} and the general viability and consistency of supergravity models for cosmology \cite{Hasegawa:2017hgd, Kolb:2021xfn, Kolb:2021nob,Dudas:2021njv,Antoniadis:2021jtg}.

In an expanding background, the notion of number of particles in a quantum state is intrinsically ambiguous because there is no unique definition of vacuum state \cite{Birrell:1982ix,Gorbunov:2011zzc}. The standard procedure is to consider the two initial and final stages of the evolution, the far past and future, in which the time dependence can be assumed to be turned off and a vacuum state can then be defined according to the standard rules of quantum mechanics. In quantum field theory, this is typically done through the Hamiltonian for the canonically normalized fields \cite{Zeldovich:1971mw,Kofman:1997yn}. The two ingoing and outgoing Fock spaces are, a priori, different, but can be related through the so-called Bogolyubov transformations \cite{Parker:1969au,Zeldovich:1971mw,Ford:1986sy,Birrell:1982ix,Gorbunov:2011zzc}. Schematically, these transformations act on the two sets of ladder operators $\{a_\textup{in},a^\dagger_\textup{in}\}$ and $\{a_\textup{out},a^\dagger_\textup{out}\}$ as
\beq
    \begin{pmatrix} a_\textup{out} \\ a^\dagger_\textup{out}\end{pmatrix}=\begin{pmatrix} \alpha & \beta^\dagger \\ \beta & \alpha^\dagger \end{pmatrix}\begin{pmatrix} a_\textup{in} \\ a^\dagger_\textup{in}\end{pmatrix} \ .
\label{eqn:ch2_bog_gen}
\eeq

\noindent The Bogolyubov coefficients $\alpha$ and $\beta$ are usually time-dependent and must be normalized as
\beq
    \abs{\alpha}^2-\abs{\beta}^2=1 \ ,
\label{eqn:ch2_bog_wronsk_gen}
\eeq

\noindent in order to be consistent with the algebra of the ladder operators. Then, particles are said to have been gravitationally produced in the sense that the standard number operator $N_\textup{out}=a_\textup{out}^\dagger a_\textup{out}$ associated with the outgoing vacuum has a nontrivial expectation value with respect to the ingoing one, parameterized by the off-diagonal Bogolyubov coefficient $\abs{\beta}^2$:
\beq
    \langle N_\textup{out}\rangle_\textup{in}=\abs{\beta}^2 \ .
\label{eqn:ch2_bog_number_op_gen}
\eeq

This formalism was applied in \cite{Hasegawa:2017hgd,Giudice:1999am,Giudice:1999yt,Nilles:2001fg,Nilles:2001ry,Kolb:2021xfn} to study gravitational production of gravitinos, and in \cite{Kolb:2021xfn} the orthogonal supergravity model \eqref{eqn:ch2_sugra_inflation_oc_lag} is considered. 
After decomposing the massive gravitino into its transverse and longitudinal modes, as described in eq.~\eqref{eqn:ch2_L_psi_t} and \eqref{eqn:ch2_L_theta}, the Bogolyubov coefficient $\beta$ is obtained by numerically solving the equations of motion for both polarizations. While the rate of transverse gravitino production is suppressed at high momenta, as physically expected, the rate of longitudinal gravitino production is instead found to progressively increase almost linearly with the associated momentum in the case where the sound speed parameter \eqref{eqn:ch2_grav_sound_speed} vanishes. This relation with the sound speed parameter can be understood more analytically by examining the adiabaticity parameter associated with gravitino propagation \cite{Kolb:2021xfn}. For a field $\varphi_k(\tau)$ propagating with oscillator-like dispersion relations, \textit{i.e.} of the type
\begin{equation}
    \partial_\tau^2\varphi_k+\omega_k^2\varphi_k=0 \ ,
\label{eqn:ch2_disp_rel_gen}
\end{equation}

\noindent in which $c_s$ is the sound speed, the adiabaticity parameter can be defined as
\beq   
    A_k(\tau)=\frac{\partial_\tau\omega_k}{\omega_k^2} \ .
\eeq

\noindent The longitudinal gravitino can be seen from \eqref{eqn:ch2_L_theta} to propagate with
\beq
    \omega_k^2=c_s^2k^2+a^2M^2 \ ,
\eeq

\noindent so that \cite{Kolb:2021xfn}
\beq
    A_k(\tau)=\frac{c_s\(\partial_\tau c_s\)\vk^2+a^3HM^2+a^2M\partial_\tau M}{\(c_s^2k^2+a^2M^2\)^{\frac32}} \ .
\label{eqn:ch2_long_grav_adiab_par}
\eeq   

\noindent Thus, for a non-vanishing sound speed, the adiabaticity parameter in eq.~\eqref{eqn:ch2_long_grav_adiab_par} approaches zero at high momenta, reflecting a suppression in the rate of production of the corresponding states. This is what happens for the transverse gravitino, which propagates with sound speed equal to 1, as seen in eq.~\eqref{eqn:ch2_L_psi_t}. On the contrary, if the sound speed vanishes, the adiabaticity parameter no longer depends on the momentum of the produced particles, meaning that particles with arbitrarily large momentum can be produced. Such a process clearly violates unitarity and signals the breakdown of the effective field theory.

This breakdown of the theory when the gravitino is quantized does not have a clear connection with the effective field theory setup itself. In fact, although the orthogonal constraint \eqref{eqn:ch1_oc}, upon which the model \eqref{eqn:ch2_sugra_inflation_oc_lag} is based, does not define reliable effective theories due to its higher-derivative UV origin \cite{DallAgata:2016syy, Bonnefoy:2022rcw}, the improved models we constructed in \cite{Bonnefoy:2022rcw} do have a well-defined microscopic origin, and yet they can suffer from the problematic overproduction, as the associated sound speed \eqref{eqn:ch2_sugra_sound_speed_improved} is not prevented from vanishing during the background scalar field evolution. Moreover, the same mechanism was studied in \cite{Dudas:2021njv} for several effective constrained superfield theories and it was shown that when a second fermion (the would-be inflatino) is present in the spectrum, its mixing with the longitudinal gravitino prevents the problematic unbounded production. Finally, the lagrangian \eqref{eqn:ch2_sugra_inflation_oc_lag} can also describe ---  within some assumptions --- pure supergravity coupled to a chiral multiplet,\footnote{In this last case, the chiral multiplet is the one containing the inflaton $\phi$, and its fermion superpartner is the goldstino during inflation, which can be gauged away in the unitary gauge. The scalar field is, however, complex, so that the setup \eqref{eqn:ch2_sugra_inflation_oc_lag} applies if the additional scalar is very heavy or frozen during inflation. If, on the contrary, the additional scalar mode undergoes a nontrivial dynamics, some changes are expected.} \cite{Kallosh:1999jj,Kallosh:2000ve} which is in principle defined up to the Planck scale.\\

These unclear and exceptional features of nonlinear supergravity theories motivated us to look more closely into the framework of gravitational particle production. We now report on our findings \cite{Casagrande:2023fjk} in the next sections.

\subsection{Gravitational particle production revisited -- a first example}\label{ch2_sub_gpp_scalar}

Gravitational particle production is a subtle mechanism that can reveal unexpected features of a given theory, both phenomenologically and in terms of theoretical consistency. Its richness stems from the way it intertwines particle physics, general relativity, and quantum field theory. From this point of view, the case of the gravitino discussed so far is particularly interesting because of its tight connection with gravity, which strongly persists also when supersymmetry is spontaneously broken and the gravitino becomes massive, as shown by its general equations of motion \eqref{eqn:ch2_gr_eom_1}-\eqref{eqn:ch2_gr_const}. This perspective was pushed in \cite{Kolb:2021xfn,Kolb:2021nob} by trying to make use of gravitino gravitational production to infer information about the properties of quantum gravity along the lines of the Swampland program \cite{Vafa:2005ui,Palti:2019pca}.

However, despite being at the basis of this mechanism, the time-dependence of the background in which gravitational particle production occurs also gives rise to several ambiguities in its physical interpretation. First of all, as mentioned at the end of the previous section, there is no unique, unambiguous way to identify the vacuum state of the theory \cite{Birrell:1982ix,Gorbunov:2011zzc}. Consequently, the definition of the number operator and the concept of number of produced particles can only be established relative to two chosen vacua at different times, in the sense expressed by the Bogolyubov transformation \eqref{eqn:ch2_bog_gen}. No intrinsic definitions of this type are actually possible \cite{Birrell:1982ix}. 

These ambiguities are reflected also in the hamiltonian formalism that is commonly used to quantify the gravitational particle production phenomenon, which is indeed subtle in time-dependent backgrounds \cite{Fulling:1979ac,Weiss:1986gg,Bozza:2003pr,Grain:2019vnq}. The quantization of a theory is typically performed using the hamiltonian of the canonically normalized fields. This hamiltonian --- which we will call \textit{instantaneous} in what follows --- is usually of the harmonic oscillator type and thus allows one to perform the quantization of the theory in a standard way \cite{Kofman:1997yn}. This instantaneous hamiltonian describes correctly the propagation of the associated canonically normalized field. However, it does not define the correct energy operator of the system. This is given by the energy-momentum tensor, entering Einstein equations, which is computed from the original covariant action. In hamiltonian terms, the same energy operator is obtained by working with the original, non-rescaled fields. However, these fields have a more complicated time evolution, so that the quantization of the system is less straightforward. 

The differences between the various Hamiltonian formulations are well known \cite{Fulling:1979ac,Weiss:1986gg,Bozza:2003pr,Grain:2019vnq}. While they effectively coincide for theories in flat space, they no longer align when the theory is placed in a time-dependent background. The mismatch arises from the fact that, to canonically normalize the starting covariant field, one must perform a time-dependent rescaling that breaks the field transformation properties under the original diffeomorphisms. However, the coupling to gravity selects the stress-energy tensor as unambiguous physical energy operator, being the source of the metric in the Einstein equations. Therefore, the differences with respect to the instantaneous Hamiltonian acquire physical significance once the latter is used to quantize the theory. In particular, it follows that the vacuum state defined via the instantaneous hamiltonian, from which particle production is usually computed \cite{Zeldovich:1971mw,Kofman:1997yn}, is not a state of minimum energy. On the other hand, physical processes involving such produced particles, like for example their backreaction on the spacetime geometry, should have a well-defined description in terms of the Einstein equations. From this point of view, the natural choice for the vacuum would be the state of zero energy as computed via the stress-energy tensor. The time evolution would then be governed by a different Bogolyubov transformation, so that the resulting final state cannot be interpreted as containing particles relative to the instantaneous Hamiltonian vacuum.

The physical consequences of this difference between the instantaneous hamiltonian and the stress-energy tensor are the subject of the work \cite{Casagrande:2023fjk}. In particular, we parameterize it in terms of a specific Bogolyubov transformation, which we compute analytically for massive particles of different spin propagating in an FLRW spacetime. Whereas for fields of spin 0, $\frac12$ and 1 the described difference is under control from a physical point of view, the case of the gravitino displays, again, some subtle (if not puzzling) features.

We now present the results of \cite{Casagrande:2023fjk}, and we begin with the well-known example of the real scalar field. The more interesting case of the gravitino is discussed in detail in the section that follows.

A massive scalar field $\Phi(\tau,\vx)$ in FLRW spacetime \eqref{eqn:ch2_flrw} is described by the action
\beq
\begin{aligned}
    S_\Phi=&\int d^4x{}\g{} \ \mathcal{L}_\Phi= \\
 =&-\frac12\int d^4x \g{} \ \(\partial_\mu\Phi\partial^\mu\Phi+m^2\Phi^2\)=\int d^4x \ \frac{a^2}{2}\[\(\Phi^{\prime}\)^2-\(\vec{\nabla}\Phi\)^2-a^2m^2\Phi^2\] \ .
\end{aligned}
\label{eqn:ch2_gpp_s_action}
\eeq

\noindent Canonical normalization is obtained via the time-dependent rescaling
\begin{equation}
	\Phi \equiv \frac{\varphi}{a} \ ,
\label{eqn:ch2_gpp_s_field_redef}
\end{equation}

\noindent leading to
\begin{equation}
\begin{aligned}
	S_\varphi=&\int d^4x \ {}\mathcal{L}_\varphi= \frac{1}{2} \int d^4x{} \left\{\(\varphi^{\prime}\)^2-\varphi\(-\vec{\nabla}^2+a^2m^2-\frac{a^{\prime\prime}}{a}\)\varphi\right\} \ .
\end{aligned}
\label{eqn:ch2_gpp_s_cn_action}
\end{equation}

\noindent Notice that in this way the action has the simplified form of a massive scalar field in flat space with time-dependent mass. Working in momentum space via the decomposition\footnote{The subscript $\bm{k}$ denotes dependence on the whole vector $\vec{k}$, while the subscript $k$ indicates dependence only on its modulus.}
\begin{equation}
	\varphi\(\tau,\vec{x}\)=\int \frac{d^3k}{\(2\pi\)^{\nicefrac{3}{2}}}\left\{\varphi_{\bm{k}}(\tau)e^{i\bm{k}\cdot\bm{x}}+\bar{\varphi}_{\bm{k}}(\tau)e^{-i\bm{k}\cdot\bm{x}}\right\} \ .
\end{equation}

\noindent the equation of motion becomes the one of the harmonic oscillator:
\begin{equation}
	\varphi^{\prime\prime}_{\bm{k}}+\omega^2{}\varphi_{\bm{k}}=0 \ ,
\label{eqn:ch2_gpp_s_eom} 
\end{equation}

\noindent with time-dependent frequency 
\begin{equation}
	\omega^2\equiv\vec{k}^2+a^2m^2-\frac{a^{\prime\prime}}{a}\equiv\omega_0^2-\frac{a^{\prime\prime}}{a}\ .
 \label{eqn:ch2_gpp_omega}
\end{equation}

\subsubsection{The instantaneous hamiltonian} 

Due to its simple, harmonic oscillator-like structure, the setup defined by the canonically normalized field $\varphi$ --- rather than the original $\Phi$ in \eqref{eqn:ch2_gpp_s_action} --- is the natural choice for quantizing the system, which can be done in the standard way. The conjugate momentum is given by
\beq
    \Pi_\varphi \equiv \frac{\partial \L_\varphi}{\partial \varphi^{\prime}}=\varphi^{\prime} \ ,
\eeq

\noindent so that the hamiltonian density $\H_\varphi$ associated with the canonically normalized action of eq.~(\ref{eqn:ch2_gpp_s_cn_action}) is equal to 
\begin{align}
\H_\varphi=\frac{1}{2}\left[\Pi_\varphi^2+\varphi\(-\vec{\nabla}^2+a^2m^2-\frac{a^{\prime\prime}}{a}\)\varphi\right] \ .
\label{eqn:ch2_gpp_s_H_cn}
\end{align}

\noindent This is what we called instantaneous hamiltonian. Quantization is realized upgrading the conjugate variables to operators and imposing the usual (equal-time) commutation relation
\begin{equation}
	\[\varphi(\tau,\vec{x}), \Pi_\varphi(\tau,\vec{y})\]=i{}\delta^{(3)}\(\vec{x}-\vec{y}\) \ .
\label{eq:ch2_gpp_cr1}
\end{equation} 

\noindent The variables $\varphi$ and $\Pi_\varphi$ can then be expanded in terms of ladder operators as\footnote{The mode functions $\varphi_k$ and $\Pi_k$ in eq.~\eqref{eqn:ch2_gpp_s_q_dec} depend only on $k$ because the equation of motion \eqref{eqn:ch2_gpp_s_eom} is isotropic.}
\begin{equation}
\begin{aligned}
	\varphi(\tau,\vec{x})=&\int \frac{d^3k}{\(2\pi\)^{\nicefrac{3}{2}}}\left\{\varphi_{k}(\tau)e^{i\bm{k}\cdot\bm{x}}{}a_{\bm{k}}+\bar{\varphi}_k(\tau)e^{-i\bm{k}\cdot\bm{x}}{}a_{\bm{k}}^\dagger\right\},\\
	\Pi_\varphi(\tau,\vec{x})=&\int \frac{d^3k}{\(2\pi\)^{\nicefrac{3}{2}}}\left\{\Pi_k(\tau)e^{i\bm{k}\cdot\bm{x}}{}a_{\bm{k}}+\bar{\Pi}_k(\tau)e^{-i\bm{k}\cdot\bm{x}}{}a_{\bm{k}}^\dagger\right\} \ .
\end{aligned}
\label{eqn:ch2_gpp_s_q_dec}
\end{equation}

\noindent The ladder operators $a_{\bm{k}}$ and $a_{\bm{k}}^\dagger$ satisfy as well the usual commutation relation
\begin{equation}
	\[a_{\bm{k}},a_{\bm{k^{\prime}}}^\dagger\]=\delta^{(3)}\(\vec{k}-\vec{k}^{\prime}\), \qquad \qquad \[a_{\bm{k}},a_{\bm{k^{\prime}}}\]=0=\[a_{\bm{k}}^\dagger,a_{\bm{k^{\prime}}}^\dagger\] \ ,
\label{eqn:ch2_gpp_s_ca_comm}
\end{equation}

\noindent if one imposes the additional "wronskian" condition
\begin{equation}
	\varphi_k\bar{\Pi}_k-\bar{\varphi}_k\Pi_k=i \ .
\label{eqn:ch2_gpp_s_wronsk}
\end{equation}

\noindent Applying this decomposition to the instantaneous hamiltonian (\ref{eqn:ch2_gpp_s_H_cn}) yields \cite{Nilles:2001fg}
\begin{equation}
\begin{aligned}
	H_\varphi=&\int d^3k \ {}\H_\varphi=\frac{1}{2}\int d^3k {} \begin{pmatrix}a^\dagger_{\bm{k}} & a_{-\bm{k}}\end{pmatrix}\begin{pmatrix}\E_k & \F^\dagger_k\\ \F_k & \E_k\end{pmatrix}\begin{pmatrix}a_{\bm{k}} \\ a^\dagger_{-\bm{k}}\end{pmatrix} \ , 
\end{aligned}
\label{eqn:ch2_gpp_s_H_EF}
\end{equation}

\noindent where we introduced the functions
\begin{align}
	\mathcal{E}_k=& |\Pi_k|^2+\omega^2|\varphi_k|^2 \ , & \mathcal{F}_k=\Pi_k^2+\omega^2\varphi_k^2 \ .
\label{eqn:ch2_gpp_s_E_F}
\end{align}

This hamiltonian can be brought in diagonal form by means of a Bogolyubov decomposition
\begin{align}
    \varphi_k=\frac{1}{\sqrt{2\omega}}\(\alpha_k+\beta_k\) \ , &&
	\Pi_k=-\frac{i\omega}{\sqrt{2\omega}}\(-\alpha_k+\beta_k\) \ . 
\label{eqn:ch2_gpp_s_bog_dec}
\end{align}

\noindent The Bogolyubov coefficients $\alpha_k$ and $\beta_k$ are constrained by the wronskian condition (\ref{eqn:ch2_gpp_s_wronsk}) to be such that
\begin{equation}
    \abs{\alpha_k}^2-\abs{\beta_k}^2=1 \ ,
\end{equation}

\noindent while the $\varphi_k$ equation of motion \eqref{eqn:ch2_gpp_s_eom} turns into the system of equations
\begin{align}
    \alpha^{\prime}_k=-i\omega\alpha_k+\frac{\omega^{\prime}}{2\omega}\beta_k \ , &&  \beta^{\prime}_k= i\omega\beta_k +\frac{\omega^{\prime}}{2\omega}\alpha_k \ .
\label{eqn:ch2_gpp_s_bog_eom}
\end{align}

\noindent As introduced schematically in eq.~\eqref{eqn:ch2_bog_gen}, the Bogolyubov transformations are maps between different sets of ladder operators. They in fact leave invariant, by construction, the commutation relations \eqref{eqn:ch2_gpp_s_ca_comm}. In the case at hand, the Bogolyubov decomposition \eqref{eqn:ch2_gpp_s_bog_dec} defines the basis of ladder operators
\begin{equation}
	\begin{pmatrix} \tilde{a}_{\bm{k}}\\\tilde{a}_{-\bm{k}}^\dagger\end{pmatrix}=\begin{pmatrix} \alpha_k & \beta_k^\dagger \\ \beta_k & \alpha_k^\dagger\\\end{pmatrix}\begin{pmatrix} a_{\bm{k}}\\ a_{-\bm{k}}^\dagger\end{pmatrix} \ ,
\label{eqn:s_bog_tr_1}
\end{equation}

\noindent in which the instantaneous hamiltonian \eqref{eqn:ch2_gpp_s_H_EF}) becomes diagonal  \cite{Nilles:2001fg}:
\begin{equation}
\begin{aligned}
	H_\varphi=&\frac{1}{2}\int d^3k{} \begin{pmatrix}a^\dagger_{\bm{k}} & a_{-\bm{k}}\end{pmatrix}\begin{pmatrix} \alpha_k^\dagger & \beta_k^\dagger \\ \beta_k & \alpha_k \\\end{pmatrix}\begin{pmatrix}\omega & 0\\ 0 & \omega\end{pmatrix}\begin{pmatrix} \alpha_k & \beta_k^\dagger \\ \beta_k & \alpha_k^\dagger\\\end{pmatrix}\begin{pmatrix}a_{\bm{k}} \\ a^\dagger_{-\bm{k}}\end{pmatrix}=\\
 =&\int d^3k{}\frac{\omega}{2}\(\tilde{a}_{\bm{k}}^\dagger\tilde{a}_{\bm{k}}+\tilde{a}_{\bm{k}}\tilde{a}_{\bm{k}}^\dagger\) \ ,
\end{aligned}
\label{eqn:ch2_gpp_s_H_diag}
\end{equation}

\noindent In particular, the resulting diagonal hamiltonian is the one of a harmonic oscillator with (time-dependent) frequency $\omega$. 

The diagonalization of the hamiltonian is performed because in this form it has a direct interpretation in terms of number of particles and associated occupation numbers. This is computed, at a given (conformal) time $\tau$ of the evolution, through the number operator $N_{\varphi}(\tau) = \tilde{a}_{\bm{k}}^\dagger\tilde{a}_{\bm{k}}$, as in eq.~\eqref{eqn:ch2_bog_number_op_gen}. Following this perspective, it is natural to define the vacuum of the theory at initial time $\tau_0$ as a state with no particles. This is possible if the initial conditions for the Bogolyubov coefficients \eqref{eqn:ch2_gpp_s_bog_dec}--\eqref{eqn:ch2_gpp_s_bog_eom} are set to be
\begin{align}
    \alpha_l\(\tau_0\)=1 \ , && \beta_k\(\tau_0\)= 0 \ ,
\end{align}

\noindent so that the vacuum state is the one annihilated by all the operators $a_k$:
\begin{equation}
    a_{\bm{k}} | 0 \rangle_{\rm BD} \equiv {\tilde a}_{\bm{k}} (\tau_0)| 0 \rangle_{\rm BD} \equiv 0 \ . \label{eqn:ch2_BD}
\end{equation}

\noindent This is the so-called Bunch--Davies vacuum \cite{Birrell:1982ix}. Thus, although initially diagonal, the time evolution of the system, governed by the Bogolyubov coefficients through the equations of motion \eqref{eqn:ch2_gpp_s_bog_eom}, causes the Hamiltonian to evolve into a non-diagonal form. Therefore, the analogue of the Bunch--Davies vacuum \eqref{eqn:ch2_BD} at time $\tau>\tau_0$ has a non-vanishing occupation number with respect to the initial one, proportional to the off-diagonal Bogolyubov coefficient
\beq
    \left\langle N_\varphi(\tau)\right\rangle_\textup{BD}=\abs{\beta_k(\tau)}^2 \ .
\label{eqn:ch2_gpp_s_N_beta_2}
\eeq

\noindent The interpretation of this dynamics is that the time-dependent background sources the production of particles during the evolution of the system.

\subsubsection{The stress-energy tensor} 

Defined in this way, the Bunch--Davies vacuum is the ground-state of the instantaneous hamiltonian \eqref{eqn:ch2_gpp_s_H_cn}. In standard flat space quantum field theory, the above discussion would lead to the interpretation of the excited states with respect to this vacuum as states with $\langle N_\varphi\rangle_\textup{BD}$ particles, of total energy given by the (continuous) sum of all the single-particle energies $\sim \omega_k\abs{\beta_k}^2$, according to eq.~\eqref{eqn:ch2_gpp_s_H_diag}-\eqref{eqn:ch2_gpp_s_N_beta_2}. However, this is not true anymore for curved backgrounds. In this case, the energy must be described by the energy-momentum tensor, in order to be consistent with Einstein equations and the general curved spacetime setup. The instantaneous hamiltonian \eqref{eqn:ch2_gpp_s_H_cn} quantifies correctly the time evolution of the field $\varphi$ and its quantization, but it does not coincide with the energy operator given by the stress-energy tensor and therefore does not describe the energy of the system.

We now see this explicitly. The stress-energy tensor of the scalar field --- computed from the starting covariant action \eqref{eqn:ch2_gpp_s_action} --- is
\begin{equation}
	T_{\mu\nu}=\partial_\mu\Phi\partial_\nu\Phi-\frac{1}{2}g_{\mu\nu}\[\(\partial\Phi\)^2+m^2\Phi^2\] \ .
\label{eqn:ch2_gpp_s_T}
\end{equation}

\noindent The energy operator is encoded in the $00$ component of this tensor. In terms of the canonically normalized field defined in eq.~\eqref{eqn:ch2_gpp_s_field_redef}, it is equal to 
\begin{equation}
\begin{aligned}
	E_\varphi = \int d^3x \g{} \ T^0{}_0=&\int d^3x{}\frac{a^2}{2}\left\{\(\Phi^{\prime}\)^2+\(\vec{\nabla}\Phi\)^2+a^2m^2\Phi^2\right\}=\\
 =&\frac{1}{2}\int d^3x{}\left\{\(\varphi^{\prime}-aH\varphi\)^2+\varphi\(-\vec{\nabla}{}^2+a^2m^2\)\varphi\right\} \ ,
\end{aligned}
\label{eqn:ch2_gpp_s_E}
\end{equation}

\noindent which is clearly different from the instantaneous hamiltonian \eqref{eqn:ch2_gpp_s_H_cn}. Performing then the canonical quantization (\ref{eqn:ch2_gpp_s_q_dec}) and applying the Bogolyubov decomposition (\ref{eqn:ch2_gpp_s_bog_dec}), one obtains
\begin{equation}
    E_\varphi=\frac{1}{2}\int d^3k{} \begin{pmatrix}\tilde{a}^\dagger_{\bm{k}} & \tilde{a}_{-\bm{k}}\end{pmatrix}\begin{pmatrix}\mathscr{E}_k & \mathscr{F}_k^\dagger\\ \mathscr{F}_k & \mathscr{E}_k\end{pmatrix}\begin{pmatrix}\tilde{a}_{\bm{k}} \\ \tilde{a}^\dagger_{-\bm{k}}\end{pmatrix} \ ,
\label{eqn:ch2_gpp_s_E_nd}
\end{equation}

\noindent where $\{\tilde{a}_{\bm{k}},\tilde{a}^\dagger_{\bm{k}}\}$ is the Fock space associated with the diagonalized instantaneous hamiltonian (\ref{eqn:ch2_gpp_s_H_diag}), whereas the functions $\mathscr{E}_k$ and $\mathscr{F}_k$ are defined to be
\begin{equation}
\begin{aligned}
    \mathscr{E}_k=&\omega_0^2+\omega^2+a^2H^2 \ ,\\ \mathscr{F}_k=&\omega_0^2-\omega^2+a^2H^2-2iaH\omega \ .
\end{aligned}
\label{eqn:ch2_gpp_s_E_EF}
\end{equation}

Thus, the energy operator $E_\varphi$ is not diagonal in the basis where the instantaneous hamiltonian \eqref{eqn:ch2_gpp_s_H_diag} is. To put also $E_\varphi$ in diagonal form, one needs a further Bogolyubov transformation, which can be computed explicitly to be \cite{Casagrande:2023fjk}
\begin{equation}
	\begin{pmatrix} \tilde{a}_{\bm{k}} \\ \tilde{a}_{-\bm{k}}^\dagger\end{pmatrix}\equiv	\begin{pmatrix} \halpha_k & \hbeta_k^\dagger \\ \hbeta_k & \halpha_k^\dagger \end{pmatrix}\begin{pmatrix} \hat{a}_{\bm{k}} \\ \hat{a}_{-\bm{k}}^\dagger\end{pmatrix} \ ,
\label{eqn:ch2_gpp_s_2_bog}
\end{equation}

\noindent with 
\begin{equation}
	\halpha_k=\frac{\omega+\omega_0+iaH}{2\sqrt{\omega\omega_0}} \ , \qquad \qquad \hbeta_k=\frac{\omega-\omega_0-iaH}{2\sqrt{\omega\omega_0}} \ .
\label{eqn:ch2_gpp_s_2_bog_coeff}
\end{equation}

\noindent The resulting diagonal energy operator is therefore
\begin{equation}
E_\varphi=\int d^3k{}\frac{\omega_0}{2}\(\hat{a}_{\bm{k}}^\dagger\hat{a}_{\bm{k}}+\hat{a}_{\bm{k}}\hat{a}_{\bm{k}}^\dagger\) \ .
\label{eqn:ch2_gpp_s_E_diag}
\end{equation}

\noindent Notice that whereas the diagonal instantaneous hamiltonian \eqref{eqn:ch2_gpp_s_H_diag} involved the shifted frequency $\omega$, the diagonal energy involves instead the simpler $\omega_0$, the two frequencies being related according to eq.~\eqref{eqn:ch2_gpp_omega}.

\subsubsection{On the difference between the two diagonal Fock spaces}

This computation shows explicitly the claimed difference between the instantaneous hamiltonian and the energy operator computed from the stress-energy tensor and, with them, their associated Fock spaces. Because of the curved spacetime setting, the hamiltonian and the energy-momentum tensor are formally identified only as long as the covariant formulation is employed. This is clearly not the case for the instantaneous hamiltonian \eqref{eqn:ch2_gpp_s_H_cn} of the canonically normalized field \eqref{eqn:ch2_gpp_s_field_redef}-\eqref{eqn:ch2_gpp_s_cn_action}. The time-dependent rescaling \eqref{eqn:ch2_gpp_s_field_redef} breaks by construction the original covariance, and the resulting instantaneous hamiltonian is not connected anymore in a simple way to the starting ("covariant") one \cite{Grain:2019vnq}. 

The immediate physical consequence of this mismatch is that the zero-particle ground state \eqref{eqn:ch2_BD} of the instantaneous hamiltonian is not a state of minimal energy
\begin{equation}
    \left\langle E_{\varphi} (\tau_0) \right \rangle_\textup{BD} =V_3 \[\omega_0 \(  \frac{1}{2}+ |{\hat \beta}_k|^2 \)\](\tau_0)  = V_3\[\frac{\omega^2 + \omega_0^2 + a^2 H^2}{4 \omega}\](\tau_0) \ ,  \label{eq:ch2_gpp_energyBD}     
\end{equation}

\noindent where we have introduced the space volume $V_3$ as the momentum space delta function $V_3 = \delta^3 ({\bm k}=0)$. In the same way, the state of minimum energy that can be defined through \eqref{eqn:ch2_gpp_s_E_diag} does not have a definite interpretation in terms of number of particles from the point of view of the Fock space of the instantaneous hamiltonian \eqref{eqn:ch2_gpp_s_H_diag}. Actually, the total number of such particles, obtained by integrating $\abs{\hbeta_k}^2$ \eqref{eqn:ch2_gpp_s_2_bog_coeff} over the whole momentum space, can be seen to be divergent. Nevertheless, at the level of the energy-momentum tensor, this divergence is not a problem \textit{per se}, since the energy requires renormalization. This is a further indication that one should always refer to the proper energy operator as defined by the stress-energy tensor, rather then the instantaneous hamiltonian, for all physical applications.

From the physical point of view, the two different diagonal Fock spaces  $\{\tilde{a}_{\bm{k}},\tilde{a}^\dagger_{\bm{k}}\}$ and  $\{\hat{a}_{\bm{k}},\hat{a}^\dagger_{\bm{k}}\}$ are expected to align in the large momentum limit $k\to\infty$, where the gravitational effects are suppressed. This physical requirement is realized if the Bogolyubov transformation \eqref{eqn:ch2_gpp_s_2_bog}-\eqref{eqn:ch2_gpp_s_2_bog_coeff} that connects the two Fock spaces becomes trivial in this limit, namely
\beq
    \lim_{k\to\infty}\hbeta_k=0 \ .
\label{eqn:ch2_gpp_s_uv_lim}
\eeq

\noindent From the explicit expression in eq.~\eqref{eqn:ch2_gpp_s_2_bog_coeff} we easily see that this condition is indeed satisfied, since for $k\to\infty$ one has $\omega\to\omega_0$, so that $\hbeta_k\to 0$.

\subsection{Gravitational particle production revisited -- the massive gravitino}\label{ch2_sub_gpp_grav}

The simple case of the scalar field already exemplifies the subtleties of the gravitational particle production mechanisms. In \cite{Casagrande:2023fjk} we analyze in the same fashion also the fields of spin-$\frac12$, 1 and $\frac32$. The gravitational particle production of the spin-$\frac12$ Dirac fermion was originally discussed, from the instantaneous hamiltonian point of view, in \cite{Nilles:2001fg, Peloso:2000hy, Chung:2011ck}. We find that the instantaneous hamiltonian and the energy operator actually coincide, so that there is no difference between the two respective diagonal Fock spaces. The production of the massive spin-1 Proca gauge field was originally discussed in \cite{Himmetoglu:2008zp, Graham:2015rva, Ahmed:2020fhc}. Similarly to the spin-$\frac12$ case, the transverse Proca gauge field is found to be free of ambiguities in the different Fock spaces. In contrast, the longitudinal polarization of the Proca field closely follows the case of the real scalar field and the Fock space of the energy operator differs from the one of the standard instantaneous hamiltonian by a Bogolyubov transformation of the type outlined in eq.~\eqref{eqn:ch2_gpp_s_2_bog_coeff}. This correspondence can be understood consistently by means of the equivalence theorem for massive gauge fields. 

Instead, the analysis of the physical differences between the energy operator and the instantaneous Hamiltonian of the massive spin-$\frac32$ field --- which is the central element of \cite{Casagrande:2023fjk} --- yields nontrivial and intriguing results, which we now present in detail.

\subsubsection{The instantaneous hamiltonian} 

We begin, as before, with the instantaneous hamiltonian and the associated canonical quantization \cite{Kallosh:1999jj,Kallosh:2000ve,Nilles:2001fg,Nilles:2001ry,Giudice:1999am,Giudice:1999yt,Kolb:2021xfn}. The canonically normalized action for the transverse and longitudinal components of the massive gravitino field in momentum space are given respectively in eq.~\eqref{eqn:ch2_L_psi_t} and \eqref{eqn:ch2_L_theta}. The associated instantaneous hamiltonian densities are then
\begin{align}
    	\mathcal{H}_\textup{t}=&\frac{1}{2}\sum_{i=1}^3\bpsi_{i,k}^\textup{t}\[i\kg+am\]\psi_{i,k}^\textup{t} \ , &&
	   \mathcal{H}_\vartheta=\frac{1}{2}\bar{\vartheta}_k\[-i\kg C +aM\]\vartheta_k \ . 
\label{eqn:ch2_gpp_H_theta}
\end{align}

\noindent The quantization can be performed following the prescriptions for the spin-$\frac12$ fermions discussed in \cite{Nilles:2001fg,Peloso:2000hy,Chung:2011ck}. From this point of view, the transverse gravitino $\psi_{i,k}^\textup{t}$ has the same exact structure of the standard spin-$\frac12$ fermion, for which, as we said, there are no ambiguities related to the different energy formulations \cite{Casagrande:2023fjk}. We therefore focus on the longitudinal component $\vartheta_k$. Its conjugate momentum is given by
\beq
    \Pi_\vartheta \equiv \frac{\partial\L_\vartheta}{\partial\partial_0\vartheta}=\frac{i}{2}\vartheta^\dagger \ ,
\eeq

\noindent so that canonical quantization is realized by imposing the anticommutation relations
\begin{align}
	\left\{\vartheta(\tau,\vx),\Pi_\vartheta(\tau,\vec{y})\right\}=\delta^{(3)}\(\vx-\vec{y}\) \ , &&
 \left\{\vartheta(\tau,\vx),\vartheta(\tau,\vec{y})\right\}=\left\{\Pi_\vartheta(\tau,\vx),\Pi_\vartheta(\tau,\vec{y})\right\}=0 \ .
\label{eqn:ch2_gpp_psi_comm_rel_ps}
\end{align}

\noindent These conditions can be met by expanding the field $\vartheta$ in momentum space in terms of creation and annihilation operators as
\begin{equation}
	\vartheta(\tau,\vec{x})=\int \frac{d^3k}{(2\pi)^{\nicefrac{3}{2}}}\sum_r e^{i\vk\cdot\vx}\[\vartheta_{r,\bm{k}}a_{r,\bm{k}}+v_{r,\bm{k}}a_{r,-\bm{k}}^\dagger\] \ .
\label{eqn:ch2_gpp_theta_ladder_dec}
\end{equation}

\noindent Note that the mode functions and ladder operators now depend on the whole momentum vector $\bm{k}$ and on the specific $\pm \nicefrac{1}{2}$ longitudinal polarization, labeled by $r=\pm1$. Moreover, the mode functions $\vartheta_{r,\bm{k}}$ and $v_{r,\bm{k}}$ have to satisfy a series of constructional requirements. They are normalized as
\begin{equation}
	\vartheta_{r,\bm{k}}^\dagger\vartheta_{s,\bm{k}}=\delta_{rs}=v_{r,\bm{k}}^\dagger v_{s,\bm{k}} \ ,
\label{eqn:ch2_gpp_theta_mf_norm}
\end{equation}

\noindent and are related by the Majorana condition 
\begin{equation}
	\bar{v}_{r,\bm{k}}=\vartheta_{r,-\bm{k}}^\textup{T}C_4 \ , 
\label{eqn:maj_cond_mod_f}
\end{equation}

\noindent where $C_4$ is the charge conjugation matrix (see Appendix \ref{app_conv_dirac}). It is then convenient to express these mode functions using two-component notation \cite{Nilles:2001fg}. Working in the reference frame where $\vk=\(0,0,k\)$, the mode function $\vartheta_{r,\bm{k}}$ is written as
\begin{equation}
	\vartheta_{r,\bm{k}}=\begin{pmatrix}\vartheta_A(r,\bm{k})\\\vartheta_B(r,\bm{k})\end{pmatrix}=\frac{1}{2}\begin{pmatrix}\(\vartheta_+-r\vartheta_-\)\psi_r\\ \(\vartheta_+ + r\vartheta_-\)\psi_r\end{pmatrix} \ ,
\label{eqn:ch2_gpp_theta_dec}
\end{equation}

\noindent where the $\bm{k}$-dependence in $\vartheta_{\pm}$ is left implicit and the two-component spinors $\psi_r\equiv (\psi_+,\psi_-)$ are the eigenvectors of the helicity operator $\vec{\sigma}\cdot\hat{k}=\sigma^3$:
\begin{equation}
	\psi_+=\begin{pmatrix}1\\0\\\end{pmatrix} \ , \quad \psi_-=\begin{pmatrix}0\\1\\\end{pmatrix} \ , \qquad \qquad \sigma^3\psi_r=r\psi_r \ .
\label{eqn:ch2_gpp_psi_r_def}
\end{equation}

With this parameterization, the ladder operators entering the decomposition \eqref{eqn:ch2_gpp_theta_ladder_dec} satisfy the canonical commutation relations 
\begin{equation}
	\left\{a_{r,\bm{k}},a_{s,\bm{q}}^\dagger\right\}=\delta_{rs}\delta^{(3)}\(\vk-\vec{q}\) \ , \qquad \qquad \left\{a_{r,\bm{k}},a_{s,\bm{q}}\right\}=0=\left\{a_{r,\bm{k}}^\dagger,a_{s,\bm{q}}^\dagger\right\} \ ,
\label{eqn:ch2_gpp_a_ad_cr}
\end{equation}

\noindent if the two-component mode functions $\vartheta_\pm$ are normalized as
\begin{equation}
	\abs{\vartheta_+}^2+\abs{\vartheta_-}^2=2 \ .
\label{eqn:ch2_gpp_theta_pm_norm}
\end{equation}

\noindent Moreover, the longitudinal gravitino equations of motion, computed from the lagrangian \eqref{eqn:ch2_L_theta}, turn into a coupled system of equations for $\vartheta_\pm$:
\begin{equation}
	{\vartheta}^{\prime}_\pm = -i\(C_\textup{R}\mp i C_\textup{I}\)k \vartheta_{\mp}\mp i aM \vartheta_\pm \ .
\label{eqn:ch2_gpp_theta_pm_eom}
\end{equation}

\noindent Thus, the instantaneous hamiltonian \eqref{eqn:ch2_gpp_H_theta} becomes
\begin{equation}
	\H_\vartheta=\frac{1}{2}\sum_r\[\begin{pmatrix} a_{r,\bm{k}}^\dagger&a_{-r,-\bm{k}}\end{pmatrix}\begin{pmatrix}\mathcal{E}_k & \mathcal{F}_k^\dagger \\ \mathcal{F}_k&-\mathcal{E}_k\end{pmatrix}\begin{pmatrix} a_{r,\bm{k}}\\a_{-r,-\bm{k}}^\dagger\end{pmatrix}\] \ , 
\label{eqn:ch2_gpp_H_theta_ladder}
\end{equation}

\noindent where
\begin{equation}
	\begin{aligned}
		\mathcal{E}_k=&-\frac{\cR k}{2}\(\vartheta_+^*\vartheta_-+\vartheta_+\vartheta_-^*\)+\frac{i}{2}\cI k \(\vartheta_+^*\vartheta_- - \vartheta_+\vartheta_-^*\)+\frac{aM}{2}\(\abs{\vartheta_+}^2-\abs{\vartheta_-}^2\) \ , \\
		\mathcal{F}_k=&-\frac{\cR k}{2}\(\vartheta_+^2-\vartheta_-^2\)-\frac{i}{2}\cI k \(\vartheta_+^2+\vartheta_-^2\)-aM\vartheta_+\vartheta_- \ .
	\end{aligned}
\label{eqn:ch2_gpp_theta_E_F}
\end{equation}

Following the same steps as for the scalar field case, we now diagonalize the hamiltonian \eqref{eqn:ch2_gpp_H_theta_ladder} by means of a Bogolyubov transformation. This is implemented through the following decomposition of the mode functions $\vartheta_\pm$ \cite{Nilles:2001fg, Peloso:2000hy, Chung:2011ck}
\begin{equation}
\begin{aligned}
	\vartheta_+ \ = \ &c_+(k) \ \alpha_{\bm{k}}-c_-(k) \ \beta_{\bm{k}} \ , \\
	\vartheta_- \ = \ &c_-(k) \ \alpha_{\bm{k}}+c_+(k) \ \beta_{\bm{k}} \ .
\end{aligned}
\label{eqn:ch2_gpp_theta_bog_tr}
\end{equation}

\noindent Consistency with the mode functions equations of motion \eqref{eqn:ch2_gpp_theta_pm_eom} requires the Bogolyubov coefficients $\alpha_{\bm{k}}$ and $\beta_{\bm{k}}$ to be respectively even and odd in $\bm{k}$:
\begin{align}
    \alpha_{-\bm{k}}=\alpha_{\bm{k}} \ , &&  \beta_{-\bm{k}}=-\beta_{\bm{k}} \ ,
\label{eqn:ch2_gpp_theta_bog_parity}
\end{align}

\noindent should satisfy 
\begin{equation}
	\abs{\alpha_{\bm{k}}}^2+\abs{\beta_{\bm{k}}}^2=1 \ ,
\label{eqn:ch2_gpp_bog_coeff_norm}
\end{equation}

\noindent in order to preserve the commutation relations \eqref{eqn:ch2_gpp_psi_comm_rel_ps}-\eqref{eqn:ch2_gpp_a_ad_cr}. Moreover, the real coefficients $c_\pm$ are such that
\begin{align}
    c^2_+({k})+c^2_-({k})=2 \ , &&  c_+(-k)=c_+(k) \ , && c_-(-k)=-c_-(k) \ .
\label{eqn:ch2_gpp_c_pm_norm_parity}
\end{align}

\noindent These are the necessary requirements for the Bogolyubov transformation \eqref{eqn:ch2_gpp_theta_bog_tr} to be well-defined. The new set of ladder operators is then given by
\begin{equation}
	\begin{pmatrix} \ta_{r,\bm{k}}\\ \ta^\dagger_{-r,-\bm{k}},\end{pmatrix}= \begin{pmatrix}\alpha_{\bm{k}}& -\beta_{\bm{k}}^\dagger\\\beta_{\bm{k}}&\alpha_{\bm{k}}^\dagger\end{pmatrix}\begin{pmatrix} a_{r,\bm{k}}\\a_{-r,-\bm{k}}^\dagger\end{pmatrix} \ ,
\label{eqn:ch2_gpp_theta_bog_tr_1}
\end{equation}

\noindent and the instantaneous hamiltonian \eqref{eqn:ch2_gpp_H_theta_ladder} in this basis is equal to
\begin{equation}
    \H_\vartheta=\frac{1}{2}\sum_r\begin{pmatrix} \ta_{r,\bm{k}}^\dagger&\ta_{-r,-\bm{k}}\end{pmatrix}\begin{pmatrix}A_k & B_k\\ B_k^\dagger &-A_k\end{pmatrix}\begin{pmatrix} \ta_{r,\bm{k}}\\ \ta^\dagger_{-r,-\bm{k}},\end{pmatrix} \ ,
\label{eqn:ch2_gpp_H_theta_1}
\end{equation}

\noindent where the functions $A_k$ and $B_k$ are given by
\begin{equation}
\begin{aligned}
	A_k=&-\[c_+(k)c_-(k)\]\cR k+\[\frac{c_+^2(k)-c_-^2(k)}{2}\] aM \ ,\\
	B_k=&-\[\frac{c_+^2(k)-c_-^2(k)}{2}\]\cR k+i\cI k-\[c_+(k)c_-(k)\] aM \ .
\end{aligned}
\label{eqn:ch2_gpp_theta_AB}
\end{equation}

Contrary to the previous scalar field case, the Bogolyubov decomposition \eqref{eqn:ch2_gpp_theta_bog_tr} is not enough to diagonalize the instantaneous hamiltonian of the longitudinal gravitino. In fact, the non-vanishing imaginary part of the sound speed operator $\cI$ of eq.~\eqref{eqn:ch2_const_coeff} does not allow one to solve for $B_k=0$ in eq.~\eqref{eqn:ch2_gpp_theta_AB} in terms of the real coefficients $c_\pm$.\footnote{It is though possible to reabsorb the complex phase of the sound speed operator $C(\tau)$ into a field redefinition of $\vartheta$. In this case, since the resulting sound speed parameter is real, there exist a choice of $c_\pm$ in \eqref{eqn:ch2_gpp_theta_bog_tr} that directly diagonalizes the hamiltonian. However, the final physical result does not depend on this specific phase choice. \label{fn:theta_ham}} A further Bogolyubov transformation is actually needed to reach the desired diagonal hamiltonian. This transformation is given by
\begin{equation}
    \begin{pmatrix} \hat{b}_{r,\bm{k}}\\ \hat{b}^\dagger_{-r,-\bm{k}}\end{pmatrix}= \begin{pmatrix}\halpha_{\bm{k}}& -\hbeta_{\bm{k}}^\dagger\\ \hbeta_{\bm{k}}&\halpha_{\bm{k}}^\dagger\end{pmatrix}\begin{pmatrix} \ta_{r,\bm{k}}\\ \ta_{-r,-\bm{k}}^\dagger\end{pmatrix} \ , 
\label{eqn:ch2_gpp_theta_bog_tr_2}
\end{equation}

\noindent with\footnote{In (\ref{eqn:ch2_gpp_H_hbog}) we write only the modulus of the Bogolyubov coefficients because the equation defining them can only be solved up to an unphysical phase.} 
\begin{align}
     \halpha_{\bm{k}}^\dagger=&-\frac{B_k}{\omega-A_k}\hbeta_{\bm{k}}, & \abs{\halpha_{\bm{k}}}^2=&\frac{\omega+A_k}{2\omega},& \abs{\hbeta_{\bm{k}}}^2=\frac{\omega-A_k}{2\omega} \ .
\label{eqn:ch2_gpp_H_hbog}
\end{align}

\noindent In these coefficients, the parameter $\omega$ is the frequency defined by
\begin{align}
    \omega^2=&c_s^2k^2+a^2M^2 \ , & c_s^2=\cR^2+\cI^2 \ 
\label{eqn:ch2_gpp_theta_omega}
\end{align}

\noindent and $c_s$ is the sound speed of the longitudinal gravitino. The resulting diagonal instantaneous hamiltonian is therefore
\begin{equation}
    \H_\vartheta=\frac{\omega}{2}\sum_r\(\hat{b}_{r,\bm{k}}^\dagger \hat{b}_{r,\bm{k}}-\hat{b}_{r,\bm{k}} \hat{b}_{r,\bm{k}}^\dagger\) \ .
\label{eqn:ch2_gpp_H_theta_diag}
\end{equation}

\noindent We remark that this final result is actually independent of the specific choice of the coefficients $c_\pm$ in \eqref{eqn:ch2_gpp_theta_bog_tr}, since they turn out to appear only in the combination fixed by the normalization \eqref{eqn:ch2_gpp_c_pm_norm_parity}.

\subsubsection{The stress-energy tensor} 

The instantaneous hamiltonian \eqref{eqn:ch2_gpp_H_theta_diag} has to be compared with the energy operator for the longitudinal gravitino. The massive gravitino energy-momentum tensor, computed in generality from the starting covariant action \eqref{eqn:ch2_m_grav_lag}, is equal (on-shell) to \cite{Casagrande:2023fjk}
\begin{equation}
\begin{aligned}
	T^{\mu\nu}=&\frac{1}{2}\bPsi_\rho\gamma^{\rho\sigma(\mu}\({\nabla}_\sigma\Psi^{\nu)}-{\nabla}^{\nu)}\Psi_\sigma\)-\frac{m}{2}\bPsi_\rho\gamma^{\rho(\mu}\Psi^{\nu)}\\
		&-\frac{1}{2}\nabla_\rho\(\bPsi^\rho\gamma^{(\mu}\Psi^{\nu)}\)-\frac{1}{2}\nabla^{(\mu}\left(\bPsi^{\nu)}\gamma^\rho\Psi_\rho\)-\frac{1}{2}g^{\mu\nu}\,\nabla_\rho\left(\bPsi_\sigma\gamma^\sigma\Psi^\rho\right) \ .
\end{aligned}
\label{eqn:ch2_gpp_emt_full}
\end{equation}

\noindent The terms in the first line are the ones coming from the variation of the explicit vierbeins in \eqref{eqn:ch2_m_grav_lag}, the second line is instead the contribution from the spin-connection in the covariant derivative. The proof of the symmetry and the covariant conservation of this tensor can be found in Appendix \ref{app_emt_grav}. Starting from this general formula, the energy operator for the canonically normalized gravitino in FLRW spacetime, defined in eq.~\eqref{eqn:ch2_psi_cn}, results in
\begin{equation}
    E_\psi=\int d^3x \g \ T^0{}_0=\frac{1}{2}\int d^3x\left\{\bpsi_i\fgamma^{i j l}{\partial}_j\psi_l-am\bpsi_i\fgamma^{ij}\psi_j\right\} \ .
\label{eqn:gr_eom_flrw_0_comp}
\end{equation}

\noindent In order to properly identify the transverse and longitudinal components, one has to go to momentum space via \eqref{eqn:ch2_psi_mom_space} and expand the field $\psi_{i,k}$ according to the decomposition \eqref{eqn:ch2_psi_dec}. The resulting energy operator is given by
\beq
    E_\psi=\int d^3k\[\E_\textup{t}+\E_\theta\] \ ,
\label{eqn:ch2_gpp_psi_E_flrw_tot}
\eeq

\noindent with
\begin{align}
    \E_\textup{t}&=\frac{1}{2}\sum_{j=1}^3\[\bpsi_{j,k}^\textup{t}\kg\psi_{j,k}^\textup{t}+am\bpsi_{j,k}^\textup{t}\psi_{j,k}^\textup{t}\] \ ,  \label{eqn:ch2_gpp_psi_ET}\\
	\E_\vartheta&=\frac{1}{2}\bar{\vartheta}_k\[\kg D+am\]\vartheta_k \ .
\label{eqn:ch2_gpp_E_theta_D} 
\end{align}

\noindent where the operator $D$ is equal to
\begin{equation}
	D=d_\textup{R}+id_\textup{I}\fgamma^0\equiv\frac{3m^2+H^2}{3\(m^2+H^2\)}+\frac{2mH}{3\(m^2+H^2\)}\fgamma_0 \ ,
\label{eqn:ch2_gpp_theta_D}
\end{equation}

Thus, for transverse gravitino the energy operator and the instantaneous hamiltonian of eq.~\eqref{eqn:ch2_gpp_H_theta} match exactly, as anticipated previously. On the contrary, the resulting energy operator for the longitudinal gravitino --- which agrees with the energy density given in \cite{Kallosh:1999jj} --- does not coincide with the corresponding instantaneous hamiltonian \eqref{eqn:ch2_gpp_H_theta}. They have though the same structure: the latter is characterized by the sound speed parameter $C(\tau)$ of eq.~\eqref{eqn:ch2_const_coeff} and non-minimal time-dependent mass parameter $M$ \eqref{eqn:ch2_grav_mass_M}, the former by sound speed parameter $D\(\tau\)$ of eq.~\eqref{eqn:ch2_gpp_theta_D} and minimal mass parameter $m$. This similarity allows one to perform the quantization and diagonalization of the longitudinal energy operator \eqref{eqn:ch2_gpp_E_theta_D} in a simple way, adapting the steps performed for the instantaneous hamiltonian. The resulting diagonal energy of the longitudinal gravitino is 
\begin{equation}
	\E_\vartheta=\frac{\omega_d}{2}\sum_r\(\hat{a}_{r,\bm{k}}^\dagger \hat{a}_{r,\bm{k}}-\hat{a}_{r,\bm{k}} \hat{a}_{r,\bm{k}}^\dagger\) \ ,
\label{eqn:ch2_gpp_E_theta_diag}
\end{equation}

\noindent where the diagonal frequency is equal to
\begin{align}
	\omega_d^2=d_s^2k^2+a^2m^2 \ ,  && d_s^2=d_\textup{R}^2+d_\textup{I}^2=\frac{m^2+\nicefrac{H^2}{9}}{m^2+H^2} \ . 
\label{eqn:ch2_gpp_theta_wd}
\end{align}

\noindent The Bogolyubov transformation between the diagonal ladder operators and the original ones in eq.~\eqref{eqn:ch2_gpp_theta_ladder_dec} is
\begin{equation}
    \begin{pmatrix} \hat{a}_{r,\bm{k}}\\ \hat{a}^\dagger_{-r,-\bm{k}},\end{pmatrix}=\begin{pmatrix}x_{\bm{k}}& -y_{\bm{k}}^\dagger\\ y_{\bm{k}}&x_{\bm{k}}^\dagger\end{pmatrix}\begin{pmatrix} \ta_{r,\bm{k}}\\ \ta_{-r,-\bm{k}}^\dagger\end{pmatrix} =\begin{pmatrix}x_{\bm{k}}& -y_{\bm{k}}^\dagger\\ y_{\bm{k}}&x_{\bm{k}}^\dagger\end{pmatrix}\begin{pmatrix}\alpha_{\bm{k}}& -\beta_{\bm{k}}^\dagger\\\beta_{\bm{k}}&\alpha_{\bm{k}}^\dagger\end{pmatrix}\begin{pmatrix} a_{r,\bm{k}}\\a_{-r,-\bm{k}}^\dagger\end{pmatrix} \ .
\label{eqn:ch2_gpp_E_diag_lad}
\end{equation}

\noindent The Bogolyubov coefficients $x_k$ and $y_k$ are the energy analogue of the coefficients $\halpha_k$ and $\hbeta_k$ of eq.~\eqref{eqn:ch2_gpp_theta_bog_tr_2}-\eqref{eqn:ch2_gpp_H_hbog} for the instantaneous hamiltonian, and are\footnote{As for the coefficients in eq.~\eqref{eqn:ch2_gpp_H_hbog}, the Bogolyubov coefficients $x_k$ and $y_k$ in eq.~\eqref{eqn:ch2_gpp_E_hbog} are defined up to an unphysical phase.}
\begin{align}
    x_{\bm{k}}^\dagger=-\frac{B_k^\textup{D}}{\omega_d-A_k^\textup{D}}y_{\bm{k}} \ , && \abs{x_{\bm{k}}}^2=\frac{\omega_d+A_k^\textup{D}}{2\omega_d} \ , && \abs{y_{\bm{k}}}^2=\frac{\omega_d-A_k^\textup{D}}{2\omega_d} \ , 
\label{eqn:ch2_gpp_E_hbog}
\end{align}

\noindent with 
\begin{align}
    A_k^\textup{D} = &[c_+(k)c_-(k)]\dr k+\(\frac{c^2_+(k)-c^2_-(k)}{2}\)am \ , \\
    B_k^\textup{D} = &\(\frac{c^2_+(k)-c^2_-(k)}{2}\)\dr k-[c_+(k)c_-(k)]am -i\di k \ .
\end{align}

\subsubsection{On the difference between the two diagonal Fock spaces and the equivalence theorem}

Therefore, as in the scalar field case, the diagonal energy operator and instantaneous hamiltonian are different, and are associated with distinct diagonal Fock spaces. The relation between the two is again a Bogolyubov transformation, that can be worked out explicitly, combining eq.~\eqref{eqn:ch2_gpp_theta_bog_tr_2}-\eqref{eqn:ch2_gpp_H_hbog} with eq.~\eqref{eqn:ch2_gpp_E_diag_lad}-\eqref{eqn:ch2_gpp_E_hbog}. The resulting transformation is \cite{Casagrande:2023fjk}
\begin{equation}
    \begin{pmatrix} \hat{a}_{r,\bm{k}}\\ \hat{a}_{-r,-\bm{k}},^\dagger\end{pmatrix}=\begin{pmatrix}\mathscr{A}_{\bm{k}}& -\mathscr{B}_{\bm{k}}^\dagger\\ \mathscr{B}_{\bm{k}}&\mathscr{A}_{\bm{k}}^\dagger\end{pmatrix}\begin{pmatrix} \hat{b}_{r,\bm{k}}\\ \hat{b}_{-r,-\bm{k}},^\dagger\end{pmatrix} \ , 
\label{eqn:ch2_gpp_psi_comp_fs}
\end{equation}

\noindent where the composite Bogolyubov coefficients
\begin{align}
    \mathscr{A}_{\bm{k}}=x_{\bm{k}}\halpha_{\bm{k}}^\dagger+y_{\bm{k}}^\dagger\hbeta_{\bm{k}} \ , && \mathscr{B}_{\bm{k}}=y_{\bm{k}}\halpha_{\bm{k}}^\dagger-x_{\bm{k}}^\dagger\hbeta_{\bm{k}} \ .
\label{eqn:psi_fs_bog}
\end{align}

\noindent are such that

\beq
\begin{aligned}
    \abs{\mathscr{A}_{\bm{k}}}^2=&\frac{1}{2}\[1-\frac{\(\cR\dr+\cI\di\)}{\omega\omega_d}k^2+\frac{a^2mM}{\omega\omega_d}\],\\
    \abs{\mathscr{B}_{\bm{k}}}^2=&\frac{1}{2}\[1+\frac{\(\cR\dr+\cI\di\)}{\omega\omega_d}k^2-\frac{a^2mM}{\omega\omega_d}\] \ . 
\end{aligned}
\label{eq:calB2}
\eeq

\noindent Note that, once again, these formulas hold independently of the particular choice of coefficients $c_\pm$ in eq.~\eqref{eqn:ch2_gpp_theta_bog_tr}. 

Equipped with this correspondence, we can address the physical consequences on the dynamics of the massive gravitino in FLRW. First of all, we go back to the "catastrophic" production of longitudinal gravitinos for vanishing sound speed in the orthogonal supergravity model \eqref{eqn:ch2_sugra_inflation_oc_lag}, that was put forward in \cite{Kolb:2021xfn} and discussed here in Section \ref{ch2_sec_grav_flrw}. Despite the conditions for the sound speed of eq.~\eqref{eqn:ch2_sound_speed_flrw} to vanish in this particular model are in tension with the corresponding effective field theory approximation, physical processes associated with the produced particles before the breakdown --- like for instance their backreaction on the geometry for small but non-vanishing sound speed --- should have an unambiguous description. The gravitino overproduction discussed in \cite{Kolb:2021xfn} is computed by means of the instantaneous diagonal hamiltonian given in eq.~\eqref{eqn:ch2_gpp_H_theta_diag}, following the standard prescription of eq.~\eqref{eqn:ch2_bog_number_op_gen}. This hamiltonian density is indeed proportional to the frequency $\omega$ of eq.~\eqref{eqn:ch2_gpp_theta_omega}, which directly depends on the sound speed $c_s$ of eq.~\eqref{eqn:ch2_sound_speed_flrw}. However, we highlighted and proved in this section that describing physical processes in curved background with the instantaneous hamiltonian is misleading, since the associated ground state is not the state of minimal energy. We showed this explicitly in eq.~\eqref{eq:ch2_gpp_energyBD} for the case of the scalar field, but it holds as long as there is a mismatch between the two Fock spaces. Taking therefore the non-ambiguous perspective of the energy-momentum tensor \eqref{eqn:ch2_gpp_emt_full} we computed in \cite{Casagrande:2023fjk}, one has that energy operator of the longitudinal gravitino \eqref{eqn:ch2_gpp_E_hbog} is proportional to the frequency $\omega_d$ of eq.~\eqref{eqn:ch2_gpp_theta_wd} rather than $\omega$, which means that it depends on the parameter $d_s^2$ of eq.~\eqref{eqn:ch2_gpp_theta_wd} rather than on the troublesome sound speed $c_s^2$. This means that the energy of the longitudinal gravitino is not associated with any divergence or non-unitary behavior in the limit $c_s\to0$ \cite{Casagrande:2023fjk}. Moreover, even if divergences are found at this level they should not be an obstacle a priori, since energy requires anyway renormalization, and clearly the renormalization procedure concerns the energy-momentum tensor.
 
In addition, the correspondence \eqref{eqn:ch2_gpp_psi_comp_fs} between the diagonal Fock spaces highlights a further puzzling feature. As we discussed for the scalar field in eq.~\eqref{eqn:ch2_gpp_s_uv_lim}, the energy operator and the instantaneous hamiltonian are physically expected to align in the UV limit $k\to\infty$, where the effect of the coupling to gravity are suppressed. However, contrary to this expectation, this does not happen, since the off-diagonal Bogolyubov coefficient $\mathscr{B}_{\bm k}$ of eq.~\eqref{eqn:ch2_gpp_psi_comp_fs} does not vanish in the UV limit \cite{Casagrande:2023fjk}:
\begin{equation}
    \lim_{k\to\infty}\abs{\mathscr{B}_{\bm{k}}}^2=\frac{1}{2}\(1+\frac{\cR\dr+\cI\di}{c_sd_s}\) \ . 
\label{eqn:ch2_gpp_calB3} 
\end{equation}

\noindent This peculiar misalignment is specific of the spin-$\frac32$ field case and does not affect the particles of lower spins \cite{Casagrande:2023fjk}.

The same problem can be seen from a second point of view, with respect to the gravitino-goldstino equivalence theorem \eqref{eqn:ch2_sugra_eq_theorem} \cite{Fayet:1986zc,Casalbuoni:1988kv,Casalbuoni:1988qd}. According to this theorem, the dynamics of the longitudinal gravitino should be described in terms of the goldstino in the limit of high momenta. In the case of the orthogonal supergravity model \eqref{eqn:ch2_sugra_inflation_oc_lag}, one can check that the equivalence theorem is explicitly realized at the lagrangian level, without canonical normalization. For the longitudinal gravitino, this is given by the field $\theta_k\equiv\fgamma^i\psi_{i,k}$ before the redefinition \eqref{eqn:ch2_theta_cn} is applied. The scalar field background of this model is defined by the Einstein equations \eqref{eqn:ch2_EE_flrw}, and in such a background the sound speed takes the form given in eq.~\eqref{eqn:ch2_sound_speed_flrw}. Writing the longitudinal gravitino  field $\theta_k$ in two-component notation as
\begin{equation}
	\theta(t,\vk)=\begin{pmatrix}\chi_{\bm{k}}(t)\\\bar{\chi}_{-\bm{k}}(t)\end{pmatrix} \ ,
\label{eqn:ch2_gpp_theta_2c_k}
\end{equation}

\noindent the longitudinal gravitino lagrangian is found from eq.~\eqref{eqn:ch2_grav_lag_cn} to be
\begin{equation}
\begin{aligned}
	\mathcal{L}_\theta= &\frac{a^2f^2}{2\vk^2}\left\{\(1+\frac{\dot{\phi}^2}{2f^2}\)\frac{i}{2}\partial_0\chi_{\bm{k}}\sigma^0\bar{\chi}_{\bm{k}}+\(1-\frac{\dot{\phi}^2}{2f^2}\)\frac{1}{2}\chi_{\bm{k}}\(\vk\cdot\vec{\sigma}\)\bar{\chi}_{\bm{k}}\right.+\\
	&\left. +\frac{g^{\prime}\,\dot{\phi}}{f^2}\chi_{\bm{k}}\(\vk\cdot\vec{\sigma}\)\bar{\sigma}^0\chi_{-\bm{k}}+\frac{1}{2}\[\(\frac{\dot{\phi}^2}{f^2}-1\)am+\(\frac{3g^{\prime}\,\dot{\phi}}{f^2}\)aH\]\chi_{\bm{k}}\chi_{-\bm{k}}+h.c.\right\} \ .
\end{aligned}
\label{eqn:ch2_gpp_L_theta_bkgr}
\end{equation}

\noindent On the goldstino side, the non-linear supersymmetry goldstino lagrangians were computed in \cite{Bonnefoy:2022rcw} and reported in detail in Section \ref{ch1_sec_nls}. The relevant one in this case is the quadratic sector of the orthogonal constraint lagrangian of eq.~\eqref{eqn:ch1_oc_lag_1}.  Denoting with $G_{\bm k}$ the goldstino in the same momentum space parameterization as $\chi_{\bm k}$, this lagrangian is found to be
\begin{equation}
	\begin{aligned}
		\mathcal{L}_G=&\(1+\frac{\dot{\phi}^2}{2f^2}\)\frac{i}{2}\partial_0 G_{\bm{k}}\sigma^0\bar{G}_{\bm{k}}+\(1-\frac{\dot{\phi}^2}{2f^2}\)\frac{1}{2}G_{\bm{k}}\(\vk\cdot\vec{\sigma}\)\bar{G}_{\bm{k}} + \\
		& + i\frac{g^{\prime}\,\dot{\phi}}{f^2}G_{\bm{k}}\(\vk\cdot\vec{\sigma}\)\bar{\sigma}^0G_{-\bm{k}}+h.c.\ .
	\end{aligned}
\label{eqn:ch2_gpp_L_G}
\end{equation}

Therefore, in the large momentum limit $k\gg am$ and $k\gg aH$, the lagrangians \eqref{eqn:ch2_gpp_L_theta_bkgr} and \eqref{eqn:ch2_gpp_L_G} match via the identification
\begin{equation}
    \chi_{\bm{k}}=\frac{2k }{af}G_{\bm{k}} \ ,
\label{eqn:ch2_gpp_G_theta_id}
\end{equation}

\noindent which suggests the following prescription for the equivalence theorem:
\begin{equation}
	\psi_\mu\longrightarrow\frac{M_\textup{P}}{f}\partial_\mu G \ ,
\label{eq:equiv}
\end{equation}

\noindent along the lines of eq.~\eqref{eqn:ch2_sugra_eq_theorem}.\footnote{Notice that during inflation the relation $f= \sqrt{3} m \MP$ is valid only on a flat space background.} However, the same alignment of the lagrangian is not realized at the level of the energy operators. The goldstino hamiltonian computed from the relative lagrangian \eqref{eqn:ch2_gpp_L_G} is in fact equal, in the UV limit, to the instantaneous hamiltonian of eq.~\eqref{eqn:ch2_gpp_H_theta}, and therefore remains disconnected from the actual energy operator \eqref{eqn:ch2_gpp_E_theta_D}, according to eq.~\eqref{eqn:ch2_gpp_calB3} \cite{Casagrande:2023fjk}. 

We still do not have a detailed understanding of the puzzling behavior of the massive gravitino in non-trivial, time-dependent backgrounds. This, once again, provides further motivation to stick to the stress-energy tensor operator \eqref{eqn:ch2_gpp_E_theta_D} to describe the dynamics of the massive gravitino, in any limit. 

\subsection{Summary of results}

In this section we analyzed the dynamics of the massive gravitino \eqref{eqn:ch2_m_grav_lag} in the cosmological scenario of the time-dependent FLRW spacetime. These models are embedded in supergravity as effective theories with nonlinearly realized supersymmetry, like the orthogonal supergravity model presented in eq.~\eqref{eqn:ch2_sugra_inflation_oc_lag} \cite{Ferrara:2015tyn,Carrasco:2015iij,Hasegawa:2017hgd,Kolb:2021xfn,Dudas:2021njv}. In time-dependent backgrounds, the dynamics of the longitudinal gravitino --- which is the goldstino of spontaneous supersymmetry breaking, absorbed through the super-Higgs mechanism, discussed in Section \ref{ch2_sec_sugra_Higgs} --- is governed by the sound-speed parameter defined in eq.~\eqref{eqn:ch2_const_coeff}-\eqref{eqn:ch2_grav_sound_speed} \cite{Hasegawa:2017hgd,Kolb:2021xfn,Dudas:2021njv}. Depending on the specific background solution, this sound speed exposes two major inconsistencies related to the gravitino propagation. The first is the possibility for the gravitino to propagate with superluminal sound speed. We proved in \cite{Bonnefoy:2022rcw} that this is a consequence of the ill-defined orthogonal constraint \eqref{eqn:ch1_oc}, which does not define reliable effective field theories. This constraint originates in fact from a higher-derivative theory in the UV, which turns into causality constraints like eq.~\eqref{eqn:ch2_sugra_caus_const}. The issue disappears once the improved setup of eq.~\eqref{sub_ch1_imp_const} proposed in \cite{Bonnefoy:2022rcw} is employed. This yields a lagrangian analogous to the one in eq.~\eqref{eqn:ch2_sugra_inflation_oc_lag} but free of causality problems, as shown in eq.~\eqref{eqn:ch2_sugra_sound_speed_improved}. The second issue is more subtle and related to the phenomenon of gravitational particle production \cite{Parker:1969au,Zeldovich:1971mw,Ford:1986sy,Kofman:1997yn,Kallosh:1999jj,Kallosh:2000ve,Giudice:1999am,Giudice:1999yt,Nilles:2001fg,Nilles:2001ry,Kolb:2021nob,Kolb:2021xfn,Ahmed:2020fhc,Graham:2015rva,Chung:2011ck, Birrell:1982ix,Gorbunov:2011zzc}. It was argued in \cite{Kolb:2021xfn} that a vanishing sound speed parameter during the background scalar field evolution is related to an unbounded production of gravitinos. More specifically, a vanishing sound speed allows gravitinos to be gravitationally produced with arbitrary large momentum, which is clearly a non-unitary process.

This inconsistent behavior signals that the theory breaks down once the gravitino is quantized. This breakdown can be related only partially to the effective field theory setup under consideration, and therefore motivated us to look more closely into the gravitational particle production mechanism. This is usually computed based on the ground state and the Fock space of what we called instantaneous hamiltonian, which is the hamiltonian for the canonically normalized fields \cite{Zeldovich:1971mw,Kofman:1997yn,Ford:1986sy,Parker:1969au,Birrell:1982ix,Gorbunov:2011zzc}. However, despite describing correctly the field propagation, the instantaneous hamiltonian does not represent the physical energy of the system when placed in curved spacetime. This is rather given by the stress-energy tensor, sourcing the metric field in the Einstein equations. In fact, the time-dependent rescaling that one performs to canonically normalize the fields breaks their transformation properties under diffeomorphism, so that the corresponding instantaneous hamiltonian is not connected in a simple way to the energy operator encoded in the stress-energy tensor \cite{Fulling:1979ac,Weiss:1986gg,Bozza:2003pr,Grain:2019vnq}.

In \cite{Casagrande:2023fjk} we investigate the consequences of this difference between the Fock spaces of the instantaneous hamiltonian and the energy-momentum tensor, parameterizing it in terms of a Bogolyubov transformation. In Section \ref{ch2_sub_gpp_scalar} we have reported explicitly this analysis for the simple case of the real massive scalar field. The two relative Fock spaces are indeed different --- see eq.~\eqref{eqn:ch2_gpp_s_2_bog}-\eqref{eqn:ch2_gpp_s_2_bog_coeff} --- but they align in the UV limit of large momentum. This is physically consistent, since the gravitational effects are suppressed in such a limit. On the contrary, the analysis on the massive gravitino is more involved and yields a further puzzle. The energy operator \eqref{eqn:ch2_gpp_E_theta_D}, computed from the general stress-energy tensor \eqref{eqn:ch2_gpp_emt_full}, and the instantaneous hamiltonian \eqref{eqn:ch2_gpp_H_theta} are related through the Bogolyubov transformation \eqref{eqn:ch2_gpp_psi_comp_fs}--\eqref{eq:calB2}. However, this transformation does not trivialize in the limit of high momenta, meaning that the operators are completely disconnected, even in the limit in which gravity should be suppressed. The counterpart of this puzzling behavior is related to the gravitino equivalence theorem \eqref{eqn:ch2_sugra_eq_theorem} \cite{Fayet:1986zc,Casalbuoni:1988qd,Casalbuoni:1988kv}, which we prove at the lagrangian level in eq.~\eqref{eqn:ch2_gpp_theta_2c_k} and eq.~\eqref{eqn:ch2_gpp_L_G}, but it is clearly not realized at the level of the energy operators, since the goldstino hamiltonian corresponds to the instantaneous hamiltonian of the longitudinal gravitino. On the other hand, a further consequence of the difference between the two Fock spaces is that the actual energy operator \eqref{eqn:ch2_gpp_E_theta_D} of the longitudinal gravitino is not associated with any specific behavior in the vanishing sound speed limit discussed in \cite{Kolb:2021xfn}. Moreover, the energy in quantum field theory usually requires renormalization, which must be taken into account in the presence of divergences in the energy operator of the gravitino. 

The underlying reason for this mismatch in the gravitino case remains unknown, and we plan on further investigating it in future work. Nevertheless, it clearly suggests that one should rely on the gravitino stress-energy tensor \eqref{eqn:ch2_gpp_emt_full} for any physical application, even in setups that are not expected to be sensitive to gravitational coupling. More generally, the results of \cite{Casagrande:2023fjk} point to the need for a re-analysis of non-thermal gravitino production in cosmology.

\section{Off-shell supergravity and supercurrent superfields}\label{ch2_sec_off_shell_sugra}

We now leave the framework of spontaneous supersymmetry breaking and return to the pure supergravity multiplet $\{g_{\mu\nu},\psi_\mu\}$ of eq.~\eqref{eqn:ch2_pure_sugra_susy}-\eqref{eqn:ch2_sugra_lag_1}. In Chapter \ref{ch_susy} we saw that rigid supersymmetry can be conveniently formulated in terms of off-shell degrees of freedom. This description is particularly natural from the superspace point of view\footnote{The superspace formalism can indeed be extended to the local supersymmetry case and used to construct supergravity theories. We will not discuss this approach further in the manuscript, and refer for example to \cite{Wess:1977fn, Grimm:1977kp, Wess:1978bu,Siegel:1978mj,Wess:1992cp} for more information.} of Section \ref{ch1_sec_ss_sf}. In a similar fashion, supergravity also admits different off-shell formulations \cite{Stelle:1978ye,Ferrara:1978em,Siegel:1978mj,Sohnius:1981tp}, which we discuss in this section. Although not as convenient as in the rigid case, these formulations have many interesting features and applications. In particular, as will become clear early in this section, they allow one to investigate the leading-order coupling to supergravity of rigid theories (see for example \cite{Ferrara:1977mv,Komargodski:2010rb,Festuccia:2011ws}). This is the setup and perspective adopted in the paper \cite{ms2}, which we will present in detail in Section \ref{ch2_sec_ms2}.

To understand how off-shell formulations of the pure supergravity multiplet $\left\{e_\mu{}^a,\psi_\mu\right\}$ \eqref{eqn:ch2_pure_sugra_susy} can be constructed, we go back to the Noether procedure of eq.~\eqref{eqn:ch2_noether_psi_Xi}--\eqref{eqn:ch2_delta_sugra_noether}. In particular, we saw in eq.~\eqref{eqn:ch2_noether_psi_Xi} that the current of supersymmetry $\Xi^\mu$ associated with a given multiplet transforms under supersymmetry into the stress-energy tensor $T^{\mu\nu}$. This is a general relation imposed by the Super-Poincar\'e algebra \eqref{eqn:ch1_SP_alg}, which has the major consequence that the current of supersymmetry and the stress-energy tensor are themselves part of a supersymmetric multiplet, the so-called supercurrent superfield \cite{Ferrara:1974pz,Sohnius:1981tp,Clark:1995bg,Komargodski:2010rb}. As shown in the original paper by Ferrara and Zumino \cite{Ferrara:1974pz}, the supercurrent multiplet is a vector superfield $\S^\mu$ that satisfies the so-called linear multiplet condition, which in superspace language is
\beq
    D^2\S^\mu=0 \ ,
\label{eqn:ch2_lin_sf}
\eeq

\noindent along with an additional conservation-like equation that holds on-shell and determines its components. This equation can take different forms, hence leading to multiple supercurrent superfields \cite{Ferrara:1974pz,Sohnius:1981tp,Clark:1995bg,Komargodski:2010rb}. They all contain the supersymmetry current and the stress-energy tensor, but differ in the additional superpartners that complete the multiplet.

Then, going back to the coupling to supergravity, which at leading order is given by eq.~\eqref{eqn:ch2_noether_psi_Xi}--\eqref{eqn:ch2_delta_sugra_noether}, the supercurrent superfield is naturally interpreted as the multiplet of currents of the supergravity gauge fields. Thus, following this path, the supercurrent superfield can be used to construct an off-shell extension of the pure supergravity multiplet \eqref{eqn:ch2_pure_sugra_susy}: as the metric $g_{\mu\nu}$ and the gravitino $\psi_\mu$ couple respectively to the stress-energy tensor $T^{\mu\nu}$ and the current of supersymmetry $\Xi^\mu$, the additional off-shell degrees of freedom for the supergravity multiplet can be identified based on the structure of the additional elements of the supercurrent multiplet \cite{Stelle:1978ye,Ferrara:1978em,Sohnius:1981tp,Siegel:1978mj}.

Let us now present the different supercurrent superfields and the corresponding off-shell formulation of supergravity they define.

\subsection{The Ferrara--Zumino supercurrent and old-minimal off-shell supergravity}

We begin with the so-called Ferrara--Zumino multiplet $\S^\mu_\textup{FZ}$ \cite{Ferrara:1974pz}. It is defined by the superspace equation
\beq
    \(\sigma_\mu\)_{\alpha\dalpha}\overline{D}^{\dalpha}\S_\textup{FZ}^\mu=D_\alpha X \ ,
\label{eqn:ch2_FZ_sc_eq}
\eeq    

\noindent where $X$ is a chiral multiplet $\overline{D}_{\dalpha}X=0$, as in eq.~\eqref{eqn:ch1_chiral_sf_def}. This equation fixes the components of the Ferrara--Zumino supercurrent superfield $\S^\mu_\textup{FZ}$ to be
\beq
    \S^\mu_\textup{FZ}=\left\{J^\mu,\Xi^\mu, T^{\mu\nu}, x\right\} \ ,
\label{eqn:ch2_FZ_sc_comp}
\eeq

\noindent where $J^\mu$ is a non-conserved axial current and $x=A+iB$ is a complex scalar field. Together with the stress-energy tensor $T^{\mu\nu}$, they account for 12 bosonic degrees of freedom, which match the 12 fermionic ones of the supersymmetry current $\Xi^\mu$. The chiral superfield $X$ contains the complex scalar $x$ as lowest component, together with the divergence $\partial_\mu J^\mu$ and the traces $T^\mu{}_\mu$ and $\(\sigma^\mu\Xi_\mu\)_\alpha$.\footnote{For simplicity of notation, we use the same name to denote the supersymmetry current both in 2-component and 4-component notation.} In the presence of the full superconformal symmetry, the superfield $X$ vanishes and $J^\mu$ becomes the superconformal R-symmetry current \cite{Ferrara:1974pz}. The supersymmetry transformations characterizing the $\S^\mu_\textup{FZ}$ superfield are 
\beq
\begin{aligned}
    &\delta J^\mu= -i\bepsilon\gamma_5\Xi^\mu \ ,\\
    &\delta\Xi^\mu=T^{\mu\nu}\gamma_\nu\epsilon+\frac{1}{8}\varepsilon^{\mu\nu\rho\sigma}\partial_\rho J_\sigma \gamma_\nu\epsilon+\frac{i}{4}\partial_\rho J^\mu\gamma_5\gamma^\rho\epsilon +\frac{i}{4}\partial^\mu\(A-iB\gamma_5\)\gamma_5\epsilon \ , \\
    &\delta T^{\mu\nu}=-\frac{1}{4}\bepsilon\gamma_{(\mu}{}^{\lambda}\partial_\lambda \Xi_{\nu)} \ ,\\
    &\delta x=2i\bepsilon_R\gamma^\rho\Xi_{\rho L}\ .
\end{aligned}
\label{eqn:ch2_FZ_sc_susy}
\eeq

This multiplet was used to construct the first off-shell formulation of pure supergravity, known as old-minimal \cite{Stelle:1978ye,Ferrara:1978em}. The additional off-shell fields of this old-minimal supergravity multiplet are an axial vector $V_\mu$ and a complex scalar $Z=M+iN$. These couple respectively to the current $J^\mu$ and the complex scalar $x$ of the Ferrara--Zumino supercurrent $\S^\mu_\textup{FZ}$. In terms of off-shell degrees of freedom, the bosonic ones are 6 from the graviton, 4 from the axial vector $V_\mu$ and two from the complex scalar $Z$, which match the 12 off-shell fermionic degrees of freedom of the gravitino. The linearised supersymmetry transformations of this old-minimal off-shell supergravity multiplet are  
\beq
\begin{aligned}
    &\delta g_{\mu\nu}= \bepsilon\gamma_{(\mu}\psi_{\nu)} \ , \\
    &\delta\psi_\mu=\partial_\mu \epsilon - \frac14 \partial_\nu g_{\rho\mu} \gamma^{\nu\rho} \epsilon -\frac{1}{6}\(M+iN\gamma_5\)\gamma_\mu\epsilon+\frac{i}{2}\(V_\mu-\frac13\gamma_\mu\slashed{V}\)\gamma_5\epsilon \ , \\
    &\delta V_\mu=\frac{i}{4}\bepsilon\gamma_5\(\gamma_{\mu}{}^{\rho\sigma}-4\delta_{\mu}{}^{[\rho}\gamma^{\sigma]}\)\partial_\rho\psi_\sigma  \ ,\\
    &\delta Z=\bepsilon_R\gamma^{\mu\nu}\partial_\mu\psi_{\nu R} \ ,
\end{aligned}
\label{eqn:ch2_old_min_susy}
\eeq

\noindent and the full nonlinear lagrangian is given by \cite{Stelle:1978ye,Ferrara:1978em}
\beq
    e^{-1}\L_\textup{OM}=\L_\textup{SUGRA}+\frac13\(\abs{Z}^2+V_\mu V^\mu\) \ ,
\eeq

\noindent where $\L_\textup{sugra}$ is the pure supergravity action of eq.~\eqref{eqn:ch2_sugra_lag_1}. 

An example of multiplet that can be coupled to supergravity in this old-minimal off-shell formulation through the Ferrara--Zumino supercurrent \eqref{eqn:ch2_FZ_sc_comp} is the Wess--Zumino model of eq.~\eqref{eqn:ch1_wz_KW}-\eqref{eqn:ch1_L_wz}, which is the case studied in the original Ferrara--Zumino paper \cite{Ferrara:1974pz}.

\subsection{The \texorpdfstring{$R$}{R}-multiplet and new-minimal off-shell supergravity}

A second type of supercurrent superfield can be defined for theories with a conserved R-symmetry. This multiplet contains the R-symmetry current $J^\mu_\textup{R}$ as lowest component and it is then called the R-multiplet $\S^\mu_\textup{R}$. Its characteristic equation is \cite{Sohnius:1981tp,Komargodski:2010rb}
\beq
    \(\sigma_\mu\)_{\alpha\dalpha}\overline{D}^{\dalpha}\S^\mu_\textup{R}=\chi_\alpha \ ,
\label{eqn:ch2_R_sc_eq}
\eeq

\noindent where the superfield $\chi_\alpha$ is such that
\begin{align}
    \overline{D}_{\dalpha}\chi_\alpha=0 \ , && D^\alpha\chi_\alpha=\overline{D}_{\dalpha}\bar{\chi}^{\dalpha} \ ,
\label{eqn:ch2_R_sc_chi_def}
\end{align}

\noindent namely it is a chiral superfield with a field strength-like structure. These two equations imply that $\partial_\mu \S^\mu_\textup{R}=0$, which means that the components of the R-multiplet are conserved currents. These are fixed by \eqref{eqn:ch2_R_sc_eq}-\eqref{eqn:ch2_R_sc_chi_def} to be
\beq
    \S^\mu_\textup{R}=\left\{J^\mu_\textup{R},\Xi^\mu,T^{\mu\nu},X^{\mu\nu}\right\} \ ,
\label{eqn:ch2_R_sc_comp}
\eeq

\noindent where $J^\mu_\textup{R}$ is indeed the R-symmetry current and $X^{\mu\nu}$ is a totally antisymmetric tensor. This field $X^{\mu\nu}$ can be written as the dual of a curl:
\beq
    X^{\mu\nu}=\varepsilon^{\mu\nu\rho\sigma}\partial_\rho X_\sigma \ ,
\label{eqn:ch2_R_sc_X_dual}
\eeq

\noindent which implies that it is trivially conserved and hence ensures the correct matching of bosonic and fermionic degrees of freedom. 
The supersymmetry transformation rules of the R-multiplet \eqref{eqn:ch2_R_sc_comp} are \cite{Sohnius:1981tp}
\beq
\begin{aligned}
    &\delta J_\mu=-i\bar{\epsilon}\gamma_5\Xi_\mu \ , \\
    &\delta \Xi_\mu=  T_{\mu\nu}\gamma^\nu\epsilon+X_{\mu\nu}\gamma^\nu\epsilon+\frac{1}{8}\varepsilon_{\mu\nu\rho\sigma}\partial^\rho J^\sigma\gamma^\nu \epsilon+\frac{i}{4}\partial_\rho J_\mu\gamma_5\gamma^\rho\epsilon \ , \\
    &\delta T_{\mu\nu}=-\frac{1}{4}\bepsilon\gamma_{(\mu}{}^{\lambda}\partial_\lambda \Xi_{\nu)} \ , \\
    &\delta X_{\mu\nu}=-\frac{i}{8}\varepsilon_{\mu\nu\rho\sigma}\bepsilon\gamma_5\gamma^\rho\gamma^\lambda\partial^\sigma \Xi_\lambda \ . 
\end{aligned}
\label{eqn:ch2_R_sc_susy}
\eeq

This R-multiplet of currents defines an alternative off-shell formulation of supergravity, called new-minimal, in which the supergravity multiplet auxiliary fields are now an axial vector field $V_\mu$ and an antisymmetric tensor field $C_{\mu\nu}$ of field strength $C^\mu =\frac12  \varepsilon^{\mu\nu\rho\sigma} \partial_\nu C_{\rho\sigma} $. The linearised supersymmetry transformations of this new-minimal off-shell supergravity multiplet are given by \cite{Sohnius:1981tp}
\beq
\begin{aligned}
    &\delta g_{\mu\nu} = \bar \epsilon \gamma_{(\mu} \psi_{\nu)}\ , \\
    &\delta \psi_\mu =  \partial_\mu \epsilon - \frac14 \partial_\nu g_{\rho\mu} \gamma^{\nu\rho} \epsilon + i \gamma_5 V_\mu \epsilon - \frac{i}{4} \gamma_\mu G_\nu \gamma_5 \gamma^\nu \epsilon \ ,  \\
    &\delta V_\mu = - \frac{i}{4} \bar \epsilon \gamma_5 \gamma_\mu \gamma^{\nu \rho} \partial_\nu \psi_\rho + \frac{3}{8} \varepsilon_{\mu\nu}{}^{\rho\sigma} \bar \epsilon \gamma^\nu \partial_\rho \psi_\sigma \ , \\
    &\delta C_{\mu\nu} =\bar\epsilon \gamma_{[\mu} \psi_{\nu]} \ .
\end{aligned}
\label{eqn:ch2_new_min_susy}
\eeq

\noindent and the full nonlinear off-shell lagrangian of pure supergravity is equal to \cite{Sohnius:1981tp}
\beq
    e^{-1}\L_\textup{NM}=\L_\textup{SUGRA}-3V_\mu V^\mu+2V^\mu C_\mu \ ,
\eeq

\noindent where again $\L_\textup{SUGRA}$ is the lagrangian given in eq.~\eqref{eqn:ch2_sugra_lag_1}.

We employed explicitly this new-minimal setup in the original work \cite{ms2}, where we studied the coupling to supergravity of the massive spin-2 field. We will discuss in detail the results of this paper in Section \ref{ch2_sec_ms2}.

\subsection{The non-minimal formulations}

The two previous supercurrent multiplets $\S^\mu_\textup{FZ}$ \eqref{eqn:ch2_FZ_sc_eq}-\eqref{eqn:ch2_FZ_sc_comp} and $\S^\mu_\textup{R}$ \eqref{eqn:ch2_R_sc_eq}-\eqref{eqn:ch2_R_sc_comp} admit a generalization into a further supercurrent superfield $\S^\mu$ interpolating between them. There are in fact instances in which neither the R-multiplet nor the Ferrara--Zumino multiplet can be defined consistently \cite{Komargodski:2010rb}. This concerns in particular the Ferrara--Zumino supercurrent, which is not well-defined in the presence of Fayet--Iliopoulos terms \cite{Komargodski:2009pc} and for specific classes of K\"ahler geometries \cite{Komargodski:2010rb}. In cases of this type, one has to resort to such a generalized multiplet $\S^\mu$ \eqref{eqn:ch2_sc_gen_eq}. This generalized supercurrent superfield $\S^\mu$ is defined by the set of equations \cite{Komargodski:2010rb}
\beq
\begin{aligned}
    \(\sigma_\mu\)_{\alpha\dalpha}\overline{D}^{\dalpha}\S^\mu=D_\alpha X+\chi_\alpha &\ , \\
    \overline{D}_{\dalpha} X=\overline{D}_{\dalpha} \chi_\alpha = 0& \ , \\
    D^\alpha\chi_\alpha=\overline{D}_{\dalpha} \bar{\chi}^{\dalpha}&\ ,
\end{aligned}
\label{eqn:ch2_sc_gen_eq}
\eeq

\noindent and contains the fields of $\S^\mu_\textup{FZ}$ and the 2-form field of $\S^\mu_\textup{R}$, together with an additional Majorana fermion and a real scalar, yielding a total of sixteen bosonic and sixteen fermionic degrees of freedom. 

With respect to off-shell supergravity, this multiplet $\S^\mu$ defines what are known as non-minimal off-shell formulations \cite{Siegel:1978mj}.


\section{The coupling to supergravity of the massive spin-2 field}\label{ch2_sec_ms2}

The off-shell formulations of supergravity and the supercurrent superfield formalism introduced in Section \ref{ch2_sec_off_shell_sugra} set the stage for presenting our findings in \cite{ms2}. As we remarked at the beginning of this chapter when discussing in Section \ref{ch2_sec_gauged_susy} the explicit construction of supergravity through the Noether procedure, and then further emphasized in Section \ref{ch2_sec_off_shell_sugra}, the supercurrent superfield encodes, by construction, the leading-order coupling to pure supergravity of the given set of fields under consideration. It contains in fact the current of supersymmetry and the stress-energy tensor, which are the currents of the local Super-Poincar\'e algebra, and define therefore the minimal coupling to the gravitino and the metric, as in eq.~\eqref{eqn:ch2_noether_psi_Xi} and \eqref{eqn:ch2_g_emt_noether}. Therefore, constructing explicitly the supercurrent superfield associated with a supersymmetric multiplet allows one to probe directly its coupling to supergravity at leading order in the Planck mass $\MP$, without relying on a specific lagrangian (see for example \cite{Komargodski:2010rb,Festuccia:2011ws}). We take this perspective in \cite{ms2} and use the formalism of the supercurrent superfield to investigate the coupling to (off-shell) supergravity of the massive spin-2 field.

Constructing reliable effective theories of a single massive spin-2 field coupled to gravity is well known to be a difficult task. First of all, being a massive field of higher-spin, the massive spin-2 theories are likely to propagate ghosts. At the quadratic order, the only ghost-free theory in flat space for massive spin-2 fields is the famous Fierz--Pauli lagrangian\footnote{In eq.~\eqref{eqn:ch2_ms2_L_FP}, the symbol $h_{\mu\nu}$ denotes the massive spin-2 field, of mass $m$, and we employed the standard definition $h\equiv h^\mu{}_\mu$.}
\cite{Fierz:1939ix}
\beq
    \L_\textup{FP}=\partial_\mu h_{\nu\rho}\partial^\nu h^{\mu\rho}-\frac12\partial_\rho h_{\mu\nu}\partial^\rho h^{\mu\nu}-\partial_\mu h^{\mu\nu}\partial_\nu h+\frac12\partial_\mu h\partial^\mu h-\frac{m^2}{2}\(h^{\mu\nu}h_{\mu\nu}-h^2\)\ .
\label{eqn:ch2_ms2_L_FP}
\eeq    

\noindent Further ghost-like instabilities arise when interactions and non-linearities are added, especially in the presence of a nontrivial spacetime background. In massive gravity, this is known as the Boulware--Deser ghost \cite{Boulware:1972yco}. In addition, aside from the ghost problem,\footnote{For more details on the problems of ghosts, we refer to \cite{Bouatta:2004kk,Hinterbichler:2011tt}.} the couplings to gravity of the massive spin-2 field are inherently associated with low unitarity cutoffs. This has been proven in several settings of increasing generality, by studying the behavior of the elastic scattering amplitudes $\A(s)$ for the massive spin-2 field in the limit of large center-of-mass energy $s$ \cite{Arkani-Hamed:2002bjr, Arkani-Hamed:2003roe, Schwartz:2003vj, Bonifacio:2018vzv, Bonifacio:2018aon, Bonifacio:2019mgk, Kundu:2023cof}. Gravitational theories in which this amplitude has a power growth of the type $\A(s)\sim s^\lambda$ will become strongly coupled at a scale
\begin{equation}
    \Lambda_\lambda= \(m^{\lambda-1}M_{\scalebox{0.6}{P}}\)^{\frac1\lambda},
\label{eqn:ch2_ms2_cutoff}
\end{equation}

\noindent so that the higher the power $\lambda$, the narrower will be the window of validity of the effective field theory approximation. It has been shown that, while an arbitrary interacting theory will generically scale with $\Lambda_5$ --- this is for example the case of the Fierz--Pauli theory \eqref{eqn:ch2_ms2_L_FP} coupled to gravity \cite{Arkani-Hamed:2002bjr} --- , specific choices of interactions, usually nonlinear and of the higher-derivative type, can raise the cutoff to higher energy scales. However, the cutoff of the theory is never greater than the scale $\Lambda_3$, which appears to be an insurmountable threshold for the couplings of a single massive spin-2 field. This has been shown to hold for nonlinear deformations of the Fierz–Pauli theory \eqref{eqn:ch2_ms2_L_FP}, as well as for interactions with gravity \cite{Bonifacio:2018aon} and with fields of lower spins \cite{Bonifacio:2019mgk, Kundu:2023cof}. Theories that saturate the $\Lambda_3$ threshold are therefore somewhat special. Indeed, as shown in \cite{Bonifacio:2018vzv} and \cite{Bonifacio:2018aon}, this applies to the well-known ghost-free nonlinear theories of de Rham--Gabadadze--Tolley (dRGT) massive gravity \cite{deRham:2010ik, deRham:2010kj} and Hassan--Rosen bigravity \cite{Hassan:2011zd}. 

We emphasize that these results for the scattering amplitudes and the corresponding strong coupling scales apply to theories with a single massive spin-2 field. By contrast, interactions involving additional massive spin-2 fields can raise the cutoff beyond the $\Lambda_3$ scale \cite{Schwartz:2003vj,Bonifacio:2019mgk}. For example, this is the case for compactified theories of gravity, in which the Kaluza--Klein interactions are tuned in such a way that the cutoff of the theory is, consistently, the higher-dimensional Planck scale. In order to reproduce this behavior, one must include the whole tower of Kaluza--Klein massive spin-2 modes \cite{SekharChivukula:2019yul,Chivukula:2020hvi,Bonifacio:2019ioc}. It is therefore interesting to ask whether the only consistent effective field theories describing a single massive spin-2 field coupled to gravity are those that arise from higher-dimensional theories, or whether additional constructions are possible.

In \cite{ms2} we investigate such a question in the case of supersymmetric theories. The supersymmetric $\N=1$ multiplet of the massive spin-2 field includes, as superpartners, a massive spin-1 field and two massive Majorana spin-$\frac32$ fields,\footnote{This is a $\N=1$ massive multiplets of the type discussed in Section \ref{ch1_sec_m_rep}.} and the associated free theory was first constructed in \cite{Buchbinder:2002gh,Zinoviev:2002xn}. Further studies on the self-interactions of this multiplet can be found in \cite{DelMonte:2016czb, Zinoviev:2018juc, Engelbrecht:2022aao}, but less is known about explicit couplings to (super)gravity. A theory of this type that certainly exists is, as we emphasized, the one arising from a higher-dimensional theory through a proper Kaluza--Klein reduction. In this case, this theory is pure 5D supergravity \cite{Gunaydin:1983bi, Ceresole:2000jd} compactified on the orbifold $S^1/\mathbb{Z}_2$, which yields, in four dimensions, pure supergravity coupled to a massless chiral multiplet and a tower of massive spin-2 multiplets. We explicitly perform this reduction in Appendix \ref{app_ms2_kk}. From the supersymmetric point of view, the question we address in \cite{ms2} becomes whether the Kaluza--Klein coupling of a single massive spin-2 field to supergravity is universal or if more supersymmetric couplings are instead possible. 

In fact, supersymmetry further restricts the spectrum of allowed couplings and, at the same time, provides a framework in which to identify them. This is given, as we pointed out at the beginning of the section, by the supercurrent superfield, which we construct explicitly for the massive spin-2 supermultiplet and use to probe its leading-order coupling to supergravity. The result is that supersymmetry and diffeomorphism invariance imply the existence of a \textit{unique} consistent coupling of the massive spin-2 multiplet to new-minimal off-shell supergravity. The resulting theory includes non-minimal couplings to the gravitino and, in particular, to the metric field, both of which involve higher-derivative terms. Moreover, the interactions of the massive spin-2 field with gravity described by the theory are found to have an improved effective field theory cutoff, equal to $\Lambda_4$ instead of $\Lambda_5$. Additional possible couplings can then be obtained by deforming this theory into one involving a coupling to non-minimal off-shell supergravity, as in eq.~\eqref{eqn:ch2_sc_gen_eq}. We also argue that this class of couplings does not contain, by construction, the Kaluza--Klein theory. We now present this analysis and its results in full detail.

\subsection{The massive spin-2 supermultiplet and its supercurrent superfield}\label{ch2_sec_ms2_mult}

We begin with the rigid $\N=1$ supermultiplet of the massive spin-2 field, which we denote as
\beq
    \left\{h_{\mu\nu},\lambda_\mu,\chi_\mu,A_\mu\right\} \ ,
\label{eqn:ch2_ms2_mult_names}
\eeq

\noindent where $h_{\mu\nu}$ is the massive spin-2, $\lambda_\mu$ and $\chi_\mu$ are the Majorana spin-$\frac32$, and $A_\mu$ is the massive spin-1. At the linear level, they are related by supersymmetry as\footnote{The supersymmetry transformations \eqref{eqn:ch2_ms2_ms2_susy} were first determined in \cite{Buchbinder:2002gh,Zinoviev:2002xn}. Alternatively, one can extract them from the ones of pure 5D supergravity by compactification on the orbifold $S^1/\mathbb{Z}_2$. We perform this computation explicitly in Appendix \ref{app_ms2_kk}.}
\begin{equation}
\begin{aligned}
    &\delta A_\mu=\frac{1}{2}\bepsilon\mpsi_\mu \ , \\
    &\delta\mpsi_\mu=-\frac{i}{4}\(mh_{\mu\rho}+\tilde{F}_{\mu\rho}+2i\partial_\rho A_\mu\gamma_5\)\gamma_5\gamma^\rho\epsilon \ ,  \\
    &\delta h_{\mu\nu}=\frac{i}{m}\bepsilon\gamma_5\(\partial_{(\mu}\mpsi_{\nu)}-m\gamma_{(\mu}\chi_{\nu)}\) \ , \\
    &\delta\chi_\mu=\frac{i}{4}\partial_\rho h_{\mu\sigma}\gamma^{\rho\sigma}\gamma_5\epsilon+\frac{1}{8m}\partial_\mu F_{\rho\sigma}\gamma^{\rho\sigma}\epsilon+\frac{m}{4}A_\rho\(2\delta^\rho_\mu+\gamma^\rho{}_\mu\)\epsilon \ ,
\end{aligned}
\label{eqn:ch2_ms2_ms2_susy}
\end{equation}

\noindent where $m$ is the mass of the multiplet. The theory associated with these transformations is the supersymmetric extension of the Fierz--Pauli lagrangian \eqref{eqn:ch2_ms2_L_FP}, given by\footnote{For convenience, we employ a non-canonical normalization for the bosonic fields. The normalization of the spin-1 field $A_\mu$ differs from the canonical one by a factor $\sqrt{\frac{2}{3}}$, whereas the one of the spin-2 field $h_{\mu\nu}$ by a factor $\frac14$.}
\beq
\begin{aligned}
	\mathcal{L}_\textup{SFP}=&-\frac{3}{8}F^{\mu\nu}F_{\mu\nu}-\frac{3}{4}m^2A^{\mu}A_\mu -\frac{1}{2}\bar{\mpsi}_\mu\gamma^{\mu\nu\rho}\partial_\nu\mpsi_\rho-\frac{1}{2}\bchi_\mu\gamma^{\mu\nu\rho}\partial_\nu\chi_\rho-m\bar{\mpsi}_\mu\gamma^{\mu\nu}\chi_\nu\\
	&+\frac{1}{4}\partial_\mu h_{\nu\rho}\partial^\nu h^{\mu\rho}-\frac18\partial_\rho h_{\mu\nu}\partial^\rho h^{\mu\nu}-\frac14\partial_\mu h^{\mu\nu}\partial_\nu h+\frac18\partial_\mu h\partial^\mu h-\frac{m^2}{8}\(h^{\mu\nu}h_{\mu\nu}-h^2\) \ .
\label{eqn:ch2_ms2_L_SFP}
\end{aligned}
\eeq

\noindent This lagrangian yields the equations of motion
\beq
\begin{aligned}
	\begin{aligned}  \partial_\mu F^{\mu\nu}-m^2 A^\nu&=0,  &  \partial_{[\mu}F_{\nu\rho]}&=0, \\
	\gamma^{\mu\nu\rho}\partial_\nu\mpsi_\rho+m\gamma^{\mu\nu}\chi_\nu&=0,  &  \gamma^{\mu\nu\rho}\partial_\nu\chi_\rho+m\gamma^{\mu\nu}\mpsi_\nu&=0, \end{aligned} \\
	 \Box h_{\mu\nu}-2\partial_\rho\partial_{(\mu} h^\rho{}_{\nu)}+\eta_{\mu\nu}\partial_\rho\partial_\sigma h^{\rho\sigma}+\partial_\mu\partial_\nu h-\eta_{\mu\nu}\Box h-m^2\(h_{\mu\nu}-\eta_{\mu\nu }h\)=0, 
\label{eqn:ch2_ms2_eom_full}
\end{aligned}
\eeq

\noindent which can be expressed as the following set of dynamical equations and constraints:
\beq
\begin{aligned}
								 &		& 	&\Box A_\mu =m^2 A_\mu \ ,   		&	 \partial^\mu A_\mu &=0 \ ,   		& 	& \\
								 &		& 	&\Box h_{\mu\nu}=m^2 h_{\mu\nu} \ ,  	&	 \partial^\rho h_{\rho\mu}&=0 \ ,  	&  	h&=0 \ ,  \\
	\cancel{\partial}\mpsi_\mu&=m\chi_\mu \ ,   		& 	&\Box\mpsi_\mu=m^2\mpsi_\mu \ ,  	& 	\partial^\mu\mpsi_\mu&=0 \ , 	& 	 \gamma^\mu \mpsi_\mu&=0 \ ,     \\
	 \cancel{\partial}\chi_\mu&=m\mpsi_\mu \ ,  		& 	&\Box\chi_\mu=m^2\chi_\mu \ ,   		&	 \partial^\mu\chi_\mu&=0 \ , 		& 	\gamma^\mu \chi_\mu&=0 \ .    \\
\label{eqn:ch2_ms2_eom}
\end{aligned}
\eeq

\noindent Using these equations, one can check that the supersymmetry transformations \eqref{eqn:ch2_ms2_ms2_susy} close, accordingly to the Super-Poincar\'e algebra \eqref{eqn:ch1_SP_alg}-\eqref{eqn:ch1_comm_susy_derivative}, on
\begin{equation}
	\[\delta_1,\delta_2\]=-\frac{1}{2}\(\bepsilon_1\gamma^\rho\epsilon_2\) \partial_\rho \ .
\label{eqn:ch2_ms2_susy_cl}
\end{equation}

We can now compute the supercurrent superfield \cite{Ferrara:1974pz,Sohnius:1981tp,Komargodski:2010rb} associated with this massive spin-2 field. For this purpose, we observe that the supersymmetry transformations \eqref{eqn:ch2_ms2_ms2_susy} enjoy the following R-symmetry:
\begin{align}
    \mpsi_\mu \to e^{i \alpha \gamma_5}\mpsi_\mu \ , && \chi_\mu\to e^{-i \alpha \gamma_5}\chi_\mu \ , && \epsilon\to e^{-i \alpha \gamma_5}\epsilon \ .
\label{eqn:ch2_ms2_axial_sym}
\end{align}

\noindent Thus, the massive spin-2 supermultiplet \eqref{ch2_sec_ms2_mult}-\eqref{eqn:ch2_ms2_ms2_susy} is naturally associated with a supercurrent superfield of the R-multiplet type, given in eq.~\eqref{eqn:ch2_R_sc_eq}-\eqref{eqn:ch2_R_sc_susy} \cite{Sohnius:1981tp}. As described in section \ref{ch2_sec_off_shell_sugra}, this supercurrent superfield contains the R-symmetry conserved current as bottom component and closes on an antisymmetric two-form field. Therefore, we can explicitly construct this multiplet by starting with the Noether current of the R-symmetry \eqref{eqn:ch2_ms2_axial_sym} and applying successive supersymmetry transformations \eqref{eqn:ch2_ms2_ms2_susy}. These generate the remaining currents step by step, as dictated by the R-multiplet structure in \eqref{eqn:ch2_R_sc_susy}. The R-symmetry current associated with eq.~\eqref{eqn:ch2_ms2_axial_sym} is
\begin{equation}
    J^\mu_0=\varepsilon^{\mu\nu\rho\sigma}\(\bar{\mpsi}_\rho\gamma_\nu\mpsi_\sigma-\bchi_\rho\gamma_\nu\chi_\sigma\) \ .
\label{eqn:ch2_ms2_axial_c}
\end{equation}

\noindent This R-symmetry current is indeed conserved on-shell and one can check that it satisfies the linear superfield condition \eqref{eqn:ch2_lin_sf}. Following \eqref{eqn:ch2_R_sc_susy}, its supersymmetry transformation \eqref{eqn:ch2_ms2_ms2_susy} yields the current of supersymmetry $\Xi^\mu_0$, which results in
\begin{equation}
\begin{aligned}
\Xi^\mu_0=\frac{1}{2}\varepsilon^{\mu\nu\rho\sigma}&\biggl\{\(mh_{\rho\tau}+\tilde{F}_{\rho\tau}-2i\partial_\tau A_\rho\gamma_5\)\gamma^\tau\gamma_\nu\mpsi_\sigma \\
&\quad+\[\partial_\lambda h_{\tau\rho}\gamma^{\lambda\tau}\gamma_5-\frac{i}{2m}\partial_\rho F_{\lambda\tau}\gamma^{\lambda\tau}+imA_\tau\(2\delta^\tau_\rho+\gamma_\rho{}^\tau\)\]\gamma_\nu\gamma_5\chi_\sigma\biggr\} \  ,
\end{aligned}
\label{eqn:ch2_ms2_susy_c}
\end{equation}

\noindent which is also conserved on-shell. Then, a further supersymmetry transformations yields the antisymmetric field $X^{\mu\nu}_0=\varepsilon^{\mu\nu\rho}{}_{\sigma}\partial_\rho X_{0}^\sigma$, as in eq.~\eqref{eqn:ch2_R_sc_X_dual}, resulting in
\begin{equation}
    X^{\mu}_0=\frac{1}{2}A_\nu\(mh^{\nu\mu}+\tilde{F}^{\nu\mu}\)+\frac{i}{4}\bar{\mpsi}^\nu\gamma_5 \gamma^\mu\mpsi_\nu \ ,
\label{eqn:ch2_ms2_sc_X}
\end{equation}

\noindent as well the stress-energy tensor
\beq
    T^{\mu\nu}_0=T^{\mu\nu}_\textup{K}+\partial_\rho B^{\rho\mu\nu}_0+\partial_\rho\partial_\sigma G^{\mu\sigma\nu\rho}_0 \ .
\label{eqn:ch2_ms2_sc_T_1}
\eeq

Let us now focus on this energy-momentum tensor and describe the three terms into which it is organized. The term $T^{\mu\nu}_\textup{K}$ is the universal contribution from the kinetic terms of the various fields, associated the Proca, Rarita--Schwinger and Fierz--Pauli lagrangians:
\beq
    T^{\mu\nu}_\textup{K}=\frac{3}{2}T^{\mu\nu}_\textup{P}+T^{\mu\nu}_\textup{RS}+\frac14T^{\mu\nu}_\textup{FP} \ ,
\eeq

\noindent with
\beq
\begin{aligned}
    &T_\textup{P}^{\mu\nu}=F^{\mu\rho}F^\nu{}_\rho-\frac{1}{4}\eta^{\mu\nu}F^2+m^2A^\mu A^\nu-\frac{m^2}{2}\eta^{\mu\nu}A^2 \ , \\
    &T^{\mu\nu}_\textup{RS}=\frac{1}{2}\(\bar{\mpsi}^\sigma\gamma^{(\mu}\partial^{\nu)}\mpsi_\sigma+\bchi^\sigma\gamma^{(\mu}\partial^{\nu)}\chi_\sigma\)-\(\bar{\mpsi}^\sigma\gamma^{(\mu}\partial_\sigma\psi^{\nu)}+\bchi^\sigma\gamma^{(\mu}\partial_\sigma\chi^{\nu)}\) \ ,\\
    &\begin{aligned}
    T^{\mu\nu}_\textup{FP}=& \partial^\mu h^{\rho\sigma}\partial^\nu h_{\rho\sigma}+\partial^\rho h^{\mu\sigma}\partial_\rho h^{\nu}{}_{\sigma}-2\partial^{(\mu|}h^{\rho\sigma}\partial_\rho h^{|\nu)}{}_{\sigma}+m^2h^{\mu\rho}h^{\nu}{}_{\rho}\\
    &+\frac{\eta^{\mu\nu}}{2}\(-\partial^\lambda h^{\rho\sigma}\partial_\lambda h_{\rho\sigma}+\partial^\lambda h^{\rho\sigma}\partial_\rho h_{\lambda\sigma}-m^2h^{\rho\sigma}h_{\rho\sigma}\) \ .
    \end{aligned}
\end{aligned}
\label{eqn:ch2_ms2_sc_T_K}
\eeq

\noindent We remark that these tensors correspond to the Minkowski background limit of the standard, fully covariant tensors of the corresponding fields.\footnote{The complete stress-energy tensor of the covariantized Fierz--Pauli action can be found in \cite{Petrov:2017bdx}. The one for the Rarita--Schwinger field was instead reported here in eq.~\eqref{eqn:ch2_gpp_emt_full}, referring to \cite{Casagrande:2023fjk}.} Whereas for the Proca and Fierz--Pauli fields the Minkowski limit of the complete tensor coincide with the one obtained directly in flat space  --- namely, from the lagrangian \eqref{eqn:ch2_ms2_L_SFP} --- via the standard Belinfante symmetrization procedure, for the Rarita--Schwinger one the two tensors differ by terms which come from the spin-connection of the original covariant derivatives. This can be checked explicitly working out the limit from the stress-energy tensor given in eq.~\eqref{eqn:ch2_gpp_emt_full}. Despite being improvements from the flat space point of view \cite{Callan:1970ze}, these terms are crucial correctly interpreting the stress-energy tensor \eqref{eqn:ch2_ms2_sc_T_1} and the associated coupling to gravity.

The terms $B^{\rho\mu\nu}_0$ and $G^{\mu\sigma\nu\rho}_0$ in eq.~\eqref{eqn:ch2_ms2_sc_T_1} are instead equal to
\begin{equation}
\begin{aligned}
B^{\rho\mu\nu}_0=&\frac{1}{4}\(h^{(\nu}{}_{\sigma}\partial^{\mu)}h^{\rho\sigma}-\partial^{(\mu}h^{\nu)}{}_{\sigma}h^{\rho\sigma}\)+\frac{5}{4}\(A^{(\nu}\partial^{\mu)} A^\rho-A^\rho\partial^{(\mu}A^{\nu)}\) \\
&  +\frac{1}{4m}\(\partial^{(\mu}\tilde{F}^{\rho\sigma}h^{\nu)}{}_\sigma-\tilde{F}^{\rho\sigma}\partial^{(\mu}h^{\nu)}{}_\sigma+\tilde{F}^{(\nu}{}_\sigma\partial^{\mu)}h^{\rho\sigma}-\partial^{(\mu}\tilde{F}^{\nu)}{}_\sigma h^{\rho\sigma}\)\\
& +\frac{1}{2m}\(\partial^{(\mu}\bchi^{\nu)}\mpsi^\rho-\bchi^{(\nu}\partial^{\mu)}\mpsi^\rho+\partial^{(\mu}\bar{\mpsi}^{\nu)}\chi^\rho-\bar{\mpsi}^{(\nu}\partial^{\mu)}\chi^\rho\)\; , 
\end{aligned}
\label{eqn:ch2_ms2_sc_B}
\end{equation}

\noindent and\footnote{See Appendix \ref{app_conventions} for the notation on the antisymmetrized indices used in eq.~\eqref{eqn:ch2_ms2_sc_G} and in the following.}
\beq
\begin{aligned}
    G^{\mu\sigma\nu\rho}_0=&\frac{1}{2m^2}\eta^{\nu][\sigma}F^{\mu]\lambda}F_{\lambda}{}^{[\rho}+\frac{1}{8m^2}\eta^{\nu[\sigma}\eta^{\mu]\rho}F^2+3\eta^{\nu][\sigma}A^{\mu]}A^{[\rho}+\frac{1}{4}\eta^{\nu[\mu}\eta^{\sigma]\rho}A^2\\
    &+\frac{1}{2m}\eta^{\rho][\mu}\(\tilde{F}^{\sigma]}{}_{\lambda}h^{\lambda[\nu}-h^{\sigma]}{}_{\lambda}\tilde{F}^{\lambda[\nu}\)-h^{\mu[\rho}h^{\nu]\sigma}+\frac{1}{2}\eta^{\rho][\mu}h^{\sigma]}{}_\lambda h^{\lambda[\nu}\\
    &+\frac{1}{m}\(\bar{\mpsi}^{[\sigma}\gamma^{\mu][\rho}\chi^{\nu]}+\bar{\mpsi}^{[\rho}\gamma^{\nu][\sigma}\chi^{\mu]}\)\; . 
\end{aligned}
\label{eqn:ch2_ms2_sc_G}
\eeq

\noindent These two terms are improvements with respect to $T^{\mu\nu}_\textup{K}$, namely deformations of the current that are trivially conserved and thus do not modify the conservation of the associated charges. \cite{Callan:1970ze}. They have in fact the following symmetry structures:\footnote{While the symmetry properties (\ref{eqn:ch2_ms2_imp_sym}) are manifest for $\partial_\rho\partial_\sigma G^{\mu\sigma\nu\rho}_0$ in eq.~\eqref{eqn:ch2_ms2_sc_G}, those of $\partial_\rho B^{\rho\mu\nu}_0$ in eq.~\eqref{eqn:ch2_ms2_sc_B} require an on-shell manipulation to be manifest. Taking the Proca sector of eq.~(\ref{eqn:ch2_ms2_sc_B}) as an example, one has
\begin{equation}
\begin{aligned}
\frac{4}{5}\partial_\rho B_A^{\rho\mu\nu}=&\partial_\rho \(A^{(\nu}\partial^{\mu)} A^\rho-A^\rho\partial^{(\mu}A^{\nu)}\)=\\
=&\partial_\rho \(A^\nu\partial^{[\mu}A^{\rho]}+A^{[\mu}\partial^{\rho]}A^\nu+A^{[\mu|}\partial^\nu A^{|\rho]}\)+\underbrace{\partial_\rho\(A^{[\nu|}\partial^\rho A^{|\mu]}\)}_{=0}=\frac{4}{5}\partial_\rho B_A^{[\rho\mu]\nu}\; . 
\end{aligned}
\end{equation}

\noindent Analogous manipulations yield also $\partial_\rho B^{\rho\mu\nu}=\partial_\rho B^{[\rho|\mu|\nu]}$.}
\begin{equation}
\begin{aligned}
\partial_\rho B^{\rho\mu\nu}_0=\partial_\rho B^{\rho(\mu\nu)}_0=\partial_\rho B^{[\rho\mu]\nu}_0=\partial_\rho B^{[\rho|\mu|\nu]}_0 \ ,&\\
\partial_\rho\partial_\sigma G^{\mu\sigma\nu\rho}_0= \partial_\rho\partial_\sigma G^{(\mu|\sigma|\nu)\rho}_0=\partial_\rho\partial_\sigma G^{[\mu\sigma]\nu\rho}_0=\partial_\rho\partial_\sigma G^{\mu\sigma[\nu\rho]}_0 \ , &
\end{aligned}
\label{eqn:ch2_ms2_imp_sym}
\end{equation}

\noindent

\noindent   The presence of improvements in the currents of the multiplet is rather standard \cite{Ferrara:1974pz,Komargodski:2010rb} and they are perfectly consistent from the rigid theory point of view \cite{Callan:1970ze}. The peculiar property of the ones found in eq.~\eqref{eqn:ch2_ms2_sc_B}-\eqref{eqn:ch2_ms2_sc_G} for the massive spin-2 supermultiplet is that they contain higher-derivative terms.

Therefore, we obtained this way the supercurrent superfield of the massive spin-2 supermultiplet \eqref{eqn:ch2_ms2_ms2_susy}, which is the fist result of \cite{ms2}. This is given by
\beq
	\S^\mu_0=\left\{J^\mu_0,\Xi^\mu_0,T^{\mu\nu}_0,X^{\mu\nu}_0\right\},
\label{eqn:ch2_ms2_sc_0}
\eeq

\noindent where $J^\mu_0$ is the $R$-symmetry current \eqref{eqn:ch2_ms2_axial_c} from which the multiplet was constructed, and the other currents defined respectively in \eqref{eqn:ch2_ms2_susy_c}, \eqref{eqn:ch2_ms2_sc_T_1} and \eqref{eqn:ch2_ms2_sc_X}. The current multiplet $\S^\mu_0$ \eqref{eqn:ch2_ms2_sc_0} is, by definition, a R-multiplet like eq.~\eqref{eqn:ch2_R_sc_eq}-\eqref{eqn:ch2_R_sc_susy}, and one can check that the supersymmetry transformations \eqref{eqn:ch2_R_sc_susy} of this multiplet close, on-shell, on \eqref{eqn:ch2_ms2_susy_cl}.

\subsection{The coupling to supergravity and its uniqueness}\label{ch2_sec_ms2_gravity}

The motivation for constructing the supercurrent superfield $\S^\mu_0$ \eqref{eqn:ch2_ms2_sc_0} was that it can be used to define the coupling to supergravity of the massive spin-2 multiplet \eqref{eqn:ch2_ms2_mult_names}-\eqref{eqn:ch2_ms2_ms2_susy}, at leading order in the Planck mass. In particular, as we discussed in Section \ref{ch2_sec_off_shell_sugra}, supercurrent superfields are dual to the off-shell formulations of supergravity. As an R-symmetry multiplet, the supercurrent $\S^\mu_0$ \eqref{eqn:ch2_ms2_sc_0} defines a coupling to the so-called new-minimal off-shell supergravity, introduced in eq.~\eqref{eqn:ch2_new_min_susy} \cite{Sohnius:1981tp}. The corresponding leading-order coupling lagrangian is given by
\begin{equation}
\mathcal{L}\sim-\frac{1}{2}(g_{\mu\nu}-\eta_{\mu\nu})T^{\mu\nu}-\frac12 \bar{\psi}_\mu \Xi^\mu - \frac12 V_\mu J^\mu + C_\mu X^\mu \;,
\label{eqn:ch2_ms2_new_min_min_coupling}
\end{equation}

\noindent where $V_\mu$ is an axial vector and $C_\mu$ is the field strength of antisymmetric tensor field $C_\mu=\frac12\varepsilon_{\mu\nu\rho\sigma}\partial^\nu C^{\rho\sigma}$, as in eq.~\eqref{eqn:ch2_new_min_susy}, while $g_{\mu\nu}$ denotes here the fluctuation of the metric field around the Minkowski background.

When going from rigid to local supersymmetry, the currents of the supercurrent superfield are promoted to couplings to the gauge fields of supergravity. This is expressed by the leading-order coupling lagrangian \eqref{eqn:ch2_ms2_new_min_min_coupling}. This means that also the improvements to such currents are not arbitrary anymore, but rather acquire physical significance as non-minimal couplings to the associated gauge fields \cite{Callan:1970ze}. Focusing on the stress-energy tensor $T^{\mu\nu}_0$ of eq.~\eqref{eqn:ch2_ms2_sc_T_1}, the improvement terms $B^{\rho\mu\nu}_0$ of eq.~\eqref{eqn:ch2_ms2_sc_B} and $G^{\mu\sigma\nu\rho}_0$ of eq.~\eqref{eqn:ch2_ms2_sc_G} should correspond to non-minimal couplings to the metric field of the new-minimal off-shell supergravity, as in eq.~\eqref{eqn:ch2_ms2_new_min_min_coupling}. Their specific form can be reconstructed from the symmetries  \eqref{eqn:ch2_ms2_imp_sym} of the two improvement terms. Focusing on the linearised coupling of lagrangian \eqref{eqn:ch2_ms2_new_min_min_coupling}, it follows that $G^{\mu\sigma\nu\rho}_0$ defines a non-minimal coupling to the linearised Riemann tensor:
\begin{equation}
    g_{\mu\nu}\partial_\rho\partial_\sigma G^{\mu\sigma\nu\rho}_0\approx\(\partial_{\rho]}\partial_{[\sigma} g_{\mu][\nu}\)G^{[\mu\sigma][\nu\rho]}_0\sim   - \frac12  R_{\mu\sigma\nu\rho}G^{\mu\sigma\nu\rho}_0\, ,
\label{eqn:ch2_ms2_riem_L}
\end{equation}

\noindent according to
\beq
    R_{\mu\sigma\nu\rho}\sim  -2 \partial_{\rho]}\partial_{[\sigma}g_{\mu][\nu} \ .
\label{eqn:ch2_ms2_riem_lin}
\eeq

\noindent These non-minimal couplings are indeed peculiar, since the operator $G^{\mu\sigma\nu\rho}$ of eq.~\eqref{eqn:ch2_ms2_sc_G} contains higher-derivative terms, but they are perfectly consistent with the invariance under diffeomorphisms of the theory coupled to gravity. In contrast, the improvement $B^{\rho\mu\nu}_0$ corresponds to a non-minimal coupling to the linearised Levi--Civita connection:
\begin{equation}
    g_{\mu\nu}\partial_\rho B^{\rho\mu\nu}_0\approx\(\partial_{[\rho} g_{\mu]\nu}+\frac{1}{2}\partial_\nu g_{\mu\rho}\)B^{[\rho\mu]\nu}_0\sim\Gamma_{\rho\mu\nu}B^{\rho\mu\nu}_0 \ ,
\label{eqn:ch2_ms2_lc_L}
\end{equation}

\noindent as
\begin{align}
    g_{\rho\sigma} \Gamma^\sigma_{\mu\nu}\sim\partial_{[\mu}g_{\rho]\nu}+\frac{1}{2}\partial_\nu g_{\mu\rho} \ .
\label{eqn:ch2_ms2_lc_lin}
\end{align}

\noindent Since these terms are couplings to the bare Levi--Civita connections --- namely, not part of any covariant derivative/operator --- they are in open contrast with diffeomorphism invariance. Therefore, because of these $B^{\rho\mu\nu}_0$-improvements in the stress-energy tensor $T^{\mu\nu}_0$ \eqref{eqn:ch2_ms2_sc_T_1}, the supercurrent superfield $\S^\mu_0$ \eqref{eqn:ch2_ms2_sc_0} does not define a consistent coupling to gravity of the massive spin-2 multiplet \eqref{eqn:ch2_ms2_ms2_susy} \cite{ms2}.

Although naturally associated with the massive spin-2 multiplet through the R-symmetry current $J^\mu_0$ \eqref{eqn:ch2_ms2_axial_c}, it is necessary to deform the starting supercurrent $\S^\mu_0$ \eqref{eqn:ch2_ms2_sc_0} in order to look for consistent couplings to supergravity. Since the source of inconsistency are improvement terms in the stress-energy tensor $T^{\mu\nu}_0$, the strategy we employ in \cite{ms2} is to study superfield improvements to the whole supercurrent $\S^\mu_0$, of the form 
\begin{align}
    \S^\mu_0\,\,\longrightarrow\,\,\S^\mu=\S^\mu_0+\Delta\S^\mu \ ,
\label{eqn:ch2_ms2_imp_names}
\end{align}

\noindent with 
\begin{align}
  \Delta\S^\mu=\left\{\Delta J^\mu,\Delta\Xi^\mu,\Delta T^{\mu\nu},\Delta X^{\mu\nu}\right\} \ .
\label{eqn:ch2_ms2_DJ_def}
\end{align}

\noindent The idea is to find the most general supercurrent improvement $\Delta\S^\mu$ such that the resulting stress-energy tensor $T^{\mu\nu}=T^{\mu\nu}_0+\Delta T^{\mu\nu}$ does not contain any diffeomorphism breaking improvement terms of the $B^{\rho\mu\nu}$ type \eqref{eqn:ch2_ms2_sc_B}. In superspace notation, this general improvement is written as
\beq  
    \Delta\S^\mu=\partial_\nu \mathcal{L}^{\mu\nu} + \sigma^{\mu\, \alpha\dot{\beta}} [ D_\alpha , \bar D_{\dot{\beta}} ] \mathcal{U}   ,
\label{GeneralImproved} 
\eeq

\noindent where $\mathcal{L}^{\mu\nu}  =- \mathcal{L}^{\nu\mu}$ is taken to be a linear superfield, namely satisfying the condition \eqref{eqn:ch2_lin_sf},
\beq
    D^2\L^{\mu\nu}=0 \ .
\label{eqn:ch2_ms2_Dj_lin_m}
\eeq

\noindent The superfield $\mathcal{U}$ is instead real but unconstrained. Notice that the $\L^{\mu\nu}$-improvement still defines, by construction, a coupling to new-minimal off-shell supergravity \eqref{eqn:ch2_new_min_susy}. On the contrary, the $\mathcal{U}$-improvement breaks the conservation of the bottom component of $\S^\mu$ \cite{Komargodski:2010rb}. Such a supercurrent $\S^\mu$ would be an example of the generalized supercurrent superfield introduced in eq.~\eqref{eqn:ch2_sc_gen_eq}, which defines a coupling to a non-minimal formulation of off-shell supergravity \cite{Siegel:1978mj}.

\subsubsection{The unique coupling to new-minimal off-shell supergravity}

We begin by analyzing the $\mathcal{U}=0$ case, \textit{i.e.} the improved couplings to new-minimal off-shell supergravity. As in the previous section constructing $\S^\mu_0$, we begin with the bottom component of the linear improvement $\L^{\mu\nu}$, namely the axial improvement $\Delta J^\mu$ of eq.~\eqref{eqn:ch2_ms2_imp_names}. As for its superfield embedding, this improvement current has the form
\beq
    \Delta J^\mu=\frac{1}{m}\partial_\nu L^{\mu\nu} \ ,
\label{eqn:ch2_ms2_Dj_gen_2}
\eeq

\noindent with $L^{\mu\nu}$ being a real and antisymmetric composite operator. We restrict these operators to have a limited number of derivatives, in such a way as to produce contributions to the stress-energy tensor with at most four derivatives, as the ones characterizing $T^{\mu\nu}_0$ in eq.~\eqref{eqn:ch2_ms2_sc_T_1}, \eqref{eqn:ch2_ms2_sc_B} and \eqref{eqn:ch2_ms2_sc_G}. With this working assumption, a general ansatz for $L^{\mu\nu}$ can be written based on its consistency with the setup under consideration. First, the current $\Delta J^\mu$ has to be an axial vector, and thus the operators entering $L^{\mu\nu}$ have to be consistent with such a parity. This requirement is already quite stringent, and constraints $L^{\mu\nu}$ to the following linear combination of independent operators:
\beq
\begin{aligned}
    L^{\mu\nu}=&c_\textup{B1} h^{[\mu}{}_\lambda \partial^\lambda A^{\nu]}+c_\textup{B2}h^{[\mu}{}_\lambda\partial^{\nu]}A^\lambda+\frac{c_\textup{B3}}{m}\varepsilon^{\mu\nu\rho\sigma}\partial_\lambda A_\rho\partial_\sigma A^\lambda\\
    &+c_\textup{F1}\varepsilon^{\mu\nu\rho\sigma}\bchi_\rho\mpsi_\sigma+ic_\textup{F2}\bchi^{[\mu}\gamma_5\mpsi^{\nu]}.
\end{aligned}
\label{eqn:ch2_ms2_Dj_1}
\eeq

\noindent Then, the second, essential requirement on $L^{\mu\nu}$ is that it has to define, by construction, a linear multiplet, like the R-symmetry current of eq.~\eqref{eqn:ch2_ms2_axial_c}. The linear multiplet constraint on the superfield $\L^{\mu\nu}$ of eq.~\eqref{eqn:ch2_ms2_Dj_lin_m} translates on its bottom component to 
\beq
	\delta_\textup{L}^2 L^{\mu\nu} =0\; ,
\label{eqn:ch2_ms2_linear_m_comp}
\eeq

\noindent where $\delta_\textup{L}$ is the supersymmetry transformation with the left-handed spinor $\epsilon_\textup{L}$ as parameter. This condition restricts further the linear combination of eq.~\eqref{eqn:ch2_ms2_Dj_1}. Using the Fierz identity
\begin{equation}
    \epsilon_\textup{L}\bepsilon_\textup{L}=-\frac{1}{2}\left(\bepsilon_\textup{L}\epsilon_\textup{L}\right)P_\textup{L},
\end{equation}

\noindent the various operators of $L^{\mu\nu}$ in eq.~\eqref{eqn:ch2_ms2_Dj_1} transform as
\beq
\begin{aligned}
    \bullet& & &\delta_\textup{L}^2\left(h^{[\mu}{}_\sigma \partial^\sigma A^{\nu]}\)=\frac{i}{4}\(\bepsilon_\textup{L}\epsilon_\textup{L}\)\[-\frac{1}{m}\partial_\sigma\bar{\mpsi}^{[\mu}_\textup{L}\partial^{\nu]}\mpsi^\sigma_{\textup{L}}+\bchi^\sigma_\textup{R}\gamma^{[\mu}\partial^{\nu]}\mpsi_{\sigma\textup{L}}\] \ , \\
    \bullet& & &\delta_\textup{L}^2\left(h^{[\mu}{}_\sigma\partial^{\nu]}A^\sigma\right)=\frac{i}{4}\(\bepsilon_\textup{L}\epsilon_\textup{L}\)\[\frac{1}{m}\partial_\sigma\bar{\mpsi}^{[\mu}_\textup{L}\partial^{\nu]}\mpsi^\sigma_{\textup{L}}+\bchi^\sigma_\textup{R}\gamma^{[\mu}\partial^{\nu]}\mpsi_{\sigma\textup{L}}\] \ , \\
    \bullet& & &\delta_\textup{L}^2\left(\frac{1}{m}\varepsilon^{\mu\nu\rho\sigma}\partial_\lambda A_\rho\partial_\sigma A^\lambda\right)=\frac{i}{4}\(\bepsilon_\textup{L}\epsilon_\textup{L}\)\[\frac{2}{m}\partial_\sigma\bar{\mpsi}^{[\mu}_\textup{L}\partial^{\nu]}\mpsi^\sigma_{\textup{L}}-2\bchi^\sigma_\textup{R}\gamma^{[\mu}\partial^{\nu]}\mpsi_{\sigma\textup{L}}\] \ , \\
    \bullet& & &\delta_\textup{L}^2\(\varepsilon^{\mu\nu\rho\sigma}\bchi_\rho\mpsi_\sigma\)=0 \ , \\
\bullet& & &\delta_\textup{L}^2\(\bchi^{[\mu}\gamma_5\mpsi^{\nu]}\)=0.
\end{aligned}
\label{eqn:ch2_ms2_d2L_Dj}
\eeq

\noindent Thus, while the fermionic operators are consistent on their own with the linear multiplet structure, the bosonic ones are restricted to the combination given by
\begin{align}
\quad c_\textup{B2}=0 \ ,  && c_\textup{B1}=2c_\textup{B3}\equiv 2c_\textup{B}.
\end{align}

\noindent Thus, the most general ansatz --- at the four-derivative order in the associated stress-energy tensor --- for the axial current \eqref{eqn:ch2_ms2_Dj_gen_2} takes the compact form
\begin{equation}
	L^{\mu\nu}=c_\textup{B}\(2h^{[\mu|}{}_\lambda\partial^\lambda A^{|\nu]}+\frac{1}{m}\varepsilon^{\mu\nu\rho\sigma}\partial_\lambda A_\rho\partial_\sigma A^\lambda\)+c_\textup{F1}\varepsilon^{\mu\nu\rho\sigma}\bchi_\rho\mpsi_\sigma+ic_\textup{F2}\bchi^{[\mu}\gamma_5\mpsi^{\nu]} \ .
\label{eqn:ch2_ms2_Dj_fin}
\end{equation}

The linear superfield improvements $\L^{\mu\nu}$ of eq.~\eqref{GeneralImproved} are therefore spanned by the three real coefficients $\left\{c_\textup{B},c_\textup{F1},c_\textup{F2}\right\}$. The goal is to find the subset that corresponds to a diffeomorphism-invariant coupling of the stress-energy tensor $T^{\mu\nu}=T^{\mu\nu}_0+\Delta T^{\mu\nu}$. Applying, as before, the R-multiplet supersymmetry algebra \eqref{eqn:ch2_R_sc_susy}, one finds that the three axial improvements in eq.~\eqref{eqn:ch2_ms2_Dj_fin} give the following $B^{\rho\mu\nu}$-like contributions to the associated energy-momentum tensors:
\begin{align}
    &\begin{aligned}
        \partial_\rho\Delta B^{\rho\mu\nu}_\textup{B}=&\frac{1}{8}\partial_\rho\(A^{(\mu}\partial^{\nu)}A^\rho-A^\rho \partial^{(\mu}A^{\nu)}\)\\
        	&+\frac{1}{8m^2}\partial_\rho\(\partial^\sigma A^{(\mu}\partial_\sigma\partial^{\nu)}A^\rho-\partial_\sigma A^\rho \partial^\sigma\partial^{(\mu}A^{\nu)}\),
    \end{aligned} \label{eqn:ch2_ms2_DB_B} \\
     &\begin{aligned}
        \partial_\rho \Delta B^{\rho\mu\nu}_\textup{F1}=&\frac{1}{2m}\partial_\rho\(\partial^{(\mu}\bchi^{\nu)}\mpsi^\rho-\bchi^{(\nu}\partial^{\mu)}\mpsi^\rho+\partial^{(\mu}\bar{\mpsi}^{\nu)}\chi^\rho-\bar{\mpsi}^{(\nu}\partial^{\mu)}\chi^\rho\)\\
        &+\frac{1}{2m}\partial_\rho\(\partial^{(\mu|}\bar{\mpsi}_\sigma\gamma^\rho\gamma^{|\nu)}\chi^\sigma-\partial^\rho\bar{\mpsi}_\sigma\gamma^{(\mu}\gamma^{\nu)}\chi^\sigma\), 
    \end{aligned} \label{eqn:ch2_ms2_DB_F1} \\
    &\partial_\rho\Delta B^{\rho\mu\nu}_\textup{F2}=-\frac{1}{4}\partial_\rho B^{\rho\mu\nu}_0. \label{eqn:ch2_ms2_DB_F2}
\end{align}

\noindent From these equations, we see that the axial improvements dubbed with B and F1 in \eqref{eqn:ch2_ms2_Dj_fin} lead to \textit{new} non-diffeomorphism invariant improvements to the stress-energy tensor, and thus must be discarded. On the contrary, the improvement F2 is exactly proportional to the $B^{\rho\mu\nu}$-improvement \eqref{eqn:ch2_ms2_sc_B} that affects the tensor $T^{\mu\nu}_0$ of eq.~\eqref{eqn:ch2_ms2_sc_T_1}. Hence, there is a solution for the coefficients in \eqref{eqn:ch2_ms2_Dj_fin}, given by
\beq
	c_\textup{B}=c_\textup{F1}=0\;, \qquad c_\textup{F2}=4\; ,
\label{eqn:ch2_ms2_scu_coeff_sol}
\eeq

\noindent that satisfies the original requirement, namely that removes the non-diffeomorphism invariant terms from the R-symmetry current superfield $\S_0^\mu$ \eqref{eqn:ch2_ms2_sc_0} without introducing additional terms of the same type. This solution is \textit{unique}, meaning that there exists only a single three-point coupling of the massive spin-2 supermultiplet \eqref{eqn:ch2_ms2_ms2_susy} to new minimal off-shell supergravity \eqref{eqn:ch2_new_min_susy} that is consistent with diffeomorphism invariance and supersymmetry at the same time. This result holds up to four derivatives operators, which are actually essential for realizing such a consistent coupling \cite{ms2}.

The supercurrent supermultiplet that defines this unique coupling is 
\begin{equation}
	\S^\mu=\left\{J^\mu,\Xi^\mu,T^{\mu\nu},X^{\mu\nu} \right\},
\label{eqn:ch2_ms2_scu_tot}
\end{equation}

\noindent with 
\begin{align}
    &J^\mu=\varepsilon^{\mu\nu\rho\sigma}\(\bar{\mpsi}_\rho\gamma_\nu\mpsi_\sigma-\bchi_\rho\gamma_\nu\chi_\sigma\)+\frac{4i}{m}\partial_\nu\(\bchi^{[\mu}\gamma_5\mpsi^{\nu]}\)\; ,  \label{eqn:ch2_ms2_scu_j} \\
    &\begin{aligned}
        \Xi^\mu=&\Xi^\mu_0-\frac{i}{m}\partial_\nu\biggl[\(mh^{[\nu}{}_\lambda+\tilde{F}^{[\nu}{}_\lambda-2i\partial_\lambda A^{[\nu}\gamma_5\)\gamma^{\lambda}\gamma_5\chi^{\mu]} \\
        &+\(\partial_\rho h^{[\mu}{}_\sigma\gamma^{\rho\sigma}\gamma_5-\frac{i}{2m}\partial^{[\mu}F_{\rho\sigma}\gamma^{\rho\sigma}+2imA^{[\mu}-imA_\rho\gamma^{\rho[\mu}\)\mpsi^{\nu]}\biggr]\; ,   
    \end{aligned}\label{eqn:ch2_ms2_scu_Xi} \\
    &T^{\mu\nu}=T^{\mu\nu}_\textup{K}+\partial_\rho\partial_\sigma \G^{\mu\sigma\nu\rho}\; ,  \label{eqn:ch2_ms2_scu_T} \\
    &\begin{aligned}
        X^{\mu}=&\frac{i}{4}\bar{\mpsi}^\nu\gamma_5 \gamma^\mu\mpsi_\nu+\frac{3m}{4}A_{\nu}h^{\nu\mu}+\frac{9}{8}A_{\nu}\tilde{F}^{\nu\mu}+\frac{1}{4m^2}\partial_\rho\partial_\sigma\(\tilde{F}^{\mu\rho}A^\sigma\)   \\
        & +\frac{1}{8m}\partial_\nu\(F^{\nu\rho}h^\mu{}_{\rho}-F^{\mu\rho}h^{\nu}{}_{\rho}\)-\frac{3}{4m}\partial_\nu\(\partial^\mu A^\rho h^{\nu}{}_{\rho}+\partial_\rho A^{\nu} h^{\rho\mu}\)  \\
        & +\frac{1}{4}\varepsilon^{\mu\nu\rho\sigma}\partial_\nu h^\lambda{}_{\rho} h_{\sigma\lambda}-\frac{1}{4m}\varepsilon^{\mu\nu\rho\sigma}\(\bchi_\nu \partial_\rho\mpsi_\sigma+\bar{\mpsi}_\nu \partial_\rho\chi_\sigma\)\; ,  
    \end{aligned}\label{eqn:ch2_ms2_scu_X}
\end{align}

\noindent where $\Xi_0^\mu$ and $T^{\mu\nu}_\textup{K}$ are, respectively, the current of supersymmetry of the minimal R-multiplet \eqref{eqn:ch2_ms2_susy_c}-\eqref{eqn:ch2_ms2_sc_0} and the Hilbert stress-energy tensor \eqref{eqn:ch2_ms2_sc_T_K}. The main feature of the resulting coupling to gravity --- namely, the stress-energy tensor in eq.~\eqref{eqn:ch2_ms2_scu_T} --- is the following non-minimal coupling to the Riemann tensor:
\begin{equation}
    \begin{aligned}
        \mathcal{G}^{\mu\sigma\nu\rho}=&-\frac{1}{2}h^{\rho[\mu}h^{\sigma]\nu}+\frac{1}{2}\eta^{\rho][\mu}h^{\sigma]}{}_\lambda h^{\lambda[\nu}+2\eta^{\rho][\mu}A^{\sigma]}A^{[\nu}-\frac{3}{4m^2}F^{\mu\sigma}F^{\nu\rho}+\frac{1}{2m^2}\eta^{\rho][\mu}F^{\sigma]}{}_\lambda F^{\lambda[\nu}\\
        &+\frac{1}{m^2}\(\eta^{\sigma][\nu}F^{\rho]}{}_\lambda \partial^\lambda A^{[\mu}+\eta^{\rho][\mu}F^{\sigma]}{}_\lambda \partial^\lambda A^{[\nu}\)+\frac{1}{2m}\eta^{\rho][\mu}\(\tilde{F}^{\sigma]}{}_{\lambda}h^{\lambda[\nu}-h^{\sigma]}{}_{\lambda}\tilde{F}^{\lambda[\nu}\)\\
        &+\frac{1}{2m}\(\varepsilon^{\mu\sigma\lambda\tau} A_\lambda \partial^{[\nu}h^{\rho]}{}_\tau+\varepsilon^{\nu\rho\lambda\tau} A_\lambda \partial^{[\mu}h^{\sigma]}{}_\tau\)+\frac{1}{m}\(\bar{\mpsi}^{[\sigma}\gamma^{\mu][\rho}\chi^{\nu]}+\bar{\mpsi}^{[\rho}\gamma^{\nu][\sigma}\chi^{\mu]}\)\; .
    \end{aligned}
\label{eqn:ch2_ms2_scu_G}
\end{equation}

\noindent As stressed, this coupling contains higher-derivative terms, which are required by the consistency of the theory with supersymmetry --- as operators of this type occurred already in the minimal setup \eqref{eqn:ch2_ms2_sc_T_1}-\eqref{eqn:ch2_ms2_sc_G} --- as well as with diffeomorphism invariance \cite{ms2}.

Starting from the supercurrent \eqref{eqn:ch2_ms2_scu_tot} and the linear order coupling to supergravity \eqref{eqn:ch2_ms2_new_min_min_coupling}, we can extrapolate a nonlinear, covariant lagrangian that reduces to the described couplings in the proper limit. This is given by the terms
\begin{equation}
    \mathcal{L}=\mathcal{L}_\textup{SUGRA}+\mathcal{L}_\text{SFP}+\mathcal{L}_{\psi}+\mathcal{L}_\text{non-min}+\mathcal{L}_\text{quartic}+\dots\;,
 \label{full lagrangian}
\end{equation}

\noindent which we now proceed to describe. The term $\mathcal{L}_\textup{SUGRA}$ is the standard pure supergravity lagrangian of eq.~\eqref{eqn:ch2_sugra_lag_1}. The term $\mathcal{L}_\text{SFP}$ is the minimal covariantization of the supersymmetric Fierz--Pauli lagrangian of eq.~\eqref{eqn:ch2_ms2_L_SFP}, equal to
\begin{equation}
    \begin{aligned}
            e^{-1}\mathcal{L}_\textup{SFP}=&-\frac{3}{8}F^{\mu\nu}F_{\mu\nu}-\frac{3}{4}m^2A^{\mu}A_\mu -\frac{1}{2}\bar{\mpsi}_\mu\gamma^{\mu\nu\rho}D_\nu\mpsi_\rho-\frac{1}{2}\bchi_\mu\gamma^{\mu\nu\rho}D_\nu\chi_\rho-m\bar{\mpsi}_\mu\gamma^{\mu\nu}\chi_\nu\\
			&\hspace{-2mm}+\frac{1}{4}\nabla_\mu h_{\nu\rho}\nabla^\nu h^{\mu\rho}-\frac18\nabla_\rho h_{\mu\nu}\nabla^\rho h^{\mu\nu}-\frac14\nabla_\mu h^{\mu\nu}\nabla_\nu h+\frac18\nabla_\mu h\nabla^\mu h-\frac{m^2}{8}\(h^{\mu\nu}h_{\mu\nu}-h^2\)\;. 
    \end{aligned}
\label{covariant SFP lagrangian}
\end{equation}

\noindent Next, the term $\L_\psi$ encodes all the couplings to the gravitino, namely both the minimal coupling to the supersymmetry current $\Xi^\mu$ of eq.~\eqref{eqn:ch2_ms2_susy_c} and the non-minimal one defined by the improvement term in eq.~\eqref{eqn:ch2_ms2_scu_Xi}:
\begin{equation}
\begin{aligned}
    e^{-1}\mathcal{L}_{\psi}=-\frac{e^{-1}}{4}\varepsilon^{\mu\nu\rho\sigma}\bpsi_\mu&\biggl[\(mh_{\rho\tau}+\tilde{F}_{\rho\tau}-2i\nabla_\tau A_\rho\gamma_5\)\gamma^\tau\gamma_\nu\mpsi_\sigma \\
    &\quad+\(\nabla_\lambda h_{\tau\rho}\gamma^{\lambda\tau}\gamma_5-\frac{i}{2m}\nabla_\rho F_{\lambda\tau}\gamma^{\lambda\tau}+imA_\tau\(2\delta^\tau_\rho+\gamma_\rho{}^\tau\)\)\gamma_\nu\gamma_5\chi_\sigma\biggr]\\
    +\frac{i}{2m}\bar{\rho}_{\mu\nu} \biggl[&\(mh^{\nu}{}_\lambda+\tilde{F}^{\nu}{}_\lambda-2i\nabla_\lambda A^{\nu}\gamma_5\)\gamma^{\lambda}\gamma_5\chi^{\mu}  \\
    &+\(\nabla_\rho h^{\mu}{}_\sigma\gamma^{\rho\sigma}\gamma_5-\frac{i}{2m}\nabla^{\mu}F_{\rho\sigma}\gamma^{\rho\sigma}+2imA^{\mu}-imA_\rho\gamma^{\rho\mu}\)\mpsi^{\nu} \biggr]\;,
\end{aligned}
\label{gravitino couplings lagrangian}
\end{equation}

\noindent where we introduced the gravitino field-strength $\rho_{\mu\nu}=D_\mu\psi_\nu-D_\nu\psi_\mu$. Instead, the non-minimal couplings to the metric field are all cast into the term $\mathcal{L}_\text{non-min}$, equal to
\beq
\begin{aligned}
    e^{-1}\mathcal{L}_\text{non-min}=&\frac{1}4 R_{\mu\sigma\nu\rho}\[-\frac{1}{2}h^{\rho\mu}h^{\sigma\nu}+\frac{1}{2}g^{\rho\mu}h^{\sigma}{}_\lambda h^{\lambda\nu}+2g^{\rho\mu}A^{\sigma}A^{\nu}-\frac{3}{4m^2}F^{\mu\sigma}F^{\nu\rho}+\frac{1}{2m^2}g^{\rho\mu}F^{\sigma}{}_\lambda F^{\lambda\nu}\right. \\
	&\left.+\frac{2}{m^2} g^{\sigma \nu}F^{\rho}{}_\lambda \nabla^\lambda A^{\mu}+\frac{e^{-1}}{m} \varepsilon^{\mu\sigma\lambda\tau} A_\lambda \nabla^{\nu}h^{\rho}{}_\tau +\frac{1}{m}g^{\rho\mu} \tilde{F}^{\sigma}{}_{\lambda}h^{\lambda\nu} +\frac{2}{m} \bar{\mpsi}^{\sigma}\gamma^{\mu\rho}\chi^{\nu}\]\; .    
\end{aligned}
\label{non-minimal couplings lagrangian}
\eeq

\noindent Finally, the term $\L_\textup{quartic}$ contains terms quartic in the fields, arising from integrating out the auxiliary fields $V_\mu$ and $C_{\mu\nu}$ in the minimal coupling lagrangian \eqref{eqn:ch2_ms2_new_min_min_coupling}, which yields a leftover term equal to $- \frac{1}{4} X_\mu J^\mu$. The associated eq.~\eqref{eqn:ch2_ms2_scu_X} and \eqref{eqn:ch2_ms2_scu_j} give
\beq
\begin{aligned}
    e^{-1}\mathcal{L}_\text{quartic}=&-\frac{1}{4}\bigg(e^{-1}\varepsilon_{\mu}{}^{\nu\rho\sigma}\(\bar{\mpsi}_\rho\gamma_\nu\mpsi_\sigma-\bchi_\rho\gamma_\nu\chi_\sigma\)+\frac{4i}{m}\nabla^\nu\(\bchi_{[\mu}\gamma_5\mpsi_{\nu]}\)\bigg)\bigg[\frac{i}{4}\bar{\mpsi}^\tau\gamma_5 \gamma^\mu\mpsi_\tau+\frac{3m}{4}A_{\tau}h^{\tau\mu}  \\
    &+\frac{9}{8}A_{\tau}\tilde{F}^{\tau\mu} +\frac{1}{8m}\nabla_\tau\(F^{\tau\lambda}h^\mu{}_{\lambda}-F^{\mu\lambda}h^{\tau}{}_{\lambda}\)-\frac{3}{4m}\nabla_\tau\(\nabla^\mu A^\lambda h^{\tau}{}_{\lambda}+\nabla_\lambda A^{\tau} h^{\lambda\mu}\) \\
    &+\frac{1}{4m^2}\nabla_\lambda\nabla_\pi\(\tilde{F}^{\mu\lambda}A^\pi\)  +\frac{e^{-1}}{4}\varepsilon^{\mu\tau\lambda\pi}\nabla_\tau h^\alpha{}_{\lambda} h_{\pi\alpha}-\frac{e^{-1}}{4m}\varepsilon^{\mu\tau\lambda\pi}\(\bchi_\tau D_\lambda\mpsi_\pi+\bar{\mpsi}_\tau D_\lambda\chi_\pi\)\bigg]\; .
\end{aligned}
\label{quartic lagrangian}
\eeq

\noindent The dots in the total lagrangian of eq.~\eqref{full lagrangian} denote further terms of higher-order in the number of fields which the analysis we carried out does not capture. Indeed, the supercurrent superfield $\S^\mu$ of eq.~\eqref{eqn:ch2_ms2_scu_tot} define a leading-order coupling, in which the theory is exact, but is insensitive to higher-order ones. The only couplings of this type that we can determine are those collected in the $\mathcal{L}_\text{quartic}$ term, which still arise from the quadratic off-shell lagrangian.

\subsubsection{The non-minimal deformations}

The analysis of the previous section showed that there exists a unique coupling of the massive spin-2 supermultiplet \eqref{eqn:ch2_ms2_ms2_susy} to new-minimal off-shell supergravity \eqref{eqn:ch2_new_min_susy}. This is given by the supercurrent superfield $\S^\mu$ of eq.~\eqref{eqn:ch2_ms2_scu_tot}, which was obtained by properly improving the minimal supercurrent $\S^\mu_0$ \eqref{eqn:ch2_ms2_sc_0} via the linear improvement superfield $\L^{\mu\nu}$, given by eq.~\eqref{GeneralImproved}-\eqref{eqn:ch2_ms2_Dj_gen_2} and \eqref{eqn:ch2_ms2_Dj_fin}-\eqref{eqn:ch2_ms2_scu_coeff_sol}. As we stressed, this is a coupling to new-minimal off-shell supergravity, since the supercurrent $\S^\mu$ is, by construction, an R-multiplet \eqref{eqn:ch2_R_sc_eq}-\eqref{eqn:ch2_R_sc_susy}, and it is unique at the four-derivative order.

Possible extensions of such a coupling can be found by deforming the new-minimal setup through the $\mathcal{U}$-term in the general improvement superfield of eq.~\eqref{GeneralImproved}. The bottom component of this $\mathcal{U}$-improvement term yields a contribution to the axial current that is not conserved \cite{Komargodski:2010rb}, so that the coupling to supergravity defined by the resulting supercurrent multiplet is of the non-minimal type \cite{Siegel:1978mj}. 

This non-minimal improvement yields the following contributions to the stress energy tensor and supersymmetry current:
\beq
    T^{\mu\nu} \rightarrow T^{\mu\nu} + ( \partial^\mu \partial^\nu - \eta^{\mu\nu} \Box ) U \; , \quad  \Xi^\mu \rightarrow \Xi^\mu - 2 \gamma^{\mu\nu} \partial_\nu \Upsilon \;
\label{eqn:ch2_ms2_U_imp_T_Xi}
\eeq

\noindent where 
\begin{align}
    U=\mathcal{U}|_{\vartheta = 0} \ , && \delta U  = \bar \epsilon \Upsilon \; .
\label{eqn:ch2_ms2_U_comp}
\end{align}

\noindent Hence, the $\mathcal{U}$-improvement contributes to the lagrangian with non-minimal couplings of the type \cite{Komargodski:2010rb}
\beq
    e^{-1}\L_{\mathcal{U}}=-\frac{1}{2} R\,  U + \bar \rho_{\mu\nu} \gamma^{\mu\nu} \Upsilon \; ,
\label{eqn:ch2_ms2_non_min_imp_lag}
\eeq

\noindent Similarly to the new-minimal case, the operator $U$ is a polynomial of composite operators quadratic in the number of fields and containing at most two derivatives. However, it cannot be used to remove the non-diffeomorphism invariant operator $B^{\rho\mu\nu}$ of eq.~\eqref{eqn:ch2_ms2_sc_B} that affects the minimal supercurrent superfield $\S^\mu_0$ \eqref{eqn:ch2_ms2_sc_0}. 

We therefore conclude that the unique new-minimal coupling defined by the supercurrent superfield $S^\mu$ found in eq.~\eqref{eqn:ch2_ms2_scu_tot} is the core, characteristic element of a class of couplings of the massive spin-2 field \eqref{eqn:ch2_ms2_ms2_susy} to supergravity. The various elements of this set are obtained by non-minimal improvements to the new-minimal supercurrent $\S^\mu$, and the corresponding lagrangians differ only in the couplings to the Ricci scalar and gravitino field strength as in eq.~\eqref{eqn:ch2_ms2_non_min_imp_lag} \cite{ms2}.

\subsection{The Kaluza--Klein supergravity and the St\"uckelberg symmetry}\label{ch2_sec_stuck}

The outcome of the above analysis --- which is the main result of \cite{ms2} --- is that the coupling to supergravity of the massive spin-2 supermultiplet \eqref{eqn:ch2_ms2_ms2_susy} is unique (up to non-minimal improvement terms \eqref{eqn:ch2_ms2_non_min_imp_lag}) and contains 4-derivative operators. On the other hand, as mentioned at the beginning of this chapter, one such coupling that certainly exists is the one originating from pure supergravity in five dimensions, compactified on the orbifold $S^1/\mathbb{Z}_2$. From the point of view of the off-shell formulations of supergravity, the action of the orbifold projection on the auxiliary fields of the minimal off-shell supergravity in five dimensions \cite{Zucker:1999ej} predicts that the four-dimensional compactified theory should be realized as a non-minimal off-shell supergravity \cite{Siegel:1978mj}. However, this Kaluza--Klein theory is clearly different from the one that could be obtained from the non-minimal $\mathcal{U}$-type deformations \eqref{eqn:ch2_ms2_Dj_gen_2}-\eqref{eqn:ch2_ms2_non_min_imp_lag} of the supercurrent $\S^\mu$ \eqref{eqn:ch2_ms2_scu_tot}, since the latter necessarily contains higher-derivative operators, whereas the former does not. The source of this apparent contradiction lies in the underlying assumptions of the analysis described so far, which we now discuss. 

The unique coupling of the massive spin-2 supermultiplet to new-minimal off-shell supergravity defined by the supercurrent $\S^\mu$ of eq.~\eqref{eqn:ch2_ms2_scu_tot} --- and similarly for its non-minimal deformations ---  was constructed step by step, following the transformation rules of the R-symmetry supercurrent superfield \eqref{eqn:ch2_R_sc_susy}. Performing such a computation implicitly assumes that the supersymmetry transformations of the massless supergravity multiplet, dual to the supermultiplet of currents, are the ones of the new-minimal off-shell formulations \eqref{eqn:ch2_new_min_susy} and do not get modified. However, this assumption is nontrivial, since it is not guaranteed a priori that the coupling of a massive multiplet to supergravity does not require that the supersymmetry transformations of the massless fields be modified, even at leading order. 

This is precisely what happens to the coupling of the massive spin-2 multiplet of eq.~\eqref{eqn:ch2_ms2_mult_names} to gravity in the context of the Kaluza--Klein theory. The reason is that in the compactified theories, the degrees of freedom making up the massive gauge fields of the spin-2 supermultiplet of eq.~\eqref{eqn:ch2_ms2_mult_names} naturally appear in  St\"uckelberg form,\footnote{For a review on the St\"uckelberg formulation of massive spin-2 fields, see for example \cite{Hinterbichler:2011tt}. For the case of massive spin-$\frac32$, see instead \cite{Gherghetta:2002nr,Zinoviev:2002xn}.} since the St\"uckelberg symmetry relating the massive gauge fields' degrees of freedom arises as the massive Kaluza--Klein modes of the higher-dimensional gauge symmetries. The formulation in which the fields organize as in eq.~\eqref{eqn:ch2_ms2_mult_names} is the unitary gauge of the St\"uckelberg symmetry, in which the pure gauge degrees of freedom are effectively set to zero. However, in order for the supersymmetry transformations of the system to preserve this unitary gauge, the transformations of the massless Kaluza--Klein fields must be modified at the linearised level. 

Let us show this explicitly for the pure 5D supergravity algebra compactified on the orbifold $S^1/\mathbb{Z}_2$. This compactification is performed in Appendix \ref{app_ms2_kk}, to which we refer for all details. The pure 5D supergravity multiplet contains the f\"unfbein $E_M{}^A$, a vector field $\A_M$ and a symplectic Majorana gravitino $\Psi_M$. To study the dimensional reduction on $S^1/\mathbb{Z}_2$, we start from the compactification ansatz, in the language of differential forms,
\beq
\begin{aligned}
    E^a&=\phi^{-\frac{1}{2}}e^a\; , &&&  \A&=a\(dy+B\)+A\; , \\
    E^4&=\phi\(dy+B\)\; , &&&    \Psi&=\phi^{\frac{5}{4}}\,\zeta\(dy+B\)+\phi^{-\frac{1}{4}}\(\psi-\frac{1}{2}e^a\gamma_a\gamma_5\zeta\)\; . 
\end{aligned}
\label{KK_ansatz} 
\eeq

\noindent Under the $\mathbb{Z}_2$ parity on the $S^1$ coordinate $y$, the bosonic fields $e_\mu{}^a$, $\phi$ and $a$ are even and $A_\mu$ and $B_\mu$ are odd, while the fermions are such that 
\beq
\psi{}_\mu(-y)=\sigma_3\gamma_5\psi{}_\mu(y)\; , \qquad  \zeta(-y)=-\sigma_3\gamma_5\zeta(y)\; ,
\eeq

\noindent where the action of the Pauli matrix $\sigma_3$ is related to the 5D R-symmetry. The massive Kaluza--Klein modes of the fields $B_\mu$, $\phi$, $a$ and $\zeta$ are the pure St\"uckelberg degrees of freedom. The unitary gauge is effectively imposed by setting these fields to zero, namely
\begin{align}
    B_\mu(x,y) = 0 \; ,  && \phi(x,y) = \phi(x) \; , && a(x,y) = a(x) \; , && \zeta(x,y) = \tfrac12\bigl( 1-\sigma_3\gamma_5\bigr) \zeta(x)\; .
\label{eqn:ch2_stuck_ug}
\end{align}

\noindent On the other hand, these fields transform nontrivially under supersymmetry. Focusing on the fields $B_\mu$ and $\phi$ --- related to the massive spin-2 --- and including a 5D diffeomorphism of parameter $\xi^M(x,y)$, these are given by 
\beq
\begin{aligned} 
    \delta B_\mu(x,y) &=\frac{\phi(x)^{-\frac{3}{2}}}{2}\bar{\epsilon}(x)  \gamma_5\psi_\mu(x,y)   + \phi(x)^{-3} g_{\mu\nu}(x,y) \partial_y \xi^\nu(x,y) + \partial_\mu \xi^5(x,y)  \ , \\
    \delta \phi(x,y) &= \frac{\phi(x)}{2} \bar \epsilon(x) \gamma_5 \zeta(x) + \xi^\mu(x,y) \partial_\mu \phi(x) + \partial_y \xi^5(x,y) \phi(x) \; .
\end{aligned}
\label{eqn:ch2_delta_stuck}
\eeq

\noindent Thus, in order for the unitary gauge \eqref{eqn:ch2_stuck_ug} and these supersymmetry transformations to be consistent with each other, one must set the diffeomorphism parameters to be
\beq
\begin{aligned}
    \partial_y \xi^5(x,y) &= - \xi^\mu(x,y) \partial_\mu \log \phi(x)\; ,  \\
    \partial_y \xi^\mu(x,y) + \phi(x)^3 g^{\mu\nu}(x,y) \partial_\nu \xi^5(x,y) &= - \frac{\phi(x)^{\frac{3}{2}}}{2} g^{\mu\nu}(x,y) \bar{\epsilon}(x)\gamma_5\psi_\nu(x,y) \ .
\end{aligned}
\eeq

\noindent In turn, the supersymmetry variation of the vierbein $e_\mu{}^a$ becomes
\beq 
    \delta e_\mu{}^a(x,y) = \frac{1}{2}\bar{\epsilon}(x)\gamma^a\psi_\mu(x,y) + D_\mu \bigl(  \xi^\nu(x,y)e_\nu{}^a(x,y) \bigr) + \xi^5(x,y) \partial_y e_\mu{}^a(x,y)\; , 
\eeq

\noindent from which one sees that the massless vierbein transformation is modified at leading order, with terms quadratic in the massive fields, involving additional derivatives. This deformation cannot be reabsorbed by a local redefinition in the unitary gauge, since also the closure of the massless supersymmetry algebra gets modified, with respect to eq.~\eqref{eqn:ch2_ms2_susy_cl}, with terms quartic in the massive modes:
\beq 
    \xi^\mu(\epsilon_1,\epsilon_2) \sim - \frac12  \bar \epsilon_1 \gamma^\mu \epsilon_2 + \frac{1}{4m^2}  \bar \epsilon_1 h^{\nu\rho} \psi_\rho  \partial_\nu \bigl(  \bar \epsilon_2 h^{\mu\sigma} \psi_\sigma \bigr) - \frac{1}{4m^2}  \bar \epsilon_2 h^{\nu\rho} \psi_\rho  \partial_\nu \bigl(  \bar \epsilon_1 h^{\mu\sigma} \psi_\sigma \bigr)  \; . 
\eeq

\noindent Note that repeating a similar analysis for the spin-$\frac32$ field and the massive modes of the supersymmetry parameter, the massless transformations of the 4D gravitino would also be modified, but this time by terms that can be reabsorbed by field redefinitions in the unitary gauge. The nontrivial deformation of the massless supersymmetry transformation is a feature and a consequence of the spin-2 field only \cite{ms2}. 

Therefore, the Kaluza--Klein theory cannot be retrieved assuming both the St\"uckelberg unitary gauge and the undeformed supersymmetry transformations of the massless off-shell supergravity multiplet. One can either work outside the unitary gauge, keeping the St\"uckelberg degrees of freedom as additional fields, and the undeformed massless supersymmetry algebra, or impose the unitary gauge and work in a setup in which the whole supersymmetry transformations are deformed. The analysis performed in the previous sections is clearly based on the use of the unitary gauge for the massive spin-2 multiplet gauge fields, as in eq.~\eqref{eqn:ch2_ms2_mult_names}-\eqref{eqn:ch2_ms2_ms2_susy} and the undeformed supersymmetry transformations of the massless off-shell supergravity. This implies, on one hand, that the Kaluza--Klein theory is not captured by such a setup. On the other hand, it also means that the unique coupling to new-minimal supergravity described by the supercurrent superfield $\S^\mu$ of eq.~\eqref{eqn:ch2_ms2_scu_tot} is a genuinely additional coupling of the massive spin-2 multiplet to supergravity \cite{ms2}.

\subsection{The \texorpdfstring{${2\to2}$}{2->2} massive spin-2 scattering amplitude}

We now analyze in more detail the unique class of couplings of the massive spin-2 supermultiplet to off-shell supergravity discussed in the previous sections. From the leading-order coupling described by the supercurrent superfield of eq.~\eqref{eqn:ch2_ms2_imp_names}--\eqref{GeneralImproved}, we can compute the $2\to2$ elastic scattering amplitude of the massive spin-2 field $h_{\mu\nu}$. In the theory under consideration, this amplitude receives contributions only from the stress-energy tensor. In general, this is given by the new-minimal coupling tensor $T^{\mu\nu}$ of eq.~\eqref{eqn:ch2_ms2_scu_T}, plus the possible non-minimal extensions like eq.~\eqref{eqn:ch2_ms2_U_imp_T_Xi}--\eqref{eqn:ch2_ms2_non_min_imp_lag}. We emphasize that this amplitude is calculated at the three-point level, \textit{i.e.} at leading order in the Planck mass $\MP$, where the defining supercurrent is exact. The amplitude is generally sensitive to corrections from higher-order interactions, such as four-point couplings of the massive spin-2. These couplings are, however, not captured by the current setup.

We initially focus, as before, on the unique coupling to new-minimal supergravity, given by the supercurrent $\S^\mu$ of eq.~\eqref{eqn:ch2_ms2_scu_tot}. From the associated stress-energy tensor $T^{\mu\nu}$ of eq.~\eqref{eqn:ch2_ms2_scu_T}, one reads the following three-point function:
\beq\begin{aligned} \includegraphics[scale=0.5]{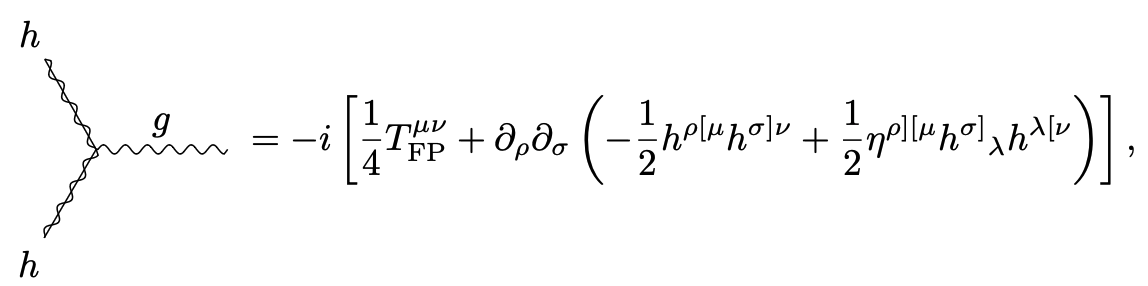} \end{aligned}\label{eqn:ch2_ms2_3pc_fig}\eeq

\noindent This three-point function defines the tree-level amplitude 
\beq
  i  (2\pi)^4 \delta^{\scalebox{0.6}{(4)}}(p_1+p_2-p_1^{\prime}-p_2^{\prime}) \A=-\frac{\kappa^2}{2}  \langle {\rm out} |  \int  d^4x d^4 y :T^{\mu\nu}_{h}(x)D_{\mu\nu\rho\sigma}(x-y)T^{\rho\sigma}_{h}(y):  |{\rm in}\rangle  \; ,
\label{eqn:ch2_ms2_amp_ch}
\eeq

\noindent where we have reinstated the Planck mass factor as $\kappa^{2}=8\pi\MP^{-2}$ and $D_{\mu\nu\rho\sigma}$ is the graviton propagator, which we define in momentum space as 
\beq
    D_{\mu\nu\rho\sigma}(p)=-\frac{i}{2p^2}\(\eta_{\mu\rho}\eta_{\nu\sigma}+\eta_{\mu\sigma}\eta_{\nu\rho}-\eta_{\mu\nu}\eta_{\rho\sigma}\) \ .
\eeq

\noindent The general amplitude \eqref{eqn:ch2_ms2_amp_ch} results in the standard sum of the $s$, $t$ and $u$-channel 
\beq \begin{aligned}\includegraphics[scale=0.6]{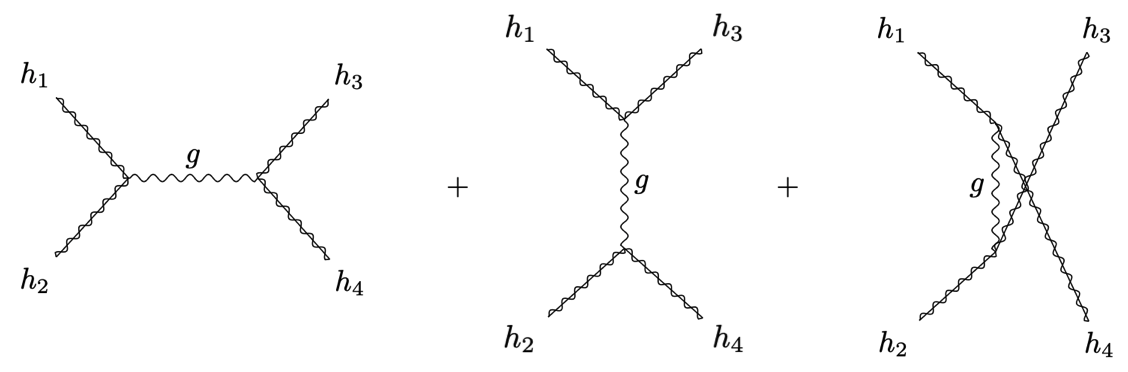}\end{aligned}. \label{eqn:ch2_ms2_amp_fig}\eeq

\noindent We now compute this amplitude explicitly. We work in momentum space, where the momentum $p^{\mu}=\(E_{\boldsymbol{p}},\boldsymbol{p}\)$ is parameterized as
\beq
\begin{aligned}
	 E_{\boldsymbol{p}}=\sqrt{\boldsymbol{p}^2+m^2 }, &&&& \boldsymbol{p}=\abs{\boldsymbol{p}}\(\sin\theta\cos\phi,\sin\theta\sin\phi,\cos\theta\)\equiv\abs{\boldsymbol{p}}\hat{\bp} \ ,
\end{aligned}
\eeq

\noindent with $\(\theta,\phi\)$ being the angles with respect to the $z$-axis. The expansion of the massive spin-2 field in momentum space is\footnote{The factor of $2$ in the expansion of eq.~\eqref{eqn:ch2_ms2_h_exp_mom} reinstates the canonical normalization of the spin-2 field $h_{\mu\nu}$ with respect to the non-canonical one chosen in eq.~\eqref{eqn:ch2_ms2_L_SFP}-\eqref{covariant SFP lagrangian}.} 
\beq
    h_{\mu\nu}(x)=2 \int \frac{d^3p}{\(2\pi\)^{\frac32}\sqrt{2E_{{\boldsymbol p}}}}\sum_{{\sigma}}\(\epsilon_{\mu\nu}\({\sigma},\boldsymbol{p}\)e^{ip\cdot x}a_{{\sigma},\boldsymbol{p}}+\bepsilon_{\mu\nu}\({\sigma},\boldsymbol{p}\)e^{-ip\cdot x}a_{{\sigma},\boldsymbol{p}}^\dagger\) \ ,
\label{eqn:ch2_ms2_h_exp_mom}
\eeq

\noindent where $\{a_{{\sigma},\boldsymbol{p}},a_{{\sigma},\boldsymbol{p}}^\dagger\}$ are the ladder operators and $\epsilon_{\mu\nu}({\sigma},\boldsymbol{p})$ are the polarization functions. They depend on the space momentum $\boldsymbol p$ and on the five polarization states ${\sigma}=\left\{\pm2,\pm1,0\right\}$ of the massive spin-2 field. The polarization functions satisfy the following orthogonality and completeness relations:
\beq
\begin{aligned}
    &\epsilon^{\mu\nu}({\sigma},\bp)\bepsilon_{\mu\nu}(\sigma^\prime,\bp)=\delta_{\sigma\sigma^\prime},\\
    &\sum_{\sigma}\epsilon^{\mu\nu}({\sigma},\bp)\bepsilon^{\rho\sigma}({\sigma},\bp)=\frac{1}{2}\(P^{\mu\rho}P^{\nu\sigma}+P^{\nu\rho}P^{\nu\sigma}\)-\frac{1}{3}P^{\mu\nu}P^{\rho\sigma} \ ,
\end{aligned}
\eeq

\noindent where $\textstyle{P^{\mu\nu}=\eta^{\mu\nu}+\frac{p^\mu p^\nu}{m^2}}$. Moreover, the equations of motion \eqref{eqn:ch2_ms2_eom_full}-\eqref{eqn:ch2_ms2_eom} impose
\begin{align}
    p^\mu\epsilon_{\mu\nu}({\sigma},\bp)=0, && \epsilon^\mu{}_\mu({\sigma},\bp)=0 \ .
\end{align}

\noindent According to these general properties, we choose the polarization functions to be \cite{SekharChivukula:2019yul,Chivukula:2020hvi}
\beq
\begin{aligned}
    \epsilon^{\mu\nu}\({\pm2},\bp\)&=\epsilon^\mu_{\pm 1}(\bp)\epsilon^\nu_{\pm 1}(\bp), \\
    \epsilon^{\mu\nu}\(\pm1,\bp\)&=\frac{1}{\sqrt{2}}\(\epsilon^\mu_{\pm1}(\bp)\epsilon^\nu_{0}(\bp)+\epsilon^\nu_{\pm1}(\bp)\epsilon^\mu_{0}(\bp)\), \\
    \epsilon^{\mu\nu}(0,\bp)&=\frac{1}{\sqrt{6}}\(\epsilon^\mu_{1}(\bp)\epsilon^\nu_{-1}(\bp)+\epsilon^\nu_{1}(\bp)\epsilon^\mu_{-1}(\bp)+2\epsilon^\mu_{0}(\bp)\epsilon^\nu_{0}(\bp)\),
\end{aligned}
\eeq

\noindent with
\beq
\begin{aligned}
    \epsilon_{\pm1}^\mu(\bp)=&\pm\frac{e^{\pm i\phi}}{\sqrt{2}}\(0,-\cos\theta\cos\phi\pm i\sin\phi,-\cos\theta\sin\phi\mp i\cos\phi,\sin\theta\), \\
    \epsilon_0^\mu(\bp)=&\frac{E_{\bp}}{m}\(\sqrt{1-\frac{m^2}{E_{\bp}^2}},\hat{\bp}\).
\end{aligned}
\eeq

\noindent These are all the tools needed to compute the scattering amplitude \eqref{eqn:ch2_ms2_amp_ch}-\eqref{eqn:ch2_ms2_amp_fig}. This involves in general four arbitrary helicity states for the external massive spin-2 fields, of the form $\A\(\sigma_1,\sigma_2,\sigma_3,\sigma_4\)$. In the absence of the improvement term produced by supersymmetry, the  most divergent of the combinations is the elastic helicity-0 scattering \cite{SekharChivukula:2019yul,Chivukula:2020hvi}. In the present case, this scattering amplitude is found to be \cite{ms2}
\begin{equation}
    \begin{aligned}
         \A(0,0,0,0) &= \frac{\kappa^2}{16 m^4 s (s - 4 m^2)} \[ 
  s^5 \sin^2\theta \(3 + \cos^2\theta\) - \frac{4 m^2s^4}{3}  (21 - 29 \cos^2\theta)\right. \\
  &\hspace{20mm}+   2 m^4 s^3 \( \frac{ 32}{\sin^2\theta} + \frac{37 - 231  \cos^2\theta + 18  \cos^4\theta}{3}  \)\\
  &\hspace{20mm}- 
   16 m^6 s^2 \(\frac{ 16}{\sin^2\theta} - 7 + 2  \cos^2\theta  - \cos^4\theta\) \\
   &\hspace{20mm}\left.+   32 m^8 s\,  \(\frac{ 4}{\sin^2\theta} + \frac{11 + 26  \cos^2\theta}{3}  \) +128 m^{10}\cos(2\theta) \]
\end{aligned}
\label{eqn:ch2_ms2_amp_us}
\end{equation}

\noindent In the Regge limit of large $s$ at fixed $t$, this amplitude grows as $\A\(0,0,0,0\)\sim s^2$, while in the high-energy limit of large $s$ and fixed angle $\theta$ it scales as $\A\(0,0,0,0\)\sim s^3$. Thus, this amplitude saturates the $\Lambda_3$ threshold discussed in \cite{Arkani-Hamed:2002bjr,Arkani-Hamed:2003roe,Schwartz:2003vj,Bonifacio:2018vzv,Bonifacio:2018aon,Bonifacio:2019mgk,Kundu:2023cof} for the interactions of the massive spin-2 field with gravity. However, this is not the cutoff scale of the whole theory, because the other helicty states scattering actually exhibit worse behavior. More specifically, this concerns the helicity states $\sigma=\pm1$, whose scattering with the helicity-0 ones grows as $s^4$ both in the Regge and high-energy limit \cite{ms2}. For instance, one has 
\beq
\begin{aligned}
     \A(1,1,0,0) &= \frac{\kappa^2}{384 m^6 s (s - 4 m^2)}\times \\
     &\hspace{5mm}\times\[ s^6 \(5 - \cos^2\theta\)- m^2 s^5 \(49 - 9 \cos^4\theta\) +2 m^4 s^4 \(217 - 113 \cos^2\theta + 26\cos^4\theta\)\right.\\
     &\hspace{6mm} - 
  4 m^6 s^3 \(480 - 265  \cos^2\theta + 69 \cos^4\theta \) +  32 m^8 s^2 \(103 + 96 \cos^2\theta - 9 \cos^4\theta \)\\
  & \hspace{6mm}\left.- 
  64 m^{10} s \(56 + 119 \cos^2\theta + \cos^4\theta\) 
- 3072 m^{12}\cos(2\theta) \] \ ,
\end{aligned}
\label{eqn:ch2_ms2_amp_us_2}
\eeq
for which one can check that $\A\(1,1,0,0\)\sim s^4$ in both limits.\footnote{In addition to the $\A\(1,1,0,0\)$ amplitude of eq.~\eqref{eqn:ch2_ms2_amp_us_2}, other amplitudes with the same exact behavior are  $\A(0,0,\pm 1,\pm 1)$, $\A(\pm 1,\pm 1,0,0)$, $\A(0,\pm 1,\pm 1,0)$, $\A(\pm 1,0,0,\pm 1)$, $ \A(\pm 1,0,\mp 1 ,0)$ and $\A(0,\pm 1,0,\mp 1 )$.} Therefore, following \cite{Arkani-Hamed:2002bjr,Arkani-Hamed:2003roe,Schwartz:2003vj,Bonifacio:2018vzv,Bonifacio:2018aon,Bonifacio:2019mgk,Kundu:2023cof} and eq.~\eqref{eqn:ch2_ms2_cutoff}, the coupling of the massive spin-2 field to gravity defined by the new-minimal supercurrent $\S^\mu$ of eq.~\eqref{eqn:ch2_ms2_scu_tot} is associated with the cutoff scale $\Lambda_4$. This is slightly improved with respect to generic behavior in energy of the interactions between gravity and the massive spin-2 field, which is typically $\Lambda_5$ \cite{Arkani-Hamed:2002bjr,Arkani-Hamed:2003roe,Schwartz:2003vj,Bonifacio:2018vzv}, thanks to local supersymmetry invaraiance. However, it is necessary to include self-interactions of the massive spin-2 field in order to raise the cutoff of the theory up to the $\Lambda_3$ threshold \cite{Bonifacio:2018aon}.

Let us now deform this new-minimal setup and include the couplings to non-minimal off-shell supergravity \eqref{GeneralImproved}-\eqref{eqn:ch2_ms2_non_min_imp_lag}. In this formulation, the stress-energy tensor is modified according to eq.~\eqref{eqn:ch2_ms2_U_imp_T_Xi}, \textit{i.e.} in terms of a non-minimal coupling to the Ricci scalar $R$ to the composite operator $U= \mathcal{U}|_{\vartheta = 0}$ of the non-minimal improvement of eq.~\eqref{GeneralImproved}. This operator is, in general, quadratic in fields and contains up to two derivatives. Therefore, the most general operator of this type that contributes nontrivially to the $2\to2$ scattering amplitude of the massive spin-2 fields $h_{\mu\nu}$ has the form
\beq
	T^{\mu\nu}_{U}=\(\partial^\mu\partial^\nu-\eta^{\mu\nu}\Box\)\(c_1 h^{\rho\sigma}h_{\rho\sigma}+c_2 \partial_\lambda  h^{\rho\sigma}\partial^\lambda h_{\rho\sigma} + c_3 \partial_\rho \partial_\sigma ( h^{\rho\lambda} h^{\sigma}{}_\lambda ) \) \ .
\label{eqn:ch2_ms2_U_terms_amp}
\eeq

\noindent One can verify that this deformation worsens the growth of the $2 \to 2$ scattering amplitude with energy, causing it to scale as $\A(s,t) \sim s^5$ or higher in the large $s$ limit at fixed angle. 

Hence, the supersymmetric coupling to gravity of the massive spin-2 in the new-minimal setup, defined by the supercurrent $\S^\mu$ of eq.~\eqref{eqn:ch2_ms2_scu_tot} and reported in eq.~\eqref{eqn:ch2_ms2_3pc_fig}, proves to be the one with best behavior in the energy growth of the scattering amplitude, so that the associated effective field theory is the one defined with the highest degree of reliability. All the other couplings of the spin-2 field to gravity in the non-minimal off-shell setup are associated with lower cutoff scales, as just showed. A priori, one could consider non-minimal deformations which do not modify the massive spin-2 scattering amplitude which we are analyzing. This is achieved by any operator $U$ that does not contain any of the terms appearing in eq.~\eqref{eqn:ch2_ms2_U_terms_amp}, but involves only the rest of the fields of the massive multiplet \eqref{eqn:ch2_ms2_mult_names}. However, although we did not compute them explicitly, it is reasonable to expect that, by supersymmetry, also the elastic scattering  amplitudes involving the other massive gauge fields, the superpartners of $h_{\mu\nu}$ in eq.~\eqref{eqn:ch2_ms2_mult_names}, should scale in the same way, so that any non-minimal deformation $U$ \eqref{eqn:ch2_ms2_U_imp_T_Xi}-\eqref{eqn:ch2_ms2_non_min_imp_lag} would inevitably raise the cutoff of the theory.

From this perspective, the theory defined by the supercurrent superfield $\mathcal{S}^\mu$ in eq.~\eqref{eqn:ch2_ms2_scu_tot} is therefore not only the unique coupling to new-minimal off-shell supergravity but is also the unique coupling to any off-shell supergravity with improved cutoff   $\Lambda_4$ at tree-level instead of $\Lambda_5$.

\subsection{On the relation between the unique coupling to new-minimal supergravity and string theory}

The main result of \cite{ms2} we have been discussing in this section is that the coupling of the massive spin-2 multiplet in the unitary gauge \eqref{eqn:ch2_ms2_ms2_susy} to off-shell supergravity is unique at leading-order in the Planck Mass. This is in particular a coupling to new-minimal off-shell supergravity \eqref{eqn:ch2_new_min_susy}, given by the supercurrent superfield $\S^\mu$ of eq.~\eqref{eqn:ch2_ms2_scu_tot}, and contains non-minimal, higher-derivative couplings to the massless supergravity fields, as in eq.~\eqref{eqn:ch2_ms2_scu_Xi}-\eqref{gravitino couplings lagrangian} and \eqref{eqn:ch2_ms2_scu_T}-\eqref{non-minimal couplings lagrangian}. This non-minimal coupling term involves all the fields in the massive spin-2 supermultiplet and contains in particular the massive spin-1 higher-derivative operator
\beq
    -\frac{3}{16m^2}R_{\mu\sigma\nu\rho}F^{\mu\sigma}F^{\nu\rho} \ .
\label{RFF}
\eeq
It was shown in \cite{Camanho:2014apa} that in the case of a massless gauge field, this operator leads to causality violations in $D>4$ dimensions. We expect this result to hold also in the present four-dimensional setup involving a massive gauge field if the mass is such that $m\ll\MP$. In \cite{Camanho:2014apa} it is also argued that acausality problems of this type can be resolved through the introduction of a tower of higher-spin fields of increasing mass. This resembles clearly the typical Regge tower of states that occurs in string theory, so that effectively proving that the operator \eqref{RFF} does lead to violations of causality would hint towards a UV completion of the string theory type directly from a bottom-up point of view.

Actually, the coupling defined by $\S^\mu$ \eqref{eqn:ch2_ms2_scu_tot} has further connections with superstring theory. Expanding the metric field around Minkowski space as $g_{\mu\nu} = \eta_{\mu\nu} + 2 \kappa g_{\mu\nu}^{\scalebox{0.6}{(1)}}$, the on-shell three-point coupling of two massive spin-2 fields and a massless graviton, given previously in eq.~\eqref{eqn:ch2_ms2_3pc_fig}, can be written in the form
\begin{equation}
    \begin{aligned}
        \frac{\kappa}{2} &\[  g_{\mu\nu}^{\scalebox{0.6}{(1)}} \( \partial^\mu h^{\rho\sigma} \partial^\nu h_{\rho\sigma} - 4 \partial^\mu h^{\rho\sigma} \partial_\rho h^{\nu}{}_\sigma \) + h^{\mu\nu} \( \partial_\mu g_{\rho\sigma}^{\scalebox{0.6}{(1)}}  \partial_\nu h^{\rho\sigma} -\partial_\rho g_{\mu\sigma}^{\scalebox{0.6}{(1)}}  \partial_\nu h^{\rho\sigma} \)\right. \\
        &\qquad+\left. \partial_\mu \partial^\rho g_{\nu\rho}^{\scalebox{0.6}{(1)}}  \(2 h^{\mu}{}_\sigma h^{\nu\sigma}- \tfrac14 \eta^{\mu\nu} h_{\sigma\lambda} h^{\sigma\lambda} \)\] \;  .
    \end{aligned}
\end{equation} 
This term exactly matches the coupling of a massive open string state of mass $m = \frac{1}{\sqrt{\alpha^\prime}}$ to the graviton in superstring theory \cite{Lust:2021jps}, as well as in bosonic string theory \cite{Buchbinder:1999ar}.\footnote{The massive open string theory states also have self-interactions, which are, however, beyond the limit of the analysis carried out in \cite{ms2}.} Therefore, the coupling $\S^\mu$ \eqref{eqn:ch2_ms2_scu_tot} has a natural interpretation in superstring theory as the coupling to gravity of a non-trivial spin-2 Regge excitation of the string. This is a nontrivial top-down match of the UV completion of the theory based on the causality violation related to the operator of eq.~\eqref{RFF}. Moreover, string theory brings a whole Regge tower of particles of mass of the same order $\sqrt{\frac{n}{\alpha^{\prime}}}$ for $n\in \mathbb{N}$, and thus the $\Lambda_3$ threshold for the interactions of the massive spin-2 field and gravity is far beyond the scale of new physics for $m\ll M_{\scalebox{0.6}{P}}$, according to this picture.

Starting from type~I superstring theory,\footnote{A general introduction to string theory is given in Chapter \ref{ch3_st}.} a non-BPS massive multiplet for the massive spin-2 field is obtained by compactifying the theory on $T^6$ from the first oscillator mode of the open string, of mass $m= \frac{1}{\sqrt{\alpha^{\prime}}}$ \cite{Lust:2021jps}. In $\N=4$ supergravity, the massive spin-2 field is contained in a long multiplet, describing $128+128$ degrees of freedom. On the other hand, the Kaluza--Klein reduction along $T^6$ of the ten-dimensional massless fields associated with the closed string yields the massive spin-2 field within $\frac12$-BPS spin-2 multiplets, comprising $24+24$ degrees of freedom. The two compactifications lead to the same tree-level interactions once an appropriate orbifold projection is included, breaking supersymmetry down to $\N=1$ in four dimensions. One can then take the bottom-up perspective of the unique coupling found in \cite{ms2} and extrapolate the connection with extended supersymmetry in higher dimensions. The supercurrent $\S^\mu$ of eq.~\eqref{eqn:ch2_ms2_scu_tot} describes the unique coupling to supergravity of the massive spin-2 multiplet of the long type, like the one arising from the first string oscillator mode \cite{Lust:2021jps}. By supersymmetry, we also expect the gravitational coupling of the $\frac12$-BPS multiplet to be unique and equal to the Kaluza--Klein one, since it is associated with the only nontrivial deformation of the massless supersymmetry algebra discussed in Section \ref{ch2_sec_stuck}. This picture fits within the swampland infinite distance paradigm \cite{Ooguri:2006in}. For an asymptotic point in the moduli space of vacua in which the mass of the $\frac12$-BPS spin-2 multiplet vanishes, we must get an infinite tower of massless spin-2 multiplets that will be interpreted as a decompactification limit. On the other hand, an asymptotic point in moduli space where the mass of a long multiplet vanishes will lead to an infinite tower of higher-spin fields becoming massless, reproducing the tensionless limit of a perturbative string theory. 

This picture is in agreement with the string lamppost principle, proposing that all theories in the quantum gravity landscape are realized as string theories \cite{Adams:2010zy,Kim:2019ths,Kim:2019vuc,Bedroya:2021fbu,Montero:2020icj}. Our results in \cite{ms2} actually provide evidence in favor of this principle within a four-dimensional setup and minimal supersymmetry.

\subsection{Summary of results}

Let us conclude by summarizing the results of \cite{ms2} presented in this section. In this work we investigate the couplings to supergravity of the massive spin-2 supermultiplet \eqref{eqn:ch2_ms2_ms2_susy}. We do so by means of the formalism of the supercurrent superfields, which encode the leading-order couplings to off-shell supergravity, as discussed in Section \ref{ch2_sec_off_shell_sugra}. We construct this multiplet explicitly for the massive multiplet \eqref{eqn:ch2_ms2_ms2_susy} and we find that the essential requirements of invariance under supersymmetry and general diffeomorphisms, imposed in full generality through the superfield improvement of eq.~\eqref{GeneralImproved}, yield a unique class of couplings to off-shell supergravity. The core, fundamental element of this set of couplings is the supercurrent $\S^\mu$ of eq.~\eqref{eqn:ch2_ms2_scu_tot}, which is the unique consistent coupling to new-minimal off-shell supergravity, with the highest possible cutoff scale $\Lambda_4$. It is characterized by non-minimal, higher-derivative couplings to the massless supergravity fields. Extensions of this coupling can be obtained by deforming the new-minimal setup associated to $\S^\mu$ to a non-minimal off-shell formulation of supergravity, by means of the $\mathcal{U}$-type improvement of eq.~\eqref{GeneralImproved}. These additional theories only differ by non-minimal couplings to the Ricci scalar $R$ and the gravitino field strength $\rho_{\mu\nu}$, as described in eq.~\eqref{eqn:ch2_ms2_U_imp_T_Xi}-\eqref{eqn:ch2_ms2_non_min_imp_lag}. Moreover, all couplings to non-minimal supergravity have a lower cutoff scale than the new-minimal one defined by $\S^\mu$.

We also remark that this class of couplings is a genuine alternative to the Kaluza--Klein theory arising from pure 5D supergravity on $S^1/\mathbb{Z}_2$. As we argue in Section \ref{ch2_sec_stuck}, this theory does not fall within the framework considered in \cite{ms2}, since working with the massive spin-2 supermultiplet in the unitary gauge of the St\"uckelberg symmetry, as the one in eq.~\eqref{eqn:ch2_ms2_ms2_susy}, leads to deformed supersymmetry transformations of the massless fields in the Kaluza--Klein reduction, so that they are not described by any of the off-shell supergravities considered in this analysis. Moreover, the unique new-minimal coupling defined by the supercurrent $\S^\mu$ of eq.~\eqref{eqn:ch2_ms2_scu_tot} has a natural interpretation in superstring theory, as it reproduces exactly, from the bottom-up, the coupling of an open string state of mass $\frac{1}{\sqrt{\alpha^\prime}}$ to the graviton \cite{Lust:2021jps}.

Further questions that we plan on investigating in the future are the following. First, the presence of higher-derivative operators, as in eq.~\eqref{eqn:ch2_ms2_scu_G} and \eqref{eqn:ch2_ms2_non_min_imp_lag}, require checking their compatibility with unitarity and causality, which is a further and strong constraint on the spectrum of allowed such couplings. In particular, this concerns the massive spin-1 operator of eq.~\eqref{RFF}. If this operator were found to be the source of causality violations along the lines of \cite{Camanho:2014apa}, it would provide even stronger motivation for a string-like UV completion, grounded in the effective field theory setup itself. On the other hand, it would be interesting to investigate even higher-derivative extensions of the theories determined in \cite{ms2}, which are restricted by construction to the 4-derivatives level. This could be done by properly extending the ansatz \eqref{eqn:ch2_ms2_Dj_fin} with higher-derivative axial improvements, starting for example from the 6-derivative case. This would be an interesting test of the uniqueness property of the new-minimal coupling $\S^\mu$. Higher-derivative deformations could in fact lead, on-shell, to modifications of the single 4-derivative consistent solution of eq.~\eqref{eqn:ch2_ms2_scu_coeff_sol}, or rather be also constrained to a unique solution, pointing in the direction of a full perturbative resummation of the theory. Finally, the study of the supersymmetric couplings involving a deformation of the massless supersymmetry algebra to prove the uniqueness of the Kaluza--Klein deformation discussed in Section \ref{ch2_sec_stuck} is a further subject of future investigations.





\chapter{String theory and orientifolds}\label{ch3_st}

The last chapter of the present manuscript is devoted to string theory. The main focus will be on orientifold projections and the string vacuum amplitudes in the presence of the non-dynamical objects known as orientifold planes. In particular, we will present the novel orientifold construction that we put forward in \cite{Bossard:2024mls}. The main properties of the resulting string theory are that it lives in nine dimensions, has no open-string background sector, does not preserve supersymmetry, and that the orientifold planes couple only to the twisted states in the vacuum amplitudes. Because of this feature, we refer to them as \textit{twisted O-planes}. Moreover, we argue that this string theory is invariant under S-duality.

The content of the chapter is organized as follows. We provide a general review of the bosonic string in Section \ref{ch3_sec_BS} and superstring theory in Section \ref{ch3_sec_sst}. We discuss the dynamics of these theories, their light-cone quantization, the resulting spectra of excitations, and, in particular,  we describe the formalism of the string vacuum amplitudes and their physical interpretation. Orientifold projections are introduced in Section \ref{ch3_sec_type_I}, while in Section \ref{ch3_sec_ss_dp_orientifolds} we analyze in detail the orientifold compactifications of the type IIB string in nine dimensions. These are the three Scherk--Schwarz orientifolds previously known in the literature, and the supersymmetric orientifold of Dabholkar and Park. In Section \ref{ch3_sec_new_sso} we present the detailed construction of the new orientifold of \cite{Bossard:2024mls}. This includes its definition, its interpretation in the supergravity limit, and the computation of the one-loop effective potential and of the D-brane spectrum. We also provide evidence that the resulting nine-dimensional string theory is invariant under S-duality and argue for its realization as an F-theory compactification. \\

In addition to the standard books \cite{Polchinski:1998rq,Polchinski:1998rr,Green:1987mn,Green:1987sp,Becker:2006dvp}, the introductory part of this chapter is mostly based on the reviews \cite{reviews_1,reviews_2,reviews_4}.

\section{The bosonic string}\label{ch3_sec_BS}

Strings are one-dimensional objects, arising as the generalization of the concept of point-particle to extended dimensions. Their basic description follows from this analogy. The propagation of strings in spacetime spans a 2-dimensional surface, known as the \textit{worldsheet}, embedded in spacetime. It is parameterized by two coordinates $\xi^a=(\tau,\sigma)$, where $\tau\in(-\infty,+\infty)$ is the proper time, as for point-particles, and $\sigma$ describes instead the specific point along the string. Strings can be either closed or open. In the former case, one has $\sigma\in(0,2\pi)$ with proper periodicity condition $\sigma=0\simeq\sigma=2\pi$. In the latter, one has  $\sigma\in(0,\pi)$, corresponding to two boundary points. The relativistic action describing this worldsheet evolution has the natural form
\beq
    S=-T\int dA=-T\int d^2\xi\sqrt{-\gamma} \ ,
\label{eqn:ch3_action_ws_0}
\eeq

\noindent where $T$ is the string tension --- namely, the string analogue of the mass of a particle --- and $dA$ is the infinitesimal worldsheet area element, which is written in terms of the induced metric on the worldsheet $\gamma_{ab}$. The position of the string in spacetime is given by the coordinates $X^{\mu}=X^\mu(\tau,\sigma)$, known as the \textit{target space} coordinates. Calling $G_{\mu\nu}$ the target space metric --- which we take to be of general dimensions $D$ ---  the induced metric on the worldsheet is equal to
\beq
    \gamma_{ab}=\frac{\partial X^\mu}{\partial \xi^a}\frac{\partial X^\nu}{\partial \xi^b}G_{\mu\nu}=\begin{pmatrix}
        \dot{X}^\mu\dot{X}_\mu & \dot{X}^\mu {X^\prime}_\mu \\ {X^\prime}^\mu\dot{X}_\mu & {X^\prime}^\mu {X^\prime}_\mu
    \end{pmatrix} \ ,
\eeq

\noindent where we employed the standard notation $\dot{X}^\mu\equiv\partial_\tau X^\mu$ and ${X^\prime}^\mu\equiv\partial_\sigma X^\mu$. With this formula, the action \eqref{eqn:ch3_action_ws_0} becomes
\beq
    S_\textup{NG}=-T\int d^2\xi\sqrt{\(\dot{X}\cdot {X^\prime}\)^2-\(\dot{X}\cdot\dot{X}\)\({X^\prime}\cdot {X^\prime}\)} \ ,
\label{eqn:ch3_action_ws_NG}
\eeq

\noindent which is known as the Nambu--Goto action. As in the case of the relativistic particle, an alternative action can be written by introducing an auxiliary metric $g_{ab}$ on the worldsheet. This is known as the Polyakov action, and it is equal to
\beq
    S_\textup{P}=-\frac{T}{2}\int d^2\xi \sqrt{-g}\,\,g^{ab}\partial_a X^\mu\partial_b X^\nu G_{\mu\nu} \ .
\label{eqn:ch3_action_ws_P}
\eeq

\noindent The variation of the Polyakov action with respect to its auxiliary worldsheet metric yields a vanishing condition for its associated stress-energy tensor:
\beq
    T_{ab}=\partial_a X^\mu\partial_b X_\mu-\frac12g_{ab}\(g^{cd}\partial_c X^\mu\partial_d X_\mu\)=0 \ .
\label{eqn:ch3_emt_=_0}
\eeq

\noindent This constraint is explicitly solved by $g_{ab}=f\partial_a X^\mu\partial_b X_\mu$, with $f$ being a proportionality constant. When plugged back into the Polyakov action \eqref{eqn:ch3_action_ws_P}, this solution gives back the Nambu--Goto action \eqref{eqn:ch3_action_ws_NG}. On the other hand, the Polyakov action is best suited to discuss the dynamics and especially the quantization of the strings, in which the vanishing condition of the stress-energy tensor \eqref{eqn:ch3_emt_=_0} plays a central role.

\subsection{The string dynamics}

We thus focus on the Polyakov action \eqref{eqn:ch3_action_ws_P}. We observe that this action enjoys a Weyl symmetry $g_{ab}\to e^\omega g_{ab}$. This is reflected in the stress-energy tensor \eqref{eqn:ch3_emt_=_0}, which is traceless: $T^a{}_a=0$. Exploiting this symmetry, along with the reparameterization and Lorentz invariance of the action \eqref{eqn:ch3_action_ws_P}, the worldsheet metric can be set to the two-dimensional Minkowski metric $g_{ab}=\eta_{ab}$, in what is known as the conformal gauge. Then, the string action is conveniently parameterized in terms of the light-cone coordinates
\begin{align}
    \xi^{\pm}=\tau\pm\sigma \ , &&  ds^2=-d\xi^+d\xi^- \ , && \eta=\begin{pmatrix}
        0 & -\frac12 \\ -\frac12 & 0   \end{pmatrix} \ .
\label{eqn:ch3_lc_coord}
\end{align}

\noindent Choosing a flat target space $G_{\mu\nu}=\eta_{\mu\nu}$ as well, the Polyakov action \eqref{eqn:ch3_action_ws_P} takes the form
\beq
    S_\textup{P}=2T\int d\xi^+d\xi^-\,\,\partial_+ X^\mu\partial_-X_\mu \ .
\label{eqn:ch3_action_ws_P_lc}
\eeq

The equations of motion for the string are computed from the variation of this action. This is equal to
\begin{equation}
    \begin{aligned}
        \delta S_\textup{P}&=2T\int d\xi^+d\xi^-\(\partial_+ \delta X^\mu\partial_-X_\mu+\partial_+  X^\mu\partial_- \delta X_\mu \) = \\
         &=2T\int d\xi^+d\xi^-\[\partial_+\(\delta X^\mu\partial_- X_\mu\)+\partial_-\(\partial_+  X^\mu \delta X_\mu \)-2\(\partial_+\partial_- X^\mu\)\delta X_\mu\] = \\
        &= -4T\int d\xi^+d\xi^-\(\partial_+\partial_- X^\mu\)\delta X_\mu -T\int d^2\xi\,\,\partial_\sigma\(\delta X^\mu\partial_\sigma X_\mu\)\ .
    \end{aligned}
\label{eqn:ch3_eom_string_action}
\end{equation}

\noindent Since strings are extended objects, the boundary term in the $\sigma$ coordinate is, a priori, nontrivial. For closed strings, the periodicity condition $\sigma=2\pi\simeq\sigma=0$ is such that the boundary term automatically vanishes, since $X^\mu(\sigma=2\pi)=X^{\mu}(\sigma=0)$ must be satisfied by consistency. In contrast, the open string has, by construction, two boundary points $\sigma=0$ and $\sigma=\pi$, on which the boundary term in eq.~\eqref{eqn:ch3_eom_string_action} depends explicitly:
\beq
    \int_0^\pi d\sigma\partial_\sigma\(\delta X^\mu\partial_\sigma X_\mu\)= \(\delta X^\mu\partial_\sigma X_\mu\)|_{\sigma=\pi}-\(\delta X^\mu\partial_\sigma X_\mu\)|_{\sigma=0} \ .
\label{eqn:ch3_os_bt}
\eeq

\noindent Boundary conditions are thus required for the boundary term of open strings to vanish. These conditions can be either of the Neumann or Dirichlet type, imposed at the two ends of the open string:
\begin{align}
    \mathrm{NN:}\quad \partial_\sigma X^\mu|_{\sigma=0}=\partial_\sigma X^\mu|_{\sigma=\pi}=0 \ , && \mathrm{DD:}\quad \delta X^\mu|_{\sigma=0}=\delta X^\mu|_{\sigma=\pi}=0 \ .
\label{eqn:ch3_open_bc}
\end{align}

\noindent In addition, also mixed ND/DN boundary conditions can be defined accordingly.

Therefore, the string equations of motion that one finds from the Polyakov action \eqref{eqn:ch3_action_ws_P}-\eqref{eqn:ch3_action_ws_P_lc} are, with these assumptions,
\beq
    \partial_+\partial_-X^\mu=0 \ .
\label{eqn:ch3_eom_string}
\eeq

\noindent They have the form of a D'Alembertian equation,  resembling the dynamics of a system of $D$ free scalar fields. The solution has then the form of a superposition of waves, left-movers and right-movers:
\beq
    X^\mu(\tau,\sigma)=X^\mu_\textup{L}\(\tau+\sigma\)+X^\mu_\textup{R}\(\tau-\sigma\) \ ,
\label{eqn:ch3_LR_movers_tot}
\eeq

\noindent which have the general form of an expansion in constant, linear, and oscillating terms in the progressive/regressive wave coordinates:
\beq
\begin{aligned}
    X^\mu_\textup{L}\(\tau+\sigma\)&=x^\mu_\textup{L}+\frac{\alpha^{\prime}}{2}p^\mu_\textup{L}\(\tau+\sigma\)+i\sqrt{\frac{\alpha^{\prime}}{2}}\sum_{n\ne 0}\frac{\alpha_n^\mu}{n}e^{-in\(\tau+\sigma\)} \ , \\
    X^\mu_\textup{R}\(\tau-\sigma\)&=x^\mu_\textup{R}+\frac{\alpha^{\prime}}{2}p^\mu_\textup{R}\(\tau-\sigma\)+i\sqrt{\frac{\alpha^{\prime}}{2}}\sum_{n\ne 0}\frac{\balpha_n^\mu}{n}e^{-in\(\tau-\sigma\)} \ ,
\end{aligned}
\label{eqn:ch3_LR_movers_sing}
\eeq

\noindent where $\alpha^{\prime}=\(2\pi T\)^{-1}$. This general solution must be supplemented with the appropriate periodicity/boundary conditions for closed/open strings. In addition, this solution is further constrained by the requirement that the stress-energy tensor must vanish on-shell \eqref{eqn:ch3_emt_=_0}. In light-cone coordinates \eqref{eqn:ch3_lc_coord}, the nontrivial components of the stress-energy tensor \eqref{eqn:ch3_emt_=_0} are
\begin{align}
    T_{++}=\partial_+ X^\mu\partial_+X_\mu \ , && T_{--}=\partial_- X^\mu\partial_-X_\mu \ ,
\label{eqn:ch3_emt_lc}
\end{align}

\noindent and its conservation implies $T_{++}=T_{++}\(\xi^+\)$ and $T_{--}=T_{--}\(\xi^-\)$.

\subsubsection{Closed string dynamics}

The closed string solution must satisfy the periodicity condition $X^{\mu}(\sigma=2\pi)=X^{\mu}(\sigma=0)$. This implies that $p^\mu_\textup{L}=p^\mu_\textup{R}\equiv p^\mu$ and $n\in\mathbb{Z}$ in eq.~\eqref{eqn:ch3_LR_movers_tot}-\eqref{eqn:ch3_LR_movers_sing}. Calling $x^\mu\equiv x^\mu_\textup{L}+x^\mu_\textup{R}$, eq.~\eqref{eqn:ch3_LR_movers_tot} becomes
\beq
    X^\mu(\tau,\sigma)=x^\mu+\alpha^{\prime}p^\mu\tau+i\sqrt{\frac{\alpha^{\prime}}{2}}\sum_{n\in\mathbb{Z}\setminus\{0\}}\frac{1}{n}\[\alpha_n^\mu e^{-in\(\tau+\sigma\)}+\balpha_n^\mu e^{-in\(\tau-\sigma\)}\] \ .
\label{eqn:ch3_cs_X}
\eeq

\noindent The reality condition $\(X^\mu\)^*=X^\mu$ yields then $\(\alpha^\mu_n\)^*=\alpha_{-n}^\mu$ and $\(\balpha^\mu_n\)^*=\balpha_{-n}^\mu$. Defining, as customary,
\beq
    \alpha_0^\mu\equiv \sqrt{\frac{\alpha^{\prime}}{2}}p^\mu \ ,
\label{eqn:ch3_osc_mom}
\eeq
the stress-energy tensor \eqref{eqn:ch3_emt_lc} takes the form
\begin{align}
    T_{++}=\alpha^{\prime}\sum_n L_n e^{-in\(\tau+\sigma\)} \ , && T_{--}=\alpha^{\prime}\sum_n \bL_n e^{-in\(\tau-\sigma\)} \ ,
\label{eqn:ch3_cs_emt_lc}
\end{align}

\noindent where $L_n$ and $\bL_n$ are the Virasoro generators
\begin{align}
    L_n=\frac12\sum_m\alpha_{n-m}^\mu\alpha_m{}_\mu \ , &&  \bL_n=\frac12\sum_m\balpha_{n-m}^\mu\balpha_m{}_\mu \ ,
\label{eqn:ch3_cs_virasoro_gen_1}
\end{align}

\noindent satisfying the Virasoro algebra
\begin{align}
    \left\{L_n,L_m\right\}=-i\(n-m\)L_{n+m} \ , && \left\{\bL_n,\bL_m\right\}=-i\(n-m\)\bL_{n+m} \ .
\label{eqn:ch3_cs_virasoro_alg_1}
\end{align}

\noindent Therefore, the vanishing condition on the stress-energy tensor \eqref{eqn:ch3_emt_=_0} turns into the infinite set of constraints
\beq
    L_n=\bL_n=0 \quad \forall\,\, n\in\mathbb{Z} \ .
\label{eqn:ch3_Vir_gen_=_0}
\eeq

\noindent The $n=0$ constraint yields the mass formula for the propagating states:
\beq
	 M^2=\frac{2}{\alpha^{\prime}}\sum_{m\ne0}\alpha_{-m}^\mu\alpha_m{}_\mu \ ,
\label{eqn:cs_mass_1}
\eeq

\noindent along with the so-called \textit{level matching condition}
\begin{align}
    \sum_{m\ne0}\alpha_{-m}^\mu\alpha_m{}_\mu=\sum_{m\ne0}\balpha_{-m}^\mu\balpha_m{}_\mu \  ,
\label{eqn:cs_lmc_1}
\end{align}

\noindent relating the progressive and regressive oscillators.


\subsubsection{Open string dynamics}

For open strings one proceeds in the same way as for closed ones, but enforcing the boundary conditions \eqref{eqn:ch3_open_bc}. These imply the main feature of open strings: that left and right oscillators are now related to each other.

\paragraph{NN open strings} The NN boundary conditions in eq.~\eqref{eqn:ch3_open_bc} are satisfied by $p^\mu_\textup{L}=p^\mu_\textup{R}$ and $n\in\mathbb{Z}$, as for the closed strings, together with
\begin{align}
    \alpha_n^\mu=\balpha_n^\mu \ .
\end{align}

\noindent This yields
\beq 
    X^\mu(\tau,\sigma)=x^\mu+2\alpha^{\prime}p^\mu\tau+i\sqrt{2\alpha^{\prime}}\sum_{n\in\mathbb{Z}\setminus\{0\}}\frac{\alpha_n^\mu}{n}\cos(n\sigma)e^{-in\tau} \ .
\label{eqn:ch3_os_nn_X}
\eeq

\noindent In this case, one has only one set of Virasoro generators $L_n$, defined as in eq.~\eqref{eqn:ch3_cs_virasoro_gen_1}, where one now takes $\alpha^\mu_0\equiv\sqrt{2\alpha^{\prime}}p^\mu$. These generators are constrained to identically vanish by the energy-momentum tensor condition \eqref{eqn:ch3_emt_=_0}. The constraint on $L_0$ yields the mass of the open-string states, which is found to be
\beq
    M^2=\frac{1}{2\alpha^{\prime}}\sum_{m\ne0}\alpha^\mu_{-m}\alpha_m{}_\mu \ .
\eeq


\paragraph{DD open strings} The DD boundary conditions in eq.~\eqref{eqn:ch3_open_bc} are such that the open string endpoints are completely fixed, so that the string can only have vibration modes and no time translation one. The correspondent solution reads
\beq
   X^\mu(\tau,\sigma)=x^\mu+d^\mu\sigma+\sqrt{2\alpha^{\prime}}\sum_{n\in\mathbb{Z}\setminus\{0\}}\frac{\alpha_n^\mu}{n}\sin(n\sigma)e^{-in\tau} \ .
\label{eqn:ch3_os_dd_X}
\eeq

\subsection{String quantization}

Once the solutions to the equations of motion \eqref{eqn:ch3_eom_string} for closed and open strings are obtained, one can perform their quantization in the standard way. This amounts to the promotion of the target space coordinates $X^\mu(\tau, \sigma)$ and their conjugate momenta $\Pi_\mu(\tau,\sigma)=\dot{X}_\mu(\tau,\sigma)$ to quantum operators, satisfying  the canonical equal-time commutation relations
\beq
    \[X^\mu\(\tau,\sigma\),\Pi^\nu\(\tau,\sigma'\)\]=i\eta^{\mu\nu}\delta\(\sigma-\sigma'\) \ .
\label{eqn:ch3_quant_cov}
\eeq

\noindent Combining this quantum condition with the explicit solutions \eqref{eqn:ch3_cs_X} and \eqref{eqn:ch3_os_nn_X}-\eqref{eqn:ch3_os_dd_X} and the vanishing stress-energy tensor condition \eqref{eqn:ch3_emt_=_0}-\eqref{eqn:ch3_Vir_gen_=_0} leads to the rich and powerful features of string theory.

We illustrate this explicitly in the case of open strings with NN boundary conditions, given in eq.~\eqref{eqn:ch3_os_nn_X}. The DD open string and the closed string cases follow in complete analogy. We do so by following the paradigm of the light-cone quantization \cite{Goddard:1973qh}. This method exploits a residual gauge freedom of the Polyakov action --- which reflects the invariance of the D'Alembert equations \eqref{eqn:ch3_eom_string} under a reparameterization of the light-cone coordinates of the type $\xi^{\pm}{}'\to\tilde{\xi}^{\pm}\(\xi^{\pm}\)$ --- to align the worldsheet light-cone with the one in target space. This is given by
\begin{align}
    X^\pm\equiv\frac{1}{\sqrt{2}}\(X^0\pm X^{D-1}\) \ , && ds^2=-2dX^+ dX^-+\sum_{i=1}^{D-2}dX^idX_i \ .
\label{eqn:ch3_lc_gauge_coord}
\end{align}

\noindent The identification condition, known as the light-cone gauge, that can be fixed with the residual gauge invariance is
\beq
    X^+\(\tau,\sigma\)=x^++2\alpha^{\prime}p^+\tau \ .
\label{eqn:ch3_lc_q_X+}
\eeq
The rest of the target space coordinates are written in this setup as
\beq
\begin{aligned}
    X^-\(\tau,\sigma\)&=x^-+2\alpha^{\prime}p^-\tau+i\sqrt{2\alpha^{\prime}}\sum_{n\ne0}\frac{\alpha_n^-}{n}\cos(n\sigma)e^{-in\tau} \ , \\ X^i\(\tau,\sigma\)&=x^i+2\alpha^{\prime}p^i\tau+i\sqrt{2\alpha^{\prime}}\sum_{n\ne0}\frac{\alpha_n^i}{n}\cos(n\sigma)e^{-in\tau} \ .
\end{aligned}
\eeq

\noindent This light-cone gauge parameterization breaks explicitly Lorentz-invariance, so that one has to restore it at the end. Its main advantage is that it allows one to solve very easily the stress-energy tensor constraint \eqref{eqn:ch3_emt_=_0}. In particular, it implies that also the coordinate $X^-(\tau,\sigma)$ is constrained, as
\begin{align}
    T_{++}=T_{--}=0 \quad \implies \quad  \partial_\pm X^{-}=\frac{1}{2\alpha^{\prime}p^+}\partial_\pm X^i\partial_\pm X_i \ ,
\end{align}

\noindent which turns into a condition on the $\alpha_n^-$ oscillators:
\beq
    \alpha_n^-=\sqrt{\frac{2}{\alpha^{\prime}}}\frac{1}{4p^+}\sum_m\alpha_{n-m}^i\alpha_m{}_i \ . 
\eeq

\noindent Evaluated for the $n=0$ oscillator, this relation yields the mass formula
\beq
    M^2=\frac{1}{2\alpha^{\prime}}\sum_{m\ne0}\alpha^i_{-m}\alpha_m{}_i \ ,
\label{eqn:ch3_M_os_lc}
\eeq

\noindent depending only on the physical transverse oscillators.

Quantization is then performed following the standard eq.~\eqref{eqn:ch3_quant_cov}. The light-cone gauge is such that the canonical commutation relations act nontrivially only on a restricted, simplified set of operators. This concerns, in particular, the transverse oscillators, which take the form of creation and annihilation operators:
\beq
    \[\alpha_n^i,\alpha_{m}^j\]=n\delta_{n+m,0}\delta^{ij}.
\label{eqn:ch3_osc_lad}
\eeq

\noindent Therefore, the transverse oscillators define a Fock space for the string quantum excitations, which is obtained starting from the vacuum $\ket{0}$ --- such that $\alpha_m\ket{0}=0$ --- and then acting iteratively with $\alpha_{-m}^i$, where $m$ is taken positive. The mass in eq.~\eqref{eqn:ch3_M_os_lc} becomes a quantum operator as well, and being quadratic in the ladder operators it requires to be normal-ordered. A proper renormalization of the zero-point contributions yields
\beq
     M^2=\frac{1}{2\alpha^{\prime}}\sum_{m\ne0}:\alpha^i_{-m}\alpha_m{}_i:=\frac{1}{\alpha^{\prime}}\sum_{m>0}\alpha^i_{-m}\alpha_m{}_i -\frac{D-2}{24\alpha^{\prime}}\equiv\frac{1}{\alpha^{\prime}}\(N-\frac{D-2}{24}\) \ .
\label{eqn:ch3_M_os_lc_q}
\eeq

\noindent We can now fully characterize the spectrum of the open string quantum states:
\vspace{0.5cm}
\begin{itemize}
    \item beginning with the vacuum $\ket{0}$, the above mass formula immediately tells that this state is tachyonic:
\beq
    M^2\ket{0}=-\frac{D-2}{24\alpha^{\prime}}\ket{0} \ ,
\label{eqn:ch3_os_vac_tachyon}
\eeq

\noindent denoting a fundamental instability of bosonic string theory. 

\vspace{0.3cm}
\item The first excited state is given by $\alpha_{-1}^i\ket{0}$. It contains $D-2$ degrees of freedom and its mass is equal to
\beq
    M^2\(\alpha_{-1}^i\ket{0}\)=-\frac{D-26}{24\alpha^{\prime}}\(\alpha_{-1}^i\ket{0}\) \ . 
\label{eqn:ch3_os_massless_vec}
\eeq

\noindent This state transforms, by construction, in the vectorial representation of the little group $\mathrm{SO}(D-2)$, and since this state contains $D-2$ degrees of freedom, this representation must be massless. Thus, consistency with Lorentz symmetry implies that the bosonic string can be defined only in the critical dimension
\beq
    D=26 \ .
\eeq

\vspace{0.3cm}
\item All the higher excitations are massive. The first such massive level $M^2=\frac{1}{\alpha^{\prime}}$ is given by the states
\begin{align}
    \alpha_{-2}^i\ket{0} \ , && \alpha_{-1}^i\alpha_{-1}^j\ket{0} \ .
\label{eqn:ch3_os_first_massive_states}
\end{align}

\noindent These two states organize into a symmetric representation of the little group $\mathrm{SO}(D-1)$ corresponding to the degrees of freedom of a massive spin-2 field:
\beq
\begin{aligned}
    \(D-2\)+\frac{\(D-2\)\(D-1\)}{2}&=\frac{D\(D-1\)}{2}-1 \\ 
    &=\[\frac{\(D-2\)\(D-1\)}{2}-1\]+\(D-2\)+1 \ .
\end{aligned}
\eeq

\noindent The second line represents in fact a massless spin-2 tensor plus a vector and a scalar field, which correspond to the polarization states of a massive spin-2 field (see for example eq.~\eqref{eqn:ch2_ms2_h_exp_mom} in Chapter \ref{ch2_sugra}).

\end{itemize}

\vspace{0.5cm}
\noindent Following the same pattern, the higher massive states correspond to higher-spin representations. Therefore, the string excitations spectrum is characterized by a tower of states of increasing mass and spin: this is the famous Regge trajectory, which is a consequence of the non-canonical normalization of the commutation relations among the ladder operators of eq.~\eqref{eqn:ch3_osc_lad}.

\subsubsection{The closed-string spectrum} As remarked at the beginning of this section, the quantization procedure is analogous for closed strings. The main difference is that now the left and right oscillators define two distinct set of ladder operators, related by the level-matching condition \eqref{eqn:cs_lmc_1}, so that the resulting spectrum is richer. The mass operator in eq.~\eqref{eqn:cs_mass_1} becomes
\beq
    M^2=\frac{2}{\alpha^{\prime}}\[\sum_{n>0}\(\alpha^i_{-m}\alpha_m{}_i+\balpha^i_{-m}\balpha_m{}_i\)-\frac{D-2}{12}\]\equiv\frac{2}{\alpha^{\prime}}\(N+\bN-\frac{D-2}{12}\),
\label{eqn:ch3_cs_masses}
\eeq

\noindent and the level-matching condition enforces $N=\bN$ as a relation on the action of these quantum operators on physical states. 

The closed string vacuum state $\ket{0,\bar{0}}$ is also tachyonic, as the open string one in eq.~\eqref{eqn:ch3_os_vac_tachyon}. Then, the first excited state that is allowed by the level-matching condition is the tensor state
\beq
    \alpha_{-1}^i\balpha_{-1}^j\ket{0,\bar{0}} \ .
\label{eqn:ch3_cs_massless_state}
\eeq

\noindent Its $\(D-2\)\times\(D-2\)$ degrees of freedom can be organized into
\beq
    \(D-2\)\(D-2\)=\[\frac{\(D-2\)\(D-1\)}{2}-1\]+\frac{\(D-2\)\(D-3\)}{2}+1 \ ,
\label{eqn:ch3_cs_massless_count}
\eeq

\noindent namely, into the symmetric traceless, antisymmetric and singlet representations of the little group $\mathrm{SO}(D-2)$. As before, these states must be massless in order to be consistent with Lorentz invariance, which is indeed the case for the critical dimension $D=26$. In the field theory limit, these states are identified, respectively, with the graviton $G_{\mu\nu}$, a rank-2 antisymmetric tensor field $B_{\mu\nu}$, known as the Kalb--Ramond field, and the dilaton $\phi$. Therefore, the quantization of the bosonic string yields, in a natural way, the gravitational interaction. This string realization is the only known consistent quantum description of gravity.

\subsection{The field theory limit and background fields}

We have seen that the spectrum of excitations of the quantized string is built on a tachyonic vacuum and includes a massless state and a tower of higher-spin states of increasing mass. Physically, there should exist an energy threshold below which this massive tower of states can be neglected. This corresponds to the field theory limit of the string, where the theory reduces to an effective field theory describing the massless degrees of freedom of the string. For the closed string, this field theory limit consists of the graviton, the Kalb–Ramond antisymmetric field, and the dilaton, as discussed above.

This field theory limit of closed strings can be understood also from an alternative perspective, in terms of a background-field description. To see this, we return to the Polyakov action of eq.~\eqref{eqn:ch3_action_ws_P}, in which the string coordinates couple to the auxiliary worldsheet metric $g_{ab}$ and to the target space metric $G_{\mu\nu}$. Taking a nontrivial background metric $G_{\mu\nu}$ is already a generalization of the above discussion on the string dynamics and quantization, which was carried out in flat Minkowski space. Nevertheless, further generalizations of the Polyakov action \eqref{eqn:ch3_action_ws_P} are actually possible. Two additional nontrivial terms can be introduced through appropriate background fields. The first such term is the Einstein--Hilbert action for the worldsheet metric $g_{ab}$. Since the worldsheet metric is two-dimensional, the Einstein--Hilbert action reduces to the Euler characteristic of the worldsheet, \textit{i.e.} a constant. This can be made nontrivial through a coupling to a background scalar field, which we call $\phi$. The second term is the antisymmetric version of the string coordinates kinetic term, in which the contraction of the worldsheet indices is made with the two-dimensional Levi--Civita tensor $\varepsilon^{ab}$ (instead of the worldsheet metric), and is introduced in the action via coupling to an antisymmetric background field $B_{\mu\nu}$ (instead of the target space metric). The resulting generalization of the Polyakov action is then
\beq
    S_\textup{P}=-\frac{1}{4\pi\alpha^{\prime}}\int d^2\xi\left\{g^{ab}\partial_a X^\mu\partial_b X^\nu G_{\mu\nu}(X)+\varepsilon^{ab}\partial_a X^\mu\partial_b X^\nu B_{\mu\nu}(X)-\alpha^{\prime}R(g)\phi(X)\right\} \ .
\eeq

However, this background is not completely arbitrary. In particular, it must be consistent with the conformal symmetry of the worldsheet and not source any Weyl anomaly at the quantum level. The conformal symmetry can be imposed on the background fields by thinking of them as coupling constants and requiring them to define a fixed point of the renormalization group flow, as fixed points indeed correspond to conformal field theories. This condition is met if the $\beta$-functions associated with the background coupling fields identically vanish. These equations can be computed to be \cite{Callan:1985ia}
\beq
\begin{aligned}
    \beta{}_{\mu \nu}^{(G)}&=R_{\mu \nu}+\frac{1}{4} H_{\mu}{ }^{\lambda \rho} H_{\nu \lambda \rho}-2 \nabla_{\mu} \nabla_{\nu} \phi+\mathcal{O}\left(\alpha^{\prime}\right)=0 \ , \\
    \beta_{\mu \nu}^{(B)}&=\nabla_{\lambda} H^{\lambda}{ }_{\mu \nu}-2 \nabla_{\lambda} \phi H^{\lambda}{ }_{\mu \nu}+\mathcal{O}\left(\alpha^{\prime}\right)=0 \ , \\
    \beta^{(\phi)}&=4 \nabla_{\lambda} \phi \nabla^{\lambda} \phi-4 \nabla_{\lambda} \nabla^{\lambda} \phi+R+\frac{1}{12} H_{\lambda \rho \sigma} H^{\lambda \rho \sigma}+\mathcal{O}\left(\alpha^{\prime}\right)=0 \ ,
\end{aligned}
\label{eqn:ch3_bkgr_beta_eq}
\eeq

\noindent in which $H_{\mu\nu\rho}=3\partial_{[\mu}H_{\nu\rho]}$ is the antisymmetric tensor field strength and $\nabla_\mu$ is the covariant derivative associated with the target space metric. The striking feature of these equations is that they exactly match the equations of motion associated with the action
\beq
    S=\frac12\int d^{26}x\sqrt{-G}\,\,e^{-2\phi}\(R-\frac{1}{12}H^{\mu\nu\rho}H_{\mu\nu\rho}+4\partial_\mu\phi\partial^\mu\phi\) \ ,
\eeq

\noindent which is the field theory limit of the closed bosonic string. Consequently,  $G_{\mu\nu}$, $B_{\mu\nu}$ and $\phi$ are identified with the graviton, the Kalb--Ramond field and the dilaton introduced in eq.~\eqref{eqn:ch3_cs_massless_state}-\eqref{eqn:ch3_cs_massless_count}. Higher-order corrections in $\alpha^{\prime}$ in the $\beta$-functions equations \eqref{eqn:ch3_bkgr_beta_eq} correspond to higher-derivative, stringy  corrections to the low-energy effective field theory action.

\subsection{The bosonic string partition function and 1-loop vacuum amplitudes}\label{ch3_sec_bs_va}

A quantity of crucial importance in understanding the physics described by a given string theory is the string \textit{partition function} $\Z$. Schematically, it has the form
\beq
    \Z=\int DX(\tau,\sigma)e^{-S_\textup{P}} \ ,
\label{eqn:ch3_Z_0}
\eeq

\noindent where $S_\textup{P}$ is the Polyakov action \eqref{eqn:ch3_action_ws_P}. The meaning of this formula can be understood in analogy with the path integral of the one-particle state in quantum mechanics. This is given in terms of the position operator $q(t)$ by
\beq
    \braket{q_\textup{F},t_\textup{F}}{q_\textup{I},t_\textup{I}}=\int_{q_\textup{I}}^{q_\textup{F}} Dq(t)e^{-\frac{1}{\hbar}S(q,\dot{q})} \ .
\label{eqn:ch3_path_int_qm}
\eeq

\noindent The interpretation of this formula is that the transition amplitude for the quantum one-particle state to go from the position $q_\textup{I}$ at time $t_\textup{I}$ to the position $q_\textup{F}$ at time $t_\textup{F}$ is given by the infinite sum over all possible trajectories between $q_\textup{I}$ and $q_\textup{F}$, weighted by a Boltzmann-like factor $e^{-\frac{1}{\hbar}S}$. When going from the point-particle to the extended string, the trajectories in spacetime become two-dimensional manifolds, \textit{i.e.} the worldsheet, so that a partition function like eq.~\eqref{eqn:ch3_Z_0} can be interpreted as the sum over all possible geometries that the worldsheet can describe. Actually, the set of symmetries enjoyed by the Polyakov action are such that this sum can be effectively performed over the possible \textit{topologies} of the worldsheet. This is made precise and manifest by the Einstein--Hilbert term of the (generalized) Polyakov action:
\beq
    \frac{\phi}{4\pi}\int d^2\xi\sqrt{-g}\,\,R(g)=\phi\chi \ ,
\eeq

\noindent Here $\phi$ is the dilaton background field and $\chi$ is the Euler characteristic of the worldsheet. Assuming a vacuum expectation value $\langle \phi\rangle$, the infinite sum over the worldsheet geometries described by the partition function \eqref{eqn:ch3_Z_0} is actually a sum over topologies, each one associated with a different Euler characteristic $\chi$ and weighted by a factor $e^{-\langle \phi\rangle\chi}\equiv g_s^{-\chi}$, with $g_s$ being the string coupling. The Euler characteristic of a Riemann surface is equal to
\begin{align}
    \chi=2-2h-b-c \ , 
\label{eqn:ch3_euler_ch_gen}
\end{align}
where $h$ is the number of handles, $b$ the number of boundaries and $c$ the number of crosscaps. Thus, the partition function \eqref{eqn:ch3_Z_0} is given by an expansion of the form
\beq
    \Z\sim\frac{1}{g_s^2}\[\chi=2\]+\frac{1}{g_s}\[\chi=1\]+\[\chi=0\]+\sum_{n>0}g_s^n\[\chi=n\] \ .
\label{eqn:ch3_poly_exp_gen}
\eeq

\noindent This is known as Polyakov expansion \cite{Polyakov:1981rd,Polyakov:1981re}, and it is the string analogue of the loop expansion of amplitudes in quantum field theory. In contrast to the field theory case, we see that in string theory the dominant term of the expansion \eqref{eqn:ch3_poly_exp_gen} is a loop contribution, given by the topologies of Euler characteristic $\chi=0$. There are four Riemann surfaces with vanishing Euler characteristic:
\begin{equation*}
    \begin{array}{rcccc}
        \text{Torus}  & \T: & h=1, & b=0, & c=0, \\
        \text{Cylinder}  & \A: & h=0, & b=2, & c=0, \\
        \text{Klein bottle}  & \K: & h=0, & b=0, & c=2, \\
        \text{M\"obius strip}  & \M: & h=0, & b=1, & c=1. \\
    \end{array}
\end{equation*}
%
%
%

\noindent For the bosonic strings at hand, the worldsheet of closed strings must be a closed (oriented) Riemann surface with no boundaries, so that the corresponding partition function is dominated by the torus amplitude $\T$. On the contrary, the open strings worldsheet is, by definition, a Riemann surface with nontrivial boundaries, so that the leading contribution in the corresponding partition function is the cylinder amplitude $\A$. The Klein bottle and the M\"obius strip --- and, in general, Riemann surfaces with crosscaps $c\ne0$ --- appear instead in the case of unoriented strings, which we will discuss in Section \ref{ch3_sec_type_I}. 

\subsubsection{The torus amplitude}

We now focus on the case of closed strings and their corresponding torus amplitude $\T$. In order to derive its analytic expression, we start again from an analogy, as for eq.~\eqref{eqn:ch3_Z_0}-\eqref{eqn:ch3_path_int_qm}, this time drawing from quantum field theory \cite{reviews_1}. The vacuum energy $\Gamma$ in quantum field theory is obtained from the trace over the particles masses as
\beq
    \Gamma=-\log\Z=-\frac{V_D}{2\(4\pi\)^{\frac{D}{2}}}\int_\epsilon^\infty\frac{dt}{t^{\frac{D}{2}+1}}\mathrm{Str}\(e^{-tM^2}\) \ ,
\label{eqn:ch3_ve_field_theory}
\eeq

\noindent in which $\epsilon$ is a UV cutoff, $V_D$ is the total spacetime volume, $t$ is a Schwinger time parameter and $\mathrm{Str}$ is the super-trace operator counting the masses of bosonic and fermionic fields with appropriate multiplicity. A first extension of this formula to string theory is obtained by inserting the masses of the closed string excitations, given in eq.~\eqref{eqn:ch3_cs_masses}. In the critical dimension $D=26$, these are given by
\beq
    M^2=\frac{2}{\alpha^{\prime}}\(N+\bN-2\) \ .
\eeq

\noindent In addition, one must also account for the level-matching condition $N=\bN$ of eq.~\eqref{eqn:cs_lmc_1}. We do this by supplementing the formula \eqref{eqn:ch3_ve_field_theory} with a Dirac $\delta$-function of the type
\beq
    \delta\(N-\bN\)=\int_{-\frac12}^{\frac12}ds\,\,e^{2\pi i\(N-\bN\)s} \ ,
\eeq

\noindent  yielding
\beq
    \Gamma=-\frac{V_{26}}{2\(4\pi\)^{13}}\int_{-\frac12}^{\frac12}ds\int_\epsilon^\infty\frac{dt}{t^{14}}\mathrm{Tr}\[e^{-\frac{2t}{\alpha^{\prime}}\(N+\bN-2\)}e^{2\pi i\(N-\bN\)s}\] \ .
\eeq

\noindent This formula can be rewritten in compact form introducing
\begin{align}
    \tau=\tau_1+i\tau_2\equiv s+\frac{i}{\pi\alpha^{\prime}}t \ , && q=e^{2\pi i \tau} \ ,
\end{align}

\noindent as
\beq
    \Gamma=-\frac{V_{26}}{2\(4\pi^2\alpha^{\prime}\)^{13}}\int_{\mathcal{D}}\frac{d^2\tau}{\tau_2^{14}}\mathrm{Tr}\(q^{N-1}\bar{q}^{\bN-1}\) \ .
\label{eqn:ch3_torus_gamma_step}
\eeq

\noindent The domain of integration is, by now, $\mathcal{D}=\left\{\tau\in\mathbb{H}^+,\abs{\tau_1}\le\frac12\right\}$, corresponding to the strip of width one centered in the complex upper-half plane $\mathbb{H}^+$. We first focus on the integrand. The trace in eq.~\eqref{eqn:ch3_torus_gamma_step} factorizes into contributions from left- and right-moving modes. The former is computed explicitly to be
\beq
\begin{aligned}
    \mathrm{Tr}\(q^{N-1}\)&=\frac{1}{q}\mathrm{Tr}\(q^{\sum_{n>0}\alpha^i_{-n}\alpha_n{}_i}\)=\frac{1}{q}\prod_{i=1}^{24}\prod_{n=1}^{\infty}\mathrm{Tr}\(q^{\alpha^i_{-n}\alpha_n^i}\)= \\
    &=\frac{1}{q}\prod_{i=1}^{24}\prod_{n=1}^{\infty}\sum_{k_n}\bra{k_n}q^{\alpha^i_{-n}\alpha_n^i}\ket{k_n}=\frac{1}{q}\prod_{i=1}^{24}\prod_{n=1}^{\infty}\sum_{k_n}q^{k_nn}=\\
    &=\frac{1}{q}\prod_{i=1}^{24}\prod_{n=1}^{\infty}\frac{1}{1-q^n}=\[\frac{1}{q^{\frac{1}{24}}\prod_{n=1}^\infty\(1-q^n\)}\]^{24}=\frac{1}{\eta(\tau)^{24}} \ ,
\end{aligned}
\eeq

\noindent where we introduced the famous Dedekind function $\eta(\tau)$. The right-movers give the conjugate contribution $\bar{\eta}(\bar{\tau})$, so that eq.~\eqref{eqn:ch3_torus_gamma_step} becomes
\beq
    \Gamma=-\frac{V_{26}}{2\(4\pi^2\alpha^{\prime}\)^{13}}\int_{\mathcal{D}}\frac{d^2\tau}{\tau_2^{14}}\frac{1}{\abs{\eta(\tau)}^{48}} \ .
\eeq

However, this is not yet the final result. Defined in this way, this integral is actually divergent. The reason is that the integration domain $\mathcal D$, which was drawn from the analogy with the field theory case \eqref{eqn:ch3_ve_field_theory}, is redundant in the case of the string. In fact, the string worldsheet enjoys a significant number of symmetries, which must be taken into account in order to properly count the physical contributions to the vacuum energy. In the present case, the topology of the worldsheet is that of the torus, and the integration in eq.~\eqref{eqn:ch3_torus_gamma_step} sums over all torus geometries, parameterized by $\tau\in\mathcal{D}$. This integral turns out to be divergent because this integration overcounts the number of inequivalent tori. In fact, equivalent tori are identified by modular transformations of $\mathrm{PSL}(2,\mathbb{Z})$, acting on the complex parameter $\tau$ as
\begin{align}
    \tau\to\frac{a\tau+b}{c\tau+d} \ ,  
\label{eqn:ch3_mod_tr}
\end{align}

\noindent with $\left\{a,b,c,d\right\}\in\mathbb{Z}$ and $ad-bc=1$. This group is generated by the transformations
\begin{align}
    T=\begin{pmatrix}1 & 1 \\ 0& 1\end{pmatrix} \ , && S=\begin{pmatrix}0 & -1 \\ 1& 0\end{pmatrix} \ , && \(S\)^2=\(ST\)^3=-\mathbb{1}\ ,
\label{eqn:ch3_mod_gen}
\end{align}

\noindent acting respectively as the shift $\tau\to\tau+1$ and the reflection $\tau\to-\tau^{-1}$. The inequivalent tori are therefore identified by the so-called fundamental domain
\beq
    \mathscr{F}=\mathbb{H}^+\setminus\mathrm{PSL}(2,\mathbb{Z})=\left\{-\frac12<\tau_1\le\frac12,\abs{\tau}\ge 1,\tau_2>0\right\} \ .
\label{eqn:ch3_torus_fd}
\eeq

\noindent Therefore, the resulting torus amplitude, describing the leading vacuum energy contribution from closed oriented strings, is found to be\footnote{With respect to the above formulae, the torus amplitude in eq.~\eqref{eqn:ch3_cs_torus_amp} is rescaled to canonical normalization.}
\beq
    \T=\frac{1}{\(4\pi^2\alpha^{\prime}\)^{13}}\int_{\mathscr{F}}\frac{d^2\tau}{\tau_2^{14}}\frac{1}{\abs{\eta(\tau)}^{48}} \ .
\label{eqn:ch3_cs_torus_amp}
\eeq

\noindent The emergence of the Dedekind $\eta$-function in the combination of eq.~\eqref{eqn:ch3_cs_torus_amp} has now a nice interpretation in terms of modular invariance of the whole amplitude, consistently with the symmetry properties of the worldsheet and the Polyakov action. This can be verified using the following transformation properties of the $\eta$-function:
\begin{align}
    T\text{:}\quad \eta(\tau+1)=e^{\frac{i\pi}{12}}\,\eta(\tau) \ , &&  S\text{:}\quad \eta\(-\textstyle{\frac{1}{\tau}}\)=\sqrt{-i\tau}\,\eta(\tau) \ .
\end{align}

Vacuum amplitudes such as the torus have the interesting property that they encode the entirety of the degrees of freedom of the quantized string. The degrees of freedom associated with each mass level can be identified through a Taylor expansion around $q=0$ of the integrand of eq.~\eqref{eqn:ch3_cs_torus_amp}. Recalling its definition \eqref{eqn:ch3_ve_field_theory} as an (infinite) sum over mass states, the expansion has the form
\beq
\begin{aligned}
    \frac{1}{\eta(q)^{24}\bar{\eta}(\bar{q})^{24}}&=\sum_{n,m}c(n,m)q^n\bar{q}^m \ ,
\end{aligned}
\eeq

\noindent and the physical states are identified by the level-matching condition, which selects only the states with equal powers of $q$ and $\bar{q}$, \textit{i.e.} with $n=m$ in the sum. The relative coefficient $c(n,n)$ gives the number of degrees of freedom of the states of mass $M=\frac{n}{\alpha^{\prime}}$. Restricting the expansion above to the physical states, one obtains
\beq
\begin{aligned}
    \frac{1}{\eta(q)^{24}\bar{\eta}(\bar{q})^{24}}\Big|_\textup{physical}&=\sum_nc(n,n)\(q\bar{q}\)^n=\frac{1}{q\bar{q}}+576+104976 q\bar{q}+\dots \ ,
\end{aligned}
\label{eqn:ch3_bs_torus_state_counting}
\eeq

\noindent in which we recognize the tachyonic vacuum, the 576 degrees of freedom given by the 26-dimensional massless graviton, Kalb--Ramond field and dilaton, and so on with the massive tower of states.

\subsubsection{The cylinder amplitude and the double channel interpretation}

The cylinder amplitude $\A$ is the open string analogue of the torus amplitude for closed strings. Following this analogy, one obtains
\beq
    \A=\frac{1}{2\(4\pi^2\alpha^{\prime}\)^{24}}\int_0^\infty \frac{d\tau_2}{\tau_2^{14}}\frac{1}{\eta\(\frac{i\tau_2}{2}\)^{24}} \ .
\label{eqn:ch3_os_cylinder_loop_ch}
\eeq

\noindent As for the torus in eq.~\eqref{eqn:ch3_bs_torus_state_counting}, we can count the degrees of freedom of the quantized open string by Taylor expanding this cylinder amplitude around $q=0$, obtaining:
\beq
    \frac{1}{\eta(q)^{24}}=\frac{1}{q}+24+324q\dots \ .
\label{eqn:ch3_bs_cylinder_state_counting}
\eeq

\noindent In this expansion, one clearly recognizes the tachyonic vacuum, the massless vector, and the massive spin-2 state of the open-string spectrum discussed in eq.~\eqref{eqn:ch3_os_vac_tachyon}, \eqref{eqn:ch3_os_massless_vec} and \eqref{eqn:ch3_os_first_massive_states}.

Because of the open string setting, only one power of the Dedekind $\eta$-function appears in the amplitude \eqref{eqn:ch3_os_cylinder_loop_ch}, depending on the parameter $\tau=\frac{i\tau_2}{2}$. This is the so-called modulus of the doubly covering torus, describing the cylinder in the real projective plane \cite{reviews_1}. As for the torus in eq.~\eqref{eqn:ch3_cs_torus_amp}, it parameterizes the cylinder shapes, and the associated fundamental domain --- which must be the domain of integration of the vacuum amplitude --- defines all inequivalent cylinders. This is the general structure of vacuum string amplitudes. 


\noindent With this choice of the "time" integration parameter, the amplitude is called the loop (or direct) channel. In this picture, one looks at the cylinder vertically, from the point of view of an open string that sweeps it closing a loop. However, one can also move to a "horizontal time" integration parameter $l=\frac{2}{\tau_2}$. The amplitude obtained with this choice is the so-called transverse channel, equal to 
\beq
    \tilde{\A}=\frac{2^{-13}}{2\(4\pi^2\alpha^{\prime}\)^{24}}\int_0^\infty d\ell\frac{1}{\eta\(i\ell\)^{24}} \ .
\label{eqn:ch3_os_cylinder_trans_ch}
\eeq

\noindent Formally, it can be obtained from the loop channel \eqref{eqn:ch3_os_cylinder_loop_ch} via an $S$ transformation \eqref{eqn:ch3_mod_gen}. In this basis the cylinder is parameterized horizontally, as a hollow tube. While in the direct channel \eqref{eqn:ch3_os_cylinder_loop_ch} the amplitude describes the 1-loop propagation of an open string, the transverse channel \eqref{eqn:ch3_os_cylinder_trans_ch} allows for a dual interpretation in terms of tree-level propagation of a closed string.

\section{Superstring theory}\label{ch3_sec_sst}

The bosonic string discussed in Section \ref{ch3_sec_BS} already displays a number of striking features, the main one being achieving a quantum description of gravity. However, this theory has two important drawbacks. First, its vacuum is tachyonic, meaning that the theory is inherently unstable. Second, it does not describe any fermionic degrees of freedom.

These two problems are solved at once introducing fermionic degrees of freedom on the worldsheet \cite{Polyakov:1981re,Brink:1976sc,Deser:1976rb}. We do so following the Ramond--Neveu--Schwarz formalism \cite{Ramond:1971gb,Neveu:1971fz} and introduce the fermionic coordinate $\psi^\mu\(\tau,\sigma\)$, which is a vector in target space --- like the bosonic coordinates $X^\mu$ --- and a spinor on the worldsheet. The two-dimensional Clifford algebra is described by the matrices
\begin{align}
    \rho^0=\begin{pmatrix}  0 & -1 \\ 1 & 0  \end{pmatrix} \ , && \rho^1=\begin{pmatrix}  0 & 1 \\ 1 & 0  \end{pmatrix} \ , && \left\{\rho^a,\rho^b\right\}=2\eta^{ab} \ ,
\end{align}

\noindent and the worldsheet spinor $\psi^\mu$ is a Majorana fermion, which one writes as 
\begin{align}
    \psi^\mu=\begin{pmatrix} \psi^\mu_- \\ \psi^\mu_+ \end{pmatrix} \ , && \psi_\pm^\mu{}^*=\psi_\pm^\mu\ , &&\bpsi^\mu=i\psi^\mu{}^\dagger\rho^0 \ .
\end{align}

Working in the conformal gauge, the extended Polyakov action is
\beq
    S_\textup{P}=-\frac{1}{4\pi\alpha^{\prime}}\int d^2\xi\(\partial_a X^\mu\partial^a X_\mu+\bpsi^\mu\rho^a\partial_a\psi_\mu\) \ .
\label{eqn:ch3_ss_action_P}
\eeq

\noindent The crucial property of this action is that it enjoys worldsheet supersymmetry:
\begin{align}
    \delta X^\mu=\bepsilon\psi^\mu \ , && \delta\psi^\mu=\rho^a\partial_a X^\mu \epsilon \ .
\label{eqn:ch3_susy_string}
\end{align}

\noindent As a consequence, this theory has now two conserved currents, the stress-energy tensor $T_{ab}$, as in the bosonic case, and the supersymmetry current $\Xi_a$. In the light-cone gauge, their nontrivial components are
\begin{align}
    T_{\pm\pm}=\partial_\pm X^\mu\partial_\pm X_\mu+\frac{i}{2}\psi^\mu_\pm\partial_\pm\psi_\pm{}_\mu \ , && \Xi_\pm=\partial_\pm X^\mu\psi_\pm{}_\mu \ .
\label{eqn:ch3_ss_emt_sc}
\end{align}

\noindent As in a chiral multiplet like eq.~\eqref{eqn:ch1_wz_free_susy}, the dependence on the left- and right-moving components factorizes in the currents, as well as in the supersymmetry transformations, as can be seen from eq.~\eqref{eqn:ch3_susy_string}. Because of the supersymmetry relation with the stress-energy tensor, also the supersymmetry current $\Xi_a$ must vanish. This can be seen also by writing the action \eqref{eqn:ch3_ss_action_P} in an arbitrary gauge: in addition to the worldsheet metric $g_{ab}$, supersymmetry requires the introduction of an auxiliary gravitino, and its equation of motion implies the vanishing of the supersymmetry current.

\subsection{Superstring dynamics and quantization}

The analysis of the theory now proceeds in the same way as for the bosonic string, although the new supersymmetric structure will imply important differences. First, we need to understand which periodicity/boundary conditions are satisfied by the fermionic string. The equations of motion boundary term are found to be
\beq
    \left.\int d\tau\(\psi_-^\mu\delta\psi_-{}_\mu-\psi_+^\mu\delta\psi_+{}_\mu\)\right\rvert_{\sigma=0}^{\sigma=\bfrac{2\pi}{\pi}}=0 \ ,
\label{eqn:ch3_ss_eom_bt}
\eeq

\noindent For closed strings, this amounts to imposing periodic or antiperiodic boundary conditions: $\psi_\pm\(\sigma=2\pi\)=\pm\psi_\pm\(\sigma=0\)$. For open strings, eq.~\eqref{eqn:ch3_ss_eom_bt} must be satisfied independently at the two endpoints, so that the inequivalent choices are given by $\psi_+^\mu\(\sigma=0\)=\psi_-^\mu\(\sigma=0\)$ and $\psi_+^\mu\(\sigma=\pi\)=\pm\psi_-^\mu\(\sigma=\pi\)$. The corresponding expansions involve, for both closed and open strings, a sum over integer oscillators when the boundary conditions are periodic, over half-integer when they are antiperiodic. The former are referred to as the Ramond sector (R), the latter as the Neveu--Schwarz sector (NS) \cite{Ramond:1971kx,Ramond:1971gb,Neveu:1971rx,Neveu:1971iv,Neveu:1971iw}. The corresponding expansions are
\vspace{0.3cm}
\begin{itemize}
    \item closed strings:
    \begin{align}
        \text{R:}&& \psi_+^\mu&=\sqrt{2\pi\alpha^{\prime}}\sum_{n\in\mathbb{Z}}d_n^\mu e^{-2i\pi n\(\tau+\sigma\)} \ ,  &&& \psi_-^\mu&=\sqrt{2\pi\alpha^{\prime}}\sum_{n\in\mathbb{Z}}\td_n^\mu e^{-2i\pi n\(\tau-\sigma\)} \ ,\\
        \text{NS:}&& \psi_+^\mu&=\sqrt{2\pi\alpha^{\prime}}\sum_{r\in\mathbb{Z}+\frac12}b_r^\mu e^{-2i\pi r\(\tau+\sigma\)} \ , &&& \psi_-^\mu&=\sqrt{2\pi\alpha^{\prime}}\sum_{r\in\mathbb{Z}+\frac12}\tb_r^\mu e^{-2i\pi r\(\tau-\sigma\)}  \ ,
    \end{align}

    \noindent where the Majorana condition fixes $\(d_n^\mu\)^\dagger=d_{-n}^\mu$ and similarly for all the other oscillators.

    \vspace{0.3cm}
    \item Open strings:
    \begin{align}
        \text{R:}\quad \psi_\pm^\mu=\sqrt{\pi\alpha^{\prime}}\sum_{n\in\mathbb{Z}}d_n^\mu e^{-i\pi n\(\tau\pm\sigma\)} \ , && \text{NS:}\quad \psi_\pm^\mu&=\sqrt{\pi\alpha^{\prime}}\sum_{r\in\mathbb{Z}+\frac12}\tb_r^\mu e^{-i\pi r\(\tau\pm\sigma\)} \ .
    \end{align}
\label{eqn:ch3_ss_os_R_NS_exp}
\end{itemize}

\begin{table}[]
    \centering
    \begin{tabular}{crlrl}
        \toprule
         &  \multicolumn{2}{c}{Closed strings} & \multicolumn{2}{c}{Open strings} \\
         \midrule
       \multirow{2}*{Bosonic} &&  \multirow{2}*{$X^\mu\(\tau,\sigma+2\pi\)=X^\mu\(\tau,\sigma\)$}  & NN: & $\partial_\sigma X^\mu\(\tau,0\)=\partial_\sigma X^\mu\(\tau,\pi\)=0$ \\
       &&& DD: & $\delta X^\mu\(\tau,0\)=\delta X^\mu\(\tau,\pi\)=0$ \\
       \midrule
       \multirow{6}*{Fermionic} & R-R: & $\psi_\pm^\mu\(\tau,\sigma+2\pi\)=+\psi_\pm^\mu\(\tau,\sigma\)$ & & \multirow{2}*{$\psi_+^\mu\(\tau,0\)=\psi_-^\mu\(\tau,0\)$} \\
       & NS-NS:& $\psi_\pm^\mu\(\tau,\sigma+2\pi\)=-\psi_\pm^\mu\(\tau,\sigma\)$ &&\\
       & \multirow{2}*{R-NS:} & $\psi_+^\mu\(\tau,\sigma+2\pi\)=+\psi_+^\mu\(\tau,\sigma\)$ & \multirow{2}*{R:} &\multirow{2}*{$\psi_+^\mu\(\tau,\pi\)=+\psi_-^\mu\(\tau,\pi\)$} \\
       && $\psi_-^\mu\(\tau,\sigma+2\pi\)=-\psi_+^\mu\(\tau,\sigma\)$ & &\\
       & \multirow{2}*{NS-R:} & $\psi_+^\mu\(\tau,\sigma+2\pi\)=-\psi_+^\mu\(\tau,\sigma\)$ & \multirow{2}*{NS:} & \multirow{2}*{$\psi_+^\mu\(\tau,\pi\)=-\psi_-^\mu\(\tau,\pi\)$} \\
       && $\psi_-^\mu\(\tau,\sigma+2\pi\)=+\psi_-^\mu\(\tau,\sigma\)$ && \\
       \bottomrule
    \end{tabular}
    \caption{Summary of the various boundary conditions and corresponding sectors of superstrings.}
    \label{tab:ss_sectors}
\end{table}

\noindent All the boundary conditions and the associated sectors of the theory are summarized in Table \ref{tab:ss_sectors}. As we will see, the R-R and NS-NS sectors of the closed string and the R sector of the open strings describe bosonic excitations, while the closed R-NS and NS-R sectors and the open NS sector describe fermionic ones.

We now turn to the quantization, which we perform in the light-cone gauge. As in the bosonic case \eqref{eqn:ch3_lc_gauge_coord}-\eqref{eqn:ch3_lc_q_X+}, one can use the residual worldsheet symmetry to set
\begin{align}
X^+=x^++2\pi\alpha^{\prime}p^+\tau \ , && \psi^+_\pm=0 \ ,
\end{align}

\noindent and the vanishing conditions on the stress-energy tensor and the current of supersymmetry \eqref{eqn:ch3_ss_emt_sc} are solved by
\begin{align}
    \partial_\pm X^-=\frac{1}{2\pi\alpha^{\prime}p^+}\(\partial_\pm X^i\partial_\pm X_i+\frac{i}{2}\psi_\pm^i\partial_\pm\psi_\pm{}_i\) \ , && \psi_\pm^-=\frac{1}{\pi\alpha^{\prime}p^+}\psi^i_\pm\partial_\pm X_i \ .
\end{align}

\noindent This constraint identifies the transverse oscillators as the physical ones, on which one imposes canonical commutation relations to quantize the theory. The fermionic number operators result in
\begin{align}
    \text{R:}\quad N_\psi=\sum_{n>0}nd_{-n}^id_n{}_i \ , && \text{NS:} \quad N_\psi=\sum_{r\ge\frac12}rb_{-r}^ib_r{}_i \ .
\end{align}

\noindent Denoting with $N_X$ the bosonic string number operators found in the previous section in eq.~\eqref{eqn:ch3_M_os_lc_q}-\eqref{eqn:ch3_cs_masses}, the mass operators for the various sectors --- resulting from the proper regularized normal ordering of both bosonic and fermionic oscillators --- are given by
\vspace{0.5cm}
\begin{itemize}
    \item open strings:
    \begin{align}
        \text{R:}\quad M^2=\frac{1}{\alpha^{\prime}}\(N_X+N_\psi\) \ , && \text{NS:}\quad M^2=\frac{1}{\alpha^{\prime}}\(N_X+N_\psi-\frac{D-2}{16}\) \ .
    \label{eqn:ch3_ss_os_masses}
    \end{align} \\

    \item closed strings left-movers:
    \begin{align}
        \text{R:}\quad M^2_L=\frac{4}{\alpha^{\prime}}\(N_X+N_\psi\) \ , && \text{NS:}\quad M^2_L=\frac{4}{\alpha^{\prime}}\(N_X+N_\psi-\frac{D-2}{16}\) \ ,
    \label{eqn:ch3_ss_cs_masses_L}
    \end{align}

    \noindent accompanied by analogous expressions for the right-movers and, most importantly, the level-matching condition $M^2_L=M^2_R$.
\end{itemize}

\vspace{0.5cm}
\noindent These mass formulae allow us to discuss the spectrum of excitations of the quantized superstring. We proceed, again, sector by sector:
\vspace{0.5cm}
\begin{description}
    \item[Open NS sector] the vacuum $\ket{0}_\textup{NS}$ of the NS open strings is a tachyonic scalar, of mass 
    \beq
        M^2\ket{0}_\textup{NS}=-\frac{\(D-2\)}{16\alpha^\prime}\ket{0}_\textup{NS} \ . 
    \label{eqn:ch3_ss_os_NS_tachyon}
    \eeq
    
    \noindent The first excited state is obtained acting on $\ket{0}_\textup{NS}$ with the fermionic string creation operator $b^i_{-\frac12}$, and its mass is equal to
    \beq
        M^2\(b^i_{-\frac12}\ket{0}_\textup{NS}\)=\frac{10-D}{2\alpha^\prime}\(b^i_{-\frac12}\ket{0}_\textup{NS}\) \ .
    \label{eqn:ch3_ss_os_NS_first_state_massless}
    \eeq

    This state transforms in the vectorial representation of the little group $\mathrm{SO}(D-2)$ and contains $\(D-2\)$ degrees of freedom. Lorentz invariance requires therefore this state to be massless, and thus fixes the superstring theory critical dimension
    \beq
        D=10 \ .
    \eeq

    \noindent The excited states are then built acting with $b_{-r}^i$ and $\alpha_{-n}^i$ and represent massive bosonic higher-spin fields.\\

    \item[Open R sector] In the case of the R open sector, the vacuum state is massless, as it is clear from eq.~\eqref{eqn:ch3_ss_os_masses}, but it carries a spinorial representation of the little group, now fixed to $\mathrm{SO}(8)$. This spinorial structure can be understood from the level-zero oscillators $d_0^i$, which satisfy the Clifford algebra $\left\{d_0^i,d_0^j\right\}=2\eta^{ij}$. More specifically, the vacuum $\ket{0}_\textup{R}$ transforms in the Majorana--Weyl $\boldsymbol{8}_s\oplus\boldsymbol{8}_c$ representation of $\mathrm{SO}(8)$, where $s$ and $c$ denote two different chirality. Therefore, the open R sector contains fermionic excitations, which were missing in the bosonic string case. \\

    \item[Closed NS-NS sector] In the closed string case, there are two possible choices for each left and right-mover, so that four different sectors arise. We begin from the NS-NS one. The vacuum $\ket{0,\bar{0}}_\textup{NSNS}$ is again a tachyonic scalar, of mass 
    \beq
        M^2\ket{0,\bar{0}}_\textup{NSNS}=-\frac{\(D-2\)}{4\alpha^\prime}\ket{0,\bar{0}}_\textup{NSNS}\ . 
    \label{eqn:ch3_ss_cs_NSNS_tachyon}
    \eeq
    
    \noindent The first excited state is instead
    \begin{align}
        b_{-\frac12}^i\bar{b}_{-\frac12}^j\ket{0,\bar{0}}_\textup{NSNS} \ , && M^2\(b_{-\frac12}^i\bar{b}_{-\frac12}^j\ket{0,\bar{0}}_\textup{NSNS}\)=\frac{10-D}{2\alpha^\prime}\(b_{-\frac12}^i\bar{b}_{-\frac12}^j\ket{0,\bar{0}}_\textup{NSNS}\) \ .
    \label{eqn:ch3_ss_cs_NSNS_bb_state}
    \end{align}

    \noindent In the critical dimension $D=10$ this state is indeed massless, consistently with Lorentz symmetry, and its degrees of freedom organize in the symmetric traceless plus antisymmetric plus singlet representations of $\mathrm{SO}(8)$: these are the graviton, the Kalb--Ramond field and the dilaton degrees of freedom. Gravity ---  which belonged to the closed-string sector of the bosonic string --- is therefore described by the NS-NS sector of closed superstrings.\\

    \item[Closed R-R sector] The R-R sector also yields bosonic excitations, originating from the tensor composition of the left and right movers vacuum representations of the type $\boldsymbol{8}_s\oplus\boldsymbol{8}_c$, as for the open string case. This takes the form
\beq
    \begin{aligned}
        \ket{0,\bar{0}}_\textup{RR}&\simeq\(\boldsymbol{8}_s\oplus\boldsymbol{8}_c\)\otimes\(\boldsymbol{8}_s\oplus\boldsymbol{8}_c\)=\\
        &=\(\boldsymbol{8}_s\otimes\boldsymbol{8}_s\)\oplus\(\boldsymbol{8}_c\otimes\boldsymbol{8}_c\)\oplus\[\(\boldsymbol{8}_s\otimes\boldsymbol{8}_c\)\oplus\(\boldsymbol{8}_c\otimes\boldsymbol{8}_s\)\]=\\
        &=\(\boldsymbol{1}\oplus\boldsymbol{28}\oplus\boldsymbol{35}_+\)\oplus\(\boldsymbol{1}\oplus\boldsymbol{28}\oplus\boldsymbol{35}_-\)\oplus\[\(\boldsymbol{8}_\textup{v}\oplus\boldsymbol{56}\)\oplus\(\boldsymbol{8}_\textup{v}^\prime\oplus\boldsymbol{56}^\prime\)\] \ .
    \end{aligned}
    \label{eqn:ch3_ss_cs_RR_full_spectrum}
    \eeq
    \noindent Thus, this spectrum comprises, in ten dimensions, two real scalars ($\boldsymbol{1}$), two antisymmetric 2-form fields ($\boldsymbol{28}$) and two 4-form fields, respectively self-dual ($\boldsymbol{35}_+$) and anti self-dual ($\boldsymbol{35}_-$), together with two vectors ($\boldsymbol{8}_\textup{v}$ and $\boldsymbol{8}_\textup{v}^\prime$) and two 3-form fields ($\boldsymbol{56}$ and $\boldsymbol{56}^\prime$). These form fields are known as the Ramond--Ramond forms.\\

    \item[Closed NS-R / R-NS sectors] The mixed NS-R and R-NS sectors of the closed superstring have instead fermionic excitations. For instance, the NS-R tensor product of the massless vacuum representations for the NS-R sector yields
    \begin{equation}
        \begin{aligned}
            \ket{0,\bar{0}}_\textup{NSR}\simeq\boldsymbol{8}_\textup{v}\otimes\(\boldsymbol{8}_s\oplus\boldsymbol{8}_c\)&=\(\boldsymbol{8}_\textup{v}\otimes\boldsymbol{8}_s\)\oplus\(\boldsymbol{8}_\textup{v}\otimes\boldsymbol{8}_c\)\\
            &=\(\boldsymbol{56}_s\oplus\boldsymbol{8}_s\)\oplus\(\boldsymbol{56}_c\oplus\boldsymbol{8}_c\)\  ,
        \end{aligned}
    \end{equation}

    \noindent which correspond to two spin-$\frac32$ gravitinos, of chirality $s$ and $c$, and two spin-$\frac12$ "dilatini", of chirality $c$ and $s$. The R-NS sector yields the same type of spectrum. 
\end{description}

\subsection{The GSO projection and type~II closed superstring theories}

The superstring spectrum described above is indeed very rich. In particular, it contains fermionic excitations, which were missing in the purely bosonic string theory. However, this spectrum still contains tachyons, in the closed NS-NS sector \eqref{eqn:ch3_ss_cs_NSNS_tachyon} and in the open NS sector \eqref{eqn:ch3_ss_os_NS_tachyon}. A revealing feature of this tachyon is that it breaks supersymmetry, since it lacks a superpartner state. From the R-sector point of view, although the vacuum is massless, it lies in a spinorial representation that is too large to be supersymmetric.

According to this perspective, supersymmetry thus exhibits a kind of anomaly, since it is preserved at the level of the classical worldsheet theory but it is broken at the quantum level. Therefore, one is led to try and solve the tachyon problem by consistently imposing supersymmetry on the quantum spectrum. This is done via the Gliozzi--Scherk--Olive (GSO) projection \cite{Gliozzi:1976qd}. This is constructed from the worldsheet fermion number operator $\(-1\)^G$, which for the NS and R sectors is defined as
\begin{align}
    \text{NS:}\quad \(-1\)^{G_\textup{NS}}=\(-1\)^{\sum_{r\ge\frac12}b^i_{-r}b_r{}_i} \ , && \text{R:}\quad \(-1\)^{G_\textup{R}}=\gamma_9\(-1\)^{\sum_{n\ge 1}d^i_{-n}d_n{}_i} \ ,
\end{align}

\noindent such that states built out of an even number of oscillators have positive eigenvalue, whereas states built out of an odd number of oscillators have negative eigenvalue. The $\gamma_9$ matrix appearing in the R-sector operator encodes the chiral nature of its action on the fermionic states built out of the fermionic R-vacuum. Conventionally, one chooses the $s$ chirality to be even and the $c$ to be odd. According to this definition, the NS tachyonic vacuum is such that
\beq
    \(-1\)^{G_\textup{NS}}\ket{0}_\textup{NS}=\ket{0}_\textup{NS} \ ,
\eeq

\noindent so that it can be removed by projecting the NS spectrum with
\beq
    \mathcal{P}_\textup{GSO}^\textup{NS}=\frac{1-\(-1\)^{G_\textup{NS}}}{2} \ .
\label{eqn:ch3_gso_NS}
\eeq

\noindent In the R-sector there is instead no tachyon to remove, and there is additional freedom in choosing the chirality onto which to project:
\beq
    \mathcal{P}_\textup{GSO}^\textup{R}=\frac{1\pm\(-1\)^{G_\textup{R}}}{2} \ .
\label{eqn:ch3_gso_R}
\eeq

Projecting the quantum excitations of the superstring removes the NS tachyons, as desired, and yields a supersymmetric spectrum. Focusing now on the closed superstring case, the NS-NS sector must be projected as of eq.~\eqref{eqn:ch3_gso_NS} in both left and right-movers in order to remove the tachyon \eqref{eqn:ch3_ss_cs_NSNS_tachyon}. Instead, there exist two inequivalent projections of the R-R vacuum, corresponding to opposite or identical sign choices in the R-sector projector of eq.~\eqref{eqn:ch3_gso_R}. The former defines the so-called \textit{type~IIA} superstring theory, the latter the \textit{type~IIB}. Their massless spectrum is the following:\\
\begin{description}
    \item[Common NS-NS sector] as previously mentioned, the two theories share the same NS-NS sector. The massless vacuum is given by the state
    \beq
        b^i_{-\frac12}\bar{b}^j_{-\frac12}\ket{0,\bar{0}}_\textup{NSNS}=b^i_{-\frac12}\ket{0}_\textup{NS}\otimes \bar{b}^j_{-\frac12}\ket{\bar{0}}_\textup{NS} \ ,
    \eeq

    \noindent of eq.~\eqref{eqn:ch3_ss_cs_NSNS_bb_state}, describing the degrees of freedom corresponding to the 10D graviton $G_{\mu\nu}$, Kalb--Ramond field $B_{\mu\nu}$ and dilaton $\phi$.\\

    \item[R-R sectors] The vacua of the R-R sectors are conventionally denoted as 
    \begin{align}
        \text{type~IIA:}\quad \ket{+}_\textup{R}\otimes\ket{-}_\textup{R}\ , &&  \text{type~IIB:}\quad \ket{+}_\textup{R}\otimes\ket{+}_\textup{R} \ ,
    \end{align}
    
    \noindent referring to the choices of chiral GSO projection \eqref{eqn:ch3_gso_R}. The massless spectrum contains the so-called Ramond--Ramond forms. Following eq.~\eqref{eqn:ch3_ss_cs_RR_full_spectrum}, type~IIA contains a 1-form $C^{(1)}_\mu$ and a 3-form $C^{(3)}_{\mu\nu\rho}$, sitting in the representation $\boldsymbol{8}_s\otimes\boldsymbol{8}_c=\boldsymbol{8}_\textup{v}\oplus\boldsymbol{56}$, whereas type~IIB contains a scalar "0-form" $C^{(0)}$, a 2-form $C^{(2)}_{\mu\nu}$ and a self-dual 4-form $C^{(4)}_{\mu\nu\rho\sigma}$, associated with the representation $\boldsymbol{8}_s\otimes\boldsymbol{8}_s=\boldsymbol{1}\oplus\boldsymbol{28}\oplus\boldsymbol{35}_+$.\footnote{Furthermore, the Ramond--Ramond forms can be dualized, adding in this way a $C^{(6)}$ and a $C^{(8)}$ form to the type~IIA spectrum, and a $C^{(5)}$ and a $C^{(7)}$ form to the type~IIB one.} \\


 \item[NS-R and R-NS sectors] The NS-R and R-NS sectors contain the fermionic excitations. The massless vacua of the NS-R sector are given by
 	\beq
	\begin{aligned}
		 \text{type~IIA:} && b^i_{-\frac12}\ket{0}_\textup{NS}\otimes\ket{-}_\textup{R} \in \boldsymbol{56}_s\oplus\boldsymbol{8}_s \ , \\  
		 \text{type~IIB:} &&  b^i_{-\frac12}\ket{0}_\textup{NS}\otimes\ket{+}_\textup{R} \in\boldsymbol{56}_s\oplus\boldsymbol{8}_s \ ,
	\end{aligned}
	\eeq
	
	\noindent while those of the R-NS sector are
	\beq
	\begin{aligned}
		 \text{type~IIA:} &&  \ket{+}_\textup{R}\otimes\bar{b}^i_{-\frac12}\ket{0}_\textup{NS} \in \boldsymbol{56}_c\oplus\boldsymbol{8}_c\ , \\  
		 \text{type~IIB:} && \ket{+}_\textup{R}\otimes\bar{b}^i_{-\frac12}\ket{0}_\textup{NS}\in \boldsymbol{56}_s\oplus\boldsymbol{8}_s \ .
	\end{aligned}
	\eeq

    \noindent Thus, both type~II theories contain two gravitini, $\psi_1$ and $\psi_2$, and two dilatini $\lambda_1$ and $\lambda_2$. In the type~IIA case, the two pairs of fermions have opposite chirality, whereas in type~IIB they all have the same chirality, which is the left-handed one in these conventions. \\
\end{description}

The field theory limit of closed type~II superstrings are the type~IIA and type~IIB ten-dimensional supergravity theories, that enjoy respectively $\N=\(1,1\)$ and $\N=\(2,0\)$ supersymmetry. Note that, despite being chiral, the type~IIB theory is anomaly-free \cite{Alvarez-Gaume:1983ihn,Schellekens:1986yi,Schellekens:1986xh,Lerche:1987sg}. The spectra of the two theories are summarized in Table \ref{tab:type_II_spectra}.

\begin{table}[]
    \centering
    \begin{tabular}{crll}
    \toprule
                               		 	&				&	{Type~IIA} 					&   {Type~IIB}        \\
    \midrule
    \multirow{4}*{Bosons}   		& graviton      		& $G_{\mu\nu}$                				& $G_{\mu\nu}$           \\
                                			& dilaton       		& $\phi$ 							&  $\phi$                \\
		   				& Kalb--Ramond 	& $B_{\mu\nu}$ 					& $B_{\mu\nu}$		\\
						& R-R forms		& $C^{(1)}$, $C^{(3)}$				& $C^{(0)}$, $C^{(2)}$, $C^{(4)}_+$ \\
						&				& $C^{(5)}$, $C^{(7)}$				& $C^{(6)}$, $C^{(8)}$ \\
    \midrule
    \multirow{2}*{Fermions}		& {gravitini} 		&  {$\psi_1{}_R$, $\psi_2{}_L$}    		& {$\psi_1{}_L$, $\psi_2{}_L$} \\
                                			& {dilatini} 		&  {$\lambda_1{}_R$, $\lambda_2{}_L$}   	& {$\lambda_1{}_L$, $\lambda_2{}_L$} \\
    \bottomrule
    \end{tabular}
    \caption{Massless spectrum of type~II closed superstring theories. All the dual R-R forms are included.}
    \label{tab:type_II_spectra}
\end{table}


\subsection{The superstring vacuum amplitudes}

As discussed in Section \ref{ch3_sec_bs_va}, the string theory spectrum and properties are encoded in the corresponding 1-loop vacuum amplitude. In the case of the closed type~II superstring theories, this is the torus amplitude. To compute it, one has to supplement the supertrace formula introduced for the bosonic string with the appropriate GSO projection.

In the bosonic string case, the resulting amplitude was found to be proportional to the Dedekind $\eta$-function, as in eq.~\eqref{eqn:ch3_cs_torus_amp}, which was understood to descend from the modular invariance of the amplitude. This universal symmetry principle, inherent in all string theories, yields, in the superstring case, an additional mathematical structure, the Jacobi $\vartheta$-functions and their combinations in terms of characters of the level-one current algebra of $\mathrm{SO}(8)$ \cite{orientifolds5}. The general Jacobi $\vartheta$-function is defined as
\beq
\begin{aligned}
    \vartheta\sfrac{\alpha}{\beta}\(z \,|\, \tau\)&=\sum_{n}q^{\frac12\(n+\alpha\)^2}e^{2i\pi\(n+\alpha\)\(z+\beta\)}\\
    &=e^{2\pi i\alpha\(z+\beta\)}q^{\frac{\alpha^2}{2}}\prod_{n=1}^\infty\(1-q^n\)\(1+q^{n+\alpha-\frac12}e^{2i\pi\(z+\beta\)}\)\(1+q^{n+\alpha-\frac12}e^{-2i\pi\(z+\beta\)}\) \ .
\end{aligned}
\label{eqn:ch3_jacobi_theta_full}
\eeq

\noindent From this general definition, one picks the specific $\vartheta$-functions given by
\beq
\begin{aligned}
    \vartheta_{1}(z \,|\, \tau)&=\vartheta\sfrac{\frac12}{\frac12}(z \,|\, \tau)= 2 \sin\(\pi z\) q^{\frac18} \prod_{n=1}^{\infty}\left(1-q^{n}\right)\left(1-q^{n} e^{2 \pi i z}\right)\left(1-q^{n} e^{-2 \pi i z}\right), \\
    \vartheta_{2}(z \,|\, \tau)&=\vartheta\sfrac{\frac12}{0}(z \,|\, \tau)= 2 \cos\(\pi z\) q^{\frac18} \prod_{n=1}^{\infty}\left(1-q^{n}\right)\left(1+q^{n} e^{2 \pi i z}\right)\left(1+q^{n} e^{-2 \pi i z}\right), \\
    \vartheta_{3}(z \,|\, \tau)&=\vartheta\sfrac{0}{0}(z \,|\, \tau)=\prod_{n=1}^{\infty}\left(1-q^{n}\right)\left(1+q^{n-\frac12} e^{2 \pi i z}\right)\left(1+q^{n-\frac12} e^{-2 \pi i z}\right), \\
    \vartheta_{4}(z \,|\, \tau)&= \vartheta\sfrac{0}{\frac12}(z \,|\, \tau)=\prod_{n=1}^{\infty}\left(1-q^{n}\right)\left(1-q^{n-\frac12} e^{2 \pi i z}\right)\left(1-q^{n-\frac12} e^{-2 \pi i z}\right) \ ,
\end{aligned}
\label{eqn:ch3_jacobi_theta_1--4}
\eeq

\noindent and, in particular, the case in which these functions are evaluated at $z=0$, which we denote simply with $\vartheta_i\equiv\vartheta_i\(0\,|\, \tau\)$. Despite being identically vanishing, the function $\vartheta_1$ is needed to correctly account for the modular properties of the superstring vacuum amplitudes. The transformation of the general Jacobi $\vartheta$-function \eqref{eqn:ch3_jacobi_theta_full} under $T$ and $S$ modular transformations \eqref{eqn:ch3_mod_gen} is
\beq
\begin{aligned}
T:&& & \vartheta\sfrac{\alpha}{\beta}\(z \,|\, \tau+1\)=\mathrm{e}^{-\mathrm{i} \pi \alpha(\alpha-1)} \vartheta\sfrac{\alpha}{\beta+\alpha-\frac12}\(z \,|\, \tau\) \ , \\
S:&& & \vartheta\sfrac{\alpha}{\beta}\left(\frac{z}{\tau} \left\lvert -\frac{1}{\tau}\right.\right)=(-\mathrm{i} \tau)^{\frac12} \mathrm{e}^{2 \mathrm{i} \pi \alpha \beta+\mathrm{i} \pi z^{2} / \tau} \vartheta\sfrac{\beta}{\alpha}\(z \,|\, \tau\) \ ,
\end{aligned}
\eeq

\noindent so that the functions $\vartheta_i$ transform as
\beq
\begin{aligned}
    T:&& \vartheta_{1,2}&\to e^{\frac{i\pi}{4}}\vartheta_{1,2} \ , & \vartheta_{3,4}&\to\vartheta_{4,3} \ , \\
    S:&& \vartheta_{1,3}&\to \sqrt{-i\tau}\ \vartheta_{1,3} \ ,& \vartheta_{2,4}&\to \sqrt{-i\tau} \ \vartheta_{4,2} \ .
\end{aligned}
\label{eqn:ch3_jacobi_theta_14_TS}
\eeq

The four Jacobi $\vartheta$-functions $\vartheta_i$ define the so-called characters $\chi_{2n}=\left\{O_{2n},V_{2n},S_{2n},C_{2n}\right\}$ of the level-one current algebra of $\mathrm{SO}(2n)$, as
\beq
\begin{aligned}
    O_{2 n}(q) & =\frac{\vartheta_{3}^{n}+\vartheta_{4}^{n}}{2 \eta(q)^{n}} \ , &&&    V_{2 n}(q) & =\frac{\vartheta_{3}^{n}-\vartheta_{4}^{n}}{2 \eta(q)^{n}} \ , \\
    S_{2 n}(q) & =\frac{\vartheta_{2}^{n}+i^{-n} \vartheta_{1}^{n}}{2 \eta(q)^{n}} \ , &&&    C_{2 n}(q) & =\frac{\vartheta_{2}^{n}-i^{-n} \vartheta_{1}^{n}}{2 \eta(q)^{n}} \ .
\end{aligned}
\label{eqn:ch3_ss_characters_gen}
\eeq

\noindent The $T$ and $S$ modular transformations \eqref{eqn:ch3_mod_gen}-\eqref{eqn:ch3_jacobi_theta_14_TS} act on the character vector $\chi_{2n}$ as
\begin{align}
    T=e^{-\frac{in\pi}{12}}\begin{pmatrix}
        1 & 0 & 0 & 0 \\
        0 & -1 & 0 & 0 \\
        0 & 0 & e^{\frac{in\pi}{4}}& 0 \\
        0 & 0 & 0 & e^{\frac{in\pi}{4}}
    \end{pmatrix} \ , && 
    S=\frac12\begin{pmatrix}
        1 & 1 & 1 & 1 \\
        1 & 1 & -1 & -1 \\
        1 & -1 & i^{-n} &  -i^{-n}\\
        1 & -1 &-i^{-n}& i^{-n} 
    \end{pmatrix} \ .
\label{eqn:ch3_character_mod_tr}
\end{align}

\noindent At first order in $q$, these characters have the following expansion:
\beq
\begin{aligned}
    O_{2n}&\sim q^{-\frac{n}{24}} \ , &&& V_{2n}&\sim 2n \ q^{\frac12-\frac{n}{24}} \ , \\
    S_{2n}&\sim 2^{n-1}q^{\frac{n}{12}}\ , &&& C_{2n}&\sim 2^{n-1}q^{\frac{n}{12}}\ .
\label{eqn:ch3_characters_expansion}
\end{aligned}
\eeq

\noindent Once combined with the appropriate bosonic contributions, these characters allow one to read the spectrum of a given superstring theory off the associated partition function. As an example, the ten-dimensional combinations
\beq
\begin{aligned}
    \frac{O_8}{\eta^8}\ , && \frac{V_8}{\eta^8}\ , && \frac{S_8}{\eta^8}\ , && \frac{C_8}{\eta^8}\ ,
\end{aligned}
\label{eqn:ch3_characters_comb}
\eeq

\noindent describe respectively, at the lowest mass level, a tachyonic scalar, a massless vector and two massless Majorana fermions of opposite chirality, as one can verify by combining the expansions in eq.~\eqref{eqn:ch3_characters_expansion} and \eqref{eqn:ch3_bs_cylinder_state_counting}. Moreover, the affine characters satisfy the following set of decomposition rules:
\begin{equation}
\begin{aligned}
O_{2n} &= O_{p-1} O_{2n+1-p} + V_{p-1} V_{2n+1-p} \ , &&&  V_{2n} &= O_{p-1} V_{2n+1-p} + V_{p-1} O_{2n+1-p} \ ,\\
S_{2n} &= S_{p-1} S_{2n+1-p} + C_{p-1} C_{2n+1-p} \ , &&&  C_{2n} &= S_{p-1} C_{2n+1-p} + C_{p-1} S_{2n+1-p} \ ,
\end{aligned}
\label{eqn:ch3_characters_dec_rule}
\end{equation}

\noindent following $\mathrm{SO}(2n)\to\mathrm{SO}(p-1)\times\mathrm{SO}(2n+1-p)$.

We now have all the ingredients needed to compute the torus amplitude for the type~II superstring theories. In order to do so, one must properly include the GSO projections \eqref{eqn:ch3_gso_NS}-\eqref{eqn:ch3_gso_R} in the amplitude supertrace. Following the computation outlined explicitly in Section \ref{ch3_sec_bs_va}, one finds that the possible projections of the NS and R sectors yield precisely the $\mathrm{SO}(8)$ characters
\beq
\begin{aligned}
    \mathrm{Str}_\textup{NS}\[\frac{1+\(-1\)^{G_\textup{NS}}}{2}q^{N_X+N_\psi-\frac12}\]&= O_8(q) \ , &  \mathrm{Str}_\textup{R}\[\frac{1+\(-1\)^{G_\textup{R}}}{2}q^{N_X+N_\psi}\]&= S_8(q) \ ,  \\
    \mathrm{Str}_\textup{NS}\[\frac{1-\(-1\)^{G_\textup{NS}}}{2}q^{N_X+N_\psi-\frac12}\]&= V_8(q) \ , & \mathrm{Str}_\textup{R}\[\frac{1-\(-1\)^{G_\textup{R}}}{2}q^{N_X+N_\psi}\]&= C_8(q) \ .
\end{aligned}
\label{eqn:ch3_ss_characters_SO(8)_traces}
\eeq


\noindent Hence, the torus amplitudes for the type~II closed superstring theories are found to be
\begin{align}
    \T_\textup{IIA}&=\frac{1}{\(4\pi\alpha^\prime\)^5}\int_{\mathscr{F}}\frac{d^2\tau}{\tau_2^6}\frac{1}{\eta(q)^8\bar{\eta}(\bar q)^8}\(V_8-S_8\)\(\bar{V}_8-\bar{C}_8\) \ , \label{eqn:ch3_typeIIA_torus}\\
    \T_\textup{IIB}&=\frac{1}{\(4\pi\alpha^\prime\)^5}\int_{\mathscr{F}}\frac{d^2\tau}{\tau_2^6}\frac{1}{\eta(q)^8\bar{\eta}(\bar q)^8}\(V_8-S_8\)\(\bar{V}_8-\bar{S}_8\) \ . \label{eqn:ch3_typeIIB_torus}
\end{align}

\noindent Note that these two amplitudes do not involve the character $O_8$, which correspond to the singlet conjugacy class of $\mathrm{SO}(8)$ and carries a tachyonic scalar, but involve instead the vectorial class $V_8$ and the two spinorial ones $S_8$ and $C_8$, whose lower mass states are massless vector and fermions, consistently (see eq. \eqref{eqn:ch3_characters_expansion}-\eqref{eqn:ch3_characters_comb}). The $\mathrm{SO}(8)$ case is somewhat special, because the relative Jacobi $\vartheta$-functions entering the characters in eq.~\eqref{eqn:ch3_ss_characters_SO(8)_traces} satisfy the relation
\beq
    \vartheta_3^4-\vartheta_4^4-\vartheta_2^4=0 \ . 
\label{eqn:ch3_id_jacobi_theta}
\eeq   

\noindent Together with the fact that $\vartheta_1$ is identically vanishing, this identity imply that $V_8-S_8=~0$, so that the torus amplitudes of eq.~\eqref{eqn:ch3_typeIIA_torus}-\eqref{eqn:ch3_typeIIB_torus} are formally equal to zero. This peculiar property reflects nothing but the fact that the spectrum of the type~II theories is supersymmetric, so that the contributions to the vacuum energy of the bosonic and fermionic states cancel each other.

Moreover, the amplitudes of eq.~\eqref{eqn:ch3_typeIIA_torus}-\eqref{eqn:ch3_typeIIB_torus} are also modular invariant, as one can check by performing the $T$ and $S$ transformations of eq.~\eqref{eqn:ch3_character_mod_tr}, consistently with the symmetry properties of the worldsheet. Actually, modular invariance can be used to completely characterize the spectrum of allowed torus amplitudes, writing them in the form\footnote{We now employ a standard, simplified notation for the amplitudes, in which one leaves implicit the modular integration and the bosonic state contributions.}
\beq
    \T=\bar{\chi}^i\mathcal{N}_{ij}\chi^j \ ,
\eeq

\noindent where $\chi^i$ is the vector of characters of $\mathrm{SO}(8)$, and the $\N$ enforces the GSO projection, while being consistent with modular invariance. Then, it can be shown that the allowed closed superstring amplitudes are the two type~IIA and type~IIB amplitudes of eq.~\eqref{eqn:ch3_typeIIA_torus}-\eqref{eqn:ch3_typeIIB_torus}, together with
\beq
\begin{aligned}
    \T_{0\text{A}}&=\abs{V_8}^2+\abs{O_8}^2 +S_8\bar{C}_8+\bar{S}_8C_8\ , \\
    \T_{0\text{B}}&=\abs{V_8}^2+\abs{O_8}^2+\abs{S_8}^2+\abs{C_8}^2 \ .
\end{aligned}
\eeq

\noindent These two amplitudes belong to what are known as type~0 string theories \cite{Dixon:1986iz,Seiberg:1986by}. The spectrum of these models only contains spacetime bosonic states and includes a tachyon, signaled by the presence of the character $O_8$.

\subsection{Open superstrings and D-branes}

In the previous section, we discussed how to obtain consistent closed superstring theories by means of the GSO projection \eqref{eqn:ch3_gso_NS}-\eqref{eqn:ch3_gso_R} \cite{Gliozzi:1976qd}. These are the type~IIA and type~IIB string theories, whose spectra are tachyon-free and supersymmetric --- as summarized in Table \ref{tab:type_II_spectra} --- and their torus partition functions are given in eq.~\eqref{eqn:ch3_typeIIA_torus}-\eqref{eqn:ch3_typeIIB_torus}. 

In this section, we include open superstrings in the picture. In order to do so, one has to properly account for the boundary conditions of both bosonic and fermionic strings (see Table \ref{tab:ss_sectors}). The fermionic coordinates split the open theory into the R and NS sectors, according to eq.~\eqref{eqn:ch3_ss_os_R_NS_exp}. The bosonic ones, as discussed in Section \ref{ch3_sec_BS}, are defined either with NN or DD boundary conditions, as in eq.~\eqref{eqn:ch3_open_bc}. The NN boundary conditions are invariant under Poincar\'e symmetry and are compatible with a motion of the open string end-points in target space. On the contrary, DD boundary conditions completely fix the end-points of the string. Thus, imposing NN boundary conditions on $\(p+1\)$ coordinates and DD boundary conditions on the remaining $\(9-p\)$, ones breaks the Lorentz group according to $\mathrm{SO}(1,9)\to\mathrm{SO}(1,p)\times\mathrm{SO}(9-p)$. These open strings propagate in the $\mathrm{SO}(1,p)$ space, while their position is fixed with respect to the transverse $(9-p)$ coordinates. This $(p+1)$-dimensional hypersurface is called \textit{D$p$-brane} \cite{Dai:1989ua,Leigh:1989jq,orientifolds4,Polchinski:1995mt,Hull:1994ys,Townsend:1995kk,Witten:1995ex,Duff:1994an,Polchinski:1996na,Bachas:1998rg}. D$p$-branes are actually soliton solutions of type~II supergravities \cite{Hull:1994ys,Townsend:1995kk,Witten:1995ex} and are therefore nonperturbative, dynamical objects intrinsic to string theory, carrying tension and charges under the Ramond--Ramond forms \cite{Polchinski:1995mt}.

The splitting $\mathrm{SO}(1,9)\to\mathrm{SO}(1,p)\times\mathrm{SO}(9-p)$ introduced by a D$p$-brane is reflected on the associated Hilbert space. Calling $\mu=\(a,i\)$, with $a=0,\dots,p$ being the coordinates on the D$p$-brane and $i=p+1,\dots,9$ the coordinates transverse to it, one has that the NS sector contains the tachyonic vacuum $\ket{0}_\textup{NS}$ of eq.~\eqref{eqn:ch3_ss_os_NS_tachyon} and the massless states, following \eqref{eqn:ch3_ss_os_NS_first_state_massless}, are now a $(p+1)$-dimensional vector $b_{-\frac12}^a\ket{0}_\textup{NS}$ and $(9-p)$ scalars $b_{-\frac12}^i\ket{0}_\textup{NS}$, associated with the transverse, fixed positions of the D$p$-brane. The R-sector vacuum $\ket{0}_\textup{R}$ is still massless and fermionic, and the representations $\boldsymbol{8}_s$ and $\boldsymbol{8}_c$ also decompose accordingly. Note that the field excitations propagate along the D$p$-brane hypersurface.

As discussed in Section \ref{ch3_sec_bs_va}, the 1-loop partition function of open (oriented) strings is the cylinder amplitude. In order to understand how to properly construct this amplitude along with the closed type~II superstring ones \eqref{eqn:ch3_typeIIA_torus}-\eqref{eqn:ch3_typeIIB_torus}, we recall that the open string cylinder amplitude admits two types of descriptions. In the first one, the modular time parameter flows "vertically", for which the open string propagates while sweeping the surface of the cylinder. This is described by the direct/loop-channel amplitude, which for the bosonic string was given in eq.~\eqref{eqn:ch3_os_cylinder_loop_ch}-\eqref{eqn:ch3_os_cylinder_loop_ch}. In the second case --- formally obtained performing an $S$ modular transformation --- the modular time is "horizontal", as the open strings sweep the cylinder between its two boundaries. The associated amplitude is the so-called transverse channel, given in eq.~\eqref{eqn:ch3_os_cylinder_trans_ch} for the bosonic string, and it can be interpreted as describing the tree-level propagation of closed strings \cite{Polchinski:1987tu,orientifolds1,Bianchi:1988fr,orientifolds3}. This feature is of crucial importance because it shows that open strings always require an associated closed-string sector in order to be well-defined. This essential property is also the guideline to follow in order to properly construct the cylinder amplitudes for open superstrings and understand their structure. Focusing on the case of type~II closed superstrings, the associated open-string cylinder amplitude has then to be constructed with the same type of GSO projections \eqref{eqn:ch3_gso_NS}-\eqref{eqn:ch3_gso_R}, and the open-string states propagated in the direct channel have to be compatible with the closed string ones of the transverse channel. In the case of a D$p$-brane system --- in a ten-dimensional non-compact space ---  these requirements lead to the following direct channel cylinder amplitude:
\beq
    \A_{pp}=\frac{1}{\(8\pi\alpha^\prime\)^{\frac{p+1}{2}}}\int_0^\infty\frac{d\tau_2}{\tau_2^{\frac{p+3}{2}}}\frac{V_8-S_8}{\eta^8}\(\textstyle{\frac{i\tau_2}{2}}\) \ .
\label{eqn:ch3_Dp_brane_cylinder_direct_ch}
\eeq

\noindent An $S$ modular transformation \eqref{eqn:ch3_character_mod_tr} yields the transverse channel amplitude
\beq
    \tilde{\A}_{pp}=\frac{2^{-\frac{p+1}{2}}}{\(8\pi\alpha^\prime\)^{\frac{p+1}{2}}}\int_0^\infty\frac{d\ell}{\ell^{\frac{9-p}{2}}}\frac{V_8-S_8}{\eta^8}\(i\ell\) \ ,
\label{eqn:ch3_Dp_brane_cylinder_trans_ch}
\eeq

\noindent where $\ell=\frac{2}{\tau_2}$. The characters entering this transverse channel amplitude are interpreted, consistently with the discussion above, as the NS-NS and R-R states of the closed string, that propagates at tree-level between the two boundaries of the cylinder, which are identified with D$p$-branes. They also explicitly reveal the physical nature of the D$p$-brane, carrying tension (from the $V_8$ term) and charge under the Ramond--Ramond forms (from the $S_8$ term). Following this picture, one concludes that type~IIA string theory contains the even D$0$,~\dots,~D$8$ branes, coupling to the odd Ramond--Ramond potentials, whereas type~IIB string theory contains the odd D$(-1)$, \dots, D$9$-branes.

A more precise definition of the cylinder amplitudes for a generic D$p$-brane actually involves the decomposition of the characters of eq.~\eqref{eqn:ch3_characters_dec_rule} along the $(p+1)$-dimensional worldvolume of the brane. The direct channel amplitude takes then the form, in compact notation,
\beq
    \A_{pp}=
O_{p-1} V_{9-p} + V_{p-1} O_{9-p} -S_{p-1} S_{9-p} - C_{p-1} C_{9-p} \ .
\eeq

\subsubsection{Stack of D-branes and gauge groups}

Therefore, open strings must live on D-branes, and there are multiple configurations in which this can be realized. The simplest such configuration is the case in which the two endpoints of the open string live on the same brane. In this case, the massless spectrum contains one abelian gauge field, living on the brane worldvolume. The two endpoints of the open string can also live on two distinct D-branes, separated by a distance. In this case, there are two massless and two massive abelian gauge fields, the latter being charged under the former and with mass proportional to the distance between the two branes. 

This picture is useful to understand when multiple D-branes coincide, forming a stack. In this limit, also the gauge fields that were initially massive become massless and the gauge symmetry is enhanced to a non-abelian group \cite{orientifolds7,Witten:1995im}, from $\mathrm{U}(1)^2\to \mathrm{U}(2)$ in the example of the pair of branes under consideration. In the case of a stack of $N$ such branes, the gauge symmetry is $\mathrm{U}(N)$. The cylinder amplitude for a stack of $N$ branes is modified with respect to eq.~\eqref{eqn:ch3_Dp_brane_cylinder_direct_ch} as 
\beq
    \A_{pp}=\frac{N\bar{N}}{\(8\pi\alpha^\prime\)^{\frac{p+1}{2}}}\int_0^\infty\frac{d\tau_2}{\tau_2^{\frac{p+3}{2}}}\frac{V_8-S_8}{\eta^8}\(\textstyle{\frac{i\tau_2}{2}}\) \ .
\label{eqn:ch3_Dp_brane_cylinder_direct_ch_N_factor}
\eeq

\noindent This mechanism is the brane version of the Chan--Paton non-dynamical degrees of freedom that one can define at the endpoints of the string \cite{Paton:1969je,Marcus:1986cm}.

\subsection{The orientifold projection and the type~I string}\label{ch3_sec_type_I}

We conclude this first section by presenting the fundamental orientifold projection \cite{orientifolds1,orientifolds2,orientifolds3,orientifolds4,orientifolds5,orientifolds6,orientifolds7,Polchinski:1987tu}, which leads to type~I string theory and to a rich variety of lower-dimensional string vacua. The orientifold is a projection of the type~IIB string onto states of definite worldsheet parity, denoted $\Omega$. For closed strings, this symmetry acts as
\beq
    \Omega_\textup{closed}:\quad\sigma\to -\sigma \ .
\label{eqn:ch3_ws_parity_cs}
\eeq

\noindent and maps left-movers into right-movers and vice versa:
\begin{align}
    \alpha_m^\mu\leftrightarrow\balpha_m^\mu \ , && b^\mu_r\leftrightarrow \bar{b}^\mu_r \ , && d^\mu_n\leftrightarrow \bar{d}^\mu_m \ .
\end{align} 
For open strings, the worldsheet parity is
\beq
    \Omega_\textup{open}:\quad\sigma\to \pi-\sigma \ ,
\label{eqn:ch3_ws_parity_os}
\eeq

\noindent which exchanges the two open string end-points. Its action on the oscillators is
\begin{align}
    \alpha^\mu_m\to\(-1\)^m\alpha^\mu_m \ , && b^\mu_r\to\(-1\)^{r-\frac12}b^\mu_r \ , && d^\mu_n\to\(-1\)^n d^\mu_n \ .
\label{eqn:ch3_omega_os_osc}
\end{align}

\noindent In both cases, the outcome of this procedure is that the resulting strings are \textit{unoriented}, with their spectra projected onto a subset of the original spectra that is invariant under this worldsheet parity \eqref{eqn:ch3_ws_parity_cs}-\eqref{eqn:ch3_ws_parity_os}. Moreover, this projection introduces an additional non-dynamical object in spacetime, the \textit{orientifold plane} \cite{Polchinski:1987tu,orientifolds1,Bianchi:1988fr,orientifolds2,orientifolds3}.

The effective implementation of the orientifold projection is done at the level of the string partition function, analogously to the GSO projection of eq.~\eqref{eqn:ch3_ss_characters_SO(8)_traces}. Starting from the closed sector, the trace in the partition function takes the form
\beq
    \mathrm{Str}\[\frac{1+\Omega}{2}\,\mathcal{P}_\textup{GSO}\,q^{N_X+N_\psi-\Delta}\bar{q}^{\bar{N}_X+\bar{N}_\psi-\Delta}\] \ ,
\label{eqn:ch3_orientifold_trace_cs}
\eeq

\noindent where $\mathcal{P}_\textup{GSO}$ denotes the appropriate GSO projector for both the NS-NS and R-R sectors, while $\Delta$ is the mass shift, equal to $-\frac12$ in the NS-NS sector and $0$ in the R-R one. The first term in the trace \eqref{eqn:ch3_orientifold_trace_cs} is the torus contribution of eq.~\eqref{eqn:ch3_typeIIB_torus}. The second one is the new contribution produced by the orientifold projection, which is the Klein bottle amplitude
\beq
    \K=\frac{1}{\(4\pi\alpha^\prime\)^5}\int_0^\infty\frac{d\tau_2}{\tau_2^6}\,\,\frac{V_8-S_8}{\eta^8}\(2i\tau_2\) \ .
\label{eqn:c3_IIB_Klein_loop_ch}
\eeq

\noindent This amplitude receives contributions from the NS-NS and R-R sectors and symmetrizes the torus one \eqref{eqn:ch3_typeIIB_torus}, projecting away the Kalb--Ramond field $B_{\mu\nu}$, the $C^{(0)}$ and $C^{(4)}_+$ Ramond--Ramond forms, one gravitino and one dilatino. Therefore, the projection breaks half of the original supersymmetry and yields $\N=(1,0)$ supergravity in ten dimensions. The bosonic sector is given by the graviton, the dilaton and the R-R 2-form $C^{(2)}$, whereas the fermionic sector is given by one gravitino and one dilatino. This is understood and confirmed through the expansion of the closed-string partition function integrand:
\beq
\begin{aligned}
    \frac{1}{2}\T+\K&\sim \frac12\left\lvert{\frac{V_8-S_8}{\eta^8}}\right\rvert^2+\frac{V_8-S_8}{2\eta^8}=\\
    &=\frac{\abs{V_8}^2+\abs{S_8}^2+V_8-S_8}{2\eta^8}-\frac{V_8\bar{S}_8+\bar{V}_8S_8}{2\eta^8}\sim\(35+1+28\)-\(56+8\)\ , 
\end{aligned}
\eeq

\noindent where we recognize the degrees of freedom of the supersymmetric spectrum described above and their origin.

The Klein also bottle admits a double interpretation in terms of loop and transverse channels. The latter is obtained by the former in eq.~\eqref{eqn:c3_IIB_Klein_loop_ch} by an $S$ transformation and reads
\beq
    \tilde{\K}=\frac{2^5}{\(4\pi\alpha^\prime\)^5}\int_0^\infty d\ell\,\,\frac{V_8-S_8}{\eta^8}\(i\ell\) \ ,
\label{eqn:c3_IIB_Klein_trans_ch}
\eeq

\noindent with $\ell=\frac{1}{2\tau_2}$. Similarly for the cylinder amplitude of the open strings, while the direct channel describes the propagation of closed strings at 1-loop, the transverse channel describes instead the propagation of closed strings at tree-level, between the two crosscaps defining the Klein bottle. Following the analogy with the D-branes, the two crosscaps can be associated with new spacetime, non-dynamical objects, known as orientifold planes, corresponding to the hypersurfaces invariant with respect to the orientifold involution. From the transverse channel amplitude \eqref{eqn:c3_IIB_Klein_trans_ch} one sees that the O-planes carry tension and charge under the Ramond--Ramond forms to which they can couple (see Table \ref{tab:O-planes}). For the worldsheet parity $\Omega$ of eq.~\eqref{eqn:ch3_ws_parity_cs} at hand, the resulting orientifold planes are O9-plane, occupying the entirety of spacetime. Then, the Klein bottle amplitude in the transverse channel is interpreted as the tree-level propagation of closed strings between the orientifold planes.

\begin{table}[]
    \centering
    \begin{tabular}{ccc}
    \toprule
       O-plane  & Tension & R-R charge  \\
    \midrule
        O$p_-$  & $-$ & $-$ \\
        O$p_+$ & $+$ & $+$ \\
        $\overline{\text{O}p_-}$  & $-$ & $+$\\
        $\overline{\text{O}p_+}$ & $+$ & $-$ \\
    \bottomrule
    \end{tabular}
    \caption{Nomenclature and properties of O-planes. The $\pm$ signs refer to the tension and R-R charge of the O$p$-planes. They are given in relation to the charge of the associated D$p$-brane, according to $T_{\text{O}{p_\pm}}=\pm2^{p-5}Q_{\text{D}_p}$, $T_{\overline{\text{O}{p_\pm}}}=\pm2^{p-5}Q_{\text{D}_p}$, $Q_{\text{O}p_\pm}=\pm2^{p-5}Q_{\text{D}_p}$ and $Q_{\overline{\text{O}p_\pm}}=\mp2^{p-5}Q_{\text{D}_p}$.}
    \label{tab:O-planes}
\end{table}

Let us now move to the open-string sector. The relevant trace building the partition function is 
\beq
    \mathrm{Str}\[\frac{1+\Omega}{2}\,\mathcal{P}_\textup{GSO}\,q^{N_X+N_\psi+\Delta}\] \ .
\label{eqn:ch3_orientifold_trace_os}
\eeq

\noindent The first contribution is the cylinder amplitude we give in eq.~\eqref{eqn:ch3_Dp_brane_cylinder_direct_ch}-\eqref{eqn:ch3_Dp_brane_cylinder_direct_ch_N_factor}. Considering a stack of $N$ D9-branes, this amplitude is equal to
\beq
\begin{aligned}
    \A&=\frac{N^2}{\(8\pi\alpha^\prime\)^5}\int_0^\infty \frac{d\tau_2}{\tau_2^6}\,\,\frac{V_8-S_8}{\eta^8}\(\textstyle{\frac{i\tau_2}{2}}\) \ , \\
    \tilde{\A}&=\frac{2^{-5}N^2}{\(8\pi\alpha^\prime\)^5}\int_0^\infty d\ell\,\,\frac{V_8-S_8}{\eta^8}\(i\ell\) \ , 
\end{aligned}
\eeq

\noindent in which the Chan--Paton factor $N$ is real in order to be consistent with the orientifold projection. The $\Omega$-term in the trace of eq.~\eqref{eqn:ch3_orientifold_trace_os} yields the fourth possible $\chi=0$ amplitude, the M\"obius strip, which is a non-oriented Riemann surface with one boundary and one crosscap. In order to properly write the M\"obius strip amplitude $\M$, one has to take into account an additional subtlety. Contrary to the Klein bottle and the cylinder, the modular parameter of the M\"obius strip has a non-vanishing constant real part $\tau_{\M}=\frac{1}{2}+\frac{i\tau_2}{2}$, produced by the action of the worldsheet parity on the oscillators as in eq.~\eqref{eqn:ch3_omega_os_osc}, which makes its partition function integral not obviously real. In order to properly fix this complex phase, one must introduce the real hatted characters $\hat{\chi}_i$, which are related to the standard ones as
\beq
    \hat{\chi}_i(q)=e^{-i\pi\(h_i-\frac{c}{24}\)}\chi_i(q) \ ,
\eeq

\noindent where $h_i$ is the conformal weight of the character and $c$ the central charge of the associated conformal algebra \cite{orientifolds1}. In addition, the vacuum state of the brane system has a sign ambiguity $\epsilon=\pm1$ under the parity $\Omega$ (which will be fixed momentarily). Taking all this into account, the M\"obius amplitude $\M$ of the $N$ D9-branes stack takes the form
\beq
    \M=\frac{\epsilon N}{\(8\pi\alpha^\prime\)^5}\int_0^\infty \frac{d\tau_2}{\tau_2^6}\,\,\frac{\hat{V}_8-\hat{S}_8}{\hat{\eta}^8}\(\textstyle{\frac12+\frac{i\tau_2}{2}}\) \ .
\label{eqn:ch3_Mobius_loop_ch}
\eeq

\noindent Summing the two open string cylinder and M\"obius amplitudes, the expansion yields the massless degrees of freedom counting
\beq
    \A+\M\sim \frac{N\(N+\epsilon\)}{2}\(8-8\) \ .
\eeq

\noindent Thus, in addition to the supersymmetric condition on the bosonic and fermionic degrees of freedom, we also see that the orientifold projection modifies the gauge group associated with the brane system. While the original gauge group was $\mathrm{U}(N)$, now we see that the gauge group becomes $\mathrm{USp}(N)$ for $\epsilon=1$ and $\mathrm{SO}(N)$ for $\epsilon=-1$.

The M\"obius strip in eq.~\eqref{eqn:ch3_Mobius_loop_ch} is the loop channel amplitude, in which the open strings propagate at 1-loop exchanging their endpoints. In order to move to the transverse channel amplitude consistently, one has to perform the transformation
\begin{align}
    P=TST^2S: \quad \frac12+\frac{i\tau_2}{2}\overset{P}{\longrightarrow}\frac{1}{2}+\frac{i}{2\tau_2}\equiv \frac12+i\ell \ .
\end{align}

\noindent which yields
\beq
    \tilde{\M}=\frac{2\epsilon N}{\(8\pi\alpha^\prime\)^5}\int_0^\infty d\ell\,\,\frac{\hat{V}_8-\hat{S}_8}{\hat{\eta}^8}\(\textstyle{\frac12+i\ell}\) \ .
\label{eqn:ch3_Mobius_trans_ch}
\eeq

\noindent In this transverse channel picture, closed strings are propagated at tree-level between the boundary and the crosscap of the M\"obius strip. This is interpreted as the interaction term between the D9-branes and the O9-planes.

The O-planes and D-branes systems in orientifold constructions are actually tied together by a further consistency requirement, known as the \textit{tadpole cancellation condition} \cite{Polchinski:1987tu,Polchinski:1995mt}. In the example under consideration, the sole closed-string sector does not define a consistent theory. First, while the chiral spectrum of the original type~IIB theory (see Table \ref{tab:type_II_spectra}) is anomaly-free, the reduced chiral spectrum of the orientifold theory is, by construction, no longer so. Second, the O$p$-planes are charged under the $(p+1)$ Ramond--Ramond forms, so that the O9 of this theory couples to a $C^{(10)}$ form. However, this form is, by construction, non-dynamical, so that the associated Gauss law imposes the vanishing of the net $C^{(10)}$ charge, which is not the case for the O9-planes at hand. These issues can be resolved by properly introducing the open sector of the theory. In fact, also the D9-branes couple to the $C^{(10)}$ form, which, when combined with the O9-plane, can yield a neutral configuration. The total charge can be read from the transverse channel amplitudes $\tilde \K$, $\tilde \A$ and $\tilde \M$ --- more specifically, from their R-R sectors ---, respectively in eq.~\eqref{eqn:c3_IIB_Klein_trans_ch}, \eqref{eqn:ch3_Dp_brane_cylinder_trans_ch} and \eqref{eqn:ch3_Mobius_trans_ch}. They give the total charge
\beq
    \tilde{\K}+\tilde{\A}+\tilde{\M}\sim 2^5+2^{-5}N^2+2\epsilon N=2^{-5}\(N+2^5\epsilon\)^2 \ ,
\label{eqn:ch3_orientifold_total_trans}
\eeq

\noindent which vanish for $\epsilon=-1$ and $N=32$. This condition fixes uniquely the orientifold theory to contain 32 background D9-branes, carrying the gauge group $\mathrm{SO}(32)$. This gauge group defines indeed an anomaly-free theory, by means of the Green--Schwarz mechanism \cite{Green:1984sg}. From the point of view of the NS-NS sector, eq.~\eqref{eqn:ch3_orientifold_total_trans} also gives the total tension of the O9-plane--D9-branes system. In the field theory limit, this is related to what is known as the dilaton tadpole. The NS-NS tension and the R-R charge are equal to each other because of supersymmetry, so that the former also vanishes in the $\mathrm{SO}(32)$ theory. Assuming the tension and charge of the D9-branes to be positive, the ones of the O9-planes must both be negative, thus being identified with O$9_-$ planes (see Table \ref{tab:O-planes}).

The consistent type~IIB orientifold theory of gauge group $\mathrm{SO}(32)$ is known as the type~I string theory \cite{orientifolds1} and it is the only consistent supersymmetric theory of unoriented strings in 10D. This theory contains 32 D9-branes and one O9$_-$--plane. Its low-energy field theory limit is $\N=(1,0)$ supergravity coupled to a $\mathrm{SO}(32)$-vector multiplet.


\section{Orientifold compactifications}\label{ch3_sec_ss_dp_orientifolds}

Orientifold projections open the door to an abundant set of string vacua, with different amounts of supersymmetries and compact dimensions. Combining the basic orientifold involution $\Omega$ with target space orbifolds \cite{orientifolds2,orientifolds3,orientifolds4,orientifolds5,orientifolds6,orientifolds7,Harvey:1986bf,Ishibashi:1988tf} allows one to construct a variety of lower-dimensional descendants of the type~I string, all the way down to four dimensions \cite{Angelantonj:1996uy,Kakushadze:1997ku,Kakushadze:1997uj,Kakushadze:1998eg,Kakushadze:1998cd,Zwart:1997aj,Klein:2000qw,Blumenhagen:1999md,Cvetic:1999hb,Blumenhagen:1999ev,Pradisi:1999ii,Cvetic:2000aq,Cvetic:2000st}. The richness of such a landscape of vacua stems also from several seminal constructions uncovered thanks to the development and understanding of orientifold theories and their D-brane spectra. These include the generalized Green--Schwarz mechanism for anomaly cancellation in less than ten dimensions\cite{gss}, and novel ways of breaking supersymmetry in string theory: by compactification à la Scherk--Schwarz \cite{ss-original_1,ss-original_2,ss-original_3,blum-dienes_1,blum-dienes_2,ads1,ss_open-lower_1,ss_open-lower_2,ss_open-lower_3,ss_open-lower_4,ss_open-lower_5,aads1} --- generalizing the type~II constructions \cite{ss_closed1,ss_closed2,ss_closed3,ss_closed4,ss_closed5,ss_closed6} ---, by means of internal background magnetic fields \cite{Fradkin:1985qd,Abouelsaood:1986gd,Bachas:1995ik,Bianchi:1997gt}, and via the brane supersymmetry breaking mechanism \cite{bsb1,bsb2,bsb3,bsb4,bsb5,bsb6,bsbnl_1,bsbnl_2,bsb_rev}.

The construction of a novel nine-dimensional orientifold projection is the subject of the work \cite{Bossard:2024mls}. It is a nine-dimensional orientifold of the type~IIB string in its Scherk--Schwarz deformation, which breaks all supersymmetries. The resulting theory has no background open sector and the orientifold planes couple only to the massive twisted states, and are therefore dubbed \textit{twisted orientifold planes}. We also argue that this theory is invariant under S-duality, based on its conjectured realization as a freely-acting orbifold in F-theory.

Before its detailed presentation, we first discuss some previous constructions, which share important similarities with it. We begin by reviewing the three known supersymmetry breaking Scherk--Schwarz deformations of type~IIB string theory in nine dimensions \cite{ss_closed1,ss_closed2,ss_closed3,ss_closed4,ss_closed5,ss_closed6,blum-dienes_1,blum-dienes_2,ads1,dm1_1,dm1_2}. We then discuss a supersymmetric nine-dimensional construction, the Dabholkar--Park orientifold \cite{dp}, and describe its D-brane spectrum, which was first analyzed in \cite{Dbranes-dpGukov,Dbranes-dpGimon,Dbranes-dpGaberdiel} and completed in \cite{Bossard:2024mls}. Finally, we present six-dimensional orientifold compactifications that display the feature of twisted O-planes \cite{Dabholkar:1996zi,Gimon:1996ay}.

\subsection{The supersymmetry-breaking Scherk--Schwarz orientifolds}

The Scherk--Schwarz mechanism \cite{ss-original_1,ss-original_2,ss-original_3} is a very general procedure, both in field and string theory, that realizes the breaking of supersymmetry through dimensional compactification. The core element of this mechanism is to exploit a symmetry of the theory to impose distinct boundary conditions on bosons and fermions along the compact dimensions, which lead to a reduced theory in lower dimensions with spontaneously broken supersymmetry. This procedure can be seen as the effective gauging of the symmetry under consideration on the compact dimensions. As an example, we recall the original construction of such a mechanism \cite{ss-original_1,ss-original_2}, which was put forward in the context of a supergravity theory. The symmetry used to realize the mechanism is actually the R-symmetry of the supersymmetry algebra, which naturally acts in a different way on bosons and fermions. This R-symmetry is gauged in the compactified theory, and bosons and fermions acquire different masses.

The elements of this construction are very general and can be naturally applied to string theory as well. The basic Scherk--Schwarz mechanism of this type is a deformation of the type~IIB theory in nine dimensions \cite{ss_closed1,ss_closed2,ss_closed3,ss_closed4,ss_closed5,ss_closed6,blum-dienes_1,blum-dienes_2}. Taking the $9^\textup{th}$ spatial dimension $X^9$ to be the compact circle of radius $R$, this deformation is obtained via a freely-acting orbifold
\beq
    g=\(-1\)^F\delta \ ,
\label{eqn:ch3_ss_IIB_1_g_def}
\eeq

\noindent where $\(-1\)^F$ is the spacetime fermion number 
and $\delta$ is the shift
\beq
    \delta:\quad X^9\to X^9+\pi R \ ,
\label{eqn:ch3_ss_IIB_1_g_shift_def_1}
\eeq

\noindent along the circle. As in field theory, the momenta along compact space dimensions take discrete values and contribute to the mass of the string states, according to the Kaluza--Klein paradigm. In addition, closed strings can wrap non-trivially around the compact cycles and therefore are associated with a second discrete quantity, the winding number. These windings increase the tension of the string, which turns into a further contribution to the mass of the excitations. This is better summarized by looking at the form of the compact coordinates. In the present case of the circle compactification, the coordinate $X^9$ takes the form
\begin{align}
    X^9=x^9+2\alpha^\prime\frac{m}{R}\tau+2nR\sigma +\[\text{oscillators}\] \ ,
\end{align}
so that 
\begin{align}
    X^9_\textup{L,R}=x^9_\textup{L,R}+\alpha^\prime p^9_\textup{L,R}\(\tau\mp\sigma\) +\[\text{oscillators}\] \ , && p^9_\textup{L,R}=\frac{m}{R}\pm\frac{nR}{\alpha^\prime} \ , 
\end{align}
and the mass of the closed string becomes
\beq
    M^2=\frac{m^2}{R^2}+\frac{n^2 R^2}{{\alpha^\prime}^2}+\frac{2}{\alpha^\prime}\(N_X+N_\psi+\bar{N}_X+\bar{N}_\psi-1\) \ . 
\label{eqn:ch3_compact_dim_mass}
\eeq
In this picture, the winding numbers appear as the analogue for the string parameter $\sigma$ of the standard discrete momentum associated with the time parameter $\tau$. Note that these formulae are invariant under the transformation $R\to R^\prime=\frac{\alpha^\prime}{R}$, upon redefining the name of momentum and winding numbers. This is a simple and thus manifest example of T-duality. Then, when computing the string partition function, the integration over the compact momentum is replaced by the sum over the momentum and winding modes. In this respect, the relevant combinations appearing in the partition functions are
\begin{align}
     P_{m} \equiv q^{\frac{\alpha^\prime m^2}{R^2}}\ , && W_{n} \equiv  q^{\frac{n^2 R^2}{4 \alpha^\prime}} \ , && \Lambda_{m,n}=q^{\frac{\alpha^\prime}{4}\(\frac{m}{R}+\frac{nR}{\alpha^\prime}\)^2}\bar{q}^{\frac{\alpha^\prime}{4}\(\frac{m}{R}-\frac{nR}{\alpha^\prime}\)^2} \ .
\label{eqn:ch3_ss_P_W_terms}
\end{align}


With these tools, one can compute the torus partition function associated with the freely-acting orbifold $g$ defined in eq.~\eqref{eqn:ch3_ss_IIB_1_g_def}, which is found to be\footnote{We again employ a simplified notation for the vacuum amplitudes in which the modular integration is left implicit. Note that, since the integration over the compact dimension is replaced by a lattice sum, these modular integrations are now weighted by $\(4\pi\alpha^\prime\)^{\frac{9}{2}}\tau_2^{\frac{11}{2}}$.}
\begin{equation}
    \begin{aligned}
        \T&= \frac{1}{2|\eta|^{16}} \sum_{m,n} \Bigl( |V_8-S_8|^2 \Lambda_{m,n}  \ + \ |V_8+S_8|^2 (-1)^m \Lambda_{m,n} \Bigr. \\
    &\qquad\qquad \Bigl.   +  \ |O_8-C_8|^2 \Lambda_{m,n+1/2} \ + \
    |O_8+C_8|^2 (-1)^m \Lambda_{m,n+1/2} \Bigr)  \  .
    \end{aligned}
\label{ss4-2}
\end{equation}

\noindent The terms in the first line are the so-called \textit{untwisted sector}, and it is obtained from the standard type~IIB torus amplitude \eqref{eqn:ch3_typeIIB_torus} by requiring symmetry under the orbifold action $g$. The second line is called instead \textit{twisted sector} and must be introduced to make the orbifold symmetry compatible with modular invariance. The amplitude \eqref{ss4-2} is given in the so-called orbifold basis. A more natural basis for the spacetime interpretation of the theory is the so-called Scherk--Schwarz basis, in which one replaces $R\to 2R$ and the orbifold $g$ of eq.~\eqref{eqn:ch3_ss_IIB_1_g_def} is rewritten as
\begin{align}
    g=\(-1\)^F\delta^2 \ , && \delta^2:\quad X^9\to X^9+2\pi R \ .
\label{eqn:ch3_ss_IIB_2_g_def}
\end{align}
In this basis the shift corresponds to a full lap around the circle, so that combined with the spacetime fermion number operator $\(-1\)^{F}$ in $g$ clearly imposes different periodicity conditions on bosonic and fermionic states, as prescribed by the Scherk--Schwarz paradigm. The torus amplitude \eqref{ss4-2} in such a Scherk--Schwarz basis is equal to
\begin{equation}
\begin{aligned}
    \T &= \frac{1}{|\eta|^{16}}\sum_{m,n} \Bigl[  (|V_8|^2 + |S_8|^2) \ \Lambda_{m,2n}
    - (V_8 {\bar S}_8 + {\bar V}_8 S_8) \ \Lambda_{m+1/2,2n}  \\
    & \qquad \qquad + (|O_8|^2 + |C_8|^2) \ \Lambda_{m,2n+1}
    - (O_8 {\bar C}_8 + {\bar O}_8 C_8) \ \Lambda_{m+1/2,2n+1} \Bigr] \ ,
\end{aligned}
\label{ss4-3}
\end{equation}
\noindent and the shift between the masses of bosons and fermions and the associated breaking of supersymmetry are now manifest. 

Building upon this amplitude, we can extend this construction by including an orientifold projection. There exist three known constructions of this type, which we now review.  

\subsubsection{The first Scherk--Schwarz orientifold}

The simplest 9D orientifold construction of this type is achieved by adding the standard projection on the worldsheet parity $\Omega$ of eq.~\eqref{eqn:ch3_ws_parity_cs}. The resulting Klein bottle amplitudes, in the direct and transverse channels, are
\begin{align}
    \K_1 = \frac{1}{2} \ \frac{V_8-S_8}{\eta^8} \ \sum_m P_m \ , && {\tilde \K}_1 = \frac{2^5v}{2}\frac{V_8-S_8}{\eta^8} \ \sum_n W_{2n} \ ,
\end{align}

\noindent where $v = R/\sqrt{\alpha^{\prime}}$. The model contains a O$9_-$ plane and requires the introduction of sixteen D9-branes in order to cancel the Ramond--Ramond tadpole \cite{blum-dienes_1,blum-dienes_2}.

\subsubsection{The second Scherk--Schwarz orientifold}

A second orientifold projection is obtained by supplementing the involution $\Omega$ with a parity transformation along the circle:
\begin{align}
    \Omega_2=\Omega\Pi \ , && \Pi:\quad X^9\to2\pi-X^9 \ .
\end{align}

\noindent This orientifold was introduced in \cite{ads1} as the Scherk--Schwarz compactification of the Horava--Witten M-theory \cite{hw_1,hw_2}. The loop-channel Klein bottle amplitude is equal to
\beq
    \K_2 = \frac{1}{2\eta^8} \sum_n \[ (V_8-S_8) \ W_{2n} \ + \ (O_8-C_8) \ W_{2n+1} \]\ .
\eeq

\noindent The presence of the $O_8$ character signals a potential tachyonic scalar, which in this case is part of the spectrum. The applied projection $\Omega_2$ suggest the presence of O8-planes. The Klein bottle in the transverse channel is equal to
\beq
    {\tilde \K}_2 = \frac{2^5}{2v} \frac{1}{\eta^8} \sum_m \(V_8 P_{2m} - S_8 P_{2m+1}\) \ ,
\label{ss4-66}
\eeq
and thus the model contains a pair of O$8_-$--$\overline{\text{O}8_-}$ planes, located at the fixed points $X^9=0,\pi$ of the parity $\Pi$. Their net Ramond--Ramond charge sums to zero, so that the closed string theory is self-consistent and no open sector is needed. On the contrary, the two O8-planes have the same tension, which generate a NS-NS tadpole. A theory which is free of both R-R and NS-NS tadpoles is obtained by superposing eight pairs of D8--$\overline{\text{D}8}$ branes to the $\text{O}8_-$ and the $\overline{\text{O}8_-}$-plane respectively, generating in this way a $\mathrm{SO}(16)\times\mathrm{SO}(16)$ gauge group.

\subsubsection{The third Scherk--Schwarz orientifold}

The third orientifold projection is a more sophisticated variation of the second one, in which one adds a worldsheet left-fermion number operator \cite{dm1_1,dm1_2}:
\beq
    \Omega_3=\Omega\Pi\(-1\)^{G_L} \ .
\eeq
The Klein bottle amplitudes are equal to
\begin{equation}
\begin{aligned}
    \K_3 &= \frac{1}{2\eta^8} \sum_n \[ (V_8-S_8) W_{2n}  -  (O_8-C_8)  W_{2n+1} \] \ ,  \\ 
    {\tilde \K}_3 &= \frac{2^5}{2v} \frac{1}{\eta^8} \sum_m \(S_8 P_{2m} - V_8 P_{2m+1}\) \ .
\end{aligned}
\label{ss4-77}
\eeq
Thus, also in this case the theory contains a pair of O8-planes at the parity fixed points, but this time they are O$8_-$--$\overline{\text{O}8_+}$. The Ramond--Ramond charges now add up and the correspondent tadpole is removed by introducing sixteen D8-branes. They can be superposed either on the O$8_-$-plane or on the $\overline{\text{O}8_+}$ one. In the former case, the correspondent gauge group is $\mathrm{SO}(32)$, in the latter $\mathrm{USp}(32)$. Moreover, contrary to the previous $\Omega_2$, the projection $\Omega_3$ removes the tachyonic scalar from the spectrum.

\subsection{The supersymmetric Dabholkar--Park orientifold}

After having discussed the supersymmetry-breaking Scherk--Schwarz orientifolds, we now turn to a supersymmetric orientifold of type~IIB sting theory in nine dimensions, constructed by Dabholkar and Park in \cite{dp}. The Dabholkar--Park projection combines the worldsheet parity $\Omega$ with the half-shift $\delta$ along the circle:
\beq
    \Omega_\textup{DP}=\Omega\delta \ .
\label{eqn:ch3_dp_parity}
\eeq
Note that, contrary to the first Scherk--Schwarz orientifold, here there is no fermion number operator included in the projection, and the resulting theory  preserves in fact half of the originally supersymmetry. Moreover, despite sharing this feature with the type~I theory --- discussed here in Section \ref{ch3_sec_type_I} ---, the massive spectra of these theories are different, as the Dabholkar--Park theory is actually a deformation of the type~IIB string. The torus amplitude of this model is
\begin{equation}
    \T_\textup{DP} = \left|\frac{V_8-S_8}{\eta^8}\right|^2 \sum_{m,n} \Lambda_{m,n} \ , 
\label{dp1}
\end{equation}
whereas the Klein bottle loop and transverse amplitudes are
\begin{align}
    \K_\textup{DP} = \frac{1}{2} \ \frac{V_8-S_8}{\eta^8} \sum_m (-1)^m P_m \ , && \tilde{\K}_\textup{DP}=\frac{2^5v}{2}\frac{V_8-S_8}{\eta^8}\sum_n W_{2n+1} \ .
\label{dp2}
\end{align}
The compactified transverse amplitudes have a spacetime geometrical interpretation when they involve the sum over the momenta rather than the windings, so that auxiliary T-duality transformations are usually needed to achieve such a description. In this T-dual picture, one then recognizes, looking at the transverse amplitude \eqref{dp2}, the presence of two O8-planes, an O$8_-$-plane and a O$8_+$-plane, located respectively at $X^9=0$ and $X^9=\pi R^\prime$, with $R^\prime=\frac{\alpha^\prime}{R}$. Their net R-R charge is vanishing and thus they do not require any background open sector to be well-defined.

The spectrum of the Dabholkar--Park theory can be understood by studying the action of the symmetry operator $ \Omega_\textup{DP}$ of eq.~\eqref{eqn:ch3_dp_parity} on the type~IIB fields reduced to nine-dimensions. Starting from the bosons, one has
\beq
\begin{aligned}
    \Omega_\textup{DP}\left\lvert G_{\mu\nu},\phi,C^{(2)},C^{(6)},C^{(10)}\right\rangle^{(m)}&=\(-1\)^m\left\lvert G_{\mu\nu},\phi,C^{(2)},C^{(6)},C^{(10)}\right\rangle^{(m)} \ ,\\
    \Omega_\textup{DP}\left\lvert B_{\mu\nu},\phi,C^{(0)},C^{(4)}_+,C^{(8)}\right\rangle^{(m)}&=-\(-1\)^m\left\lvert B_{\mu\nu},\phi,C^{(0)},C^{(4)}_+,C^{(8)}\right\rangle^{(m)} \ ,
\end{aligned}
\eeq
so that only the even Kaluza--Klein modes of the first set of fields and the odd modes of the second one survive the projection. Instead, the action on the fermions is
\begin{align}
    \Omega_\textup{DP} |\psi_1{}_\mu,\lambda_1 \rangle^{(m)}
=  (-1)^m   |\psi_2{}_\mu,\lambda_2  \rangle^{(m)}\ , &&
\Omega_\textup{DP} | \psi_2{}_\mu,\lambda_2  \rangle^{(m)}
=   (-1)^m     |\psi_1{}_\mu,\lambda_1  \rangle^{(m)}\ ,
\label{dp02}
\end{align}
so that the invariant combinations are
\begin{align}
    \psi^{\prime\,(m)}_\mu{}\equiv\frac{\psi^{(m)}_{1\mu}+\(-1\)^m\psi^{(m)}_{2\mu}}{\sqrt{2}} \ , && \lambda^{\prime\,(m)}\equiv\frac{\lambda^{(m)}_1+\(-1\)^m\lambda^{(m)}_2}{\sqrt{2}} \ .
\end{align}
Thus, there is only one massless gravitino, \textit{i.e.} $\psi_{1\mu}^\prime\equiv\psi_{\mu}^{\prime\,(0)}$, and therefore the Dabholkar-Park orientifold \eqref{eqn:ch3_dp_parity} halves the original type~IIB supersymmetry. The full spectrum of the theory and relative Kaluza--Klein masses are
\beq
\begin{aligned}
    \left\{G_{\mu \nu}, \phi, C^{(2)} , C^{(6)}, C^{(10)}, {\psi^\prime}_1^{\mu}, \lambda^\prime_1 \right\}^{(2m)}& \ , &&& M^2 &= \frac{(2m)^2 }{R^2} \ , \\
    \left\{B_{\mu\nu},C^{(0)},C^{(4)}_+,C^{(8)}, {\psi^\prime}_2^{\mu}, \lambda^\prime_2\right\}^{(2m+1)}& \ , &&& M^2 &= \frac{(2m+1)^2}{R^2} \ .
\end{aligned}
\label{dp3}
\eeq
As previously mentioned, this field content is different from the nine-dimensional Kaluza--Klein spectrum of the type~I theory, since the Dabholkar--Park one contains both gravitini at the massive level, and thus the latter must be regarded as a deformation of the original type~IIB theory, which is consistently recovered in the decompactification limit $R\to\infty$. 

It can be shown that this Dabholkar--Park model is S-dual to the type~IIB asymmetric orbifold $(-1)^{F_L} \delta$, where $F_L$ is the spacetime left-fermion number \cite{dp,akp}. Further similar constructions where proposed in \cite{gepner}. 

\subsection{The D-brane spectrum of the Dabholkar--Park orientifold}\label{ch3_sec_dp_branes}

Despite the fact that background branes are not strictly required for consistency of the theory, the Dabholkar--Park orientifold model still contains a series of lower-dimensional branes, which can either be orthogonal to the compact $9^\textup{th}$ dimension or wrapping it. This brane spectrum was first studied in \cite{Dbranes-dpGukov,Dbranes-dpGimon,Dbranes-dpGaberdiel}, and we have completed its analysis in \cite{Bossard:2024mls}, following the rules described in \cite{dms}.

The stable branes are the BPS D1 and D5-branes, as dictated by the massless spectrum \eqref{dp3}, and the non-BPS D3 and D7-branes orthogonal to the circle. These are related to D4 and D8-branes wrapping the circle through a tachyonic kink transition \cite{Sen:1998tt}. These stable branes have an M-theory interpretation as M2-branes wrapping twisted cycles or M5-branes wrapping untwisted cycles \cite{Dbranes-dpGaberdiel}. The D7 and D3-branes wrapping the circle, as well as all even-dimensional branes, are unstable and we report on them in Appendix \ref{app_branes_dp}.

\subsubsection{BPS D5-branes wrapping the circle}

The BPS D5-branes can either wrap the circle or be orthogonal to it, and we begin from the former ones. For $N$ such branes, the cylinder and M\"obius strip amplitudes in the transverse channel are
\beq
\begin{aligned}
    {\tilde \A}_{55}&=\frac{2^{-3}vN^2}{2}\frac{V_4 O_4 + O_4 V_4 -S_4 S_4 - C_4 C_4}{\eta^8}\sum_n W_{n} \ , \\
    {\tilde \M}_{5}&=-N\left(\frac{4\epsilon v}{\hat{\eta}^2 \hat{\vartheta}_2^2}\right)  \left({\hat V}_4 {\hat O}_4 - {\hat O}_4 {\hat V}_4 + {\hat S}_4 {\hat S}_4 - {\hat C}_4 {\hat C}_4\right)\sum_n W_{2n+1} \ ,
\end{aligned}
\eeq
where we employed the character decomposition outlined in eq.~\eqref{eqn:ch3_characters_dec_rule}. Going in the T-dual picture, one understands that the $N$ T-dual D4-branes can either sit on top of the O$8_-$--plane or of the O$8_+$--plane, according to the sign ambiguity $\epsilon=\pm1$, respectively. The loop channel amplitudes are instead given by
\begin{equation}
    \begin{aligned}
        \A_{55}&=\frac{N^2}{2}\frac{V_4 O_4 + O_4 V_4 -S_4 S_4 - C_4 C_4}{\eta^8}\sum_m P_m  \ ,  \\
        \M_{5}&=\frac{N}{2}\left(\frac{4\epsilon}{\hat{\eta}^2 \hat{\vartheta}_2^2}\right) \left({\hat V}_4 {\hat O}_4 - {\hat O}_4 {\hat V}_4 + {\hat S}_4 {\hat S}_4 - {\hat C}_4 {\hat C}_4\right) \sum_m (-1)^m P_m \ .
    \end{aligned}
\end{equation}
Summed together, they yield the following $\N=\(1,0\)$ supersymmetric field spectrum:
\begin{equation}
    \begin{aligned}
        \A_{55}+\M_{5} &\sim  q^{- \frac{1}{3}} \[\frac{N(N+\epsilon)}{2} \( V_4 O_4 - C_4 C_4\)  + \frac{N(N-\epsilon)}{2} \( O_4 V_4  -  S_4 S_4\)  \] \ .
    \end{aligned}
\label{eq:dpd53}
\end{equation}
The characters $V_4 O_4 - C_4 C_4$ describe a vector multiplet, of gauge group $\mathrm{USp}(N)$ for $\epsilon =1$ or $\mathrm{SO}(N)$ for $\epsilon =-1$. The characters $O_4 V_4  -  S_4 S_4$ contain instead an hypermultiplet, standing either in the reducible antisymmetric representation of $\mathrm{USp}(N)$ or in the reducible symmetric one of $\mathrm{SO}(N)$.

\subsubsection{BPS D5-branes orthogonal to the circle}

For D5-branes orthogonal to the circle, the cylinder transverse amplitude is equal to
\begin{equation}
    \begin{aligned}
        {\tilde \A}_{55} &= \frac{2^{-3}}{2 v}\frac{V_8-S_8}{\eta^8} \sum_m \left[ e^{\frac{i \pi m}{2}} N + e^{-\frac{i \pi m}{2}} \overline{N}  \right]^2 P_m\\
        &= \frac{2^{-3}}{2 v} \frac{V_8-S_8}{\eta^8}\sum_m \left[ \left(N +\overline{N}\right)^2 P_{2m}-\left(N -\overline{N}\right)^2 P_{2m+1}\right] \ . 
    \end{aligned}
\label{eq:dpd54}
\end{equation}
The orientifold projection generates a doublet structure, with two copies of the D5-branes stack placed at antipodal points.  Moreover, a peculiar feature of this orthogonal D5-brane system --- which will also characterize the orthogonal D7-branes --- is that there are no closed-string states propagating between the D5-branes and the background O-planes. In fact, the cylinder amplitude does not share any character in common with the Klein bottle amplitude of eq.~\eqref{dp2}, which involves massive winding modes only, so that there is no M\"obius amplitude that can be written. As a consequence, despite the orientifold action, the cylinder amplitude must describe consistently the field spectrum by itself. This can be checked for example from the fact that the physical couplings to the NS-NS and R-R sectors are only those of the even Kaluza--Klein modes, in agreement with the orientifold projection in the closed-string sector. The corresponding loop channel amplitude is given by
\begin{equation}
    \A_{55} = \frac{V_8-S_8}{\eta^8} \sum_n \( N  \overline{N}  W_n +\frac{N^2 + \overline{N}^2}{2} W_{n+1/2} \) \ .
\label{dpd54}
\end{equation}
Given that the M\"obius amplitude vanishes identically, the spectrum described by $\A_{55}$ is maximally symmetric and of gauge group $\mathrm{U}(N)$. The states that have opposite half-integer windings, \textit{i.e.} $n+\frac12$ and $-n-\frac12$, have the same mass and are related by the shift $\delta$ of the projection symmetry $\Omega_\textup{DP}$ \eqref{eqn:ch3_dp_parity}. From the gauge group viewpoint, they are interpreted as states sitting in the complex symmetric plus antisymmetric representations $\boldsymbol{N}\otimes\boldsymbol{N}=\boldsymbol{\frac{N(N+1)}{2}}\oplus\boldsymbol{\frac{N(N-1)}{2}}$.

\subsubsection{Non-BPS D7-branes orthogonal to the circle}

The cylinder amplitudes for D7-branes orthogonal to the compact circle are equal to
\beq
\begin{aligned}
    {\tilde \A}_{77}&=\frac{2^{-4}}{2 v}\frac{1}{\eta^8} \sum_m 
\left[\left(N+{\overline N}\right)^2 \left(V_8 P_{2m}-S_8 P_{2m+1}\right)-\left(N-{\overline N}\right)^2\left(V_8 P_{2m+1}-S_8 P_{2m}\right)\right]  \\
    &=\frac{2^{-4}}{v}\frac{1}{\eta^8} \sum_m \left[ N {\overline N} \left(V_8-S_8\right) +  \frac{N^2 + \overline{N}^2}{2} (-1)^m \left(V_8+S_8\right)\right]P_m\ , \\
\end{aligned}
\label{dpd73}
\eeq
and
\beq
    {\A}_{77} =\frac{1}{\eta^8}\left[ N {\overline N} (V_8-S_8)\sum_n W_n +\frac{N^2 + \overline{N}^2}{2}(O_8-C_8)\sum_n W_{n+1/2} \right]. 
\eeq
Similarly to the orthogonal D5-branes case, also the orthogonal D7-branes do not exchange closed-string states with the background O-planes. The M\"obius amplitude vanishes identically and, consistently with the orientifold projection, only the even NS-NS momentum modes and the odd R-R ones have physical couplings in the transverse cylinder amplitudes. From the loop channel amplitude one sees that there is a supersymmetric $\mathrm{U}(N)$ vector multiplet from the $V_8-S_8$ sector, while the $O_8-C_8$ has shifted winding numbers, and states with opposite such numbers are complex conjugates, sitting in the reducible representation 
$\boldsymbol{N}\otimes\boldsymbol{N}=\boldsymbol{\frac{N(N+1)}{2}}\oplus\boldsymbol{\frac{N(N-1)}{2}}$. In addition, the scalar modes of $O_8$ are massive for $R>\sqrt{2\alpha^{\prime}}$ when the branes are at the same point on the circle.

This D7-branes system is actually stable when they are placed at antipodal points $X^9=0$ and $X^9=\pi R$, which is the critical point of a repulsion between these branes. This can be understood from the amplitude of two stacks of $N_1$ and $N_2$ D7-branes at a distance $X=2\pi a R$. This is obtained starting from the $M$ brane--$N$ antibrane amplitude in type~IIB \cite{reviews_1}
\beq 
    \tilde{\A}^{\scalebox{0.6}{IIB}}_{77} = \frac{2^{-4}}{v}\frac{1}{\eta^8} \sum_m  \[ ( M  \overline{M} + N \overline{N} ) (V_8-S_8) + ( M \overline{N} + N \overline{M})(V_8 + S_8)\] P_m \ , 
\label{eqn:ch3_dp_IIB_amp}
\eeq 

\noindent and properly incorporate the Dabholkar--Park orientifold projection under consideration. Under the orientifold involution $\Omega_\textup{DP}$ \eqref{eqn:ch3_dp_parity}, the $N_1$ branes at $X^9=0$ are mapped into $N_1$ antibranes at $X^9=\pi R$, while the $N_2$ branes at $X^9=2\pi a R$ are mapped into $N_2$ antibranes at $X^9=\(2a+1\)\pi R$. This is reproduced by the type~IIB amplitude \eqref{eqn:ch3_dp_IIB_amp} by taking
\begin{align}
    M = N_1  + e^{2\pi i a m} N_2 \ , && N = (-1)^m ( \overline{N}_1 + e^{2\pi i a m} \overline{N}_2) \ ,
\end{align}

\noindent and therefore the appropriate Dabholkar--Park amplitude is given by
\begin{equation}
    \begin{aligned}
        {\tilde \A}_{77}&=\frac{2^{-4}}{ 2 v}\frac{1}{\eta^8} \sum_m 
 \Bigl[ \bigl| N_1 + \overline{N}_1 + e^{2\pi i am} ( N_2 + \overline{N}_2) \bigr|^2    \bigl(V_8 P^{\scalebox{0.5}{even}}_{m}-S_8 P^{\scalebox{0.5}{odd}}_{m}\bigr) \\
    &\hspace{25mm} +\bigl| N_1 - \overline{N}_1 + e^{2\pi i am} ( N_2 - \overline{N}_2) \bigr|^2    \bigl(V_8 P^{\scalebox{0.5}{odd}}_{m}-S_8 P^{\scalebox{0.5}{even}}_{m}\bigr)\Bigr] \\
    &=\frac{2^{-3}}{v}\frac{1}{\eta^8} \sum_m  P_m \Biggl[ \Bigl( N_1 {\overline N}_1+N_2 {\overline N}_2+\cos(2\pi  a m) (  N_1 {\overline N}_2+ N_2 {\overline N}_1 )  \Bigr)  \bigl(V_8-S_8\bigr)  \\
    &\hspace{10mm} + \biggl(  \frac{N_1^2+\overline{N}_1^2 + N_2^2 + \overline{N}_2^2}{2}  +\cos(2\pi a m) ( N_1 N_2 + \overline{N}_1 \overline{N}_2 )\biggr) (-1)^m \bigl(V_8+S_8\bigr)\Biggr]  \ .
    \end{aligned}
\label{A77V}
\end{equation}
This amplitude involves the Chan--Paton factors in term of squared absolute values, allowed by the gauge group $\mathrm{U}(N)$. This might seem an unusual feature for an orientifold model, in which these coefficients typically appear as real perfect squares. However, as we point out in \cite{Bossard:2024mls}, this complex phase appears naturally from the exchange amplitude of states between positions $x$ and $y$ along the circle written in momentum space:
\beq 
    \langle {\mathrm{D}p}, X^9=x| {\mathrm{D}p},X^9=y  \rangle  = \sum_m e^{i\frac{m}{R} (y-x)}  \left\langle {\mathrm{D}p}, p^9 = \frac{m}{R} \left\lvert {\mathrm{D}p}, p^9 = \frac{m}{R} \right. \right\rangle \ .
\eeq
The complex phase disappears in the standard background D-branes amplitudes since they usually come in pairs built from the orientifold symmetry, which is not the case here.

The amplitude obtained in eq.~\eqref{A77V} again satisfies the consistency requirement that the physical couplings are the even momentum modes of the NS-NS sector and the odd momentum modes of the R-R sector. The contribution of this amplitude to the vacuum energy is minimized for $a=\frac12$, showing that the critical point configuration is for the two $N_1$ and $N_2$ stacks at antipodal points along the circle. On the other hand, the associated loop channel amplitude is equal to
\beq
\begin{aligned}
    {\A}_{77} &=\sum_n \biggl[ N_1 {\overline N}_1 W_n +N_2 {\overline N}_2W_n + \frac{N_1 {\overline N}_2+ N_2 {\overline N}_1}{2}\(W_{n-a} + W_{n+a}\) \biggr] \frac{V_8-S_8}{\eta^8}\\
    &+\sum_n  \biggl[ \frac{N_1^2+\overline{N}_1^2 + N_2^2 + \overline{N}_2^2}{2} W_{n+\frac12}  + \frac{N_1 N_2 + \overline{N}_1 \overline{N}_2}{2} \( W_{n+\frac12+a}  +  W_{n+\frac12-a }\)  \biggr] \frac{O_8-C_8}{\eta^8}\ ,
\end{aligned}
\eeq

\noindent which shows that for $a=\frac12$ there exists a complex tachyon in the representation $\(\boldsymbol{N_1},\boldsymbol{N_2}\)$ for all values of the radius. Therefore, a single D7-brane is stable, as it is the result of the orientifold freezing two D7-branes (before the projection) at $X^9=0$ and $X^9=\pi R$ into an invariant state, but multiple D7-branes can decay, as the brane and antibrane of each original pair can annihilate each other when they are at the same point \cite{Dbranes-dpGimon}.

In addition to this dynamics, one also expects that for $R<\sqrt{2\alpha^\prime}$ the orthogonal D7-brane system should transit into a D8-brane wrapping the circle. The transverse channel amplitudes of this D8-brane are
\begin{equation}
    \begin{aligned}
        {\tilde    \A}_{88} &= \frac{2^{-\frac{9}{2}}v}{2}\frac{N^2}{\eta^8} 
\Bigl( V_{7} O_{1} +  O_{7} V_{1}  \Bigr) \sum_n W_n \ , \\
{\tilde \M}_{8} =& - \frac{2 \, v \, N  \epsilon}{\hat{\eta}^{\frac{13}{2}} \hat{\vartheta}_2^\frac12} 
 \left( {\hat V}_{7} {\hat O}_{1} -  {\hat O}_{7} {\hat V}_{1} \right) \sum_n W_{2n+1} \ ,
    \end{aligned}
\label{dpdeven1_1}
\end{equation}
where, as before, the sign choices $\epsilon=\pm1$ are interpreted as the D7-brane being located at the position either of the $\text{O}8_-$ or the  $\text{O}8_+$ plane. In these amplitudes there is no coupling of any type of the R-R sector. Physical couplings to R-R fields are excluded by construction, since these branes are uncharged. Instead, the unphysical ones become physical couplings to the Ramond--Ramond forms in the presence of a background magnetic field, but these couplings are not possible for even D$p$-branes. The loop channel amplitudes associated with the ones in eq.~\eqref{dpdeven1_1} are
\begin{equation}
    \begin{aligned}
        {\A}_{88} =& \frac{N^2}{2}\frac{1}{\eta^8}  \Bigl[ (O_{7}+V_{7}) (O_{1}+V_{1}) - 2 S'_{7} S'_{1} \Bigr]  \sum_m P_m \ ,  \\
        \M_8 =& -   \epsilon \ \frac{N}{2} \ \frac{1}{\hat{\eta}^{7}}  \left( \frac{2 \hat{\eta}}{\hat{\vartheta}_2} \right)^{\frac{1}{2}} \left[  {\hat O}_{7} {\hat O}_{1} + {\hat V}_{7} {\hat V}_{1} -{\hat O}_{7} {\hat V}_{1} + {\hat V}_{7} {\hat O}_{1}  \right] \sum_m (-1)^m P_m \ ,
    \end{aligned}
\label{dpdeven2_1}
\end{equation}
which show that there exists a stable D8-brane configuration for $N=1$ and $\epsilon=1$, where the tachyon is absent at small radius $R\le\sqrt{2\alpha^\prime}$. For large radius $R>\sqrt{2\alpha^\prime}$, this D8-brane develops a tachyonic kink instability which makes it decay to the D7-brane \cite{Dbranes-dpGimon,Sen:1998tt}. Therefore, there is a stable brane configuration for all values of the radius.

\subsubsection{Non-BPS D3-branes orthogonal to the circle}

The physics of the orthogonal D3-branes matches exactly that of the D7-branes discussed above. The cylinder amplitudes in the transverse and loop channels are
\begin{equation}
\begin{aligned}
    {\tilde \A}_{33}&=\frac{2^{-2}}{2 v}\frac{1}{\eta^8}\sum_m\left[ \left(N+{\overline N}\right)^2\left(V_8 P_{2m}-S_8 P_{2m+1}\right)-\left(N-{\overline N}\right)^2\left(V_8 P_{2m+1}-S_8 P_{2m}\right)\right] \\
    &=\frac{2^{-4}}{v}\frac{1}{\eta^8}\sum_m \left[ N {\overline N}(V_8-S_8) +\frac{N^2 +\overline{N}^2}{2}(-1)^m (V_8+S_8)\right] P_m \ , \\
    {\A}_{33}&= N {\overline N} \left(\frac{V_8-S_8}{\eta^8}\right) \sum_n W_n+\frac{N^2 + \overline{N}^2}{2}\left(\frac{O_8-C_8}{\eta^8}\right)\sum_n W_{n+1/2} \ .
\end{aligned}
\label{dpd33}
\end{equation}

\noindent Instead, the M\"obius amplitude does not exist since the Klein bottle \eqref{dp2} contains only massive winding states and thus does not share, by construction, any sector with the cylinder amplitude \eqref{dpd33}. A single D3-brane is stable for $R>\sqrt{2\alpha^{\prime}}$, being an invariant configuration under the orientifold projection, but two such branes are repelled to antipodal points, where they develop a tachyon. 

The single D3-brane, too, is tachyonic for small radius $R < \sqrt{2\alpha^{\prime}}$, and in this limit it decays to a D4-brane wrapping the circle. The D4-brane loop channel amplitudes are
\begin{equation}
    \begin{aligned}
        {\A}_{44} =& \frac{N^2}{2}\frac{1}{\eta^8}  \left[ (O_{3}+V_{3}) (O_{5}+V_{5}) - 2 S'_{5} S'_{5} \right]  \sum_m P_m \ ,  \\
        \M_4 =&   \epsilon \ \frac{N}{2} \ \frac{1}{\hat{\eta}^{3}}  \left( \frac{2 \hat{\eta}}{\hat{\vartheta}_2} \right)^{\frac{5}{2}} \left[ {\hat O}_{3} {\hat O}_{5} + {\hat V}_{3} {\hat V}_{5} -  {\hat O}_{3} {\hat V}_{5} + {\hat V}_{3} {\hat O}_{5}   \right] \sum_m (-1)^m P_m \ . 
    \end{aligned}
\label{dpdeven2_2}
\end{equation}
As for the D8--D7-brane case, also here the tachyon at $R < \sqrt{2\alpha^{\prime}}$ disappears for $N=1$ and $\epsilon = -1$.

\subsection{Six-dimensional twisted O-planes}\label{ch3_sec_6D-twop}

In this section we discuss a six-dimensional example of orientifold compactification, the K3 supersymmetric orientifold of \cite{Dabholkar:1996zi,Gimon:1996ay}, characterized by the presence of "twisted O-planes", which is one of the main features of the new orientifold we put forward in \cite{Bossard:2024mls}, as we will describe in Section \ref{ch3_sec_new_sso}.

\subsubsection{The 6D K3 orientifolds} 

We begin by reviewing the standard 6D K3 orientifold, which combines the worldsheet parity involution $\Omega$ \eqref{eqn:ch3_ws_parity_cs} with the orbifold $T^4/\mathbb{Z}_2$. To understand the action of the reflection $\mathbb{Z}_2$, it is convenient to cast the $\mathrm{SO}(8)\to\mathrm{SO}(4)\times\mathrm{SO}(4)$ decomposition of the characters \eqref{eqn:ch3_characters_dec_rule} into the supersymmetric characters \cite{orientifolds5,orientifolds6}
\begin{equation}
\begin{aligned}
Q_o =& V_4 O_4 - C_4 C_4 \ , &&&&&&& Q_v =&  O_4 V_4  -  S_4 S_4\ , \\
Q_s =&  O_4 C_4 - S_4 O_4 \ , &&&&&&& Q_c =&  V_4 S_4 - C_4 V_4\ .
\end{aligned}
\label{6d12} 
\end{equation}
These characters encode $D=6$, $\N=(1,0)$ supermultiplets, such that $Q_o$ and $Q_s$ are even under the orbifold action, while $Q_v$ and $Q_c$ are odd. The torus amplitude for the orbifold compactification is then completely specified to be
\begin{equation}
\begin{aligned}
    \T &= \frac{1}{2} \left[ \left|\frac{Q_o + Q_v}{\eta^8}\right|^2
\sum_{m,n} q^{\frac{\alpha^{\prime}}{4} p_{{\rm L}}^{\intercal} g^{-1} p_{{\rm L}}} q^{\frac{\alpha^{\prime}}{4}  p_{{\rm R}}^{\intercal} g^{-1} p_{{\rm R}} }\right.\\
&\left. \hspace{10mm}+\frac{16}{\left|\eta\right|^4}\left( \left|\frac{Q_o - Q_v}{\vartheta_2^2} \right|^2+\left|\frac{Q_s - Q_c}{\vartheta_3^2}\right|^2+\left|\frac{Q_s + Q_c}{\vartheta_4^2} \right|^2 \right)\right], 
\end{aligned} 
\label{6d1} 
\end{equation} 

\noindent where the multiplicity 16 of the twisted contributions reflects the fixed points of the orbifold. The field spectrum described by this amplitude is summarized, in K3 language, in Table \ref{tab:K3_orientifold_spectrum_tot}. Then, taking the standard orientifold projection $\Omega$ yields the Klein bottle amplitude
\begin{equation} 
    \K = \frac{1}{4} \left[ \frac{Q_o + Q_v}{\eta^8} \left(
\sum_m q^{\frac{\alpha^{\prime}}{2} m^{\intercal} g^{-1} m} +
\sum_n q^{\frac{1}{2\alpha^{\prime}} n^{\intercal} g n}
\right) + 2 \times 16 \frac{Q_s + Q_c}{\eta^2 \vartheta^2 _4}  \right] \ ,
\label{6d4}
\end{equation} 

\noindent which shows that this orientifold projects the spectrum to  $\N=(1,0)$ 6D supergravity coupled to one tensor multiplet and 20 hyper-multiplets, as in Table \ref{tab:K3_orientifold_spectrum_tot}. The associated transverse channel is equal to 
\begin{equation}
\begin{aligned}
    \tilde{\K} &= \frac{2^5}{4} \left[ \frac{Q_o + Q_v}{\eta^8} \left(v_4 \sum_n q^{\frac{1}{\alpha^{\prime}} n^{\intercal} g n}+ \frac{1}{v_4} \sum_m q^{\alpha^{\prime} m^{\intercal} g^{-1} m}\right) +8 \frac{Q_o - Q_v}{\eta^2 \vartheta^2_2} \right] \\
    & \overset{q=0}{\sim}  8   \left[  \frac{Q_o}{\eta^8}\bigg|_{q=0} \left(\sqrt{v_4} + \frac{1}{\sqrt{v_4}} \right)^2  \ + \ \frac{Q_v}{\eta^8}\bigg|_{q=0} \left(\sqrt{v_4} - \frac{1}{\sqrt{v_4}} \right)^2 \right] \ . 
\end{aligned}
\label{6d5} 
\end{equation}

\noindent Therefore, in addition to the O9$_-$--planes, this theory contains also O5$_-$--planes, with standard negative tension and R-R charge. Tadpole cancellation is achieved by introducing 16 D9-branes and 16 D5-branes, with a maximum gauge group $\mathrm{U}(16)_9 \times \mathrm{U}(16)_5$ \cite{orientifolds5,gp}. 
\begin{table}
    \begin{minipage}[l]{12mm}
        \begin{tabular}{ccc}
       \toprule
 & tensor/ gravity  & hyper/gravitino   \\
\midrule
 \multirow{2}*{$\mathcal{Q}_o$}  & $ g_{\mu\nu}$ , $ C_{\mu\nu+} $ , $ \psi_\mu^i $ & $ B_{\mu\nu+} $ , $ C_{\mu\nu+\, I} $ , $ \psi_\mu^{\hat{\imath}} $  \\
& $ C_+$ , $ B_{\mu\nu-}$ , $ \lambda_+^i $ & $ e^{-\phi}$ , $ C_I$ , $ \lambda^{\hat{\imath}}$  \\
\midrule
\multirow{2}*{$\mathcal{Q}_v$}  & $ v_4$ , $ C_{\mu\nu-}$ , $ \lambda^i  $ & $ C_-$ , $ B_I  $ , $ \lambda^{\hat{\imath}}_- $  \\
& $ B_{\hat{I}}$ , $ C_{\mu\nu- \hat{I}} $ , $ \lambda_{\hat{I}}^i  $ & $ C_{\hat{I}}$ , $ g_{\hat{I}I} $ , $ \lambda_{\hat{I}}^{\hat{\imath}} $ \\
\midrule
$ \mathcal{Q}_s $ & $ B_A$ , $ B_{\mu\nu-A} $ , $ \lambda^i_A   $ & $ C_A $ , $ g_{A I}  $ , $ \lambda^{\hat{\imath}}_A $\\
\bottomrule
\end{tabular}
\end{minipage}
\hspace{70mm} 
\begin{minipage}[c]{12mm}
\begin{tabular}{ccc}
        \toprule
 & tensor/ gravity  & hyper   \\
\midrule
 \multirow{2}*{$\mathcal{Q}_o$}  & \multirow{2}*{$ g_{\mu\nu}$, $C_{\mu\nu+}$ , $\psi_\mu^i $} & \multirow{2}*{$ e^{-\phi}$, $C_I$, $\lambda^{\hat{\imath}}$}   \\
&  &  \\
\midrule
\multirow{2}*{$\mathcal{Q}_v$}  & \multirow{2}*{$ v_4$, $C_{\mu\nu-}$, $\lambda^i  $} & \multirow{2}*{$ C_{\hat{I}}$, $g_{\hat{I}I}$ , $\lambda_{\hat{I}}^{\hat{\imath}} $} \\
&   &  \\
\midrule
$ \mathcal{Q}_s $ &   & $ C_A$ , $g_{A I}$  , $\lambda^{\hat{\imath}}_A $\\
\bottomrule
\end{tabular}
\end{minipage}
\caption{\label{tab:K3_orientifold_spectrum_tot} Massless $\N=(1,0)$ 6D supermultiplets of the $\mathbb{Z}_2$ orbifold theory for each character, before (left panel) and after (right panel) the orientifold projection $\Omega$.}
\end{table}

\paragraph{The K3 theory description} 
The K3 surface description of this theory is given through the complex coordinates $z_1 = X_1 + i X_2$, $z_2 = X_3 + i X_4$ and the following basis of self-dual two-forms $\omega^I = \star \omega^I$ and anti self-dual two-forms $\omega^{\hat{I}} = - \star \omega^{\hat{I}}$:
\begin{equation}
    \begin{aligned}
        \omega^0 &= \frac{i}{2} \( dz_1\wedge d\bar z_1 + dz_2 \wedge d \bar z_2 \) \ , &&& \omega^+ &= dz_1\wedge d z_2 \ , &&& \omega^- &= d\bar z_1 \wedge d\bar z_2 \ , \\
       \omega^{\hat{0}} &= \frac{i}{2} \( dz_1\wedge d\bar z_1 - dz_2 \wedge d \bar z_2 \) \ , &&& \omega^{\hat{+}} &= dz_1\wedge d \bar z_2 \ , &&&  \omega^{\hat{-}} &= d\bar z_1 \wedge d z_2 \ .
    \end{aligned}
\end{equation}

\noindent The orbifold fixed points are written as $X_i = a_i \pi  R_i$ for $a_i \in \{ 0,1\}$, which we collectively denote with $A \in (\mathbb{Z}_2)^4$. Each fixed point can be blown-up to a $\mathbb{CP}^1$ cycle by turning on some twisted fields and we denote by  $\omega^A$ the dual anti self-dual two-forms on K3 \cite{Aspinwall:1996mn}. In this basis, the type~IIB Ramond--Ramond fields are written as 
\begin{equation}
\begin{aligned}
    C^{(0)} &= C_+ + C_- \ , \\ 
    C^{(2)} &= C_I \omega^I + C_{\hat{I}} \omega^{\hat{I}} + C_A \omega^A + \frac{1}{2} C_{\mu\nu} dx^\mu \wedge dx^\nu \ , \\
    C^{(4)} &=\frac14\( C_+-C_- \)  \omega^+ \wedge \bar \omega^- + \frac12 \( C_{\mu\nu+ I} \omega^I  + C_{\mu\nu- \hat{I}} \omega^{\hat{I}}  + C_{\mu\nu- A } \omega^A \) dx^\mu \wedge dx^\nu  \ .
\end{aligned}
\end{equation} 

\noindent In the NS-NS sector, the two-form field has an analogous expansion, while the internal metric components include the torus volume $v_4$ (in units of $\alpha^{\prime\, 2}$), the nine traceless components on the torus $g_{I\hat{J}}$, and the 48 components in the twisted sector $g_{AI}$. The integral basis of the K3 homology is obtained in the basis given by the $e_{ij}$ of $H_2(T^4,\mathbb{Z})$ and the exceptional divisors $e_A$ at each fixed point as the lattice generated by 22 vectors \cite{Morrison,Kummer}
\beq 
	H_2({\rm K3},\mathbb{Z})= \left\langle \frac12 e_{ij} + \frac12 \sum_{\substack{ A \in (\mathbb{Z}_2)^4\\ a_i = a_j=0}} e_A \Big|_{i<j} \ , \;  \frac12 \sum_{\substack{ A \in (\mathbb{Z}_2)^4\\ a_i = 0 }}  e_A  \Big|_{i=1}^4 \ , \;\frac12 \sum_{A \in (\mathbb{Z}_2)^4} e_A\ , \; e_{A}\Big|_{\sum_i a_i \le  2}  \right\rangle  \ ,
\label{LK3}   
\eeq
with the bilinear form 
\beq ( e_{ij},e_{kl}) = 2 \varepsilon_{ijkl}\ , \qquad (  e_A,e_B) =  - 2\delta_{AB} = - 2 \prod_{i=1}^4 \delta_{a_ib_i} \ . \eeq

\paragraph{Alternative orientifold projection}

An alternative orientifold projection of this 6D $T^4/\mathbb{Z}_2$ orbifold was proposed in \cite{Dabholkar:1996zi} and uses the orientifold projection 
$\Omega \,  \delta Z_2$, where $Z_2$ denotes the reflection on the torus coordinates $X^i$, and the shift $\delta$ acts as 
\beq 
\delta X_i = X_i + \pi R_i \ . \label{6de6}
\eeq
In this case, the Klein bottle does not receive contributions from the twisted sector of the torus \eqref{6d1} since the shift $\delta$ and the reflection $Z_2$ do not have any common fixed point. The loop channel Klein bottle amplitude is then
\begin{equation} 
    \K = \frac{1}{4} \left[ \frac{Q_o + Q_v}{\eta^8} \left(
\sum_m (-1)^m q^{\frac{\alpha^{\prime}}{2} m^{\intercal} g^{-1} m} +
\sum_n q^{\frac{1}{\alpha^{\prime}} n^{\intercal} g n}\right) + 2 \times (8-8) \frac{Q_s + Q_c}{\eta^2 \vartheta^2 _4}  \right]\ .
\label{6d4_2}
\end{equation}

\noindent The resulting massless spectrum includes nine tensor multiplets and twelve hyper-multiplets and is free of gravitational anomalies \cite{Dabholkar:1996zi}. The associated transverse channel amplitude is
\begin{equation}
\tilde{\K} = \frac{2^5}{4} \left[ \frac{Q_o + Q_v}{\eta^8} \left(
v_4 \sum_{n\; \scalebox{0.7}{odd}} q^{\frac{1}{\alpha^{\prime}} n^{\intercal} g n}
+ \frac{1}{v_4} \sum_{m} q^{\alpha^{\prime} m^{\intercal}g^{-1} m}\right)+(1-1) \frac{Q_o - Q_v}{\eta^2 \vartheta^2_2}   \right] .
\label{6d6} 
\end{equation}

\noindent Therefore, there is a neutral O9-plane, accompanied by 16 charged O5$_-$--planes. The associated R-R tadpole can be canceled by introducing an appropriate open sector with 16 D5-branes, of possible gauge groups $\mathrm{U}(8)_5 \times \mathrm{U}(8)_5$ or $\mathrm{SO}(16)_5$ \cite{Dabholkar:1996zi}. \\

This orientifold allows for many further variations, which can be chosen according to the desired shift in momentum and winding numbers \cite{reviews_1}. For instance, one can combine the shift $\delta$ of eq.~\eqref{6de6} with a reflection in a two-dimensional plane of the type $Z_2^{\text{(2D)}}(-1)^{F_L}$, under which the R-R tadpole is projected away and no open sector is required for consistency \cite{Gopakumar:1996mu}.

\subsubsection{Twisting the 6D orientifold} 
\label{6DtwistedO}

An interesting deformation of the 6D orientifold \eqref{6d1} discussed above involves a projection that is not an involution on the torus but only of the orbifold theory. An orientifold of this type was introduced in \cite{Dabholkar:1996zi,Gimon:1996ay}, where the following projection was considered:
\begin{align}
\Omega^\prime = \Omega  Z_4 \ , && \left(\Omega^\prime\right)^2 = Z_2 \ .
\label{6de1} 
\end{align}

\noindent The reflection $Z_4$ acts on the coordinates as
\begin{equation}
    Z_4 (z_1,z_2) =  (i z_1,- i z_2) \quad \longleftrightarrow \quad Z_4(X_1,X_2,X_3,X_4) = (-X_2,X_1, X_4,-X_3) \ ,
\end{equation}
while it acts on the fixed points $X_i =a_i \pi R_i $ by exchanging $a_1\leftrightarrow a_2$ and $a_3\leftrightarrow a_4$. It follows that the four fixed points of $\mathbb{Z}_4$ are such that $a_1=a_2$ and $a_3=a_4$, while the twelve other $\mathbb{Z}_2$ fixed points are exchanged by $Z_4$. The Klein bottle amplitude is
\beq  
    \K = \frac{1}{4\eta^2}\left[ 8\frac{Q_o - Q_v}{\vartheta^2_2}+  2 \times (4 + 6-6) \frac{Q_s - Q_c}{\vartheta^2_3}  \right] \ . 
\label{6de3}
\eeq

\noindent The coefficient $\(4 + 6-6\)$ in the twisted sector reflects the property that the O-planes occupy the four $\mathbb{Z}_4$ fixed points, whereas the other 12 $\mathbb{Z}_2$ fixed points are empty. 
The resulting closed-string spectrum, summarized in Table \ref{tab:6D_twist_spectrum}, contains the ${\cal N}= (1,0)$ gravitational multiplet coupled to nine tensors multiplets --- three from the untwisted sector and six from the twisted sector --- and  two untwisted and ten twisted hypermultiplets. 
\begin{table}
\centering
\begin{tabular}{ccc}
\toprule
 & tensor/ gravity  & hyper   \\
\midrule
$ \mathcal{Q}_o $ & $ g_{\mu\nu}$, $C_{\mu\nu+}$ , $\psi_\mu^i $ &  $ e^{-\phi}$, $C_I$, $\lambda^{\hat{\imath}}$  \\ 
\midrule
\multirow{2}*{$ \mathcal{Q}_v $} & $ v_4$,$ C_{\mu\nu-}$, $\lambda^i  $ & \multirow{2}*{$ C_{\hat{0}}$, $g_{\hat{0}I}$ , $\lambda_{\hat{0}}^{\hat{\imath}} $ } \\
& $ B_{\hat{\pm}}$, $C_{\mu\nu- \hat{\pm}}$ , $\lambda_{\hat{\pm}}^i  $ & \\
\midrule
$ \mathcal{Q}_s $ & $ B_{\hat{a}}$, $C_{\mu\nu-\hat{a}}$ , $\lambda^i_{\hat{a}}  $ & $ C_a$ , $g_{a I} $ , $\lambda^{\hat{\imath}}_a $\\
\bottomrule
 \end{tabular}
\caption{ Massless $\N=(1,0)$ supermultiplets of the $\mathbb{Z}_2$ orbifold theory with $\Omega^\prime=\Omega Z_4$ orientifold projection on the right. The indices $a=1,\dots,10$ denote the four $\mathbb{Z}_4$ fixed-points and the symmetric combinations of the six pairs of the remaining $\mathbb{Z}_2$ fixed points, while $\hat{a}=1,\dots,6$ denote to their antisymmetric combinations.}
\label{tab:6D_twist_spectrum}
\end{table}

The tree-level channel Klein bottle amplitude is equal to
\beq  
\tilde{\K} = \frac{16}{\eta^2} \left[\left(\frac{1}{\vartheta^2_4}-\frac{1}{\vartheta^2_3}\right)Q_s + \left(\frac{1}{\vartheta^2_4}+\frac{1}{\vartheta^2_3}\right)Q_c  \right]. 
\label{6de4}
\eeq 

\noindent We observe that the O-planes of this theory couple only to the massive states of the twisted sector of \eqref{6d1}: this is an instance of what we call twisted orientifold planes, which will characterize also the new orientifold we construct in \cite{Bossard:2024mls}. Note that the massless tadpole of $Q_s$ cancels in $ \tilde{\mathcal{K}}$, whereas $Q_c$ does not produce any tadpole. These twisted O-planes do not couple neither to the gravitational sector nor to the compactification moduli, and thus do not require any open sector to be well-defined. In fact, one can explicitly check that both the irreducible and the reducible gravitational anomalies cancel, without the need of a Green--Schwarz--Sagnotti mechanism \cite{gss}:
\beq 
\frac{1}{36(4\pi)^4} \left( \frac{n_H+ 29 n_T - 273 }{5} {\rm Tr} R^4 + \frac{n_H-7 n_T + 51}{4} {\rm Tr} R^2 {\rm Tr} R^2 \right)  \underset{\substack{n_H=12\\n_T=9}}{=}   0 \ .
\eeq

\section{The new Scherk--Schwarz orientifold}\label{ch3_sec_new_sso}

In this last section we present in detail the new nine-dimensional Scherk--Schwarz orientifold we construct in \cite{Bossard:2024mls}. It is based on the
projection 
\beq
    \Omega^\prime = \Omega (-1)^{F_L} \delta \ ,
\label{ss4-6}
\eeq

\noindent where $ (-1)^{F_L}$ is the left spacetime fermion number and $\delta$ is the half-circle shift of eq.~\eqref{eqn:ch3_ss_IIB_1_g_shift_def_1}. The square of $\Omega^\prime$ is equal to
\begin{equation}
    (\Omega^\prime)^2 = g \ ,
\label{ss4-6-2}
\end{equation}
with $g = (-1)^F \delta^2$ being the freely-acting orbifold of eq.~\eqref{eqn:ch3_ss_IIB_2_g_def} that defines the basic Scherk--Schwarz deformation of type~IIB string theory, which we discussed in Section \ref{ch3_sec_ss_dp_orientifolds}. Thus, the new projection $\Omega^\prime$ is not an involution of the type~IIB theory but it is so in its Scherk--Schwarz deformation. Similarly to the Dabholkar--Park orientifold \cite{dp} presented in Section \ref{ch3_sec_ss_dp_orientifolds}, the projection $\Omega^\prime$ defines a nine-dimensional orientifold compactification that is not related to the type~I string, but rather represents a deformation of the type~IIB theory. In contrast to the Dabholkar--Park case, this orientifold completely breaks supersymmetry, \'a la Scherk--Schwarz.

The Klein bottle amplitude in the loop channel is equal to
\begin{equation}
    \K_4 = \frac{1}{2} \frac{V_8+S_8}{\eta^8} \sum_m (-1)^m P_m \ .
\label{ss4-6-3}
\end{equation}

\noindent The orientifold action projects both the NS-NS and the R-R sectors onto their symmetric component for even momentum modes, and onto the antisymmetric one for odd momentum modes. As a result, the massless field sector is purely bosonic, containing the axio-dilaton, the metric and the R-R four-form. All fermions are instead massive, showing the breaking of supersymmetry. The associate transverse amplitude is 
\begin{equation}
    {\tilde \K}_4 = \frac{2^5v}{2}\frac{O_8-C_8}{\eta^8}\sum_n W_{2n+1} \ .
\label{ss4-7}
\end{equation}

\noindent This amplitude contains a peculiar O9-plane. It has zero tension and zero R-R charge --- so that no open-string sector is required for consistency --- and it couples only to the massive twisted states of the theory. Because of this feature, we call it \textit{twisted O-plane}.\footnote{We already presented in Section \ref{ch3_sec_ss_dp_orientifolds} a first instance of twisted O-planes, in the 6D torus orientifold of \cite{Dabholkar:1996zi,Gimon:1996ay}. O-planes with similar properties exist also in orientifolds of the type~0 string \cite{Sagnotti:1995ga,O'B_1,O'B_2,O'B_3}.} These twisted states are the massive R-R forms in $|C_8|^2$ and the real scalar field in $|O_8|^2$. For large enough radius $R \geq \sqrt{2 \alpha^{\prime}}$, all these $|O_8|^2$ scalars are massive and the theory is stable, whereas it is tachyonic at small radius $R< \sqrt{2 \alpha^{\prime}}$.

We remark that the transverse Klein bottle amplitude \eqref{ss4-7} shows explicitly that this orientifold can be interpreted only as supersymmetry-breaking deformation of the type~IIB string to nine dimensions rather than of the type~I string. In fact, the closed-string states entering $\tilde{\K}_4$ do exist only in the Scherk--Schwarz torus amplitude of eq.~\eqref{ss4-3}, whereas they do not exist in the standard amplitude of eq.~\eqref{eqn:ch3_typeIIB_torus}. In the large radius limit the Klein bottle amplitudes \eqref{ss4-6-3}-\eqref{ss4-7} vanish, since the orientifold projection $\Omega^\prime$ of eq.~\eqref{ss4-6} becomes trivial, and the torus amplitude \eqref{ss4-3} reduces to the type~IIB one, so that this new orientifold gives back the type~IIB string theory in the decompactification limit to ten dimensions.

\subsection{The supergravity realization and S-duality}\label{ch3_sec_ss_sugra}

This new orientifold theory admits a low-energy interpretation as a standard field-theoretical Scherk--Schwarz reduction \cite{ss-original_1,ss-original_2,ss-original_3} of type~IIB supergravity to nine dimensions. Supersymmetry breaking by compactification in the type~IIB string is described as a spontaneous breaking at the supergravity level \cite{ss_closed2}, with a vacuum energy proportional to $1/R^9$, without any dependence on the string scale. We will see that this is also true in the present orientifold theory. The orientifold symmetry $\Omega^\prime$ of eq.~\eqref{ss4-6} can be written in terms of the strong-weak coupling S-duality symmetry $S$. This symmetry generates, together with the axion shift, the $\mathrm{SL}(2,\mathbb{Z})$ duality group of the type~IIB theory. At the field theory level,\footnote{We recall that the field spectrum of type~IIB supergravity can be found in Table \ref{tab:type_II_spectra}.} it acts on the axio-dilaton $\tau = C^{(0)} + i e^{- \phi}$ as
\beq
    \tau \to \frac{a \tau + b}{c \tau + d} \ ,
\eeq

\noindent with $\{a,b,c,d\}\in\mathbb{Z}$ and $a d - b c =1$. The two Kalb--Ramond and Ramond--Ramond two-forms $B^{(2)}$ and $C^{(2)}$ transform as a doublet:
\beq
    \begin{pmatrix} B^{(2)} \\ C^{(2)} \end{pmatrix}=\begin{pmatrix} d B^{(2)}-c C^{(2)} \\ a C^{(2)}-b B^{(2)}\end{pmatrix}\ ,
\eeq
while the Einstein frame metric and the self-dual four-form $C^{(4)}_{+}$ are singlets. Instead, the action on the fermions is defined through the metaplectic cover $\mathrm{Mp}(2,\mathbb{Z})$ :
\beq
    \left\{ 1, (-1)^F\right\} \rightarrow  {\rm Mp}(2,\mathbb{Z}) \rightarrow  \mathrm{GL}(2,\mathbb{Z}) \  ,
\eeq

\noindent in terms of which the standard orientifold involution $\Omega$ acts on the bosons as the $\mathrm{GL}(2,\mathbb{Z})$ element $\left(\begin{smallmatrix}1 & 0 \\ 0 & -1 \end{smallmatrix}\right)$, while the operator $(-1)^{F_L}$ acts as $\left(\begin{smallmatrix} -1 & 0 \\ 0 & 1 \end{smallmatrix}\right)$. The action of these two operators on the fermions are, respectively, the $\mathrm{O}(2)$ R-symmetry reflections $\left(\begin{smallmatrix}0 & 1 \\ 1 & 0 \end{smallmatrix}\right)$ and $\left(\begin{smallmatrix}1 & 0 \\ 0 & -1 \end{smallmatrix}\right)$ \cite{Dabholkar:1997zd,Pantev:2016nze,Tachikawa:2018njr}. Putting this together, the symmetry operator $\Omega (-1)^{F_L} $ is identified with the square of an S-duality transformation
\beq
    \Omega (-1)^{F_L} = S^2 \ ,
\label{S2=OmegaFL}
\eeq

\noindent such that
\beq
    S (-1)^{F_L} = \Omega S\ , \qquad  S^4  = (-1)^F\ . 
\label{S4=F}
\eeq

Thus, the new orientifold \eqref{ss4-6} can be realized at the supergravity level in terms of a Scherk--Schwarz compactification to nine dimensions with respect to the symmetry $S^2$. We confirm this statement by matching the two pictures. On the string side, the action of the new orientifold symmetry $\Omega^\prime$ \eqref{ss4-6} on the type~IIB fields is
\begin{equation}
\begin{aligned}
    \Omega^\prime\left\lvert C^{(0)},C^{(4)}_+,C^{(8)} \right\rangle^{(m)}& =  (-1)^m \left\lvert C^{(0)},C^{(4)}_+,C^{(8)} \right\rangle^{(m)} \ , \\
    \Omega^\prime \left\lvert B^{(2)},C^{(2)},C^{(6)},C^{(10)}  \right\rangle^{(m)}& = -(-1)^m \left\lvert B^{(2)},C^{(2)},C^{(6)},C^{(10)}\right\rangle^{(m)} \ , \\
    \Omega^\prime \left\lvert\psi^\mu_1,\lambda_2 \right\rangle^{(m+\frac{1}{2})}&=-i(-1)^m \left\lvert \psi^\mu_2,\lambda_1   \right\rangle^{(m+\frac{1}{2})} \ , \\
    \Omega^\prime \left\lvert \psi_2^\mu, \lambda_1  \right\rangle^{(m+\frac{1}{2})}&=i(-1)^m \left\lvert \psi_1^\mu,\lambda_2  \right\rangle^{(m+\frac{1}{2} )} \ ,
    \end{aligned}
\label{eqn:ch3_Omega'_action_field}
\end{equation} 

\noindent so that the invariant fermionic combinations are given by
\beq
    \left\lvert \psi^\prime{}^\mu , \lambda^\prime\right\rangle^{(m+\frac{1}{2})} \equiv \frac{\left\lvert \psi^\mu_1,\lambda_2\right\rangle^{(m+\frac{1}{2})} - i (-1)^m \left\lvert \psi^\mu_2,\lambda_1  \right\rangle^{(m+\frac{1}{2})}}{\sqrt{2}} \ .
\label{ss4-07}
\eeq

\noindent The projected spectrum of the nine-dimensional theory is therefore
\beq
\begin{aligned}
\(G_{\mu \nu}, \phi, C^{(0)} \ , C^{(4)}_{+}\)^{(2m)}& \ , &&& M^2 &= \frac{(2m)^2 }{R^2} \ , \\
\(B^{(2)}, C^{(2)}\)^{(2m+1)}& \ , &&& M^2 &= \frac{(2m+1)^2 }{R^2} \ ,\\
\(\psi^\prime_\mu , \lambda^\prime\)^{\(m+\frac12\)}& \ , &&& M^2 &= \frac{\(m+\frac12\)^2 }{R^2} \ .
\end{aligned}
\label{ss4-8}
\eeq

\noindent This supergravity spectrum exactly matches the one described by the complete partition function $\Z=\frac12 \T +\K_4$ of eq.~\eqref{ss4-3} and eq.~\eqref{ss4-6-3}. 

Still working in this supergravity side, we can explicitly perform the Scherk--Schwartz compactification of type~IIB supergravity with the $S^2$ symmetry. The associated boundary conditions on the type~IIB supergravity fields are
\begin{equation}
\begin{aligned}
    \(g_{\mu \nu}, C^{(0)},C^{(4)}_+ \) (y+\pi R) =& \(g_{\mu \nu}, C^{(0)},C^{(4)}_+ \) (y) \ , \\
    \(B^{(2)},C^{(2)}\) (y+\pi R)=&-\(B^{(2)},C^{(2)}\) (y) \ , \\
    \psi_{\pm} (y+\pi R) =& \pm i \ \psi_{\pm} (y) \quad \to  \quad \psi_{\pm} (y+2 \pi R) = - \psi_{\pm} (y) \ .
\end{aligned}
\label{sugra3} 
\end{equation}

\noindent The consequent Kaluza--Klein field expansions are
\begin{equation}
\begin{aligned}
    \(g_{\mu \nu}, \tau, C^{(4)}_+ \) (y, x) =&\sum_m e^{2m i \frac{y}{R}}\(g_{\mu\nu},\tau,C^{(4)}_+ \)^{(m)} (x) \ , \\
    \(B^{(2)},C^{(2)}\) (y, x) = &\sum_m e^{(2m+1) i \frac{y}{R}} \(B^{(2)},C^{(2)}\)^{(m)} (x) \ , \\
    \psi_+ (y, x) =& \sum_m e^{(2m+ \frac{1}{2}) i \frac{y}{R}} \psi_{+}^{(m)} (x) \ , \\
    \psi_- (y, x)=& \sum_m e^{(2m- \frac{1}{2}) i \frac{y}{R}}\psi_{-}^{(m)}(x)\ . 
\end{aligned}
\label{sugra4} 
\end{equation}

\noindent We see in these formulae that the Kaluza--Klein masses of each mode precisely match those obtained from the string theory calculation in eq.~\eqref{ss4-8}. This confirms that the orientifold projection \eqref{ss4-6} is realized in supergravity as a Scherk--Schwarz compactification.

\subsubsection{S-selfduality and F-theory formulation}

Since, by definition, the transformation $S^2$ commutes with $\mathrm{SL}(2,\mathbb{Z})$, one would expect this orientifold to be invariant under S-duality. Despite one must be careful, in general, with the uplift of the definitions eq.~\eqref{S2=OmegaFL} and \eqref{S4=F} at the non-perturbative level,\footnote{For instance, the type~0B string is the orbifold of type~IIB with respect to $S^4=(-1)^F$, but it is not invariant under S-duality. Similarly, one would deduce that type~I string theory is S-dual to type~IIA instead of the $\mathrm{Spin}(32)/\mathbb{Z}_2$ heterotic string. In both examples, one either has a tachyonic state or open-string states must be added to cancel the R-R tadpole.} they should apply when the theory under consideration is a deformation of type~IIB string theory \cite{Vafa:1995gm}. This is what happens, for instance, for the S-duality between the Dabholkar--Park orientifold of eq.~\eqref{eqn:ch3_dp_parity} and the asymmetric orbifold of type~IIB by $(-1)^{F_L} \delta$ \cite{dp} (as mentioned in Section \ref{ch3_sec_ss_dp_orientifolds}), and this is also the case, as we stressed, of the new Scherk--Schwarz orientifold \eqref{ss4-6} of \cite{Bossard:2024mls}.
 
In support of the S-self duality, we argue that the new orientifold theory can be formulated as a perturbative F-theory background on the manifold $(T^2 \times S^1)/\mathbb{Z}_4$, where the $\mathbb{Z}_4$ symmetry acts as 
\begin{align}
    T^2:\quad z\,\,\overset{\mathbb{Z}_4}{\longrightarrow}\,\,-z \ ,  && S^1:\quad X^9\,\,\overset{\mathbb{Z}_4}{\longrightarrow} \,\,X^9+\frac{\pi}{2} R_9 \ .
\end{align}
In the Scherk--Schwarz basis, the manifold becomes $\mathcal{M}_3 = T^3 / \mathbb{Z}_2$, where $X^9\rightarrow X^9 + \pi  R_9$. As a first description of this model, we consider its M-theory descendant by putting the theory on an additional circle of radius $R_{\rm B}$. With respect to this circle, this M-theory model is then T-dual to the type~IIA orientifold by $\Omega^\prime = Z_2^{(8)} \Omega (-1)^{F_L} \delta$, that combines the orientifold \eqref{ss4-6} with the reflection of the eighth coordinate $X^8\longrightarrow-X^8$, with $X^8 \simeq X^8 + \frac{2\pi \alpha^{\prime}}{R_{\rm B}}$. At strong coupling, $\Omega (-1)^{F_L}$ acts on the M-theory circle as the reflection $X^{10}\rightarrow - X^{10}$. Note that the action of $\Omega$ and $(-1)^{F_L}$ individually changes the sign of the three-form, but their product acts  geometrically.  Therefore, one finds that $\Omega^\prime$ acts geometrically in M-theory as the combined reflection of $X^8$ and $X^{10}$, together with the quarter period shift of $X^9$. The rotation $z\rightarrow - z$ on the torus $(X^8,X^{10})\in T^2$ is an involution on the bosons, but it squares to minus one on the fermions. As a result, this geometric compactification of eleven-dimensional supergravity breaks all supersymmetries. 

To describe the field content, we split the indices as $\mu \ne 8,10$ and $I=8,10$, and we collect the gravitini into eight-dimensional complex spinors $\Psi$. We denote with $m$ the Kaluza--Klein mode along the coordinate $X^9 \cong X^9 + 4\pi R$, and with $\vec{n}$ the Kaluza--Klein mode along the torus of coordinates $(X^8,X^{10})$. At $\vec{n}=0$ the projected fields are 
\beq
\begin{aligned}
    (G_{\mu \nu}, G_{IJ} , C_{\mu IJ} , C_{\mu\nu\rho} )^{(2m)} \ , &&&& M^2 &= \frac{(2m)^2 }{R^2} \ ,\\
    (C_{\mu\nu I} , g_{\mu I} )^{(2m+1)} \ , &&&&  M^2 &= \frac{(2m+1)^2 }{R^2} \ , \\
    (\Psi_{\mu L} , \Psi_{I R})^{(2m+1/2)} \ , &&&& M^2 &= \frac{(2m+1/2)^2 }{R^2} \ , \\
    (\Psi_{\mu R} , \Psi_{I L})^{(2m+3/2)} \ , &&&& M^2 &= \frac{(2m+3/2)^2 }{R^2} \ .
\end{aligned}
\label{Mth}
\eeq

\noindent For ${\vec{n}}\ne 0$ they are simply identified with the states at $-\vec{n}$ with a sign depending on the parity of $m$, according to the orbifold projection. This is consistent with the spectrum \eqref{ss4-8} for $\vec{n}=0$, while the states with $\vec{n}\ne 0$ are infinitely massive in the limit $R_{\rm B}\rightarrow \infty$. 

\subsection{The vacuum energy}\label{sec:ss-energy}

Let us now compute the vacuum energy of the new orientifold $\Omega^\prime$ \eqref{ss4-6}. This is given by two contributions, that of the standard Scherk--Schwarz torus amplitude \eqref{ss4-3}, which we call $V_\textup{SS}$, and that of the Klein bottle amplitude \eqref{ss4-6-3}-\eqref{ss4-7}, which we call $V_\textup{K}$:
\beq
    V=V_\textup{SS}+V_\textup{K}=-\(\frac12\T+\K_4\) \ .
\eeq

\noindent  The complete expression of the two torus and Klein bottle contributions are
\begin{align}
    &\begin{aligned}
       V_\textup{SS}=&-\frac{1}{2(4\pi^2 \alpha^\prime)^{\frac{9}{2}}}\int_{\mathcal{F}}\frac{d^2\tau}{\tau_2^{11/2}}\frac{1}{|\eta|^{16}}\sum_{m,n} \Bigl[ \left[ V_8-S_8\right|^2 \Lambda_{m,n}+\left|V_8+S_8\right|^2 (-1)^m \Lambda_{m,n} \Bigr. \\
        &\Bigl.  \hspace{38mm}+|O_8-C_8|^2 \Lambda_{m,n+1/2} +|O_8+C_8|^2 (-1)^m \Lambda_{m,n+1/2}\Bigr] \ ,
    \end{aligned} \\
    &V_\textup{K}  =- \frac{1}{2 (4 \pi^2 \alpha^{\prime})^{\frac{9}{2}}}\int_0^{\infty} \frac{d \tau_2}{\tau_2^{\frac{11}{2}}}\frac{V_8+S_8}{\eta^8} (2 i \tau_2 )\ \sum_m (-1)^m P_m \ .
\end{align}

This energy can be computed for large radius $R>\sqrt{2\alpha^\prime}$, whereas for $R\leq\sqrt{2\alpha^\prime}$ the tachyon makes it divergent. We first look at the leading contribution for large radius. In this limit, the winding modes are suppressed with respect to the momentum ones. Thus, the twisted sector of the torus term is negligible, and using the identification $S_8\equiv V_8$ (see eq.~\eqref{eqn:ch3_ss_characters_gen}-\eqref{eqn:ch3_id_jacobi_theta}), one finds
\beq
\begin{aligned}
    V_\textup{SS} \sim&- \frac{1}{2(4 \pi^2 \alpha^{\prime})^{\frac{9}{2}}}\int_{\mathscr{F}}\frac{d^2 \tau}{{\tau}_2^{11/2}}  \left\lvert\frac{V_8+S_8}{\eta^8}\right\rvert^2\sum_{m\in\mathbb{Z}}\(-1\)^m\Lambda_{m,0}\\
    \sim&- \frac{16^2}{2(4 \pi^2 \alpha^{\prime})^{\frac{9}{2}}}\int_0^{\infty}\frac{d \tau_2}{{\tau}_2^{11/2}} \sum_{m\in\mathbb{Z}} \left[ e^{- \pi \tau_2 \frac{\alpha^{\prime} m^2}{R^2}} -  e^{- \pi \tau_2 \frac{\alpha^{\prime} \(m+\frac12\)^2}{R^2}}  \right]   \\
    \sim& -2\frac{16^2 R}{(2\pi)^9 {\alpha^{\prime}}^5} \int_0^\infty \frac{d \tau_2}{\tau_2^{\, 6}}  \ \sum_{k =0}^\infty e^{- \frac{4\pi R^2 }{\alpha^{\prime} \tau_2} \(k+\frac12\)^2} =\\
    =&- \frac{24}{\pi^{14} R^9} \sum_{k=0}^\infty \frac{1}{(2k+1)^{10}}= - \frac{31}{7560(2\pi)^4 R^9} \ .
\end{aligned}
\label{eqn:ch3_ve_T_1}
\eeq

\noindent Similarly, the Klein bottle yields
\beq
\begin{aligned}
    V_\textup{K}  &=- \frac{1}{2 (4 \pi^2 \alpha^{\prime})^{\frac{9}{2}}}\int_0^{\infty} \frac{d \tau_2}{\tau_2^{\frac{11}{2}}}\frac{V_8+S_8}{\eta^8} (2 i \tau_2 )\ \sum_m (-1)^m e^{- \pi \tau_2 \alpha^{\prime} \frac{m^2}{R^2}} \\
    &= -\frac{R}{(2\pi)^9 {\alpha^{\prime}}^5} \int_0^\infty\frac{d\tau_2}{\tau_2^{6}} \frac{ V_8+S_8}{ \eta^8}(2i\tau_2) \sum_{k =0}^\infty e^{- \frac{\pi R^2 }{\alpha^{\prime} \tau_2} \(k+\frac12\)^2}  \\
    &\sim -\frac{24 }{\pi^{14} R^9} \sum_{k=0}^\infty \frac{2^5}{(2k+1)^{10}} =  -\frac{31}{3780\pi^4 R^9} \ ,
\end{aligned}
\label{eqn:ch3_ve_K_1}
\eeq
Therefore, the two contributions have the same form,\footnote{Note that in specific constructions, the torus amplitude can contribute positively to the energy or be exponentially suppressed in the large radius limit \cite{ss-specific_1,ss-specific_2}.} so that the vacuum energy $V$ is negative and has a runaway behavior, scaling as $R^{-9}$ for large radius. In the decompactification limit $R\to \infty$ one recovers the type~IIB theory, of which this model is indeed a Scherk--Schwarz deformation. Note that in the resulting vacuum energy $V=V_\textup{SS}+V_\textup{K}\sim R^{-9}$ there is no leftover dependence on $\alpha^\prime$, meaning that it should be captured exactly by the field limit of the theory. Indeed, writing $V_\textup{SS}$ of eq.~\eqref{eqn:ch3_ve_T_1} and $V_\textup{K}$ of eq.~\eqref{eqn:ch3_ve_K_1} in terms of the Schwinger proper time $t=\pi\alpha^\prime\tau_2$, the total vacuum energy $V$ takes the form
\begin{equation}
    V \sim - \frac{1}{(4 \pi)^{9/2}} \int_0^{\infty} \frac{dt}{t^{\frac{11}{2}}}\sum_{m=-\infty}^{\infty} \left( 2 \times 36 \ e^{- \left(\frac{2m}{R}\right)^2 t} +  2\times 28 \ e^{- \left(\frac{2m+1}{R}\right)^2 t} -  64 \ e^{- \left(\frac{m+\frac{1}{2}}{R}\right)^2 t}\right) \ ,
\label{ve6}
\end{equation}
in which we recognize the field-theory vacuum energy of eq.~\eqref{eqn:ch3_ve_field_theory} given by the supergravity fields of the projected theory \eqref{ss4-8}, consistently. 

To perform these integrals, we made use of the Poisson summation formula
\begin{equation}
\sum_{n \in \mathbb{Z}} e^{- \pi n A n + 2 i \pi b n}
   = \frac{1}{{\sqrt A}}  \sum_{m \in Z} e^{- \pi (m-b) A^{-1} (m-b)} \ ,
\label{poisson}
\end{equation}
as well as the $q$-expansion
\begin{align}
     \frac{V_8+S_8}{\eta^8} =& \frac{\vartheta_2(0)^4}{\eta^{12}}= 16 \prod_{n\ge 1} \frac{(1+q^n)^8}{(1-q^n)^8} = 16 \sum_{n=0}^\infty c(n) q^n\ ,
\label{cn}
\end{align}
which at leading-order simply gives $\frac{V_8+S_8}{\eta^8}\sim16$. The large-radius computation can be made precise using the refined expansion in terms of the coefficients $c(n)$ of eq.~\eqref{cn}. Starting from
\begin{align}
    V_\textup{SS}&-\frac{2R}{(2\pi)^9 {\alpha^{\prime}}^5} \int_0^\infty \frac{d \tau_2}{\tau_2^{\, 6}}  \int_{-\frac12}^{\frac12} d\tau_1 \left|\frac{ V_8+S_8}{\eta^8}\right|^2 \sum_{k =0}^\infty e^{- \frac{4\pi R^2 }{\alpha^{\prime} \tau_2} \(k+\frac12\)^2}\ , \\
    V_\textup{K} &= -\frac{R}{(2\pi)^9 {\alpha^{\prime}}^5} \int_0^\infty\frac{d\tau_2}{\tau_2^{6}} \frac{ V_8+S_8}{ \eta^8}(2i\tau_2) \sum_{k =0}^\infty e^{- \frac{\pi R^2 }{\alpha^{\prime} \tau_2} \(k+\frac12\)^2} \ ,
\end{align}
which are obtained by means of the Poisson formula \eqref{poisson}, one obtains, using \eqref{cn},
\begin{equation}
    \begin{aligned}
        V&=-\frac{403}{189(4\pi)^4} \frac{1}{R^9} \\
        &\qquad-\frac{1 }{R^4 } \sum_{n\ge 1} \sum_{k\ge 0} \frac{(\frac{n}{\alpha^{\prime}})^{\frac52}}{(k+\frac12)^5}\left[c(n)^2 K_5 \left(8\pi R  \scalebox{1.1}{$\sqrt{\frac{n}{\alpha^{\prime}}}$}(k+\tfrac12)\right) + 2 c(n)K_5 \left(4\pi R  \scalebox{1.1}{$\sqrt{\frac{n}{\alpha^{\prime}}}$}(k+\tfrac12)\right)\right]\ , 
    \end{aligned}
\label{eqn:ch3_ve_gen}
\end{equation}

\noindent where $K_5$ is a Bessel function. Using the asymptotic expansion $c(n) \sim \frac{e^{2\pi \sqrt{2n}}}{n^{{11}/{4}}}$ at large $n$ and the behavior of the Bessel function $K_5$, one can show that this sum is absolutely convergent for $R>\sqrt{2\alpha^\prime}$, and that it diverges in the tachyonic regime. The vacuum energy is therefore strictly negative for all $R>\sqrt{2\alpha^{\prime}}$,  it evolves monotonically to zero for infinite radius, and it diverges negatively at $R=\sqrt{2\alpha^{\prime}}$ because of the tachyonic instability.  

\subsubsection{Relation to S-duality}

In the previous section we argued that the orientifold theory introduced in \cite{Bossard:2024mls} should be invariant under S-duality at the non-perturbative level. One can explore this conjecture in the context of the vacuum energy by trying to guess a non-perturbative extension of the potential of eq.~\eqref{eqn:ch3_ve_gen} that is invariant under $\mathrm{SL}(2,\mathbb{Z})$. Writing the string length $\alpha^\prime$ in terms of the $\mathrm{SL}(2,\mathbb{Z})$-invariant ten-dimensional Planck length $\ell_{10}$ as ${\alpha^\prime}=e^{-\frac{\phi}{2}}\ell_{10}^{2}$, in the Einstein frame one may write \cite{Bossard:2024mls}
\begin{equation}
    \begin{aligned}
        \hat{V}&=- \frac{403}{189(4\pi)^4} \frac{1}{R^9} \\
                &\hspace{4mm} -\frac{1}{2\ell_{10}^5 R^4 } \hspace{-2mm}\sum_{(m,n) \in \mathbb{Z}^2\smallsetminus (0,0)} \sum_{k\ge 0} \frac{ \(e^{\frac{\phi}{2}} |n+\tau m|\)^{\frac52}}{(k+\frac12)^5} \left[ + 2 c({\rm gcd}(m,n)) K_5 \left(4\pi \scalebox{1}{$\frac{ R }{\ell_{10}}$}  \scalebox{1.1}{$\sqrt{e^{\frac{\phi}{2}} |n+\tau m|}$}(k+\tfrac12)\right)\right.\\
            &\left.\qquad\qquad\qquad\qquad + c({\rm gcd}(m,n))^2 K_5 \left(8\pi \scalebox{1}{$\frac{R }{\ell_{10}}$}  \scalebox{1.1}{$\sqrt{e^{\frac{\phi}{2}} |n+\tau m|}$}(k+\tfrac12)\right)\right]\ .
    \end{aligned}\label{eqn:ch3_V_SL_full}
\end{equation}

\noindent Although the original potential is expected to receive corrections at all orders of perturbation theory, one can still try to make some predictions based on this generalized potential. The sum over $(m,n)$ is found to diverge for a radius 
\beq 
    R \, \times  {\rm min}_{m,n}   \scalebox{1.1}{$\sqrt{e^{\frac{\phi}{2}} |n+\tau m|}$} \le \sqrt{2} \ell_{10}\ ,
\eeq
which can be interpreted as the existence of a tachyonic state whenever the radius or the axio-dilaton cross this bound for some $m$ and $n$. The mass of the tachyonic closed-string scalar state can be generalized in the same $(m,n)$-dependent way, as
\beq 
    M_{\scalebox{0.6}{F1-D1}}^2 = - 2 \frac{|n + \tau m|}{\alpha^{\prime}} + \frac{R^2|n + \tau m|^2}{\alpha^{\prime}{}^2} \ .
\label{MassStrings} 
\eeq 
Expanded at weak coupling, it suggests the existence of a non-BPS D1-brane with mass $g_{\rm s} M_{\scalebox{0.6}{D1}} = \frac{R}{\sqrt{\alpha^{\prime}}}$ at tree-level: 
\beq 
    M_{\scalebox{0.6}{D1}}^2 = -2  \frac{\sqrt{ \frac{1}{g_{\rm s}^2} + C_0^2}}{\alpha^{\prime} } + \frac{R^2}{\alpha^{\prime}{}^2}\Bigl( \frac{1}{g_{\rm s}^2} + C_0^2\Bigr)  \approx  \frac{R^2}{\alpha^{\prime}{}^2 g_{\rm s}^2}  - \frac{2}{\alpha^{\prime} g_{\rm s}}  + \mathcal{O}( g_{\rm s}^0) \ .
\label{MassD1}
\eeq
The corrections are consistent with the open string expansion. However, there is no non-renormalization theorem to protect the mass of the perturbative closed-string scalar, and this interpretation holds as long as the corrections to this perturbative scalar mass modify the $(m,n)$-string mass formula \eqref{MassStrings} without affecting the leading-order D1-brane mass term \eqref{MassD1} in the weak-coupling expansion. We will find that there is indeed an unstable D1-brane with the expected tension, although one should not expect it to be stable.

One may also try to interpret these corrections as coming from D($-1$) instantons. In principle, one should use the Poisson summation formula \eqref{poisson} on the sum over $n \in \mathbb{Z}$ for $m\ne 0$ in eq.~\eqref{eqn:ch3_V_SL_full}. Nevertheless, one can anticipate the result using the relation 
\beq 
    4 e^{-2\phi} \partial_\tau \partial_{\bar \tau}  \hat{V}  =  \frac{1}{16} \Bigl(  \partial^2_{\log R}  + 22 \partial_{\log R}   +117\Bigr)  \hat{V} , 
\eeq 
and that the general solution to this differential equation with RR-charge $N$ takes the form
\beq 
    \int ds f(s) \frac{\ell_{10}^{4s}}{R^{9+4s}} e^{- \frac{\phi}{2}} K_{s-\frac12}\( 2\pi |N| e^{- \Phi} \) e^{ 2\pi i N C_0}\ .
\eeq
From these formulae, one finds that the potential $\hat{V}$ of eq.~\eqref{eqn:ch3_V_SL_full} must admit a weak-coupling expansion in Fourier modes for the D($-1$) instanton charge $N$ that has the expected exponential suppression in the classical instanton action \cite{Green:1997tv}
\beq 
    S_{{\rm D(-1)}} = 2\pi |N| e^{-\Phi}   + 2\pi i N C_0 \ . 
\eeq
 
From the supergravity point of view, the associated 1-loop potential is already S-duality invariant and gives, as we proved, the leading contribution at large radius. The loop expansion of the potential may be written as
 \beq V \sim \sum_{h=1}^\infty  d_h \frac{1}{R^9} \frac{\ell_{10}^{8(h-1)}}{R^{8(h-1)}} \ . \eeq
The supergravity loop corrections are expected to dominate the string theory ones at large radius, as for the 1-loop order. However, they are negligible at weak coupling, since one has $R\gg \ell_{10}$. Only at strong coupling $e^{\phi}\sim 1$ there could exist a minimum of the complete, non-perturbative potential in the non-tachyonic regime, that could prevent the perturbative tachyonic instability of the model.

\subsection{The D-brane spectrum of the new orientifold}\label{ch3_sec_new_ss_branes}

We now analyze the D-brane spectrum of the new orientifold \eqref{ss4-6}, whose construction can be done following the general paradigm of \cite{dms}. As mentioned, this theory does not contain any background charged D-brane. In fact, as shown by the Klein bottle amplitude \eqref{ss4-7}, the O-planes of this theory are twisted, \textit{i.e.} they only couple to the twisted sector, so that there is no R-R tadpole to cancel. Nevertheless, the theory contains, as always, D-branes of all complementary dimensions. The theory contains massless $C^{(4)}_+$ and $C^{(8)}$ Ramond--Ramond forms, as in eq.~\eqref{eqn:ch3_Omega'_action_field}, hence there will be charged D3 and D7-branes. This can be seen also from the type~IIB point of view. The orientifold symmetry $\Omega^\prime$ maps the type~IIB D3 and D7-branes into D3 and D7-branes of complex conjugate Chan--Paton factor, in such a way to form charged bound states after the orientifold projection. Similarly, the type~IIB D5 and D9-branes are mapped by $\Omega^\prime$ into $\overline{\mathrm{D}5}$ and $\overline{\mathrm{D}9}$-branes of complex conjugate Chan--Paton factor and therefore the orientifold freezes them into non-BPS D5 and D9-branes in the projected theory. The theory also contains D1-branes and even-dimensional branes. The latter are all found to be unstable, and their analysis is provided in Appendix \ref{app_branes_new_ss}.

The brane spectrum and its stability can also be understood geometrically from the perspective of the dual M-theory on $\mathcal{M}_3 = T^3 / \mathbb{Z}_2$ discussed in Section \ref{ch3_sec_ss_sugra}. In particular, the stable D-branes of the new orientifold model are in direct correspondence with the M-theory membranes wrapping cycles of $\M_3$.

\subsubsection{Homology cycles in the dual M-theory}

In order to describe the D-brane spectrum from the M-theory perspective, one needs to identify the homology cycles of the manifold $\M_3=T^3/\mathbb{Z}_2$ on which it is defined. The torus $T^3$ is quotiented over the isometry
\beq
    \sigma\( X^8,X^9,X^{1\hspace{-0.3mm}0}\) = \( - X^8,X^9 + \pi R, -X^{1\hspace{-0.3mm}0}\)\ , 
\eeq
which is freely-acting and preserves the orientation, so that $\mathcal{M}_3$ is a smooth orientable manifold. The first homology group is computed from the relation 
\beq 
    \mathcal{M}_3 = \mathbb{R}^3 / \pi_1(\mathcal{M}_3)
\label{M3quotient} \eeq
where $ \pi_1(\mathcal{M}_3)$ is generated by the isometry $\sigma$ and the three discrete translations 
\beq 
    T_{a_1,a_2,a_3} ( X^8,X^9,X^{1\hspace{-0.3mm}0}) = \( X^8 + 2\pi a_1\frac{{\alpha^{\prime}}}{R_{\rm B}},X^9 + 2\pi a_3 R, X^{1\hspace{-0.3mm}0} + 2\pi a_2 e^{\frac{2}{3}\Phi_{\rm A}} \sqrt{\alpha^{\prime}}\) \ .
\eeq
One then finds that 
\beq 
    H_1(\mathcal{M}_3, \mathbb{Z}) =  \pi_1(\mathcal{M}_3)  / [  \pi_1(\mathcal{M}_3), \pi_1(\mathcal{M}_3)] = \mathbb{Z} \oplus \mathbb{Z}_2 \oplus \mathbb{Z}_2\ , 
\eeq
where $\mathbb{Z}_2 \oplus \mathbb{Z}_2$ is the torsion subgroup of $H_1\(\mathcal{M}_3, \mathbb{Z}\)$. In the case of an orientable and compact smooth manifold of dimension $n$, the torsion subgroups of $H_{0}\(\mathbb{Z}\)$, $H_{n-1}\(\mathbb{Z}\)$ and $H_{n}\(\mathbb{Z}\)$ are trivial \cite{Hatcher}. Moreover,  Poincar\'e duality gives $H_2\(\mathcal{M}_3,\mathbb{Z}\) = \mathbb{Z}$. Thus, the homology of $\mathcal{M}_3$ is completely determined to be
\beq 
\begin{aligned}
    H_0(\mathcal{M}_3,\mathbb{Z}) &= \mathbb{Z}\ , &&&&& H_1(\mathcal{M}_3,\mathbb{Z}) &= \mathbb{Z}\oplus \mathbb{Z}_2\oplus \mathbb{Z}_2 \ , \\
    H_2(\mathcal{M}_3,\mathbb{Z}) &= \mathbb{Z} \ , &&&&& H_3(\mathcal{M}_3,\mathbb{Z}) &= \mathbb{Z}\ . 
\end{aligned}
\eeq

\noindent The cycles of  $H_0(\mathcal{M}_3,\mathbb{R})$ are associated with the coordinates as
\begin{align}
    S^1\(X^9\) \in H_1\(\mathcal{M}_3,\mathbb{Z}\)\ , && T^2\(X^8,X^{1\hspace{-0.3mm}0}\) \in H_2\(\mathcal{M}_3,\mathbb{Z}\)\ ,
\end{align}
while 1-cycles associated with the torsion $\mathbb{Z}_2$  are
\begin{align}
    S^1\(X^8\) \in H_1\(\mathcal{M}_3,\mathbb{Z}\) \ , && S^1\(X^{1\hspace{-0.3mm}0}\) \in H_1\(\mathcal{M}_3,\mathbb{Z}\)\ .
\end{align}
This geometrical structure allows one to predict that the stable membranes in this dual M-theory are M2, M5 and KK6-branes, wrapping these homology cycles. This prediction will be in perfect agreement with the stable branes of the new $\Omega^\prime$ orientifold model.

\subsubsection{Non-BPS D9-branes}

We begin to chart the D-brane spectrum in the new Scherk--Schwarz orientifold from the non-BPS uncharged D9-branes. As mentioned at the beginning of this section, they are interpreted as bound states of type~IIB
D9 and $\overline{\mathrm{D}9}$-branes under the orientifold projection. The general form of the cylinder amplitudes of a D-brane wrapping the circle are \cite{dms,reviews_1,reviews_2}
\begin{align}
    &\begin{aligned}
        \A=\frac{1}{8\eta^8} \sum_m&\Bigl[\left(\alpha^2 + \beta^2 + \gamma^2 + \delta^2\right)\left(V_8 P_m - S_8 P_{m+1/2}\right) \\
        & \left. + \left(\alpha^2 + \beta^2 - \gamma^2 - \delta^2\right)\left(V_8 P_{m+1/2} - S_8 P_{m}\right)\right.\\
        &\left. + \left(\alpha^2 - \beta^2 + \gamma^2 - \delta^2\right)\left(O_8 P_m - C_8 P_{m+1/2}\right) \right.\\
        & + \left(\alpha^2 - \beta^2 - \gamma^2 + \delta^2\right)\left(O_8 P_{m+1/2} - C_8 P_{m}\right)\Bigr]\ ,
    \end{aligned} \label{eq:ssd91}\\
    &{\tilde \A} = \frac{2^{-5}v}{2}\frac{1}{\eta^8}\sum_n \left[\left(\alpha^2 V_8- \beta^2 S_8\right) W_{2n} + \left(\gamma^2 O_8- \delta^2 C_8\right) W_{2n+1}\right] \ , \label{eq:ssd92}
\end{align}
in which the coefficients $\left\{\alpha,\beta,\gamma,\delta\right\}$ are related to the Chan--Paton factors of the branes. The M\"obius amplitudes are instead given by \cite{orientifolds1,orientifolds2,orientifolds3,orientifolds4,orientifolds5,orientifolds6,orientifolds7,reviews_1,reviews_2,dms}
\begin{equation}
\begin{aligned}
{\M}_9 &=- \frac{1}{2\eta^8} \sum_m \left[\epsilon_1 \gamma (-1)^m P_m {\hat O}_8 - i\epsilon_2 \delta (-1)^m P_{m+1/2} {\hat C}_8 \right] \ , \\
{\tilde \M}_9 &= -\frac{v}{\eta^8} \sum_n \left[ \epsilon_1 \gamma  {\hat O}_8 + (-1)^n\epsilon_2 \delta{\hat C}_8 \right] W_{2n+1} \ ,
\end{aligned}
\label{eq:ssd93}
\end{equation}
where $\epsilon_1,\epsilon_2=\pm1$ define the possible couplings of the O9-plane to the closed-string sector. Note that there is no tadpole cancellation condition fixing them. 

In the case of D9-branes, the possible combinations of Chan--Paton factors are
\begin{equation}
\begin{aligned}
    \alpha &= N_1 + \overline{N_1} + N_2 + \overline{N_2} \ , &&&&& \beta &= i ( N_1 - \overline{N_1} + N_2 - \overline{N_2}) \ , \\
    \gamma  &= N_1 + \overline{N_1} - N_2 - \overline{N_2} \ , &&&&& \delta &= i ( N_1 -\overline{N_1} - N_2 + \overline{N_2}) \ . 
\end{aligned}
\label{eq:ssd94}
\end{equation}
The two independent Chan--Paton factors $N_1$ and $N_2$ correspond to two possible type of D9-branes, which differ by the sign of the associated couplings to the closed-string tachyonic scalar $O_8$. Focusing on the case with only one type of D9-branes, say $N_1=1$ and $N_2=0$, the open spectrum is described by the loop-channel amplitudes
\begin{equation}
\begin{aligned}
    A_{99} &=\frac{1}{\eta^8}\left[ N  \overline{N}   \sum_m \(V_8 P_m - S_8 P_{m+1/2}\) + \frac{N^2 +\overline{N}^2 }{2}\sum_m \(O_8 P_m - C_8 P_{m+1/2}\) \right] \ , \\
    M_9 &= - \frac{1}{2\hat{\eta}^8} \sum_m (-1)^m \left[ \epsilon_1\( N + \overline{N}\) \ O_8 P_m -\epsilon_2 \( N -  \overline{N}\) C_8 P_{m+1/2} \right] \ .
\end{aligned}
\label{eq:ssd95}
\end{equation}
Therefore, the gauge group is $\mathrm{U}(N)$ and there are complex tachyonic scalars in the symmetric and antisymmetric representations of $\mathrm{U}(N)$. Tachyon condensation would break the gauge group to either $\mathrm{SO}(N)$ or $\mathrm{USp}(N)$ if the tachyons are in the symmetric or the antisymmetric representation. Moreover, even if these tachyons can be antisymmetrized for $\epsilon_1=1$, the first Kaluza--Klein mode of the scalar state inside $O_2O_6\supset O_8$ is tachyonic for $R>\sqrt{2\alpha^\prime}$, \textit{i.e.} at large radius, which is the limit in which one can avoid the closed-string tachyon. The only allowed value would be $R=\sqrt{2\alpha^\prime}$, where they both become massless. Thus, in the working limit of large radius, the D9-brane is unavoidably  unstable.\\

This behavior should have a geometrical interpretation, along the lines discussed in the previous paragraph. We can do so starting from F-theory compactified on one extra circle of coordinate $X^8\sim X^8 + 2\pi \alpha^{\prime} / R_{\rm B}$. After T-duality along $X^8$, the D9-brane becomes a D8-brane orthogonal to $X^8$. It has been argued that this D8-brane should correspond to an exotic M9-brane in M-theory \cite{Bergshoeff:1996ui,Bergshoeff:1997ak,Hull:1997kt,Bergshoeff:1998bs}, that couples to a $B_{10,1,1}$ form in eleven dimensions \cite{Kleinschmidt:2003mf}. Following this picture, the M9-brane should wrap a 2-cycle in the $\mathcal{M}_3$ manifold defining the dual M-theory under consideration, along the directions $X^9$ and $X^{1\hspace{-0.3mm}0}$. However, there is no such a cycle in the $\mathcal{M}_3$ manifold: this should explain the D9-brane instability from the M-theory point of view.

\subsubsection{Charged D7-branes wrapping the circle}

The D7-branes are charged and originate from a pair of type~IIB D7-branes of complex conjugate Chan--Paton factors whose relative position is frozen by the orientifold projection. The associated transverse M\"obius amplitude describes the tree-level propagation of closed-string states between the O9-plane and the D7-branes. Following the character decomposition associated with $\mathrm{SO}(8) \to \mathrm{SO}(6) \times \mathrm{SO}(2)$, as in eq.~\eqref{eqn:ch3_characters_dec_rule}, the M\"obius transverse channel is given by
\begin{equation}
    {\tilde \M}_7 = -\frac{2v}{\hat{\eta}^5 \hat{\vartheta}_2} \sum_n \left[ \epsilon_1 \gamma \left({\hat O}_6 {\hat O}_2 + {\hat V}_6 {\hat V}_2\right) (-1)^n- \epsilon_2 \delta \left({\hat S}_6 {\hat C}_2 - {\hat C}_6 {\hat S}_2\right)\right] W_{2n+1} \ .
\label{eq:ssd72}
\end{equation}

In this case, the correct parameterization of the Chan--Paton factors introduced in eq.~\eqref{eq:ssd91} is
\begin{equation}
\begin{aligned}
    \alpha &= N_1 + \overline{N_1} + N_2 + \overline{N_2}, &&&&& \gamma &= i ( N_1 -\overline{N_1} + N_2 - \overline{N_2}), \\
    \beta  &= N_1 + \overline{N_1} - N_2 - \overline{N_2}, &&&&& \delta &= i  ( N_1 -\overline{N_1} - N_2 + \overline{N_2}). 
\end{aligned}
\label{eq:ssd71}
\end{equation}

\noindent The physical couplings of the D7-branes to the R-R sector $S_8$ is given by the coefficient $\beta$. From the type~IIB point of view, if $N_1$ is associated with D7-branes, then $N_2$ is associated with $\overline{\text{D}7}$ antibranes. Considering the case with no antibranes, \textit{i.e.} $N_1= N$ and $N_2=0$, the open-string spectrum is described by the loop-channel amplitudes
\begin{equation}
\begin{aligned}
    \A_{77} &=\frac{1}{\eta^8}\left[N  \overline{N}\sum_m (V_8 P_m - S_8 P_{m+1/2}) + \frac{N^2 + \overline{N}^2 }{2} \sum_m \left(V_8 P_{m+1/2} - S_8 P_{m}\right)\right] \ ,  \\
    \M_7 &=\left( \frac{2}{\hat{\eta}^5 \hat{\vartheta}_2} \right) \frac{N- \overline{N} }{2} \sum_m  (-1)^m\left[\epsilon_1\left({\hat V}_6 {\hat O}_2 - {\hat O}_6 {\hat V}_2\right) P_{m+1/2}+\epsilon_2 \left({\hat S}_6 {\hat S}_2 - {\hat C}_6 {\hat C}_2\right) P_{m} \right] \ .
\end{aligned}
\label{eq:ssd73}
\end{equation}

\noindent The gauge group is  $\mathrm{U}(N)$ and the massless spectrum contains gauge vectors and a complex scalar in the adjoint representation, together with Weyl fermions of one chirality in the antisymmetric (or symmetric) representation and Weyl fermions of opposite chirality in the symmetric (or antisymmetric) representation of the gauge group. Note that there is no coupling to any tachyonic scalar, so that this D7-brane should be stable.\\

Let us now describe this charged brane in the dual F-theory picture. Consider the theory on one extra circle of coordinate $X^8\sim X^8 + 2\pi \alpha^{\prime} / R_{\rm B}$ and a D7-brane wrapping both circles of coordinates $(X^8,X^9)$. After T-duality along $X^8$, the D7-brane becomes a D6-brane orthogonal to $X^8$. This can be interpreted in M-theory as a KK6-brane:
\beq 
    {\rm D7}_{012345 89} \underset{{\rm T-duality}}{\rightarrow} {\rm KK6}_{0123459} \ . 
\eeq
This KK6-brane is localized at one of the fixed points $X^{10} = 0$ or $X^{10} = \pi e^{\frac{2}{3}\Phi_{\rm A}} \sqrt{\alpha^{\prime}}$ \cite{Atiyah:2001qf} and wraps the $\mathbb{Z}$ 1-cycle of $H_1\(\M_3,\mathbb{Z}\)$ along the coordinate $X^9$ in $\mathcal{M}_3$. The charge of the D7-brane is interpreted, in this geometrical picture, as the winding number of the KK6-brane along the $X^9$ 1-cycle in F-theory. A second possibility is to consider a D7-brane orthogonal to the circle subject to T-duality. This can be interpreted as a M9-brane wrapping $\mathcal{M}_3$. 

\subsubsection{Charged D7-branes orthogonal to the circle}

A similar description holds also for the charged D7-branes orthogonal to the circle. They are charged and one does not expect any tachyon-like scalar. These branes have one-point couplings to the Kaluza--Klein closed-string states along the circle. Due to the shift $\delta$ in the orientifold $\Omega^\prime$ \eqref{ss4-6}, these branes have a doublet structure, sharing the same Chan--Paton factor and being at a distance equal to $\pi R$. This will be the case for all orthogonal branes.\footnote{Similar structures are present also in other shift orientifolds \cite{aads1}.} The direct and transverse channel cylinder amplitudes are given by
\begin{align}
    & A_{77} = \frac{V_8-S_8}{\eta^8}\sum_n \left( N  \overline{N}  W_n +\frac{N^2 + \overline{N}^2}{2} W_{n+1/2} \right) \ , \label{eq:ssd75}\\
    &\begin{aligned}
    {\tilde \A}_{77} &= \frac{2^{-4}}{2 v}\left(\frac{V_8-S_8}{\eta^8}\right) \sum_m 
\left[ e^{\frac{i \pi m}{2}} N + e^{-\frac{i \pi m}{2}} \overline{N}  \right]^2 P_m  \\
    &= \frac{2^{-4}}{2 v} \left(\frac{V_8-S_8}{\eta^8}\right)\sum_m \left[ (N +\overline{N})^2 P_{2m} -(N -\overline{N})^2P_{2m+1}\right] \\
    &= \frac{2^{-4}}{v}\left(\frac{V_8-S_8}{\eta^8}\right)\sum_m \left[ N  \overline{N} + (-1)^m\frac{N^2 + \overline{N}^2}{2}\right] P_m \ .
\end{aligned} \label{eq:ssd74}
\end{align}

\noindent There are no closed-string states that couple simultaneously to the O9-plane and these D7-branes, since the transverse Klein bottle amplitude of eq.~\eqref{ss4-7} only involves massive winding states and the transverse cylinder only momentum states. Consequently, the M\"obius amplitude cannot be written. This feature already appeared in the D-brane spectrum of the Dabholkar--Park orientifold discussed in Section \ref{ch3_sec_dp_branes}. The cylinder amplitudes \eqref{eq:ssd75}-\eqref{eq:ssd74} should then be consistent by themselves. Indeed, in the tree-level channel only even Kaluza--Klein closed-string states of the NS-NS sector $V_8$ and R-R sector $S_8$ have physical couplings, while odd ones have unphysical couplings, in agreement with the physical states of the closed sector. Then, looking at the spectrum given by the loop-channel, the massless gauge group is $\mathrm{U}(N)$ and the states with shifted mass $W_{n+1/2}$ are recognized to be degenerate: states of winding numbers $n+\frac12$ and $-n-\frac12$ have the same mass and can be decomposed into even and odd combinations under the orientifold projection. The corresponding Chan--Paton representations should therefore be interpreted as symmetric $\boldsymbol{N(N{+}1)/2}$ plus antisymmetric $\boldsymbol{N(N{-}1)/2}$ representations of the gauge group $\mathrm{U}(N)$, plus their complex conjugates. This is a further similarity with the Dabholkar--Park case. Moreover, this spectrum has the surprising feature of being supersymmetric, despite the closed sector has no leftover supersymmetry. One possible explanation is that the orientifold projection $\Omega^\prime$  acts non-trivially only on momentum states, while the closed-string winding states have the same structure as in the more conventional Scherk--Schwarz compactifications.\\

As in the wrapping case, also the orthogonal D7-branes can be described in F-theory terms. Consider the theory on one extra circle of coordinate $X^8\sim X^8 + 2\pi \alpha^{\prime} / R_{\rm B}$ and a D7-brane wrapping this circle. Performing a T-duality along $X^8$, the D7-brane becomes a D6-brane orthogonal to $X^8$ and $X^9$ that can be interpreted as a KK6 brane in M-theory:
\beq 
    {\rm D7}_{0123456 8} \underset{{\rm T-duality}}{\rightarrow} {\rm KK6}_{0123456} \ . 
\eeq
This KK6 brane is localized at a point in $\mathcal{M}_3$ and it has an associated $H_0(\mathcal{M}_3,\mathbb{Z})=\mathbb{Z}$ charge. Alternatively, one can consider also a D7-brane orthogonal to the circle of T-dualization. This yields a M9-brane wrapping $T^2(X^8,X^{1\hspace{-0.3mm}}) \in \mathcal{M}_3$. 

\subsubsection{Non-BPS D5-branes orthogonal to the circle}

As for the orthogonal D7-branes, the shift $\delta$ in the orientifold $\Omega^\prime$ \eqref{ss4-6} generates a doublet structure, of equal Chan--Paton factors. The cylinder amplitudes are given by
\begin{equation}
\begin{aligned}
    {\tilde \A}_{55}&=\frac{2^{-3}}{2 v}\frac{1}{\eta^8} \sum_m  \left[\left(e^{\frac{i \pi m}{2}} N + e^{-\frac{i \pi m}{2}} \overline{N}  \right)^2 V_8 +\left(e^{\frac{i \pi m}{2}} N - e^{-\frac{i \pi m}{2}} \overline{N}  \right)^2 S_8\right]P_m \\
    &=\frac{2^{-3}}{2 v}\frac{1}{\eta^8}\sum_m \left[ (N+{\overline N})^2\left(V_8 P_{2m}-S_8 P_{2m+1}\right)-(N-{\overline N})^2\left(V_8 P_{2m+1}-S_8 P_{2m}\right)\right] \\
    &= \frac{2^{-4}}{v}\frac{1}{\eta^8} \sum_m \left[ N {\overline N} \ (V_8-S_8) +\frac{N^2 + \overline{N}^2}{2} (-1)^m (V_8+S_8)   \right] P_m \ , \\
    {\A}_{55}&=\frac{1}{\eta^8}\left[ N {\overline N} (V_8-S_8) \sum_n W_n +  \frac{N^2 + \overline{N}^2}{2}(O_8-C_8) \sum_n W_{n+1/2} \right] \ .  
\end{aligned}
\label{ssd53}
\end{equation}

\noindent Note that the physical couplings in the transverse channel of this brane are with the even modes of the closed NS-NS sector $V_8$ but with the odd modes of the R-R sector, as they are indeed uncharged. Moreover, there is no shared closed-string state with the transverse Klein bottle amplitude \eqref{ss4-7}, which contains only massive winding states, and thus the M\"obius amplitude is identically zero, since there are no closed-string states that couple simultaneously to the O9-plane and the D5-branes. The cylinder amplitude should therefore be consistent by itself, and this can be checked following the same argument as for the orthogonal D7-branes. In particular, the potentially-tachyonic open-string scalars in the loop-channel amplitude actually have positive squared masses for the same values of the radius $R > \sqrt{2 \alpha^{\prime}}$, \textit{i.e.} where there is no closed-string tachyon. If there is more than one such a brane, the brane and anti-brane will rotate by half a period to annihilate, but a single brane is stable and carries a $\mathbb{Z}_2$ charge. This is the same mechanism that occurs also for the orthogonal D3 and D7-branes in the Dabholkar--Park orientifold, as discussed in Section \ref{ch3_sec_dp_branes}. \\

This $\mathbb{Z}_2$ charge of the orthogonal D5-brane has an F-theory interpretation. Consider the theory on one extra circle of coordinate $X^8$ and a D5 brane wrapping this circle. T-duality along this circle maps this D5-brane into an M5-brane wrapping the M-theory circle:
\beq 
{\rm D5}_{01234 8} \underset{{\rm T-duality}}{\rightarrow} {\rm M5}_{01234 1\hspace{-0.3mm} 0} \ . 
\eeq
This circle is a $\mathbb{Z}_2$ torsion 1-cycle in $\mathcal{M}_3 $, leading to the $\mathbb{Z}_2$ charge. Similar conclusions are obtained considering instead  a D5-brane orthogonal to $X^8$, with 
\beq 
{\rm D5}_{012345} \underset{{\rm T-duality}}{\rightarrow} {\rm KK6}_{012345 8} \ , 
\eeq
since $X^8$ is also the coordinate of a torsion 1-cycle in $\mathcal{M}_3$. 

\subsubsection{Charged D3-branes wrapping the circle}

\begin{table}[t]
    \centering
    \begin{tabular}{ccc}
    \toprule
      KK number   &  Fields & Representation \\
      \midrule
      \multirow{2}*{$2m$}   & vectors $A^{(m)}_\mu$+6 scalars & \scalebox{0.9}{$\boldsymbol{N\overline{N}}$} \\
      & 4 fermions & $\boldsymbol{\frac{N(N-1)}{2} + \overline{\frac{N(N+1)}{2}}}$ \\
      \midrule
      \multirow{2}*{$2m+1$}   & vectors $A^{(m)}_\mu$+6 scalars & \scalebox{0.9}{$\boldsymbol{N\overline{N}}$} \\
      & 4 fermions & $\boldsymbol{\frac{N(N+1)}{2} + \overline{\frac{N(N-1)}{2}}}$ \\
      \midrule
      \multirow{3}*{$2m+\frac12$} & vectors $A^{(m)}_\mu$ & $\boldsymbol{\frac{N(N-1)}{2} + \overline{\frac{N(N+1)}{2}}}$ \\
       & 4 fermions & \scalebox{0.9}{$\boldsymbol{N\overline{N}}$} \\
       & 6 complex scalars & $\boldsymbol{\frac{N(N+1)}{2} + \overline{\frac{N(N-1)}{2}}}$ \\
       \midrule
       \multirow{3}*{$2m+\frac32$} & vectors $A^{(m)}_\mu$ & $\boldsymbol{\frac{N(N+1)}{2} + \overline{\frac{N(N-1)}{2}}}$ \\
       & 4 fermions & \scalebox{0.9}{$\boldsymbol{N\overline{N}}$} \\
       & 6 complex scalars & $\boldsymbol{\frac{N(N-1)}{2} + \overline{\frac{N(N+1)}{2}}}$ \\
      \bottomrule
    \end{tabular}
    \caption{Kaluza--Klein spectrum of D3-branes wrapping the circle defined as four-dimensional $\N=4$ super-Yang--Mills fields along the circle. The fields of mode number $m$ are complex conjugate to the fields of mode number $-m$.}
    \label{tab:D3_KK_spec}
\end{table}

Analogously to the charged D7-branes, also the D3-branes are charged and originate form a pair of type~IIB D3-branes with complex conjugate Chan--Paton factors, bounded together by the orientifold $\Omega'$ in a single charged object in the projected theory. One expects no tachyon-like scalars and, thus, a stable brane. The Chan--Paton factors organize in the same way as for the D7-brane, namely as in eq.~\eqref{eq:ssd71}. For only one type of branes, \textit{i.e.} $N_1=N$ and $N_2=0$, the loop-channel open string amplitudes are equal to
\begin{equation}
\begin{aligned}
    \A_{33} &=\frac{1}{\eta^8} \left[N  \overline{N}\sum_m (V_8 P_m - S_8 P_{m+1/2}) +\frac{N^2 + \overline{N}^2 }{2} \sum_m (V_8 P_{m+1/2} - S_8 P_{m}) \right] \ ,\\
    \M_3 &=\left( \frac{8 \hat{\eta}}{\hat{\vartheta}_2^3} \right) \frac{N- \overline{N}}{2}  \sum_m  (-1)^m\left[\epsilon_1\left({\hat O}_2 {\hat V}_6 - {\hat V}_2 {\hat O}_6\right) P_{m+1/2}+\epsilon_2\left({\hat S}_2 {\hat S}_6 - {\hat C}_2 {\hat C}_6\right)P_{m}\right] \ .
\end{aligned}
\label{eq:ssd31}
\end{equation}

\noindent In the M\"obius amplitude, $\epsilon_i = \pm 1$ and the decomposition $\mathrm{SO}(8) \to \mathrm{SO}(2) \times \mathrm{SO}(6)$ has been implemented to the characters, following eq.~\eqref{eqn:ch3_characters_dec_rule}. Thus, the massless spectrum contains $\mathrm{U}(N)$ gauge vectors and four fermions in its symmetric and antisymmetric representations. The spectrum lives in three dimensions and can be interpreted as the orbifold of $\mathcal{N}=4$ super Yang--Mills on $S^1$ of radius $R_o=2R$ with gauge group U$(2N)$. The associated Kaluza--Klein modes are summarized in Table \ref{tab:D3_KK_spec}. To see this, we define the unitary matrix 
\beq 
    \varsigma = \left(\begin{array}{cc} 0  &  i \mathbb{I}_{N\times N}\\ \mathbb{I}_{N\times N} & 0 \end{array}\right) \ , 
\eeq
along with the automorphism 
\begin{equation}
\begin{aligned}
    Z_4 A_{\mu}(X) \hspace{-2mm}&=- \varsigma A_{\mu}(X+\pi R)^\intercal \varsigma^\dagger \ , &&&  Z_4 \phi_{ij}(X) &=  \varsigma  \phi_{ij}(X+\pi R)^\intercal \varsigma^\dagger \ , \\
    Z_4 \lambda_{\alpha i}(X) \hspace{-2mm}&= i  \varsigma  \lambda_{\alpha i}(X+\pi R)^\intercal \varsigma^\dagger \ , &&& Z_4 \lambda_{\dot{\alpha}}^i(X) &= - i  \varsigma  \lambda_{\dot{\alpha}}^i(X+\pi R)^{\intercal} \varsigma^\dagger \ ,
\end{aligned}
\end{equation}
which is a symmetry of the four-dimensional super-Yang--Mills theory. The D3-brane spectrum is obtained by projecting this super-Yang--Mills theory by the $\mathbb{Z}_4$ group. This reproduces the low-energy theory on the wrapping D3-brane worldvolume. One can see that the beta-function vanishes at one-loop in the  orbifold theory, and that at energy scale $E\gg 1/R$ one recovers the $\mathcal{N}=4$ super-Yang--Mills theory, so that one expects this theory to be UV finite at all orders in perturbation theory. In addition, the conjectured $\mathrm{SL}(2,\mathbb{Z})$ duality of the new orientifold suggests that this non-supersymmetric D3-brane theory may admit $\mathrm{SL}(2,\mathbb{Z})$ Montonen--Olive duality \cite{Montonen:1977sn}. Confirming this explicitly would be, nevertheless, a nontrivial check. \\

This charged D3-brane can also be associated with a cycle in F-theory. Consider the theory on an extra circle of coordinate $X^8$. The D3 brane wrapping the circle of coordinate $X^9$ can be T-dualized to either 
\beq 
{\rm D3}_{0189} \underset{{\rm T-duality}}{\rightarrow} {\rm M2}_{01 9} \ ,
\eeq
where the M2-brane wraps the $\mathbb{Z}$ 1-cycle along $X^9$ in $\mathcal{M}_3$, or to
\beq
{\rm D3}_{012 9} \underset{{\rm T-duality}}{\rightarrow} {\rm M5}_{012 891\hspace{-0.3mm}0}  \ , 
\eeq
in which the M5-brane wraps the whole $\mathcal{M}_3$ manifold.

\subsubsection{Non-BPS D1-branes wrapping the circle}

The D1-branes are uncharged and their parameterization follows that of the D9-brane. In the minimal setup with one type of D1-brane, the transverse channel amplitudes are found to be
\begin{equation}
\begin{aligned}
    {\tilde \A}_{11} &= \frac{2^{-1}v}{2\eta^8}  \sum_n \left[ (N + {\overline N})^2 (V_8 W_{2n}+ O_8 W_{2n+1}) + (N - {\overline N})^2 (S_8 W_{2n}+ C_8 W_{2n+1})\right], \\
    {\tilde \M}_1 &= - v \left( \frac{2 \hat{\eta}}{\hat{\vartheta}_2} \right)^4 \sum_n\left[ \epsilon_1  (N + {\overline N}) \left({\hat O}_0 {\hat O}_8-{\hat V}_0 {\hat V}_8\right) + i\epsilon_2 (-1)^n  (N - {\overline N}) \left({\hat S}_0 {\hat C}_8- {\hat C}_0 {\hat S}_8 \right) \right]   W_{2n+1} \ .
\end{aligned}
\label{eq:ssd11}
\end{equation}

\noindent Instead, the loop-channel amplitudes, describing the open-string spectrum, are equal to
\begin{equation}
    \begin{aligned}
        \A_{11} &=\frac{1}{\eta^8}\left[N  \overline{N}   \sum_m \left(V_8 P_m - S_8 P_{m+1/2}\right) + \frac{N^2 + \overline{N}^2 }{2}\sum_m \left(O_8 P_m - C_8 P_{m+1/2}\right) \right], \\
        \M_1&= \left(\frac{2 \hat{\eta}}{\hat{\vartheta}_2} \right)^4 \sum_m (-1)^m\left[\epsilon_1 \frac{N+ \overline{N}}{2} \left({\hat O}_0 {\hat O}_8 - {\hat V}_0 {\hat V}_8 \right) P_{m}- \epsilon_2 \frac{N- \overline{N}}{2} \left({\hat S}_0 {\hat C}_8 - {\hat C}_0 {\hat S}_8 \right) P_{m+1/2} \right] \ .
    \end{aligned}
\label{eq:ssd12}
\end{equation}
The character decomposition along the two-dimensional D1-brane worldvolume is, according to eq.~\eqref{eqn:ch3_characters_dec_rule},
\beq
\begin{aligned}
    O_8 & = O_0 O_8 + V_0 V_8 \ , &&& V_8 &= V_0 O_8 + O_0 V_8 \ , \\
    S_8 &= S_0 S_8 + C_0 C_8 \ , &&& C_8 &= S_0 C_8 + C_0 S_8 \ ,
\end{aligned}
\label{eq:ssd13}
\eeq
in which $O_0= 1$ describes a scalar, $V_0 = 0$ describes a two-dimensional gauge boson,  that does not carry any dynamical degrees of freedom, whereas $S_0=C_0=1/2$ describe two-dimensional fermions of opposite chirality. The massless spectrum contains gauge vectors of gauge group $\mathrm{U}(N)$, eight scalars in the adjoint representation and complex tachyon-like scalars, sitting in the symmetric/antisymmetric representation for $\epsilon_1=\pm1$, respectively. Similarly to the D9-branes, the tachyon can be antisymmetrized for $\epsilon_1 = 1$, but the first Kaluza--Klein mode is tachyonic at large radius $R > \sqrt{2 \alpha^{\prime}}$ anyway. Moreover, the $\epsilon_1= -1$ configuration is energetically favored, due to the interaction with the real closed-string scalar that becomes tachyonic at small radius. Therefore, the D1-brane wrapping the circle is unstable. 

From the point of view of the $\mathrm{SL}(2,\mathbb{Z})$ duality, the D1-brane transforms in a doublet with the fundamental F1-string. One check that the mass of this D1 brane is $M_{\scalebox{0.6}{D1}}= \frac{R}{\alpha^{\prime}}$, consistently with the one in eq.~\eqref{MassD1} predicted from the proposed non-perturbative $\mathrm{SL}(2,\mathbb{Z})$-invariant potential $\hat{V}$ of eq.~\eqref{eqn:ch3_V_SL_full}. \\

The instability of the D1-brane wrapping the circle can be interpreted from the F-theory viewpoint. The naive T-dual brane in this picture would be 
\beq 
    {\rm D1}_{09} \underset{{\rm T-duality}}{\rightarrow} {\rm M2}_{0 89} \ , 
\eeq
but there is no 2-cycle along $X^8$ and $X^9$ in $\mathcal{M}_3$. Similarly, a fundamental string corresponding to M2$_{091\hspace{-0.3mm}0}$ does not wrap a cycle in $\mathcal{M}_3$ and it is expected to be unstable. 

\subsubsection{The remaining branes in the spectrum}

The D-branes that we did not yet discuss are the D5-branes  wrapping the circle, the D3 and D1-branes orthogonal to the circle, and the even branes. The wrapped D5-branes behave exactly like the D9 and the wrapping D1-branes. The gauge group is $\mathrm{U}(N)$ and the Kaluza--Klein spectrum contains a tachyonic state at large radius, rendering it unstable. Next, the orthogonal D3-branes are charged and very similar to the orthogonal D7-branes. In fact, also the theory on its worldvolume is supersymmetric, and its massless spectrum is that of $\N=4$ super-Yang--Mills. This theory is actually S-duality invariant, supporting the conjecture on the S-self duality of this orientifold  construction. Then, the orthogonal D1-branes are analogous to the orthogonal D5 ones. A single D1-brane is stable at large radius and carries a $\mathbb{Z}_2$ charge, while multiple D1-branes annihilate each other. Finally, even-dimensional branes wrapping the circle are tachyonic in the large radius limit $R>\sqrt{2\alpha^\prime}$, where the closed-string spectrum is stable. Among these branes, the D6 and D2-branes are expected to decay to the stable D5 and D1-branes through a kink transition \cite{Sen:1998tt}, as for the D4 and D8-branes of the Dabholkar--Park orientifold, discussed in Section \ref{ch3_sec_dp_branes}. Also the even branes orthogonal to the circle behave as the Dabholkar--Park ones, discussed in Appendix \ref{app_branes_dp}, and are inherently unstable for all values of the radius. A more detailed description of these branes and their behavior, including their dual F-theory interpretation, can be found in Appendix \ref{app_branes_new_ss}.

\subsection{Twisted O-planes in 5D}

We saw in Section \ref{ch3_sec_6D-twop} that twisted orientifold planes can also appear in lower-dimensional constructions, and we described the specific case of the supersymmetric six-dimensional K3 orientifold of \cite{Dabholkar:1996zi,Gimon:1996ay}. In \cite{Bossard:2024mls} we also construct, alongside the main nine-dimensional Scherk--Schwarz orientifold \eqref{ss4-6}, a related five-dimensional model with twisted orientifold planes, preserving half of the original supersymmetry. This 5D model is obtained from a freely-acting orbifold compactification of the type~IIB string on $(T^4 \times S^1 )/\mathbb{Z}_2 $, where the  $\mathbb{Z}_2$ symmetry is realized as the combination $Z_2 \delta_5$, with $Z_2$ being the reflection of all the $T^4$ coordinates:
\begin{align}
     Z_2 \ \(X^6,X^7,X^8,X^9\) = \(-X^6,-X^7,-X^8,-X^9\) \ , 
\end{align}
and $\delta_5$ being the half-circle shift in the fifth circle coordinate:
\begin{align}        
    \delta_5X^5 = X^5 + \pi R_5 \ .
\end{align}
After going to the analogue of the nine-dimensional Scherk--Schwarz basis through the replacement $R_5 \to 2 R_5$, the orbifold operation becomes $Z_2\,  \delta_5^2$, acting on the double-length circle, with $\delta_5^2 \, X_5 = X_5 + 2 \pi R_5$. Then, the orientifold action is taken to be 
\beq
    \Omega' = \Omega  Z_4  \delta_5 \ ,
\eeq
where $Z_4(X^6,X^7,X^8,X^9) = (-X^7,X^6, X^9,-X^8)$, as given in Section \ref{ch3_sec_6D-twop}. This orientifold symmetry $\Omega^\prime$ squares to the orbifold action in the "Scherk--Schwarz" basis
\beq
    \(\Omega^\prime\)^2=Z_2\delta_5^2 \ ,
\eeq
like the nine-dimensional orientifold \eqref{ss4-6}-\eqref{ss4-6-2} and the six-dimensional model of eq.~\eqref{6de1}.  

In this Scherk--Schwarz-like basis, the torus amplitude is given by
\begin{equation}
\begin{aligned}
\T =& \frac{1}{2} \left\{ \left|\frac{Q_o + Q_v}{\eta^8}\right|^2
\sum_{m,n} \left(q^{\frac{\alpha^{\prime}}{4} p_{{\rm L}}^{\intercal} g^{-1} p_{{\rm L}}} q^{\frac{\alpha^{\prime}}{4} p_{{\rm R}}^{\intercal} g^{-1} p_{{\rm R}} }\right) \sum_{m_5,n_5} \left( \Lambda_{m_5,2n_5}^{(1,1)}
+ \Lambda_{m_5+\frac{1}{2},2n_5}^{(1,1)} \right) \right .   \\ 
& \qquad \left . +\frac{16}{\left|\eta\right|^4}\left[ \left|\frac{Q_o - Q_v}{\vartheta_2^2}\right|^2 \sum_{m_5,n_5} \left( \Lambda_{m_5,2n_5}^{(1,1)} - \Lambda_{m_5+\frac{1}{2},2n_5}^{(1,1)} \right) \right.\right.\\
&\left.\left. \qquad\qquad +\left|\frac{Q_s - Q_c}{\vartheta_3^2} \right|^2  \sum_{m_5,n_5} \left( \Lambda_{m_5,2n_5+1}^{(1,1)} - \Lambda_{m_5+\frac{1}{2},2n_5+1}^{(1,1)} \right)  \right. \right. \\
&\left.\left. \qquad\qquad  + \left|\frac{Q_s + Q_c}{\vartheta_4^2}\right|^2 \sum_{m_5,n_5} \left( \Lambda_{m_5,2n_5+1}^{(1,1)}
+ \Lambda_{m_5+\frac{1}{2},2n_5+1}^{(1,1)} \right) \right] \right\} \ , 
\end{aligned}
\label{5d1} 
\end{equation}

\noindent where we employed the supersymmetric characters introduced in eq.~\eqref{6d12} \cite{orientifolds5,orientifolds6}. The Klein bottle amplitudes are found to be
\begin{equation}
\begin{aligned}
\K &= \frac{2^2}{2} \frac{Q_o-Q_v}{\eta^2\vartheta_2^2 }   \sum_{m_5} (-1)^{m_5} P_{m_5} \ ,  \\   
{\tilde \K} &= \frac{2^3}{2} \frac{R_5}{\sqrt{\alpha^{\prime}}} \frac{Q_s+Q_c}{\eta^2\vartheta_4^2}\sum_{n_5} W_{2n_5+1} \ .
\end{aligned}
\label{5d2} 
\end{equation}

\noindent The transverse Klein bottle shows that the orientifold planes of this model are indeed twisted, since they couple uniquely to the massive twisted sector, and therefore no background open strings are needed for consistency. In the decompactification limit $R_5 \to \infty$, the loop channel Klein bottle amplitude vanishes and the spectrum becomes that of the toroidal compactification of the type~IIB string on $T^4$. In full analogy with the nine-dimensional model \eqref{ss4-6}, the freely-acting orbifold operation becomes the identity in this limit, and the orientifold projection disappears.



\chapter{Summary and outlook}\label{ch_conclusions}

In this thesis I have gathered and presented the research work carried out throughout my doctoral training, which has been focused on the study of field and string theories featuring different realizations of supersymmetry. The content of the manuscript is based on the papers \cite{Bonnefoy:2022rcw,Casagrande:2023fjk,Bossard:2024mls,ms2}, which are discussed in the various chapters along with a comprehensive introduction to the theoretical frameworks in which they were developed.\\

In Chapter \ref{ch_susy} we introduced the formalism of constrained superfields to construct effective theories of nonlinearly realized rigid supersymmetry. The main element of these models is the goldstino, \textit{i.e.} the goldstone field of supersymmetry breaking, which is embedded into a nilpotent chiral superfield. This superfield allows to define further superfield constraints that effectively decouple superpartners from the multiplet they act upon, as well as to construct the associated interactions following the standard rules of supersymmetric theories. In \cite{Bonnefoy:2022rcw} we investigated the consistency of these superfield constraints and argue that those acting nontrivially on the auxiliary field of a supermultiplet are not well-defined constraints. Most notably, this is the case for the so-called orthogonal constraint. The reason is that, from the effective field theory point of view, these constraints do not descend from a microscopic two-derivative theory, but rather require higher-derivative operators already in the UV. This feature emerges in the low-energy lagrangians as non-trivial causality constraints on the goldstino scattering amplitudes, which must be imposed by hand in order to ensure a subluminal goldstino propagation. We substantiate our claim by constructing an alternative version of the ill-defined orthogonal constraint, which yields the same field content while avoiding any action on the auxiliary field. The resulting models are found to be automatically causal in the whole field space. As an application, we also formulate improved minimal models for inflation in supergravity along these lines.\\

In Chapter \ref{ch2_sugra} we move from rigid to local supersymmetry. In this context, the discussion focuses on two interesting massive fields: the massive gravitino in supergravity theories with spontaneously broken supersymmetry \cite{Casagrande:2023fjk}, and the massive spin-2 field coupled to $D=4$, $\N=1$ supergravity \cite{ms2}. The topic we address first is the dynamics of the massive gravitino in supergravity cosmological models. In the presence of local supersymmetry, the goldstino of supersymmetry breaking is absorbed by the gravitino, which becomes massive following the supersymmetric version of the Higgs mechanism. The constrained-superfield formalism presented in Chapter \ref{ch_susy} can still be used to construct effective theories of nonlinear (local) supersymmetry, which now describe (in the unitary gauge) the interactions of a massive gravitino field coupled to gravity, and are related to the massless goldstino ones through the equivalence theorem. In particular, the orthogonal constraint allows one to formulate minimal models of inflation in supergravity. However, in addition to the causality issues discussed in \cite{Bonnefoy:2022rcw}, these models were found to suffer from a further inconsistency, namely an unbounded gravitational production of massive gravitinos, occurring when the associated sound speed parameter vanishes. Motivated by this peculiar dynamics, which is only partially related to the effective field theory setup, we revisit the gravitational particle production mechanism in \cite{Casagrande:2023fjk}. The analysis we carry out in this paper stems from the fact that when discussing the gravitational particle production, the quantization of the system is usually performed with the "instantaneous hamiltonian" of the canonically normalized fields. This object correctly describes the equations of motion but not the energy of the system, which is rather encoded in the stress-energy tensor. This sources the metric in the Einstein equations and should be the quantity used to unambiguously describe any physical process in a time-dependent background, including particle production. In \cite{Casagrande:2023fjk} we parameterized the difference between the Fock spaces built from the instantaneous hamiltonian and the stress-energy tensor in terms of a Bogolyubov transformation, which we compute explicitly for massive particles of different spin, from the spin-0 scalar to the spin-$\frac32$ gravitino. Whereas for the lower-spin cases the two Fock spaces are different but align in the limit of large momenta, consistently with the physical expectation that the coupling to gravity should be suppressed in this limit, they do not align in the case of the gravitino. This ambiguity is reflected in the equivalence theorem, which is realized at the lagrangian level but not at the level of the energy operators. Moreover, contrary to the instantaneous hamiltonian, the energy operator computed from the stress-energy tensor does not display any divergent behavior in the limit of vanishing sound speed. These puzzling results of \cite{Casagrande:2023fjk} have to do with the intimate connections between the gravitino field and the spacetime geometry and require further precise studies in order to be fully understood. We intend to come back to this problem in the near future. \\

The second part of Chapter \ref{ch2_sugra} is dedicated to the study of the supersymmetric couplings of a single massive spin-2 field to supergravity. These couplings are typically difficult to construct because of an early loss of unitarity and the potential propagation of ghosts by the nonlinear interactions. In \cite{ms2} we studied the couplings to $D=4$, $\N=1$ off-shell supergravity of the massive spin-2 supermultiplet, by means of the supercurrent superfield formalism. This superfield contains the current of supersymmetry and the stress-energy tensor, and therefore defines the coupling to the supergravity massless fields at leading-order in the Planck mass. We found that consistency with local supersymmetry singles out a unique class of theories at the four-derivative level, characterized by a non-minimal coupling to the Riemann tensor. The core element of this set is a coupling to the so-called new-minimal off-shell supergravity. Additional couplings are possible in the context of non-minimal off-shell formulations, and they differ from the new-minimal one by interaction terms with the Ricci scalar $R$ and the $\gamma$-trace of the gravitino field-strength $\gamma^{\mu\nu}\rho_{\mu\nu}$. All these couplings reach strong coupling at a lower scale than the new-minimal one, according to the $2\to2$ scattering amplitude of the massive spin-2 field they describe, and can be eliminated at the three-point level by a redefinition of the massless fields. There is a further possible coupling of the massive spin-2 multiplet to supergravity, \textit{i.e.} the one arising from the Kaluza--Klein compactification of 5D pure supergravity on the orbifold $S^1/\mathbb{Z}_2$. We argue that this theory lies outside the scope of the supercurrent superfield analysis described, since preserving the St\"uckelberg symmetry --- which we used throughout the whole analysis ---  within the dimensional reduction requires the supersymmetry transformations of the massless supergravity fields to be deformed, in such a way that no (undeformed) off-shell formulation actually includes them. Along with a better energy behavior, the unique coupling to new-minimal supergravity we find in \cite{ms2} has two main features. First, the coupling between the massive spin-2 field and the graviton described by such a theory reproduces exactly the one that originates from the first oscillator mode of the open superstring. Second, it contains a specific non-minimal coupling of the Riemann tensor to the massive spin-1 field, superpartner of the massive spin-2 one, which is likely to lead to causality violations. This issue can be resolved by including a Regge trajectory of fields of higher mass and higher spin, which is precisely what happens in the string theory case. On the other hand, also the Kaluza--Klein coupling naturally comes with a tower of massive spin-2 multiplets, that are necessary to ensure unitarity. Our results of \cite{ms2} therefore suggest that string theory and Kaluza–Klein theories are the only frameworks in which the tree-level amplitudes of a single massive spin-2 multiplet coupled to supergravity are consistent with unitarity and causality. This picture provides evidence for the string lamppost principle from an explicit setup in four dimensions with minimal supersymmetry.

Future developments of this work consist in strengthening this claim by explicitly computing, on one hand, that the unique coupling found in \cite{ms2} does violate causality, and, on the other hand, that the Kaluza--Klein theory is the only possible coupling that implies a nontrivial deformation of the massless supergravity algebra. Moreover, we plan on extending and testing the uniqueness of the coupling to new-minimal supergravity to higher-derivative orders.\\

Finally, in Chapter \ref{ch3_st} we address the topics of string theory and orientifold projections. After a general introduction to string theory, D-branes and orientifold planes, we give an overview of various orientifold compactifications, with and without supersymmetry, which yield a very rich set of string vacua and allow one to probe and understand several crucial features of string theory. In this context, we then discuss in detail the new orientifold we constructed in \cite{Bossard:2024mls}. This is a Scherk--Schwarz deformation of the type~IIB string, which lives in nine dimensions, is tachyonic for values of the compactification radius smaller than $\sqrt{2\alpha^\prime}$, and does not require any background open sector. The peculiar feature of this string theory is what we called \textit{twisted O-planes}, namely orientifold planes that only couple to the massive twisted states of the theory. This model can be understood in the supergravity limit as a standard Scherk--Schwarz compactification to nine dimensions, as confirmed by the explicit computation of the string vacuum energy, although the symmetry employed in this reduction does not allow one to interpret the resulting theory in terms of a spontaneous breaking of supersymmetry. We also argue that the theory admits a formulation in terms of an F-theory compactification on a freely-acting orbifold. This, together with the analysis of the vacuum energy, provides evidence that this string theory is invariant under S-duality. In addition, we compute the D-brane spectrum of the theory and find that their stability properties exactly match the predictions that one can draw from the dual geometric formulation in F-theory. Alongside this analysis, we also describe lower dimensional orientifold compactifications, which display twisted orientifold planes with different amounts of supersymmetries.

Further studies in this context concern the conjectured S-self duality of the theory, which could give nontrivial information about its non-perturbative structure and dynamics. It would be of particular interest trying to use S-duality to investigate whether the non-perturbative potential admits a metastable vacuum.\\

I would like to conclude this thesis by mentioning two further research projects that were developed during my PhD, which I did not include in the manuscript for the sake of simplicity but which also played an important role in my doctoral formation. 

The first project concerns the so-called \textit{Scalar Weak Gravity Conjecture}, a bound on the scalar fields interactions in effective field theories that was proposed in the context of the swampland program \cite{Palti:2017elp,DallAgata:2020ino}. We propose a refined version of such conjecture based on arguments from dimensional reduction, the original Weak Gravity Conjecture and gauged supergravity, in which this scalar conjecture is realized as an identity. This project is carried out in collaboration with E.~Dudas and G.~Dall'Agata.   

The second work is about an interesting class of models of 3D gauged supergravity, dubbed \textit{RSTU models} in \cite{Arboleya:2024vnp}, that arise as flux compactifications of type~II string theories. Making use of the embedding tensor formalism and the construction of \cite{Arboleya:2024vnp}, we chart completely their rich landscape of vacua, which consists of various supersymmetric and non-supersymmetric $\mathrm{AdS}_3$ and $\mathrm{Mkw}_3$ solutions, and investigate their phenomenological properties, such as the possibility of achieving scale separation. This work is done in collaboration with A.~Arboleya, A.~Guarino and M.~Morittu.


\begin{appendices}

\chapter{Notations, Conventions, and useful formulae}\label{app_conventions}

This appendix collects the notations and conventions used throughout the manuscript. In general, we employ the following:
\vspace{0.5cm}
\begin{itemize}

    \item symmetrisation and antisymmetrisation of indices are denoted, respectively, with round and square brackets, and are always intended with weight one, \textit{e.g.}: 
    \begin{align}
        T_{(\mu\nu)}\equiv\frac12\(T_{\mu\nu}+T_{\nu\mu}\) \ , && T_{[\mu\nu]}\equiv\frac12\(T_{\mu\nu}-T_{\nu\mu}\) \ .
    \end{align}
    
    In addition, we utilize the notation
    \beq
        T_{\mu]\dots[\nu}\equiv T_{[\mu|\dots|\nu]}\ ,
    \eeq
    
    \noindent when convenient (see Section \ref{ch2_sec_ms2}).

\vspace{0.3cm}
    \item Commutation and anticommutation of operators are denoted, respectively, with square and curly brackets:
    \begin{align}
        \[O_1,O_2\]\equiv O_1 O_2-O_1O_2=2O_{[1}O_{2]} \ , && \left\{O_1,O_2\right\}\equiv O_1 O_2+O_1O_2=2O_{(1}O_{2)} \ .
    \end{align}

\vspace{0.3cm}
    \item The flat Minkowski metric is denoted with $\eta$, and it is taken with \textit{mostly-plus} convention, \textit{i.e.} $\eta=\mathrm{diag}\(-1,1,1,1,\dots\)$. When working in flat Minkowski spacetime, indices are denoted with Greek letters of the type $\(\mu,\nu,\rho,\dots\)$. When working in curved spacetime, the Greek letters denote the curved indices, while small Latin letters of the type $\(a,b,c,\dots\)$ denote the flat (local-Lorentz) indices. The curved metric is denoted with $g_{\mu\nu}$, the associated direct vierbein is denoted with $e_\mu{}^a$, its inverse with $e^\mu{}_a$, and its determinant with $e$.

\vspace{0.3cm}
    \item The definitions above are employed in any spacetime dimension. However, when different dimensions are involved in the discussion, it is useful to distinguish higher and lower-dimensional indices/quantities. In this respect, the set of rules above will generally apply to the lower-dimensional spacetime, while specific notations will be applied for the higher-dimensional one. For instance, in the dimensional reduction discussed in Appendix \ref{app_ms2_kk} (related to Section \ref{ch2_sec_stuck}), the 5D indices are denoted with capital Latin letters, of the type $(M,N,R,\dots)$ for curved ones, and of the type $(A,B,C,\dots)$ for flat ones. Instead, in the string theory setting of Chapter \ref{ch3_st}, the target space and worldsheet metric are distinguished by denoting the former with $G_{\mu\nu}$ and the latter with $g_{ab}$, where the indices $(a,b)$ label, in this context, the string coordinates. More specific notations are introduced, when needed case by case, in the main text.

 \vspace{0.3cm}
    \item The Levi--Civita connection and Lorentz spin-connection are defined as
    \beq
    \begin{aligned}
        \Gamma^{\mu}_{\nu\rho}&=\frac12g^{\mu\sigma}\(\partial_\rho g_{\nu\sigma}+\partial_\nu g_{\rho\sigma}-\partial_\sigma g_{\nu\rho}\) \ , \\
        \omega_{\mu}^{ab}&=2 e^{\nu}{}^{[a} \partial_{[\mu} e_{\nu]}{}^{b]}-e_{c}{}^{\mu} e^{\nu}{}^{[a} e^{b]}{}^{\sigma} \partial_{\nu} e_{\sigma}{}^{c} \ ,
    \end{aligned}
    \eeq

    \noindent and the Riemann, Ricci and Einstein tensors are 
    \begin{align}
        R^{\mu}_{\nu\rho\sigma}&=2\partial_{[\rho}\Gamma^\mu_{\sigma]\nu}+2\Gamma^\mu_{\lambda[\rho}\Gamma^\lambda_{\sigma]\nu} \ , && R_{\mu\nu}=R^{\rho}_{\mu\rho\nu} \ , && G_{\mu\nu}=R_{\mu\nu}-\frac12g_{\mu\nu}R \ .
    \end{align}

    \noindent The energy-momentum tensor associated with a lagrangian $\L$ is
    \beq
        T_{\mu\nu}=-\frac{2}{e}\frac{\delta\(e\L\)}{\delta g^{\mu\nu}}=-2\frac{\delta\L}{\delta g^{\mu\nu}}+\frac12g_{\mu\nu}\L \ .
    \eeq

\vspace{0.3cm}
    \item The Levi–Civita tensor is denoted, in arbitrary dimension $D$, with $\varepsilon^{\mu_1\dots\mu_D}$, and it is defined such that $\varepsilon^{012\dots(D-1)}=1$. The contraction of two Levi--Civita tensors satisfies the identity
    \beq
        \varepsilon^{\mu_1\dots \mu_p \rho_1\dots \rho_{D-p}}\varepsilon_{\nu_1\dots \nu_p\rho_1\dots \rho_{D-p}}=-\frac{1}{p!\(D-p\)!}\delta^{\mu_1\dots \mu_p}_{\nu_1\dots \nu_p} \ ,
    \label{eqn:app_conv_levi_civita_contr}
    \eeq

    \noindent in which we have introduced the completely antisymmetrized $\delta$-function: 
    \beq
    	\delta^{\mu_1\dots \mu_p}_{\nu_1\dots \nu_p}\equiv\delta^{\mu_1}_{[\nu_1}\dots \delta^{\mu_p}_{\nu_p]} \ .
    \eeq

\end{itemize}

%

\section{Clifford algebra}\label{app_conv_clifford}

This section contains the conventions and rules related to the spinor algebra. The focus will be on the four-dimensional case, unless explicitly stated otherwise. First, we present the algebra of spinors in two-component notation --- including a specific discussion on the superspace algebra --- and collect a series of identities that mainly concern the content of Chapter \ref{ch_susy}. Then, we do the same for the Dirac four-dimensional spinor algebra, which mainly refer to Chapter \ref{ch2_sugra}.

\subsection{Two-component spinors}\label{app_conv_grassmann}

Weyl spinors are two-component, grassmann-valued vectors which transform either in the $\(\frac12,0\)$ or in the $\(0,\frac12\)$ representations of the Lorentz group $\mathrm{SO}(1,3)\simeq\mathrm{SU}(2)\times\mathrm{SU}(2)^*$. The former are denoted with unbarred Greek letters and undotted indices, \textit{e.g.} $\psi_\alpha$, the latter with barred letters and dotted indices, \textit{e.g.} $\bpsi_{\dalpha}$. The two representations are related by hermitian conjugation, as
\beq
    \bpsi_{\dalpha}=\(\psi_\alpha\)^\dagger=\(\psi^\alpha\)^* \ .
\eeq    

\noindent Two-component spinor indices are raised and lowered using the 2D antisymmetric Levi--Civita tensors
\begin{align}
    \epsilon_{\alpha\beta}=\epsilon_{\dot{\alpha}\dot{\beta}}=\begin{pmatrix}0&-1\\1&0\end{pmatrix} \ , &&  \epsilon^{\alpha\beta}=\epsilon^{\dot{\alpha}\dot{\beta}}=\begin{pmatrix}0&1\\-1&0\end{pmatrix} \ ,
\end{align}

\noindent and index contractions are defined top-down for the $\psi_\alpha$ spinors and bottom-up for the $\bar{\psi}_{\dot{\alpha}}$ ones:
\begin{align}
    \psi\chi=\psi^\alpha\chi_\alpha \ , && \bpsi\bchi=\bpsi_{\dalpha}\bchi^{\dalpha} \ .
\end{align}

The Pauli matrices are taken to be
\begin{align}
     \sigma^1=\begin{pmatrix} 0 & 1 \\ 1 & 0  \end{pmatrix} \ , && \sigma^2=\begin{pmatrix} 0 & -1 \\ i & 0  \end{pmatrix} \ , && \sigma^3=\begin{pmatrix} 1 & 0 \\ 0 & -1  \end{pmatrix} \ , 
\end{align}

\noindent and they carry both dotted and undotted indices. They act through the vectors
\begin{align}
    \sigma^{m}=\(-\mathbb{1},\sigma^i\) \ , &&  \bsigma^{m}=\(-\mathbb{1},-\sigma^i\) \ , && \bsigma^{m}{}^{\dalpha\alpha}=\varepsilon^{\dalpha\dot{\beta}}\varepsilon^{\alpha\beta}\sigma^{m}_{\beta\dot{\beta}} \ .
\end{align}

\noindent and tensor combinations
\begin{align}
	\(\sigma^{{m}{n}}\)_\alpha{}^{\beta}=\frac14\(\sigma^{m}\bsigma^{n}-\sigma^{n}\bsigma^{m}\)_\alpha{}^{\beta} \ , && \(\bsigma^{{m}{n}}\)^{\dalpha}{}_{\dot{\beta}}=\frac14\(\bsigma^{m}\sigma^{n}-\bsigma^{n}\sigma^{m}\)^{\dalpha}{}_{\dot{\beta}} \ ,
\end{align}

\noindent which satisfy the following identities:
\beq
\begin{aligned}
&\mathrm{Tr}\(\sigma^{{m}} \bar{\sigma}^{{n}}\)=-2 \eta^{{m}{n}}\ , \\
&\sigma_{\alpha \dot{\alpha}}^{{m}} \bar{\sigma}_{{m}}^{\dot{\beta} \beta}=-2 \delta_{\alpha}^{\beta} \delta_{\dot{\alpha}}^{\dot{b}}\ , \\
& \sigma^{{m}} \bar{\sigma}^{{n}}+\sigma^{{n}} \bar{\sigma}^{{m}}=-2 \eta^{{m}{n}} \mathbb{1}\ , \\
&\bar{\sigma}^{{m}} \sigma^{{n}}+\bar{\sigma}^{{n}} \sigma^{{m}}=-2 \eta^{{m}{n}} \mathbb{1}\ , \\
& \sigma^{{m}} \bar{\sigma}^{{n}} \sigma^{p}=\left(-\eta^{{m}{n}} \eta^{{r}{s}}+\eta^{{m}{r}} \eta^{{n}{s}}-\eta^{{n}{r}} \eta^{{m}{s}}+i \epsilon^{{m}{n}{r}{s}}\right) \sigma_{s}\ , \\
& \bar{\sigma}^{{m}} \sigma^{{n}} \bar{\sigma}^{p}=\left(-\eta^{{m}{n}} \eta^{{r}{s}}+\eta^{{m}{r}} \eta^{{n}{s}}-\eta^{{n}{r}} \eta^{{m}{s}}-i \epsilon^{{m}{n}{r}{s}}\right) \bar{\sigma}_{s}\ . 
\end{aligned}
\eeq

Further useful identities on spinor combinations are instead
\beq
\begin{aligned}
    \theta^\alpha\theta^\beta&=-\frac{1}{2}\theta^2\epsilon^{\alpha\beta} \ , 		&&& \psi\chi=\psi^\alpha\chi_\alpha&=-\chi_\alpha\psi^\alpha=\chi\psi \ , \\
    \btheta^{\dalpha}\btheta^{\dbeta}&=\frac{1}{2}\btheta^2\epsilon^{\dalpha\dbeta} \ , &&& \bpsi\bchi=\bpsi_{\dalpha}\bchi^{\dalpha}&=-\bchi^{\dalpha}\bpsi_{\dalpha}=\bchi\bpsi \ , \\
	\theta_\alpha\theta_\beta&=\frac{1}{2}\theta^2\epsilon^{\alpha\beta} \ ,		&&&	\left(\theta\sigma^{m}\bpsi\right)\left(\theta\sigma^{n}\bchi\right)&=\frac{1}{2}\theta^2\left(\bpsi\bsigma^{m}\sigma^{n}\bchi\right) \ ,	 \\
    \btheta_{\dalpha}\btheta_{\dbeta}&=-\frac{1}{2}\btheta^2\epsilon_{\dalpha\dbeta} \ ,
     &&& \left(\psi\sigma^{m}\btheta\right)\left(\chi\sigma^{n}\btheta\right)&=\frac{1}{2}\btheta^2\left(\psi\sigma^{m}\bsigma^{n}\chi\right) \ , \\ 
	\psi\sigma^{m}\bchi&=-\bchi\bsigma^{m}\psi \ ,	&&& \left(\theta\sigma^{m}\btheta\right)\left(\theta\sigma^{n}\btheta\right)&=\frac{1}{2}\theta^2\btheta^2\eta^{{m}{n}} \ ,\\
    \theta\sigma^{m}\bsigma^{n}\theta&=\theta^2\eta^{{m}{n}} \ ,  &&& \left(\theta\psi\right)\left(\theta\chi\right)&=-\frac{1}{2}\theta^2\psi\chi \ ,  \\
    \psi\sigma^{m}\bsigma^{n}\chi&=\chi\sigma^{n}\bsigma^{m}\psi \ , &&& \left(\btheta\bpsi\right)\left(\btheta\bchi\right)&=-\frac{1}{2}\btheta^2\bpsi\bchi  \ .
\end{aligned}
\eeq

\subsection{Superspace algebra}

In the superspace formalism one introduces auxiliary fermionic coordinates $\(\theta_\alpha,\btheta_{\dalpha}\)$. These are grassmann, \textit{i.e.} anticommuting coordinates, which are, by definition, nilpotent: 
\beq
    \theta_1^2=\theta_2^2=\btheta_{\dot 1}^2=\btheta_{\dot 2}^2=0 \,\, \longleftrightarrow \,\, \theta^3=\btheta^3=0 \ .
\eeq

\noindent A function of this coordinates has therefore a finite expansion. Taking one such coordinate $\vartheta$ as an example, one has 
\beq
    F(\vartheta)=F_0+F_1\vartheta \ .
\eeq

\noindent Integration in such grassmann variable $\theta$ follows the rules
\begin{align}
     \int d\vartheta =0 \ , && \int d\vartheta\, \vartheta = 1 \ ,
\end{align}

\noindent and
\begin{align}
        \int d\vartheta\,F(\vartheta)=F_1=\partial_\vartheta F(\vartheta) \ , && \int d\vartheta\,\vartheta F(\vartheta)=F_0 \ , && \int d\vartheta\,\partial_\vartheta F(\vartheta)=0 \ .
\end{align}

\noindent In the case of superspace, \textit{i.e.} with four such grassmann coordinates, one has the integration measures
\beq
\begin{aligned}
    d^{2} \theta&=-\frac{1}{4} \epsilon_{\alpha \beta} d \theta^{\alpha} d \theta^{\beta}=\frac{1}{2} d \theta^{1} d \theta^{2}\ , \\
    d^{2} \bar{\theta}&=-\frac{1}{4} \epsilon^{\dot{\alpha} \dot{\beta}} d \bar{\theta}_{\dot{\alpha}} d \bar{\theta}_{\dot{\beta}}=-\frac{1}{2} d \bar{\theta}_{1} d \bar{\theta}_{2} \ , \\
    d^{4} \theta&=d^{2} \theta d^{2} \bar{\theta} \ ,
\end{aligned}
\eeq

\noindent and basic integration rules
\beq
    \int \theta^{2} d^{2} \theta=\int \bar{\theta}^{2} d^{2} \bar{\theta}=1 \ .
\eeq

Superspace covariant derivatives have been defined in eq.~\eqref{eqn:ch1_ss_cov_dev} to be
\begin{align}
    D_\alpha=\partial_\alpha+i\(\sigma^{m}\btheta\)_\alpha\partial_{m} \ , && \bar{D}_{\dalpha}=-\bar{\partial}_{\dalpha}-i\(\theta\sigma^{m}\)_{\dalpha}\partial_{m} \ .
\end{align}

\noindent They also satisfy a set of useful identities, namely
\beq
\begin{aligned}
D_\alpha D_\beta&=-\frac12 \varepsilon_{\alpha\beta} D^2 \ , &&& \bar{D}_{\dalpha} \bar{D}_{\dot \beta}&=\frac12 \varepsilon_{\dalpha\dot{\beta}} \bar{D}^2 \ ,	\\
 \{D_{\alpha}, \bar{D}_{\dot{\alpha}}\}&=-2 i \sigma_{\alpha \dot{\alpha}}^{m} \partial_{m} \ , &&& \{D_{\alpha}, D_{\beta}\}&=\{\bar{D}_{\dot{\alpha}}, \bar{D}_{\dot{\beta}}\}=0 \ , \\
D^{2} \theta^{2}&=-4+4 i \theta \sigma^{m} \bar{\theta} \partial_m+\theta^{2} \bar{\theta}^{2} \square \ , &&&  \left[D_{\alpha}, \bar{D}^{2}\right]&=-4 i \sigma_{\alpha \dot{\alpha}}^{m} \bar{D}^{\dot{\alpha}} \partial_{m} \ , \\
 \bar{D}^{2} \bar{\theta}^{2}&=-4-4 i \theta \sigma^{m} \bar{\theta} \partial_m+\theta^{2} \bar{\theta}^{2} \square \ , &&& \left[D^{2}, \bar{D}^{2}\right]&=-8 i D^{\alpha} \sigma_{\alpha \dot{\alpha}}^{m} \bar{D}^{\dot{\alpha}} \partial_{m}+16 \square \ ,
\end{aligned}
\eeq


\subsection{Four-component spinors and Dirac matrices}\label{app_conv_dirac}

Dirac four-component spinors are obtained by joining together two distinct Weyl fermions:
\beq
    \Psi=\begin{pmatrix}  \psi_\alpha  \\ \bchi^{\dalpha} \end{pmatrix} \ .
\label{eqn:app_cliff_dirac_spinor}
\eeq

\noindent  The associated Dirac matrices $\gamma^m$ satisfy the Clifford algebra relation
\beq
    \left\{\gamma^m, \gamma^n\right\}=2\eta^{mn} \ , 
\label{eqn:app_cliff_algebra}
\eeq

\noindent so that their hermitian conjugates are such that 
\begin{align}
    \(\gamma^\mu\)^\dagger=\gamma^0\gamma^\mu\gamma^0 && \longleftrightarrow && \(\gamma^0\)^\dagger=-\gamma^0 \ , \,\,\, \ \(\gamma^i\)^\dagger=\gamma^i  \ ,
\end{align}

\noindent and Dirac conjugation is defined as
\beq
    \bPsi=i\Psi^\dagger\gamma^0 \ .
\eeq

\noindent Throughout the manuscript, we make use of the following explicit representation:
\begin{align}
    \gamma^m=\begin{pmatrix}
        \mathbb{0} & -i\sigma^m \\ -i\bsigma^m & \mathbb{0}
    \end{pmatrix} \ , && \gamma_5=i\gamma^0\gamma^1\gamma^2\gamma^3=\begin{pmatrix}
        \mathbb{1} & \mathbb{0} \\ \mathbb{0} & -\mathbb{1}
    \end{pmatrix} \ .
\end{align}

\noindent The $\gamma_5$ matrix is hermitian, \textit{i.e.} $\gamma_5^\dagger=\gamma_5$, and allows to introduce the chiral projectors
\begin{align}
    P_L=\frac{1-\gamma_5}{2}=\begin{pmatrix} \mathbb{0} &  \mathbb{0} \\  \mathbb{0} &  \mathbb{1}\end{pmatrix} \ , && P_R=\frac{1+\gamma_5}{2}=\begin{pmatrix} \mathbb{1} &  \mathbb{0} \\  \mathbb{0} &  \mathbb{0}\end{pmatrix} \ , &&
\end{align}

\noindent so that
\begin{align}
    \Psi_{L,R}\equiv P_{L,R} \Psi  \ , && \Psi_L=\begin{pmatrix}  0  \\ \bchi^{\dalpha} \end{pmatrix} \ , && \Psi_R=\begin{pmatrix}  \psi_\alpha  \\ 0 \end{pmatrix} \ , && \bPsi_{L,R}=\bPsi P_{L,R} = \overline{\Psi_{R,L}} \ .
\end{align}

The Clifford algebra \eqref{eqn:app_cliff_algebra} is spanned by the basis
\begin{align}
    \left\{\mathbb{1},\gamma^{{m}},\gamma^{{m}{n}},\gamma^{{m}}\gamma_5,\gamma_5\right\} \ ,
\label{eqn:app_cliff_basis}
\end{align}

\noindent whose elements satisfy the identity
\beq
    \gamma^{m_1\dots m_r}=-\frac{i}{\(4-r\)!}\varepsilon^{m_r\dots m_1r_1\dots r_{4-r}}\gamma_{r_1\dots r_{4-r}}\gamma_5 \ ,
\label{eqn:app_cliff_id_basis}
\eeq

\noindent where we employ the standard definition $\gamma^{{m}_1\dots{m}_r}\equiv\gamma^{[{m}_1}\dots\gamma^{{m}_r]}$. The associated charge conjugation matrix $C_4$ is taken to be
\begin{align}
    C_4=\begin{pmatrix}
        \sigma^2& \mathbb{0} \\  \mathbb{0} & -\sigma^2
    \end{pmatrix} \ , && C_4^\textup{T}=C_4^\dagger=C_4^{-1}=-C_4 \ ,
\end{align}

\noindent and it defines the symmetry property of the Clifford algebra \eqref{eqn:app_cliff_algebra}
\begin{align}
    \(C_4\gamma^{{m}_1\dots{m}_r}\)^\textup{T}=-t_rC_4\gamma^{{m}_1\dots{m}_r} \ , 
\end{align}

\noindent With $t_r=1$ for $r=0,3$ and $t_r=-1$ for $t=1,2$ modulo 4. Moreover, a generic element $M$ of the Clifford algebra can always be written as an expansion in the basis \eqref{eqn:app_cliff_basis}. This is the basic Fierz identity, given by
\beq
\begin{aligned}
    M=&\frac14\mathrm{Tr}\(M\)\mathbb{1}+\frac14\mathrm{Tr}\(\gamma^{m} M\)\gamma_{m}-\frac18\mathrm{Tr}\(\gamma^{mn}M\)\gamma_{{m}{n}}\\
    &-\frac14\mathrm{Tr}\(\gamma^{m}\gamma_5M\)\gamma_{m}\gamma_5+\frac14\mathrm{Tr}\(\gamma_5M\)\gamma_5 \ .
\end{aligned}
\eeq

The special Dirac fermion \eqref{eqn:app_cliff_dirac_spinor} which is made by hermitian conjugated components $\bchi^{\dalpha}\equiv\bpsi^{\dalpha}$ is called a Majorana fermion:
\begin{equation}
    \lambda=\begin{pmatrix}  \psi_\alpha  \\ \bpsi^{\dalpha} \end{pmatrix} \ .
\end{equation}

\noindent From the four-component point of view, a Majorana fermion is defined by the reality condition
\beq
    \bar{\lambda}=\lambda^\textup{T}C_4 \ .
\label{eqn:app_cliff_majorana}
\eeq

In curved spacetime, the above Clifford algebra is upgraded to
\begin{align}
    \left\{\gamma^{\mu},\gamma^\nu\right\}=2g^{\mu\nu} \ , && \gamma^\mu=e^{\mu}{}_m\gamma^m \ .
\end{align}


\section{Algebra of the helicity eigenstate decomposition}\label{app:hed_algebra}

In this section we collect the detailed equations and properties that characterize the helicity eigenstate decomposition employed in Section \ref{ch2_sugra_grav_cosmology} when discussing the dynamics of the gravitino in the FLRW spacetime. We recall that the flat Dirac matrices are denoted, in the context of this discussion, with Greek indices and a bar on top, following
\beq
    \gamma^\mu=e^\mu{}_a\gamma^a=a(\tau)^{-1}\gamma^a\equiv a(\tau)^{-1}\fgamma^\mu \ .
\eeq

The indices contraction rules and notation in momentum space are
\begin{equation}
\begin{aligned} 
    \vec{k}\cdot\vec{x}=&\sum_{i=1}^{3}k^ix^i=k^ix_i \ , & 
\vfgamma\cdot\vpsi_k=&\sum_{i=1}^3\fgamma^i\psi_{i,k}=\fgamma^i\psi_{i,k} \ , \\
	\vk\cdot\vpsi_k=&\sum_{i=1}^3k^i\psi_{i,k}=k^i\psi_{i,k} \ , &
	\vec{k}\cdot\vfgamma=&\sum_{i=1}^{3}k^i\fgamma^i=k_i\fgamma^i \ ,
\end{aligned}
\label{eqn:app_hed_k_contr}
\end{equation}

\noindent and the following identities hold:
\begin{align}
    \kg\kg=\mk \ , && \kg\fgamma^i\kg = 2k^i\kg-\mk\fgamma^i \ .
\end{align}

\noindent The projectors introduced in eq.~(\ref{eqn:ch2_psi_proj}) satisfy the following algebra:
\begin{equation}
\begin{aligned}
    &\fgamma^i\left(P_\gamma\right)_i=-1 \ , &&& &\fgamma^i\left(P_k\right)_i=0 \ , \\
	&k^i\left(P_\gamma\right)_i=0 \ , &&&    &k^i\left(P_k\right)_i=-1 \ , \\
	&\left\{\left(P_\gamma\right)_i,\fgamma_0\right\}=0 \ ,	&&&		&\left\{\left(P_\gamma\right)_i,\fgamma_j\right\}=-\eta_{ij}+\frac{k_ik_j}{\mk} \ , \\
	&\left[\left(P_\gamma\right)_k,\fgamma_0\right]=0 \ ,	&&&		&\left[\left(P_k\right)_i,\fgamma_j\right]=\frac{1}{\mk}\left[\fgamma_i k_j+\fgamma_j k_i+\eta_{ij}\kg\right] \ , \\
    &\left(P_\gamma\right)_i^\dagger=\left(P_\gamma\right)_i \ , &&& &\left(P_k\right)_i^\dagger=\frac{1}{2\mk}\left[\left(\vk\cdot\vfgamma\right)\fgamma_i-3k_i\right] \ , \\
    &\left(P_\gamma\right)^i\left(P_\gamma\right)_i=\frac{1}{2} \ , &&& &\left(P_\gamma\right)^i\left(P_k\right)_i=-\frac{1}{2\mk}\kg \ ,\\
    &\left(P_k\right)^i\left(P_k\right)_i=\frac{1}{4\vec{k}{}^4} \ , &&&   &\left(P_k\right)^i\left(P_\gamma\right)_i=\frac{1}{2\mk}\kg \ ,\\
    &\left(P_k\right)^{i\,\dagger}\left(P_k\right)_i=\frac{3}{4\vec{k}{}^4} \ , &&& &\left(P_k\right)^{i\,\dagger}\left(P_\gamma\right)_i=-\frac{1}{2\mk}\kg \ .
\end{aligned}
\label{eqn:app_hed_proj_algebra}
\end{equation}

\noindent The operator $\hO_i$ introduced in eq.~(\ref{eqn:ch2_psi_dec})-(\ref{eqn:ch2_psi_dec_O}) satisfies instead the following identities:
\begin{align}
    \fgamma^0\hO_i^\dagger\fgamma^0&=\frac{1}{\mk}\left\{-k_i\kg+\frac{ia}{2}\left(m+H\fgamma^0\right)\left[\kg\fgamma_i-3k_i\right]\right\} \ , \\
    \kg\hO_i&=-\left\{k_i+\frac{ia}{2}\left[\frac{k_i}{\mk}\kg+\fgamma_i\right]\left(m-H\fgamma^0\right)\right\} \ .
\end{align}

\chapter{Symmetry and conservation of the gravitino stress-energy tensor}\label{app_emt_grav}

In this Appendix we explicitly prove the on-shell symmetry and covariant conservation of the energy-momentum tensor of the massive gravitino, computed in eq.~\eqref{eqn:ch2_gpp_emt_full} of Chapter \ref{ch2_sugra}. Starting from the covariant action for the massive Majorana spin-$\nicefrac{3}{2}$ field of eq.~\eqref{eqn:ch2_m_grav_lag}, The necessary tools are the associated equations of motion and constraints, respectively in eq.~\eqref{eqn:ch2_gr_eom_1}-\eqref{eqn:ch2_gr_eom_2} and in eq.~\eqref{eqn:ch2_gr_const}, and the Majorana reality condition \eqref{eqn:app_cliff_majorana}, along with the Levi--Civita tensor contraction rule \eqref{eqn:app_conv_levi_civita_contr} and the Clifford algebra identity \eqref{eqn:app_cliff_basis}.

The variation of the action \eqref{eqn:ch2_m_grav_lag} with respect to the vierbein yields:\footnote{In eq.~\eqref{eqn:app_emt_T_non_sym}), the equations of motion \eqref{eqn:ch2_gr_eom_1} have already been applied once.}
\begin{equation}
	\begin{aligned}
		T^{\mu\nu}=&\frac{1}{2}\bPsi_\rho\gamma^{\rho\sigma\mu}\({\nabla}_\sigma\Psi^\nu-{\nabla}^\nu\Psi_\sigma\)-\frac{m}{2}\bPsi_\rho\gamma^{\rho\mu}\Psi^\nu\\
		&-\frac{1}{4}\nabla_\rho\(\bPsi^\rho\gamma^\mu\Psi^\nu+\bPsi^\rho\gamma^\nu\Psi^\mu-\bPsi^\mu\gamma^\rho\Psi^\nu\)\\
		&-\frac{1}{2}\nabla^\nu\left(\bPsi^\mu\gamma^\rho\Psi_\rho\)-\frac{1}{2}g^{\mu\nu}\,\nabla_\rho\left(\bPsi_\sigma\gamma^\sigma\Psi^\rho\right) \ .
	\end{aligned}
\label{eqn:app_emt_T_non_sym}
\end{equation}

\noindent Making use of the equations mentioned above, one can prove that the antisymmetric part of this tensor is vanishing on-shell:
\begin{equation}
\begin{aligned}
    T^{[\mu\nu]}&=\frac{1}{2}\bPsi_\rho\gamma^{\rho\sigma[\mu|}\({\nabla}_\sigma\Psi^{|\nu]}-{\nabla}^{|\nu]}\Psi_\sigma\)-\frac{m}{2}\bPsi_\rho\gamma^{\rho[\mu|}\Psi^{|\nu]}\\
    &\quad-\frac{1}{2}\nabla^{[\nu}\(\bPsi^{\mu]}\gamma^\rho\Psi_\rho\)+\frac{1}{4}\nabla_\rho\(\bPsi^{[\mu|}\gamma^\rho\Psi^{|\nu]}\)=\\
    &=-\frac{1}{2}\(\bPsi_\rho\gamma^{\rho[\mu}-\bPsi^{[\mu}\)\[\cancel{\nabla}\psi^{\nu]}-\nabla^{\nu]}\(\gamma^\rho\Psi_\rho\)\]-\frac{m}{2}\bPsi_\rho\gamma^{\rho[\mu|}\Psi^{|\nu]}\\
    &\quad-\frac{1}{2}\bPsi^\sigma\gamma^{[\mu}\(\nabla_\sigma\Psi^{\nu]}-\nabla^{\nu]}\Psi_\sigma\)+\frac{1}{2}\bPsi_\rho\gamma^\rho\nabla^{[\mu}\Psi^{\nu]}=\\
    &=\frac{m}{2}\bPsi_\rho\gamma^{\rho[\mu}\Psi^{\nu]}-\frac{3}{2}\bPsi_\rho\gamma^{[\rho}\nabla^\nu\Psi^{\mu]}\\
    &=\frac{m}{2}\bPsi_\rho\gamma^{\rho[\mu}\Psi^{\nu]}-\frac{3}{2}\delta^{\rho\nu\mu}_{\alpha\beta\gamma}\bPsi_\rho\gamma^\alpha\nabla^\beta\Psi^\gamma=\\
    &=\frac{m}{2}\bPsi_\rho\gamma^{\rho[\mu}\Psi^{\nu]}+\frac{i}{4}\varepsilon^{\rho\nu\mu}{}_{\sigma}\bPsi_\rho\gamma_5\[\gamma^{\sigma\lambda\tau}\nabla_\lambda\Psi_\tau\]=\\
    &=\frac{m}{2}\bPsi_\rho\gamma^{\rho[\mu}\Psi^{\nu]}+\frac{m}{4}\bPsi_\rho\gamma^{\rho\nu\mu}\gamma^{\sigma}\Psi_\sigma\\
    &=\frac{m}{2}\bPsi_\rho\gamma^{\rho[\mu}\Psi^{\nu]}-\frac{m}{2}\bPsi_\rho\gamma^{\rho[\mu}\Psi^{\nu]}=0 \ .
\end{aligned}
\end{equation}

\noindent Thus, the tensor in eq.~\eqref{eqn:app_emt_T_non_sym} is equal, on-shell, to its symmetric part $T^{\mu\nu}=T^{(\mu\nu)}$, which is the tensor given in eq.~\eqref{eqn:ch2_gpp_emt_full}.

The proof of the covariant conservation goes along similar lines. Together with the equations listed at the beginning, one needs the following tensor identities:
\beq
\begin{aligned}
   & R^\mu_{[\nu\rho\sigma]}=0 \ , \\ &R_{\mu\nu\rho\sigma}\gamma^{\mu\nu\lambda}\gamma^{\rho\sigma}=4G^\lambda_\tau\gamma^\tau \ ,  \\
   &\left[\nabla_\mu,\nabla_\nu\right]\Psi^\rho=\frac{1}{4}R_{\mu\nu\lambda\tau}\gamma^{\lambda\tau}\Psi^\rho+R^\rho{}_{\sigma\mu\nu}\Psi^\sigma \ , \\
   &\[\nabla_\mu,\nabla_\nu\]V^{\rho_1\dots\rho_r}=R^{\rho_1}{}_{\sigma\mu\nu}V^{\sigma\rho_2\dots\rho_r}+\dots+R^{\rho_r}{}_{\sigma\mu\nu}V^{\rho_1\rho_2\dots\sigma} \ .
\end{aligned}
\label{eqn:app_emt_ids}
\eeq

\noindent We study the covariant divergence $\nabla_\mu T^{\mu\nu}$ piece by piece. We do so working with the original tensor \eqref{eqn:app_emt_T_non_sym} rather than the symmetric one: the two are in fact equal, but the former is simpler to deal with. We start from the first line of the tensor \eqref{eqn:app_emt_T_non_sym}, for which we have
\begin{equation}
\begin{aligned}
    \frac{1}{2}\nabla_\mu&\[\bPsi_\rho\gamma^{\rho\sigma\mu}\({\nabla}_\sigma\Psi^\nu-{\nabla}^\nu\Psi_\sigma\)-m\bPsi_\rho\gamma^{\rho\mu}\Psi^\nu\]=\\ 
	&=\frac{1}{2}\bPsi_\rho\lnabla_\mu\gamma^{\rho\sigma\mu}\(\nabla_\sigma\Psi^\nu-\nabla^\nu\Psi_\sigma\)+\frac{1}{2}\bPsi_\rho\gamma^{\rho\sigma\mu}\nabla_\mu\nabla_\sigma\Psi^\nu-\frac{1}{2}\bPsi_\rho\gamma^{\rho\mu\sigma}\nabla_\mu\nabla^\nu\Psi_\sigma\\
	&\quad -\frac{m}{2}\bPsi_\rho\lnabla_\mu\gamma^{\rho\mu}\Psi^\nu-\frac{m}{2}\bPsi_\rho\gamma^{\rho\mu}\nabla_\mu\Psi^\nu-\frac{1}{2}\(\partial_\mu m\)\bPsi_\rho\gamma^{\rho\mu}\Psi^\nu=\\
	&=-\frac{1}{4}R^\nu_{\,\rho\alpha\beta}\bPsi^{\alpha}\gamma^\rho\Psi^\beta-\frac{1}{2}R^\nu{}_{\rho}\bPsi_\alpha\gamma^\alpha\Psi^\rho \ .
\end{aligned}
\label{eqn:app_emt_cov_cons_kin}
\end{equation}

\noindent Next, all the remaining pieces are total derivatives of gravitinos bilinear. From the point of view of general diffeomorphisms, they transform as tensors of the type $V^{\mu\nu\rho}$ of eq.~\eqref{eqn:app_emt_ids}, so that there is no effective dependence on the Lorentz spin-connection in the covariant derivatives. Thus, employing the notation $V^{\mu\nu\rho}\equiv\bPsi^\mu\gamma^\nu\Psi^\rho$, we have
\begin{align}
&\begin{aligned}
	\bullet \quad -\frac{1}{4}\nabla_\mu\nabla_\rho&\[\bPsi^\rho\gamma^\nu\Psi^\mu\]=-\frac{i}{8}\[\nabla_\mu,\nabla_\rho\]V^{\rho\nu\mu}=\\
	=&-\frac{1}{8}\(R^{\rho}_{\sigma\mu\rho}V^{\sigma\nu\mu}+R^{\mu}_{\sigma\mu\rho}V^{\rho\nu\sigma}+R^{\nu}_{\sigma\mu\rho}V^{\rho\sigma\mu}\)=\frac{1}{8}R^{\nu}_{\rho\alpha\beta}\bPsi^\alpha\gamma^\rho\Psi^\beta \ ,
	\end{aligned}\\\notag\\
&\begin{aligned}
	\bullet \quad -\frac{1}{4}\nabla_\mu\nabla_\rho&\[\bPsi^\rho\gamma^\mu\Psi^\nu-\bPsi^\mu\gamma^\rho\Psi^\nu\]=-\frac{1}{4}\[\nabla_\mu,\nabla_\rho\]V^{\rho\mu\nu}=\\
	=&-\frac{1}{4}\(R^{\rho}_{\sigma\mu\rho}V^{\sigma\mu\nu}+R^{\mu}_{\sigma\mu\rho}V^{\rho\sigma\nu}+R^{\nu}_{\sigma\mu\rho}V^{\rho\mu\sigma}\)=-\frac{1}{4}R^\nu_{\beta\rho\alpha}\bPsi^\alpha\gamma^\rho\Psi^\beta \ ,
\end{aligned}\\\notag\\
&\begin{aligned}
	\bullet \quad -\frac{1}{2}\nabla_\mu&\[g^{\mu\nu}\nabla_{\rho}\(\bPsi_\alpha\gamma^\alpha\Psi^\rho\)+\nabla^\nu\(\bPsi^\mu\gamma^\rho\Psi_\rho\)\]=\\
	&=-\frac{1}{2}\[\nabla^\nu,\nabla_\mu\]\(\bPsi_\rho\gamma^\rho\Psi^\mu\)V^\mu=\frac{1}{2}R^\nu{}_\rho\bPsi_\alpha\gamma^\alpha\Psi^\rho \ .
	\end{aligned} \label{eqn:app_emt_cov_cons_vec}
\end{align}

\noindent Therefore, summing up all the terms in eq.~\eqref{eqn:app_emt_cov_cons_kin})--\eqref{eqn:app_emt_cov_cons_vec} and applying the Bianchi identity of the Riemann tensor in eq.~\eqref{eqn:app_emt_ids}, we find that the energy-momentum tensor \eqref{eqn:ch2_gpp_emt_full} is indeed covariantly conserved:
\begin{equation}
    \nabla_\mu T^{\mu\nu}=0 \ .
\end{equation}

\chapter{The Kaluza--Klein reduction of the 5D supergravity algebra on \texorpdfstring{${S^1/\mathbb{Z}_2}$}{S1/Z2}}\label{app_ms2_kk}

This appendix is devoted to the detailed Kaluza--Klein reduction of the 5D pure supergravity algebra on the orbifold $S^1/\mathbb{Z}_2$, from which it is possible to extract the $D=4$, $\N=1$ massive spin-2 supermultiplet of eq.~\eqref{eqn:ch2_ms2_ms2_susy}. The field content of 5D pure supergravity is given by the f\"unfbein $E_M{}^A$, one abelian gauge field $\mathcal{A}_M$ and one $\mathrm{SU}(2)_R$-Majorana gravitino $\Psi_M$, which is actually a doublet of the $\mathrm{SU}(2)_R$ $R$-symmetry of the theory. This type of fermions is associated with the so-called symplectic Clifford algebra, described by the following five-dimensional symplectic Dirac matrices:
\begin{align}
    \gamma^0&=\mathbb{1}\otimes\(i\sigma_1\otimes\mathbb{1}\),\qquad\gamma^i=\mathbb{1}\otimes\(\sigma_2\otimes\sigma^i\),\qquad\gamma^4=\mathbb{1}\otimes\(\sigma_3\otimes\mathbb{1}\) \ ,
\label{eqn:5D_symp_gamma}
\end{align}

\noindent where the elements inside the parenthesis correspond to the usual Dirac indices, whereas the elements in front correspond to the $\mathrm{SU}(2)_R$ indices. In this setup, the appropriate charge conjugation matrix $\mathcal{C}_5$ is given by
\begin{equation}
    \mathcal{C}_5=\sigma_2\otimes\(\mathbb{1}\otimes\sigma_2\)\equiv \sigma_2\otimes \(\gamma_5C_4\) \ , 
\label{eqn:5D_C_mat}
\end{equation}

\noindent and it is such that
\beq
    \mathcal{C}_5^{-1}=\mathcal{C}_5^\textup{T}=\mathcal{C}_5^\dagger=\mathcal{C}_5 \ .
\eeq

\noindent By construction, this symplectic Clifford algebra possess the symmetry property
\begin{align}
    \(\mathcal{C}_5\gamma^{M_1\dots M_r}\)^\textup{T}=&-t_r\mathcal{C}_5\gamma^{M_1\dots M_r} \ ,
\label{eqn:5D_Cgamma_sym}
\end{align}

\noindent with $t_r=-1$ for $r=0,1\,[4]$ and $t_r=1$ for $r=2,3\,[4]$. Thus, a five-dimensional, symplectic-Majorana fermion $\lambda$ is defined by the condition
\begin{equation}
\bar\lambda=\lambda^\textup{T} \mathcal{C}_5 \ .
\label{eqn:symp_maj}
\end{equation}

\noindent In this section we will work directly with symplectic-Majorana spinors like \eqref{eqn:symp_maj} as long as five-dimensional quantities are concerned. We will then open the $\mathrm{SU}(2)_R$ components, which we denote with $\lambda^i$, with $i=1,2$, when needed, in particular when performing the dimensional reduction on the orbifold $S^1/\mathbb{Z}_2$.

We can now go back to the discussion of such a compactification, which we begin from the supersymmetry transformations of pure 5D supergravity. At the linearized level, they are \cite{Gunaydin:1983bi, Ceresole:2000jd}
\beq
\begin{aligned}
&\delta E^A=\frac{1}{2}\bar{\varepsilon}\gamma^A\Psi,  \\
&\delta \A = -\frac{i}{2}\bar{\varepsilon}\Psi , \\
&\delta \Psi= \mathcal{D}\varepsilon +i E^A\(\frac{1}{8}\gamma_{A}{}^{BC}\F_{BC}-\frac{1}{2}\F_{AB}\gamma^B\)\varepsilon,
\end{aligned}
\label{eqn:ch2_ms2_5D_susy}
\eeq

\noindent where $\F_{MN}=2\partial_{[M}\A_{N]}$ is the abelian gauge field strength, $\mathcal{D}_M$ is the five-dimensional Lorentz-covariant derivative and $\varepsilon$ is the 5D supersymmetry parameter. To perform the desired dimensional reduction, we start from the compactification ansatz
\beq
\begin{aligned}
&E^a=\phi^{-\frac{1}{2}}e^a, \\
&E^4=\phi\(dy+B\),  \\
&\A=a\(dy+B\)+A, \\
&\Psi=\phi^{\frac{5}{4}}\,\zeta\(dy+B\)+\phi^{-\frac{1}{4}}\(\psi-\frac{1}{2}e^a\gamma_a\gamma_5\zeta\),
\end{aligned}
\label{eqn:ch2_ms2_comp_ansatz}
\eeq

\noindent in which $y$ is the compact dimension and $e^a$ is the vierbein, $A$ and $B$ are 4D abelian gauge fields, of field strengths $F=dA$ and $G=dB$, $a$ and $\phi$ are two real scalar fields, and $\psi$ and $\zeta$ are two $\mathrm{SU}(2)_R$-Majorana fermions. Alongside this ansatz, it is convenient to redefine also the five dimensional supersymmetry parameter $\varepsilon$ as
\beq
\varepsilon=\phi^{-\frac{1}{4}}\epsilon.
\label{eqn:ch2_ms2_4D_susy_param}
\eeq

\noindent with $\epsilon$ that will give the supersymmetry parameter in four dimensions.

The next step of this procedure is to plug the compactification ansatz \eqref{eqn:ch2_ms2_comp_ansatz} into the 5D algebra \eqref{eqn:ch2_ms2_5D_susy}, which yields its decomposition in terms of the 4D parameterization of \eqref{eqn:ch2_ms2_comp_ansatz}. In order to do this, we need the 4D decomposition along the ansatz \eqref{eqn:ch2_ms2_comp_ansatz} of the components of the 5D gauge field strength $\F_{MN}$ and spin-connection $\(\omega_M\)^A{}_B$. The former is given by
\begin{align}
    \F_{ab}=&\phi e_a{}^\mu e_b{}^\nu F_{\mu\nu}, & \F_{a4}=&\phi^{-\frac{1}{2}}e_a{}^\mu\(\partial_\mu a-\partial_yA_\mu\) .
\end{align}

\noindent The latter is more conveniently expressed through the anholonomy coefficients $\Omega_{AB}{}^C$, according to 
\begin{align}
  dE^A=\frac{1}{2}\Omega_{BC}{}^A{}E^B\wedge E^C,  && \(\omega_C\)_{AB}=&\frac{1}{2}\(\Omega_{CAB}-\Omega_{CBA}+\Omega_{BAC}\),&&
\end{align}

\noindent which decompose as
\begin{equation}
    \begin{aligned}
        \Omega_{bc}{}^a=&\phi^{\frac{1}{2}}\Omega^{(4)}_{bc}{}^a-\phi^{-\frac{1}{2}}\partial_\mu\phi e_{[b}{}^\mu \delta^a_{c]}, &&&&& \Omega_{ab}{}^4=&\phi^2\,e_{[a}{}^\mu e_{b]}{}^\nu G_{\mu\nu}, \\
	   \Omega_{b4}{}^a=&\frac{1}{2}\phi^{-2}\partial_y\phi\,\delta^a_b-\phi^{-1}e_b{}^\mu\partial_y e_\mu{}^a, &&&&&	\Omega_{a4}{}^4=&\phi^{-\nicefrac{1}{2}}\,e_a{}^\mu\(\partial_\mu\phi-\phi\partial_y B_\mu\).
    \end{aligned}
\end{equation}

\noindent where $\Omega^{(4)}_{bc}$ are the anholonomy coefficients in 4D. Moreover, consistency between the compactification ansatz \eqref{eqn:ch2_ms2_comp_ansatz} and the structure of the original 5D supergravity algebra \eqref{eqn:ch2_ms2_5D_susy} requires, when performing the reduction, to implement the latter with an additional 5D local Lorentz transformation. At the linearised level, such transformations affects only the f\"unfbein, on which it acts as
\begin{equation}
    \delta_\textup{LL}E^A=\Lambda^A{}_B E^B,
\end{equation}

\noindent and the explicit transformation required is given by
\begin{align}
    \Lambda^a{}_b=&\frac{1}{4}\phi^{\frac14}\(\bar{\varepsilon}\gamma^a{}_b\gamma_5\zeta\), & \Lambda^a{}_4 =&\frac{1}{4}\phi^{\frac14}\(\bar{\varepsilon}\gamma^a\zeta\)
\end{align}

The 4D supersymmetry transformations resulting from this procedure are then
\beq
\begin{aligned}
&\delta e_\mu{}^a=\frac{1}{2}\bar{\epsilon}\gamma^a\psi_\mu, \\
&\delta\phi=\frac{\phi}{2}\bar{\epsilon}\gamma_5\zeta, \\
&\delta a=-\frac{i}{2}\phi\bar{\epsilon}\zeta,  \\
&\delta A_\mu=-\frac{i}{2}\phi^{-\frac{1}{2}}\bar{\epsilon}\(\psi_\mu-\frac{1}{2}\gamma_\mu\gamma_5\zeta\), \\
&\delta B_\mu=\frac{\phi^{-\frac{3}{2}}}{2}\bar{\epsilon}\(\gamma_5\psi_\mu+\frac{3}{2}\gamma_\mu\zeta\), \\
&\begin{aligned}\delta\zeta=&-\frac{\phi^{-\frac{5}{2}}}{4}\partial_y\phi\epsilon+\frac{\phi^{-\frac{3}{2}}}{4}Q_{ab}\gamma^{ab}\epsilon+\frac{i}{8}\phi^{\frac{1}{2}}\(F_{\rho\sigma}+i\phi G_{\rho\sigma}\gamma_5\)\gamma^{\rho\sigma}\gamma_5\epsilon\\
&+\frac{i}{2}\phi^{-1}\[\partial_\rho a-\partial_yA_\rho+i\(\phi\partial_yB_\rho-\partial_\rho\phi\)\gamma_5\]\gamma^\rho \epsilon,
\end{aligned}\\
&\begin{aligned}\delta\psi_\mu=& D_\mu\epsilon-\frac{3}{8}\phi^{-\frac{5}{2}}\partial_y\phi\gamma_\mu\gamma_5\epsilon+\frac{\phi^{-\frac{3}{2}}}{2}e^a_{\mu}\(P_{ab}\gamma^b+\frac{1}{4}Q_{bc}\gamma_a\gamma^{bc}\)\gamma_5\epsilon \\
&-\frac{1}{4}\partial_yB_\rho\gamma_\mu\gamma^\rho\epsilon+\frac{3i}{4}\phi^{-1}\(\partial_yA_\mu-\partial_\mu a\)\gamma_5\epsilon-\frac{i}{8}\phi^{\frac{1}{2}}\(3F_{\mu\rho}^{+,5}-i\phi G_{\mu\rho}^{+,5}\gamma_5\)\gamma^\rho\epsilon.
\end{aligned}
\end{aligned}
\label{eqn:ch2_ms2_4d_5d_susy}
\eeq

\noindent In these formulae, $D_\mu$ is the 4D Lorentz-covariant derivative, while $\epsilon$ is the 4D supersymmetry parameter defined in \eqref{eqn:ch2_ms2_4D_susy_param}. The quantities $P_{ab}$ and $Q_{ab}$ are defined to be
\begin{align}
P_{ab}\equiv e_{(a}{}^\lambda\partial_y e_{b)}{}_{\lambda}, && Q_{ab}\equiv e_{[a}{}^\lambda\partial_y e_{b]}{}_{\lambda},
\end{align}

\noindent while $F_{\mu\nu}^{\pm,5}$ and $G_{\mu\nu}^{\pm,5}$ are the self-dual field strengths dressed by a $\gamma_5$, defined as
\begin{align}
F_{\mu\nu}^{\pm,5}\equiv F_{\mu\nu}\pm i\tilde{F}_{\mu\nu}\gamma_5 , && \tilde{F}^{\mu\rho}\equiv\frac{1}{2}\varepsilon^{\mu\nu\rho\sigma}F_{\rho\sigma}.
\end{align}

The transformations \eqref{eqn:ch2_ms2_4d_5d_susy} still carry the full dependence on the $5^\textup{th}$ compact coordinate $y$. Thus, the next step is to properly expand each field in modes along the $S^1/\mathbb{Z}_2$ interval spanned by $y$, which will further decompose the transformations \eqref{eqn:ch2_ms2_4d_5d_susy} into massless and massive transformation rules, properly organized in $\N=1$ multiplets. This mode expansion has to be consistent with the $\mathbb{Z}_2$ parity along $y$ modded out by the orbifold projection under consideration, namely with the transformation of each field under such $\mathbb{Z}_2$, fixed by the invariance of the lagrangian\footnote{For the explicit expression of the lagrangian of pure 5D supergravity, see \cite{Gunaydin:1983bi,Ceresole:2000jd}}. Starting from the bosons, the vierbein $e_\mu{}^a$ and the scalars $\phi$ and $a$ are even under $\mathbb{Z}_2$, while the gauge vectors $A_\mu$ and $B_\mu$ are odd: this fixes their Kaluza--Klein expansion to be\footnote{The expansion of the field $\phi$ is fixed to the one in \eqref{eqn:ch2_ms2_kk_bosons} by the additional requirement of being an exponential function.}
\begin{equation}
\begin{aligned}
e_\mu{}^a(y)=&e^{\scalebox{0.6}{$(0)$}}_\mu{}^a(x)+\sum_{n=1}^{\infty}e^{\scalebox{0.6}{$(n)$}}_\mu{}^a(x)\cos\(\frac{ny}{R}\), &  B_\mu(y)=&\sum_{n=1}^{\infty}B_\mu^{(n)}(x)\sin\(\frac{ny}{R}\),  \\
\phi(y)=&\phi^{(0)}(x)\[1+\sum_{n=1}^{\infty}\phi^{(n)}(x)\cos\(\frac{ny}{R}\)\], &
A_\mu(y)=&\sum_{n=1}^{\infty}A_\mu^{(n)}(x)\sin\(\frac{ny}{R}\), \\
a(y)=&a^{(0)}(x)+\sum_{n=1}^{\infty}a^{(n)}(x)\cos\(\frac{ny}{R}\). 
\end{aligned}
\label{eqn:ch2_ms2_kk_bosons}
\end{equation}

\noindent The projection on the fermions is instead more involved. The action of $\mathbb{Z}_2$ is in fact chiral and breaks therefore the original $\mathrm{SU}(2)_R$ symmetry:
\begin{align}
\Psi_i{}_\mu(-y)=\sigma^3_{ij}\gamma_5\Psi_j{}_\mu(y), && \Psi_i{}_5(-y)=-\sigma^3_{ij}\gamma_5\Psi_j{}_5(y),
\end{align}

\noindent so that
\beq
\begin{aligned}
\psi_1{}_\mu(-y)=&\gamma_5\psi_1{}_\mu(y), & \zeta_1{}_\mu(-y)=&-\gamma_5\zeta_1{}_\mu(y), \\
\psi_2{}_\mu(-y)=&-\gamma_5\psi_2{}_\mu(y), & \zeta_2{}_\mu(-y)=&\gamma_5\zeta_2{}_\mu(y).
\end{aligned}
\eeq

\noindent Then, the Kaluza--Klein mode expansion for the fermions result to be
\beq
\begin{aligned}
\psi_{1\mu}=&\psi_{1\mu \text{R}}+\psi_{1\mu \text{L}}=\psi^{(0)}_{1\mu \text{R}}+\sum_{n=1}^{\infty}\[\psi^{(n)}_{1\mu \text{R}}\cos\(\frac{ny}{R}\)+i\psi^{(n)}_{1\mu \text{L}}\sin\(\frac{ny}{R}\)\], \\
\psi_{2\mu}=&\psi_{1\mu \text{L}}+\psi_{1\mu \text{R}}=\psi^{(0)}_{2\mu \text{L}}+\sum_{n=1}^{\infty}\[\psi^{(n)}_{2\mu \text{L}}\cos\(\frac{ny}{R}\)+i\psi^{(n)}_{2\mu \text{R}}\sin\(\frac{ny}{R}\)\],\\
\zeta_1=&\zeta_{1\text{L}}+\zeta_{1\text{R}}=\zeta^{(0)}_{1\text{L}}+\sum_{n=1}^{\infty}\[\zeta^{(n)}_{1\text{L}}\cos\(\frac{ny}{R}\)+i\zeta^{(n)}_{1\text{R}}\sin\(\frac{ny}{R}\)\],  \\
\zeta_2=&\zeta_{2\text{R}}+\zeta_{2\text{L}}=\zeta^{(0)}_{2\text{R}}+\sum_{n=1}^{\infty}\[\zeta^{(n)}_{2\text{R}}\cos\(\frac{ny}{R}\)+i\zeta^{(n)}_{2\text{L}}\sin\(\frac{ny}{R}\)\]. 
\end{aligned}
\label{eqn:ch2_ms2_kk_fermions}
\eeq

\noindent This decomposition is consistent with the original symplectic Majorana condition, which at the level of the modes fixes, for every $n\ge 0$,
\begin{align}
    \zeta^{(n)}_{2\text{R}}=-\(\zeta^{(n)}_{1\text{L}}\)^*, && \zeta^{(n)}_{2\text{L}}=-\(\zeta^{(n)}_{1\text{R}}\)^*, &&\psi^{(n)}_{2\mu \text{R}}=\(\psi^{(n)}_{1\mu \text{L}}\)^*, && \psi^{(n)}_{2\mu \text{L}}=\(\psi^{(n)}_{1\mu \text{R}}\)^*.
\label{eqn:ch2_ms2_simp_maj}
\end{align}

Also the (local) supersymmetry parameter undergoes an analogous Kaluza--Klein decomposition and the associated zero mode correspond to the actual 4D supersymmetry parameters. The action of the $\mathbb{Z}_2$ symmetry is
\begin{equation}
    \epsilon_i(-y)=\sigma^3_{ij}\gamma_5\epsilon_j(y),
\label{eqn:ch2_ms2_susy_p_z2}
\end{equation}

\noindent where $\sigma^3$ is the third Pauli matrix. The actual 4D supersymmetry parameters correspond to the zero modes of $\epsilon_i(y)$ (the massive ones are discarded), which the $\mathbb{Z}_2$ parity \eqref{eqn:ch2_ms2_susy_p_z2} constrains to be
\begin{equation}
\begin{aligned}
\epsilon_1=\epsilon_{1\text{R}}, &&&&&& \epsilon_2=\epsilon_{2\text{L}},
\end{aligned}
\label{eqn:ch2_ms2_epsilon_chirality}
\end{equation}

\noindent with $\epsilon_{2\text{L}}=\(\epsilon_{1\text{R}}\)^*$, consistently with the five dimensional symplectic-Majorana condition. This indicates therefore that the combination
\beq
\epsilon\equiv\epsilon_{1\text{R}}+\epsilon_{2\text{L}},
\label{eqn:ch2_ms2_4d_susy_p_def}
\eeq

\noindent that defines a 4D Majorana fermion, is the actual supersymmetry parameter in four dimension, showing explicitly how the orbifold projection breaks the original $\N=2$ supersymmetry in 5D to $\N=1$ in 4D.

Plugging the expansions \eqref{eqn:ch2_ms2_kk_bosons} and \eqref{eqn:ch2_ms2_kk_fermions} into \eqref{eqn:ch2_ms2_4d_5d_susy} yields the transformation rules associated to every Kaluza--Klein mode of the various fields, both massless and massive, which we now discuss separately.

\section{The massless sector}

The supersymmetry transformations of the massless modes result to be
\begin{equation}
    \begin{aligned}
        &\delta e^{\scalebox{0.6}{$(0)$}}_\mu{}^a=\frac{1}{2}\(\bar{\epsilon}_1\gamma^a\psi^{(0)}_{1\mu}+\bar{\epsilon}_2\gamma^a\psi^{(0)}_{2\mu}\), \\
        &\delta\phi^{(0)}=\frac{\phi^{(0)}}{2}\(\bepsilon_2\zeta^{(0)}_2-\bepsilon_1\zeta^{(0)}_1\), \\
        &\delta a^{(0)}=-\frac{i}{2}\phi^{(0)}\(\bepsilon_1\zeta^{(0)}_1+\bepsilon_2\zeta^{(0)}_2\), \\
        &\delta\psi^{(0)}_{1\mu R}=D_\mu\epsilon_1-\frac{3i}{4} {\phi^{\scalebox{0.6}{$(0)$}}}^{-1}  \partial_\mu a^{(0)} \epsilon_1,  \\
        &\delta\psi^{(0)}_{2\mu L}=D_\mu\epsilon_2+\frac{3i}{4}{\phi^{\scalebox{0.6}{$(0)$}}}^{-1}\partial_\mu a^{(0)} \epsilon_2,\\
        &\delta\zeta^{(0)}_{2R}=\frac{{\phi^{\scalebox{0.6}{$(0)$}}}^{-1}}{2}\partial_\rho\(\phi^{(0)}+ia^{(0)}\)\gamma^\rho\epsilon_2, \\
        &\delta\zeta^{(0)}_{1L}=-\frac{{\phi^{\scalebox{0.6}{$(0)$}}}^{-1}}{2}\partial_\rho\(\phi^{(0)}-ia^{(0)}\)\gamma^\rho\epsilon_1,
    \end{aligned}
\label{eqn:ch2_ms2_kk_m0_susy_1}
\end{equation}

\noindent and they correspond to pure $\N=1$ supergravity coupled to a chiral multiplet. This structure becomes manifest once the fields are redefined in a proper way. Equations \eqref{eqn:ch2_ms2_simp_maj} and \eqref{eqn:ch2_ms2_4d_susy_p_def} suggest to recombine the Majorana massless fermions as
\begin{align}
    \psi_{\mu}\equiv \psi^{(0)}_{1\mu R}+ \psi^{(0)}_{2\mu L}, && \zeta \equiv \phi^{(0)}\(\zeta^{(0)}_{2R}-\zeta^{(0)}_{1L}\).
\end{align}

\noindent These redefinitions, together with
\begin{equation}
    T\equiv \phi^{(0)}+ia^{(0)},
\end{equation}

\noindent and the renaming $e^{\scalebox{0.6}{$(0)$}}_{\mu}{}^a\to e_{\mu}{}^a$, fix the transformations of the massless sector to be
\beq
\begin{aligned}
&\delta e_{\mu}{}^a=\frac{1}{2}\bepsilon\gamma^a\psi_{\mu}, \\
&\delta\psi_{\mu}=D_\mu\epsilon+\frac{i}{2}Q_\mu\gamma_5\epsilon, \\
&\delta T=\bepsilon_\text{R}\zeta_\text{R},  \\
&\delta\zeta_\text{R}=\frac{1}{2}\cancel{\partial} T\epsilon_\text{L},
\end{aligned}
\eeq

\noindent which are the transformations of pure $D=4$, $\N=1$ supergravity $\left\{e_{\mu}{}^a, \psi_{\mu}\right\}$ coupled to chiral multiplet $\left\{T, \zeta \right\}$ \cite{Freedman:2012zz,DallAgata:2021uvl}, with K\"ahler potential and associated K\"ahler-covariantisation given by
\begin{align}
K=&-3\log\(T+\overline{T}\), & Q_\mu=&-\frac{3}{2i\(T+\overline{T}\)}\(\partial_\mu T-\partial_\mu\overline{T}\).
\end{align}

\noindent This is a no-scale K\"ahler potential expected in Kaluza--Klein reductions of this type.

\section{The massive sector}

We then move to the massive sector. We define the Kaluza--Klein mass of every mode as
\begin{align}
m_n={\phi^{\scalebox{0.6}{$(0)$}}}^{-\frac{3}{2}}\frac{n}{R},
\end{align}

\noindent with $R$ being the length of the compact dimension. Using then the formulae
\begin{equation}
\begin{aligned}
\partial_y\phi=&-{\phi^{\scalebox{0.6}{$(0)$}}}^{\frac52}\sum_nm_n\phi^{(n)}\sin\(\frac{ny}{R}\), &&& P_{ab}=&-{\phi^{\scalebox{0.6}{$(0)$}}}^{\frac32}\sum_n m_n e^{\scalebox{0.6}{$(0)$}}_{(a}{}^\rho e^{\scalebox{0.6}{$(n)$}}_{b)}{}_{\rho} \sin\(\frac{ny}{R}\),\\
\partial_yB_\mu=&{\phi^{\scalebox{0.6}{$(0)$}}}^{\frac32}\sum_n m_nB^{(n)}_\mu\cos\(\frac{ny}{R}\),  &&& Q_{ab}=&-{\phi^{\scalebox{0.6}{$(0)$}}}^{\frac32}\sum_n m_n e^{\scalebox{0.6}{$(0)$}}_{[a}{}^\rho e^{\scalebox{0.6}{$(n)$}}_{b]}{}_{\rho} \sin\(\frac{ny}{R}\), \\
\partial_yA_\mu=&{\phi^{\scalebox{0.6}{$(0)$}}}^{\frac32}\sum_n m_nA^{(n)}_\mu\cos\(\frac{ny}{R}\), &&& \omega^{\scalebox{0.6}{$(n)$}}_\mu{}^{ab}=&2e^{\scalebox{0.6}{$(0)$}}{}^{[a|}{}^{\alpha}\partial^{\scalebox{0.6}{\color{white}$(n)$}}_{[\mu}e^{\scalebox{0.6}{$(n)$}}_{\alpha]}{}^{|b]}-e^{\scalebox{0.6}{$(0)$}}_\alpha{}^{[a}e^{\scalebox{0.6}{$(0)$}}_\beta{}^{ b]}e^{\scalebox{0.6}{$(0)$}}_\mu{}^c\partial^\alpha e^{\scalebox{0.6}{$(n)$}}_c{}^{\beta}, 
\end{aligned}
\end{equation}

\noindent one obtains that the massive modes of the set of supersymmetry transformations \eqref{eqn:ch2_ms2_4d_5d_susy}. The bosonic transformation rules are  
\beq
\begin{aligned}
&\delta e_n{}^a_\mu=\frac{1}{2}\(\bar{\epsilon}_1\gamma^a\psi^{(n)}_{1\mu}+\bar{\epsilon}_2\gamma^a\psi^{(n)}_{2\mu }\), \\
&\delta\phi^{(n)}=\frac{1}{2}\(\bepsilon_2\zeta^{(n)}_2-\bepsilon_1\zeta^{(n)}_1\), \\
&\delta a^{(n)}=-\frac{i}{2}\phi^{(0)}\(\bepsilon_1\zeta^{(n)}_1+\bepsilon_2\zeta^{(n)}_2\),  \\
&\delta A^{(n)}_\mu=\frac{{\phi^{\scalebox{0.6}{$(0)$}}}^{-\frac12}}{2}\[\bepsilon_1\(\psi^{(n)}_{1\mu}-\frac{1}{2}\gamma_\mu\zeta^{(n)}_1\)+\bepsilon_2\(\psi^{(n)}_{2\mu}-\frac{1}{2}\gamma_\mu\zeta^{(n)}_2\)\], \\
&\delta B^{(n)}_\mu=\frac{i}{2}{\phi^{\scalebox{0.6}{$(0)$}}}^{-\frac{3}{2}}\[\bepsilon_2\(\psi^{(n)}_{2\mu}+\frac{3}{2}\gamma_\mu\zeta^{(n)}_2\)-\bepsilon_1\(\psi^{(n)}_{1\mu}-\frac{3}{2}\gamma_\mu\zeta^{(n)}_1\)\],
\end{aligned}
\label{eqn:ch2_ms2_susy_bare_m_B}
\end{equation}

\noindent while the transformations on the massive spin-$\textstyle{\frac12}$ and spin-$\textstyle{\frac32}$ are respectively
\beq
\begin{aligned}
&\delta\zeta^{(n)}_{1\text{L}}=\frac{m_n}{2}\[-i\({\phi^{\scalebox{0.6}{$(0)$}}}^{\frac12}A^{(n)}_\rho-\frac{{\phi^{\scalebox{0.6}{$(0)$}}}^{-1}}{m_n}\partial_\rho a^{(n)}\)+\({\phi^{\scalebox{0.6}{$(0)$}}}^{\frac32}B^{(n)}_\rho-\frac{1}{m_n}\partial_\rho \phi^{(n)}\)\]\gamma^\rho\epsilon_1, \\
&\delta\zeta^{(n)}_{1\text{R}}=\frac{{\phi^{\scalebox{0.6}{$(0)$}}}^{\frac12}}{8}\(F^{(n)}_{\rho\sigma}+i\phi^{(0)}G^{(n)}_{\rho\sigma}\)\gamma^{\rho\sigma}\epsilon_1+\frac{i}{4}m_n \(e^{\scalebox{0.6}{$(0)$}}_a{}^\rho e^{\scalebox{0.6}{$(n)$}}_b{}_{\rho}\gamma^{ab}-\phi^{(n)}\)\epsilon_1, \\
&\delta\zeta^{(n)}_{2\text{L}}=-\frac{{\phi^{\scalebox{0.6}{$(0)$}}}^{\frac12}}{8}\(F^{(n)}_{\rho\sigma}-i\phi^{(0)}G^{(n)}_{\rho\sigma}\)\gamma^{\rho\sigma}\epsilon_2+\frac{i}{4}m_n\(e^{\scalebox{0.6}{$(0)$}}_a{}^\rho e^{\scalebox{0.6}{$(n)$}}_b{}_{\rho}\gamma^{ab}-\phi^{(n)}\)\epsilon_2, \\
&\delta\zeta^{(n)}_{2\text{R}}=\frac{m_n}{2}\[-i\({\phi^{\scalebox{0.6}{$(0)$}}}^{\frac12}A^{(n)}_\rho-\frac{{\phi^{\scalebox{0.6}{$(0)$}}}^{-1}}{m_n}\partial_\rho a^{(n)}\)-\({\phi^{\scalebox{0.6}{$(0)$}}}^{\frac32}B^{(n)}_\rho-\frac{1}{m_n}\partial_\rho \phi^{(n)}\)\]\gamma^\rho\epsilon_2,
\end{aligned}
\label{eqn:ch2_ms2_susy_bare_m_12}
\eeq

\noindent and 
\beq
\begin{aligned}
&\begin{aligned}
\delta\psi^{(n)}_{1\mu \text{R}}=&\frac{1}{4}\(2e^{\scalebox{0.6}{$(0)$}}{}^{a}{}^{\alpha}\partial^{\scalebox{0.6}{\color{white}$(n)$}}_{[\mu}e^{\scalebox{0.6}{$(n)$}}_{\alpha]}{}^{b}-e^{\scalebox{0.6}{$(0)$}}_\alpha{}^{a}e^{\scalebox{0.6}{$(0)$}}_\beta{}^{ b}e^{\scalebox{0.6}{$(0)$}}_\mu{}^c\partial^\alpha e^{\scalebox{0.6}{$(n)$}}_c{}^{\beta}\)\gamma_{ab}\epsilon_1-\frac{m_n}{4}{\phi^{\scalebox{0.6}{$(0)$}}}^{\frac32}B^{(n)}_\rho\gamma_\mu\gamma^\rho\epsilon_1\\
&+\frac{3i}{4}m_n\({\phi^{\scalebox{0.6}{$(0)$}}}^{\frac12}A^{(n)}_\mu-\frac{{\phi^{\scalebox{0.6}{$(0)$}}}^{-1}}{m_n}\partial_\mu a^{(n)}\)\epsilon_1,
\end{aligned}  \\
&\begin{aligned}
\delta\psi^{(n)}_{1\mu \text{L}}=&-\frac{{\phi^{\scalebox{0.6}{$(0)$}}}^{\frac12}}{8}\(3F^{(n)+,5}_{\mu\rho}+i\phi^{(0)}G^{(n)+,5}_{\mu\rho}\)\gamma^\rho\epsilon_1+\frac{i}{4}m_ng^{(n)}_{\mu\rho}\gamma^\rho\epsilon_1\\
&+\frac{i}{4}m_n\(e^{\scalebox{0.6}{$(0)$}}_a{}^\rho e^{\scalebox{0.6}{$(n)$}}_b{}_{\rho}\gamma_\mu\gamma^{ab}-\phi^{(n)}\gamma_\mu\)\epsilon_1,
\end{aligned} \\
&\begin{aligned}
\delta\psi^{(n)}_{2\mu \text{R}}=&-\frac{{\phi^{\scalebox{0.6}{$(0)$}}}^{\frac12}}{8}\(3F^{(n)+,5}_{\mu\rho}-i\phi^{(0)}G^{(n)+,5}_{\mu\rho}\)\gamma^\rho\epsilon_2-\frac{i}{4}m_ng^{(n)}_{\mu\rho}\gamma^\rho\epsilon_2\\
&-\frac{i}{4}m_n\(e^{\scalebox{0.6}{$(0)$}}_a{}^\rho e^{\scalebox{0.6}{$(n)$}}_b{}_{\rho}\gamma_\mu\gamma^{ab}-\phi^{(n)}\gamma_\mu\)\epsilon_2,
\end{aligned}  \\
&\begin{aligned}
\delta\psi^{(n)}_{2\mu \text{L}}=&\frac{1}{4}\(2e^{\scalebox{0.6}{$(0)$}}{}^{a}{}^{\alpha}\partial^{\scalebox{0.6}{\color{white}$(n)$}}_{[\mu}e^{\scalebox{0.6}{$(n)$}}_{\alpha]}{}^{b}-e^{\scalebox{0.6}{$(0)$}}_\alpha{}^{a}e^{\scalebox{0.6}{$(0)$}}_\beta{}^{ b}e^{\scalebox{0.6}{$(0)$}}_\mu{}^c\partial^\alpha e^{\scalebox{0.6}{$(n)$}}_c{}^{\beta}\)\gamma_{ab}\epsilon_2-\frac{m_n}{4}{\phi^{\scalebox{0.6}{$(0)$}}}^{\frac32}B^{(n)}_\rho\gamma_\mu\gamma^\rho\epsilon_2\\
&-\frac{3i}{4}m_n\({\phi^{\scalebox{0.6}{$(0)$}}}^{\frac12}A^{(n)}_\mu-\frac{{\phi^{\scalebox{0.6}{$(0)$}}}^{-1}}{m_n}\partial_\mu a^{(n)}\)\epsilon_2.
\end{aligned}
\end{aligned}
\label{eqn:ch2_ms2_susy_bare_m_32}
\eeq

The whole set massive degrees of freedom recombines, for each mode $n$, to a massive spin-1, two massive Majorana spin-$\textstyle{\frac32}$ and a massive spin-2 fields, according to the St\"uckelberg parameterisation, as customary for massive gauge fields. The St\"uckelberg symmetry comes naturally in dimensional compactifications, as the massive modes of the higher-dimensional gauge transformations\footnote{See \cite{Hinterbichler:2011tt} for an explicit analysis of this property in the case of the massive spin-2 field.}. To determine the combinations of the massive degrees of freedom of eqs. \eqref{eqn:ch2_ms2_kk_bosons} and \eqref{eqn:ch2_ms2_kk_fermions} associated to the unitary gauge of the correspondent St\"uckelberg symmetry, we start from the simplest (and more standard) of such combinations, the one for the massive spin-1 field, which we take to be
\beq
    {A_\mu^{\scalebox{0.6}{$(n)$}}}{}^\textup{S}={\phi^{\scalebox{0.6}{$(0)$}}}^{\frac12}A^{(n)}_\mu-\frac{{\phi^{\scalebox{0.6}{$(0)$}}}^{-1}}{m_n}\partial_\mu a^{(n)},
\eeq

\noindent as suggested also by the massive spin-$\textstyle{\frac32}$ supersymmetry transformations \eqref{eqn:ch2_ms2_susy_bare_m_32}. Acting with a supersymmetry transformation \eqref{eqn:ch2_ms2_susy_bare_m_B}, one finds
\begin{equation}
    \delta  {A_\mu^{\scalebox{0.6}{$(n)$}}}{}^\textup{S}=\frac{1}{2}\[\bepsilon_1\(\psi^{(n)}_{1\mu \text{L}}-\frac{1}{2}\gamma_\mu\zeta^{(n)}_{1\text{R}}+\frac{i}{m_n}\partial_\mu\zeta^{(n)}_{1\text{L}}\)+\bepsilon_2\(\psi^{(n)}_{2\mu \text{R}}+\frac{1}{2}\gamma_\mu\zeta^{(n)}_{2\text{L}}+\frac{i}{m_n}\partial_\mu\zeta^{(n)}_{2\text{R}}\)\],
\end{equation}

\noindent which indicates that the correct St\"uckelberg combinations for $\psi^{(n)}_{1\mu \text{L}}$ and $\psi^{(n)}_{2\mu \text{R}}$ are
\begin{equation}
    \begin{aligned}
        {{\psi_{1\mu\text{L}}^{\scalebox{0.6}{$(n)$}}}}{}^\textup{S}=&\psi^{(n)}_{1\mu \text{L}}-\frac{1}{2}\gamma_\mu\zeta^{(n)}_{1\text{R}}+\frac{i}{m_n}\partial_\mu\zeta^{(n)}_{1\text{L}},\\ 
         {{\psi_{2\mu\text{R}}^{\scalebox{0.6}{$(n)$}}}}{}^\textup{S}=&\psi^{(n)}_{2\mu \text{R}}+\frac{1}{2}\gamma_\mu\zeta^{(n)}_{2\text{L}}+\frac{i}{m_n}\partial_\mu\zeta^{(n)}_{2\text{R}},
    \end{aligned}
\label{eqn:ch2_ms2_psi_1L_S}
\end{equation}

\noindent where $ {{\psi_{2\mu\text{R}}^{\scalebox{0.6}{$(n)$}}}}{}^\textup{S}=\( {{\psi_{1\mu\text{L}}^{\scalebox{0.6}{$(n)$}}}}{}^\textup{S}\)^*$ consistently, so that they can be combined in a Majorana spin-$\textstyle{\frac32}$ field
\beq
\begin{aligned}
    {\lambda_\mu^{\scalebox{0.6}{$(n)$}}}{}^\textup{S}\equiv& {{\psi_{1\mu\text{L}}^{\scalebox{0.6}{$(n)$}}}}{}^\textup{S}+ {{\psi_{2\mu\text{R}}^{\scalebox{0.6}{$(n)$}}}}{}^\textup{S}=\(\psi^{(n)}_{1\mu \text{L}}+\psi^{(n)}_{2\mu \text{R}}\)-\frac{1}{2}\gamma_\mu\(\zeta^{(n)}_{1\text{R}}-\zeta^{(n)}_{2\text{L}}\)+\frac{i}{m_n}\partial_\mu\(\zeta^{(n)}_{1\text{L}} +\zeta^{(n)}_{2\text{R}}\),
\end{aligned}
\eeq

\noindent Proceeding in this way, one finds the correct St\"uckelberg combinations for the remaining spin-$\textstyle{\frac32}$ and spin-2 fields ${\chi_\mu^{\scalebox{0.6}{$(n)$}}}{}^\textup{S}$ and ${g_{\mu\nu}^{\scalebox{0.6}{$(n)$}}}{}^\textup{S}$, which are 
\begin{align}
&{\chi_\mu^{\scalebox{0.6}{$(n)$}}}{}^\textup{S}=-i\left[\(\psi^{(n)}_{1\mu \text{R}}-\psi^{(n)}_{2\mu \text{L}}\)-\frac{1}{2}\gamma_\mu\(\zeta^{(n)}_{1\text{L}}+\zeta^{(n)}_{2\text{R}}\)+\frac{i}{m_n}\partial_\mu\(\zeta^{(n)}_{1\text{R}}-\zeta^{(n)}_{2\text{L}}\)\right], \\
&{g_{\mu\nu}^{\scalebox{0.6}{$(n)$}}}{}^\textup{S}=2e^{\scalebox{0.6}{$(0)$}}_{(\mu}{}^a e^{\scalebox{0.6}{$(n)$}}_{\nu)}{}_a-\phi^{(n)}g^{(0)}_{\mu\nu}+\frac{{\phi^{\scalebox{0.6}{$(0)$}}}^{\frac32}}{m_n}\(\partial_\mu B^{(n)}_\nu+\partial_\nu B^{(n)}_\mu\)-\frac{2}{m_n^2}\partial_\mu\partial_\nu\phi^{(n)}.
\end{align}

\noindent which together with  ${A_\mu^{\scalebox{0.6}{$(n)$}}}{}^\textup{S}$ and $ {\lambda_\mu^{\scalebox{0.6}{$(n)$}}}{}^\textup{S}$ form the massive spin-2 supermultiplet we were looking for. The linearised supersymmetry transformations of the above combinations, where one retains only the coupling to the massless modes of the vierbein $e_\mu{}^a$ and the scalar $\phi$ (but not their derivatives) and discards all other couplings, give us the algebra of a rigid massive spin-2 supermultiplet of mass $m_n$, for each $n$ mode of the Kaluza--Klein tower. These are the transformations given in eq.~\eqref{eqn:ch2_ms2_ms2_susy}, which can be thought of independently from the Kaluza--Klein reduction used to obtain them.

\chapter{D-branes in orientifold constructions}\label{app_branes}.

In this final appendix, we complete the analysis of the brane spectra in the orientifold models discussed in Chapter \ref{ch3_st}. We begin by examining the unstable lower-dimensional branes in the Dabholkar–Park orientifold introduced in Section \ref{ch3_sec_ss_dp_orientifolds}. We then turn to the new Scherk–Schwarz orientifold proposed in \cite{Bossard:2024mls} and discuss those branes which were omitted from the analysis in Section \ref{ch3_sec_new_sso}.

\section{The unstable branes of the Dabholkar--Park orientifold}\label{app_branes_dp}

The unstable D-branes in the Dabholkar--Park orientifold are non-BPS D2, D3, D6 and D7-branes wrapping the compact circle, and all even-dimensional branes orthogonal to the circle. We now present all of them, referring to the discussion of Section \ref{ch3_sec_ss_dp_orientifolds}.

\subsection{Non-BPS D7-branes wrapping the circle}

The cylinder and M\"obius amplitudes in the transverse channel for the wrapping D7-branes are equal to
\begin{equation}
    \begin{aligned}
        {\tilde \A}_{77} =& \frac{2^{-4}v}{2}\frac{1}{\eta^8} 
\left[ \left(N+{\overline N}\right)^2 V_8 + \left(N-{\overline N}\right)^2 S_8\right] \sum_n W_n \ , \\
    {\tilde \M}_{7} =& \left( \frac{2\epsilon v}{\hat{\eta}^5 \hat{\vartheta}_2} \right)
\left[ \left(N+{\overline N}\right) \left({\hat O}_6 {\hat V}_2 -
{\hat V}_6 {\hat O}_2\right) - \left(N-{\overline N}\right)
\left({\hat S}_6 {\hat S}_2 - {\hat C}_6 {\hat C}_2\right) \right] \sum_n W_{2n+1} \ ,
    \end{aligned}
\label{dpd71}
\end{equation}

\noindent where $\epsilon = \pm 1$ reflects the position of the T-dual D6-branes on top of the $\text{O}8_-$--plane or on the $\text{O}8_+$--plane, respectively. There are no physical couplings to the R-R sector, and therefore these branes are uncharged. The loop-channel open amplitudes are instead equal to
\begin{equation}
    \begin{aligned}
        {\A}_{77} =&\frac{1}{\eta^8}\left[ N {\overline N} \left(V_8-S_8\right) +  \frac{N^2 + \overline{N}^2}{2} \left(O_8-C_8\right) \right] \sum_m P_m \ ,  \\
        \M_7 =& - \left( \frac{2\epsilon}{\hat{\eta}^5 \hat{\vartheta}_2} \right)\left[ \frac{N+{\overline N}}{2} \left({\hat O}_6 {\hat O}_2 + {\hat V}_6 {\hat V}_2\right) +   \frac{N-{\overline N}}{2}\left({\hat S}_6 {\hat C}_2 - {\hat C}_6 {\hat S}_2\right)\right] \sum_m (-1)^m P_m \ . 
    \end{aligned}
\label{dpd72}
\end{equation}
The gauge group of these branes is $\mathrm{U}(N)$, associated with a complete ${\cal N}=4$ super Yang--Mills massless multiplet. In addition, there are complex tachyons in the symmetric plus antisymmetric representations of $\mathrm{U}(N)$, so that, for $\epsilon=1$, the tachyon disappears from a single $N=1$ D7-brane at radius $R \leq \sqrt{2 \alpha'}$.\footnote{Note that this dynamics is different with respect to what happens in the type~I string, where the existemce of background D9-branes -- absent in the Dabholkar--Park orientifold under consideration -- implies the presence of tachyons in the D7--D9 sector.}
However, the transverse M\"obius amplitude \eqref{dpd71} gives, in this setup, a positive contribution to the potential energy. This is interpreted, in the T-dual picture, as the attraction of the D6-brane from the $\text{O}8_-$--plane, where it is stable, towards the O$8_+$--plane, where instead it develops a tachyon \cite{Dbranes-dpGimon}. This process is indeed allowed, given that the appropriate brane positions moduli on the circle do exist. This implies that no D7-brane wrapping the circle can be stable.

\subsection{Non-BPS D3-branes wrapping the circle}

The non-BPS D3-branes wrapping the circle are uncharged and their behavior exactly matches that of the wrapping D7-branes discussed above. The transverse channel amplitudes are equal to
\beq
\begin{aligned}
    {\tilde \A}_{33} =& \frac{2^{-2}v}{2}\frac{1}{\eta^8}\left[\left(N+{\overline N}\right)^2 V_8+\left(N-{\overline N}\right)^2 S_8 \right] \sum_n W_n \ , \\
    {\tilde \M}_{3}=&\left(\frac{8\epsilon v \hat{\eta}}{\hat{\vartheta}_2^3}\right)\left[ \left(N+{\overline N}\right)\left({\hat O}_2 {\hat V}_6 -{\hat V}_2 {\hat O}_6\right)-\left(N-{\overline N}\right)\left({\hat S}_2 {\hat S}_6 - {\hat C}_2 {\hat C}_6\right)\right]\sum_n W_{2n+1} \ ,
\end{aligned}
\label{dpd31}
\eeq

\noindent where $\epsilon = \pm 1$ signals that the branes sit on top of the O$8\mp$--plane. The loop-channel open amplitudes are
\begin{equation}
    \begin{aligned}
        \A_{33}=&\left[ N {\overline N} \left(\frac{V_8-S_8}{\eta^8} \right)+  \frac{N^2 + \overline{N}^2}{2}\left(\frac{O_8-C_8}{\eta^8}\right)\right] \sum_m P_m \ ,  \\
        {\M}_{3}=&\left(\frac{8\epsilon \hat{\eta}}{\hat{\vartheta}_2^3}\right)\left[ \frac{N+{\overline N}}{2} \left({\hat O}_2 {\hat O}_6 + {\hat V}_2 {\hat V}_6\right) +\frac{N-{\overline N}}{2}\left({\hat S}_2 {\hat C}_6 - {\hat C}_2 {\hat S}_6\right)\right] \sum_m (-1)^m P_m \ .
    \end{aligned}
\label{dpd32}
\end{equation}

\noindent The gauge group is $\mathrm{U}(N)$ and there is a complete ${\cal N}=4$ super Yang--Mills multiplet at the massless level. There are complex scalar tachyons in all configurations but that with a single $N=1$ D3-brane and $\epsilon=-1$, where the tachyon disappears at small radius $R<\sqrt{2\alpha'}$. On the other hand, the associated transverse M\"obius amplitude \eqref{dpd31} gives a positive contribution to the potential energy, so that the $\epsilon=1$ setup is energetically favored, but the corresponding D3-brane has a symmetric $\bm{\frac{N(N+1)}{2}}$ tachyon for all $N$. 

\subsection{Even-dimensional branes} 

Except for the D4 and D8-branes wrapping the circle -- which are related to the orthogonal D3 and D7-branes through a kink transition \cite{Sen:1998tt}, as we discuss in Section \ref{ch3_sec_ss_dp_orientifolds} -- all the other even D$p$ brane are found to be unstable. We begin with those wrapping the circle. The transverse channel amplitudes are given by
\begin{equation}
    \begin{aligned}
        {\tilde \A}_{pp} =& \frac{2^{-\frac{p+1}{2}}v}{2}\frac{N^2}{\eta^8} 
\left( V_{p-1} O_{9-p} +  O_{p-1} V_{9-p}  \right) \sum_n W_n \ , \\
{\tilde \M}_{p} =& -\frac{\sqrt{2}vN  \epsilon}{\hat{\eta}^{p-1}} \left( \frac{2 \hat{\eta}}{\hat{\vartheta}_2} \right)^{\frac{9-p}{2}}
 \left( {\hat V}_{p-1} {\hat O}_{9-p} -  {\hat O}_{p-1} {\hat V}_{9-p} \right) \sum_n W_{2n+1} \ ,
    \end{aligned}
\label{dpdeven1_2}
\end{equation}

\noindent where $\epsilon = \pm 1$ identifies the position of the T-dual D($p-1$)-brane on either the $\text{O}8_-$ or to the  $\text{O}8_+$--plane. As for the D8-brane discussed in Section \ref{ch3_sec_ss_dp_orientifolds}, there are neither physical nor unphysical couplings to R-R fields. The former are not allowed by construction, whereas the presence of the latter would be related to background magnetic fields, which will turn them into physical couplings to R-R forms that, however, do not exist for the even D$p$ branes under consideration. 

The loop-channel open amplitudes are equal to
\begin{equation}
\begin{aligned}
        {\A}_{pp} &= \frac{N^2}{2}\frac{1}{\eta^8}  \left[ (O_{p-1}+V_{p-1}) (O_{9-p}+V_{9-p}) - 2 S'_{p-1} S'_{9-p} \right]  \sum_m P_m \ ,  \\
        \M_p &= - \frac{\epsilon N}{\sqrt{2}} \ \frac{1}{\hat{\eta}^{p-1}}  \left( \frac{2 \hat{\eta}}{\hat{\vartheta}_2} \right)^{\frac{9-p}{2}} \left[ \sin \tfrac{(p-5) \pi}{4} \left({\hat O}_{p-1} {\hat O}_{9-p} + {\hat V}_{p-1} {\hat V}_{9-p} \right) \right.\\
        &\left.\hspace{42mm}+ \cos \tfrac{(p-5) \pi}{4}  \left({\hat O}_{p-1} {\hat V}_{9-p} - {\hat V}_{p-1} {\hat O}_{9-p} \right)   \right] \sum_m (-1)^m P_m\; . 
\end{aligned}
\label{dpdeven2_3}
\end{equation}
The gauge group is either $\mathrm{SO}(N)$ or $\mathrm{USp}(N)$, and there is a tachyon in either the symmetric or the antisymmetric representation. The massless tachyon can be removed in the $\mathrm{SO}(1)$ configuration if in the antisymmetric representation, which is the case for $p=0$ mod 4 and $\epsilon = (-1)^{\frac{p}{4}}$. This confirms that D4 and D8 are the only stable even branes wrapping the circle.

Let us now turn to the even D-branes orthogonal to the circle.   The cylinder amplitudes are given by
\begin{equation}
\begin{aligned}
    {\tilde \A}_{pp} &= \frac{2^{-\frac{p+1}{2}}}{2v \ \eta^8} \left( e^{\frac{i \pi m}{2}} N + e^{-\frac{i \pi m}{2}} {\overline N} \right)^2 
\left( V_{p-1} O_{9-p} +  O_{p-1} V_{9-p}  \right) \sum_m P_m \ , \\
    {\A}_{pp} &= \frac{1}{\eta^8}  
\left( N {\overline N} \sum_n W_n + \frac{N^2 +{\overline N}^2 }{2} \sum_n W_{n+\frac{1}{2}} \right)  
 \left[ (O_{p-1}+V_{p-1}) (O_{9-p}+V_{9-p}) - 2 S'_{p-1} S'_{9-p} \right] \ .
\end{aligned}
\end{equation}
There is no M\"obius amplitude, as the Klein bottle \eqref{dp2} only involves non-vanishing winding numbers. The gauge group is $\mathrm{U}(N)$, as signaled by the complex Chan-Paton factors. Tachyons are present for any value of the radius, and they transform in the adjoint representation of $\mathrm{U}(N)$. There exist also charged scalars, which are tachyonic for $R < \sqrt{2 \alpha'}$. Hence, orthogonal, even-dimensional branes are unstable for any value of the radius.

\section{Additional branes of the new Scherk--Schwarz orientifold}\label{app_branes_new_ss}

The D-branes of the new Scherk--Schwarz orientifold \eqref{ss4-6} constructed in \cite{Bossard:2024mls} that we report here are the non-BPS D5-branes wrapping the circle, the charged D3-branes and the non-BPS D1-branes orthogonal to the circle, along with all even-dimensional branes. The wrapping D5-branes resembles the behavior of the D9-branes, whereas the charged orthogonal D3-branes are similar to the orthogonal D7-branes and the orthogonal D1-branes are similar to the orthogonal D5-branes. These analogue branes are discussed in Section \ref{ch3_sec_new_ss_branes}. Instead, the even-dimensional branes behave like those of the Dabholkar--Park orientifold, discussed in Section \ref{ch3_sec_dp_branes} and in Appendix \ref{app_branes_dp}.

\subsection{Non-BPS D5-branes wrapping the circle}

The wrapping D5-branes are uncharged and are parameterized as the D9-branes discussed in Section \ref{ch3_sec_new_ss_branes}. In the case of a single type of branes, the cylinder and M\"obius amplitudes in the transverse channel are
\begin{equation}
\begin{aligned}
    {\tilde \A}_{55} &= \frac{2^{-3}v}{2\eta^8}  \sum_n \left[ (N + {\overline N})^2 (V_8 W_{2n}+ O_8 W_{2n+1}) + (N - {\overline N})^2 (S_8 W_{2n}+ C_8 W_{2n+1})\right] \ , \\
    {\tilde \M}_5 &= - \left( \frac{4v}{\hat{\eta}^2 \hat{\vartheta}_2^2} \right) \sum_n\left[ \epsilon_1  (N + {\overline N}) \left({\hat O}_4 {\hat O}_4-{\hat V}_4 {\hat V}_4\right) + i (-1)^n\epsilon_2  (N - {\overline N}) \left({\hat S}_4 {\hat C}_4- {\hat C}_4 {\hat S}_4\right) \right]   W_{2n+1} \ .
\end{aligned}
\label{eq:ssd51}
\end{equation}

\noindent The direct channel amplitudes are instead
\begin{equation}
    \begin{aligned}
        \A_{55} &=\frac{1}{\eta^8}\left[N  \overline{N}   \sum_m \left(V_8 P_m - S_8 P_{m+1/2}\right) + \frac{N^2 + \overline{N}^2 }{2}\sum_m \left(O_8 P_m - C_8 P_{m+1/2}\right) \right] \ , \\
        \M_5&=\left(\frac{2}{\hat{\eta}^2 \hat{\vartheta}_2^2} \right)\sum_m (-1)^m\left[\epsilon_1 (N+ \overline{N})\left({\hat O}_4 {\hat O}_4 - {\hat V}_4 {\hat V}_4\right) P_{m}+\epsilon_2 (N- \overline{N})\left({\hat S}_4 {\hat C}_4 - {\hat C}_4 {\hat S}_4 \right) P_{m+1/2} \right] \ .
    \end{aligned}
\label{eq:ssd52}
\end{equation}
  
\noindent The gauge group is $\mathrm{U}(N)$ and the massless spectrum contains, in addition to the gauge vectors, four adjoint scalars and complex tachyonic scalars in the symmetric/antisymmetric representation for $\epsilon_1=\pm1$, respectively. These tachyons can be antisymmetrized for $\epsilon_1 = -1$, but even in this case the first Kaluza--Klein mode of the scalar is still tachyonic for $R > \sqrt{2 \alpha'}$. Hence, there is no value of the radius for which both the closed string and the D5-brane tachyons are eliminated, except for $R = \sqrt{2 \alpha'}$ where both scalars are massless. At large radius the closed string tachyon is massive, but the D5-brane is unstable. \\

The instability of the D5-brane has the following interpretation in F-theory. Consider the theory on one extra circle of coordinate $X^8$ and a D5-brane wrapping both $X^8$ and $X^9$. After T-duality along this circle, the D5-brane is mapped into an M5-brane wrapped over the M-theory circle:
\beq 
    {\rm D5}_{0123 89} \underset{{\rm T-duality}}{\rightarrow} {\rm M5}_{0123 91\hspace{-0.3mm} 0} \; . 
\eeq
However, there is no 2-cycle in $\mathcal{M}_3$ along $(X^9,X^{10})$, which suggests that the brane is unstable. Similar conclusions are reached considering a D5-brane orthogonal to $X^8$, with 
\beq 
{\rm D5}_{01234 9} \underset{{\rm T-duality}}{\rightarrow} {\rm KK6}_{01234 89} \; , 
\eeq
since the there is no 2-cycle in $\mathcal{M}_3$ along $(X^8,X^{9})$. 

\subsection{Charged D3-branes orthogonal to the circle}

These D3-branes are charged and follow the same behavior as the charged D7-branes orthogonal to the circle discussed in Section \ref{ch3_sec_new_ss_branes}. The loop and transverse cylinder amplitudes are equal to
\begin{align}
    &A_{33} = \sum_n \left( N  \overline{N}  W_n +\frac{N^2 + \overline{N}^2}{2}W_{n+1/2}\right)\frac{V_8-S_8}{\eta^8}\; , \label{eq:ssd34} \\
    &\begin{aligned}
        {\tilde \A}_{33} &= \frac{2^{-2}}{2 v}\left (\frac{V_8-S_8}{\eta^8}\right) \sum_m \left[ e^{\frac{i \pi m}{2}} N + e^{-\frac{i \pi m}{2}} \overline{N}  \right]^2 P_m \\
        &=\frac{2^{-2}}{v} \left (\frac{V_8-S_8}{\eta^8}\right)\sum_m \left[ N  \overline{N}+ (-1)^m\frac{N^2 + \overline{N}^2}{2}\right] P_m\; .
    \end{aligned} \label{eq:ssd33}
\end{align}

\noindent The transverse amplitude tells that there is no closed-string state that can be exchanged between the D3-brane and the O9-plane, since the associated Klein bottle transverse amplitude of eq.~\eqref{ss4-7} only involves massive winding states. The M\"obius amplitude is therefore absent, and the cylinder one alone describes the dynamics of the brane even in the presence of the orientifold. The theory living on the worldvolume of the orthogonal D3-brane is found to be ${\cal N}=4$ super-Yang--Mills. It is known that this theory has a strong-weak coupling S-duality, thus supporting the conjectured S-duality of the full non-supersymmetric orientifold construction \eqref{ss4-6}.\\

From the F-theory perspective, this orthogonal charged D3-brane is mapped by T-duality to 
\begin{align}
    {\rm D3}_{0128} \underset{{\rm T-duality}}{\rightarrow} {\rm M2}_{012} \ , && {\rm D3}_{0123} \underset{{\rm T-duality}}{\rightarrow} {\rm M5}_{0123 81\hspace{-0.3mm}0}  \ ,
\end{align}
where the M2-brane is point-like in $\mathcal{M}_3$, while the M5-brane wraps the $\mathbb{Z}$ 2-cycle in $\mathcal{M}_3$.

\subsection{Non-BPS D1-branes orthogonal to the circle}

D1-branes orthogonal to the circle resemble the behavior of the orthogonal D5-branes discussed in Section \ref{ch3_sec_new_ss_branes}. A single D1-brane is stable at large radius
$R \geq \sqrt{2 \alpha'}$, while multiple D1-branes annihilate each other.\\

In the F-theory picture, the corresponding T-dual branes would be 
\begin{align} 
    {\rm D1}_{01} \underset{{\rm T-duality}}{\rightarrow} {\rm M2}_{01 8} \; , &&  {\rm D1}_{08} \underset{{\rm T-duality}}{\rightarrow} {\rm KK}(X^{1\hspace{-0.3mm}0}) \ ,
\end{align}
where the M2-brane wraps the torsion $\mathbb{Z}_2$ 1-cycle along $X^8$ in $\mathcal{M}_3$, while the KK-brane is along the circle $X^{10}$ that is only conserved modulo two. In both cases one obtains that a single D1 brane orthogonal to the circle naturally carries a conserved $\mathbb{Z}_2$ charge, associated with its stability. 

\subsection{Even-dimensional branes} 

Finally, we describe the even-dimensional D$p$-branes, \textit{i.e.} with $p=0,2,4,6,8$. These branes come in two different configurations, related to the sign of their couplings to the closed-string tachyonic-like scalar. 

Let us start with the branes wrapping the circle. The transverse channel open string amplitudes are found to be
\begin{equation}
    \begin{aligned}
        {\tilde \A}_{pp} &= \frac{2^{-\frac{p+1}{2}}v}{2 \eta^8}  
\left[ (N_1{+}N_2)^2V_8\sum_n W_{2n} + (N_1{-}N_2)^2 O_8 \sum_n W_{2n+1} \right]  \ , \\
{\tilde \M}_{p} &= - \sqrt{2} \ v \ (N_1{-}N_2)  \ \frac{1}{\hat{\eta}^{p-1}} \left( \frac{2 \hat{\eta}}{\hat{\vartheta}_2} \right)^{\frac{9-p}{2}}
 \left( {\hat O}_{p-1} {\hat O}_{9-p} + {\hat V}_{p-1} {\hat V}_{9-p} \right) \sum_n W_{2n+1} \ ,
    \end{aligned}
\label{ss4deven1}
\end{equation}
\noindent where the usual sign $\epsilon = \pm 1$ can be absorbed in the definition of $N_1$ and $N_2$. Similarly to the supersymmetric Dabholkar--Park orientifold (see Appendix \ref{ch3_sec_dp_branes}), there are neither physical nor unphysical couplings to the R-R sector. The loop-channel open amplitudes are then
\begin{equation}
    \begin{aligned}
        {\A}_{pp} &= \frac{N_1^2+N_2^2}{2\eta^8}  \left[ (O_{p-1}+V_{p-1}) (O_{9-p}+V_{9-p}) \sum_m P_m  - 2 S'_{p-1} S'_{9-p} \sum_m P_{m+\frac{1}{2}}  \right]   \\
        &\qquad + \frac{N_1 N_2}{\eta^8} \left[ (O_{p-1}+V_{p-1}) (O_{9-p}+V_{9-p}) \sum_m P_{m+\frac{1}{2}}   - 2 S'_{p-1} S'_{9-p} \sum_m   P_m \right] \ , \\ 
        \M_p &= -   \ \frac{N_1-N_2}{\sqrt{2}} \ \frac{1}{\hat{\eta}^{p-1}}  \left( \frac{2 \hat{\eta}}{\hat{\vartheta}_2} \right)^{\frac{9-p}{2}}\left[ \sin \tfrac{(p-5) \pi}{4} \left({\hat V}_{p-1} {\hat O}_{9-p} - {\hat O}_{p-1} {\hat V}_{9-p} \right) \right. \\
       &\left. \hspace{40mm}   + \cos \tfrac{(p-5) \pi}{4}  \left({\hat O}_{p-1} {\hat O}_{9-p} + {\hat V}_{p-1} {\hat V}_{9-p} \right)   \right]\sum_m (-1)^m P_m \ . 
    \end{aligned}
\label{ss4deven2}
\end{equation}

\noindent The gauge group is either $\mathrm{SO}$ or $\mathrm{USp}$ and the spectrum contains brane-position moduli, Dirac fermions in the symmetric and the antisymmetric representation, and potentially tachyonic real scalars, also in the antisymmetric or the symmetric representation of the gauge group. The tachyons are removed for a single D6-brane with $(N_1,N_2)=(1,0)$ and for a single D2-brane with $(N_1,N_2)=(0,1)$. The gauge group is $\mathrm{SO}(1)=\mathbb{Z}_2$ in both cases, and the corresponding branes have no Wilson-line moduli (T-dual to brane positions) on the circle. However, these open-string tachyons disappear in the small radius regime $R \leq \sqrt{2 \alpha'}$, where the closed-string scalar is instead tachyonic. Thus, there are no stable even-dimensional D$p$-brane wrapping the circle in the perturbative regime. In particular, in the large radius limit $R > \sqrt{2 \alpha'}$ the unstable D6 and D2-branes are expected to decay, respectively, into a stable D5 and D1-brane orthogonal to the circle, following the tachyonic kink transition discussed in \cite{Sen:1998tt}. This dynamics takes place also for the D8 and D4-branes in the Dabholkar--Park orientifold, as described in Section \ref{ch3_sec_dp_branes} and Appendix \ref{app_branes_dp}.

Finally, the behavior of the even branes orthogonal to the circle matches that of the supersymmetric Dabholkar--Park case, and are indeed unstable for all values of the radius.

\end{appendices}

\backmatter

\bibliographystyle{JHEP}
\bibliography{bibliography}\addcontentsline{toc}{chapter}{References}

\providecommand{\href}[2]{#2}\begingroup\raggedright\begin{thebibliography}{100}

\bibitem{Bonnefoy:2022rcw}
Q.~Bonnefoy, G.~Casagrande and E.~Dudas, \emph{{Causality constraints on
  nonlinear supersymmetry}},
  \href{http://dx.doi.org/10.1007/JHEP11(2022)113}{\emph{JHEP} {\bfseries 11}
  (2022) 113}, [\href{https://arxiv.org/abs/2206.13451}{{\ttfamily
  2206.13451}}].

\bibitem{Casagrande:2023fjk}
G.~Casagrande, E.~Dudas and M.~Peloso, \emph{{On energy and particle production
  in cosmology: the particular case of the gravitino}},
  \href{http://dx.doi.org/10.1007/JHEP06(2024)003}{\emph{JHEP} {\bfseries 06}
  (2024) 003}, [\href{https://arxiv.org/abs/2310.14964}{{\ttfamily
  2310.14964}}].

\bibitem{Bossard:2024mls}
G.~Bossard, G.~Casagrande and E.~Dudas, \emph{{Twisted orientifold planes and
  S-duality without supersymmetry}},
  \href{http://dx.doi.org/10.1007/JHEP02(2025)062}{\emph{JHEP} {\bfseries 02}
  (2025) 062}, [\href{https://arxiv.org/abs/2411.00955}{{\ttfamily
  2411.00955}}].

\bibitem{ms2}
G.~Bossard, G.~Casagrande, E.~Dudas and A.~Loty, \emph{{A unique coupling of
  the massive spin-2 field to supergravity}},
  \href{https://arxiv.org/abs/2502.09599}{{\ttfamily 2502.09599}}.

\bibitem{Glashow:1961tr}
S.~L. Glashow, \emph{{Partial Symmetries of Weak Interactions}},
  \href{http://dx.doi.org/10.1016/0029-5582(61)90469-2}{\emph{Nucl. Phys.}
  {\bfseries 22} (1961) 579--588}.

\bibitem{Weinberg:1967tq}
S.~Weinberg, \emph{{A Model of Leptons}},
  \href{http://dx.doi.org/10.1103/PhysRevLett.19.1264}{\emph{Phys. Rev. Lett.}
  {\bfseries 19} (1967) 1264--1266}.

\bibitem{Salam:1968rm}
A.~Salam, \emph{{Weak and Electromagnetic Interactions}},
  \href{http://dx.doi.org/10.1142/9789812795915_0034}{\emph{Conf. Proc. C}
  {\bfseries 680519} (1968) 367--377}.

\bibitem{Englert:1964et}
F.~Englert and R.~Brout, \emph{{Broken Symmetry and the Mass of Gauge Vector
  Mesons}}, \href{http://dx.doi.org/10.1103/PhysRevLett.13.321}{\emph{Phys.
  Rev. Lett.} {\bfseries 13} (1964) 321--323}.

\bibitem{Higgs:1964pj}
P.~W. Higgs, \emph{{Broken Symmetries and the Masses of Gauge Bosons}},
  \href{http://dx.doi.org/10.1103/PhysRevLett.13.508}{\emph{Phys. Rev. Lett.}
  {\bfseries 13} (1964) 508--509}.

\bibitem{ATLAS:2012yve}
{\scshape ATLAS} collaboration, G.~Aad et~al., \emph{{Observation of a new
  particle in the search for the Standard Model Higgs boson with the ATLAS
  detector at the LHC}},
  \href{http://dx.doi.org/10.1016/j.physletb.2012.08.020}{\emph{Phys. Lett. B}
  {\bfseries 716} (2012) 1--29},
  [\href{https://arxiv.org/abs/1207.7214}{{\ttfamily 1207.7214}}].

\bibitem{CMS:2012qbp}
{\scshape CMS} collaboration, S.~Chatrchyan et~al., \emph{{Observation of a New
  Boson at a Mass of 125 GeV with the CMS Experiment at the LHC}},
  \href{http://dx.doi.org/10.1016/j.physletb.2012.08.021}{\emph{Phys. Lett. B}
  {\bfseries 716} (2012) 30--61},
  [\href{https://arxiv.org/abs/1207.7235}{{\ttfamily 1207.7235}}].

\bibitem{Aoyama:2012wj}
T.~Aoyama, M.~Hayakawa, T.~Kinoshita and M.~Nio, \emph{{Tenth-Order QED
  Contribution to the Electron g-2 and an Improved Value of the Fine Structure
  Constant}},
  \href{http://dx.doi.org/10.1103/PhysRevLett.109.111807}{\emph{Phys. Rev.
  Lett.} {\bfseries 109} (2012) 111807},
  [\href{https://arxiv.org/abs/1205.5368}{{\ttfamily 1205.5368}}].

\bibitem{Fan:2022eto}
X.~Fan, T.~G. Myers, B.~A.~D. Sukra and G.~Gabrielse, \emph{{Measurement of the
  Electron Magnetic Moment}},
  \href{http://dx.doi.org/10.1103/PhysRevLett.130.071801}{\emph{Phys. Rev.
  Lett.} {\bfseries 130} (2023) 071801},
  [\href{https://arxiv.org/abs/2209.13084}{{\ttfamily 2209.13084}}].

\bibitem{Muong-2:2004fok}
{\scshape Muon g-2} collaboration, G.~W. Bennett et~al., \emph{{Measurement of
  the negative muon anomalous magnetic moment to 0.7 ppm}},
  \href{http://dx.doi.org/10.1103/PhysRevLett.92.161802}{\emph{Phys. Rev.
  Lett.} {\bfseries 92} (2004) 161802},
  [\href{https://arxiv.org/abs/hep-ex/0401008}{{\ttfamily hep-ex/0401008}}].

\bibitem{Goroff:1985sz}
M.~H. Goroff and A.~Sagnotti, \emph{{QUANTUM GRAVITY AT TWO LOOPS}},
  \href{http://dx.doi.org/10.1016/0370-2693(85)91470-4}{\emph{Phys. Lett. B}
  {\bfseries 160} (1985) 81--86}.

\bibitem{Goroff:1985th}
M.~H. Goroff and A.~Sagnotti, \emph{{The Ultraviolet Behavior of Einstein
  Gravity}}, \href{http://dx.doi.org/10.1016/0550-3213(86)90193-8}{\emph{Nucl.
  Phys. B} {\bfseries 266} (1986) 709--736}.

\bibitem{Veneziano:1968yb}
G.~Veneziano, \emph{{Construction of a crossing - symmetric, Regge behaved
  amplitude for linearly rising trajectories}},
  \href{http://dx.doi.org/10.1007/BF02824451}{\emph{Nuovo Cim. A} {\bfseries
  57} (1968) 190--197}.

\bibitem{Virasoro:1969me}
M.~A. Virasoro, \emph{{Alternative constructions of crossing-symmetric
  amplitudes with regge behavior}},
  \href{http://dx.doi.org/10.1103/PhysRev.177.2309}{\emph{Phys. Rev.}
  {\bfseries 177} (1969) 2309--2311}.

\bibitem{Shapiro:1969km}
J.~A. Shapiro, \emph{{Narrow-resonance model with regge behavior for pi pi
  scattering}}, \href{http://dx.doi.org/10.1103/PhysRev.179.1345}{\emph{Phys.
  Rev.} {\bfseries 179} (1969) 1345--1353}.

\bibitem{Scherk:1974ca}
J.~Scherk and J.~H. Schwarz, \emph{{Dual Models for Nonhadrons}},
  \href{http://dx.doi.org/10.1016/0550-3213(74)90010-8}{\emph{Nucl. Phys. B}
  {\bfseries 81} (1974) 118--144}.

\bibitem{Yoneya:1974jg}
T.~Yoneya, \emph{{Connection of Dual Models to Electrodynamics and
  Gravidynamics}}, \href{http://dx.doi.org/10.1143/PTP.51.1907}{\emph{Prog.
  Theor. Phys.} {\bfseries 51} (1974) 1907--1920}.

\bibitem{Green:1981yb}
M.~B. Green and J.~H. Schwarz, \emph{{Supersymmetrical String Theories}},
  \href{http://dx.doi.org/10.1016/0370-2693(82)91110-8}{\emph{Phys. Lett. B}
  {\bfseries 109} (1982) 444--448}.

\bibitem{Gross:1984dd}
D.~J. Gross, J.~A. Harvey, E.~J. Martinec and R.~Rohm, \emph{{The Heterotic
  String}}, \href{http://dx.doi.org/10.1103/PhysRevLett.54.502}{\emph{Phys.
  Rev. Lett.} {\bfseries 54} (1985) 502--505}.

\bibitem{Witten:1995ex}
E.~Witten, \emph{{String theory dynamics in various dimensions}},
  \href{http://dx.doi.org/10.1201/9781482268737-32}{\emph{Nucl. Phys. B}
  {\bfseries 443} (1995) 85--126},
  [\href{https://arxiv.org/abs/hep-th/9503124}{{\ttfamily hep-th/9503124}}].

\bibitem{Candelas:1985en}
P.~Candelas, G.~T. Horowitz, A.~Strominger and E.~Witten, \emph{{Vacuum
  configurations for superstrings}},
  \href{http://dx.doi.org/10.1016/0550-3213(85)90602-9}{\emph{Nucl. Phys. B}
  {\bfseries 258} (1985) 46--74}.

\bibitem{Douglas:2003um}
M.~R. Douglas, \emph{{The Statistics of string / M theory vacua}},
  \href{http://dx.doi.org/10.1088/1126-6708/2003/05/046}{\emph{JHEP} {\bfseries
  05} (2003) 046}, [\href{https://arxiv.org/abs/hep-th/0303194}{{\ttfamily
  hep-th/0303194}}].

\bibitem{Wess:1973kz}
J.~Wess and B.~Zumino, \emph{{A Lagrangian Model Invariant Under Supergauge
  Transformations}},
  \href{http://dx.doi.org/10.1016/0370-2693(74)90578-4}{\emph{Phys. Lett. B}
  {\bfseries 49} (1974) 52}.

\bibitem{Wess:1974tw}
J.~Wess and B.~Zumino, \emph{{Supergauge Transformations in Four-Dimensions}},
  \href{http://dx.doi.org/10.1016/0550-3213(74)90355-1}{\emph{Nucl. Phys. B}
  {\bfseries 70} (1974) 39--50}.

\bibitem{Coleman:1967ad}
S.~R. Coleman and J.~Mandula, \emph{{All Possible Symmetries of the S Matrix}},
  \href{http://dx.doi.org/10.1103/PhysRev.159.1251}{\emph{Phys. Rev.}
  {\bfseries 159} (1967) 1251--1256}.

\bibitem{Haag:1974qh}
R.~Haag, J.~T. Lopuszanski and M.~Sohnius, \emph{{All Possible Generators of
  Supersymmetries of the s Matrix}},
  \href{http://dx.doi.org/10.1016/0550-3213(75)90279-5}{\emph{Nucl. Phys. B}
  {\bfseries 88} (1975) 257}.

\bibitem{Freedman:1976xh}
D.~Z. Freedman, P.~van Nieuwenhuizen and S.~Ferrara, \emph{{Progress Toward a
  Theory of Supergravity}},
  \href{http://dx.doi.org/10.1103/PhysRevD.13.3214}{\emph{Phys. Rev. D}
  {\bfseries 13} (1976) 3214--3218}.

\bibitem{Deser:1976eh}
S.~Deser and B.~Zumino, \emph{{Consistent Supergravity}},
  \href{http://dx.doi.org/10.1016/0370-2693(76)90089-7}{\emph{Phys. Lett. B}
  {\bfseries 62} (1976) 335}.

\bibitem{Bern:2006kd}
Z.~Bern, L.~J. Dixon and R.~Roiban, \emph{{Is N = 8 supergravity ultraviolet
  finite?}},
  \href{http://dx.doi.org/10.1016/j.physletb.2006.11.030}{\emph{Phys. Lett. B}
  {\bfseries 644} (2007) 265--271},
  [\href{https://arxiv.org/abs/hep-th/0611086}{{\ttfamily hep-th/0611086}}].

\bibitem{Wess:1992cp}
J.~Wess and J.~Bagger, \emph{{Supersymmetry and supergravity}}.
\newblock Princeton University Press, Princeton, NJ, USA, 1992.

\bibitem{Bertolini:2024xny}
M.~Bertolini, \emph{{Supersymmetry - From the basics to exact results in gauge
  theories}}.
\newblock World Scientific, 12, 2024,
  \href{http://dx.doi.org/10.1142/14026}{10.1142/14026}.

\bibitem{Bilal:2001nv}
A.~Bilal, \emph{{Introduction to supersymmetry}},
  \href{https://arxiv.org/abs/hep-th/0101055}{{\ttfamily hep-th/0101055}}.

\bibitem{Martin:1997ns}
S.~P. Martin, \emph{{A Supersymmetry primer}},
  \href{http://dx.doi.org/10.1142/9789812839657_0001}{\emph{Adv. Ser. Direct.
  High Energy Phys.} {\bfseries 18} (1998) 1--98},
  [\href{https://arxiv.org/abs/hep-ph/9709356}{{\ttfamily hep-ph/9709356}}].

\bibitem{Gates:1983nr}
S.~J. Gates, M.~T. Grisaru, M.~Rocek and W.~Siegel, \emph{{Superspace Or One
  Thousand and One Lessons in Supersymmetry}}, vol.~58 of \emph{Frontiers in
  Physics}.
\newblock 1983.

\bibitem{Freedman:2012zz}
D.~Z. Freedman and A.~Van~Proeyen, \emph{{Supergravity}}.
\newblock Cambridge Univ. Press, Cambridge, UK, 5, 2012,
  \href{http://dx.doi.org/10.1017/CBO9781139026833}{10.1017/CBO9781139026833}.

\bibitem{DallAgata:2021uvl}
G.~Dall\textquoteright{}Agata and M.~Zagermann, \emph{{Supergravity: From First
  Principles to Modern Applications}}, vol.~991 of \emph{Lecture Notes in
  Physics}.
\newblock 7, 2021,
  \href{http://dx.doi.org/10.1007/978-3-662-63980-1}{10.1007/978-3-662-63980-1}.

\bibitem{Antoniadis:2024hvw}
I.~Antoniadis, E.~Dudas, F.~Farakos and A.~Sagnotti, \emph{{Non-Linear
  Supergravity and Inflationary Cosmology}}.
\newblock 9, 2024.
\newblock \href{https://arxiv.org/abs/2409.14943}{{\ttfamily 2409.14943}}.

\bibitem{Salam:1974yz}
A.~Salam and J.~A. Strathdee, \emph{{Supergauge Transformations}},
  \href{http://dx.doi.org/10.1016/0550-3213(74)90537-9}{\emph{Nucl. Phys. B}
  {\bfseries 76} (1974) 477--482}.

\bibitem{Salam:1974jj}
A.~Salam and J.~A. Strathdee, \emph{{On Superfields and Fermi-Bose Symmetry}},
  \href{http://dx.doi.org/10.1103/PhysRevD.11.1521}{\emph{Phys. Rev. D}
  {\bfseries 11} (1975) 1521--1535}.

\bibitem{Zumino:1979et}
B.~Zumino, \emph{{Supersymmetry and Kahler Manifolds}},
  \href{http://dx.doi.org/10.1016/0370-2693(79)90964-X}{\emph{Phys. Lett. B}
  {\bfseries 87} (1979) 203}.

\bibitem{Fayet:1974jb}
P.~Fayet and J.~Iliopoulos, \emph{{Spontaneously Broken Supergauge Symmetries
  and Goldstone Spinors}},
  \href{http://dx.doi.org/10.1016/0370-2693(74)90310-4}{\emph{Phys. Lett. B}
  {\bfseries 51} (1974) 461--464}.

\bibitem{ORaifeartaigh:1975nky}
L.~O'Raifeartaigh, \emph{{Spontaneous Symmetry Breaking for Chiral Scalar
  Superfields}},
  \href{http://dx.doi.org/10.1016/0550-3213(75)90585-4}{\emph{Nucl. Phys. B}
  {\bfseries 96} (1975) 331--352}.

\bibitem{Intriligator:2007py}
K.~A. Intriligator, N.~Seiberg and D.~Shih, \emph{{Supersymmetry breaking,
  R-symmetry breaking and metastable vacua}},
  \href{http://dx.doi.org/10.1088/1126-6708/2007/07/017}{\emph{JHEP} {\bfseries
  07} (2007) 017}, [\href{https://arxiv.org/abs/hep-th/0703281}{{\ttfamily
  hep-th/0703281}}].

\bibitem{Coleman:1973jx}
S.~R. Coleman and E.~J. Weinberg, \emph{{Radiative Corrections as the Origin of
  Spontaneous Symmetry Breaking}},
  \href{http://dx.doi.org/10.1103/PhysRevD.7.1888}{\emph{Phys. Rev. D}
  {\bfseries 7} (1973) 1888--1910}.

\bibitem{Nambu:1961tp}
Y.~Nambu and G.~Jona-Lasinio, \emph{{Dynamical Model of Elementary Particles
  Based on an Analogy with Superconductivity. 1.}},
  \href{http://dx.doi.org/10.1103/PhysRev.122.345}{\emph{Phys. Rev.} {\bfseries
  122} (1961) 345--358}.

\bibitem{Goldstone:1961eq}
J.~Goldstone, \emph{{Field Theories with Superconductor Solutions}},
  \href{http://dx.doi.org/10.1007/BF02812722}{\emph{Nuovo Cim.} {\bfseries 19}
  (1961) 154--164}.

\bibitem{Goldstone:1962es}
J.~Goldstone, A.~Salam and S.~Weinberg, \emph{{Broken Symmetries}},
  \href{http://dx.doi.org/10.1103/PhysRev.127.965}{\emph{Phys. Rev.} {\bfseries
  127} (1962) 965--970}.

\bibitem{Volkov:1973ix}
D.~V. Volkov and V.~P. Akulov, \emph{{Is the Neutrino a Goldstone Particle?}},
  \href{http://dx.doi.org/10.1016/0370-2693(73)90490-5}{\emph{Phys. Lett. B}
  {\bfseries 46} (1973) 109--110}.

\bibitem{Deser:1977uq}
S.~Deser and B.~Zumino, \emph{{Broken Supersymmetry and Supergravity}},
  \href{http://dx.doi.org/10.1103/PhysRevLett.38.1433}{\emph{Phys. Rev. Lett.}
  {\bfseries 38} (1977) 1433--1436}.

\bibitem{Fayet:1986zc}
P.~Fayet, \emph{{Lower Limit on the Mass of a Light Gravitino from e+ e-
  Annihilation Experiments}},
  \href{http://dx.doi.org/10.1016/0370-2693(86)90626-X}{\emph{Phys. Lett. B}
  {\bfseries 175} (1986) 471--477}.

\bibitem{Casalbuoni:1988kv}
R.~Casalbuoni, S.~De~Curtis, D.~Dominici, F.~Feruglio and R.~Gatto, \emph{{A
  GRAVITINO - GOLDSTINO HIGH-ENERGY EQUIVALENCE THEOREM}},
  \href{http://dx.doi.org/10.1016/0370-2693(88)91439-6}{\emph{Phys. Lett. B}
  {\bfseries 215} (1988) 313--316}.

\bibitem{Casalbuoni:1988qd}
R.~Casalbuoni, S.~De~Curtis, D.~Dominici, F.~Feruglio and R.~Gatto,
  \emph{{High-Energy Equivalence Theorem in Spontaneously Broken
  Supergravity}}, \href{http://dx.doi.org/10.1103/PhysRevD.39.2281}{\emph{Phys.
  Rev. D} {\bfseries 39} (1989) 2281}.

\bibitem{Rocek:1978nb}
M.~Rocek, \emph{{Linearizing the Volkov-Akulov Model}},
  \href{http://dx.doi.org/10.1103/PhysRevLett.41.451}{\emph{Phys. Rev. Lett.}
  {\bfseries 41} (1978) 451--453}.

\bibitem{Lindstrom:1979kq}
U.~Lindstrom and M.~Rocek, \emph{{CONSTRAINED LOCAL SUPERFIELDS}},
  \href{http://dx.doi.org/10.1103/PhysRevD.19.2300}{\emph{Phys. Rev. D}
  {\bfseries 19} (1979) 2300--2303}.

\bibitem{Komargodski:2009rz}
Z.~Komargodski and N.~Seiberg, \emph{{From Linear SUSY to Constrained
  Superfields}},
  \href{http://dx.doi.org/10.1088/1126-6708/2009/09/066}{\emph{JHEP} {\bfseries
  09} (2009) 066}, [\href{https://arxiv.org/abs/0907.2441}{{\ttfamily
  0907.2441}}].

\bibitem{DallAgata:2015zxp}
G.~Dall'Agata and F.~Farakos, \emph{{Constrained superfields in Supergravity}},
  \href{http://dx.doi.org/10.1007/JHEP02(2016)101}{\emph{JHEP} {\bfseries 02}
  (2016) 101}, [\href{https://arxiv.org/abs/1512.02158}{{\ttfamily
  1512.02158}}].

\bibitem{DallAgata:2016syy}
G.~Dall'Agata, E.~Dudas and F.~Farakos, \emph{{On the origin of constrained
  superfields}}, \href{http://dx.doi.org/10.1007/JHEP05(2016)041}{\emph{JHEP}
  {\bfseries 05} (2016) 041},
  [\href{https://arxiv.org/abs/1603.03416}{{\ttfamily 1603.03416}}].

\bibitem{Ivanov:1978mx}
E.~A. Ivanov and A.~A. Kapustnikov, \emph{{General Relationship Between Linear
  and Nonlinear Realizations of Supersymmetry}},
  \href{http://dx.doi.org/10.1088/0305-4470/11/12/005}{\emph{J. Phys. A}
  {\bfseries 11} (1978) 2375--2384}.

\bibitem{Ivanov:1982bpa}
E.~A. Ivanov and A.~A. Kapustnikov, \emph{{THE NONLINEAR REALIZATION STRUCTURE
  OF MODELS WITH SPONTANEOUSLY BROKEN SUPERSYMMETRY}},
  \href{http://dx.doi.org/10.1088/0305-4616/8/2/004}{\emph{J. Phys. G}
  {\bfseries 8} (1982) 167--191}.

\bibitem{Kuzenko:2010ef}
S.~M. Kuzenko and S.~J. Tyler, \emph{{Relating the Komargodski-Seiberg and
  Akulov-Volkov actions: Exact nonlinear field redefinition}},
  \href{http://dx.doi.org/10.1016/j.physletb.2011.03.020}{\emph{Phys. Lett. B}
  {\bfseries 698} (2011) 319--322},
  [\href{https://arxiv.org/abs/1009.3298}{{\ttfamily 1009.3298}}].

\bibitem{Adams:2006sv}
A.~Adams, N.~Arkani-Hamed, S.~Dubovsky, A.~Nicolis and R.~Rattazzi,
  \emph{{Causality, analyticity and an IR obstruction to UV completion}},
  \href{http://dx.doi.org/10.1088/1126-6708/2006/10/014}{\emph{JHEP} {\bfseries
  10} (2006) 014}, [\href{https://arxiv.org/abs/hep-th/0602178}{{\ttfamily
  hep-th/0602178}}].

\bibitem{Bellazzini:2016xrt}
B.~Bellazzini, \emph{{Softness and amplitudes\textquoteright{} positivity for
  spinning particles}},
  \href{http://dx.doi.org/10.1007/JHEP02(2017)034}{\emph{JHEP} {\bfseries 02}
  (2017) 034}, [\href{https://arxiv.org/abs/1605.06111}{{\ttfamily
  1605.06111}}].

\bibitem{Ferrara:2015tyn}
S.~Ferrara, R.~Kallosh and J.~Thaler, \emph{{Cosmology with orthogonal
  nilpotent superfields}},
  \href{http://dx.doi.org/10.1103/PhysRevD.93.043516}{\emph{Phys. Rev. D}
  {\bfseries 93} (2016) 043516},
  [\href{https://arxiv.org/abs/1512.00545}{{\ttfamily 1512.00545}}].

\bibitem{Carrasco:2015iij}
J.~J.~M. Carrasco, R.~Kallosh and A.~Linde, \emph{{Minimal supergravity
  inflation}}, \href{http://dx.doi.org/10.1103/PhysRevD.93.061301}{\emph{Phys.
  Rev. D} {\bfseries 93} (2016) 061301},
  [\href{https://arxiv.org/abs/1512.00546}{{\ttfamily 1512.00546}}].

\bibitem{Kolb:2021xfn}
E.~W. Kolb, A.~J. Long and E.~McDonough, \emph{{Catastrophic production of slow
  gravitinos}},
  \href{http://dx.doi.org/10.1103/PhysRevD.104.075015}{\emph{Phys. Rev. D}
  {\bfseries 104} (2021) 075015},
  [\href{https://arxiv.org/abs/2102.10113}{{\ttfamily 2102.10113}}].

\bibitem{Dudas:2021njv}
E.~Dudas, M.~A.~G. Garcia, Y.~Mambrini, K.~A. Olive, M.~Peloso and S.~Verner,
  \emph{{Slow and Safe Gravitinos}},
  \href{http://dx.doi.org/10.1103/PhysRevD.103.123519}{\emph{Phys. Rev. D}
  {\bfseries 103} (2021) 123519},
  [\href{https://arxiv.org/abs/2104.03749}{{\ttfamily 2104.03749}}].

\bibitem{Kahn:2015mla}
Y.~Kahn, D.~A. Roberts and J.~Thaler, \emph{{The goldstone and goldstino of
  supersymmetric inflation}},
  \href{http://dx.doi.org/10.1007/JHEP10(2015)001}{\emph{JHEP} {\bfseries 10}
  (2015) 001}, [\href{https://arxiv.org/abs/1504.05958}{{\ttfamily
  1504.05958}}].

\bibitem{Dine:2009sw}
M.~Dine, G.~Festuccia and Z.~Komargodski, \emph{{A Bound on the
  Superpotential}},
  \href{http://dx.doi.org/10.1007/JHEP03(2010)011}{\emph{JHEP} {\bfseries 03}
  (2010) 011}, [\href{https://arxiv.org/abs/0910.2527}{{\ttfamily 0910.2527}}].

\bibitem{Bellazzini:2014waa}
B.~Bellazzini, L.~Martucci and R.~Torre, \emph{{Symmetries, Sum Rules and
  Constraints on Effective Field Theories}},
  \href{http://dx.doi.org/10.1007/JHEP09(2014)100}{\emph{JHEP} {\bfseries 09}
  (2014) 100}, [\href{https://arxiv.org/abs/1405.2960}{{\ttfamily 1405.2960}}].

\bibitem{Trott:2020ebl}
T.~Trott, \emph{{Causality, unitarity and symmetry in effective field theory}},
  \href{http://dx.doi.org/10.1007/JHEP07(2021)143}{\emph{JHEP} {\bfseries 07}
  (2021) 143}, [\href{https://arxiv.org/abs/2011.10058}{{\ttfamily
  2011.10058}}].

\bibitem{Benakli:2014bpa}
K.~Benakli, L.~Darm\'e and Y.~Oz, \emph{{The Slow Gravitino}},
  \href{http://dx.doi.org/10.1007/JHEP10(2014)121}{\emph{JHEP} {\bfseries 10}
  (2014) 121}, [\href{https://arxiv.org/abs/1407.8321}{{\ttfamily 1407.8321}}].

\bibitem{Hasegawa:2017hgd}
F.~Hasegawa, K.~Mukaida, K.~Nakayama, T.~Terada and Y.~Yamada, \emph{{Gravitino
  Problem in Minimal Supergravity Inflation}},
  \href{http://dx.doi.org/10.1016/j.physletb.2017.02.030}{\emph{Phys. Lett. B}
  {\bfseries 767} (2017) 392--397},
  [\href{https://arxiv.org/abs/1701.03106}{{\ttfamily 1701.03106}}].

\bibitem{Kolb:2021nob}
E.~W. Kolb, A.~J. Long and E.~McDonough, \emph{{Gravitino Swampland
  Conjecture}},
  \href{http://dx.doi.org/10.1103/PhysRevLett.127.131603}{\emph{Phys. Rev.
  Lett.} {\bfseries 127} (2021) 131603},
  [\href{https://arxiv.org/abs/2103.10437}{{\ttfamily 2103.10437}}].

\bibitem{Terada:2021rtp}
T.~Terada, \emph{{Minimal supergravity inflation without slow gravitino}},
  \href{http://dx.doi.org/10.1103/PhysRevD.103.125022}{\emph{Phys. Rev. D}
  {\bfseries 103} (2021) 125022},
  [\href{https://arxiv.org/abs/2104.05731}{{\ttfamily 2104.05731}}].

\bibitem{Antoniadis:2021jtg}
I.~Antoniadis, K.~Benakli and W.~Ke, \emph{{Salvage of too slow gravitinos}},
  \href{http://dx.doi.org/10.1007/JHEP11(2021)063}{\emph{JHEP} {\bfseries 11}
  (2021) 063}, [\href{https://arxiv.org/abs/2105.03784}{{\ttfamily
  2105.03784}}].

\bibitem{Alberte:2020jsk}
L.~Alberte, C.~de~Rham, S.~Jaitly and A.~J. Tolley, \emph{{Positivity Bounds
  and the Massless Spin-2 Pole}},
  \href{http://dx.doi.org/10.1103/PhysRevD.102.125023}{\emph{Phys. Rev. D}
  {\bfseries 102} (2020) 125023},
  [\href{https://arxiv.org/abs/2007.12667}{{\ttfamily 2007.12667}}].

\bibitem{Tokuda:2020mlf}
J.~Tokuda, K.~Aoki and S.~Hirano, \emph{{Gravitational positivity bounds}},
  \href{http://dx.doi.org/10.1007/JHEP11(2020)054}{\emph{JHEP} {\bfseries 11}
  (2020) 054}, [\href{https://arxiv.org/abs/2007.15009}{{\ttfamily
  2007.15009}}].

\bibitem{Alberte:2020bdz}
L.~Alberte, C.~de~Rham, S.~Jaitly and A.~J. Tolley, \emph{{QED positivity
  bounds}}, \href{http://dx.doi.org/10.1103/PhysRevD.103.125020}{\emph{Phys.
  Rev. D} {\bfseries 103} (2021) 125020},
  [\href{https://arxiv.org/abs/2012.05798}{{\ttfamily 2012.05798}}].

\bibitem{Caron-Huot:2021rmr}
S.~Caron-Huot, D.~Mazac, L.~Rastelli and D.~Simmons-Duffin, \emph{{Sharp
  boundaries for the swampland}},
  \href{http://dx.doi.org/10.1007/JHEP07(2021)110}{\emph{JHEP} {\bfseries 07}
  (2021) 110}, [\href{https://arxiv.org/abs/2102.08951}{{\ttfamily
  2102.08951}}].

\bibitem{Arkani-Hamed:2021ajd}
N.~Arkani-Hamed, Y.-t. Huang, J.-Y. Liu and G.~N. Remmen, \emph{{Causality,
  unitarity, and the weak gravity conjecture}},
  \href{http://dx.doi.org/10.1007/JHEP03(2022)083}{\emph{JHEP} {\bfseries 03}
  (2022) 083}, [\href{https://arxiv.org/abs/2109.13937}{{\ttfamily
  2109.13937}}].

\bibitem{Ferrara:1974pz}
S.~Ferrara and B.~Zumino, \emph{{Transformation Properties of the
  Supercurrent}},
  \href{http://dx.doi.org/10.1016/0550-3213(75)90063-2}{\emph{Nucl. Phys. B}
  {\bfseries 87} (1975) 207}.

\bibitem{Cremmer:1982wb}
E.~Cremmer, S.~Ferrara, L.~Girardello and A.~Van~Proeyen, \emph{{Coupling
  Supersymmetric Yang-Mills Theories to Supergravity}},
  \href{http://dx.doi.org/10.1016/0370-2693(82)90332-X}{\emph{Phys. Lett. B}
  {\bfseries 116} (1982) 231--237}.

\bibitem{Cremmer:1982en}
E.~Cremmer, S.~Ferrara, L.~Girardello and A.~Van~Proeyen, \emph{{Yang-Mills
  Theories with Local Supersymmetry: Lagrangian, Transformation Laws and
  SuperHiggs Effect}},
  \href{http://dx.doi.org/10.1016/0550-3213(83)90679-X}{\emph{Nucl. Phys. B}
  {\bfseries 212} (1983) 413}.

\bibitem{Bagger:1982ab}
J.~A. Bagger, \emph{{Coupling the Gauge Invariant Supersymmetric Nonlinear
  Sigma Model to Supergravity}},
  \href{http://dx.doi.org/10.1016/0550-3213(83)90411-X}{\emph{Nucl. Phys. B}
  {\bfseries 211} (1983) 302}.

\bibitem{Elvang:2006jk}
H.~Elvang, D.~Z. Freedman and B.~Kors, \emph{{Anomaly cancellation in
  supergravity with Fayet-Iliopoulos couplings}},
  \href{http://dx.doi.org/10.1088/1126-6708/2006/11/068}{\emph{JHEP} {\bfseries
  11} (2006) 068}, [\href{https://arxiv.org/abs/hep-th/0606012}{{\ttfamily
  hep-th/0606012}}].

\bibitem{DeRydt:2007vg}
J.~De~Rydt, J.~Rosseel, T.~T. Schmidt, A.~Van~Proeyen and M.~Zagermann,
  \emph{{Symplectic structure of N=1 supergravity with anomalies and
  Chern-Simons terms}},
  \href{http://dx.doi.org/10.1088/0264-9381/24/20/017}{\emph{Class. Quant.
  Grav.} {\bfseries 24} (2007) 5201--5220},
  [\href{https://arxiv.org/abs/0705.4216}{{\ttfamily 0705.4216}}].

\bibitem{Binetruy:2004hh}
P.~Binetruy, G.~Dvali, R.~Kallosh and A.~Van~Proeyen, \emph{{Fayet-Iliopoulos
  terms in supergravity and cosmology}},
  \href{http://dx.doi.org/10.1088/0264-9381/21/13/005}{\emph{Class. Quant.
  Grav.} {\bfseries 21} (2004) 3137--3170},
  [\href{https://arxiv.org/abs/hep-th/0402046}{{\ttfamily hep-th/0402046}}].

\bibitem{Ferrara:2016ntj}
S.~Ferrara and A.~Van~Proeyen, \emph{{Mass Formulae for Broken Supersymmetry in
  Curved Space-Time}},
  \href{http://dx.doi.org/10.1002/prop.201600100}{\emph{Fortsch. Phys.}
  {\bfseries 64} (2016) 896--902},
  [\href{https://arxiv.org/abs/1609.08480}{{\ttfamily 1609.08480}}].

\bibitem{Ivanov:1984hs}
E.~A. Ivanov and A.~A. Kapustnikov, \emph{{On a Model Independent Description
  of Spontaneously Broken $N=1$ Supergravity in Superspace}},
  \href{http://dx.doi.org/10.1016/0370-2693(84)91486-2}{\emph{Phys. Lett. B}
  {\bfseries 143} (1984) 379--383}.

\bibitem{Ivanov:1989bh}
E.~A. Ivanov and A.~A. Kapustnikov, \emph{{Geometry of Spontaneously Broken
  Local $N=1$ Supersymmetry in Superspace}},
  \href{http://dx.doi.org/10.1016/0550-3213(90)90046-G}{\emph{Nucl. Phys. B}
  {\bfseries 333} (1990) 439--470}.

\bibitem{Samuel:1982uh}
S.~Samuel and J.~Wess, \emph{{A Superfield Formulation of the Nonlinear
  Realization of Supersymmetry and Its Coupling to Supergravity}},
  \href{http://dx.doi.org/10.1016/0550-3213(83)90622-3}{\emph{Nucl. Phys. B}
  {\bfseries 221} (1983) 153--177}.

\bibitem{Farakos:2013ih}
F.~Farakos and A.~Kehagias, \emph{{Decoupling Limits of sGoldstino Modes in
  Global and Local Supersymmetry}},
  \href{http://dx.doi.org/10.1016/j.physletb.2013.06.001}{\emph{Phys. Lett. B}
  {\bfseries 724} (2013) 322--327},
  [\href{https://arxiv.org/abs/1302.0866}{{\ttfamily 1302.0866}}].

\bibitem{Antoniadis:2014oya}
I.~Antoniadis, E.~Dudas, S.~Ferrara and A.~Sagnotti, \emph{{The
  Volkov\textendash{}Akulov\textendash{}Starobinsky supergravity}},
  \href{http://dx.doi.org/10.1016/j.physletb.2014.04.015}{\emph{Phys. Lett. B}
  {\bfseries 733} (2014) 32--35},
  [\href{https://arxiv.org/abs/1403.3269}{{\ttfamily 1403.3269}}].

\bibitem{Ferrara:2014kva}
S.~Ferrara, R.~Kallosh and A.~Linde, \emph{{Cosmology with Nilpotent
  Superfields}}, \href{http://dx.doi.org/10.1007/JHEP10(2014)143}{\emph{JHEP}
  {\bfseries 10} (2014) 143},
  [\href{https://arxiv.org/abs/1408.4096}{{\ttfamily 1408.4096}}].

\bibitem{Kallosh:2014via}
R.~Kallosh and A.~Linde, \emph{{Inflation and Uplifting with Nilpotent
  Superfields}},
  \href{http://dx.doi.org/10.1088/1475-7516/2015/01/025}{\emph{JCAP} {\bfseries
  01} (2015) 025}, [\href{https://arxiv.org/abs/1408.5950}{{\ttfamily
  1408.5950}}].

\bibitem{DallAgata:2014qsj}
G.~Dall'Agata and F.~Zwirner, \emph{{On sgoldstino-less supergravity models of
  inflation}}, \href{http://dx.doi.org/10.1007/JHEP12(2014)172}{\emph{JHEP}
  {\bfseries 12} (2014) 172},
  [\href{https://arxiv.org/abs/1411.2605}{{\ttfamily 1411.2605}}].

\bibitem{Kallosh:2014hxa}
R.~Kallosh, A.~Linde and M.~Scalisi, \emph{{Inflation, de Sitter Landscape and
  Super-Higgs effect}},
  \href{http://dx.doi.org/10.1007/JHEP03(2015)111}{\emph{JHEP} {\bfseries 03}
  (2015) 111}, [\href{https://arxiv.org/abs/1411.5671}{{\ttfamily 1411.5671}}].

\bibitem{Dudas:2015eha}
E.~Dudas, S.~Ferrara, A.~Kehagias and A.~Sagnotti, \emph{{Properties of
  Nilpotent Supergravity}},
  \href{http://dx.doi.org/10.1007/JHEP09(2015)217}{\emph{JHEP} {\bfseries 09}
  (2015) 217}, [\href{https://arxiv.org/abs/1507.07842}{{\ttfamily
  1507.07842}}].

\bibitem{Bergshoeff:2015tra}
E.~A. Bergshoeff, D.~Z. Freedman, R.~Kallosh and A.~Van~Proeyen, \emph{{Pure de
  Sitter Supergravity}},
  \href{http://dx.doi.org/10.1103/PhysRevD.93.069901}{\emph{Phys. Rev. D}
  {\bfseries 92} (2015) 085040},
  [\href{https://arxiv.org/abs/1507.08264}{{\ttfamily 1507.08264}}].

\bibitem{Hasegawa:2015bza}
F.~Hasegawa and Y.~Yamada, \emph{{Component action of nilpotent multiplet
  coupled to matter in 4 dimensional $ \mathcal{N}=1 $ supergravity}},
  \href{http://dx.doi.org/10.1007/JHEP10(2015)106}{\emph{JHEP} {\bfseries 10}
  (2015) 106}, [\href{https://arxiv.org/abs/1507.08619}{{\ttfamily
  1507.08619}}].

\bibitem{Ferrara:2015gta}
S.~Ferrara, M.~Porrati and A.~Sagnotti, \emph{{Scale invariant
  Volkov\textendash{}Akulov supergravity}},
  \href{http://dx.doi.org/10.1016/j.physletb.2015.08.066}{\emph{Phys. Lett. B}
  {\bfseries 749} (2015) 589--591},
  [\href{https://arxiv.org/abs/1508.02939}{{\ttfamily 1508.02939}}].

\bibitem{Kuzenko:2015yxa}
S.~M. Kuzenko, \emph{{Complex linear Goldstino superfield and supergravity}},
  \href{http://dx.doi.org/10.1007/JHEP10(2015)006}{\emph{JHEP} {\bfseries 10}
  (2015) 006}, [\href{https://arxiv.org/abs/1508.03190}{{\ttfamily
  1508.03190}}].

\bibitem{Antoniadis:2015ala}
I.~Antoniadis and C.~Markou, \emph{{The coupling of Non-linear Supersymmetry to
  Supergravity}},
  \href{http://dx.doi.org/10.1140/epjc/s10052-015-3783-0}{\emph{Eur. Phys. J.
  C} {\bfseries 75} (2015) 582},
  [\href{https://arxiv.org/abs/1508.06767}{{\ttfamily 1508.06767}}].

\bibitem{Kallosh:2015tea}
R.~Kallosh and T.~Wrase, \emph{{De Sitter Supergravity Model Building}},
  \href{http://dx.doi.org/10.1103/PhysRevD.92.105010}{\emph{Phys. Rev. D}
  {\bfseries 92} (2015) 105010},
  [\href{https://arxiv.org/abs/1509.02137}{{\ttfamily 1509.02137}}].

\bibitem{Kallosh:2015sea}
R.~Kallosh, \emph{{Matter-coupled de Sitter Supergravity}},
  \href{http://dx.doi.org/10.1134/S0040577916050068}{\emph{Theor. Math. Phys.}
  {\bfseries 187} (2016) 695--705},
  [\href{https://arxiv.org/abs/1509.02136}{{\ttfamily 1509.02136}}].

\bibitem{DallAgata:2015pdd}
G.~Dall'Agata, S.~Ferrara and F.~Zwirner, \emph{{Minimal scalar-less
  matter-coupled supergravity}},
  \href{http://dx.doi.org/10.1016/j.physletb.2015.11.066}{\emph{Phys. Lett. B}
  {\bfseries 752} (2016) 263--266},
  [\href{https://arxiv.org/abs/1509.06345}{{\ttfamily 1509.06345}}].

\bibitem{Bandos:2015xnf}
I.~Bandos, L.~Martucci, D.~Sorokin and M.~Tonin, \emph{{Brane induced
  supersymmetry breaking and de Sitter supergravity}},
  \href{http://dx.doi.org/10.1007/JHEP02(2016)080}{\emph{JHEP} {\bfseries 02}
  (2016) 080}, [\href{https://arxiv.org/abs/1511.03024}{{\ttfamily
  1511.03024}}].

\bibitem{Bandos:2016xyu}
I.~Bandos, M.~Heller, S.~M. Kuzenko, L.~Martucci and D.~Sorokin, \emph{{The
  Goldstino brane, the constrained superfields and matter in $ \mathcal{N}=1 $
  supergravity}}, \href{http://dx.doi.org/10.1007/JHEP11(2016)109}{\emph{JHEP}
  {\bfseries 11} (2016) 109},
  [\href{https://arxiv.org/abs/1608.05908}{{\ttfamily 1608.05908}}].

\bibitem{Kallosh:1999jj}
R.~Kallosh, L.~Kofman, A.~D. Linde and A.~Van~Proeyen, \emph{{Gravitino
  production after inflation}},
  \href{http://dx.doi.org/10.1103/PhysRevD.61.103503}{\emph{Phys. Rev. D}
  {\bfseries 61} (2000) 103503},
  [\href{https://arxiv.org/abs/hep-th/9907124}{{\ttfamily hep-th/9907124}}].

\bibitem{Kallosh:2000ve}
R.~Kallosh, L.~Kofman, A.~D. Linde and A.~Van~Proeyen, \emph{{Superconformal
  symmetry, supergravity and cosmology}},
  \href{http://dx.doi.org/10.1088/0264-9381/17/20/308}{\emph{Class. Quant.
  Grav.} {\bfseries 17} (2000) 4269--4338},
  [\href{https://arxiv.org/abs/hep-th/0006179}{{\ttfamily hep-th/0006179}}].

\bibitem{Giudice:1999yt}
G.~F. Giudice, I.~Tkachev and A.~Riotto, \emph{{Nonthermal production of
  dangerous relics in the early universe}},
  \href{http://dx.doi.org/10.1088/1126-6708/1999/08/009}{\emph{JHEP} {\bfseries
  08} (1999) 009}, [\href{https://arxiv.org/abs/hep-ph/9907510}{{\ttfamily
  hep-ph/9907510}}].

\bibitem{Giudice:1999am}
G.~F. Giudice, A.~Riotto and I.~Tkachev, \emph{{Thermal and nonthermal
  production of gravitinos in the early universe}},
  \href{http://dx.doi.org/10.1088/1126-6708/1999/11/036}{\emph{JHEP} {\bfseries
  11} (1999) 036}, [\href{https://arxiv.org/abs/hep-ph/9911302}{{\ttfamily
  hep-ph/9911302}}].

\bibitem{Nilles:2001fg}
H.~P. Nilles, M.~Peloso and L.~Sorbo, \emph{{Coupled fields in external
  background with application to nonthermal production of gravitinos}},
  \href{http://dx.doi.org/10.1088/1126-6708/2001/04/004}{\emph{JHEP} {\bfseries
  04} (2001) 004}, [\href{https://arxiv.org/abs/hep-th/0103202}{{\ttfamily
  hep-th/0103202}}].

\bibitem{Nilles:2001ry}
H.~P. Nilles, M.~Peloso and L.~Sorbo, \emph{{Nonthermal production of
  gravitinos and inflatinos}},
  \href{http://dx.doi.org/10.1103/PhysRevLett.87.051302}{\emph{Phys. Rev.
  Lett.} {\bfseries 87} (2001) 051302},
  [\href{https://arxiv.org/abs/hep-ph/0102264}{{\ttfamily hep-ph/0102264}}].

\bibitem{Parker:1969au}
L.~Parker, \emph{{Quantized fields and particle creation in expanding
  universes. 1.}},
  \href{http://dx.doi.org/10.1103/PhysRev.183.1057}{\emph{Phys. Rev.}
  {\bfseries 183} (1969) 1057--1068}.

\bibitem{Zeldovich:1971mw}
Y.~B. Zeldovich and A.~A. Starobinsky, \emph{{Particle production and vacuum
  polarization in an anisotropic gravitational field}}, {\emph{Zh. Eksp. Teor.
  Fiz.} {\bfseries 61} (1971) 2161--2175}.

\bibitem{Ford:1986sy}
L.~H. Ford, \emph{{Gravitational Particle Creation and Inflation}},
  \href{http://dx.doi.org/10.1103/PhysRevD.35.2955}{\emph{Phys. Rev. D}
  {\bfseries 35} (1987) 2955}.

\bibitem{Birrell:1982ix}
N.~D. Birrell and P.~C.~W. Davies, \emph{{Quantum Fields in Curved Space}}.
\newblock Cambridge Monographs on Mathematical Physics. Cambridge University
  Press, Cambridge, UK, 1982,
  \href{http://dx.doi.org/10.1017/CBO9780511622632}{10.1017/CBO9780511622632}.

\bibitem{Gorbunov:2011zzc}
D.~S. Gorbunov and V.~A. Rubakov, \emph{{Introduction to the theory of the
  early universe: Cosmological perturbations and inflationary theory}}.
\newblock 2011, \href{http://dx.doi.org/10.1142/7873}{10.1142/7873}.

\bibitem{Kofman:1994rk}
L.~Kofman, A.~D. Linde and A.~A. Starobinsky, \emph{{Reheating after
  inflation}}, \href{http://dx.doi.org/10.1103/PhysRevLett.73.3195}{\emph{Phys.
  Rev. Lett.} {\bfseries 73} (1994) 3195--3198},
  [\href{https://arxiv.org/abs/hep-th/9405187}{{\ttfamily hep-th/9405187}}].

\bibitem{Kofman:1997yn}
L.~Kofman, A.~D. Linde and A.~A. Starobinsky, \emph{{Towards the theory of
  reheating after inflation}},
  \href{http://dx.doi.org/10.1103/PhysRevD.56.3258}{\emph{Phys. Rev. D}
  {\bfseries 56} (1997) 3258--3295},
  [\href{https://arxiv.org/abs/hep-ph/9704452}{{\ttfamily hep-ph/9704452}}].

\bibitem{Vafa:2005ui}
C.~Vafa, \emph{{The String landscape and the swampland}},
  \href{https://arxiv.org/abs/hep-th/0509212}{{\ttfamily hep-th/0509212}}.

\bibitem{Palti:2019pca}
E.~Palti, \emph{{The Swampland: Introduction and Review}},
  \href{http://dx.doi.org/10.1002/prop.201900037}{\emph{Fortsch. Phys.}
  {\bfseries 67} (2019) 1900037},
  [\href{https://arxiv.org/abs/1903.06239}{{\ttfamily 1903.06239}}].

\bibitem{Fulling:1979ac}
S.~A. Fulling, \emph{{REMARKS ON POSITIVE FREQUENCY AND HAMILTONIANS IN
  EXPANDING UNIVERSES}}, \href{http://dx.doi.org/10.1007/BF00756661}{\emph{Gen.
  Rel. Grav.} {\bfseries 10} (1979) 807--824}.

\bibitem{Weiss:1986gg}
N.~Weiss, \emph{{Consistency of Hamiltonian Diagonalization for Field Theories
  in a Robertson-walker Background}},
  \href{http://dx.doi.org/10.1103/PhysRevD.34.1768}{\emph{Phys. Rev. D}
  {\bfseries 34} (1986) 1768}.

\bibitem{Bozza:2003pr}
V.~Bozza, M.~Giovannini and G.~Veneziano, \emph{{Cosmological perturbations
  from a new physics hypersurface}},
  \href{http://dx.doi.org/10.1088/1475-7516/2003/05/001}{\emph{JCAP} {\bfseries
  05} (2003) 001}, [\href{https://arxiv.org/abs/hep-th/0302184}{{\ttfamily
  hep-th/0302184}}].

\bibitem{Grain:2019vnq}
J.~Grain and V.~Vennin, \emph{{Canonical transformations and squeezing
  formalism in cosmology}},
  \href{http://dx.doi.org/10.1088/1475-7516/2020/02/022}{\emph{JCAP} {\bfseries
  02} (2020) 022}, [\href{https://arxiv.org/abs/1910.01916}{{\ttfamily
  1910.01916}}].

\bibitem{Peloso:2000hy}
M.~Peloso and L.~Sorbo, \emph{{Preheating of massive fermions after inflation:
  Analytical results}},
  \href{http://dx.doi.org/10.1088/1126-6708/2000/05/016}{\emph{JHEP} {\bfseries
  05} (2000) 016}, [\href{https://arxiv.org/abs/hep-ph/0003045}{{\ttfamily
  hep-ph/0003045}}].

\bibitem{Chung:2011ck}
D.~J.~H. Chung, L.~L. Everett, H.~Yoo and P.~Zhou, \emph{{Gravitational Fermion
  Production in Inflationary Cosmology}},
  \href{http://dx.doi.org/10.1016/j.physletb.2012.04.066}{\emph{Phys. Lett. B}
  {\bfseries 712} (2012) 147--154},
  [\href{https://arxiv.org/abs/1109.2524}{{\ttfamily 1109.2524}}].

\bibitem{Himmetoglu:2008zp}
B.~Himmetoglu, C.~R. Contaldi and M.~Peloso, \emph{{Instability of anisotropic
  cosmological solutions supported by vector fields}},
  \href{http://dx.doi.org/10.1103/PhysRevLett.102.111301}{\emph{Phys. Rev.
  Lett.} {\bfseries 102} (2009) 111301},
  [\href{https://arxiv.org/abs/0809.2779}{{\ttfamily 0809.2779}}].

\bibitem{Graham:2015rva}
P.~W. Graham, J.~Mardon and S.~Rajendran, \emph{{Vector Dark Matter from
  Inflationary Fluctuations}},
  \href{http://dx.doi.org/10.1103/PhysRevD.93.103520}{\emph{Phys. Rev. D}
  {\bfseries 93} (2016) 103520},
  [\href{https://arxiv.org/abs/1504.02102}{{\ttfamily 1504.02102}}].

\bibitem{Ahmed:2020fhc}
A.~Ahmed, B.~Grzadkowski and A.~Socha, \emph{{Gravitational production of
  vector dark matter}},
  \href{http://dx.doi.org/10.1007/JHEP08(2020)059}{\emph{JHEP} {\bfseries 08}
  (2020) 059}, [\href{https://arxiv.org/abs/2005.01766}{{\ttfamily
  2005.01766}}].

\bibitem{Wess:1977fn}
J.~Wess and B.~Zumino, \emph{{Superspace Formulation of Supergravity}},
  \href{http://dx.doi.org/10.1016/0370-2693(77)90015-6}{\emph{Phys. Lett. B}
  {\bfseries 66} (1977) 361--364}.

\bibitem{Grimm:1977kp}
R.~Grimm, J.~Wess and B.~Zumino, \emph{{Consistency Checks on the Superspace
  Formulation of Supergravity}},
  \href{http://dx.doi.org/10.1016/0370-2693(78)90753-0}{\emph{Phys. Lett. B}
  {\bfseries 73} (1978) 415--417}.

\bibitem{Wess:1978bu}
J.~Wess and B.~Zumino, \emph{{Superfield Lagrangian for Supergravity}},
  \href{http://dx.doi.org/10.1016/0370-2693(78)90057-6}{\emph{Phys. Lett. B}
  {\bfseries 74} (1978) 51--53}.

\bibitem{Siegel:1978mj}
W.~Siegel and S.~J. Gates, Jr., \emph{{Superfield Supergravity}},
  \href{http://dx.doi.org/10.1016/0550-3213(79)90416-4}{\emph{Nucl. Phys. B}
  {\bfseries 147} (1979) 77--104}.

\bibitem{Stelle:1978ye}
K.~S. Stelle and P.~C. West, \emph{{Minimal Auxiliary Fields for
  Supergravity}},
  \href{http://dx.doi.org/10.1016/0370-2693(78)90669-X}{\emph{Phys. Lett. B}
  {\bfseries 74} (1978) 330--332}.

\bibitem{Ferrara:1978em}
S.~Ferrara and P.~van Nieuwenhuizen, \emph{{The Auxiliary Fields of
  Supergravity}},
  \href{http://dx.doi.org/10.1016/0370-2693(78)90670-6}{\emph{Phys. Lett. B}
  {\bfseries 74} (1978) 333}.

\bibitem{Sohnius:1981tp}
M.~F. Sohnius and P.~C. West, \emph{{An Alternative Minimal Off-Shell Version
  of N=1 Supergravity}},
  \href{http://dx.doi.org/10.1016/0370-2693(81)90778-4}{\emph{Phys. Lett. B}
  {\bfseries 105} (1981) 353--357}.

\bibitem{Ferrara:1977mv}
S.~Ferrara and B.~Zumino, \emph{{Structure of Conformal Supergravity}},
  \href{http://dx.doi.org/10.1016/0550-3213(78)90548-5}{\emph{Nucl. Phys. B}
  {\bfseries 134} (1978) 301--326}.

\bibitem{Komargodski:2010rb}
Z.~Komargodski and N.~Seiberg, \emph{{Comments on Supercurrent Multiplets,
  Supersymmetric Field Theories and Supergravity}},
  \href{http://dx.doi.org/10.1007/JHEP07(2010)017}{\emph{JHEP} {\bfseries 07}
  (2010) 017}, [\href{https://arxiv.org/abs/1002.2228}{{\ttfamily 1002.2228}}].

\bibitem{Festuccia:2011ws}
G.~Festuccia and N.~Seiberg, \emph{{Rigid Supersymmetric Theories in Curved
  Superspace}}, \href{http://dx.doi.org/10.1007/JHEP06(2011)114}{\emph{JHEP}
  {\bfseries 06} (2011) 114},
  [\href{https://arxiv.org/abs/1105.0689}{{\ttfamily 1105.0689}}].

\bibitem{Clark:1995bg}
T.~E. Clark and S.~T. Love, \emph{{The Supercurrent in supersymmetric field
  theories}}, \href{http://dx.doi.org/10.1142/S0217751X9600136X}{\emph{Int. J.
  Mod. Phys. A} {\bfseries 11} (1996) 2807--2823},
  [\href{https://arxiv.org/abs/hep-th/9506145}{{\ttfamily hep-th/9506145}}].

\bibitem{Komargodski:2009pc}
Z.~Komargodski and N.~Seiberg, \emph{{Comments on the Fayet-Iliopoulos Term in
  Field Theory and Supergravity}},
  \href{http://dx.doi.org/10.1088/1126-6708/2009/06/007}{\emph{JHEP} {\bfseries
  06} (2009) 007}, [\href{https://arxiv.org/abs/0904.1159}{{\ttfamily
  0904.1159}}].

\bibitem{Fierz:1939ix}
M.~Fierz and W.~Pauli, \emph{{On relativistic wave equations for particles of
  arbitrary spin in an electromagnetic field}},
  \href{http://dx.doi.org/10.1098/rspa.1939.0140}{\emph{Proc. Roy. Soc. Lond.
  A} {\bfseries 173} (1939) 211--232}.

\bibitem{Boulware:1972yco}
D.~G. Boulware and S.~Deser, \emph{{Can gravitation have a finite range?}},
  \href{http://dx.doi.org/10.1103/PhysRevD.6.3368}{\emph{Phys. Rev. D}
  {\bfseries 6} (1972) 3368--3382}.

\bibitem{Bouatta:2004kk}
N.~Bouatta, G.~Compere and A.~Sagnotti, \emph{{An Introduction to free
  higher-spin fields}},  in \emph{{1st Solvay Workshop on Higher Spin Gauge
  Theories}}, pp.~79--99, 9, 2004.
\newblock \href{https://arxiv.org/abs/hep-th/0409068}{{\ttfamily
  hep-th/0409068}}.

\bibitem{Hinterbichler:2011tt}
K.~Hinterbichler, \emph{{Theoretical Aspects of Massive Gravity}},
  \href{http://dx.doi.org/10.1103/RevModPhys.84.671}{\emph{Rev. Mod. Phys.}
  {\bfseries 84} (2012) 671--710},
  [\href{https://arxiv.org/abs/1105.3735}{{\ttfamily 1105.3735}}].

\bibitem{Arkani-Hamed:2002bjr}
N.~Arkani-Hamed, H.~Georgi and M.~D. Schwartz, \emph{{Effective field theory
  for massive gravitons and gravity in theory space}},
  \href{http://dx.doi.org/10.1016/S0003-4916(03)00068-X}{\emph{Annals Phys.}
  {\bfseries 305} (2003) 96--118},
  [\href{https://arxiv.org/abs/hep-th/0210184}{{\ttfamily hep-th/0210184}}].

\bibitem{Arkani-Hamed:2003roe}
N.~Arkani-Hamed and M.~D. Schwartz, \emph{{Discrete gravitational dimensions}},
  \href{http://dx.doi.org/10.1103/PhysRevD.69.104001}{\emph{Phys. Rev. D}
  {\bfseries 69} (2004) 104001},
  [\href{https://arxiv.org/abs/hep-th/0302110}{{\ttfamily hep-th/0302110}}].

\bibitem{Schwartz:2003vj}
M.~D. Schwartz, \emph{{Constructing gravitational dimensions}},
  \href{http://dx.doi.org/10.1103/PhysRevD.68.024029}{\emph{Phys. Rev. D}
  {\bfseries 68} (2003) 024029},
  [\href{https://arxiv.org/abs/hep-th/0303114}{{\ttfamily hep-th/0303114}}].

\bibitem{Bonifacio:2018vzv}
J.~Bonifacio and K.~Hinterbichler, \emph{{Bounds on Amplitudes in Effective
  Theories with Massive Spinning Particles}},
  \href{http://dx.doi.org/10.1103/PhysRevD.98.045003}{\emph{Phys. Rev. D}
  {\bfseries 98} (2018) 045003},
  [\href{https://arxiv.org/abs/1804.08686}{{\ttfamily 1804.08686}}].

\bibitem{Bonifacio:2018aon}
J.~Bonifacio and K.~Hinterbichler, \emph{{Universal bound on the strong
  coupling scale of a gravitationally coupled massive spin-2 particle}},
  \href{http://dx.doi.org/10.1103/PhysRevD.98.085006}{\emph{Phys. Rev. D}
  {\bfseries 98} (2018) 085006},
  [\href{https://arxiv.org/abs/1806.10607}{{\ttfamily 1806.10607}}].

\bibitem{Bonifacio:2019mgk}
J.~Bonifacio, K.~Hinterbichler and R.~A. Rosen, \emph{{Constraints on a
  gravitational Higgs mechanism}},
  \href{http://dx.doi.org/10.1103/PhysRevD.100.084017}{\emph{Phys. Rev. D}
  {\bfseries 100} (2019) 084017},
  [\href{https://arxiv.org/abs/1903.09643}{{\ttfamily 1903.09643}}].

\bibitem{Kundu:2023cof}
S.~Kundu, E.~Palti and J.~Quirant, \emph{{Regge growth of isolated massive
  spin-2 particles and the Swampland}},
  \href{http://dx.doi.org/10.1007/JHEP05(2024)139}{\emph{JHEP} {\bfseries 05}
  (2024) 139}, [\href{https://arxiv.org/abs/2311.00022}{{\ttfamily
  2311.00022}}].

\bibitem{deRham:2010ik}
C.~de~Rham and G.~Gabadadze, \emph{{Generalization of the Fierz-Pauli Action}},
  \href{http://dx.doi.org/10.1103/PhysRevD.82.044020}{\emph{Phys. Rev. D}
  {\bfseries 82} (2010) 044020},
  [\href{https://arxiv.org/abs/1007.0443}{{\ttfamily 1007.0443}}].

\bibitem{deRham:2010kj}
C.~de~Rham, G.~Gabadadze and A.~J. Tolley, \emph{{Resummation of Massive
  Gravity}},
  \href{http://dx.doi.org/10.1103/PhysRevLett.106.231101}{\emph{Phys. Rev.
  Lett.} {\bfseries 106} (2011) 231101},
  [\href{https://arxiv.org/abs/1011.1232}{{\ttfamily 1011.1232}}].

\bibitem{Hassan:2011zd}
S.~F. Hassan and R.~A. Rosen, \emph{{Bimetric Gravity from Ghost-free Massive
  Gravity}}, \href{http://dx.doi.org/10.1007/JHEP02(2012)126}{\emph{JHEP}
  {\bfseries 02} (2012) 126},
  [\href{https://arxiv.org/abs/1109.3515}{{\ttfamily 1109.3515}}].

\bibitem{SekharChivukula:2019yul}
R.~Sekhar~Chivukula, D.~Foren, K.~A. Mohan, D.~Sengupta and E.~H. Simmons,
  \emph{{Scattering amplitudes of massive spin-2 Kaluza-Klein states grow only
  as ${\cal O}(s)$}},
  \href{http://dx.doi.org/10.1103/PhysRevD.101.055013}{\emph{Phys. Rev. D}
  {\bfseries 101} (2020) 055013},
  [\href{https://arxiv.org/abs/1906.11098}{{\ttfamily 1906.11098}}].

\bibitem{Chivukula:2020hvi}
R.~S. Chivukula, D.~Foren, K.~A. Mohan, D.~Sengupta and E.~H. Simmons,
  \emph{{Massive Spin-2 Scattering Amplitudes in Extra-Dimensional Theories}},
  \href{http://dx.doi.org/10.1103/PhysRevD.101.075013}{\emph{Phys. Rev. D}
  {\bfseries 101} (2020) 075013},
  [\href{https://arxiv.org/abs/2002.12458}{{\ttfamily 2002.12458}}].

\bibitem{Bonifacio:2019ioc}
J.~Bonifacio and K.~Hinterbichler, \emph{{Unitarization from Geometry}},
  \href{http://dx.doi.org/10.1007/JHEP12(2019)165}{\emph{JHEP} {\bfseries 12}
  (2019) 165}, [\href{https://arxiv.org/abs/1910.04767}{{\ttfamily
  1910.04767}}].

\bibitem{Buchbinder:2002gh}
I.~L. Buchbinder, S.~J. Gates, Jr., W.~D. Linch, III and J.~Phillips,
  \emph{{New 4-D, N=1 superfield theory: Model of free massive superspin 3/2
  multiplet}},
  \href{http://dx.doi.org/10.1016/S0370-2693(02)01772-0}{\emph{Phys. Lett. B}
  {\bfseries 535} (2002) 280--288},
  [\href{https://arxiv.org/abs/hep-th/0201096}{{\ttfamily hep-th/0201096}}].

\bibitem{Zinoviev:2002xn}
Y.~M. Zinoviev, \emph{{Massive spin two supermultiplets}},
  \href{https://arxiv.org/abs/hep-th/0206209}{{\ttfamily hep-th/0206209}}.

\bibitem{DelMonte:2016czb}
F.~Del~Monte, D.~Francia and P.~A. Grassi, \emph{{Multimetric Supergravities}},
  \href{http://dx.doi.org/10.1007/JHEP09(2016)064}{\emph{JHEP} {\bfseries 09}
  (2016) 064}, [\href{https://arxiv.org/abs/1605.06793}{{\ttfamily
  1605.06793}}].

\bibitem{Zinoviev:2018juc}
Y.~M. Zinoviev, \emph{{On massive super(bi)gravity in the constructive
  approach}}, \href{http://dx.doi.org/10.1088/1361-6382/aad1fb}{\emph{Class.
  Quant. Grav.} {\bfseries 35} (2018) 175006},
  [\href{https://arxiv.org/abs/1805.01650}{{\ttfamily 1805.01650}}].

\bibitem{Engelbrecht:2022aao}
L.~Engelbrecht, C.~R.~T. Jones and S.~Paranjape, \emph{{Supersymmetric Massive
  Gravity}}, \href{http://dx.doi.org/10.1007/JHEP10(2022)130}{\emph{JHEP}
  {\bfseries 10} (2022) 130},
  [\href{https://arxiv.org/abs/2205.12982}{{\ttfamily 2205.12982}}].

\bibitem{Gunaydin:1983bi}
M.~Gunaydin, G.~Sierra and P.~K. Townsend, \emph{{The Geometry of N=2
  Maxwell-Einstein Supergravity and Jordan Algebras}},
  \href{http://dx.doi.org/10.1016/0550-3213(84)90142-1}{\emph{Nucl. Phys. B}
  {\bfseries 242} (1984) 244--268}.

\bibitem{Ceresole:2000jd}
A.~Ceresole and G.~Dall'Agata, \emph{{General matter coupled N=2, D = 5 gauged
  supergravity}},
  \href{http://dx.doi.org/10.1016/S0550-3213(00)00339-4}{\emph{Nucl. Phys. B}
  {\bfseries 585} (2000) 143--170},
  [\href{https://arxiv.org/abs/hep-th/0004111}{{\ttfamily hep-th/0004111}}].

\bibitem{Petrov:2017bdx}
A.~N. Petrov, S.~M. Kopeikin, R.~R. Lompay and B.~Tekin, \emph{{Metric Theories
  of Gravity: Perturbations and Conservation Laws}}, vol.~38 of \emph{De
  Gruyter Studies in Mathematical Physics}.
\newblock De Gruyter, 4, 2017,
  \href{http://dx.doi.org/10.1515/9783110351781}{10.1515/9783110351781}.

\bibitem{Callan:1970ze}
C.~G. Callan, Jr., S.~R. Coleman and R.~Jackiw, \emph{{A New improved energy -
  momentum tensor}},
  \href{http://dx.doi.org/10.1016/0003-4916(70)90394-5}{\emph{Annals Phys.}
  {\bfseries 59} (1970) 42--73}.

\bibitem{Zucker:1999ej}
M.~Zucker, \emph{{Minimal off-shell supergravity in five-dimensions}},
  \href{http://dx.doi.org/10.1016/S0550-3213(99)00750-6}{\emph{Nucl. Phys. B}
  {\bfseries 570} (2000) 267--283},
  [\href{https://arxiv.org/abs/hep-th/9907082}{{\ttfamily hep-th/9907082}}].

\bibitem{Gherghetta:2002nr}
T.~Gherghetta and A.~Pomarol, \emph{{A Stuckelberg formalism for the gravitino
  from warped extra dimensions}},
  \href{http://dx.doi.org/10.1016/S0370-2693(02)01874-9}{\emph{Phys. Lett. B}
  {\bfseries 536} (2002) 277--282},
  [\href{https://arxiv.org/abs/hep-th/0203120}{{\ttfamily hep-th/0203120}}].

\bibitem{Camanho:2014apa}
X.~O. Camanho, J.~D. Edelstein, J.~Maldacena and A.~Zhiboedov, \emph{{Causality
  Constraints on Corrections to the Graviton Three-Point Coupling}},
  \href{http://dx.doi.org/10.1007/JHEP02(2016)020}{\emph{JHEP} {\bfseries 02}
  (2016) 020}, [\href{https://arxiv.org/abs/1407.5597}{{\ttfamily 1407.5597}}].

\bibitem{Lust:2021jps}
D.~Lust, C.~Markou, P.~Mazloumi and S.~Stieberger, \emph{{Extracting bigravity
  from string theory}},
  \href{http://dx.doi.org/10.1007/JHEP12(2021)220}{\emph{JHEP} {\bfseries 12}
  (2021) 220}, [\href{https://arxiv.org/abs/2106.04614}{{\ttfamily
  2106.04614}}].

\bibitem{Buchbinder:1999ar}
I.~L. Buchbinder, D.~M. Gitman, V.~A. Krykhtin and V.~D. Pershin,
  \emph{{Equations of motion for massive spin-2 field coupled to gravity}},
  \href{http://dx.doi.org/10.1016/S0550-3213(00)00389-8}{\emph{Nucl. Phys. B}
  {\bfseries 584} (2000) 615--640},
  [\href{https://arxiv.org/abs/hep-th/9910188}{{\ttfamily hep-th/9910188}}].

\bibitem{Ooguri:2006in}
H.~Ooguri and C.~Vafa, \emph{{On the Geometry of the String Landscape and the
  Swampland}},
  \href{http://dx.doi.org/10.1016/j.nuclphysb.2006.10.033}{\emph{Nucl. Phys. B}
  {\bfseries 766} (2007) 21--33},
  [\href{https://arxiv.org/abs/hep-th/0605264}{{\ttfamily hep-th/0605264}}].

\bibitem{Adams:2010zy}
A.~Adams, O.~DeWolfe and W.~Taylor, \emph{{String universality in ten
  dimensions}},
  \href{http://dx.doi.org/10.1103/PhysRevLett.105.071601}{\emph{Phys. Rev.
  Lett.} {\bfseries 105} (2010) 071601},
  [\href{https://arxiv.org/abs/1006.1352}{{\ttfamily 1006.1352}}].

\bibitem{Kim:2019ths}
H.-C. Kim, H.-C. Tarazi and C.~Vafa, \emph{{Four-dimensional
  $\mathbf{\mathcal{N}=4}$ SYM theory and the swampland}},
  \href{http://dx.doi.org/10.1103/PhysRevD.102.026003}{\emph{Phys. Rev. D}
  {\bfseries 102} (2020) 026003},
  [\href{https://arxiv.org/abs/1912.06144}{{\ttfamily 1912.06144}}].

\bibitem{Kim:2019vuc}
H.-C. Kim, G.~Shiu and C.~Vafa, \emph{{Branes and the Swampland}},
  \href{http://dx.doi.org/10.1103/PhysRevD.100.066006}{\emph{Phys. Rev. D}
  {\bfseries 100} (2019) 066006},
  [\href{https://arxiv.org/abs/1905.08261}{{\ttfamily 1905.08261}}].

\bibitem{Bedroya:2021fbu}
A.~Bedroya, Y.~Hamada, M.~Montero and C.~Vafa, \emph{{Compactness of brane
  moduli and the String Lamppost Principle in d \ensuremath{>} 6}},
  \href{http://dx.doi.org/10.1007/JHEP02(2022)082}{\emph{JHEP} {\bfseries 02}
  (2022) 082}, [\href{https://arxiv.org/abs/2110.10157}{{\ttfamily
  2110.10157}}].

\bibitem{Montero:2020icj}
M.~Montero and C.~Vafa, \emph{{Cobordism Conjecture, Anomalies, and the String
  Lamppost Principle}},
  \href{http://dx.doi.org/10.1007/JHEP01(2021)063}{\emph{JHEP} {\bfseries 01}
  (2021) 063}, [\href{https://arxiv.org/abs/2008.11729}{{\ttfamily
  2008.11729}}].

\bibitem{Polchinski:1998rq}
J.~Polchinski, \emph{{String theory. Vol. 1: An introduction to the bosonic
  string}}.
\newblock Cambridge Monographs on Mathematical Physics. Cambridge University
  Press, 12, 2007,
  \href{http://dx.doi.org/10.1017/CBO9780511816079}{10.1017/CBO9780511816079}.

\bibitem{Polchinski:1998rr}
J.~Polchinski, \emph{{String theory. Vol. 2: Superstring theory and beyond}}.
\newblock Cambridge Monographs on Mathematical Physics. Cambridge University
  Press, 12, 2007,
  \href{http://dx.doi.org/10.1017/CBO9780511618123}{10.1017/CBO9780511618123}.

\bibitem{Green:1987mn}
M.~B. Green, J.~H. Schwarz and E.~Witten, \emph{{SUPERSTRING THEORY. VOL. 2:
  LOOP AMPLITUDES, ANOMALIES AND PHENOMENOLOGY}}.
\newblock 7, 1988.

\bibitem{Green:1987sp}
M.~B. Green, J.~H. Schwarz and E.~Witten, \emph{{SUPERSTRING THEORY. VOL. 1:
  INTRODUCTION}}.
\newblock Cambridge Monographs on Mathematical Physics. 7, 1988.

\bibitem{Becker:2006dvp}
K.~Becker, M.~Becker and J.~H. Schwarz, \emph{{String theory and M-theory: A
  modern introduction}}.
\newblock Cambridge University Press, 12, 2006,
  \href{http://dx.doi.org/10.1017/CBO9780511816086}{10.1017/CBO9780511816086}.

\bibitem{reviews_1}
C.~Angelantonj and A.~Sagnotti, \emph{{Open strings}},
  \href{http://dx.doi.org/10.1016/S0370-1573(02)00273-9}{\emph{Phys. Rept.}
  {\bfseries 371} (2002) 1--150},
  [\href{https://arxiv.org/abs/hep-th/0204089}{{\ttfamily hep-th/0204089}}].

\bibitem{reviews_2}
E.~Dudas, \emph{{Theory and phenomenology of type I strings and M theory}},
  \href{http://dx.doi.org/10.1088/0264-9381/17/22/201}{\emph{Class. Quant.
  Grav.} {\bfseries 17} (2000) R41--R116},
  [\href{https://arxiv.org/abs/hep-ph/0006190}{{\ttfamily hep-ph/0006190}}].

\bibitem{reviews_4}
C.~Angelantonj and I.~Florakis, \emph{{A Lightning Introduction to String
  Theory}},  \href{https://arxiv.org/abs/2406.09508}{{\ttfamily 2406.09508}}.

\bibitem{Goddard:1973qh}
P.~Goddard, J.~Goldstone, C.~Rebbi and C.~B. Thorn, \emph{{Quantum dynamics of
  a massless relativistic string}},
  \href{http://dx.doi.org/10.1016/0550-3213(73)90223-X}{\emph{Nucl. Phys. B}
  {\bfseries 56} (1973) 109--135}.

\bibitem{Callan:1985ia}
C.~G. Callan, Jr., E.~J. Martinec, M.~J. Perry and D.~Friedan, \emph{{Strings
  in Background Fields}},
  \href{http://dx.doi.org/10.1016/0550-3213(85)90506-1}{\emph{Nucl. Phys. B}
  {\bfseries 262} (1985) 593--609}.

\bibitem{Polyakov:1981rd}
A.~M. Polyakov, \emph{{Quantum Geometry of Bosonic Strings}},
  \href{http://dx.doi.org/10.1016/0370-2693(81)90743-7}{\emph{Phys. Lett. B}
  {\bfseries 103} (1981) 207--210}.

\bibitem{Polyakov:1981re}
A.~M. Polyakov, \emph{{Quantum Geometry of Fermionic Strings}},
  \href{http://dx.doi.org/10.1016/0370-2693(81)90744-9}{\emph{Phys. Lett. B}
  {\bfseries 103} (1981) 211--213}.

\bibitem{Brink:1976sc}
L.~Brink, P.~Di~Vecchia and P.~S. Howe, \emph{{A Locally Supersymmetric and
  Reparametrization Invariant Action for the Spinning String}},
  \href{http://dx.doi.org/10.1016/0370-2693(76)90445-7}{\emph{Phys. Lett. B}
  {\bfseries 65} (1976) 471--474}.

\bibitem{Deser:1976rb}
S.~Deser and B.~Zumino, \emph{{A Complete Action for the Spinning String}},
  \href{http://dx.doi.org/10.1016/0370-2693(76)90245-8}{\emph{Phys. Lett. B}
  {\bfseries 65} (1976) 369--373}.

\bibitem{Ramond:1971gb}
P.~Ramond, \emph{{Dual Theory for Free Fermions}},
  \href{http://dx.doi.org/10.1103/PhysRevD.3.2415}{\emph{Phys. Rev. D}
  {\bfseries 3} (1971) 2415--2418}.

\bibitem{Neveu:1971fz}
A.~Neveu and J.~H. Schwarz, \emph{{Tachyon-free dual model with a
  positive-intercept trajectory}},
  \href{http://dx.doi.org/10.1016/0370-2693(71)90669-1}{\emph{Phys. Lett. B}
  {\bfseries 34} (1971) 517--518}.

\bibitem{Ramond:1971kx}
P.~Ramond, \emph{{An Interpretation of Dual Theories}},
  \href{http://dx.doi.org/10.1007/BF02731370}{\emph{Nuovo Cim. A} {\bfseries 4}
  (1971) 544--548}.

\bibitem{Neveu:1971rx}
A.~Neveu and J.~H. Schwarz, \emph{{Factorizable dual model of pions}},
  \href{http://dx.doi.org/10.1016/0550-3213(71)90448-2}{\emph{Nucl. Phys. B}
  {\bfseries 31} (1971) 86--112}.

\bibitem{Neveu:1971iv}
A.~Neveu and J.~H. Schwarz, \emph{{Quark Model of Dual Pions}},
  \href{http://dx.doi.org/10.1103/PhysRevD.4.1109}{\emph{Phys. Rev. D}
  {\bfseries 4} (1971) 1109--1111}.

\bibitem{Neveu:1971iw}
A.~Neveu, J.~H. Schwarz and C.~B. Thorn, \emph{{Reformulation of the Dual Pion
  Model}}, \href{http://dx.doi.org/10.1016/0370-2693(71)90391-1}{\emph{Phys.
  Lett. B} {\bfseries 35} (1971) 529--533}.

\bibitem{Gliozzi:1976qd}
F.~Gliozzi, J.~Scherk and D.~I. Olive, \emph{{Supersymmetry, Supergravity
  Theories and the Dual Spinor Model}},
  \href{http://dx.doi.org/10.1016/0550-3213(77)90206-1}{\emph{Nucl. Phys. B}
  {\bfseries 122} (1977) 253--290}.

\bibitem{Alvarez-Gaume:1983ihn}
L.~Alvarez-Gaume and E.~Witten, \emph{{Gravitational Anomalies}},
  \href{http://dx.doi.org/10.1016/0550-3213(84)90066-X}{\emph{Nucl. Phys. B}
  {\bfseries 234} (1984) 269}.

\bibitem{Schellekens:1986yi}
A.~N. Schellekens and N.~P. Warner, \emph{{Anomalies and Modular Invariance in
  String Theory}},
  \href{http://dx.doi.org/10.1016/0370-2693(86)90760-4}{\emph{Phys. Lett. B}
  {\bfseries 177} (1986) 317--323}.

\bibitem{Schellekens:1986xh}
A.~N. Schellekens and N.~P. Warner, \emph{{Anomalies, Characters and Strings}},
  \href{http://dx.doi.org/10.1016/0550-3213(87)90108-8}{\emph{Nucl. Phys. B}
  {\bfseries 287} (1987) 317}.

\bibitem{Lerche:1987sg}
W.~Lerche, B.~E.~W. Nilsson and A.~N. Schellekens, \emph{{Heterotic String Loop
  Calculation of the Anomaly Cancelling Term}},
  \href{http://dx.doi.org/10.1016/0550-3213(87)90397-X}{\emph{Nucl. Phys. B}
  {\bfseries 289} (1987) 609}.

\bibitem{orientifolds5}
M.~Bianchi and A.~Sagnotti, \emph{{On the systematics of open string
  theories}}, \href{http://dx.doi.org/10.1016/0370-2693(90)91894-H}{\emph{Phys.
  Lett. B} {\bfseries 247} (1990) 517--524}.

\bibitem{Dixon:1986iz}
L.~J. Dixon and J.~A. Harvey, \emph{{String Theories in Ten-Dimensions Without
  Space-Time Supersymmetry}},
  \href{http://dx.doi.org/10.1016/0550-3213(86)90619-X}{\emph{Nucl. Phys. B}
  {\bfseries 274} (1986) 93--105}.

\bibitem{Seiberg:1986by}
N.~Seiberg and E.~Witten, \emph{{Spin Structures in String Theory}},
  \href{http://dx.doi.org/10.1016/0550-3213(86)90297-X}{\emph{Nucl. Phys. B}
  {\bfseries 276} (1986) 272}.

\bibitem{Dai:1989ua}
J.~Dai, R.~G. Leigh and J.~Polchinski, \emph{{New Connections Between String
  Theories}}, \href{http://dx.doi.org/10.1142/S0217732389002331}{\emph{Mod.
  Phys. Lett. A} {\bfseries 4} (1989) 2073--2083}.

\bibitem{Leigh:1989jq}
R.~G. Leigh, \emph{{Dirac-Born-Infeld Action from Dirichlet Sigma Model}},
  \href{http://dx.doi.org/10.1142/S0217732389003099}{\emph{Mod. Phys. Lett. A}
  {\bfseries 4} (1989) 2767}.

\bibitem{orientifolds4}
P.~Horava, \emph{{Background Duality of Open String Models}},
  \href{http://dx.doi.org/10.1016/0370-2693(89)90209-8}{\emph{Phys. Lett. B}
  {\bfseries 231} (1989) 251--257}.

\bibitem{Polchinski:1995mt}
J.~Polchinski, \emph{{Dirichlet Branes and Ramond-Ramond charges}},
  \href{http://dx.doi.org/10.1103/PhysRevLett.75.4724}{\emph{Phys. Rev. Lett.}
  {\bfseries 75} (1995) 4724--4727},
  [\href{https://arxiv.org/abs/hep-th/9510017}{{\ttfamily hep-th/9510017}}].

\bibitem{Hull:1994ys}
C.~M. Hull and P.~K. Townsend, \emph{{Unity of superstring dualities}},
  \href{http://dx.doi.org/10.1201/9781482268737-24}{\emph{Nucl. Phys. B}
  {\bfseries 438} (1995) 109--137},
  [\href{https://arxiv.org/abs/hep-th/9410167}{{\ttfamily hep-th/9410167}}].

\bibitem{Townsend:1995kk}
P.~K. Townsend and P.~V. Landshoff, \emph{{The eleven-dimensional supermembrane
  revisited}}, \href{http://dx.doi.org/10.1201/9781482268737-25}{\emph{Phys.
  Lett. B} {\bfseries 350} (1995) 184--187},
  [\href{https://arxiv.org/abs/hep-th/9501068}{{\ttfamily hep-th/9501068}}].

\bibitem{Duff:1994an}
M.~J. Duff, R.~R. Khuri and J.~X. Lu, \emph{{String solitons}},
  \href{http://dx.doi.org/10.1016/0370-1573(95)00002-X}{\emph{Phys. Rept.}
  {\bfseries 259} (1995) 213--326},
  [\href{https://arxiv.org/abs/hep-th/9412184}{{\ttfamily hep-th/9412184}}].

\bibitem{Polchinski:1996na}
J.~Polchinski, \emph{{Tasi lectures on D-branes}},  in \emph{{Theoretical
  Advanced Study Institute in Elementary Particle Physics (TASI 96): Fields,
  Strings, and Duality}}, pp.~293--356, 11, 1996.
\newblock \href{https://arxiv.org/abs/hep-th/9611050}{{\ttfamily
  hep-th/9611050}}.

\bibitem{Bachas:1998rg}
C.~P. Bachas, \emph{{Lectures on D-branes}},  in \emph{{A Newton Institute
  Euroconference on Duality and Supersymmetric Theories}}, pp.~414--473, 6,
  1998.
\newblock \href{https://arxiv.org/abs/hep-th/9806199}{{\ttfamily
  hep-th/9806199}}.

\bibitem{Polchinski:1987tu}
J.~Polchinski and Y.~Cai, \emph{{Consistency of Open Superstring Theories}},
  \href{http://dx.doi.org/10.1016/0550-3213(88)90382-3}{\emph{Nucl. Phys. B}
  {\bfseries 296} (1988) 91--128}.

\bibitem{orientifolds1}
A.~Sagnotti, \emph{{Open Strings and their Symmetry Groups}},  in \emph{{NATO
  Advanced Summer Institute on Nonperturbative Quantum Field Theory (Cargese
  Summer Institute)}}, 9, 1987.
\newblock \href{https://arxiv.org/abs/hep-th/0208020}{{\ttfamily
  hep-th/0208020}}.

\bibitem{Bianchi:1988fr}
M.~Bianchi and A.~Sagnotti, \emph{{The Partition Function of the SO(8192)
  Bosonic String}},
  \href{http://dx.doi.org/10.1016/0370-2693(88)91884-9}{\emph{Phys. Lett. B}
  {\bfseries 211} (1988) 407--416}.

\bibitem{orientifolds3}
P.~Horava, \emph{{Strings on World Sheet Orbifolds}},
  \href{http://dx.doi.org/10.1016/0550-3213(89)90279-4}{\emph{Nucl. Phys. B}
  {\bfseries 327} (1989) 461--484}.

\bibitem{orientifolds7}
M.~Bianchi, G.~Pradisi and A.~Sagnotti, \emph{{Toroidal compactification and
  symmetry breaking in open string theories}},
  \href{http://dx.doi.org/10.1016/0550-3213(92)90129-Y}{\emph{Nucl. Phys. B}
  {\bfseries 376} (1992) 365--386}.

\bibitem{Witten:1995im}
E.~Witten, \emph{{Bound states of strings and p-branes}},
  \href{http://dx.doi.org/10.1016/0550-3213(95)00610-9}{\emph{Nucl. Phys. B}
  {\bfseries 460} (1996) 335--350},
  [\href{https://arxiv.org/abs/hep-th/9510135}{{\ttfamily hep-th/9510135}}].

\bibitem{Paton:1969je}
J.~E. Paton and H.-M. Chan, \emph{{Generalized veneziano model with isospin}},
  \href{http://dx.doi.org/10.1016/0550-3213(69)90038-8}{\emph{Nucl. Phys. B}
  {\bfseries 10} (1969) 516--520}.

\bibitem{Marcus:1986cm}
N.~Marcus and A.~Sagnotti, \emph{{Group Theory from Quarks at the Ends of
  Strings}}, \href{http://dx.doi.org/10.1016/0370-2693(87)90705-2}{\emph{Phys.
  Lett. B} {\bfseries 188} (1987) 58--64}.

\bibitem{orientifolds2}
G.~Pradisi and A.~Sagnotti, \emph{{Open String Orbifolds}},
  \href{http://dx.doi.org/10.1016/0370-2693(89)91369-5}{\emph{Phys. Lett. B}
  {\bfseries 216} (1989) 59--67}.

\bibitem{orientifolds6}
M.~Bianchi and A.~Sagnotti, \emph{{Twist symmetry and open string Wilson
  lines}}, \href{http://dx.doi.org/10.1016/0550-3213(91)90271-X}{\emph{Nucl.
  Phys. B} {\bfseries 361} (1991) 519--538}.

\bibitem{Green:1984sg}
M.~B. Green and J.~H. Schwarz, \emph{{Anomaly Cancellation in Supersymmetric
  D=10 Gauge Theory and Superstring Theory}},
  \href{http://dx.doi.org/10.1016/0370-2693(84)91565-X}{\emph{Phys. Lett. B}
  {\bfseries 149} (1984) 117--122}.

\bibitem{Harvey:1986bf}
J.~A. Harvey and J.~A. Minahan, \emph{{OPEN STRINGS ON ORBIFOLDS}},
  \href{http://dx.doi.org/10.1016/0370-2693(87)90703-9}{\emph{Phys. Lett. B}
  {\bfseries 188} (1987) 44}.

\bibitem{Ishibashi:1988tf}
N.~Ishibashi and T.~Onogi, \emph{{OPEN STRING MODEL BUILDING}},
  \href{http://dx.doi.org/10.1016/0550-3213(89)90054-0}{\emph{Nucl. Phys. B}
  {\bfseries 318} (1989) 239--280}.

\bibitem{Angelantonj:1996uy}
C.~Angelantonj, M.~Bianchi, G.~Pradisi, A.~Sagnotti and Y.~S. Stanev,
  \emph{{Chiral asymmetry in four-dimensional open string vacua}},
  \href{http://dx.doi.org/10.1016/0370-2693(96)00869-6}{\emph{Phys. Lett. B}
  {\bfseries 385} (1996) 96--102},
  [\href{https://arxiv.org/abs/hep-th/9606169}{{\ttfamily hep-th/9606169}}].

\bibitem{Kakushadze:1997ku}
Z.~Kakushadze and G.~Shiu, \emph{{A Chiral N=1 type I vacuum in four-dimensions
  and its heterotic dual}},
  \href{http://dx.doi.org/10.1103/PhysRevD.56.3686}{\emph{Phys. Rev. D}
  {\bfseries 56} (1997) 3686--3697},
  [\href{https://arxiv.org/abs/hep-th/9705163}{{\ttfamily hep-th/9705163}}].

\bibitem{Kakushadze:1997uj}
Z.~Kakushadze and G.~Shiu, \emph{{4-D chiral N=1 type one vacua with and
  without D5-branes}},
  \href{http://dx.doi.org/10.1016/S0550-3213(98)00056-X}{\emph{Nucl. Phys. B}
  {\bfseries 520} (1998) 75--92},
  [\href{https://arxiv.org/abs/hep-th/9706051}{{\ttfamily hep-th/9706051}}].

\bibitem{Kakushadze:1998eg}
Z.~Kakushadze, \emph{{On four-dimensional N=1 type I compactifications}},
  \href{http://dx.doi.org/10.1016/S0550-3213(98)00643-9}{\emph{Nucl. Phys. B}
  {\bfseries 535} (1998) 311--334},
  [\href{https://arxiv.org/abs/hep-th/9806008}{{\ttfamily hep-th/9806008}}].

\bibitem{Kakushadze:1998cd}
Z.~Kakushadze, G.~Shiu and S.~H.~H. Tye, \emph{{Type IIB orientifolds, F
  theory, type I strings on orbifolds and type I - Heterotic duality}},
  \href{http://dx.doi.org/10.1016/S0550-3213(98)00491-X}{\emph{Nucl. Phys. B}
  {\bfseries 533} (1998) 25--87},
  [\href{https://arxiv.org/abs/hep-th/9804092}{{\ttfamily hep-th/9804092}}].

\bibitem{Zwart:1997aj}
G.~Zwart, \emph{{Four-dimensional N=1 Z(N) x Z(M) orientifolds}},
  \href{http://dx.doi.org/10.1016/S0550-3213(98)00288-0}{\emph{Nucl. Phys. B}
  {\bfseries 526} (1998) 378--392},
  [\href{https://arxiv.org/abs/hep-th/9708040}{{\ttfamily hep-th/9708040}}].

\bibitem{Klein:2000qw}
M.~Klein and R.~Rabadan, \emph{{Z(N) x Z(M) orientifolds with and without
  discrete torsion}},
  \href{http://dx.doi.org/10.1088/1126-6708/2000/10/049}{\emph{JHEP} {\bfseries
  10} (2000) 049}, [\href{https://arxiv.org/abs/hep-th/0008173}{{\ttfamily
  hep-th/0008173}}].

\bibitem{Blumenhagen:1999md}
R.~Blumenhagen, L.~Gorlich and B.~Kors, \emph{{Supersymmetric orientifolds in
  6-D with D-branes at angles}},
  \href{http://dx.doi.org/10.1016/S0550-3213(99)00795-6}{\emph{Nucl. Phys. B}
  {\bfseries 569} (2000) 209--228},
  [\href{https://arxiv.org/abs/hep-th/9908130}{{\ttfamily hep-th/9908130}}].

\bibitem{Cvetic:1999hb}
M.~Cvetic, M.~Plumacher and J.~Wang, \emph{{Three family type IIB orientifold
  string vacua with nonAbelian Wilson lines}},
  \href{http://dx.doi.org/10.1088/1126-6708/2000/04/004}{\emph{JHEP} {\bfseries
  04} (2000) 004}, [\href{https://arxiv.org/abs/hep-th/9911021}{{\ttfamily
  hep-th/9911021}}].

\bibitem{Blumenhagen:1999ev}
R.~Blumenhagen, L.~Gorlich and B.~Kors, \emph{{Supersymmetric 4-D orientifolds
  of type IIA with D6-branes at angles}},
  \href{http://dx.doi.org/10.1088/1126-6708/2000/01/040}{\emph{JHEP} {\bfseries
  01} (2000) 040}, [\href{https://arxiv.org/abs/hep-th/9912204}{{\ttfamily
  hep-th/9912204}}].

\bibitem{Pradisi:1999ii}
G.~Pradisi, \emph{{Type I vacua from diagonal Z(3) orbifolds}},
  \href{http://dx.doi.org/10.1016/S0550-3213(00)00089-4}{\emph{Nucl. Phys. B}
  {\bfseries 575} (2000) 134--150},
  [\href{https://arxiv.org/abs/hep-th/9912218}{{\ttfamily hep-th/9912218}}].

\bibitem{Cvetic:2000aq}
M.~Cvetic and P.~Langacker, \emph{{D = 4 N=1 type IIB orientifolds with
  continuous Wilson lines, moving branes, and their field theory realization}},
  \href{http://dx.doi.org/10.1016/S0550-3213(00)00414-4}{\emph{Nucl. Phys. B}
  {\bfseries 586} (2000) 287--302},
  [\href{https://arxiv.org/abs/hep-th/0006049}{{\ttfamily hep-th/0006049}}].

\bibitem{Cvetic:2000st}
M.~Cvetic, A.~M. Uranga and J.~Wang, \emph{{Discrete Wilson lines in N=1 D = 4
  type IIB orientifolds: A Systematic exploration for Z(6) orientifold}},
  \href{http://dx.doi.org/10.1016/S0550-3213(00)00669-6}{\emph{Nucl. Phys. B}
  {\bfseries 595} (2001) 63--92},
  [\href{https://arxiv.org/abs/hep-th/0010091}{{\ttfamily hep-th/0010091}}].

\bibitem{gss}
A.~Sagnotti, \emph{{A Note on the Green-Schwarz mechanism in open string
  theories}}, \href{http://dx.doi.org/10.1016/0370-2693(92)90682-T}{\emph{Phys.
  Lett. B} {\bfseries 294} (1992) 196--203},
  [\href{https://arxiv.org/abs/hep-th/9210127}{{\ttfamily hep-th/9210127}}].

\bibitem{ss-original_1}
J.~Scherk and J.~H. Schwarz, \emph{{Spontaneous Breaking of Supersymmetry
  Through Dimensional Reduction}},
  \href{http://dx.doi.org/10.1016/0370-2693(79)90425-8}{\emph{Phys. Lett. B}
  {\bfseries 82} (1979) 60--64}.

\bibitem{ss-original_2}
J.~Scherk and J.~H. Schwarz, \emph{{How to Get Masses from Extra Dimensions}},
  \href{http://dx.doi.org/10.1016/0550-3213(79)90592-3}{\emph{Nucl. Phys. B}
  {\bfseries 153} (1979) 61--88}.

\bibitem{ss-original_3}
E.~Cremmer, J.~Scherk and J.~H. Schwarz, \emph{{Spontaneously Broken N=8
  Supergravity}},
  \href{http://dx.doi.org/10.1016/0370-2693(79)90654-3}{\emph{Phys. Lett. B}
  {\bfseries 84} (1979) 83--86}.

\bibitem{blum-dienes_1}
J.~D. Blum and K.~R. Dienes, \emph{{Strong / weak coupling duality relations
  for nonsupersymmetric string theories}},
  \href{http://dx.doi.org/10.1016/S0550-3213(97)00803-1}{\emph{Nucl. Phys. B}
  {\bfseries 516} (1998) 83--159},
  [\href{https://arxiv.org/abs/hep-th/9707160}{{\ttfamily hep-th/9707160}}].

\bibitem{blum-dienes_2}
J.~D. Blum and K.~R. Dienes, \emph{{Duality without supersymmetry: The Case of
  the SO(16)$\times$ SO(16) string}},
  \href{http://dx.doi.org/10.1016/S0370-2693(97)01172-6}{\emph{Phys. Lett. B}
  {\bfseries 414} (1997) 260--268},
  [\href{https://arxiv.org/abs/hep-th/9707148}{{\ttfamily hep-th/9707148}}].

\bibitem{ads1}
I.~Antoniadis, E.~Dudas and A.~Sagnotti, \emph{{Supersymmetry breaking, open
  strings and M theory}},
  \href{http://dx.doi.org/10.1016/S0550-3213(98)00806-2}{\emph{Nucl. Phys. B}
  {\bfseries 544} (1999) 469--502},
  [\href{https://arxiv.org/abs/hep-th/9807011}{{\ttfamily hep-th/9807011}}].

\bibitem{ss_open-lower_1}
I.~Antoniadis, G.~D'Appollonio, E.~Dudas and A.~Sagnotti, \emph{{Partial
  breaking of supersymmetry, open strings and M theory}},
  \href{http://dx.doi.org/10.1016/S0550-3213(99)00232-1}{\emph{Nucl. Phys. B}
  {\bfseries 553} (1999) 133--154},
  [\href{https://arxiv.org/abs/hep-th/9812118}{{\ttfamily hep-th/9812118}}].

\bibitem{ss_open-lower_2}
A.~L. Cotrone, \emph{{A $\mathbb{Z}_2\times \mathbb{Z}_2$ orientifold with
  spontaneously broken supersymmetry}},
  \href{http://dx.doi.org/10.1142/S0217732399002595}{\emph{Mod. Phys. Lett. A}
  {\bfseries 14} (1999) 2487--2497},
  [\href{https://arxiv.org/abs/hep-th/9909116}{{\ttfamily hep-th/9909116}}].

\bibitem{ss_open-lower_3}
C.~Angelantonj and I.~Antoniadis, \emph{{Suppressing the cosmological constant
  in nonsupersymmetric type I strings}},
  \href{http://dx.doi.org/10.1016/j.nuclphysb.2003.09.047}{\emph{Nucl. Phys. B}
  {\bfseries 676} (2004) 129--148},
  [\href{https://arxiv.org/abs/hep-th/0307254}{{\ttfamily hep-th/0307254}}].

\bibitem{ss_open-lower_4}
C.~Angelantonj, M.~Cardella and N.~Irges, \emph{{An Alternative for Moduli
  Stabilisation}},
  \href{http://dx.doi.org/10.1016/j.physletb.2006.08.072}{\emph{Phys. Lett. B}
  {\bfseries 641} (2006) 474--480},
  [\href{https://arxiv.org/abs/hep-th/0608022}{{\ttfamily hep-th/0608022}}].

\bibitem{ss_open-lower_5}
S.~Abel, E.~Dudas, D.~Lewis and H.~Partouche, \emph{{Stability and vacuum
  energy in open string models with broken supersymmetry}},
  \href{http://dx.doi.org/10.1007/JHEP10(2019)226}{\emph{JHEP} {\bfseries 10}
  (2019) 226}, [\href{https://arxiv.org/abs/1812.09714}{{\ttfamily
  1812.09714}}].

\bibitem{aads1}
I.~Antoniadis, G.~D'Appollonio, E.~Dudas and A.~Sagnotti, \emph{{Open
  descendants of $\mathbb{Z}_2\times \mathbb{Z}_2$ freely acting orbifolds}},
  \href{http://dx.doi.org/10.1016/S0550-3213(99)00616-1}{\emph{Nucl. Phys. B}
  {\bfseries 565} (2000) 123--156},
  [\href{https://arxiv.org/abs/hep-th/9907184}{{\ttfamily hep-th/9907184}}].

\bibitem{ss_closed1}
R.~Rohm, \emph{{Spontaneous Supersymmetry Breaking in Supersymmetric String
  Theories}}, \href{http://dx.doi.org/10.1016/0550-3213(84)90007-5}{\emph{Nucl.
  Phys. B} {\bfseries 237} (1984) 553--572}.

\bibitem{ss_closed2}
C.~Kounnas and M.~Porrati, \emph{{Spontaneous Supersymmetry Breaking in String
  Theory}}, \href{http://dx.doi.org/10.1016/0550-3213(88)90153-8}{\emph{Nucl.
  Phys. B} {\bfseries 310} (1988) 355--370}.

\bibitem{ss_closed3}
S.~Ferrara, C.~Kounnas, M.~Porrati and F.~Zwirner, \emph{{Superstrings with
  Spontaneously Broken Supersymmetry and their Effective Theories}},
  \href{http://dx.doi.org/10.1016/0550-3213(89)90048-5}{\emph{Nucl. Phys. B}
  {\bfseries 318} (1989) 75--105}.

\bibitem{ss_closed4}
C.~Kounnas and B.~Rostand, \emph{{Coordinate Dependent Compactifications and
  Discrete Symmetries}},
  \href{http://dx.doi.org/10.1016/0550-3213(90)90543-M}{\emph{Nucl. Phys. B}
  {\bfseries 341} (1990) 641--665}.

\bibitem{ss_closed5}
I.~Antoniadis and C.~Kounnas, \emph{{Superstring phase transition at high
  temperature}},
  \href{http://dx.doi.org/10.1016/0370-2693(91)90442-S}{\emph{Phys. Lett. B}
  {\bfseries 261} (1991) 369--378}.

\bibitem{ss_closed6}
E.~Kiritsis and C.~Kounnas, \emph{{Perturbative and nonperturbative partial
  supersymmetry breaking: $N=4\rightarrow N=2 \rightarrow N=1$}},
  \href{http://dx.doi.org/10.1016/S0550-3213(97)00430-6}{\emph{Nucl. Phys. B}
  {\bfseries 503} (1997) 117--156},
  [\href{https://arxiv.org/abs/hep-th/9703059}{{\ttfamily hep-th/9703059}}].

\bibitem{Fradkin:1985qd}
E.~S. Fradkin and A.~A. Tseytlin, \emph{{Nonlinear Electrodynamics from
  Quantized Strings}},
  \href{http://dx.doi.org/10.1016/0370-2693(85)90205-9}{\emph{Phys. Lett. B}
  {\bfseries 163} (1985) 123--130}.

\bibitem{Abouelsaood:1986gd}
A.~Abouelsaood, C.~G. Callan, Jr., C.~R. Nappi and S.~A. Yost, \emph{{Open
  strings in background gauge fields}},
  \href{http://dx.doi.org/10.1016/0550-3213(87)90164-7}{\emph{Nucl. Phys. B}
  {\bfseries 280} (1987) 599--624}.

\bibitem{Bachas:1995ik}
C.~Bachas, \emph{{A Way to break supersymmetry}},
  \href{https://arxiv.org/abs/hep-th/9503030}{{\ttfamily hep-th/9503030}}.

\bibitem{Bianchi:1997gt}
M.~Bianchi and Y.~S. Stanev, \emph{{Open strings on the Neveu-Schwarz
  penta-brane}},
  \href{http://dx.doi.org/10.1016/S0550-3213(98)00281-8}{\emph{Nucl. Phys. B}
  {\bfseries 523} (1998) 193--210},
  [\href{https://arxiv.org/abs/hep-th/9711069}{{\ttfamily hep-th/9711069}}].

\bibitem{bsb1}
S.~Sugimoto, \emph{{Anomaly cancellations in type I D9 -- anti-D9 system and
  the USp(32) string theory}},
  \href{http://dx.doi.org/10.1143/PTP.102.685}{\emph{Prog. Theor. Phys.}
  {\bfseries 102} (1999) 685--699},
  [\href{https://arxiv.org/abs/hep-th/9905159}{{\ttfamily hep-th/9905159}}].

\bibitem{bsb2}
I.~Antoniadis, E.~Dudas and A.~Sagnotti, \emph{{Brane supersymmetry breaking}},
  \href{http://dx.doi.org/10.1016/S0370-2693(99)01023-0}{\emph{Phys. Lett. B}
  {\bfseries 464} (1999) 38--45},
  [\href{https://arxiv.org/abs/hep-th/9908023}{{\ttfamily hep-th/9908023}}].

\bibitem{bsb3}
C.~Angelantonj, \emph{{Comments on open string orbifolds with a nonvanishing
  B(ab)}}, \href{http://dx.doi.org/10.1016/S0550-3213(99)00662-8}{\emph{Nucl.
  Phys. B} {\bfseries 566} (2000) 126--150},
  [\href{https://arxiv.org/abs/hep-th/9908064}{{\ttfamily hep-th/9908064}}].

\bibitem{bsb4}
G.~Aldazabal and A.~M. Uranga, \emph{{Tachyon free nonsupersymmetric type IIB
  orientifolds via Brane - anti-brane systems}},
  \href{http://dx.doi.org/10.1088/1126-6708/1999/10/024}{\emph{JHEP} {\bfseries
  10} (1999) 024}, [\href{https://arxiv.org/abs/hep-th/9908072}{{\ttfamily
  hep-th/9908072}}].

\bibitem{bsb5}
C.~Angelantonj, I.~Antoniadis, G.~D'Appollonio, E.~Dudas and A.~Sagnotti,
  \emph{{Type I vacua with brane supersymmetry breaking}},
  \href{http://dx.doi.org/10.1016/S0550-3213(00)00052-3}{\emph{Nucl. Phys. B}
  {\bfseries 572} (2000) 36--70},
  [\href{https://arxiv.org/abs/hep-th/9911081}{{\ttfamily hep-th/9911081}}].

\bibitem{bsb6}
C.~Angelantonj, C.~Condeescu, E.~Dudas and G.~Leone, \emph{{Rigid vacua with
  Brane Supersymmetry Breaking}},
  \href{http://dx.doi.org/10.1007/JHEP04(2024)103}{\emph{JHEP} {\bfseries 04}
  (2024) 103}, [\href{https://arxiv.org/abs/2403.02392}{{\ttfamily
  2403.02392}}].

\bibitem{bsbnl_1}
E.~Dudas and J.~Mourad, \emph{{Consistent gravitino couplings in
  nonsupersymmetric strings}},
  \href{http://dx.doi.org/10.1016/S0370-2693(01)00777-8}{\emph{Phys. Lett. B}
  {\bfseries 514} (2001) 173--182},
  [\href{https://arxiv.org/abs/hep-th/0012071}{{\ttfamily hep-th/0012071}}].

\bibitem{bsbnl_2}
G.~Pradisi and F.~Riccioni, \emph{{Geometric couplings and brane supersymmetry
  breaking}},
  \href{http://dx.doi.org/10.1016/S0550-3213(01)00441-2}{\emph{Nucl. Phys. B}
  {\bfseries 615} (2001) 33--60},
  [\href{https://arxiv.org/abs/hep-th/0107090}{{\ttfamily hep-th/0107090}}].

\bibitem{bsb_rev}
J.~Mourad and A.~Sagnotti, \emph{{An Update on Brane Supersymmetry Breaking}},
  \href{https://arxiv.org/abs/1711.11494}{{\ttfamily 1711.11494}}.

\bibitem{dm1_1}
E.~Dudas and J.~Mourad, \emph{{D-branes in nontachyonic 0B orientifolds}},
  \href{http://dx.doi.org/10.1016/S0550-3213(00)00781-1}{\emph{Nucl. Phys. B}
  {\bfseries 598} (2001) 189--224},
  [\href{https://arxiv.org/abs/hep-th/0010179}{{\ttfamily hep-th/0010179}}].

\bibitem{dm1_2}
E.~Dudas, J.~Mourad and C.~Timirgaziu, \emph{{Time and space dependent
  backgrounds from nonsupersymmetric strings}},
  \href{http://dx.doi.org/10.1016/S0550-3213(03)00248-7}{\emph{Nucl. Phys. B}
  {\bfseries 660} (2003) 3--24},
  [\href{https://arxiv.org/abs/hep-th/0209176}{{\ttfamily hep-th/0209176}}].

\bibitem{dp}
A.~Dabholkar and J.~Park, \emph{{Strings on orientifolds}},
  \href{http://dx.doi.org/10.1016/0550-3213(96)00395-1}{\emph{Nucl. Phys. B}
  {\bfseries 477} (1996) 701--714},
  [\href{https://arxiv.org/abs/hep-th/9604178}{{\ttfamily hep-th/9604178}}].

\bibitem{Dbranes-dpGukov}
S.~Gukov, \emph{{K theory, reality, and orientifolds}},
  \href{http://dx.doi.org/10.1007/s002200050793}{\emph{Commun. Math. Phys.}
  {\bfseries 210} (2000) 621--639},
  [\href{https://arxiv.org/abs/hep-th/9901042}{{\ttfamily hep-th/9901042}}].

\bibitem{Dbranes-dpGimon}
O.~Bergman, E.~G. Gimon and P.~Horava, \emph{{Brane transfer operations and T
  duality of nonBPS states}},
  \href{http://dx.doi.org/10.1088/1126-6708/1999/04/010}{\emph{JHEP} {\bfseries
  04} (1999) 010}, [\href{https://arxiv.org/abs/hep-th/9902160}{{\ttfamily
  hep-th/9902160}}].

\bibitem{Dbranes-dpGaberdiel}
M.~R. Gaberdiel and S.~Schafer-Nameki, \emph{{NonBPS D branes and M theory}},
  \href{http://dx.doi.org/10.1088/1126-6708/2001/09/028}{\emph{JHEP} {\bfseries
  09} (2001) 028}, [\href{https://arxiv.org/abs/hep-th/0108202}{{\ttfamily
  hep-th/0108202}}].

\bibitem{Dabholkar:1996zi}
A.~Dabholkar and J.~Park, \emph{{An Orientifold of type IIB theory on K3}},
  \href{http://dx.doi.org/10.1016/0550-3213(96)00199-X}{\emph{Nucl. Phys. B}
  {\bfseries 472} (1996) 207--220},
  [\href{https://arxiv.org/abs/hep-th/9602030}{{\ttfamily hep-th/9602030}}].

\bibitem{Gimon:1996ay}
E.~G. Gimon and C.~V. Johnson, \emph{{K3 orientifolds}},
  \href{http://dx.doi.org/10.1016/0550-3213(96)00356-2}{\emph{Nucl. Phys. B}
  {\bfseries 477} (1996) 715--745},
  [\href{https://arxiv.org/abs/hep-th/9604129}{{\ttfamily hep-th/9604129}}].

\bibitem{hw_1}
P.~Horava and E.~Witten, \emph{{Eleven-dimensional supergravity on a manifold
  with boundary}},
  \href{http://dx.doi.org/10.1016/0550-3213(96)00308-2}{\emph{Nucl. Phys. B}
  {\bfseries 475} (1996) 94--114},
  [\href{https://arxiv.org/abs/hep-th/9603142}{{\ttfamily hep-th/9603142}}].

\bibitem{hw_2}
P.~Horava and E.~Witten, \emph{{Heterotic and Type I string dynamics from
  eleven dimensions}},
  \href{http://dx.doi.org/10.1201/9781482268737-35}{\emph{Nucl. Phys. B}
  {\bfseries 460} (1996) 506--524},
  [\href{https://arxiv.org/abs/hep-th/9510209}{{\ttfamily hep-th/9510209}}].

\bibitem{akp}
O.~Aharony, Z.~Komargodski and A.~Patir, \emph{{The Moduli space and M(atrix)
  theory of 9d N=1 backgrounds of M/string theory}},
  \href{http://dx.doi.org/10.1088/1126-6708/2007/05/073}{\emph{JHEP} {\bfseries
  05} (2007) 073}, [\href{https://arxiv.org/abs/hep-th/0702195}{{\ttfamily
  hep-th/0702195}}].

\bibitem{gepner}
C.~Angelantonj, M.~Bianchi, G.~Pradisi, A.~Sagnotti and Y.~S. Stanev,
  \emph{{Comments on Gepner models and type I vacua in string theory}},
  \href{http://dx.doi.org/10.1016/0370-2693(96)01124-0}{\emph{Phys. Lett. B}
  {\bfseries 387} (1996) 743--749},
  [\href{https://arxiv.org/abs/hep-th/9607229}{{\ttfamily hep-th/9607229}}].

\bibitem{dms}
E.~Dudas, J.~Mourad and A.~Sagnotti, \emph{{Charged and uncharged D-branes in
  various string theories}},
  \href{http://dx.doi.org/10.1016/S0550-3213(01)00552-1}{\emph{Nucl. Phys. B}
  {\bfseries 620} (2002) 109--151},
  [\href{https://arxiv.org/abs/hep-th/0107081}{{\ttfamily hep-th/0107081}}].

\bibitem{Sen:1998tt}
A.~Sen, \emph{{SO(32) spinors of type I and other solitons on brane --
  anti-brane pair}},
  \href{http://dx.doi.org/10.1088/1126-6708/1998/09/023}{\emph{JHEP} {\bfseries
  09} (1998) 023}, [\href{https://arxiv.org/abs/hep-th/9808141}{{\ttfamily
  hep-th/9808141}}].

\bibitem{gp}
E.~G. Gimon and J.~Polchinski, \emph{{Consistency conditions for orientifolds
  and D-manifolds}},
  \href{http://dx.doi.org/10.1103/PhysRevD.54.1667}{\emph{Phys. Rev. D}
  {\bfseries 54} (1996) 1667--1676},
  [\href{https://arxiv.org/abs/hep-th/9601038}{{\ttfamily hep-th/9601038}}].

\bibitem{Aspinwall:1996mn}
P.~S. Aspinwall, \emph{{K3 surfaces and string duality}},  in
  \emph{{Theoretical Advanced Study Institute in Elementary Particle Physics
  (TASI 96): Fields, Strings, and Duality}}, pp.~421--540, 11, 1996.
\newblock \href{https://arxiv.org/abs/hep-th/9611137}{{\ttfamily
  hep-th/9611137}}.

\bibitem{Morrison}
D.~R. Morrison, \emph{{On K3 surfaces with large Picard number}},
  \href{http://dx.doi.org/10.1007/BF01403093}{\emph{Invent Math} {\bfseries 75}
  (1984) 105--121}.

\bibitem{Kummer}
A.~Garbagnati and A.~Sarti, \emph{{Kummer surfaces and K3 surfaces with
  $(\mathbb{Z}/2\mathbb{Z})^4$ symplectic action}},
  \href{http://dx.doi.org/10.1216/RMJ-2016-46-4-1141}{\emph{Rocky Mountain
  J.Math.} {\bfseries 46 (4)} (2016) 1141--1205}.

\bibitem{Gopakumar:1996mu}
R.~Gopakumar and S.~Mukhi, \emph{{Orbifold and orientifold compactifications of
  F - theory and M - theory to six-dimensions and four-dimensions}},
  \href{http://dx.doi.org/10.1016/0550-3213(96)00460-9}{\emph{Nucl. Phys. B}
  {\bfseries 479} (1996) 260--284},
  [\href{https://arxiv.org/abs/hep-th/9607057}{{\ttfamily hep-th/9607057}}].

\bibitem{Sagnotti:1995ga}
A.~Sagnotti, \emph{{Some properties of open string theories}},  in
  \emph{{International Workshop on Supersymmetry and Unification of Fundamental
  Interactions (SUSY 95)}}, pp.~473--484, 9, 1995.
\newblock \href{https://arxiv.org/abs/hep-th/9509080}{{\ttfamily
  hep-th/9509080}}.

\bibitem{O'B_1}
A.~Sagnotti, \emph{{Surprises in open string perturbation theory}},
  \href{http://dx.doi.org/10.1016/S0920-5632(97)00344-7}{\emph{Nucl. Phys. B
  Proc. Suppl.} {\bfseries 56} (1997) 332--343},
  [\href{https://arxiv.org/abs/hep-th/9702093}{{\ttfamily hep-th/9702093}}].

\bibitem{O'B_2}
C.~Angelantonj, \emph{{Nontachyonic open descendants of the 0B string theory}},
  \href{http://dx.doi.org/10.1016/S0370-2693(98)01430-0}{\emph{Phys. Lett. B}
  {\bfseries 444} (1998) 309--317},
  [\href{https://arxiv.org/abs/hep-th/9810214}{{\ttfamily hep-th/9810214}}].

\bibitem{O'B_3}
R.~Blumenhagen, A.~Font and D.~Lust, \emph{{Tachyon free orientifolds of type
  0B strings in various dimensions}},
  \href{http://dx.doi.org/10.1016/S0550-3213(99)00381-8}{\emph{Nucl. Phys. B}
  {\bfseries 558} (1999) 159--177},
  [\href{https://arxiv.org/abs/hep-th/9904069}{{\ttfamily hep-th/9904069}}].

\bibitem{Dabholkar:1997zd}
A.~Dabholkar, \emph{{Lectures on orientifolds and duality}},  in \emph{{ICTP
  Summer School in High-Energy Physics and Cosmology}}, pp.~128--191, 6, 1997.
\newblock \href{https://arxiv.org/abs/hep-th/9804208}{{\ttfamily
  hep-th/9804208}}.

\bibitem{Pantev:2016nze}
T.~Pantev and E.~Sharpe, \emph{{Duality group actions on fermions}},
  \href{http://dx.doi.org/10.1007/JHEP11(2016)171}{\emph{JHEP} {\bfseries 11}
  (2016) 171}, [\href{https://arxiv.org/abs/1609.00011}{{\ttfamily
  1609.00011}}].

\bibitem{Tachikawa:2018njr}
Y.~Tachikawa and K.~Yonekura, \emph{{Why are fractional charges of orientifolds
  compatible with Dirac quantization?}},
  \href{http://dx.doi.org/10.21468/SciPostPhys.7.5.058}{\emph{SciPost Phys.}
  {\bfseries 7} (2019) 058},
  [\href{https://arxiv.org/abs/1805.02772}{{\ttfamily 1805.02772}}].

\bibitem{Vafa:1995gm}
C.~Vafa and E.~Witten, \emph{{Dual string pairs with N=1 and N=2 supersymmetry
  in four-dimensions}},
  \href{http://dx.doi.org/10.1016/0920-5632(96)00025-4}{\emph{Nucl. Phys. B
  Proc. Suppl.} {\bfseries 46} (1996) 225--247},
  [\href{https://arxiv.org/abs/hep-th/9507050}{{\ttfamily hep-th/9507050}}].

\bibitem{ss-specific_1}
C.~Kounnas and H.~Partouche, \emph{{Super no-scale models in string theory}},
  \href{http://dx.doi.org/10.1016/j.nuclphysb.2016.10.001}{\emph{Nucl. Phys. B}
  {\bfseries 913} (2016) 593--626},
  [\href{https://arxiv.org/abs/1607.01767}{{\ttfamily 1607.01767}}].

\bibitem{ss-specific_2}
S.~Abel, K.~R. Dienes and E.~Mavroudi, \emph{{Towards a nonsupersymmetric
  string phenomenology}},
  \href{http://dx.doi.org/10.1103/PhysRevD.91.126014}{\emph{Phys. Rev. D}
  {\bfseries 91} (2015) 126014},
  [\href{https://arxiv.org/abs/1502.03087}{{\ttfamily 1502.03087}}].

\bibitem{Green:1997tv}
M.~B. Green and M.~Gutperle, \emph{{Effects of D instantons}},
  \href{http://dx.doi.org/10.1016/S0550-3213(97)00269-1}{\emph{Nucl. Phys. B}
  {\bfseries 498} (1997) 195--227},
  [\href{https://arxiv.org/abs/hep-th/9701093}{{\ttfamily hep-th/9701093}}].

\bibitem{Hatcher}
A.~Hatcher, \emph{Algebraic Topology}.
\newblock Cambridge University Press, 2002.

\bibitem{Bergshoeff:1996ui}
E.~Bergshoeff, M.~de~Roo, M.~B. Green, G.~Papadopoulos and P.~K. Townsend,
  \emph{{Duality of type II 7 branes and 8 branes}},
  \href{http://dx.doi.org/10.1016/0550-3213(96)00171-X}{\emph{Nucl. Phys. B}
  {\bfseries 470} (1996) 113--135},
  [\href{https://arxiv.org/abs/hep-th/9601150}{{\ttfamily hep-th/9601150}}].

\bibitem{Bergshoeff:1997ak}
E.~Bergshoeff, Y.~Lozano and T.~Ortin, \emph{{Massive branes}},
  \href{http://dx.doi.org/10.1016/S0550-3213(98)00045-5}{\emph{Nucl. Phys. B}
  {\bfseries 518} (1998) 363--423},
  [\href{https://arxiv.org/abs/hep-th/9712115}{{\ttfamily hep-th/9712115}}].

\bibitem{Hull:1997kt}
C.~M. Hull, \emph{{Gravitational duality, branes and charges}},
  \href{http://dx.doi.org/10.1016/S0550-3213(97)00501-4}{\emph{Nucl. Phys. B}
  {\bfseries 509} (1998) 216--251},
  [\href{https://arxiv.org/abs/hep-th/9705162}{{\ttfamily hep-th/9705162}}].

\bibitem{Bergshoeff:1998bs}
E.~Bergshoeff and J.~P. van~der Schaar, \emph{{On M nine-branes}},
  \href{http://dx.doi.org/10.1088/0264-9381/16/1/002}{\emph{Class. Quant.
  Grav.} {\bfseries 16} (1999) 23--39},
  [\href{https://arxiv.org/abs/hep-th/9806069}{{\ttfamily hep-th/9806069}}].

\bibitem{Kleinschmidt:2003mf}
A.~Kleinschmidt, I.~Schnakenburg and P.~C. West, \emph{{Very extended Kac-Moody
  algebras and their interpretation at low levels}},
  \href{http://dx.doi.org/10.1088/0264-9381/21/9/021}{\emph{Class. Quant.
  Grav.} {\bfseries 21} (2004) 2493--2525},
  [\href{https://arxiv.org/abs/hep-th/0309198}{{\ttfamily hep-th/0309198}}].

\bibitem{Atiyah:2001qf}
M.~Atiyah and E.~Witten, \emph{{M theory dynamics on a manifold of $G_2$
  holonomy}}, \href{http://dx.doi.org/10.4310/ATMP.2002.v6.n1.a1}{\emph{Adv.
  Theor. Math. Phys.} {\bfseries 6} (2003) 1--106},
  [\href{https://arxiv.org/abs/hep-th/0107177}{{\ttfamily hep-th/0107177}}].

\bibitem{Montonen:1977sn}
C.~Montonen and D.~I. Olive, \emph{{Magnetic Monopoles as Gauge Particles?}},
  \href{http://dx.doi.org/10.1016/0370-2693(77)90076-4}{\emph{Phys. Lett. B}
  {\bfseries 72} (1977) 117--120}.

\bibitem{Palti:2017elp}
E.~Palti, \emph{{The Weak Gravity Conjecture and Scalar Fields}},
  \href{http://dx.doi.org/10.1007/JHEP08(2017)034}{\emph{JHEP} {\bfseries 08}
  (2017) 034}, [\href{https://arxiv.org/abs/1705.04328}{{\ttfamily
  1705.04328}}].

\bibitem{DallAgata:2020ino}
G.~Dall'Agata and M.~Morittu, \emph{{Covariant formulation of BPS black holes
  and the scalar weak gravity conjecture}},
  \href{http://dx.doi.org/10.1007/JHEP03(2020)192}{\emph{JHEP} {\bfseries 03}
  (2020) 192}, [\href{https://arxiv.org/abs/2001.10542}{{\ttfamily
  2001.10542}}].

\bibitem{Arboleya:2024vnp}
A.~Arboleya, A.~Guarino and M.~Morittu, \emph{{Type II orientifold flux vacua
  in 3D}}, \href{http://dx.doi.org/10.1007/JHEP12(2024)087}{\emph{JHEP}
  {\bfseries 12} (2024) 087},
  [\href{https://arxiv.org/abs/2408.01403}{{\ttfamily 2408.01403}}].

\end{thebibliography}\endgroup

\includepdf[pages=-]{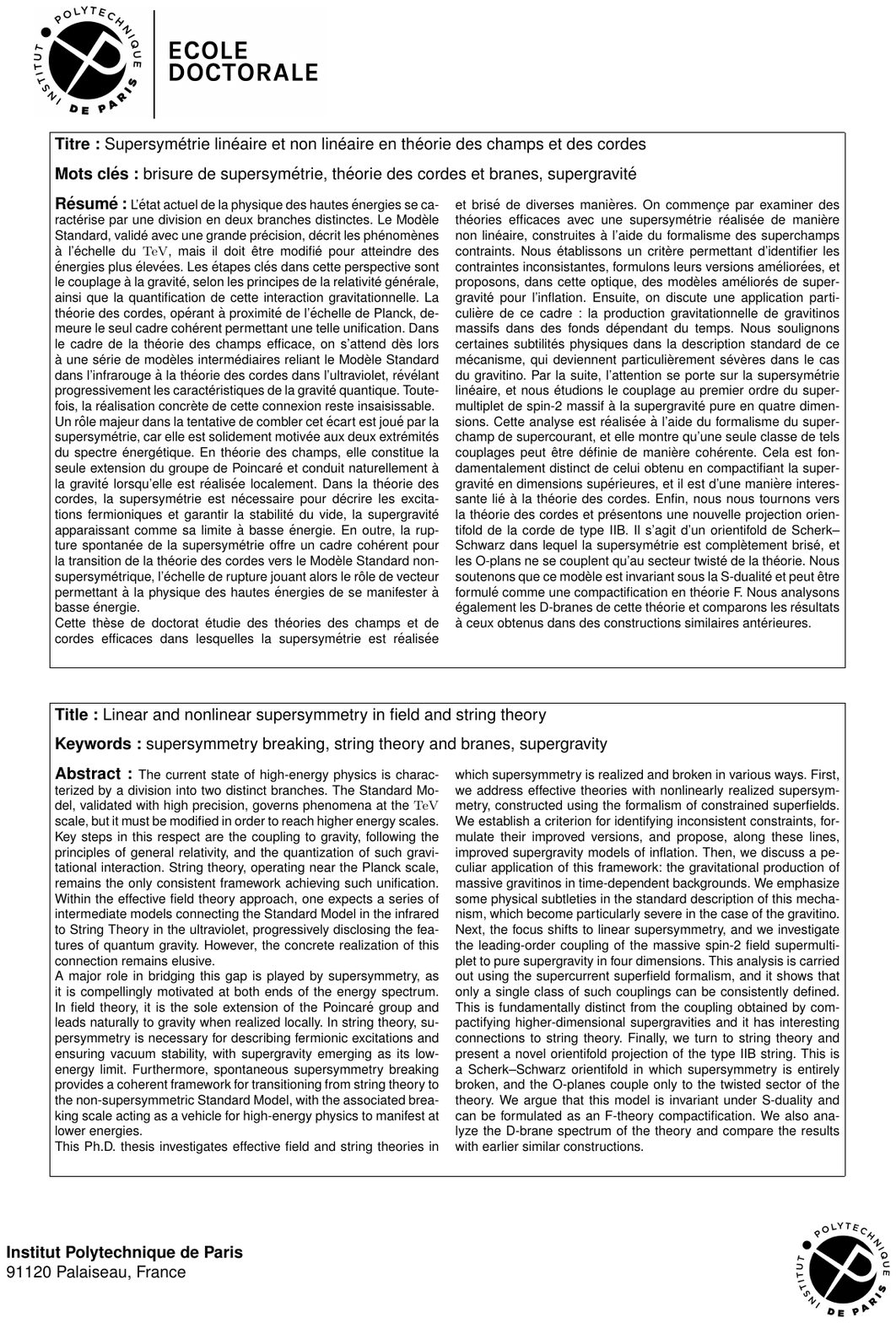}

\end{document}